\documentclass[10pt]{article}

\makeatletter
\renewcommand\paragraph{\@startsection{paragraph}{4}{\z@}%
            {-2.5ex\@plus -1ex \@minus -.25ex}%
            {1.25ex \@plus .25ex}%
            {\normalfont\normalsize\bfseries}}
\makeatother
\usepackage{amsmath}
\usepackage{amssymb}
\usepackage{graphics}
\usepackage{tensor}
\usepackage{amsbsy} 
\usepackage{mathrsfs} 

\usepackage[all]{xy}
\usepackage{compozition}
\usepackage{bm}
\usepackage{accents} 

\usepackage{makeidx}

\makeindex

\newcommand{\negs }{\hspace{-1pt}}
\newcommand{\smallspace}{\hspace{1pt}}

\usepackage{mathtools} 

\newcommand\presub[1]{ \prescript{}{#1} }
\newcommand\presup[1]{ \prescript{#1}{} \negs }

\DeclareMathAlphabet{\mathpzc}{OT1}{pzc}{m}{it}  

\newcommand{\mathbnd}{\mathpzc}   
\newcommand{\mathbpro}{\mathbb}  
\newcommand\mathlbnd[1]{{\bar{#1}}} 
\newcommand{\mathlbpro}{\mathpzc}   
\newcommand{\mathqbnd}{\mathpzc}  

\newcommand\ubar[1]{%
  \underaccent{\bar}{#1}}

\DeclareMathOperator*{\SumInt}{%
\mathchoice%
  {\ooalign{$\displaystyle\sum$\cr\hidewidth$\displaystyle\int$\hidewidth\cr}}
  {\ooalign{\raisebox{.14\height}{\scalebox{.7}{$\textstyle\sum$}}\cr\hidewidth$\textstyle\int$\hidewidth\cr}}
  {\ooalign{\raisebox{.2\height}{\scalebox{.6}{$\scriptstyle\sum$}}\cr$\scriptstyle\int$\cr}}
  {\ooalign{\raisebox{.2\height}{\scalebox{.6}{$\scriptstyle\sum$}}\cr$\scriptstyle\int$\cr}}
}

\title{\textbf{Operational General Relativity: Possibilistic, Probabilistic, and Quantum}}
\author{Lucien Hardy\\
\textit{Perimeter Institute,}\\
\textit{31 Caroline Street North,}\\
\textit{Waterloo, Ontario N2L 2Y5, Canada}}
\date{}

\begin{document}

\pagestyle{empty}

\begin{titlepage}

\maketitle

\vfill

{\center{\includegraphics[width=\textwidth]{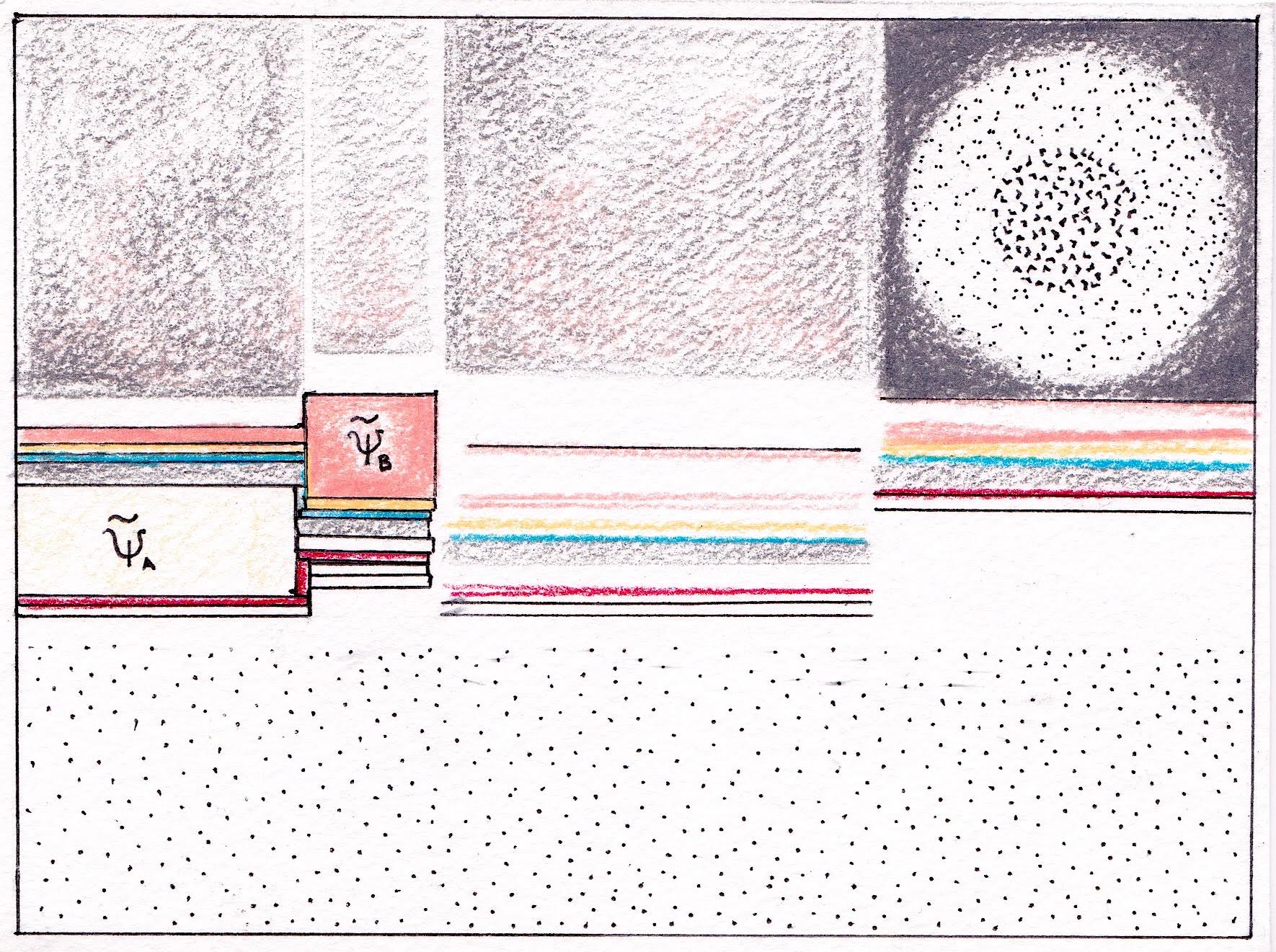}}}

\vfill
{.}
\thispagestyle{empty}
\end{titlepage}

\newpage

\begin{abstract}
In this paper we develop an operational formulation of General Relativity similar in spirit to existing operational formulations of Quantum Theory.   To do this we introduce an operational space (or op-space) built out of scalar fields. A point in op-space corresponds to some nominated set of scalar fields taking some given values in coincidence.  We assert that op-space is the space in which we observe the world.  We introduce also a notion of agency (this corresponds to the ability to set knob settings just like in Operational Quantum Theory).  The effects of agents' actions should only be felt to the future so we introduce also a time direction field.  Agency and time direction can be understood as effective notions.

We consider regions, $\mathtt{A}$, of op-space. We show how to write down solutions, $\Psi_\mathtt{A}$, corresponding to such regions.  Further, we show how to join together such solutions at boundaries in op-space.  We find that there is a curious kind of non-separability in General Relativity in which fully specifying the solutions, $\Psi_\mathtt{A}$ and $\Psi_\mathtt{B}$, in each of two regions, $\mathtt{A}$ and $\mathtt{B}$ respectively, of op-space, can be insufficient to fully specify the solution, $\Psi_\mathtt{A\cup B}$, in the composite region, $\mathtt{A\cup B}$.

We show how to formulate General Relativity as a possibilistic theory and as a probabilistic theory.  In the possibilistic case we provide a compositional framework for calculating whether some operationally described situation is possible or not.  In the probabilistic version we introduce probabilities and provide a compositional framework for calculating the probability of some operationally described situation.

Finally we look at the quantum case.  We review the operator tensor formulation of Quantum Theory and use it to set up an  approach to Quantum Field Theory that is both operational and compositional.  Then we consider three different strategies for solving the problem of Quantum Gravity.  One approach, which attempts to combine the operator tensor formulation of Quantum Field Theory with the probabilistic operational formulation of General Relativity, is developed further than the other two.

By referring only to operational quantities we are able to provide formulations for the possibilistic, probabilistic, and (the nascent) quantum cases that are manifestly invariant under diffeomorphisms.  These formulations, further, satisfy the principle of general compositionality - that the laws of physics should be written in such a way that they apply to any compositional description of any object and in terms of calculations having the same compositional structure as this description.
\end{abstract}

\vfill

\noindent Cover art by Ka\'ca Bradonji\'c: Impression of Lucien Hardy's \lq\lq Operational road to Quantum Gravity"

\newpage

\pagestyle{plain}

\pagenumbering{roman} 
\setcounter{page}{1}

{~}





\section*{Contents}     

\contentsline {part}{I\hspace {1em}Introduction and Outline}{1}
\contentsline {section}{\numberline {1}Introduction}{1}
\contentsline {section}{\numberline {2}This research program and other work}{5}
\contentsline {section}{\numberline {3}Basics}{8}
\contentsline {section}{\numberline {4}Possibilistic Operational General Relativity}{12}
\contentsline {section}{\numberline {5}Probabilistic General Relativity}{15}
\contentsline {section}{\numberline {6}Operator Tensor Quantum Theory}{18}
\contentsline {section}{\numberline {7}Operator Tensor Quantum Field Theory}{20}
\contentsline {section}{\numberline {8}Quantum Gravity}{22}
\contentsline {subsection}{\numberline {8.1}Abstract approach}{22}
\contentsline {subsection}{\numberline {8.2}Ontological approach}{23}
\contentsline {subsection}{\numberline {8.3}Principled approach}{24}
\contentsline {part}{II\hspace {1em}Ontological and Operational Elements}{25}
\contentsline {section}{\numberline {9}Introductory material on General Relativity}{25}
\contentsline {subsection}{\numberline {9.1}Manifolds, chartable spaces, and diffeomorphisms}{25}
\contentsline {subsubsection}{\numberline {9.1.1}Manifolds}{25}
\contentsline {subsubsection}{\numberline {9.1.2}Chartable spaces}{26}
\contentsline {subsubsection}{\numberline {9.1.3}Diffeomorphisms}{27}
\contentsline {subsubsection}{\numberline {9.1.4}Multiple chartable spaces}{28}
\contentsline {subsection}{\numberline {9.2}Candidate solutions, field equations, and beables}{29}
\contentsline {subsection}{\numberline {9.3}Fields are everything}{31}
\contentsline {subsection}{\numberline {9.4}The Westman Sonego approach}{32}
\contentsline {section}{\numberline {10}Operationalism and ontology}{33}
\contentsline {subsection}{\numberline {10.1}Experiments and their purview}{33}
\contentsline {subsection}{\numberline {10.2}Observables}{35}
\contentsline {subsection}{\numberline {10.3}Operational space}{36}
\contentsline {subsection}{\numberline {10.4}Chartable space for an op-space region}{39}
\contentsline {subsection}{\numberline {10.5}Solutions for an op-space region}{39}
\contentsline {subsubsection}{\numberline {10.5.1}Solutions, $\tilde{\Psi }_\mathtt {A}$}{39}  
\contentsline {subsubsection}{\numberline {10.5.2}Solutions, $\Psi _\mathtt {A}$}{40}
\contentsline {subsubsection}{\numberline {10.5.3}Beables for a region of op-space}{40}
\contentsline {subsubsection}{\numberline {10.5.4}Observables for a region of op-space}{41}
\contentsline {subsubsection}{\numberline {10.5.5}The restriction operation}{41}
\contentsline {subsubsection}{\numberline {10.5.6}Turning a solution inside out}{42}
\contentsline {subsection}{\numberline {10.6}Degeneracy in Op-Space}{43}
\contentsline {subsection}{\numberline {10.7}Blobs and hidden variables}{44}
\contentsline {subsection}{\numberline {10.8}Spaces of solutions}{46}
\contentsline {subsubsection}{\numberline {10.8.1}Types of solution}{46}
\contentsline {subsubsection}{\numberline {10.8.2}Pure solutions}{46}
\contentsline {subsubsection}{\numberline {10.8.3}Mixed solutions}{47}
\contentsline {subsubsection}{\numberline {10.8.4}Special solutions}{48}
\contentsline {paragraph}{\numberline {10.8.4.1}The empty solution:}{48}
\contentsline {paragraph}{\numberline {10.8.4.2}The null solution}{48}
\contentsline {paragraph}{\numberline {10.8.4.3}Deterministic solution}{49}
\contentsline {paragraph}{\numberline {10.8.4.4}Given $\Gamma _\mathtt {A}$ solution}{49}
\contentsline {paragraph}{\numberline {10.8.4.5}Operational solutions}{50}
\contentsline {section}{\numberline {11}Composition and non-separability}{51}
\contentsline {subsection}{\numberline {11.1}Typing surfaces}{51}
\contentsline {subsection}{\numberline {11.2}Boundary conditions}{52}
\contentsline {subsubsection}{\numberline {11.2.1}Manifold patch at typing surface}{53}
\contentsline {subsubsection}{\numberline {11.2.2}Manifold boundary conditions}{53}
\contentsline {paragraph}{\numberline {11.2.2.1}The problem}{53}
\contentsline {paragraph}{\numberline {11.2.2.2}Identification condition}{53}
\contentsline {paragraph}{\numberline {11.2.2.3}Direction condition}{53}
\contentsline {paragraph}{\numberline {11.2.2.4}Smoothness condition}{55}
\contentsline {paragraph}{\numberline {11.2.2.5}Resulting manifold}{55}
\contentsline {subsubsection}{\numberline {11.2.3}Field boundary conditions}{56}
\contentsline {subsubsection}{\numberline {11.2.4}The full boundary conditions}{56}
\contentsline {subsubsection}{\numberline {11.2.5}Composite typing surfaces}{57}
\contentsline {subsubsection}{\numberline {11.2.6}Joining solutions}{57}
\contentsline {subsubsection}{\numberline {11.2.7}The local matching assumption}{58}
\contentsline {subsubsection}{\numberline {11.2.8}The null join}{59}
\contentsline {subsubsection}{\numberline {11.2.9}Special cases}{60}
\contentsline {subsubsection}{\numberline {11.2.10}Sets of boundary conditions}{61}
\contentsline {subsubsection}{\numberline {11.2.11}Special boundary conditions}{62}
\contentsline {subsection}{\numberline {11.3}Composite boundary condition space}{63}
\contentsline {subsection}{\numberline {11.4}Given $\mathpzc {a}$ solutions}{63}
\contentsline {subsection}{\numberline {11.5}Curious nonseparability}{63}
\contentsline {section}{\numberline {12}Agency}{66}
\contentsline {subsection}{\numberline {12.1}Introducing Agency}{66}
\contentsline {subsection}{\numberline {12.2}An example: a vast fleet of spaceships}{68}
\contentsline {subsection}{\numberline {12.3}Agency fields}{69}
\contentsline {subsection}{\numberline {12.4}Agent choices}{70}
\contentsline {subsection}{\numberline {12.5}Time direction field}{70}
\contentsline {subsection}{\numberline {12.6}Representing solutions with agency}{72}
\contentsline {subsection}{\numberline {12.7}Turning a solution with agency inside out}{73}
\contentsline {subsection}{\numberline {12.8}Solving field equations given agent strategies}{73}
\contentsline {subsection}{\numberline {12.9}Boundary conditions}{74}
\contentsline {subsection}{\numberline {12.10}Causality}{74}
\contentsline {subsubsection}{\numberline {12.10.1}No influence regions}{74}
\contentsline {subsubsection}{\numberline {12.10.2}Causality condition}{75}
\contentsline {subsubsection}{\numberline {12.10.3}Deterministic input oriented boundary conditions}{76}
\contentsline {subsubsection}{\numberline {12.10.4}Causal completeness}{76}
\contentsline {subsubsection}{\numberline {12.10.5}Further discussion}{77}
\contentsline {subsection}{\numberline {12.11}Special solutions}{77}
\contentsline {subsection}{\numberline {12.12}Agent jitter}{78}
\contentsline {subsection}{\numberline {12.13}Joining solutions when we have agency}{79}
\contentsline {part}{III\hspace {1em}Possibilistic formulation: PoAGeR}{80}
\contentsline {section}{\numberline {13}Introduction}{80}
\contentsline {section}{\numberline {14}Operations}{80}
\contentsline {subsection}{\numberline {14.1}The idea of an operation}{80}
\contentsline {subsection}{\numberline {14.2}Symbolic and diagrammatic notation}{81}
\contentsline {subsection}{\numberline {14.3}Joining operations}{81}
\contentsline {section}{\numberline {15}Encapsulated propositions}{83}
\contentsline {subsection}{\numberline {15.1}Propositions in General Relativity}{83}
\contentsline {subsection}{\numberline {15.2}Joining of propositions}{83}
\contentsline {subsection}{\numberline {15.3}Encapsulated propositions}{84}
\contentsline {subsection}{\numberline {15.4}Reversing and detyping}{85}
\contentsline {subsection}{\numberline {15.5}Joining encapsulated propositions}{86}
\contentsline {section}{\numberline {16}Boundary propositions}{87}
\contentsline {subsection}{\numberline {16.1}Introducing boundary propositions}{87}
\contentsline {subsection}{\numberline {16.2}Composition of boundary propositions}{88}
\contentsline {section}{\numberline {17}Duotensors, fiducials, and generalized states}{90}
\contentsline {subsection}{\numberline {17.1}Possibilistic equivalence}{90}
\contentsline {subsection}{\numberline {17.2}Boundary fiducials}{91}
\contentsline {subsection}{\numberline {17.3}The hopping metric}{93}
\contentsline {subsection}{\numberline {17.4}Simple case}{95}
\contentsline {subsection}{\numberline {17.5}General case}{96}
\contentsline {section}{\numberline {18}Operational possibilistic formulation}{98}
\contentsline {subsection}{\numberline {18.1}Operational fiducial boundary conditions}{98}
\contentsline {subsection}{\numberline {18.2}Operational generalized possibilistic states}{100}
\contentsline {subsection}{\numberline {18.3}Operational hopping metric}{101}
\contentsline {subsection}{\numberline {18.4}Choices of $\Upsilon $ sets}{101}
\contentsline {subsection}{\numberline {18.5}Manifestly diffeomorphism invariant calculations}{103}
\contentsline {subsection}{\numberline {18.6}Operational manifestly invariant formulation}{103}
\contentsline {subsection}{\numberline {18.7}Comments on PoAGeR}{105}
\contentsline {subsection}{\numberline {18.8}Black and white dots: external and internal points of view}{106}
\contentsline {subsection}{\numberline {18.9}Formalism locality}{106}
\contentsline {subsection}{\numberline {18.10}Causality in possibilistic formulation}{108}
\contentsline {part}{IV\hspace {1em}Probabilistic formulation: PAGeR}{109}
\contentsline {section}{\numberline {19}Introduction}{109}
\contentsline {section}{\numberline {20}Objects in formalism}{109}
\contentsline {subsection}{\numberline {20.1}Loaded Operations}{109}
\contentsline {subsection}{\numberline {20.2}Probabilities for closed loaded operations}{111}
\contentsline {subsection}{\numberline {20.3}loaded encapsulated propositions}{112}
\contentsline {subsection}{\numberline {20.4}Probabilities for closed loaded encapsulated propositions}{112}
\contentsline {subsection}{\numberline {20.5}Map from loaded operations to loaded encapsulated propositions}{113}
\contentsline {subsection}{\numberline {20.6}Loaded boundary propositions}{114}
\contentsline {subsection}{\numberline {20.7}Probabilities for closed loaded boundary propositions}{114}
\contentsline {section}{\numberline {21}The simple case}{114}
\contentsline {subsection}{\numberline {21.1}Composite region: simple case}{114}
\contentsline {subsection}{\numberline {21.2}Introducing fiducials}{115}
\contentsline {subsection}{\numberline {21.3}Equivalence, equality, and the $p(\cdot )$ function}{116}
\contentsline {subsection}{\numberline {21.4}The simple case again}{117}
\contentsline {subsection}{\numberline {21.5}The hopping metric}{118}
\contentsline {subsection}{\numberline {21.6}Possible choices of fiducials}{120}
\contentsline {section}{\numberline {22}The General Case}{120}
\contentsline {subsection}{\numberline {22.1}Decomposition locality}{120}
\contentsline {subsection}{\numberline {22.2}Loading for loaded operations}{121}
\contentsline {subsection}{\numberline {22.3}Calculation for general case}{121}
\contentsline {section}{\numberline {23}Operational probabilistic formulation}{123}
\contentsline {subsection}{\numberline {23.1}Operational loaded boundary propositions}{123}
\contentsline {subsection}{\numberline {23.2}Fiducial loaded operations}{124}
\contentsline {subsection}{\numberline {23.3}Operational hopping metric}{125}
\contentsline {subsection}{\numberline {23.4}Decomposition locality for loaded operations}{125}
\contentsline {subsection}{\numberline {23.5}Manifest invariance of PAGeR}{127}
\contentsline {subsection}{\numberline {23.6}Black and white dots}{127}
\contentsline {subsection}{\numberline {23.7}Formalism locality in PAGeR}{127}
\contentsline {section}{\numberline {24}Causality in PAGeR}{129}
\contentsline {part}{V\hspace {1em}Operator Tensor Quantum Theory}{131}
\contentsline {section}{\numberline {25}Introduction}{131}
\contentsline {section}{\numberline {26}Finite dimensional Hilbert spaces}{131}
\contentsline {subsection}{\numberline {26.1}Operations}{131}
\contentsline {subsection}{\numberline {26.2}Decomposition locality}{132}
\contentsline {subsection}{\numberline {26.3}Hopping metric}{133}
\contentsline {subsection}{\numberline {26.4}Operator tensors}{133}
\contentsline {subsection}{\numberline {26.5}Operator decomposition locality}{135}
\contentsline {subsection}{\numberline {26.6}Operator Circuits}{135}
\contentsline {subsection}{\numberline {26.7}Physicality}{136}
\contentsline {subsection}{\numberline {26.8}Causality}{137}
\contentsline {section}{\numberline {27}Continuous dimensional Hilbert spaces}{138}
\contentsline {section}{\numberline {28}Operator Tensor Quantum Field Theory}{141}
\contentsline {subsection}{\numberline {28.1}Introduction}{141}
\contentsline {subsection}{\numberline {28.2}The discrete case}{142}
\contentsline {subsection}{\numberline {28.3}Physicality conditions}{143}
\contentsline {subsection}{\numberline {28.4}Change in notation convention}{144}
\contentsline {subsection}{\numberline {28.5}The continuous limit}{146}
\contentsline {subsection}{\numberline {28.6}Calculations}{148}
\contentsline {subsection}{\numberline {28.7}Discussion}{150}
\contentsline {part}{VI\hspace {1em}Quantum Gravity: QuAGeR}{152}
\contentsline {section}{\numberline {29}The problem of Quantum Gravity}{152}
\contentsline {section}{\numberline {30}An abstract approach - Quantization}{153}
\contentsline {subsection}{\numberline {30.1}Quantization and GRization}{153}
\contentsline {subsection}{\numberline {30.2}Square root and square approach to quantization}{154}
\contentsline {subsection}{\numberline {30.3}Classical level of description}{155}
\contentsline {subsection}{\numberline {30.4}Loaded operations}{155}
\contentsline {subsection}{\numberline {30.5}Choosing the Hilbert space: taking the ``square root"}{156}
\contentsline {subsection}{\numberline {30.6}Choosing the fiducials: taking the ``square"}{157}
\contentsline {subsection}{\numberline {30.7}Operational Quantum Gravity}{158}
\contentsline {subsection}{\numberline {30.8}A calculation in this framework}{158}
\contentsline {subsection}{\numberline {30.9}What are the physicality conditions?}{160}
\contentsline {section}{\numberline {31}Ontological approach to Quantum Gravity}{162}
\contentsline {subsection}{\numberline {31.1}Ontology}{162}
\contentsline {subsection}{\numberline {31.2}Maybe PAGeR is Quantum Gravity}{162}
\contentsline {subsection}{\numberline {31.3}More general ontological models}{163}
\contentsline {subsection}{\numberline {31.4}Discussion}{164}
\contentsline {section}{\numberline {32}Principles, axioms and postulates}{164}
\contentsline {subsection}{\numberline {32.1}Introduction}{164}
\contentsline {subsection}{\numberline {32.2}Postulates for Quantum Theory}{164}
\contentsline {subsection}{\numberline {32.3}Postulates for General Relativity?}{166}
\contentsline {subsection}{\numberline {32.4}Postulates for Quantum Gravity?}{168}
\contentsline {part}{VII\hspace {1em}Discussions}{170}
\contentsline {section}{\numberline {33}Other Tools}{170}
\contentsline {subsection}{\numberline {33.1}Free operations}{170}
\contentsline {subsection}{\numberline {33.2}Infinitesimal approach}{172}
\contentsline {section}{\numberline {34}Conclusions}{174}
\contentsline {part}{VIII\hspace {1em}Appendices}{177}
\contentsline {section}{\numberline {A}Standard formulation of General Relativity}{177}
\contentsline {subsection}{\numberline {A.1}The principle of equivalence and the principle of general covariance}{178}
\contentsline {subsection}{\numberline {A.2}Manifolds}{178}
\contentsline {subsection}{\numberline {A.3}Tensors}{179}
\contentsline {subsection}{\numberline {A.4}Tensor fields}{180}
\contentsline {subsection}{\numberline {A.5}The metric}{182}
\contentsline {subsection}{\numberline {A.6}The covariant derivative}{183}
\contentsline {subsection}{\numberline {A.7}Locally flat and local inertial reference frames}{185}
\contentsline {subsection}{\numberline {A.8}Derivatives of the metric at a point}{186}
\contentsline {subsubsection}{\numberline {A.8.1}Zeroth derivative and the Minkowski metric}{187}
\contentsline {subsubsection}{\numberline {A.8.2}First derivatives and local flatness}{188}
\contentsline {subsubsection}{\numberline {A.8.3}Second derivatives and the curvature tensor}{189}
\contentsline {subsubsection}{\numberline {A.8.4}Third derivatives and the Bianchi identity}{191}
\contentsline {subsection}{\numberline {A.9}Matter}{191}
\contentsline {subsubsection}{\numberline {A.9.1}Matter in Special Relativity}{192}
\contentsline {subsubsection}{\numberline {A.9.2}Minimal substitution technique}{194}
\contentsline {subsubsection}{\numberline {A.9.3}Matter field equations in General Relativity}{195}
\contentsline {subsubsection}{\numberline {A.9.4}Small fluid blobs move along geodesics}{196}
\contentsline {subsection}{\numberline {A.10}The Einstein field equations}{197}
\contentsline {subsection}{\numberline {A.11}General Relativity as a system of coupled field equations}{198}
\contentsline {subsection}{\numberline {A.12}The hole argument and diffeomorphism invariance}{200}
\contentsline {subsubsection}{\numberline {A.12.1}The argument}{201}
\contentsline {subsubsection}{\numberline {A.12.2}An example}{202}
\contentsline {paragraph}{\numberline {A.12.2.1}Smooth transformations.}{202}
\contentsline {paragraph}{\numberline {A.12.2.2}Vacuum solutions.}{202}
\contentsline {paragraph}{\numberline {A.12.2.3}Add a little matter.}{203}
\contentsline {paragraph}{\numberline {A.12.2.4}Intersection graph.}{203}
\contentsline {paragraph}{\numberline {A.12.2.5}A solution in the $x$ coordinate system.}{203}
\contentsline {paragraph}{\numberline {A.12.2.6}A solution in the $x'$ coordinate system.}{203}
\contentsline {paragraph}{\numberline {A.12.2.7}Another solution in the $x$ coordinate system.}{204}
\contentsline {subsubsection}{\numberline {A.12.3}The resolution}{204}
\contentsline {subsubsection}{\numberline {A.12.4}Diffeomorphisms}{205}
\contentsline {paragraph}{\numberline {A.12.4.1}Push forward and pull back.}{205}
\contentsline {paragraph}{\numberline {A.12.4.2}Push forward with a diffeomorphism.}{206}
\contentsline {paragraph}{\numberline {A.12.4.3}The hole argument using diffeomorphisms.}{207}
\contentsline {subsubsection}{\numberline {A.12.5}Role of the Bianchi identities}{208}
\contentsline {subsection}{\numberline {A.13}Causality}{208}
\contentsline {subsection}{\numberline {A.14}General Relativity is about fields}{210}
\contentsline {section}{\numberline {B}The substitution operator}{211}

\newpage

\pagenumbering{arabic} 
\setcounter{page}{1}

\part{Introduction and Outline}

\section{Introduction}\label{sec:introductionatbeginning}

The theoretical, experimental, and metaphysical foundations of physics are respectively rooted in mathematical formalism, operational procedures, and ontological description.  In the physics of Newton and Maxwell, these three great \lq\lq continental plates" were unified.  The objects in the mathematical formalism describe exactly the underlying ontological entities and delimit what we can observe at the operational level.  However, the tectonic forces unleashed by General Relativity and Quantum Theory broke these continental plates apart.  This is the conceptual landscape from which a solution to the problem of Quantum Gravity must be formed.

A common perception is that General Relativity is pretty much the same as the classical theories that precede it and that we must look to Quantum Theory for conceptual novelty.  But this is not so.  In some respects, Quantum Theory is conceptually pedestrian compared with its cousin General Relativity.  At the very least we need to treat the two theories on an equal footing when developing a theory of Quantum Gravity.

One tool that can be used to shed light on physical theories is reformulation.  There exist many formulations of both Quantum Theory and of General Relativity.  These different formulations often serve as a starting point for this or that attack on the problem of Quantum Gravity.  In particular, if we can formulate the two theories in a similar spirit then we have a possible route to their unification.  For example, we can formulate both Quantum Theory and General Relativity with a Hamiltonian and this provides the basis for the canonical approach to Quantum Gravity (the modern version of this is loop Quantum Gravity).  Similarly, the string theory approach to Quantum Gravity starts from a Lagrangian formulation of these two theories.

There is a formulation of Quantum Theory that is very successful but has not generally been embraced by that part of the theoretical physics community interested in more fundamental questions (such as the problem of Quantum Gravity).  This is \emph{Operational Quantum Theory}.   This unsung hero is, nevertheless, the workhorse that powers much of the progress in the new subjects of Quantum Information and Quantum Computing.   The version of Operational Quantum Theory you will find in textbooks uses \emph{density matrices}, \emph{completely positive maps}, and \emph{positive operator valued measures} (POVMs) to calculate probabilities for certain measurement outcomes.  Operational Quantum Theory was developed in parallel with other developments in Quantum Theory.  The idea of density matrices goes back to von Neuman in 1927 \cite{vonNeumann1927wahrscheinlichkeitstheoretischer}, completely positive maps were first discovered by Sudarshan in 1961 \cite{Sudarshan1961stochastic}. Such maps can be simply understood in terms of Kraus operators introduced by Kraus in 1971 \cite{kraus1971general}.  In 1967 Jauch and Piron introduced POVMs \cite{jauch1967generalized} (see also the paper by Davies and Lewis \cite{davies1970operational}).  An early book on Operational Quantum Theory is Davies's 1976 publication \cite{davies1976quantum}.  The grander project of operational quantum theory seems to go back to von Neuman (see his 1932 book translated in to English in \cite{john1955mathematical}).  Other architects of this framework include Segal \cite{segal1947postulates}, Mackey \cite{mackey1963mathematical}, Haag and Kastler \cite{haag1964algebraic}, and Ludwig \cite{ludwig1954grundlagen}.  A more comprehensive discussion of the history can be found in the article by Busch and Grabowski \cite{busch1989some}.  A modern treatment of Operational Quantum Theory, as used in Quantum Information, can be found in Chapter 8 of the book by Nielsen and Chuang \cite{nielsen2000quantum}.

In recent years alternative operational formulations of Quantum Theory have been developed that are more suitable, in some respects, for a relativistic setting and put all the aforementioned objects on an equal footing.  The causaloid approach \cite{hardy2005probability, hardy2007towards, hardy2009quantum3} is one such framework which provides a formulation of Quantum Theory in the context of a general probability theory framework suitable for theories with indefinite causal structure. Also there is the quantum combs approach \cite{chiribella2009theoretical} of Chiribella, D'Ariano, and Perinotti, and the operator tensor approach \cite{hardy2011reformulating, hardy2012operator} of the present author (these approaches associate operators with arbitrary parts of a circuit).  In Part \ref{part:operatortensorQT} I will we present the operator tensor formulation of Quantum Theory (first presented in \cite{hardy2011reformulating}).  In this approach, we can calculate the probability for a circuit by means of an operator expression having the same compositional form. We can represent this diagrammatically. For example
\begin{equation}\label{introcircuitexample}
\text{prob}\left(
\begin{Compose}{0}{-2} \setdefaultfont{\mathsf}\setsecondfont{\mathsf}\setthirdfont{\mathsf}
\Ucircle{A}{0,0} \Ucircle{B}{-5,9} \Ucircle{C}{3,4} \Ucircle{D}{-3, 15} \Ucircle{E}{2,13}
\joincc[below left]{A}{115}{B}{-65} \csymbolalt{a}
\joincc[below]{A}{80}{C}{-90} \csymbolalt{a}
\joincc[below]{C}{170}{B}{-10} \csymbolalt{c}
\joincc[above left]{B}{25}{E}{-110} \csymbolalt{d}
\joincc[left]{B}{80}{D}{-100} \csymbolalt{a}
\joincc[above right]{E}{170}{D}{-15}\csymbolalt{c}
\joincc[right]{C}{100}{E}{-80} \csymbolalt{g}
\end{Compose}
\right)
~~=~~
\begin{Compose}{0}{-2} \setdefaultfont{\hat}\setsecondfont{\mathsf}\setthirdfont{\mathsf}
\Ucircle{A}{0,0} \Ucircle{B}{-5,9} \Ucircle{C}{3,4} \Ucircle{D}{-3, 15} \Ucircle{E}{2,13}
\joincc[below left]{A}{115}{B}{-65} \csymbolalt{a}
\joincc[below]{A}{80}{C}{-90} \csymbolalt{a}
\joincc[below]{C}{170}{B}{-10} \csymbolalt{c}
\joincc[above left]{B}{25}{E}{-110} \csymbolalt{d}
\joincc[left]{B}{80}{D}{-100} \csymbolalt{a}
\joincc[above right]{E}{170}{D}{-15}\csymbolalt{c}
\joincc[right]{C}{100}{E}{-80} \csymbolalt{g}
\end{Compose}
\end{equation}
I will also present a preliminary version of operator tensor Quantum Field Theory (this was outlined in \cite{hardy2011reformulating} but not developed).  A similar formulation has been developed by Oeckl \cite{oeckl2013positive} (though there are some important differences).

Given the success of Operational Quantum Theory it seems pertinent to attempt to reformulate General Relativity along similar operational lines. If we can do this then we have the basis for an attack on the problem of Quantum Gravity.  This is the purpose of the present paper.

A desideratum is that such theories satisfy the \emph{principle of general compositionality} \index{general compositionality} (based on the \emph{composition principle} outlined in \cite{hardy2013theory}).
\begin{quote} {\bf The principle of general compositionality}: The laws of physics should be written in such a way that they apply to any compositional description of any object and in terms of calculations having the same compositional structure as this description.
\end{quote}
This principle is written in similar language to the \emph{principle of general covariance} (see Appendix \ref{Sec:principleequivcovar}).  The latter played a very important role in the construction of General Relativity.  In fact, as pointed out by Kretchmann, any physical theory can be formulated in accord with this principle \cite{kretschmann1917uber}.  However, as Einstein responded \cite{Einstein1918prinzipielles} once we attempt to formulate theories in this way, our notions of simplicity and elegance are modified.  General Relativity looks like a very natural theory in the resulting calculus of tensor fields.  Newtonian mechanics, on the other hand, appears unnatural.  We might hope for something similar for the principle of general compositionality.

We will use operational and compositional techniques, and we will be interested in possibilistic and probabilistic theories.

An example of a probabilistic calculation is given in \eqref{introcircuitexample}.  In a possibilistic theory we want to determine whether some given situation is possible or not.  An interesting example \cite{hardy2013theory} to illustrate this idea is Penrose's impossible triangle \index{impossible triangle}
\[
\begin{Compose}{0}{0}
\draw[thick] (0,0) coordinate(A) -- (10,0) coordinate (B) -- (5, 8.66) coordinate (C) -- cycle;
\draw[thick] (A) -- ++ (-1, -1.732) coordinate (AA); \draw[thick] (B) -- ++ (2, 0) coordinate (BB); \draw[thick] (C) -- ++(-1, 1.732) coordinate (CC);
\draw[thick] (AA) -- ++(16, 0) coordinate (BBB) -- ++ (-120:2) coordinate (BBBB);
\draw[thick] (BB) -- ++ (120: 16) coordinate (CCC) -- ++(2,0) coordinate (CCCC) -- (BBB) ;
\draw[thick] (CC) -- ++(-120: 16) coordinate (AAA)-- ++(120:2) coordinate(AAAA) -- (CCC);
\draw[thick] (BBBB) -- (AAA);
\draw[thin] (BB) -- (intersection cs: first line={(CCC)--(BB)}, second line={(AA)--(BBB)} ) coordinate (B5) -- ++($(BBBB) - (BBB)$);
\draw[thin] (AA) -- (intersection cs: first line={(BBB)--(AA)}, second line={(CC)--(AAA)} ) coordinate (A5) -- ++($(AAAA) - (AAA)$);
\draw[thin] (CC) -- (intersection cs: first line={(AAA)--(CC)}, second line={(BB)--(CCC)} ) coordinate (C5) -- ++($(CCCC) - (CCC)$);
\end{Compose}
\]
Each join is legal but the whole object is impossible.  We can represent this triangle abstractly as
\begin{equation}\label{penrosetriangleopdescr}
\begin{Compose}{0}{0}
\ucircle{A}{0,0}\csymbol{B} \ucircle{B}{-4,5} \csymbol{B} \ucircle{C}{4,5} \csymbol{B}
\joincc[below left]{A}{150}{B}{210} \csymbol{a}  \joincc[above]{B}{30}{C}{210} \csymbol{a} \joincc[below right]{C}{30}{A}{-30} \csymbol{a}
\end{Compose}
\end{equation}
A general theory of Penrose objects (of beams joined at their ends) would allow us to calculate whether such objects are possible or not.  If the principle of general compositionality is satisfied, these calculations will have the same compositional structure as the operational description (as in \eqref{penrosetriangleopdescr}).

In Part \ref{part:ontologicalandoperational} we show how to think of General Relativity in operational terms.  First we build an \emph{operational space} as corresponding to the space in which we observe the world.  This is built out of scalar fields.  The choice of scalar fields for the operational space is motivated by work of Westman and Sonego \cite{westman2008events, westman2009coordinates} (described in Sec.\ \ref{sec:theWestmanSonegoapproach}). We show how to write down solutions pertaining to parts of operational space and how to combine such solutions.  We also introduce agency (analogous to knob settings in Operational Quantum Theory).

The approach taken in this paper is actually suitable for the more general situation in which we have field equations that are invariant under diffeomorphisms and whose solutions are take the form of specifying tensor fields at each point on a manifold.  General Relativity is one such theory and the one we are most interested in, but everything we do is actually also suited to other such physical theories.

In Part \ref{part:possibilisticformulation} we show how to formulate General Relativity as a \emph{possibilistic} operational theory that is composite in nature. The idea is that, given operational descriptions of various regions of operational space, we have a way of determining whether this constitutes a possible state of affairs (i.e.\ does there exist at least one solution having these operational descriptions).  We provide both symbolic and diagrammatic ways to do these calculations.  We call this formulation PoAGeR (for Possibilistic General Relativity with Agency).  \index{PoAGeR}

In Part \ref{part:probabilisticformulation} we show how to include probabilities in the picture.  These are just ignorance probabilities of the kind that we can have in any theory (deterministic or not).  Again we develop symbolic and diagrammatic ways to do the calculations.  We call this formulation PAGeR (for Probabilistic General Relativity with Agency). 
\index{PAGeR} 

In Part \ref{part:operatortensorQT} we review the operator tensor approach to Quantum Theory.  We show how it can be adapted for infinite dimensional Hilbert spaces and Quantum Field Theory.  The calculations for these quantum situations can also be represented symbolically and diagrammatically and they have a similar structure to the possibilistic and probabilistic formulations of General Relativity.

In Part \ref{part:QuantumGravity} we look at the problem of Quantum Gravity.   This is to find a theory that reduces, in appropriate limits to Quantum Theory on the one hand and to General Relativity on the other hand.  We suggest three approaches. First we consider an abstract approach in which we attempt to \lq\lq quantize" the probabilistic formulation of General Relativity. We dub the approach taken to quantization as the \lq\lq square root and square" approach.  This is an attempt to take the classical probability simplices of probabilistic General Relativity and turn them into quantum like spaces.  We develop this approach quite a long way (though not into a fully fledged theory of Quantum Gravity).  The second approach is the ontological approach.  The idea is that General Relativity, once understood as an operational theory, may actually reproduce Quantum Theory in appropriative circumstances. Or, failing that, an ontological model that extends General Relativity, may also give rise to Quantum Theory.  The third approach is to apply principles to frameworks that are rich enough to contain Quantum Theory and General Relativity as special cases.  We call such possible formulations of Quantum Gravity, QuAGeR (for Quantum General Relativity with Agency).  \index{QuAGeR}

This paper will, I hope, be appreciated by people from Quantum Foundations, Quantum Information, and Computer Science backgrounds.  Since General Relativity is not generally an active area of research for people with this background I have included an extensive review of the standard formalism of General Relativity in Appendix \ref{appendix:standarformulationofGR}.  This covers all the basics of the subject: manifolds; tensor fields; the metric; the covariant derivative; the curvature tensor; the Bianchi identities; matter field equations in Special Relativity and how to convert them to matter field equations for General Relativity; the Einstein field equations; the theory of diffeomorphisms.  There is also an extensive discussion of the hole argument.

In this Part we will provide a brief sketch of the main ideas in the paper glossing over many details for the time being.

\section{This research program and other work}\label{sec:thisresearchprogramandotherwork}

Many people have appreciated that we need to think in operational terms when confronting the problem of quantum gravity.  For example, in his famous 1967 paper setting up the canonical approach to Quantum Gravity \cite{dewitt1967quantum} de Witt says
\begin{quote}
Perhaps the most impressive fact which emerges
from a study of the quantum theory of gravity is that
it is an extraordinarily economical theory. It gives one
just exactly what is needed in order to analyze a particular
physical situation, but not a bit more. Thus it
will say nothing about time unless a clock to measure
time is provided, and it will say nothing about geometry
unless a device (either a material object, gravitational
waves, or some other form of radiation) is introduced to
tell when and where the geometry is to be measured. "
In view of the strongly operational foundations of
both the quantum theory and general relativity this
is to be expected. When the two theories are united the
result is an operational theory {\it par excellence}.
\end{quote}
The approach here is also operational though very different from the canonical approach to Quantum Gravity.

In broad terms can imagine two kinds of operational approach to General Relativity (and, ultimately, to Quantum Gravity).  The first operational approach (not the one adopted here) is where we take instruments (rods, clocks, light rays, GPS satelites...) as primitives.  Special Relativity, as invented by Einstein, is in this tradition. In the context of General Relativity there is a long history of such approaches. Ehlers, Pirani, Schild  (originally published in 1972, republished in \cite{ehlers2012republication}) presented an axiomatic operational approach to space-time where light rays and test particles under free fall are taken as primitives. De Felice and Bini \cite{de2010classical} look at how to model measurements in General Relativity. Hartmann defines some basic primitives such as comparison allowing him to operationalize basic observables in relativity and mechanics \cite{hartmann2012operationalization, hartmann2015operationalization} Rovelli \cite{rovelli2007quantum} considers using GPS coordinates to parameterize space-time.  Lloyd has made similar proposals \cite{lloyd2012quantum} and investigated the accuracy limits imposed by quantum theory for such GPS coordinates. The second operational approach is where we build observables directly out of the fields that appear in the fundamental theory. This is the approach we adopt in this paper.

This paper represents another step in the authors ongoing research program with the aim of bringing operational probabilistic techniques to the problem of Quantum Gravity.   There are many parallel research programs bringing these operational techniques and related techniques to bear on various aspects of physics (mostly Quantum Theory).  These approaches have in common that they are \emph{linear in probability}. Rather than working with amplitudes they work with objects, such as density matrices, that are linear in probability.  This distinguishes this approach from much of the rest of the theoretical physics community where amplitudes are used.

In \cite{hardy2001quantum} I showed how to reconstruct quantum theory from operational axioms in the context of a framework for general probability theories.  There has been significant progress on this subject in recent years (some papers are \cite{dakic2009quantum, masanes2010derivation, chiribella2010probabilistic, chiribella2010informational, hardy2011reformulating, wilce2009four, wilce2016royal, goyal2008information, appleby2011properties, barnum2014higher, hoehn2014toolbox, hoehn2015quantum}, for a more comprehensive list see \cite{chiribella2016quantum}).  This recent work has been strongly influenced by the quantum information way of thinking (motivated by Fuchs \cite{fuchs2002quantum}).  Fuchs also suggests an interesting approach to quantum cosmology (in particular, see Fig.\ 6 of \cite{fuchs2010qbism} and surrounding discussion) that has strong resonance with the approach taken in this paper.  An earlier tradition of deriving quantum theory from axioms, often in the tradition of quantum logic, goes back to von Neumann \cite{von1996mathematical}, Mackey \cite{mackey1963mathematical}, and others \cite{birkhoff1936logic, zierler1975axioms, piron1964axiomatique,  fivel1994interference}.  Here we develop the generalized probabilistic  theories framework. Early work on this framework includes \cite{mackey1963mathematical, ludwig1985axiomatic, davies1970operational, gunson1967algebraic, mielnik1969theory, araki1980characterization, gudder1999convex, foulis1979empirical}.

In \cite{hardy2005probability} I presented the causaloid formalism as a possible framework for Quantum Gravity - this is a framework for operational probabilistic theories that does not assume definite causal structure.  Indefinite causal structure has become a subject of much study in recent years buoyed up by the work of Chirbella, D'Ariano, Perinotti, and Valiron \cite{chiribella2009beyond, chiribella2013quantum} and of Oreshkov, Costa, Brukner \cite{oreshkov2012quantum}. See the review by Brukner \cite{brukner2014quantum}.

In \cite{hardy2013formalism} I set up the duotensor framework for treating probabilistic circuits.  Duotensors are basically tensors but with both prescripts and postscripts (raised and lowered) and a \emph{hopping metric} which can be used to hop indices over (take prescripts to postscripts and vice versa).  This framework associates a duotensor with operations.  We make much use of these in the present paper for describing generalized states.  The duotensor approach was adapted to quantum theory in the operator tensor formulation of Quantum Theory \cite{hardy2011reformulating, hardy2012operator}. In this operator tensors (operators equipped with certain tensorial type structure) are associated with operations in a quantum circuit.  In \cite{hardy2013theory} I proposed a compositional approach to physics.  The key idea is the \emph{composition principle} that the calculation for a particular physical situation should have the same compositional form as the operational description it is a calculation for.  Here we take this a step further with the principle of general compositionality (see Sec.\ \ref{sec:introductionatbeginning}).

The causaloid approach and the operator tensor approach to quantum theory puts arbitrary fragments of a circuit (including preparations, transformations, and measurements) on an equal footing.  The operator tensor approach is more specifically tailored to quantum theory than the causaloid approach.  There is a dictionary between the operator tensor approach and the quantum combs approach (due to Chiribella, D'Ariano, and Perinotti \cite{chiribella2009theoretical}.  Gutoski and Watrous developed a similar framework \cite{gutoski2007toward}.   The process matrix approach of Oreshkov, Costa, and Brukner \cite{oreshkov2012quantum}, and other more recent approaches \cite{portmann2015causal, pollock2015complete}, are also similar.

There are a number of quantum foundations groups around the world working on related techniques.  The following is a very incomplete list.

Coecke and Abramsky in Oxford have pioneered a category theoretic approach to Quantum Theory \cite{abramsky2004categorical, abramsky2007physics, coecke2010quantum, abramsky2009categorical}.  Such category theoretic calculations can be written in diagrammatic form (these diagrams motivated the present author to use a diagrammatic approach).

D'Ariano in Pavia initiated his own approach to finding principles for Quantum Theory \cite{d2008probabilistic}.  Chiribella, D'Ariano, and Perinotti set up the quantum combs approach \cite{chiribella2009theoretical}, set up an operational diagrammatic approach to convex probability theories \cite{chiribella2010probabilistic}, and provided a new reconstruction of Quantum Theory from physical axioms \cite{chiribella2010informational}.  In recent years D'Ariano, Perinotti and co-workers have shifted their attention to Quantum Field Theory \cite{d2014derivation, bisio2015quantum}.

Brukner and collaborators in Vienna have made significant contributions.  Dakic and Brukner \cite{dakic2009quantum} provided a reconstruction of Quantum Theory which went a long way to removing the \lq\lq simplicity axiom" of \cite{hardy2001quantum}.  Oreshkov, Costa, and Brukner developed inequalities for classical mixtures causal order and also showed that a certain framework (called the process matrix theory) is capable of violating these inequalities \cite{oreshkov2012quantum}.  Violation of these inequalities are a potential witness for true quantum gravitational effects.  Oreshkov, now in Brussels, working with Cerf, has continued to build framework for theories with indefinite causal structure \cite{oreshkov2014operational, oreshkov2015operational}.

Barrett \cite{barrett2007information}, in an important paper, developed the generalized probabilities theory approach, gave a simple account of the tensor product structure, and looked at the landscape of such theories.  Barnum, Wilce, Leifer, Barrett, Mueller, and many others have further developed this framework in various papers (here is a very incomplete selection \cite{barnum2011information, barnum2007generalized, barnum2010entropy, barnum2012teleportation, short2010strong}).

One aspect of the operator tensor approach to Quantum Theory is that it attempts to be causally neutral.  There is much other work in this tradition in the Quantum Foundations community.  Aharonov and collaborators have the time-symmetric framework \cite{aharonov1964time} from 1964 and, more recently, the multi-time framework \cite{aharonov2009multiple}. Leifer and Spekkens have proposed formulating Quantum Theory as a causally neutral theory of Bayesian inference.  Henson, Lal, and Pusey \cite{henson2014theory} have looked at general probability theories in the context of Bayesian networks.

There are, of course, many people round the world working on Quantum Gravity.  Some of the more related approaches are listed here.

The approach coming from within the Quantum Gravity community most related to the present program is Oeckl 's general boundary formalism.  He initiated this in 2003 \cite{oeckl2003general}.  He has also applied this approach to Quantum Field Theory \cite{oeckl2008general}.  In Oeckl's formulism states are associated with the boundary of arbitrary regions of a manifold.  When two such regions are joined at parts of their boundaries there are gluing rules from which the new state associated with boundary of the composite region is deduced. Originally this was restricted to pure states. In recent years, influenced by work in the Quantum Foundations community, Oeckl has pursued a more operational approach in which operators on complex Hilbert spaces (rather than elements of such complex Hilbert spaces) are associated with boundaries \cite{oeckl2013positive, oeckl2014first, oeckl2015quantum}.  He calls this the \emph{positive formulation} because he associated positive operators with such general boundaries.  There is a certain convergence between Oeckl's general boundary approach and the operator tensor approach.  In Part \ref{part:operatortensorQT} of this paper we develop an operator tensor formulation of Quantum Field Theory (along lines originally suggested in \cite{hardy2011reformulating}).  In this case the operators are positive under input transpose (rather than simply being positive).  This extra qualification is essential here because we have input/output structure.

Stochastic Gravity is a semiclassical approach to General Relativity (see the Living Reviews of Relativity article by Hu and Verdaguer \cite{hu2008stochastic}). This is a very different approach from the approach taken here but does attempt to combine probability and General Relativity.

John Baez has pursued a categorical approach to Quantum Theory, General Relativity, and Quantum Gravity \cite{baez2006quantum}.

Seth Lloyd has taken ideas from Quantum Information and applied them to the problem of Quantum Gravity \cite{lloyd2005theory, lloyd2012quantum, lloyd2013universe}

I will not attempt to survey in any detail the more established programs pursuing Quantum Gravity.  These include the string theory approach (see for example \cite{polchinski1998string}), the loop quantum gravity approach (see \cite{smolin2004invitation, rovelli2007quantum, rovelli2014covariant}), the causal set approach (see \cite{sorkin2005causal}), and the dynamical triangulations approach (see \cite{ambjorn2004emergence}).  Many researches in the loop quantum gravity approach have pursued the spin-foam approach  (see \cite{perez2013spin} for a review).  The basic idea goes back to Penrose's spin network approach to space-time \cite{penrose1971angular} which has some overlap with the current approach.  Further, Rovelli's approach to Quantum Gravity places a lot of emphasis on what might be called operational structure (see his talk entitled \lq\lq loop quantum gravity and time" in the discussion recorded at \cite{Rovelli2016emergent} and also \cite{rovelli2007quantum, rovelli2014covariant}).

\section{Basics}

In Part \ref{part:ontologicalandoperational} we provide the elements for such an operational reformulation of General Relativity.  We can represent a solution \index{solution} to the field equations of General Relativity by
\begin{equation}
\tilde{\Psi}=\left\{ (p, \pmb{\Phi}(p)): \forall p\in\mathscr{M} \right\}
\end{equation}
where $\pmb{\Phi}=(\mathbf{g}, \text{matter fields})$ - it is a list of physical fields (gravitational and matter) at play in the physical situation being studied.  Such solutions are only meaningful up to diffeomorphisms. \index{diffeomorphism} A diffeomorphism, $\varphi$, maps the point $p$ to the point $q=\varphi(p)$ and the set of tensor fields, $\pmb{\Phi}(p)$, to $\varphi^*\pmb{\Phi}(q)$. Hence we map the solution $\tilde{\Psi}$ to
\begin{equation}
\tilde{\Psi}=\left\{ (q, \varphi^*\pmb{\Phi}(q)): \forall q\in\varphi(\mathscr{M}) \right\}
\end{equation}
This solution looks different but actually represents the same physical situation.  Beables (i.e.\ physically real properties) are given by functions, $B(\cdot)$, that are invariant under diffeomorphisms such that $B(\tilde{\Psi})= B(\varphi^*\tilde{\Psi})$ for all diffeomorphisms.  We see from this that we cannot define beables to be local on the manifold - if we attempt to define some beable in terms of the fields in some region, $\mathscr{A}\subset\mathscr{M}$, on the manifold then we can use a diffeomorphism to replace the fields in $\mathscr{A}$ with fields from elsewhere.  In more picturesque terms - the coefficient of friction between reality and the manifold is zero.

We can represent the actual physical situation by
\begin{equation}
\Psi=\left\{ \varphi^*\tilde{\Psi}:\forall \varphi \right\}
\end{equation}
where $\varphi$ is a diffeomorphism.  This is a useful object for formal manipulations, but it does not provide a very clear picture of what is actually happening.

\begin{figure}
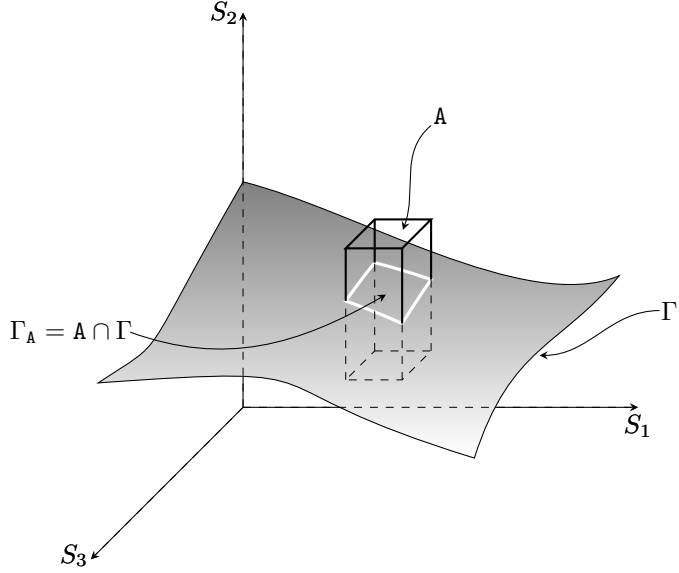
\label{fig:opspaceintro}
\begin{Compose}{0}{0}
\Cxaxis{S_1}{0,0}{21} \Cyaxis{S_2}{0,0}{21} \Czaxis{S_3}{0,0}{21}
\csurface[0.7]{0,0}{1}
\ccuboid[0.05]{7,3}{3}{7}{4}{4.7}{3.8}{3}{4.2}
\thispoint{GS}{20, 7, 11}
\thispoint{G}{26, 9, 10}
\pointingarrow{G}{-180}{GS}{0} \csymbolalt[1,0]{\Gamma}
\thispoint{AO}{8.5,9.44} \thispoint{A}{10,15} \pointingarrow[above right]{A}{-140}{AO}{70} \csymbolalt{\mathtt{A}}
\thispoint{GAS}{7.6, 6} \thispoint{GA}{-6, 4} \pointingarrow[left]{GA}{-20}{GAS}{-150} \csymbolalt[-75,0]{\Gamma_\mathtt{A}=\mathtt{A}\cap\Gamma}
\Cxaxis[ultra thin, dashed]{S_1}{0,0}{21} \Cyaxis[ultra thin, dashed]{S_2}{0,0}{21} \Czaxis[ultra thin, dashed]{S_3}{0,0}{21}
\end{Compose}
\caption{The surface $\Gamma$ is plotted into operational space.  An arbitrary region, $\mathtt{A}$, of operational space is shown (this need not be cuboid).  The intersection of $\mathtt{A}$ and $\Gamma$ is shown.}
\end{figure}

To make progress, we adapt ideas from Westman and Sonego \index{Westman and Sonego} \cite{westman2008events, westman2009coordinates} (see Sec.\ \ref{sec:theWestmanSonegoapproach}) for our own operational purposes.  In particular, we make the assertion that \emph{observables} \index{observables}(which we take here to be properties that can be directly observed) are functions of coincidences in the values of scalar fields.  To elaborate a little, consider a set of scalars $\mathbf{S}=(S_1, S_2, \dots S_K)$ constructed from the metric and matter fields in $\pmb{\Phi}$.   For each point $p$ in the solution, $\Psi$, we can plot a point $\mathbf{S}$ into a space whose axes are $S_k$.  This is the operational space (or op-space). \index{operational space}\index{op-space}  The solution $\tilde{\Psi}$ will induce a surface $\Gamma$ in operational space (see Fig.\ \ref{fig:opspaceintro}).  If we plot $\varphi^*\tilde{\Psi}$ into op-space we get the same surface, $\Gamma$, as for $\tilde{\Psi}$ (i.e.\ $\Gamma$ is invariant under diffeomorphisms).   We can, consequently, consider observables which are local on op-space.

Consider a region, $\mathtt{A}$, of op-space.  Let $\mathscr{M}_\mathtt{A}\subseteq\mathscr{M}$ be the set of points in $\mathscr{M}$ that are plotted to points in $\mathtt{A}$ for some given solution, $\tilde{\Psi}$.  We define
\begin{equation}
\tilde{\Psi}_\mathtt{A}=\left\{ (p, \pmb{\Phi}(p)): \forall p\in\mathscr{M}_\mathtt{A} \right\}
\end{equation}
and
\begin{equation}
\Psi_\mathtt{A}=\left\{ \varphi^*\tilde{\Psi}_\mathtt{A}:\forall \varphi \right\}
\end{equation}
This describes the actual situation in region $\mathtt{A}$.  Observables associated with the region $\mathtt{A}$ are functions of $\Gamma_\mathtt{A}=\Gamma\cap\mathtt{A}$.  We can write
\begin{equation}
O_\mathtt{A} (\Gamma_\mathtt{A})
\end{equation}
for observables.

\begin{figure}
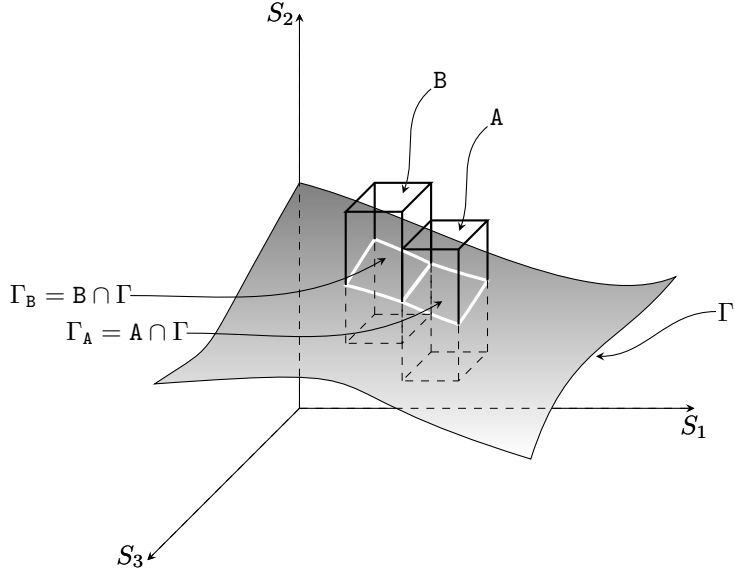
\label{fig:AandB}
\begin{Compose}{0}{0}
\Cxaxis{S_1}{0,0}{21} \Cyaxis{S_2}{0,0}{21} \Czaxis{S_3}{0,0}{21}
\csurface[0.7]{0,0}{1}
\ccuboid[-0.05]{7,3}{3}{7}{4}{4.7}{3.8}{3}{4.2}
\ccuboid[0.05]{4,5}{3}{7}{4}{4}{2.7}{2.2}{3.1}
\thispoint{GS}{20, 7, 11} \thispoint{G}{26, 9, 10} \pointingarrow{G}{-180}{GS}{0} \csymbolalt[1,0]{\Gamma}
\thispoint{AO}{8.5,9.44} \thispoint{A}{10,15} \pointingarrow[above right]{A}{-140}{AO}{70} \csymbolalt{\mathtt{A}}
\thispoint{BO}{5.5,11.44} \thispoint{B}{7,17} \pointingarrow[above right]{B}{-140}{BO}{70} \csymbolalt{\mathtt{B}}
\thispoint{GAS}{7.6, 5.7} \thispoint{GA}{-6, 4} \pointingarrow[left]{GA}{0}{GAS}{-150} \csymbolalt[-75,0]{\Gamma_\mathtt{A}=\mathtt{A}\cap\Gamma}
\thispoint{GBS}{4.6, 8} \thispoint{GB}{-9, 6} \pointingarrow[left]{GB}{0}{GBS}{-150} \csymbolalt[-75,0]{\Gamma_\mathtt{B}=\mathtt{B}\cap\Gamma}
\Cxaxis[ultra thin, dashed]{S_1}{0,0}{21} \Cyaxis[ultra thin, dashed]{S_2}{0,0}{21} \Czaxis[ultra thin, dashed]{S_3}{0,0}{21}
\end{Compose}
\caption{This shows two regions, $\mathtt{A}$ and $\mathtt{B}$, joined at a typing surface $\mathtt{a}$ (not marked).}
\end{figure}

We can consider joining two solutions $\Psi_\mathtt{A}$ and $\Psi_\mathtt{B}$ for two regions $\mathtt{A}$ and $\mathtt{B}$ of op-space (see Fig.\ \ref{fig:AandB}).  Let these two regions meet at the \emph{typing surface} \index{typing surface} $\mathtt{a}$ (a typing surface is a set of points in op-space of dimension $K-1$ with a direction indicated also).  The natural way of joining two solutions is
\begin{equation}
\Psi_\mathtt{A}\Cup_\mathtt{a}\Psi_\mathtt{B}
= \left\{ \varphi_\mathtt{A}^*\tilde{\Psi}_\mathtt{A} \cup_\mathtt{a} \varphi_\mathtt{B}^*\tilde{\Psi}_\mathtt{B}: \forall \varphi_\mathtt{A}, \varphi_\mathtt{B} \right\}
\end{equation}
Here $\cup_\mathtt{a}$ checks to see if the appropriate boundary conditions are met at the typing surface (such that we have a solution for the composite region). If the boundary conditions are met then it takes the union. If not, it returns nothing.  Thus, $\Psi_\mathtt{A}\Cup_\mathtt{a}\Psi_\mathtt{B}$ collects all cases where the boundary conditions are met.  It can happen that there is more than one equivalence class in the entries in $\Psi_\mathtt{A}\Cup_\mathtt{a}\Psi_\mathtt{B}$ (i.e.\ this is a mixed rather than pure solution).  This means that, by specifying the solutions fully in each of the subregions, $\mathtt{A}$ and $\mathtt{B}$ we do not necessarily specify fully the solution for the composite region, $\mathtt{A}\cup\mathtt{B}$.  Put another way, we have examples where
\begin{equation}
B_{\mathtt{A}\cup\mathtt{B}} (\Psi_{\mathtt{A}\cup\mathtt{B}}) \not= f(B_\mathtt{A}(\Psi_\mathtt{A}), B_\mathtt{B}(\Psi_\mathtt{B}))
\end{equation}
for any function, $f$.  Here $B_\mathtt{A}$ are the beables for region $\mathtt{A}$.   This curious kind of non-separability is a consequence of diffeomorphism invariance.

We can, generally, allow solutions, $\Psi_\mathtt{A}$, to be mixed in that they contain solutions that cannot be mapped to one another by diffeomorphisms.  Solutions that are not mixed are pure.

Observables, on the other hand, are separable since we can write
\begin{equation}
O_{\mathtt{A}\cup\mathtt{B}}(\Gamma_{\mathtt{A}\cup\mathtt{B}}) = O_{\mathtt{A}\cup\mathtt{B}}(\Gamma_\mathtt{A}\cup\Gamma_\mathtt{B})
\end{equation}
and so we can write the observables of a composite region as a function of the observables for the component regions.

In Operational Quantum Theory we have knobs on the apparatuses that can be adjusted to vary the operation implemented.  We can do something similar in General Relativity.  For example, we can imagine varying the interaction between two fluids.   In the text we give the example of a vast fleet of space ships that are powered by interacting with another fluid (providing the wind). The fleet is so vast that we can think of it as a dust fluid.  The interaction can be controlled by some tensor fields, $\pmb{\chi}$, which we call \emph{agency fields} (this is in addition to the fields in $\pmb{\Phi}$).  From the agency fields, $\pmb{\chi}$,  and the tensors in $\pmb{\Phi}$ we can extract a list of scalars, $\mathbf{Q}$ (similar to the way we extracted $\mathbf{S}$).  We can imagine agents setting $\mathbf{Q}$ as a function of $\mathbf{S}$ according to some strategy, $\mathbf{Q}(\mathbf{S})$.  The choice in region $\mathtt{A}$ of op-space is given by
\begin{equation}
\mathbf{Q}_\mathtt{A}= \left\{ (\mathbf{S}, \mathbf{Q}(\mathbf{S})):\forall \mathbf{S}\in\mathtt{A}\right\}
\end{equation}
This is the \lq\lq knob setting" for $\mathtt{A}$.

Agency fields can be understood as an effective notion - they arise because we do not solve the equations below a certain resolution.  For example, we do not have a model of the brains of the captains steering the space ships but rather take the choices the captains make as an input. When we introduce this effective notion we need to introduce another effective notion - a time direction field, $\pmb{\tau}$ (this is a vector pointing into the forward light cone).  This is so we can implement the idea that choices influence the future but not the past.  This allows us to have a more operational understanding of causality than is usually possible in General Relativity.  There is some gauge degree of freedom associated with the time direction field as any vector pointing into the same light cone will do.  We combine this gauge freedom with the diffeomorphism symmetry to have a bigger gauge group $G^+$.

Once we have agency and scalar fields in place we have a new specification of solution for region $\mathtt{A}$ as
\begin{equation}
\tilde{\Psi}_\mathtt{A} = \left\{ (p, \pmb{\Phi}, \pmb{\chi}, \pmb{\tau}): \forall p\in\mathscr{M}_\mathtt{A} \right\}
\end{equation}
and
\begin{equation}
\Psi_\mathtt{A}= \left\{ \vartheta^*\tilde{\Psi}\mathtt{A}:\forall \vartheta\in G^+\right\}
\end{equation}
This object now specifies the physical situation in region $\mathtt{A}$ of op-space.

\section{Possibilistic Operational General Relativity}

The usual picture in General Relativity is that some initial conditions are supplied then we evolve a state forward in time to find the full solution.  Here we take a different attitude that is more tuned to people living in a universe described by General Relativity.  We seek a framework for calculating whether certain operationally described observations are possible or not (given some General Relativistic field equations).

First we define the notion of an operation.  Operations, such as
\begin{equation}
\mathsf{A}^\mathtt{a}_\mathtt{bc} ~~~ \Longleftrightarrow
\begin{Compose}{0}{0} \setdefaultfont{\mathsf}\setsecondfont{\mathtt}
\Ucircle{A}{0,0}\thispoint{DL}{-120:4} \thispoint{DR}{-60:4} \thispoint{UC}{90:4}
\joincc[above left]{DL}{60}{A}{-120} \csymbolalt{b}
\joincc[above right]{DR}{120}{A}{-60} \csymbolalt{c}
\joincc[left]{A}{90}{UC}{-90} \csymbolalt{a}
\end{Compose}
\end{equation}
can be associated with any region, $\mathtt{A}$ of op-space.  Superscripts correspond to outward pointing typing surfaces and subscripts correspond to inward pointing typing surfaces.  These typing surfaces can meet at their edges but do not overlap more than this. The typing surfaces can, but to not have to, cover the entire boundary of $\mathtt{A}$.  The operation has some has some agency setting, $\mathbf{Q}_\mathtt{A}$, and some set of outcomes, $O_\mathtt{A}$, associated with it.  Here $O_\mathtt{A}$ is a set of possible $\Gamma_\mathtt{A}$'s and represents a course-grained observable for region $\mathtt{A}$.  The set $O_\mathtt{A}$ can, but does not have to, contain all possible $\Gamma_\mathtt{A}$.  We say that the operation \lq\lq happens" if the outcome is in the associated outcome set.

Next we define an encapsulated proposition.  An encapsulated proposition,
\begin{equation}
\mathcal{A}^\mathtt{a}_\mathtt{bc} ~~~ \Longleftrightarrow
\begin{Compose}{0}{0} \setdefaultfont{\mathcal}\setsecondfont{\mathtt}
\Ucircle{A}{0,0}\thispoint{DL}{-120:4} \thispoint{DR}{-60:4} \thispoint{UC}{90:4}
\joincc[above left]{DL}{60}{A}{-120} \csymbolalt{b}
\joincc[above right]{DR}{120}{A}{-60} \csymbolalt{c}
\joincc[left]{A}{90}{UC}{-90} \csymbolalt{a}
\end{Compose}
\end{equation}
can be associated with any region, $\mathtt{A}$, of op-space.  The typing surface convention is the same as for an operation.  The encapsulated proposition has a proposition, $\text{prop}(\mathcal{A})$, associated with it.  We can represent the proposition by a solution, $\Psi_\mathtt{A}$ that may be mixed. The proposition can then be read as the proposition that the actual (i.e.\ pure) solution is contained in the mixture $\Psi_\mathtt{A}$.   The encapsulated proposition also has a agent strategy, $\mathbf{Q}_\mathtt{A}$, associated with it.

We can form a map from an operation with outcome set $O_\mathtt{A}$ to an encapsulated proposition where the solutions associated with the proposition is $\Psi_\mathtt{A}[O_\mathtt{A}]$.  This is the mixed solution containing all pure solutions that might give rise to any $\Gamma_\mathtt{A}$ in $O_\mathtt{A}$.

Next we use encapsulated propositions to induce \emph{boundary propositions}
\begin{equation}
\mathbpro{A}^\mathtt{a}_\mathtt{bc} ~~~ \Longleftrightarrow
\begin{Compose}{0}{0} \setdefaultfont{\mathbpro}\setsecondfont{\mathtt}
\Ucircle{A}{0,0}\thispoint{DL}{-120:4} \thispoint{DR}{-60:4} \thispoint{UC}{90:4}
\joincc[above left]{DL}{60}{A}{-120} \csymbolalt{b}
\joincc[above right]{DR}{120}{A}{-60} \csymbolalt{c}
\joincc[left]{A}{90}{UC}{-90} \csymbolalt{a}
\end{Compose}
\end{equation}
This is the logical XOR of all the propositions concerning boundary conditions that are induced at the typing surfaces $\mathtt{a}$, $\mathtt{b}$, \dots by the solution $\Psi_\mathtt{A}[O_\mathtt{A}]$ associated with the encapsulated proposition.

We can expand a boundary proposition in terms of some fiducial set of boundary propositions (represented by $\presub{\mathbnd{a}}{\mathbpro{X}}^\mathtt{a}$, $\mathbpro{X}_\mathtt{b}^\mathbnd{b}$, \dots).  An example of such an expansion is
\begin{equation}\label{introdecomploc}
\mathbpro{A}^\mathtt{a}_\mathtt{bc} \equiv  \presup{\mathbnd{a}}A_\mathbnd{bc} ~  \presub{\mathbnd{a}}{\mathbpro{X}}^\mathtt{a} \mathbpro{X}_\mathtt{b}^\mathbnd{b} \mathbpro{X}_\mathtt{c}^\mathbnd{c}
~~~~~ \Longleftrightarrow ~~~~~
\begin{Compose}{0}{0} \setdefaultfont{\mathbpro}\setsecondfont{\mathtt}\setthirdfont{\mathbnd}
\Ucircle{A}{0,0}\thispoint{DL}{-120:4} \thispoint{DR}{-60:4} \thispoint{UC}{90:4}
\joincc[above left]{DL}{60}{A}{-120} \csymbolalt{b}
\joincc[above right]{DR}{120}{A}{-60} \csymbolalt{c}
\joincc[left]{A}{90}{UC}{-90} \csymbolalt{a}
\end{Compose}
~~~ \equiv ~~~
\begin{Compose}{0}{0} \setdefaultfont{\mathbpro}\setsecondfont{\mathtt}\setthirdfont{\mathbnd}
\ucircle{A}{0,0}\csymbolfourth{A}
\scircle{DLX}{-120:5}\csymbol{X} \scircle{DRX}{-60:5}\csymbol{X} \scircle{UCX}{90:5} \csymbol{X}
\thispoint{DL}{-120:8} \thispoint{DR}{-60:8} \thispoint{UC}{90:8}
\joincc[above left]{DL}{60}{DLX}{-120} \csymbolalt{b}
\joincc[above right]{DR}{120}{DRX}{-60} \csymbolalt{c}
\joincc[left]{UCX}{90}{UC}{-90} \csymbolalt{a}
\joinccbwsq[above left]{DLX}{60}{A}{-120} \csymbolthird{b}
\joinccbwsq[above right]{DRX}{120}{A}{-60} \csymbolthird{c}
\joinccwbsq[left]{A}{90}{UCX}{-90} \csymbolthird{a}
\end{Compose}
\end{equation}
where $\mathbnd{a}\in\Lambda_\mathtt{a}[\text{fid}]$ are boundary conditions at $\mathtt{a}$.  The fiducial proposition, $\presub{\mathbnd{a}}{\mathbpro{X}}^\mathtt{a}$, is the proposition that the boundary condition at $\mathtt{a}$ is $\mathbnd{a}$.  The expression
$\presub{\mathbnd{a}}{\mathbpro{X}}^\mathtt{a} \mathbpro{X}_\mathtt{b}^\mathbnd{b} \mathbpro{X}_\mathtt{c}^\mathbnd{c}$ should be interpreted as the logical AND of the three factors.

The expansion coefficients, $\presup{\mathbnd{a}}A_\mathbnd{bc}$, are equal to 1 for $\mathbnd{abc}$ that are induced by $\Psi_\mathtt{A}[O_\mathtt{A}]$ and 0 otherwise.  The expansion in \eqref{introdecomploc} is to be interpreted as taking the logical XOR of all $\mathbnd{abc}$ for which this expansion coefficient is equal to 1.  We call $\presup{\mathbnd{a}}A_\mathbnd{bc}$ the \emph{generalized possibilistic state}.

An important object that crops up in this approach is the \emph{hopping metric}.  This is defined as
\begin{equation}
\presub{\mathbnd{a}'}h^\mathbnd{a}= \text{poss} (\presub{\mathbnd{a}'}X^\mathtt{a} X_\mathtt{a}^\mathbnd{a} )
~~~~~~\Leftrightarrow~~~~~~
\begin{Compose}{0}{-0.1}\setthirdfont{\mathbnd}
\vbbmatrixsq{h}{0,0}\csymbolthird{a}
\end{Compose}
~=~
\text{poss}\left(
\begin{Compose}{0}{-0.5} \setdefaultfont{\mathbpro}\setsecondfont{\mathtt}\setthirdfont{\mathbnd}
\blackdotsq{d1}{0,-3} \scircle{X1}{0,0}\csymbol{X}\scircle{X2}{0,4} \csymbol{X} \blackdotsq{d2}{0,7}
\joincc[left]{d1}{90}{X1}{-90} \csymbolthird{a} \joincc[left]{X1}{90}{X2}{-90}\csymbolalt{a} \joincc[left]{X2}{90}{d2}{-90} \csymbolthird{a}
\end{Compose}
\right)
\end{equation}
where the function, $\text{poss}(\cdot)$ returns 1 if there are any matches in the given boundary conditions and 0 if not.

The boundary conditions, $\mathbnd{a}\in\Lambda_\mathtt{a}[\text{fid}]$, are complicated functions of the variables appearing in $\Psi_\mathtt{A}[O_\mathtt{A}]$ depending on tensor fields and the like.  While formally invariant under diffeomorphisms, they are unwieldy.  Under certain circumstances we can replace these variables with a set of operational descriptions pertaining to the region $\mathtt{A}$ (these operational descriptions, $(\mathbf{Q}_\mathtt{A}, O_\mathtt{A})$ in $\mathtt{A}$ induce boundary conditions at $\mathtt{a}$).  We represent these as $a_\mathtt{A}\in\Upsilon[\mathtt{A}]_\mathtt{a}$.  Under this substitution we can expand boundary propositions as
\begin{equation}
\mathbpro{A}^\mathtt{a}_\mathtt{bc} \equiv  \presup{a_\mathtt{A}}A_{b_\mathtt{A}c_\mathtt{A}} ~
\presub{a_\mathtt{A}}{\mathbpro{X}}^\mathtt{a} \mathbpro{X}_\mathtt{b}^{b_\mathtt{A}} \mathbpro{X}_\mathtt{c}^{c_\mathtt{A}}
~~~~~ \Longleftrightarrow ~~~~~
\begin{Compose}{0}{0} \setdefaultfont{\mathbpro}\setsecondfont{\mathtt}
\Ucircle{A}{0,0}\thispoint{DL}{-120:4} \thispoint{DR}{-60:4} \thispoint{UC}{90:4}
\joincc[above left]{DL}{60}{A}{-120} \csymbolalt{b}
\joincc[above right]{DR}{120}{A}{-60} \csymbolalt{c}
\joincc[left]{A}{90}{UC}{-90} \csymbolalt{a}
\end{Compose}
~~~ \equiv ~~~
\begin{Compose}{0}{0} \setdefaultfont{\mathbpro}\setsecondfont{\mathtt}
\ucircle{A}{0,0}\csymbolthird{A}
\scircle{DLX}{-120:5}\csymbol{X} \scircle{DRX}{-60:5}\csymbol{X} \scircle{UCX}{90:5} \csymbol{X}
\thispoint{DL}{-120:8} \thispoint{DR}{-60:8} \thispoint{UC}{90:8}
\joincc[above left]{DL}{60}{DLX}{-120} \csymbolalt{b}
\joincc[above right]{DR}{120}{DRX}{-60} \csymbolalt{c}
\joincc[left]{UCX}{90}{UC}{-90} \csymbolalt{a}
\joinccbw[above left]{DLX}{60}{A}{-120} \csymbolthird{b_\mathtt{A}}
\joinccbw[above right]{DRX}{120}{A}{-60} \csymbolthird{c_\mathtt{A}}
\joinccwb[left]{A}{90}{UCX}{-90} \csymbolthird{a_\mathtt{A}}
\end{Compose}
\end{equation}
The object, $\presup{a_\mathtt{A}}A_{b_\mathtt{A}c_\mathtt{A}}$, is the operational generalized possibilistic state.  We have hopping metric
\begin{equation}
\presub{a_\mathtt{A}}h^{a_\mathtt{B}}= \text{poss} (\presub{a_\mathtt{A}}{\mathbpro{X}}^\mathtt{a} \mathbpro{X}_\mathtt{a}^{a_\mathtt{B}} )
~~~~~~\Leftrightarrow~~~~~~
\begin{Compose}{0}{-0.1}
\vbbmatrix{h}{0,0}\csymbolthird[32,-62]{a_\mathtt{A}} \csymbolthird[32,62]{a_\mathtt{B}}
\end{Compose}
~=~
\text{poss}\left(
\begin{Compose}{0}{-0.5} \setdefaultfont{\mathbpro}\setsecondfont{\mathtt}
\blackdot{d1}{0,-3} \scircle{X1}{0,0}\csymbol{X}\scircle{X2}{0,4} \csymbol{X} \blackdot{d2}{0,7}
\joincc[left]{d1}{90}{X1}{-90} \csymbolthird[32,-60]{a_\mathtt{A}} \joincc[left]{X1}{90}{X2}{-90}\csymbolalt{a}
\joincc[left]{X2}{90}{d2}{-90} \csymbolthird[32,60]{a_\mathtt{B}}
\end{Compose}
\right)
\end{equation}
Going from black and white squares to black and white circles represents substituting $a_\mathtt{A}$ for $\mathbnd{a}$ (and so on with the other instances).

We finally have a manifestly diffeomorphism invariant way of formulating General Relativity.  A calculation looks like the following
\begin{equation}
\begin{array}{lccc}
{} &
\begin{Compose}{0}{-1.5} \setdefaultfont{\mathsf}\setsecondfont{\mathtt}
\Ucircle{A}{0,0} \Ucircle{B}{-5,5} \Ucircle{C}{3,4} \Ucircle{D}{-3, 11} \Ucircle{E}{2,9}
\joincc[below left]{B}{-65}{A}{115} \csymbolalt{a}
\joincc[below]{A}{80}{C}{-90} \csymbolalt{b}
\joincc[below]{C}{170}{B}{-10} \csymbolalt{c}
\joincc[above left]{B}{25}{E}{-110} \csymbolalt{d}
\joincc[left]{B}{80}{D}{-100} \csymbolalt{e}
\joincc[above right]{D}{-15}{E}{170}
\joincc[right]{C}{100}{E}{-80} \csymbolalt{g}
\end{Compose}
& \longrightarrow &
\begin{Compose}{0}{-1.5} \setdefaultfont{\mathcal}\setsecondfont{\mathtt}
\Ucircle{A}{0,0} \Ucircle{B}{-5,5} \Ucircle{C}{3,4} \Ucircle{D}{-3, 11} \Ucircle{E}{2,9}
\joincc[below left]{B}{-65}{A}{115} \csymbolalt{a}
\joincc[below]{A}{80}{C}{-90} \csymbolalt{b}
\joincc[below]{C}{170}{B}{-10} \csymbolalt{c}
\joincc[above left]{B}{25}{E}{-110} \csymbolalt{d}
\joincc[left]{B}{80}{D}{-100} \csymbolalt{e}
\joincc[above right]{D}{-15}{E}{170}
\joincc[right]{C}{100}{E}{-80} \csymbolalt{g}
\end{Compose}
\\
\longrightarrow &
\begin{Compose}{0}{-1.5} \setdefaultfont{\mathbpro}\setsecondfont{\mathtt}
\Ucircle{A}{0,0} \Ucircle{B}{-5,5} \Ucircle{C}{3,4} \Ucircle{D}{-3, 11} \Ucircle{E}{2,9}
\joincc[below left]{B}{-65}{A}{115} \csymbolalt{a}
\joincc[below]{A}{80}{C}{-90} \csymbolalt{b}
\joincc[below]{C}{170}{B}{-10} \csymbolalt{c}
\joincc[above left]{B}{25}{E}{-110} \csymbolalt{d}
\joincc[left]{B}{80}{D}{-100} \csymbolalt{e}
\joincc[above right]{D}{-15}{E}{170}
\joincc[right]{C}{100}{E}{-80} \csymbolalt{g}
\end{Compose}
& \longrightarrow &
\begin{Compose}{0}{-2} \setdefaultfont{\mathnormal}\setsecondfont{\mathnormal}
\Ucircle{A}{-2,-2} \Ucircle{B}{-9,5} \Ucircle{C}{6,4} \Ucircle{D}{-6, 16} \Ucircle{E}{3,13}
\joinccwbbw[below left]{B}{-65}{A}{115}  \csymbolwbunder{a_\mathtt{B}} \csymbolbwunder{a_\mathtt{A}}
\joinccwbbw[below]{A}{80}{C}{-90} \csymbolwbunder{b_\mathtt{A}}\csymbolbwunder{b_\mathtt{C}}
\joinccwbbw[below]{C}{170}{B}{-10} \csymbolwbunder[0,-6.5]{c_\mathtt{C}}\csymbolbwunder[0,-6.5]{c_\mathtt{B}}
\joinccwbbw[above left]{B}{25}{E}{-110} \csymbolwb{d_\mathtt{B}}\csymbolbw{d_\mathtt{E}}
\joinccwbbw[left]{B}{80}{D}{-100} \csymbolwb{e_\mathtt{B}}\csymbolbw{e_\mathtt{D}}
\joinccwbbw[above right]{D}{-15}{E}{170}\csymbolwb{k_\mathtt{D}}\csymbolbw{k_\mathtt{E}}
\joinccwbbw[right]{C}{100}{E}{-80} \csymbolwb{g_\mathtt{C}}\csymbolbwunder{g_\mathtt{E}}
\end{Compose}
\end{array}
\end{equation}
In the last step we have matched black and white dots.  Where these black and white dots meet we do possibilistic summation (such that $0+0=0$, $0+1=1$, $1+1=1$, and so on). This models the underling logical aspect of the calculation.  Hence the final calculation gives an answer equal to 0 or 1.  If the answer is 0 then there are no solutions in General Relativity that are consistent with the given operational specifications.  If the answer is 1 then there exists at least one solution consistent with the given operational specification.

To specify General Relativity in this context we need (i) to provide the space of possible operational generalized possibilistic states for each region of op-space of interest and (ii) the hopping metric for each typing surface of interest.  These are constrained by the field equations of General Relativity.

\section{Probabilistic General Relativity}

General Relativity is a deterministic theory (sufficient boundary conditions will determine what happens).  However, we can still have probabilistic ignorance represented by a probability distribution over beables.  In Part \ref{part:probabilisticformulation} we show how to set up probabilistic general relativity in an operational and compositional way.  This will look very similar to the possibilistic case.  The main difference is that we have to introduce \emph{loading}.  This is some extra probabilistic information associated with the various elements.

First we have \emph{loaded operations}
\begin{equation}
\mathsf{A}^\mathtt{a}_\mathtt{bc} ~~~ \Longleftrightarrow
\begin{Compose}{0}{0} \setdefaultfont{\mathsf}\setsecondfont{\mathtt}
\Ucircle{A}{0,0}\thispoint{DL}{-120:4} \thispoint{DR}{-60:4} \thispoint{UC}{90:4}
\joincc[above left]{DL}{60}{A}{-120} \csymbolalt{b}
\joincc[above right]{DR}{120}{A}{-60} \csymbolalt{c}
\joincc[left]{A}{90}{UC}{-90} \csymbolalt{a}
\end{Compose}
\end{equation}
In addition to agency setting ($\mathbf{Q}_\mathtt{A}$) and outcome set ($O_\mathtt{A}$) these have some loading, $L_\mathtt{A}$.  This is something that can be specified by the agent using the theory (it may correspond to beliefs, be provided by some statistical considerations, or come from elsewhere).

Next we have \emph{loaded encapsulated propositions}
\begin{equation}
\mathcal{A}^\mathtt{a}_\mathtt{bc} ~~~ \Longleftrightarrow
\begin{Compose}{0}{0} \setdefaultfont{\mathcal}\setsecondfont{\mathtt}
\Ucircle{A}{0,0}\thispoint{DL}{-120:4} \thispoint{DR}{-60:4} \thispoint{UC}{90:4}
\joincc[above left]{DL}{60}{A}{-120} \csymbolalt{b}
\joincc[above right]{DR}{120}{A}{-60} \csymbolalt{c}
\joincc[left]{A}{90}{UC}{-90} \csymbolalt{a}
\end{Compose}
\end{equation}
These also have a load, $\text{load}(\mathcal{A})$. The way in which this load is specified will emerge as the theory is developed.  However, if we have a closed loaded encapsulated proposition (one with no typing surfaces) then we demand that there exists a subnormalized probability distribution over the possible solutions, $\Psi_\mathtt{A}$, in $\mathtt{A}$ that are consistent with the given $\mathbf{Q}_\mathtt{A}$ and $O_\mathtt{A}$.

Then we consider \emph{loaded boundary propositions}
\begin{equation}
\mathlbpro{A}^\mathtt{a}_\mathtt{bc} ~~~ \Longleftrightarrow
\begin{Compose}{0}{0} \setdefaultfont{\mathlbpro}\setsecondfont{\mathtt}
\Ucircle{A}{0,0}\thispoint{DL}{-120:4} \thispoint{DR}{-60:4} \thispoint{UC}{90:4}
\joincc[above left]{DL}{60}{A}{-120} \csymbolalt{b}
\joincc[above right]{DR}{120}{A}{-60} \csymbolalt{c}
\joincc[left]{A}{90}{UC}{-90} \csymbolalt{a}
\end{Compose}
\end{equation}
These are built up by weighting over fiducial loaded boundary propositions ($\presub{\mathlbnd{a}}{\mathlbpro{X}}^\mathtt{a}$, $\mathlbpro{X}_\mathtt{b}^\mathlbnd{b}$, \dots)  at each of the typing surfaces as follows
\begin{equation}
\mathlbpro{A}^\mathtt{a}_\mathtt{bc} = \presup{\mathlbnd{a}}A_{\mathlbnd{b}\mathlbnd{c}} ~  \presub{\mathlbnd{a}}{\mathlbpro{X}}^\mathtt{a} \mathlbpro{X}_\mathtt{b}^\mathlbnd{b} \mathlbpro{X}_\mathtt{c}^\mathlbnd{c}
~~~~~ \Longleftrightarrow ~~~~~
\begin{Compose}{0}{0} \setdefaultfont{\mathlbpro}\setsecondfont{\mathtt}\setthirdfont{\mathlbnd}
\Ucircle{A}{0,0}\thispoint{DL}{-120:4} \thispoint{DR}{-60:4} \thispoint{UC}{90:4}
\joincc[above left]{DL}{60}{A}{-120} \csymbolalt{b}
\joincc[above right]{DR}{120}{A}{-60} \csymbolalt{c}
\joincc[left]{A}{90}{UC}{-90} \csymbolalt{a}
\end{Compose}
~~~ = ~~~
\begin{Compose}{0}{0} \setdefaultfont{\mathlbpro}\setsecondfont{\mathtt}\setthirdfont{\mathlbnd}
\ucircle{A}{0,0}\csymbolfourth{A}
\scircle{DLX}{-120:5}\csymbol{X} \scircle{DRX}{-60:5}\csymbol{X} \scircle{UCX}{90:5} \csymbol{X}
\thispoint{DL}{-120:8} \thispoint{DR}{-60:8} \thispoint{UC}{90:8}
\joincc[above left]{DL}{60}{DLX}{-120} \csymbolalt{b}
\joincc[above right]{DR}{120}{DRX}{-60} \csymbolalt{c}
\joincc[left]{UCX}{90}{UC}{-90} \csymbolalt{a}
\joinccbwsq[above left]{DLX}{60}{A}{-120} \csymbolthird{b}
\joinccbwsq[above right]{DRX}{120}{A}{-60} \csymbolthird{c}
\joinccwbsq[left]{A}{90}{UCX}{-90} \csymbolthird{a}
\end{Compose}
\end{equation}
The object, $\presup{\mathlbnd{a}}A_{\mathlbnd{b}\mathlbnd{c}}$ is the \emph{generalized probabilistic state}.  This state consists of real numbers (not just equal to 0 and 1 as in the possibilistic case).

We can also define a hopping metric using the fiducials
\begin{equation}
\presub{\mathlbnd{a}'}h^{\mathlbnd{a}}= \text{prob} (\presub{\mathlbnd{a}'}X^\mathtt{a} X_\mathtt{a}^\mathlbnd{a} )
~~~~~~\Leftrightarrow~~~~~~
\begin{Compose}{0}{-0.14}\setthirdfont{\mathlbnd}
\vbbmatrixsq{h}{0,0}\csymbolthird{a}
\end{Compose}
~=~
\text{prob}\left(
\begin{Compose}{0}{-0.55} \setdefaultfont{\mathlbpro}\setsecondfont{\mathtt}\setthirdfont{\mathlbnd}
\blackdotsq{d1}{0,-3} \scircle{X1}{0,0}\csymbol{X}\scircle{X2}{0,4} \csymbol{X} \blackdotsq{d2}{0,7}
\joincc[left]{d1}{90}{X1}{-90} \csymbolthird{a} \joincc[left]{X1}{90}{X2}{-90}\csymbolalt{a} \joincc[left]{X2}{90}{d2}{-90} \csymbolthird{a}
\end{Compose}
\right)
\end{equation}
where the function, $\text{prob}(\cdot)$ returns the probability.

Finally, we can substitute a loaded operational description, $a_\mathtt{A}$ (associated with region $\mathtt{A}$) for the unwieldy $\mathlbnd{a}$ descriptions. The idea is that the loaded operational description $a_\mathtt{A}$ (corresponding to some $(\mathbf{Q}_\mathtt{A}, O_\mathtt{A}, L_\mathtt{A})$) induces a certain fiducial loaded boundary proposition.  We obtain operational loaded boundary propositions
\begin{equation}
\mathlbpro{A}^\mathtt{a}_\mathtt{bc} =  \presup{a_\mathtt{A}}A_{b_\mathtt{A}c_\mathtt{A}} ~
\presub{a_\mathtt{A}}{\mathlbpro{X}}^\mathtt{a} \mathlbpro{X}_\mathtt{b}^{b_\mathtt{A}} \mathlbpro{X}_\mathtt{c}^{c_\mathtt{A}}
~~~~~ \Longleftrightarrow ~~~~~
\begin{Compose}{0}{0} \setdefaultfont{\mathlbpro}\setsecondfont{\mathtt}
\Ucircle{A}{0,0}\thispoint{DL}{-120:4} \thispoint{DR}{-60:4} \thispoint{UC}{90:4}
\joincc[above left]{DL}{60}{A}{-120} \csymbolalt{b}
\joincc[above right]{DR}{120}{A}{-60} \csymbolalt{c}
\joincc[left]{A}{90}{UC}{-90} \csymbolalt{a}
\end{Compose}
~~~ = ~~~
\begin{Compose}{0}{0} \setdefaultfont{\mathlbpro}\setsecondfont{\mathtt}
\ucircle{A}{0,0}\csymbolthird{A}
\scircle{DLX}{-120:5}\csymbol{X} \scircle{DRX}{-60:5}\csymbol{X} \scircle{UCX}{90:5} \csymbol{X}
\thispoint{DL}{-120:8} \thispoint{DR}{-60:8} \thispoint{UC}{90:8}
\joincc[above left]{DL}{60}{DLX}{-120} \csymbolalt{b}
\joincc[above right]{DR}{120}{DRX}{-60} \csymbolalt{c}
\joincc[left]{UCX}{90}{UC}{-90} \csymbolalt{a}
\joinccbw[above left]{DLX}{60}{A}{-120} \csymbolthird{b_\mathtt{A}}
\joinccbw[above right]{DRX}{120}{A}{-60} \csymbolthird{c_\mathtt{A}}
\joinccwb[left]{A}{90}{UCX}{-90} \csymbolthird{a_\mathtt{A}}
\end{Compose}
\end{equation}
The object, $\presup{a_\mathtt{A}}A_{b_\mathtt{A}c_\mathtt{A}}$, is the \emph{operational generalized possibilistic state}.

We can form a map from loaded operations to operational generalized probabilistic states and hence
\begin{equation}
\text{prob}\left(
\begin{Compose}{0}{-1.3} \setdefaultfont{\mathsf}\setsecondfont{\mathtt}
\Ucircle{A}{0,0} \Ucircle{B}{-5,5} \Ucircle{C}{3,4} \Ucircle{D}{-3, 11} \Ucircle{E}{2,9}
\joincc[below left]{B}{-65}{A}{115} \csymbolalt{a}
\joincc[below]{A}{80}{C}{-90} \csymbolalt{b}
\joincc[below]{C}{170}{B}{-10} \csymbolalt{c}
\joincc[above left]{B}{25}{E}{-110} \csymbolalt{d}
\joincc[left]{B}{80}{D}{-100} \csymbolalt{e}
\joincc[above right]{D}{-15}{E}{170}
\joincc[right]{C}{100}{E}{-80} \csymbolalt{g}
\end{Compose}
\right) ~~~=~~~
\begin{Compose}{0}{-1.3} \setdefaultfont{\mathnormal}\setsecondfont{\mathnormal}
\Ucircle{A}{-2,-2} \Ucircle{B}{-9,5} \Ucircle{C}{6,4} \Ucircle{D}{-6, 16} \Ucircle{E}{3,13}
\joinccwbbw[below left]{B}{-65}{A}{115}  \csymbolwbunder{a_\mathtt{B}} \csymbolbwunder{a_\mathtt{A}}
\joinccwbbw[below]{A}{80}{C}{-90} \csymbolwbunder{b_\mathtt{A}}\csymbolbwunder{b_\mathtt{C}}
\joinccwbbw[below]{C}{170}{B}{-10} \csymbolwbunder[0,-6.5]{c_\mathtt{C}}\csymbolbwunder[0,-6.5]{c_\mathtt{B}}
\joinccwbbw[above left]{B}{25}{E}{-110} \csymbolwb{d_\mathtt{B}}\csymbolbw{d_\mathtt{E}}
\joinccwbbw[left]{B}{80}{D}{-100} \csymbolwb{e_\mathtt{B}}\csymbolbw{e_\mathtt{D}}
\joinccwbbw[above right]{D}{-15}{E}{170}\csymbolwb{k_\mathtt{D}}\csymbolbw{k_\mathtt{E}}
\joinccwbbw[right]{C}{100}{E}{-80} \csymbolwb{g_\mathtt{C}}\csymbolbwunder{g_\mathtt{E}}
\end{Compose}
\end{equation}
Given a loaded operational description for each of the relevant regions of operational space we are able to calculate a probability.  For this we need to know the hopping metric for each typing surface of interest and the space of operational generalized probabilistic states for each region of op-space of interest.

\section{Operator Tensor Quantum Theory}

In Part \ref{part:operatortensorQT} we review the operator tensor formulation of Quantum Theory, show how it can be applied to infinite dimensional Hilbert spaces and provide a preliminary version of operator tensor Quantum Field Theory.

An operation corresponds to a single use of an apparatus and has associated with it some given knob settings and some given set of outcomes (detector clicks, lights flashing, meter readings, \dots).  It also has inputs and outputs (through which quantum systems pass).  We can wire together such operations to form circuits
\begin{equation}\label{QTcircuit}
\mathsf{A^{a_1a_2} B_{a_1c_3}^{a_5d_4} C_{a_2}^{c_3b_6} D_{a_5c_7} E_{d_4b_6}^{c_7}} ~~~~~~\Leftrightarrow~~~~~~
\begin{Compose}{0}{-2.3} \setdefaultfont{\mathsf}\setsecondfont{\mathsf}\setthirdfont{\mathsf}
\Ucircle{A}{0,0} \Ucircle{B}{-5,9} \Ucircle{C}{3,4} \Ucircle{D}{-3, 15} \Ucircle{E}{2,13}
\joincc[below left]{A}{115}{B}{-65} \csymbolalt{a}
\joincc[below]{A}{80}{C}{-90} \csymbolalt{a}
\joincc[below]{C}{170}{B}{-10} \csymbolalt{c}
\joincc[above left]{B}{25}{E}{-110} \csymbolalt{d}
\joincc[left]{B}{80}{D}{-100} \csymbolalt{a}
\joincc[above right]{E}{170}{D}{-15}\csymbolalt{c}
\joincc[right]{C}{100}{E}{-80} \csymbolalt{g}
\end{Compose}
\end{equation}
These circuits must be directed acyclic graphs (there are no closed loops).  Here $\mathsf{a}$, $\mathsf{b}$, \dots label the system types.  The integer subscripts in the symbolic notation label the individual wires.

An operation can be written in terms of an equivalent (from the point of view of probabilities) decomposition local form in terms of fiducial operations
\begin{equation}
\begin{Compose}{0}{0}\setdefaultfont{\mathsf}
\Ucircle{A}{0,0}
\thispoint{ALL}{-3,-3}\thispoint{ALR}{3,-3} \thispoint{AUL}{-3,3}\thispoint{AUR}{3,3}
\joincc{ALL}{90}{A}{-110} \csymbol{a} \joincc{ALR}{90}{A}{-70} \csymbol{b}
\joincc{A}{110}{AUL}{-90} \csymbol{a} \joincc{A}{70}{AUR}{-90}\csymbol{c}
\end{Compose}
~\equiv~
\begin{Compose}{0}{0}\setdefaultfont{\mathsf}\setsecondfont{\mathnormal}
\Ucircle{A}{0,0}
\scircle{XLL}{-4,-4}\csymbol{X}\thispoint{ALL}{-7,-7}\joincc{ALL}{90}{XLL}{-135} \csymbol{a} \joinccbw{XLL}{45}{A}{-110}\csymbolalt{a}
\scircle{XLR}{4,-4}\csymbol{X}\thispoint{ALR}{7,-7}  \joincc{ALR}{90}{XLR}{-45} \csymbol{b}
\joinccbw[above right]{XLR}{135}{A}{-70}\csymbolalt{b}
\scircle{XUL}{-4,4}\csymbol{X}\thispoint{AUL}{-7,7}  \joinccwb{A}{110}{XUL}{-45}  \csymbolalt{a}   \joincc{XUL}{135}{AUL}{-90} \csymbol{a}
\scircle{XUR}{4,4}\csymbol{X}\thispoint{AUR}{7,7}    \joinccwb{A}{70}{XUR}{-135} \csymbolalt{c} \joincc{XUR}{45}{AUR}{-90}\csymbol{c}
\end{Compose}
\end{equation}
where $a$, $b$, \dots label the fiducial elements.

We define a hopping metric
\begin{equation}
\begin{Compose}{0}{-0.1}
\vbbmatrix{h}{0,0}\csymbolthird{a}
\end{Compose}
~=~
\text{prob}\left(
\begin{Compose}{0}{-0.5} \setdefaultfont{\mathsf}\setsecondfont{\mathsf}\setthirdfont{\mathnormal}
\blackdot{d1}{0,-3} \scircle{X1}{0,0}\csymbol{X}\scircle{X2}{0,4} \csymbol{X} \blackdot{d2}{0,7}
\joincc[left]{d1}{90}{X1}{-90} \csymbolthird{a} \joincc[left]{X1}{90}{X2}{-90}\csymbolalt{a} \joincc[left]{X2}{90}{d2}{-90} \csymbolthird{a}
\end{Compose}
\right)
\end{equation}
in terms of the fiducial elements.  We can find a fiducial set of operators (acting on complex Hilbert spaces with dimension determined by the system type) such that
\begin{equation}
\begin{Compose}{0}{-0.5} \setdefaultfont{\hat}\setsecondfont{\mathsf}\setthirdfont{\mathnormal}
\blackdot{d1}{0,-3} \hatcircle{X1}{0,0}\csymbol{X}\hatcircle{X2}{0,4} \csymbol{X} \blackdot{d2}{0,7}
\joincc[left]{d1}{90}{X1}{-90} \csymbolthird{a} \joincc[left]{X1}{90}{X2}{-90}\csymbolalt{a} \joincc[left]{X2}{90}{d2}{-90} \csymbolthird{a}
\end{Compose}
~=~
\begin{Compose}{0}{-0.1}
\vbbmatrix{h}{0,0}\csymbolthird{a}
\end{Compose}
\end{equation}
This means we can write
\begin{equation}
\begin{Compose}{0}{-2} \setdefaultfont{\mathsf}\setsecondfont{\mathsf}\setthirdfont{\mathsf}
\Ucircle{A}{0,0} \Ucircle{B}{-5,9} \Ucircle{C}{3,4} \Ucircle{D}{-3, 15} \Ucircle{E}{2,13}
\joincc[below left]{A}{115}{B}{-65} \csymbolalt{a}
\joincc[below]{A}{80}{C}{-90} \csymbolalt{a}
\joincc[below]{C}{170}{B}{-10} \csymbolalt{c}
\joincc[above left]{B}{25}{E}{-110} \csymbolalt{d}
\joincc[left]{B}{80}{D}{-100} \csymbolalt{a}
\joincc[above right]{E}{170}{D}{-15}\csymbolalt{c}
\joincc[right]{C}{100}{E}{-80} \csymbolalt{g}
\end{Compose}
\equiv
\begin{Compose}{0}{-2} \setdefaultfont{\mathsf}\setsecondfont{\mathsf}\setthirdfont{\mathsf}
\Ucircle{A}{0,0} \Ucircle{B}{-7,11} \Ucircle{C}{5,6} \Ucircle{D}{-5, 18.5} \Ucircle{E}{2,15}
\joinccwbbw[below left]{A}{115}{B}{-65} \csymbolalt{a}
\joinccwbbw[below]{A}{80}{C}{-90} \csymbolalt{a}
\joinccwbbw[below]{C}{170}{B}{-10} \csymbolalt{c}
\joinccwbbw[above left]{B}{25}{E}{-110} \csymbolalt{d}
\joinccwbbw[left]{B}{80}{D}{-100} \csymbolalt{a}
\joinccwbbw[above right]{E}{170}{D}{-15}\csymbolalt{c}
\joinccwbbw[right]{C}{100}{E}{-80} \csymbolalt{g}
\end{Compose}
=
\begin{Compose}{0}{-2} \setdefaultfont{\hat}\setsecondfont{\mathsf}\setthirdfont{\mathsf}
\Ucircle{A}{0,0} \Ucircle{B}{-5,9} \Ucircle{C}{3,4} \Ucircle{D}{-3, 15} \Ucircle{E}{2,13}
\joincc[below left]{A}{115}{B}{-65} \csymbolalt{a}
\joincc[below]{A}{80}{C}{-90} \csymbolalt{a}
\joincc[below]{C}{170}{B}{-10} \csymbolalt{c}
\joincc[above left]{B}{25}{E}{-110} \csymbolalt{d}
\joincc[left]{B}{80}{D}{-100} \csymbolalt{a}
\joincc[above right]{E}{170}{D}{-15}\csymbolalt{c}
\joincc[right]{C}{100}{E}{-80} \csymbolalt{g}
\end{Compose}
\end{equation}
which gives us the probability for this circuit (as in \eqref{introcircuitexample}).

It can be shown that the necessary and sufficient conditions on the operators in this circuit are
\begin{equation}\label{physicalityconditionsA}
0\leq  \hat{B}_\mathsf{a^T_1}^\mathsf{b_2}    ~~~~\text{and}~~~  \hat{B}_\mathsf{a_1}^\mathsf{b_2} \hat{I}_\mathsf{b_2}\leq \hat{I}_\mathsf{a_1}
\end{equation}
where the superscript $\mathsf{T}$ indicates taking the partial transpose in the part of the Hilbert space associated with the given label.

We also show how to implement this for continuous dimensional Hilbert spaces.

\section{Operator Tensor Quantum Field Theory}

To go to Quantum Field Theory we consider first a grid consisting of many left moving and right moving qubits.
\begin{equation}
\begin{Compose}{0}{0}
\Cgrid{0.5}{11}{11}{0,0}
\end{Compose}
\end{equation}
At each vertex is an operation with setting $\mathbf{Q}(x)$ and outcome set $O(x)$.  Next we consider arbitrary regions
\begin{equation}
\begin{Compose}{0}{0}
\setdefaultfont{\mathsf}
\Cobject{\thetooth}{C}{1}{1}{1,-0.07}
\Cgrid[->, ultra thin]{0.5}{11}{11}{0,0}
\end{Compose}
\end{equation}
The collection of operations inside the boundary constitute an operation whose typing is determined by the wires crossed by the boundary curve.  We can associate an operator with this collection of operations.

We can associate operators with arbitrary regions of this grid.
\begin{equation}
\begin{Compose}{0}{0}
\cobjectwhite{\theblob}{A}{2}{2}{-0.2,-5} \csymbol[0,-90]{A}
\cobjectwhite{\theflag}{B}{2}{1}{-4.75,0} \csymbol[-50,0]{B}
\cobjectwhite{\theblob}{E}{1.5}{1.5}{2.15,3} \csymbol[0,50]{E}
\cobjectwhite{\thetooth}{C}{1}{1}{1.4,-1.75} \csymbol{C}
\cobjectwhite{\theblob}{D}{0.95}{0.75}{-3.8,8} \csymbol{D}
\Cgrid[->, ultra thin]{0.5}{20}{20}{0,-1}
\end{Compose}
\end{equation}
This is schematically represented by the diagram
\begin{equation}
\begin{Compose}{0}{0} \setdefaultfont{\mathsf}\setsecondfont{\mathtt}
\Ucircle{A}{1,-1} \Ucircle{B}{-5,5} \Ucircle{C}{-0.3,4} \Ucircle{D}{-3, 11} \Ucircle{E}{2,9}
\joincc[below left]{B}{-65}{A}{130} \csymbolalt{a}
\joincc[right]{A}{100}{C}{-90} \csymbolalt{b}
\joincc[below]{C}{170}{B}{-10} \csymbolalt{c}
\joincc[above left]{B}{25}{E}{-140} \csymbolalt{d}
\joincc[left]{B}{80}{D}{-100} \csymbolalt{e}
\joincc[above right]{D}{-15}{E}{170}\csymbolalt{l}
\joincc[below right]{C}{70}{E}{-95} \csymbolalt{g}
\joincc[right]{E}{-60}{A}{40}\csymbolalt{k}
\thispoint{nA}{-2,-2} \joincc[above left]{A}{-135}{nA}{45} \csymbolalt{f}
\thispoint{nB}{-8,5} \joincc[above]{B}{180}{nB}{0} \csymbolalt{h}
\thispoint{nD}{-1,13} \joincc[above left]{D}{45}{nD}{-135} \csymbolalt{i}
\thispoint{nE}{4,11} \joincc[above left]{nE}{-145}{E}{45} \csymbolalt{j}
\end{Compose}
\end{equation}
The boundary crosses some set of wires (some are inputs ($-$) into the given region and some are outputs ($+$)).  We define a typing surface, $\mathtt{a}$, to be a set of wires with $\pm$ (output or input) indicated for each wire.
\begin{equation}
\mathtt{a} = \left\{ (x, \pm): \forall x\in \text{set}(\mathtt{c}) \right\}
\end{equation}
where $x$ are the positions of the midpoints of the wires.  If this typing surface is presented as a subscript the sign of each wire is reversed (but not if it is a superscript).

We can write an operation as
\begin{equation}
\mathsf{A}_\mathtt{ab}^\mathtt{c}
\end{equation}
where $\mathtt{a}^R\cup \mathtt{b}^R\cup \mathtt{c}$ is the full set of wires in the boundary (the $R$ superscript reverses the direction of the wires).  Associated with an operation is a set of knob settings and settings
\begin{equation}
\mathbf{Q}_\mathtt{A} = \{ (x, \mathbf{Q}(x)): \forall x\in \mathtt{A} \}, ~~~~~~O_\mathtt{A} = \{ (x, O(x)): \forall x\in \mathtt{A} \}
\end{equation}
We can associate operators,
\begin{equation}
\hat{A}_\mathtt{ab}^\mathtt{c}
\end{equation}
with such operations. These act on a Hilbert space determined by the tying surfaces.

In the continuous limit (as the grid spacing goes to zero) we have an setting field, $\mathbf{Q}(x)$, and an outcome field, $O(x)$, defined over the region in question. Further, the typing surfaces become continuous surfaces.   A calculation is given by an expression such as
\begin{equation}
\begin{Compose}{0}{0} \setdefaultfont{\hat}\setsecondfont{\mathtt}
\Ucircle{A}{1,-1} \Ucircle{B}{-5,5} \Ucircle{C}{-0.3,4} \Ucircle{D}{-3, 11} \Ucircle{E}{2,9}
\joincc[below left]{B}{-65}{A}{130} \csymbolalt{a}
\joincc[right]{A}{100}{C}{-90} \csymbolalt{b}
\joincc[below]{C}{170}{B}{-10} \csymbolalt{c}
\joincc[above left]{B}{25}{E}{-140} \csymbolalt{d}
\joincc[left]{B}{80}{D}{-100} \csymbolalt{e}
\joincc[above right]{D}{-15}{E}{170} \csymbolalt{l}
\joincc[below right]{C}{70}{E}{-95} \csymbolalt{g}
\joincc[right]{E}{-60}{A}{40}\csymbolalt{k}
\thispoint{nA}{-2,-2} \joincc[above left]{A}{-135}{nA}{45} \csymbolalt{f}
\thispoint{nB}{-8,5} \joincc[above]{B}{180}{nB}{0} \csymbolalt{h}
\thispoint{nD}{-1,13} \joincc[above left]{D}{45}{nD}{-135} \csymbolalt{i}
\thispoint{nE}{4,11} \joincc[above left]{nE}{-145}{E}{45} \csymbolalt{j}
\end{Compose}
\end{equation}
This provides an operator tensor version of Quantum Field Theory.

\section{Quantum Gravity}

The main reason for pursuing an operational approach to General Relativity is to find a way to address the problem of Quantum Gravity.  In Part \ref{part:QuantumGravity} we suggest three approaches: an abstract approach, an ontological approach, and a principled approach.   We do not actually arrive at a fully fledged theory of Quantum Gravity. However we make some progress, especially on the abstract approach.

\subsection{Abstract approach}

The abstract approach is motivated by the following diagram
\begin{displaymath}
\xymatrix{ \text{OpQT} \ar[r]^{\text{GRize}} & \text{QuAGeR}  \\
\text{CProbT} \ar[u]^{\text{quantize}} \ar[r]_{\text{GRize}} & \text{PAGeR} \ar[u]_{\text{quantize}} }
\end{displaymath}
We can attempt to obtain a theory of Quantum Gravity by modifying the space of states in PAGeR in an analogous way to the modification made as we go from classical probability theory to Operational Quantum Theory.
First, since we are taking an operational approach, we need an operational space (op-space). We assume that we can nominate some scalars as corresponding to the classical level of description (this is the Heisenberg cut).  We can talk about regions, $\mathtt{A}$, $\mathtt{B}$, \dots of op-space in the same way as in the classical case (as illustrated in Fig.\ \ref{fig:opspaceintro}).

For quantization we propose a \lq\lq square root and square" approach.  We can think of the classical state space as being seeded by points, $(q,p)$, in phase space while the quantum state space is \lq\lq seeded" by points in Hilbert space. We can gain insight into this by thinking about the discrete case (though the results are not restricted to this case).  If $q$ and $p$ each only take a finite number of values, $N$ say, then $(q,p)$ takes $N^2$ possible values.  Hence, a classical mixture requires $N^2$ probabilities.  A basis for this Hilbert space is formed only from the $q$ part of the phase space.  This constitutes \lq\lq taking the square root" as $q$ only takes $N$ values.  Then we need to \lq\lq square" because we want the space of Hermitian operators acting on this Hilbert space (we require $N^2$ probabilities to specify such an operator the discrete case).  In this abstract approach to Quantum Gravity we nominate part of the boundary conditions as corresponding to the $q$ degrees of freedom - this accomplishes the \lq\lq square root" part of the quantization.  We show how to set up a space of operators associated with arbitrary parts of the op-space.  It remains to find what the correct physicality conditions are on these operators.  However, we are motivated by the following analogy
\begin{equation}
\begin{array}{lcl}
g_{\bar{\mu}\bar{\nu}}=\eta_{\bar{\mu}\bar{\nu}} ~~\text{in SR} & \longrightarrow & G_{\mu\nu} = 8\pi T_{\mu\nu}  \\
\text{QFT Physicality cond.} & \longrightarrow & \text{missing QG cond.}
\end{array}
\end{equation}
Solving this problem would give us an actual theory of quantum gravity.   A calculation in this framework looks like the following:
\begin{equation}
\begin{Compose}{0}{-1.5} \setdefaultfont{\mathsf}\setsecondfont{\mathtt}
\Ucircle{A}{1,-1} \Ucircle{B}{-5,5} \Ucircle{C}{-0.3,4} \Ucircle{D}{-3, 11} \Ucircle{E}{2,9}
\joincc[below left]{B}{-65}{A}{130} \csymbolalt{a}
\joincc[right]{A}{100}{C}{-90} \csymbolalt{b}
\joincc[below]{C}{170}{B}{-10} \csymbolalt{c}
\joincc[above left]{B}{25}{E}{-140} \csymbolalt{d}
\joincc[left]{B}{80}{D}{-100} \csymbolalt{e}
\joincc[above right]{D}{-15}{E}{170}\csymbolalt{l}
\joincc[below right]{C}{70}{E}{-95} \csymbolalt{g}
\joincc[right]{E}{-60}{A}{40}\csymbolalt{k}
\thispoint{nA}{-2,-2} \joincc[above left]{A}{-135}{nA}{45} \csymbolalt{f}
\thispoint{nB}{-8,5} \joincc[above]{B}{180}{nB}{0} \csymbolalt{h}
\thispoint{nD}{-1,13} \joincc[above left]{D}{45}{nD}{-135} \csymbolalt{i}
\thispoint{nE}{4,11} \joincc[above left]{nE}{-145}{E}{45} \csymbolalt{j}
\end{Compose}
\Longrightarrow
\begin{Compose}{0}{-1.5} \setdefaultfont{\hat}\setsecondfont{\mathtt}
\Ucircle{A}{1,-1} \Ucircle{B}{-5,5} \Ucircle{C}{-0.3,4} \Ucircle{D}{-3, 11} \Ucircle{E}{2,9}
\joincc[below left]{B}{-65}{A}{130} \csymbolalt{a}
\joincc[right]{A}{100}{C}{-90} \csymbolalt{b}
\joincc[below]{C}{170}{B}{-10} \csymbolalt{c}
\joincc[above left]{B}{25}{E}{-140} \csymbolalt{d}
\joincc[left]{B}{80}{D}{-100} \csymbolalt{e}
\joincc[above right]{D}{-15}{E}{170}\csymbolalt{l}
\joincc[below right]{C}{70}{E}{-95} \csymbolalt{g}
\joincc[right]{E}{-60}{A}{40}\csymbolalt{k}
\thispoint{nA}{-2,-2} \joincc[above left]{A}{-135}{nA}{45} \csymbolalt{f}
\thispoint{nB}{-8,5} \joincc[above]{B}{180}{nB}{0} \csymbolalt{h}
\thispoint{nD}{-1,13} \joincc[above left]{D}{45}{nD}{-135} \csymbolalt{i}
\thispoint{nE}{4,11} \joincc[above left]{nE}{-145}{E}{45} \csymbolalt{j}
\end{Compose}
\end{equation}

\subsection{Ontological approach}

The second approach is to assume that we can give Quantum Gravity an ontological underpinning coming from General Relativity.  In the first place, we could simply try to show that General Relativity actually gives rise to Quantum Theory in appropriate circumstances.  This strategy seems unlikely to work as the equations of General Relativity are local field equations and Quantum Theory violates Bell inequalities \cite{kent2009proposed}.  However, we have seen that, when construed operationally, locality in General Relativity is a much more subtle issue.  In particular, the manifold space (on which the field equations are local) is not the appropriate space for describing reality.  Further, General Relativity has a curious non-separability property once formulated in an operational way.

\subsection{Principled approach}

There has been a lot of work in the last fifteen years in reconstructing Quantum Theory from more reasonable sets of axioms, postulates, or principles.  This kind of approach requires a framework within which the principles apply.  Three possible frameworks are the operational framework (which sticks to descriptions in terms of op-space objects), the ontological framework (motivated by classical General Relativity formulated in an operational way with observables and hidden variables), and the operator tensor framework in which operators are associated with regions of op-space.  There are various principles we might attempt to impose in these frameworks such as causality, tomographic locality, or the constraint that probabilities are between 0 and 1.

\newpage

\part{Ontological and Operational Elements} \label{part:ontologicalandoperational}

\section{Introductory material on General Relativity}

An extensive introduction to General Relativity is given in Appendix \ref{appendix:standarformulationofGR}.  In this section we collect together a few extra notions that will play an important role in this paper.

\subsection{Manifolds, chartable spaces, and diffeomorphisms}\label{sec:diffeos}

We will consider trying to join solutions together.  To do this, we must first attempt to join the manifolds (and, only if we get a match, can we check to see if the fields also match).  In this subsection we look at how manifolds are defined (more generally, we are interested in manifolds with boundaries having corners).  Then we introduce the  notion of a chartable space, $\mathscr{W}$.  We will will associate separate chartable spaces, $\mathscr{W}_\mathtt{A}$, $\mathscr{W}_\mathtt{B}$,  with different regions, $\mathtt{A}$, $\mathtt{B}$, of our operational space (operational space will be defined in Sec.\ \ref{sec:WSspace}).  We can set up a fiducial identity map, $\varphi^I_{\mathtt{A}\leftarrow\mathtt{B}}$, from $\mathscr{W}_\mathtt{B}$ to $\mathscr{W}_\mathtt{A}$. This map identifies points having the same coordinates  between two chartable spaces. This map will later be useful testing to see if two manifolds match at their boundaries.

\subsubsection{Manifolds}\label{sec:manifoldsinchartablesec}

Roughly speaking, a \emph{manifold}, $\mathscr{M}$, is a set of points that is smooth and, everywhere, looks locally like $\mathbb{R}^N$. \index{manifolds}  Adding more technical detail, we say that a set, $\mathscr{M}$, is a manifold if:
\begin{enumerate}
\item It can be covered by a set of open subsets, $\mathscr{O}_i$, such that
\begin{equation}
\cup_i \mathscr{O}_i = \mathscr{M}
\end{equation}
\item For each $i$, there exists a bijection, $x_i(p)$ between the points in $\mathscr{O}_i$ and an open subset, $V_i$, of $\mathbb{R}^N$.
\item Consider the overlap regions $\mathscr{O}_i\cap\mathscr{O}_j\not=\varnothing$.  We require that the set $V_{i|j}:= x_i(\mathscr{O}_i\cap\mathscr{O}_j)\subseteq V_i$ is open for all $i$ and $j$ and that the transition map $x_j\circ x_i^{-1}$ which takes points from $V_{i|j}$ to $V_{j|i}$ (via points in $\mathscr{O}_i\cap\mathscr{O}_j$) is  $r$ times differentiable for all $i$ and $j$.
\end{enumerate}
This defines a $C^r$ manifold.   The sets, $V_i$, are called \emph{charts} (sometimes called \emph{coordinate systems}) and a collection of charts is called an \emph{atlas}.  Whitney \cite{whitney1936differentiable} proved that any $C^r$ manifold is $C^r$ equivalent to a $C^\infty$ manifold for all $r\geq 1$.   This means that if an atlas exists for which the transition maps where the charts overlap are $C^r$ differentiable (for $r\geq 1$), then there exists another atlas that is $C^\infty$ covering the same manifold and the transition maps between the charts in the old and new atlases are $C^r$ differentiable.  Any manifold that is at least differentiable (i.e.\ $C^1$) is in fact smooth.  Hence, there are only two classes of manifold - $C^0$ and $C^\infty$.

More generally, we are interested in \emph{manifolds with corners}.  These are manifolds with boundaries where those boundaries can have corners.  The definition for a manifold with corners, $\mathscr{M}$,  is, roughly speaking, that it can everywhere be locally covered by the points in $[0,\infty)^k\mathbb{R}^{N-k}$. This allows it to have corners.   There exist more technical definitions (see, for example, \cite{joyce2009manifolds}). Basically the same definition as that for a manifold but with $\mathbb{R}^N$ replaced by $[0,\infty)^k\mathbb{R}^{N-k}$.

Manifolds are a special case of manifolds with corners.  We will simply refer to either case as manifolds (whether or not they have corners).  Later we will introduce the notion of a \emph{manifold patch} - this is the set of point for which a given set of scalar fields defined on a bigger manifold are constrained to take values in some given set.  These manifold patches will be manifolds with corners.  We will be particularly interested in joining two or more such manifold patches together to make a bigger manifold patch.

\subsubsection{Chartable spaces}

\index{chartable space}

In General Relativity a solution is expressed by providing a set of fields on some manifold, $\mathscr{M}$. The manifold itself is also part of the solution - a different solution may live on a different manifold.  We will be considering mixtures (possibilistic and probabilistic) over multiple possible solutions to the field equations. These different solutions may be on different manifolds.  It is useful to have a place that these multiple manifolds live in.   To obtain this note that when we map points of a manifold (or manifold patch) into an atlas, some points $p$ may acquire coordinates from multiple charts when these charts overlap.  This motivates consideration of the following mathematical object
\begin{equation}
\breve{x}= ( x_i: i=1 ~\text{to}~ \infty )
\end{equation}
where, for each $i$,
\begin{equation}
x_i \in \mathbb{R}^N\cup\{-\}
\end{equation}
Each entry in the ordered set, $\breve{x}$, is either a $N$-tuple or the \lq\lq blank" entry $-$.  We denote by
\begin{equation}
\$ = \{ \breve{x}: \forall \breve{x} \}
\end{equation}
the set of all possible $\breve{x}$.   Now we imagine that we have a set of points, $\mathscr{W}$, that is in one to one correspondence with the set of points in $\$ $.  Thus, there exists an invertible function, $f(\cdot)$ such that $f(p)\in \$ $ for all $p\in\mathscr{W}$. We will call $f$ a \emph{super-atlas}.   We will stick with a given choice of super-atlas which we call the \emph{fiducial super-atlas}. This choice is arbitrary and we could have made a different choice.  The set, $\mathscr{W}$, is not a manifold (for example, the point $f^{-1}(-,-, -,\dots)\in \mathscr{W}$ is disconnected from other points).  We call $\mathscr{W}$ a \emph{chartable space} since, as we will see, it consists of points that can be charted by coordinates.

Any ($N$ dimensional) manifold can be regarded as a subset of $\mathscr{W}$ since we can use atlases to represent manifolds.  To see this, recall that an atlas for a manifold, $\mathscr{M}$, consists of a collection of charts, $V_i\subset \mathbb{R}^N$ (for $i\in Q$ where $Q$ is a subset of the integers), such that there exist functions
\begin{equation}
x_i(p) \in V_i ~~\text{for}~~ p\in\mathscr{O}_i  ~~ \text{else}~~x_i(p) = -
\end{equation}
where the sets $\mathscr{O}_i\subseteq \mathscr{M}$ are open and their union is $\mathscr{M}$.  We set $x_i=-$ for all $i\in \bar{Q}$.  Under this map, each point $p\in\mathscr{M}$ gets an $\breve{x}$ associated with it. We define
\begin{equation}
\$_\mathscr{M} = \left\{ \breve{x}(p): \forall~ p\in \mathscr{M} \right\}
\end{equation}
If $\mathscr{M}=\varnothing$ then $\$_\mathscr{M}=\{(-,-,-,\dots)\}$.  Otherwise every point in the manifold has at least one $N$-tuple associated with it (so $(-,-,-,\dots)$ is excluded). There will be some points, $p$, that are covered by multiple charts.  These points have $N$-tuples associated with them coming from more than one $V_i$.  Hence, $\mathscr{M}$ is mapped to $\$_\mathscr{M}\subset \$ $. Under $f^{-1}$ we can map $\$_\mathscr{M}$ into $f^{-1}(\$_\mathscr{M})\subset \mathscr{W}$. Now we make the move of identifying this with the original manifold - i.e. we set
\begin{equation}\label{MinW}
\mathscr{M} = f^{-1}(\$_\mathscr{M})
\end{equation}
As discussed above, certain smoothness properties are required where the sets $\mathscr{O}_i$ overlap.  These smoothness properties get mapped into smoothness properties in the set $\$_\mathscr{M}$.   We can, of course, choose any suitable atlas to map the points of a manifold. If we choose a different atlas we can compensate for this by choosing a different super-atlas so we get the same set of points according to \eqref{MinW}.  Note, however, that if we have two manifolds, $\mathscr{M}$ and $\mathscr{N}$, that have the point $p$ in common, then we need to chose atlases for these two manifolds that ascribe the same coordinate, $\breve{x}$, to $p$.  The full space of manifolds (including manifolds with corners) is the set of all subsets of $\mathscr{W}$ which are consistent with the definition of a manifold given above.

\subsubsection{Diffeomorphisms}

\index{diffeomorphisms}

A diffeomorphism, $\varphi$, is an isomorphism (a one-to-one and onto map that preserves the topology) between two manifolds, $\mathscr{M}$ and $\mathscr{M}'$ that is smooth.  We have
\begin{equation}
\varphi(\mathscr{M}) = \mathscr{M}'
\end{equation}
The map is invertible for the points in $\mathscr{M}'$ so $\varphi^{-1}$ exists and has the property
\begin{equation}
\varphi^{-1}(\mathscr{M}')= \mathscr{M}
\end{equation}
and, further, both $\varphi$ and $\varphi^{-1}$ are smooth (this can be understood in terms of the map on coordinates provided by an atlas).  It is possible to have a diffeomorphism which maps a manifold onto itself or a diffeomorphism that maps the manifold to a different set of points (possibly having some overlap with the original set of points).

We can consider moving manifolds $\mathscr{M}\subset \mathscr{W}$ around by diffeomorphisms.  First we map $\mathscr{M}$ into $\$ $ by $f(\cdot)$. We write the corresponding set of points in $\$ $ as $f(\mathscr{M})$. When restricted to these points, $f(\cdot)$ provides an atlas for $\mathscr{M}$.  We can now consider a different patch, $\mathscr{M}'\subset \mathscr{W}$, that is isomorphic to $\mathscr{M}$.   We can write
\begin{equation}
\mathscr{M}'= \varphi(\mathscr{M})
\end{equation}
where $\varphi$ is a \emph{diffeomorphism}.  We can write the function $\varphi$ as
\begin{equation}\label{diffeoonW}
\varphi = f^{-1} \circ F \circ f
\end{equation}
Here $F$ is a function that maps $\$ $ to itself.  It has the property
\begin{equation}
F(-,-,-,\dots) = (-, -, -, \dots)
\end{equation}
The composite function in \eqref{diffeoonW} first maps  $\mathscr{M}$ into $\$ $, (under $f$) then moves this set of points under $F$ before mapping them to the new set of points $\mathscr{M}'$ under $f^{-1}$.  We require that the function $F$ is invertible for $f(\mathscr{M})$.  It is reasonable to assume that $F$ can be extended to a one-to-one onto map (and therefore invertible) for all points in $\$ $ and that it has appropriate smoothness properties (whenever we restrict to its action on a manifold).  In this case, the map, $\varphi$, in \eqref{diffeoonW} is a diffeomorphism when acting on any manifold in $\mathscr{W}$.  The full set of such diffeomorphisms, $G_\text{diffeo}$, are generated by considering the full set of such functions, $F$.  We write
\begin{equation}
\varphi\in G^\text{diffeo}
\end{equation}
This can be used to generate all possible isomorphic manifolds in $\mathscr{W}$.

\subsubsection{Multiple chartable spaces}

We will later need to use multiple chartable spaces. We will indicate this by use of subscripts so we have $\mathscr{W}_\mathtt{A}$, $\mathscr{W}_\mathtt{B}$, \dots.  Associated with $\mathscr{W}_\mathtt{A}$ is a fiducial super-atlas, $f_\mathtt{A}(\cdot)$, so that $ \$_\mathtt{A} = f_\mathtt{A}(\mathscr{W}_\mathtt{A})$.  The elements of $ \$_\mathtt{A} $ are denoted
$\breve{x}_\mathtt{A} = (x_{\mathtt{A}i} : i=1 ~\text{to}~ \infty)$.  Diffeomorphisms acting in $\mathscr{W}_\mathtt{A}$ will be denoted $\varphi_\mathtt{A} \in G^{\text{diffeo}}_\mathtt{A}$.

We will only allow ourselves to associate points between pairs of chartable spaces, $\mathscr{W}_\mathtt{A}$ and $\mathscr{W}_\mathtt{B}$, having the same numerical values for the coordinates $\breve{x}_\mathtt{A}$ and $\breve{x}_\mathtt{B}$.  The map which accomplishes this is defined in terms of the fiducial super-atlases as follows
\begin{equation}
\varphi^I_\mathtt{B\leftarrow A} := f_\mathtt{B} I_\mathtt{B\leftarrow A}f_\mathtt{A}^{-1}
\end{equation}
where $I_\mathtt{B\leftarrow A}$ maps $\breve{x}_\mathtt{A}$ to $\breve{x}_\mathtt{B}$ having the same numerical values for the coordinates.  We will call $\varphi^I_\mathtt{B\leftarrow A}$ the \emph{fiducial identity map} \index{fiducial identity map} from $\mathscr{W}_\mathtt{A}$ to $\mathscr{W}_\mathtt{B}$.  We will use it for two purposes.  First, we will be interested in \lq\lq moving\rq\rq{} a manifold in one chartable space to another.  Second, we will be be interested in identifying points on the boundaries of manifolds in different chartable spaces. We will only allow identification of points that are related by the fiducial identity map.

\subsection{Candidate solutions, field equations, and beables}\label{sec:candidatesolutionsfieldequationsandbeables}

An extensive introduction to General Relativity is provided in Appendix \ref{appendix:standarformulationofGR}.  Here we collect a few ideas in preparation for the operational approach we will take.

In General Relativity a \emph{candidate solution} \index{solutions!candidate} is given by specifying the fields at each point on a manifold
\begin{equation}\label{Psioutout}
{\tilde{\Psi}}= \Big\{ \big(p, \pmb{\Phi}\big): \forall p\in \mathscr{M} \Big\}
\end{equation}
where $\pmb{\Phi}= (\pmb{\varphi}, \mathbf{g})$ is a list of the tensor fields for the matter ($\pmb{\varphi}$) and metric ($\mathbf{g}$) degrees of freedom.  The manifold, $\mathscr{M}$, is part of the solution - a different solution may be specified on a different manifold.

For a candidate solution to constitute an actual \emph{solution} \index{solutions!actual} it must satisfy a particular set of coupled partial differential equations (the field equations).  These consist of matter field equations that take some general form
\begin{equation}\label{Matterfieldequations}
f_l(\pmb{\varphi}, \pmb{\varphi}_{,\alpha}, \pmb{\varphi}_{,{\alpha\beta}}, g_{\mu\nu}, g_{\mu\nu,\gamma} ) = 0  ~~~~~l=1~\text{to}~L
\end{equation}
and the Einstein field equations
\begin{equation}\label{Einsteinfieldequations}
G^{\alpha\beta}(g_{\mu\nu}, g_{\mu\nu,\gamma}, g_{\mu\nu,\gamma\delta}) = 8\pi T^{\alpha\beta}(\pmb{\varphi}, \pmb{\varphi}_{,\gamma}, g_{\mu\nu}, g_{\mu\nu,\gamma})
\end{equation}
We have included the explicit functional form here. In particular, note that the matter field equations do not depend on second derivatives of the metric because of the equivalence principle.  These equations may depend on second and higher derivatives of the matter fields but, in all the main examples, they do not.

We will denote the full set of field equations, given in \eqref{Matterfieldequations} and \eqref{Einsteinfieldequations}, by
\begin{equation}
\text{FieldEqns}_{GR}
\end{equation}
The matter field equations \eqref{Matterfieldequations} are obtained from the matter field equations from Special Relativity in inertial coordinates, $x^{\bar{\mu}}$.  We do this by a process called minimal substitution (sometimes called the comma to semicolon rule) in which:  (i) the Minkowski metric, $\eta_{\bar{\mu}\bar{\nu}}$, is replaced by the general metric $g_{\mu\nu}$; (ii) partial differentiation, $\partial_{\bar{\mu}}$, is replaced by covariant differentiation, $\nabla_\mu$; (iii) all remaining inertial coordinate indices $\bar{\mu}$, $\bar{\nu}$, \dots are replaced by the corresponding indices, $\mu$, $\nu$, \dots for general coordinates.  Since these equations now depend on $g_{\mu\nu}$ we now have an extra ten degrees of freedom (for four dimensional spacetime).  Consequently, the matter field equations no longer constitute a complete set of equations and we need an extra ten field equations.  These are supplied by Einstein's field equations \eqref{Einsteinfieldequations}.

However, there is a twist in this tale.   The left hand side of Einstein's field equations are subject to an identity
\begin{equation}
\nabla_\mu G^{\mu\nu}= 0
\end{equation}
Consequently they only provide six independent equations rather than the ten we apparently need.

The reason for this is that the field equations are invariant under diffeomorphisms.  Hence, if ${\tilde{\Psi}}$ is a solution, then so is
\begin{equation}
\varphi^*{\tilde{\Psi}}= \Big\{ \big(\varphi(p), \varphi^*\pmb{\Phi}(\varphi(p))\big): \forall \varphi(p)\in \varphi(\mathscr{M}) \Big\}
                    = \Big\{ \big(p, \varphi^*\pmb{\Phi}(p)\big): \forall p\in \varphi(\mathscr{M}) \Big\}
\end{equation}
for any diffeomorphism $\varphi$.  Here $\mathscr{M}$ and $\varphi(\mathscr{M})$ are isomorphic manifolds.  When the diffeomorphism acts on a tensor field, $\mathbf{T}(p)$,  it \emph{pushes it forward} to $\varphi^*\mathbf{T}(\varphi(p))$.   If we use coordinates to represent this tensor field then this amounts to an active coordinate transformation on the tensor coordinates.

The usual resolution to the fact that both ${\tilde{\Psi}}$ and $\varphi^*{\tilde{\Psi}}$ are solutions to the field equations is to assert that they describe the same physical situation.   Physically real quantities are, therefore, functions of ${\tilde{\Psi}}$ that are invariant under diffeomorphisms.
\begin{equation}
B({\tilde{\Psi}}) = B(\varphi^*{\tilde{\Psi}}) ~~~\forall ~\varphi
\end{equation}
We will call such quantities \emph{beables}\index{beables} (pronounced be-ables).  This is John Bell's term taken from the foundations of Quantum Theory \cite{bell2004speakable}.  We reserve the term \emph{observables} (which is usually used in the General Relativity literature rather than the term beables) for a more restricted use.  In particular, given the assertion we will make in Sec.\ \ref{sec:observables} below, we will see that some beables in General Relativity are not directly observable and we will call them \emph{hidden variables}.
Since we only seek solutions up to diffeomorphisms, we have enough field equations to solve for the physics.

A big problem with beables is that they are not local on the manifold.  This is because there is no function of the fields on some subset, $\mathscr{A}\subset\mathscr{M}$, of the manifold that is a beable (since we can always find a diffeomorphism that will replace the fields on this patch with fields from elsewhere on the manifold). This makes it difficult to picture what the real physics is.  We will provide a solution to this by defining observables with respect to a different space (not the manifold). On this different space observables can be local.

We can represent candidate solutions in a diffeomorphism invariant fashion as
\begin{equation}
\Psi = G^\text{diffeo}(\tilde{\Psi}) := \{ \varphi^*\Psi: \forall \varphi\in G^\text{diffeo} \}
\end{equation}
where $G_\text{diffeo}$ is the full group of diffeomorphisms.  Thus, when we have a $\tilde{{}}$ on top of the symbol, it is represented by specifying the fields on a manifold (we can think of the $\tilde{{}}$ as suggestive of a curved manifold).  When we remove this symbol we go to a diffeomorphism invariant representation.  Representing a candidate solution by $\Psi$ is rather heavy-handed.  However, in representing it in this way we are forced to invent manipulations that are diffeomorphism invariant.  This will especially be the case when we consider composition of solutions.   Now we can write beables as
\begin{equation}
B(\Psi)
\end{equation}
Since $\Psi$ is a set and the elements of a set have no predefined order, it is necessarily the case that $B(\Psi)$ is invariant under diffeomorphisms.  This means that we can think of $\Psi$ as representing the full ontological situation (nothing added, nothing taken away).

\subsection{Fields are everything}\label{sec:fieldsareeverything}

Most objects we use in performing experiments, even experiments to test General Relativity, require Quantum Theory at some level in their description. For example, solids are solid because of the fermi-exclusion principle. The most accurate clocks we have use quantum theory.  However, we are interested in studying General Relativity by itself.  Thus, we have to suspend belief and consider only objects that can be built within the General Relativistic framework.

In fact, General Relativity only has fields.  Hence, we will insist that everything we consider is built out of fields.  For example, rather than test particles we will use small blobs of fluid.  It can be shown that, for sufficiently small blobs, it follows from the field equations that these blobs will follow geodesics (we do not need a bolt-on geodesic motion equation).  Similarly, clocks should be built out of fields.  It would actually be very difficult (and maybe impossible) to describe a typical mechanical or digital clock in terms of fields alone.  We can, however, construct a very simple clock from two miscible fluids as follows. \index{clock, two fluid} We form a spherical blob in which, initially, the first of the two miscible fluids is in a ball in the centre and the second miscible fluid forms a thick shell around the first.  As time elapses these two fluids will mix. So long as this two-fluid clock is moving inertially and is small enough that it is not subject to overly strong tidal forces, the two fluids will mix over time in a way that can be well characterized.  Thus, we can read the time off from the degree of mixing.  This clock can be described within the General Relativistic framework. Unlike a pendulum or quartz crystal clock, it does not measure time by counting some periodic motion. Also, interestingly, it also provides a time direction (the two fluids get more mixed in the forward time direction).   The two-fluid clock fails to work if it is accelerated too fast or subject to tidal forces that are too strong.  However, this is true of any clock.

It is worth commenting a little further on the attitude adopted here.  Einstein made much use of the notion of a test particle in setting up General Relativity. Further, he appended to the field equations an equation describing the motion of particles (the geodesic equation).  It would seem that he thought of this equation, and hence point particles, as a fundamental part of the theory. Later he tried to derive the geodesic equation (with Infeld and Hoffmann \cite{einstein1938gravitational}) from the other field equations by treating particles as singularities in the fields.  A more natural attitude is to regard test particles as a useful idealization in guiding the conceptual reasoning used to construct General Relativity but that should be abandoned once we actually have the theory.  A similar attitude should be taken to light rays, ideal clocks, ideal measuring rots, and so on.

\subsection{The Westman Sonego approach}\label{sec:theWestmanSonegoapproach}

\index{Westman and Sonego}

Westman and Sonego \cite{westman2008events, westman2009coordinates} suggested an approach to providing a manifestly diffeomorphism invariant formulation of General Relativity using the point coincidences of scalars. This works only when the universe is sufficiently \lq\lq messy" as we will explain.  There are, however, situations where we cannot describe all beables as corresponding to point coincidences scalars of scalars.  We will borrow from the approach of Westman and Sonego when we set up an operational approach in \ref{sec:WSspace}. We will do this by nominating a reduced set of scalars, point coincidences between which form our space of observables (quantities we directly measure).  This approach will work even when the universe is not messy in the above sense.

The approach of Westman and Sonego is best described by illustrating it with an example they give.  Consider a candidate solution
\begin{equation}\label{WSmodel}
\tilde{\Psi}= \left\{ (p, \pmb{\Phi}): \forall p\in\mathscr{M} \right\}
\end{equation}
where
\begin{equation}
\pmb{\Phi} = \left( g_{\pmb{\mu\nu}}, F_{\pmb{\mu\nu}}, (j^{\pmb{\mu}}[k]: k=1 ~\text{to} ~ K)  \right)
\end{equation}
Here $g_{\pmb{\mu\nu}}$ is the metric, $F_{\pmb{\mu\nu}}$ is the electromagnetic field, and $j^{\pmb{\mu}}[k]$ are the chemical currents (the currents associated with each of the elements in the periodic table).  Here we use bold indices, $\pmb{\mu}$, $\pmb{\nu}$, \dots, to indicate that these are abstract tensors in the sense of Penrose \cite{penrose1971applications} (we use this notation rather than the more conventional use of indices $a$, $b$, \dots, as we wish to use this latter notation later for a different purpose).  We can construct a multitude of scalars from the tensors in $\pmb{\Phi}$. For example,
For example,
\begin{equation}\label{WSthetascalars}
g_{\mu\nu} j^\mu[k] j^\nu[l], ~~~~
F_{\mu\nu}F^{\mu\nu}, ~~~~
\epsilon_{\mu\nu\rho\sigma} F^{\mu\nu}F^{\rho\sigma}
\end{equation}
and many more.  If we denote this collection of scalar fields by $\mathbf{X}=(X^A: A=1~\text{to}~K)$ (where $K$ is the number of scalars) then we can form further scalar fields as follows
\begin{equation}\label{WSscalartensors}
j^A[a] = \frac{\partial X^A}{\partial x^\mu} j^\mu[a],~~~~
g^{AB} = \frac{\partial X^A}{\partial x^\mu} \frac{\partial X^B}{\partial x^\nu} g^{\mu\nu}, ~~~~
F^{AB}= \frac{\partial X^A}{\partial x^\mu} \frac{\partial X^B}{\partial x^\mu} F^{\mu\nu} ~~~
\end{equation}
These objects are all scalars.  We denote the list of all scalars constructed (both the scalars in $\mathbf{X}$ and the scalars in (\ref{WSscalartensors}) by $\vec{\vartheta}=\{\vartheta[n]: n=1~\text{to}~N\}$.  We can think of these scalars as corresponding to the axes of a space.  We will call this the Westman Sonego space (or WS space).  The number of axes for the WS space will be much bigger than 4 (the dimension of space time) but it will not be that big.  We can now form $\vec{\vartheta}(\mathscr{M})$ by plotting the values of the scalars fields for each point in $\mathscr M$.  Note that $\vec{\vartheta}(\mathscr{M})$ is invariant under diffeomorphisms on $\mathscr M$.  Consequently it (and, indeed, any part of it) represents a beable (in the sense of Sec.\ \ref{sec:candidatesolutionsfieldequationsandbeables}).

Westman and Sonego argue that the non-degenerate situation in which  no two points on the manifold, $\mathscr M$, will have the same $\mathbf{X}$ is physically generic.  The reason for this is that universe is rather messy and so the scalars extracted from the chemical currents and the electromagnetic field will, generically, be different for every point in $\mathscr M$.  In this situation, the matrix $\frac{\partial X^A}{\partial x^\mu}$ will have rank $D$ (the dimension of $\mathscr{M}$) and also the surface, $\vec{\vartheta}(\mathscr{M})$, will be a manifold and have intrinsic dimension equal to $D$.   In this generic case, we can reconstruct a solution, as in (\ref{WSmodel}), from $\vec{\vartheta}(\mathscr{M})$.  To do this we first parameterize the surface with coordinate charts.  Then we can calculate $\frac{\partial X^A}{\partial x^\mu}$ everywhere.  Since these matrices are of rank $D$ everywhere we have enough independent equations in each of the equations in (\ref{WSscalartensors}) to invert and obtain the tensors, $g_{\mu\nu}$, $F_{\mu\nu}$, and $j^\mu[a]$ (the equations in (\ref{WSscalartensors}) can be thought of as an overcomplete set of simultaneous equations).  The coordinate charts for $\mathscr M$ can be thought of as parameterizing this surface in WS space.  Note that the solution we obtain by this \lq\lq reverse engineering" will be related to the one we started with by a diffeomorphism.

The approach of Westman and Sonego is arguably superior to other attempts to get a physical handle on the beables of General Relativity.  For example, Bergmann \cite{bergmann1961observables} suggested using four scalars to provide physical coordinates (see also Rovelli's article \cite{rovelli1991observable}). Another idea is to use GPS signals \cite{rovelli2007quantum, lloyd2012quantum}.  In this case four signals emitted from four satellites would provide a space-time coordinate at any point. If we think about what would be happening physically in the GPS case then we are actually comparing scalars (constructed from the electromagnetic fields emitted by the satellites).  These two examples are special cases that can be incorporated into what we have described above.  However, the approach of Westman and Sonego is more democratic in that it puts all scalars on an equal footing.

We are interested in pushing General Relativity beyond situations where it is usually applied and so will not restrict ourselves to this non-degenerate case. It is likely that there will be domains in which the degenerate case becomes physically generic.

\section{Operationalism and ontology}

\subsection{Experiments and their purview}\label{sec:purview}

Fundamentally, General Relativity says we have some solution,
\begin{equation}
{{\tilde{\Psi}}}_\text{fund}= \big\{ (p, \pmb{\Phi}_\text{fund}): \forall p\in \mathscr{M}_\text{fund} \big\}
\end{equation}
where $\pmb{\Phi}_\text{fund}$ are a full set of the fundamental fields at the most basic possible level of description and $\mathscr{M}_\text{fund}$ is a manifold covering the whole universe.  This must be a solution to the fundamental field equations,
\begin{equation}
\text{FieldEqns}_{GR}^\text{fund}
\end{equation}
which provide a complete set of equations for the fields $\pmb{\Phi}_\text{fund}$ up to diffeomorphisms.  We can write the solution in the diffeomorphism invariant fashion
\begin{equation}
\Psi_\text{fund} = \left\{ \varphi^* \tilde{\Psi}_\text{fund}: \forall \varphi\in G_\text{diffeo}  \right\}
\end{equation}
The \emph{fundamental beables} \index{beables!fundamental} are given by functions
\begin{equation}
B(\Psi_\text{fund})
\end{equation}
This ensures that they are invariant under diffeomorphisms.

However, when we describe an experiment we typically introduce a lot of effective structure and, further, an experiment only pertains to part of the manifold.  Thus, the object we use in our calculations for actual experiments is
\begin{equation}\label{solneff}
{\tilde{\Psi}}_\text{eff} = \big\{ (p, \pmb{\Phi}_\text{eff}): \forall p\in \mathscr{M}_\text{eff} \big\}
\end{equation}
where $\pmb{\Phi}_\text{eff}$ are the effective fields and $\mathscr{M}_\text{eff}$ is the effective manifold patch our experiment concerns.  We may have $\mathscr{M}_\text{eff}\subset \mathscr{M}_\text{fund}$ or $\mathscr{M}_\text{eff}$ may be some course-graining on a subset of $\mathscr{M}_\text{fund}$.
Henceforth, all our considerations will concern this object and so we will drop the $\text{eff}$ subscripts writing \index{Psitilde@\ensuremath{\tilde\Psi}}
\begin{equation}
{\tilde{\Psi}} = \big\{ (p, \pmb{\Phi}): \forall p\in \mathscr{M} \big\}
\end{equation}
for effective candidate solutions instead of \eqref{solneff}.  We write \index{Psi@\ensuremath{\Psi}}
\begin{equation}
\Psi = G_\text{diffeo}(\tilde{\Psi}) =  \left\{ \varphi^* \tilde{\Psi}:\forall \varphi\in G_\text{diffeo} \right\}
\end{equation}
for the gauge invariant form of the candidate solution.  These effective candidate solutions must satisfy field equations
\begin{equation}
\text{FieldEqns}_{GR}
\end{equation}
to count as actual solutions.

When we run an experiment we are interested in beables $B(\Psi)$ deduced from an effective candidate solution, $\Psi$.  These constitute the physics we are interested in. We will call them \emph{experimental beables}. \index{beables!experimental}  We include experimental apparatuses (clocks, rods, fluxometers, \dots) in with these. I.e.\ we demand that these apparatuses can be built out of the effective fields.  In running an experiment we use beables that are functions of ${{\tilde{\Psi}}}_\text{fund}$ but are not functions of the effective solution $\Psi$.  For example, the beables pertaining to the data carried along wires and recorded in a notebook or in a computer file are not themselves part of what we are measuring.  We will call these \emph{management beables} \index{beables!fundamental} - they pertain to the management of the experiment.  Since we are working in a gravitational context, we may still wish to insist that the bulk properties of the devices used to manage the experiment (such as the wires and computers) are described by the effective fields since these bulk properties may have a non-negliable influence on the experiment at hand.

We may wish to make inferences about places we cannot actually collect data from (for example the interior of a black hole).  This is reasonable as we can use the mathematics of General Relativity to make such inferences.  In this case, the effective manifold patch, $\mathscr{M}$, associated with our experiment should include these regions.  Thus, we can think of the manifold patch as being a union of $\mathscr{M}_I$ (the region of the manifold we are interested in making inferences about) and $\mathscr{M}_D$ (the region of the manifold we are able to collect data directly from).  Whether a particular point belongs to $\mathscr{M}_D$ or not depends on the way the experiment is set up.  If we want to write this explicitly in the solution we need to write
\begin{equation}\label{PsiDI}
\tilde{\Psi}_{DI} =\Big( \big\{ (p, \pmb{\Phi}): \forall p\in \mathscr{M}_D \big\}, \big\{ (p, \pmb{\Phi}): \forall p\in \mathscr{M}_I \big\} \Big)
\end{equation}
We can flatten this out to obtain
\begin{equation}
\tilde{\Psi} = \big\{ \big\{ (p, \pmb{\Phi}): \forall p\in \mathscr{M}_D\cup \mathscr{M}_I \big\}
\end{equation}
There will be consistency relations on how the solution is partitioned between $\mathscr{M}_D$ and $\mathscr{M}_I$.   This will depend on details of the management of the experiment and so will not be easily deducible from the field equations.  We will leave use of solutions such as \eqref{PsiDI} to a future paper.

Typically the purview of an experiment will concern only a certain range of the values of the observables.  We will discuss how to implement this constraint in Sec.\ \ref{sec:WSspace}.

\subsection{Observables}\label{sec:observables}

We experience the world we live in to be local.  Thus, the fact that beables cannot be local on the manifold suggests that the manifold is not the world we live in.  We need to find the appropriate space.  Our experience of the world is built out of the things we directly observe.  The general approach taken here is to define a subset of beables as \emph{observables}. \index{observables} We do not assume that one can necessarily directly observe all beables.  Those beables that are not observables will be regarded as \emph{hidden variables}. \index{hidden variables}

What are a good set of observables?   A clue to answering to this question comes from looking at a quote from Einstein \cite{einstein1916grundlage} in his famous 1916 paper on General Relativity
\begin{quote}
All our space-time verifications invariably amount to a determination of space-time coincidences (\dots) Moreover, the results of our measurings are nothing but verifications of such meetings of the material points of our measuring instruments with other material points, coincidences between the hands of a clock and points on the clock dial, and observed point-events happening at the same place and the same time.
\end{quote}
Since General Relativity is a theory of fields rather than of particles, we really need to talk in terms of coincidences between appropriate field properties. However, the components of tensor fields depend, in general, on the coordinate system used to express them.  Correspondingly, the tensor fields undergo transformations, $T\rightarrow\varphi^* T$ after a diffeomorphism.  The fortunate exception to this are scalar fields. We have to be careful though.  The value of a scalar field, $S(p)$ at a point, $p$, on the manifold is effected by a diffeomorphism. This is because, after diffeomorphism $\varphi$, the scalar field at $p$ is $S(\varphi^{-1}(p))$ (since the diffeomorphism replaces the fields at $p$ by the fields that were at $\varphi^{-1}(p)$.  Hence we cannot simply say that scalar fields at a given point, $p$, are beables.  Rather, we use the idea of point coincidences (as motivated by Westman and Sonego).  Thus, we can say that the property that scalar field $S_1(q)=s_1$ and $S_2(q)=s_2$ for some point $q$ in the manifold is invariant under diffeomorphisms.   This is clear as, after a diffeomorphism, we have $S_1(\varphi(q))=s_1$ and $S_2(\varphi(q))=s_2$.  Although the point on the manifold changes, the fact that we have a coincidence in which $S_1=s_1$ and $S_2=s_2$ at the same point does not change under a diffeomorphism.

This motivates the following assertion \index{Assertion 1}
\begin{quote}
{\bf Assertion 1:} We can only directly observe coincidences between scalars having specified values.
\end{quote}
Of course, some observables may be functions of such coincidences.  The important point is that beables not expressible in terms of scalar coincidences are not directly observable (according to this assertions).

This is an assertion.  The standard formalism of General Relativity does not come equipped with rules saying which beables are directly observables and which are not.  We will need to put this in by hand by making an assertion as to which beables are observables.  We could imagine making different assertions.  The general operational approach adopted here might still go through. Nevertheless, scalar coincidences do strongly suggest themselves as the most appropriate objects for building observables.

\subsection{Operational space}\label{sec:WSspace}

To form observables we borrow from the approach of Westman and Sonego and, in accord with Assertion 1, use coincidences between scalars.  We begin by \emph{nominating} an ordered set,
\begin{equation}
\mathbf{S}=(S_n: n=1 ~ \text{to}~ K )
\end{equation}                            \index{S@\ensuremath{\mathbf{S}}}
of scalars constructed from the metric and matter fields in $\pmb{\Phi}$ (recall these are the effective fields).  The key factor in deciding what set of scalars to use is that they should be sufficient to capture the observables we measure in the experiment under consideration.  Whether a particular particular set of scalars is a good choice or not is \emph{contingent} on the type of observer and the experiment they are considering.  Presumably humans, by and large, are similarly constructed and so will want use the same set of scalars for given types of experiment.  We call the space of possible $\mathbf{S}$ the \emph{operational space} \index{operational space} (op-space).

It is very important to note that we do \emph{not} think of the scalars in $\mathbf{S}$ as merely corresponding to some physical coordinate system (standing in for $x^\mu$) with respect to which we measure other physical quantities. This contrasts with the approach of Bergman \cite{bergmann1961observables} (in which four scalars are chosen to provide a physical reference frame) or the GPS approach of Rovelli (in which signals from GPS stations are used to provide a physical reference frame \cite{rovelli2007quantum}).   Rather, we should think of the scalars in $\mathbf{S}$ as corresponding to all the physical quantities we will be directly be measuring (in coincidence with one another).

While the approach here is motivated by Westman and Sonego there is an important difference.  Westman and Sonego attempt to list a sufficiently large set of scalars that they capture all ontological aspects of the physics in generic situations (where every point on the manifold is distinguished by the list of scalars). In this paper, on the other hand, we take an operational approach and nominate a set of scalars that are sufficient to capture our experience at the effective level.  This may be a rather restricted set of scalars. There is no need for these scalars to distinguish every point on the manifold and hence this method applies to non-generic situations.

We can plot any given candidate solution,$\tilde{\Psi}$, into op-space by plotting the values of the scalars $S_n$ (for $n=1$ to $N$) calculated from $\pmb{\Phi}$ for each point $p\in \mathscr{M}$.  This gives us
\begin{equation}
\Gamma=\{ \mathbf{S}(p): \forall p\in\mathscr{M} \}
\end{equation}                                            \index{Gamma@$\Gamma$}
We take the tensor fields in $\pmb{\Phi}$ to be smooth.  This means that the scalar fields in $\mathbf{S}$ are also smooth.  Hence, the surface $\Gamma$ will be smooth also.  It may be sufficient simply to demand that the fields in $\pmb{\Phi}$ are $r$ times differentiable where $r$ is big enough to ensure that $\mathbf{S}$ is continuous.

\begin{figure}
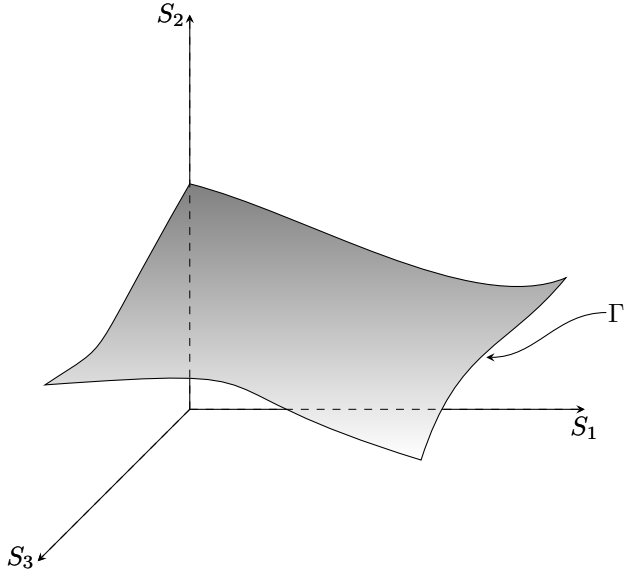
\label{fig:opspace}
\begin{Compose}{0}{0}
\Cxaxis{S_1}{0,0}{21} \Cyaxis{S_2}{0,0}{21} \Czaxis{S_3}{0,0}{21}
\csurface[0.7]{0,0}{1}
\thispoint{GS}{20, 7, 11}
\thispoint{G}{26, 9, 10}
\pointingarrow{G}{-180}{GS}{0} \csymbolalt[1,0]{\Gamma}
\Cxaxis[ultra thin, dashed]{S_1}{0,0}{21} \Cyaxis[ultra thin, dashed]{S_2}{0,0}{21} \Czaxis[ultra thin, dashed]{S_3}{0,0}{21}
\end{Compose}
\caption{The surface $\Gamma$ is plotted into operational space.}
\end{figure}

The surface, $\Gamma$ will have intrinsic dimension equal, at most, to the dimension of $\mathscr{M}$.  It may, though, pinch down to fewer dimensions in some places.  For candidate solutions, $\tilde{\Psi}$, with symmetries, $\Gamma$ will everywhere have intrinsic dimension less than that of $\mathscr{M}$.  It is important to note that $\Gamma$ is invariant under diffeomorphisms acting on $\tilde{\Psi}$.  This situation is shown in Fig.\ \ref{fig:opspace}

If we use the example of Westman and Sonego, then a possible set of scalars for observables could be those given in (\ref{WSthetascalars}).  However, we could also nominate a smaller set. For example, we could have
\begin{equation}
{\bf S} = \left( g_{\mu\nu} j^\mu[a] j^\nu[a]: \forall ~a \right)
\end{equation}
In this case the observables correspond to coincidence in the values of the rest frame energy.  Even with such a restricted set of observables, we would expect to obtain a pretty good representation of the world of chairs, tables, clocks, and apparatuses. However, we might want to use an enlarged set of observables such as
\begin{equation}
{\bf S} = \left( g_{\mu\nu} j^\mu[a] j^\nu[b]: \forall ~a, b ~\text{with}~ a\geq b \right)
\end{equation}
or even
\begin{equation}\label{Schoice3}
{\bf S} = \left( g_{\mu\nu} j^\mu[a] j^\nu[b], F_{\mu\nu}F^{\mu\nu}, \epsilon_{\mu\nu\rho\sigma} F^{\mu\nu}F^{\rho\sigma}, g^{AB}, j^A[a], F^{AB}: \forall ~a,b \right)
\end{equation}
where $g^{AB}$, $j^A[a]$ and $F^{AB}$ are the Westman Sonego scalars defined in (\ref{WSscalartensors}) defined with respect to $\mathbf{X}=\{ g_{\mu\nu} j^\mu[a] j^\nu[b]\}$ where, in this case, the index $A$ is equal to $(a,b)$ for all $a\geq b$.

If the universe is sufficiently messy then the choice in \eqref{Schoice3} allows us to directly observe Westman Sonego scalars containing all the physical information in the tensor fields in $\pmb{\Phi}$.  In the messy universe case we can formulate all of General Relativity in op-space (using the approach of Westman and Sonego).   The messy universe scenario would appear to be the physically generic case.  It is interesting, however, to explore how General Relativity looks away from this generic case.  Then we will need hidden variables in addition to observables.

As commented on in Sec.\ \ref{sec:purview}, only a certain range of values for the observables will be under the purview of any given experiment, $E$.  We incorporate this by considering a certain region, $\mathtt{U}$, of the op-space to be under the purview of the experiment $E$.  We require that $\Gamma\in\mathtt{U}$.   We will write the solution as
\begin{equation}
\tilde{\Psi}_\mathtt{U} = \left\{ (p, \pmb{\Phi}): \forall p\in \mathscr{M}_\mathtt{U} \right\}
\end{equation}
where $\mathscr{M}_\mathtt{U}$ is the part of the manifold under the purview of the experiment (and for which $\mathbf{S}$ takes only values in $\mathtt{U}$.

We will be interested in regions of op-space.  We will denote these regions by $\mathtt{A}$, $\mathtt{B}$,  $\mathtt{C}$, \dots . We must have $\mathtt{A}, \mathtt{B}, \dots \subseteq \mathtt{U}$.  Later we will consider how how to build propositions corresponding any given region, $\mathtt{A}$, of op-space.

Humans, along with the apparatuses we use in experiments, are constructed out of matter fields. Hence, it makes sense that we should construct our observables out of matter and gravity fields (as in the above examples) rather than out of purely gravitational scalars.  It is possible to define scalars just out of the metric and its derivatives (\cite{cartan1946lecons}).  Hence we can imagine creatures whose world is constructed entirely out of gravitational degrees of freedom.

\subsection{Chartable space for an op-space region}

If we have a solution,
\begin{equation}
\tilde{\Psi}_\mathtt{U} = \left\{ (p, \pmb{\Phi}): \forall p\in \mathscr{M}_\mathtt{U} \right\}
\end{equation}
then we can use it to calculate the value of $\mathbf{S}$ for each point, $p$, in the manifold.  We will write
\begin{equation}
\mathscr{M}_\mathtt{U} \subset \mathscr{W}_\mathtt{U}
\end{equation}
That is, we will associate a chartable space, $\mathscr{W}_\mathtt{U}$ with the region $\mathtt{U}$ of op-space.

For a region, $\mathtt{A}$, of op-space we have the \emph{manifold patch}
\begin{equation}
\mathscr{M}_\mathtt{A|U} = \left\{ p :\forall p\in\mathscr{M}_\mathtt{U} ~\text{s.t.} ~ \mathbf{S}(p) \in \mathtt{A} \right\}
\end{equation}
where
\begin{equation}
\mathscr{M}_\mathtt{A|U}\subseteq \mathscr{M}_\mathtt{U} \subset \mathscr{W}_\mathtt{U}
\end{equation}
If we assume that the scalar fields in $\mathbf{S}$ are continuous and the region $\mathtt{A}$ is defined by a finite set of inequalities formed from continuous functions of $\mathbf{S}$ then it is reasonable to assume that $\mathscr{M}_\mathtt{A|U}$ will be a mathematically well behaved object (a manifold with boundary having corners). This patch may consist of disjoint parts.

It turns out to be convenient (for reasons to be explained below) to associate a separate chartable space, $\mathscr{W}_\mathtt{A}$, with each region $\mathtt{A}$ we consider.
We will map $\mathscr{M}_\mathtt{A|U}$ into $\mathscr{W}_\mathtt{A}$ using the fiducial identity map $\varphi^I_\mathtt{A\leftarrow U}$.  We write
\begin{equation}
\mathscr{M}_\mathtt{A} = \varphi^I_\mathtt{A\leftarrow U} (\mathscr{M}_\mathtt{A|U})\subset \mathscr{W}_\mathtt{A}
\end{equation}          \index{MA@$\mathscr{M}_\mathtt{A}$}
Now we have a manifold patch, $\mathscr{M}_\mathtt{A}$ in the chartable space, $\mathscr{W}_\mathtt{A}$, associated with $\mathtt{A}$ that is isomorphic to $\mathscr{M}_\mathtt{A|U}$ inherited from the original solution in $\mathtt{U}$.

We will later consider how to smoothly join manifold patches associated with adjacent regions, $\mathtt{A}$ and $\mathtt{B}$, of op-space so that we get manifold patches for the composite region. We do this by identifying the appropriate points on the boundaries of $\mathscr{M}_\mathtt{A}\subset \mathscr{W}_\mathtt{A}$ and $\mathscr{M}_\mathtt{B}\subset \mathscr{W}_\mathtt{B}$.   We are guaranteed that the manifold patches will not overlap except at the points that are identified.  However, had we attempted to join $\mathscr{M}_\mathtt{A|U}\subset \mathscr{W}_\mathtt{U}$ to $\mathscr{M}_\mathtt{B|U}\subset \mathscr{W}_\mathtt{U}$ it is possible that these manifold patches could overlap in their interiors.

\subsection{Solutions for an op-space region}

\subsubsection{Solutions, $\tilde{\Psi}_\mathtt{A}$}\label{sec:solutionstildePsiA}

Op-space describes the world we observe.  It is interesting to consider candidate solutions restricted \index{restriction} to parts of op-space.  We define
\begin{equation}
\tilde{\Psi}_\mathtt{A|U} = \text{restrict}_\mathtt{A|U} (\tilde{\Psi}_\mathtt{U}) = \left\{ (p, \pmb{\Phi}(p)): \forall p\in \mathscr{M}_\mathtt{A|U} \right\}
\end{equation}
Then we define
\begin{equation}
\tilde{\Psi}_\mathtt{A} = \text{restrict}_\mathtt{A} (\tilde{\Psi}_\mathtt{U})= (\varphi^I_\mathtt{A\leftarrow U})^*\tilde{\Psi}_\mathtt{A|U}
\end{equation}
In other words, we have pushed the fields, $\pmb{\Phi}$ on $\mathscr{M}_\mathtt{A|U}$ in  $\tilde{\Psi}_\mathtt{A|U}$, forward onto $\mathscr{M}_\mathtt{A}$ using the fiducial identity map.  To avoid overly cumbersome notation, we will write this as
\begin{equation}
\tilde{\Psi}_\mathtt{A}  = \left\{ (p, \pmb{\Phi}(p)): \forall p\in \mathscr{M}_\mathtt{A} \right\}
\end{equation}                    \index{PsiAtilde@$\tilde{\Psi}_\mathtt{A}$}
it being understood from context that the fields, $\pmb{\Phi}$, have been pushed forward from $\mathscr{M}_\mathtt{A|U}$ under the fiducial identity map.

\subsubsection{Solutions, $\Psi_\mathtt{A}$}

We can write down a gauge invariant representation of a candidate solution as follows
\begin{equation}\label{PsiAdefn}
\Psi_\mathtt{A} = G^\text{diffeo}_\mathtt{A} (\tilde{\Psi}_\mathtt{A}) = \left\{  \varphi_\mathtt{A}^* \tilde{\Psi}_\mathtt{A}: \forall \varphi_\mathtt{A}\in G^\text{diffeo}_\mathtt{A} \right\}
\end{equation}               \index{PsiA@$\Psi_\mathtt{A}$}   \index{solutions!for region of op-space}
Here
\begin{align}\label{phitildePsiA}
\varphi_\mathtt{A}^* \tilde{\Psi}_\mathtt{A} &= \left\{ \left(\varphi_\mathtt{A}(p), \varphi_\mathtt{A}^*\pmb{\Phi}(\varphi(p)) \right): \forall \varphi_\mathtt{A}(p)\in \varphi_\mathtt{A}(\mathscr{M}_\mathtt{A}) \right\} \nonumber \\
&= \left\{ (p, \varphi_\mathtt{A}^*\pmb{\Phi}(p)) : \forall p\in\varphi_\mathtt{A}(\mathscr{M}_\mathtt{A}) \right\}
\end{align}
The candidate state $\Psi_\mathtt{A}$ describes the full ontological situation in region $\mathtt{A}$ of op-space.

\subsubsection{Beables for a region of op-space}

Beables pertaining to region $\mathtt{A}$ of op-space are given by functions
\begin{equation}
B_\mathtt{A}(\Psi_\mathtt{A})
\end{equation}
Note that, since we are interested in beables for region $\mathtt{A}$ of Op-space, we have the subscript $\mathtt{A}$ on the $B$.  An example of a beable is the cardinality of the set $\mathscr{M}_{\mathbf{S}|\mathtt{A}}$ (the set of points, $p\in\mathscr{M}_\mathtt{A}$, having giving rise to a particular point $\mathbf{S}$ in op-space) for some given $\mathbf{S}\in \mathtt{A}$.  If this cardinality is 0 then this particular point does not happen according to the candidate solution, $\Psi_\mathbf{A}$ and if the cardinality is 1 then the point happens only once and so on.  Another example of a beable is, given that $\mathscr{M}_{\mathbf{S}|\mathtt{A}}$ and $\mathscr{M}_{\mathbf{S}'|\mathtt{A}}$ both have cardinality of $1$, are $\mathbf{S}'$ and $\mathbf{S}$ spacelike separated or not.   The answer to this can be yes, no, or indeterminate.  If there is a spacelike path in $\mathscr{M}_\mathtt{A}$ between the point at which $\mathbf{S}$ occurs and the point at which $\mathtt{S}'$ occurs the answer is yes. If there is a forward or backward timelike path (possibly having null segments) in $\mathscr{M}_\mathtt{A}$, the answer is no. Otherwise the answer is indeterminate.

An interesting question is whether
\begin{equation}
B_\mathtt{A\cup B}(\Psi_\mathtt{A\cup B}) = f\left(B_\mathtt{A}(\Psi_\mathtt{A}), B_\mathtt{B}(\Psi_\mathtt{B}) \right)
\end{equation}
In words, can all beables for the composite region, $\mathtt{A\cup B}$, be built out of beables for the component regions $\mathtt{A}$ and $\mathtt{B}$?  If not then this would imply an interesting kind of non-separability for classical General Relativity.  We will show in Sec.\ \ref{sec:curiousnonseprability} that we do indeed see this kind of non-separability.

\subsubsection{Observables for a region of op-space}

\begin{figure}
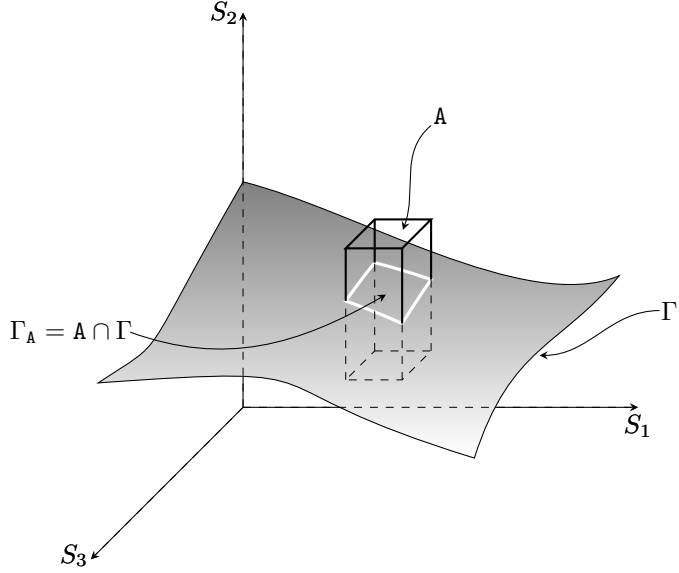
\label{fig:opspaceA}
\begin{Compose}{0}{0}
\Cxaxis{S_1}{0,0}{21} \Cyaxis{S_2}{0,0}{21} \Czaxis{S_3}{0,0}{21}
\csurface[0.7]{0,0}{1}
\ccuboid[0.05]{7,3}{3}{7}{4}{4.7}{3.8}{3}{4.2}
\thispoint{GS}{20, 7, 11}
\thispoint{G}{26, 9, 10}
\pointingarrow{G}{-180}{GS}{0} \csymbolalt[1,0]{\Gamma}
\thispoint{AO}{8.5,9.44} \thispoint{A}{10,15} \pointingarrow[above right]{A}{-140}{AO}{70} \csymbolalt{\mathtt{A}}
\thispoint{GAS}{7.6, 6} \thispoint{GA}{-6, 4} \pointingarrow[left]{GA}{-20}{GAS}{-150} \csymbolalt[-75,0]{\Gamma_\mathtt{A}=\mathtt{A}\cap\Gamma}
\Cxaxis[ultra thin, dashed]{S_1}{0,0}{21} \Cyaxis[ultra thin, dashed]{S_2}{0,0}{21} \Czaxis[ultra thin, dashed]{S_3}{0,0}{21}
\end{Compose}
\caption{An arbitrary region, $\mathtt{A}$, of operational space is shown (this need not be cuboid).  The intersection of $\mathtt{A}$ and $\Gamma$ is shown.}
\end{figure}

\index{observables!for region of op-space}

\begin{figure}
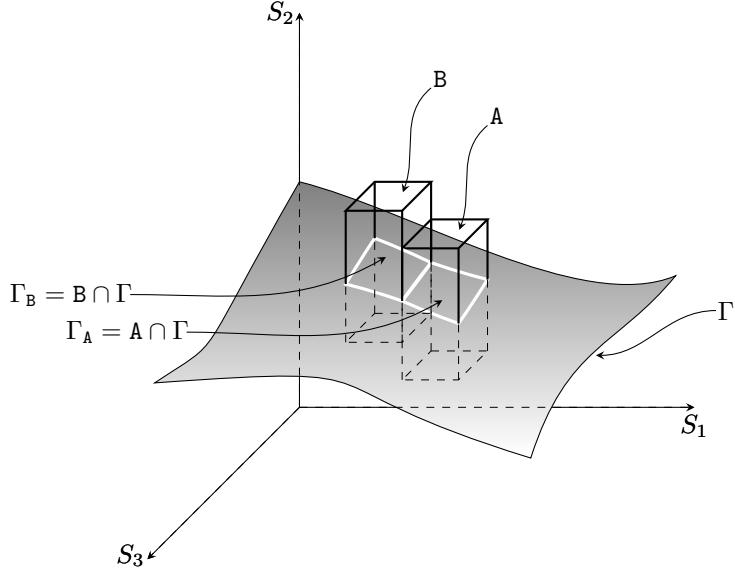
\label{fig:AandBcomposed}
\begin{Compose}{0}{0}
\Cxaxis{S_1}{0,0}{21} \Cyaxis{S_2}{0,0}{21} \Czaxis{S_3}{0,0}{21}
\csurface[0.7]{0,0}{1}
\ccuboid[-0.05]{7,3}{3}{7}{4}{4.7}{3.8}{3}{4.2}
\ccuboid[0.05]{4,5}{3}{7}{4}{4}{2.7}{2.2}{3.1}
\thispoint{GS}{20, 7, 11} \thispoint{G}{26, 9, 10} \pointingarrow{G}{-180}{GS}{0} \csymbolalt[1,0]{\Gamma}
\thispoint{AO}{8.5,9.44} \thispoint{A}{10,15} \pointingarrow[above right]{A}{-140}{AO}{70} \csymbolalt{\mathtt{A}}
\thispoint{BO}{5.5,11.44} \thispoint{B}{7,17} \pointingarrow[above right]{B}{-140}{BO}{70} \csymbolalt{\mathtt{B}}
\thispoint{GAS}{7.6, 5.7} \thispoint{GA}{-6, 4} \pointingarrow[left]{GA}{0}{GAS}{-150} \csymbolalt[-75,0]{\Gamma_\mathtt{A}=\mathtt{A}\cap\Gamma}
\thispoint{GBS}{4.6, 8} \thispoint{GB}{-9, 6} \pointingarrow[left]{GB}{0}{GBS}{-150} \csymbolalt[-75,0]{\Gamma_\mathtt{B}=\mathtt{B}\cap\Gamma}
\Cxaxis[ultra thin, dashed]{S_1}{0,0}{21} \Cyaxis[ultra thin, dashed]{S_2}{0,0}{21} \Czaxis[ultra thin, dashed]{S_3}{0,0}{21}
\end{Compose}
\caption{Composition of $\mathtt{A}$ and $\mathtt{B}$ at a typing surface $\mathtt{a}$ (not labeled in figure)}.
\end{figure}

Observables pertaining to a region $\mathtt{A}$ of op-space must be given by some function of $\Psi_\mathtt{A}$. Every observable must also be a beable (but not necessarily vice versa).  In fact, observables pertaining to $\mathtt{A}$ must be functions of $\Gamma_\mathtt{A}$ (the set of $\mathbf{S}$ calculated from $\Psi_\mathtt{A}$ - see Fig.\ \ref{fig:opspaceA}). Thus, observables have the form
\begin{equation}
O_\mathtt{A}(\Gamma_\mathtt{A})
\end{equation}
Hence, if we have a composite region, $\mathtt{A\cup B}$, of op-space (see Fig.\ \ref{fig:AandBcomposed}) then the observables are given by
\begin{equation}
O_\mathtt{A\cup B}(\Gamma_\mathtt{A\cup B})= O_\mathtt{A\cup B}(\Gamma_\mathtt{A}\cup \Gamma_\mathtt{B})
\end{equation}
Clearly, then, we can write
\begin{equation}
O_\mathtt{A\cup B}(\Gamma_\mathtt{A\cup B})= g(O_\mathtt{A}(\Gamma_\mathtt{A}), O_\mathtt{B}(\Gamma_\mathtt{B}))
\end{equation}
This means, in contrast to the situation with beables, we do have separability for observables in General Relativity.

\subsubsection{The restriction operation}

We used the restriction operation for obtaining $\tilde{\Psi}_\mathtt{A}$ from $\tilde{\Psi}_\mathtt{U}$.  We can define it more generally for obtaining
$\tilde{\Psi}_\mathtt{A}$ from $\tilde{\Psi}_\mathtt{B}$ whenever $\mathtt{A}\subseteq \mathtt{B}$.  Then
\begin{equation}
\tilde{\Psi}_\mathtt{A} = \text{restrict}_\mathtt{A}(\tilde{\Psi}_\mathtt{B}) = (\varphi_\mathtt{A\leftarrow B}^I)^* \tilde{\Psi}_\mathtt{A|B}
\end{equation}          \index{restriction}
where these definitions are obtained by replacing $\mathtt{U}$ with $\mathtt{B}$ for the definitions in Sec.\ \ref{sec:solutionstildePsiA}.

We can also write
\begin{equation}
{\Psi}_\mathtt{A} = \text{restrict}_\mathtt{A}({\Psi}_\mathtt{B})
\end{equation}
where the restrict operation acts on each $\tilde{\Psi}_\mathtt{B}\in \Psi_\mathtt{B}$.

\subsubsection{Turning a solution inside out}\label{sec:turningasolutioninsideout}

We can write a candidate solution, $\tilde{\Psi}_\mathtt{A}$, in an alternative form as
\begin{equation}\label{PsiAinsideout}
\tilde{\Psi}_\mathtt{A} = \left\{ (\mathbf{S}, \tilde{\lambda}): \forall \mathbf{S}\in \Gamma_\mathtt{A} \right\}
\end{equation}  \index{solutions!inside out form}
where $\Gamma_\mathtt{A}$ is the set of points realized by the candidate solution in op-space inside the region $\mathtt{A}$ and $\tilde{\lambda}$ are hidden variables to be defined.

To define these hidden variables first we define $\omega(p)$ such that
\begin{equation}
\pmb{\Phi} \leftrightarrow (\mathbf{S}(p), \omega(p))
\end{equation}
The idea here is that if we know $\pmb{\Phi}(p)$ then we can calculate $\mathbf{S}(p)$, but in doing so we may lose information.  We arrange so that this information goes into the variables $\omega(p)$ (these can be any kind of variables that will do this job - not necessarily a list of tensors). Hence, if we have $(\mathbf{S}(p), \omega(p))$ we can calculate $\pmb{\Phi}$.  We can trivially arrange this by putting $\omega=\pmb{\Phi}$ but, in particular examples, we may be able to make more efficient choices.

Now we define
\begin{equation}
\tilde{\lambda}= \left\{ (p, \omega): \forall p\in \mathscr{M}_{\mathbf{S}|\mathtt{A}} \right\}
\end{equation}
where $\mathscr{M}_{\mathbf{S}|\mathtt{A}}$ is the set of points $p\in\mathscr{M}_\mathtt{A}$ which map into the point $\mathbf{S}$ of Op-space for this candidate solution.  We put a tilde on top of the $\lambda$ as this object is not invariant under diffeomorphisms.

Clearly we can interconvert between the form of the solution given in \eqref{PsiAinsideout} and the outside out form
\begin{equation}
\tilde{\Psi}_\mathtt{A}= \left\{ (p, \pmb{\Phi}): \forall p\in \mathscr{M}_\mathtt{A} \right\}
\end{equation}
For this reason we will use the same symbol, $\tilde{\Psi}_\mathtt{A}$, to denote the two forms.

One great advantage of the inside out form is that diffeomorphisms do not effect $\mathbf{S}$ or $\Gamma_\mathtt{A}$. Only the hidden variables, $\tilde{\lambda}$,  in \eqref{PsiAinsideout} are effected.  The inside out form also provides an interesting way to think about the solution. It corresponds to a surface, $\Gamma_\mathtt{A}$, in Op-space in which each point, $\mathbf{S}\in\Gamma_\mathtt{A}$,  is decorated by hidden variables, $\tilde{\lambda}$.  We can go one step further and define
\begin{equation}
\lambda = \{ \varphi^*\tilde{\lambda}: \forall \varphi \}
\end{equation}
In order to calculate $\varphi^*\tilde{\lambda}$ we need to determine the effect of a diffeomorphism on $\omega(p)$.  Since $\omega{p}$ may not be a list of tensors we need to look at the effect of the diffeomorphism on $\pmb{\Phi}$ (on which $\omega(p)$ depends).  Now we can write
\begin{equation}\label{PsiAinsideoutnotildes}
\Psi_\mathtt{A} = \left\{ (\mathbf{S}, \lambda): \forall \mathbf{S}\in \Gamma_\mathtt{A} \right\}
\end{equation}
This way of representing the solution is diffeomorphism independent.  Furthermore, we have decomposed the solution into a bit that is observable and a bit that is hidden variables.

\subsection{Degeneracy in Op-Space}\label{sec:degeneracyinopspace}

For sufficiently \lq\lq messy" solutions, $\tilde{\Psi}_\mathtt{U}$, every point in the manifold will have a different value of $\mathbf{S}$ (assuming that we have an appropriate list of scalars in $\mathbf{S}$). This is the non-degenerate case.  The surface, $\Gamma_\mathtt{U}$, will have intrinsic dimension equal to that of the manifold.  At a suitable level of course-graining it is reasonable to imagine that every point on the manifold has different $\mathbf{S}$.  For example, the view from a particular point will be given by scalar quantities extracted from the electromagnetic fields at that point.   Generically the view from every point is different.  We know where we are by looking to see what is around us (see acknowledgements).  The non-degenerate case is where the Westman Sonego approach works well.  In this case, some of the scalars in $\mathbf{S}$ can act as a kind of physical coordinate system and other scalars in this list can be tensor-like (see Sec.\ \ref{sec:theWestmanSonegoapproach}).

It turns our that some particularly interesting phenomena happen in the degenerate case.  If we do not course-grain sufficiently then it is likely that there will be different points on the manifold having the same $\mathbf{S}$. This could also happen if we have a restricted set of scalars in $\mathbf{S}$ or if we have some symmetries.

The question that naturally arises is how should we understand the situation in which different points in the manifold have the same $\mathbf{S}$?  Should we think of two such points as actually corresponding to the same point?  Or should we think of them as two different places in which we happen to have the same observations?  We do not have to answer such (important) metaphysical questions as long as we know how to work operationally with such situations.

There are two points worth making that address this issue.   First, recall that we are here only talking about the experimental beables. We also have management beables. These are aspects of the management of the experiment that do not get included in the solution, $\Psi_\mathtt{U}$ (see Sec.\ \ref{sec:purview}).  Observables extracted from the management beables may act to distinguish two points on the manifold that have the same $\mathbf{S}$.  For example, we may write down the data on pieces of paper and these pieces of paper may have distinguishing marks.  Or, even if the pieces of paper do not have distinguishing marks, they may arrive at the data collection station at different times.  Hence, we then know that we have two separate instances of $\mathbf{S}$.  The problem with this is that it suggests we need to enrich our framework to include information about whether management data is distinguishing such points or not.  This brings us to the second point.  If we have degeneracy there will be at least some points in which the surface $\Gamma_\mathtt{U}$ has lower intrinsic dimension lower than of the manifold.  Hence, at any point $\mathbf{S}$, we know from the local properties of this surface whether there are multiple $\mathbf{S}$ or not. In general, of course, there can be a continuum of points in the manifold that have the same $\mathbf{S}$ and this is witnessed locally on $\Gamma_\mathtt{U}$ at any given point, $\mathbf{S}$.  Given this property, we do not need to enrich our framework to include information coming from the management beables as to whether a given $\mathbf{S}$ is happening multiple times or not as we can just measure the local properties of $\Gamma_\mathtt{U}$ at any point on its surface.

\subsection{Blobs and hidden variables}

It is interesting to explore the content of Assertion 1 (from Sec.\ \ref{sec:observables}).  Consider a solution which has four small fluid blobs, $1$, $2$, $3$, and $4$ \index{blobs} but no other matter. We assume that we form these fluids from different types of matter.  Assume that the solution for the metric is $g_{\mu\nu}=\eta_{\mu\nu}$ everywhere (this is possible since the blobs are small).   Assume that blobs $1$ and $2$ intersect once and blobs $3$ and $4$ intersect once.  Let scalars defining op-space be
\begin{equation}
{\bf S}= (\rho_1, \rho_2, \rho_3, \rho_4)
\end{equation}
where $\rho_i(p)$ is the rest frame energy of blob $i$ at point $p$.
Then, in op-space, we will witness the intersection of blobs $1$ and $2$ (since there will be points, $\mathbf{S}$, where $\rho_1$ and $\rho_2$ are both non-zero).  We will also witness the intersection of blobs $3$ and $4$ in op-space.  We can, hence, construct observables for these events.

Now we can ask what the extremal invariant distance, $s$, is between these two blob intersection events (see Sec.\ \ref{Sec:resolution}).  There is no way of reading this off the op-space.  We could attempt to supplement our observable scalar set so that we have enough scalars that we can form an observable corresponding to the distance.  However, there is no way to do this.  The only field present at points on the manifold where there are no blobs is the metric.  Since the metric is equal to $\eta_{\mu\nu}$ everywhere, all scalars we can calculate at these points (for example by forming derivatives of the metric and contracting) are also equal everywhere.  Hence, we cannot form an observable from points in the
op-space which is equal to this distance.  On the other hand, we can write down an expression for the distance (see equation (\ref{extremaldistance})) and this expression will take the same values under diffeomorphisms. Hence, the invariant distance is a beable but not an observable - it is a hidden variable. \index{hidden variables}

At this stage one might object that we have arbitrarily ruled out some beables from constituting observables.  After all, we can measure $s$.  However, if we try to imagine any experiment by which we actually attempt to measure $s$ when we will have to introduce extra physical objects.  For example, we could fire a small clock out from one intersection point so that it moved along a geodesic. If this clock happens to arrive at the other intersection point then the time elapsed would equal the invariant distance $s$.   However, if we do this, we are in a different situation than the one just described.  Nevertheless, it is interesting to see how to build an observable out an observable scalar set corresponding to such a time measurement.  To do this we first need to build a clock out of scalars.  We could imagine doing this by forming a small blob from two miscible fluids, $5$ and $6$ (of different types of matter from those of the first four blobs).  Initially the two fluids in this clock blob are arranged so that fluid $5$ forms a ball at the centre and fluid $6$ forms a thick shell round this ball. As time elapses in the local inertial reference frame of the blob the two fluids will mix in some way we can calculate.  This will be witnessed in the detail of the coincidence between the two scalars, $\rho_5(p)$ and $\rho_6(p)$ and hence we can read off a time from the detail of coincidences in $\rho_5$ and $\rho_6$.  Hence, our two-fluid blob constitutes a rather primitive clock.   We nominate our observable scalar set to be
\begin{equation}
{\bf S}= (\rho_1, \rho_2, \rho_3, \rho_4, \rho_5, \rho_6)
\end{equation}
We can deduce the time elapsed on the clock as it goes from one intersection point to the other by looking at the degree of mixing of the two clock fluids as seen in the coincidences of the scalars $\rho_5$ and $\rho_6$.

This example illustrates the content of Assertion 1.  Beables might not be readable off an op-space, but if we imagine an actual experiment to measure them then we have to introduce fields associated with the measurement apparatuses.    On doing this we will be able to read the values of these observables off the op-space.  Assertion 1 guarantees a certain locality.  While an observable (such as the example we just discussed) may pertain to different parts of the op space, it is built out of local properties in the op-space.  The claim is that all instruments (clocks, measuring rods, particle detectors, \dots ) ultimately work by looking at coincidences in the values of scalar quantities.

\subsection{Spaces of solutions}

\subsubsection{Types of solution}

We are interested in different types of solution sets associated with some arbitrary region, $\mathtt{A}$ of op-space.  For a solution, $\Psi_\mathtt{A}$, to be associated with some region, $\mathtt{A}$, of op-space, every element $\tilde{\Psi}_\mathtt{A}\in\Psi_\mathtt{A}$ must have the property that the $\Gamma_\mathtt{A}$ set calculated from it is a subset of $\mathtt{A}$.

So far the solutions, $\Psi_\mathtt{A}$, we have considered have the property that there exists a diffeomorphism between any two elements $\tilde{\Psi}_\mathtt{A}\in\Psi_\mathtt{A}$.  We will call such solutions \emph{pure} (because all elements are physically equivalent).  If we take the union of any number of distinct pure solutions for some region then we get a \emph{mixed solution}.  Mixed solutions crop up naturally when we join solutions together at some surface.

Another distinction we are interested in is between \emph{actual solutions} (which we will just call \emph{solutions}) and \emph{candidate solutions}.  We require that both candidate solutions and actual solutions have the property that, for each element, $\tilde{\Psi}_\mathtt{A}\in\Psi_\mathtt{A}$, the fields can be differentiated $r$ times where we choose $r$ to be such that the surface $\Gamma_\mathtt{U}$ is at least continuous.  This guarantees that closed regions of op-space will correspond to closed regions of the manifold (though possibly consisting of disjoint parts).  We may wish that $\Gamma_\mathtt{U}$ is, itself, differentiable some number of times.  A candidate solution is something we can check to see if it is an actual solution.

We will say that $\Psi_\mathtt{A}$ is a solution (an actual solution) to $\text{FieldEqns}_{GR}$ if all elements, ${\tilde{\Psi}}_\mathtt{A}$, of $\Psi_\mathtt{A}$ are solutions for all points $p\in \mathscr{M}_\mathtt{A}$ in the manifold patch associated with this ${\tilde{\Psi}}_\mathtt{A}$.  For pure solutions, it is sufficient to demand that any single element of $\Psi_\mathtt{A}$ to be a solution of $\text{FieldEqns}_{GR}$ since then all other elements of $\Psi_\mathtt{A}$ will also be solutions (as $\text{FieldEqns}_{GR}$ are invariant under diffeomorphisms).

We have four types of solution: pure candidate solutions, pure solutions, mixed candidate solutions, and mixed solutions.  We denote the sets of such solution types for some given region, $\mathtt{A}$, by
\begin{equation}
\Omega_\mathtt{A}[\text{spec}]
\end{equation}    \index{Omega@$\Omega_\mathtt{A}$}  \index{solutions!spaces of}
where $\text{spec}$ denotes the type of solutions set.  For pure solutions, we omit spec. For pure candidate solutions we write spec as \lq\lq cand".

\subsubsection{Pure solutions}\label{sec:puresolutions}

Let
\begin{equation}
\Omega_\mathtt{A}[\text{cand}]
\end{equation}
be the set of all pure candidate solutions, $\Psi_\mathtt{A}$, associated with region $\mathtt{A}$ of op-space.  \index{solutions!pure}

Let
\begin{equation}
\Omega_\mathtt{A} \subseteq \Omega_\mathtt{A}[\text{cand}]
\end{equation}
be the set of pure solutions, $\Psi_\mathtt{A}$, to $\text{FieldEqns}_{GR}$ for region $\mathtt{A}$.

There is an interesting subtlety here.  To illustrate this we can define
\begin{equation}
\Omega_\mathtt{A|| C}\subseteq \Omega_\mathtt{A}
\end{equation}
where $\mathtt{A}\subseteq \mathtt{C}$,
such that
\begin{equation}
\Psi_\mathtt{A}\in \Omega_\mathtt{A|| C}
\end{equation}
if and only if there exists
\begin{equation}
\Psi_\mathtt{C} \in \Omega_\mathtt{C}
\end{equation}
such that
\begin{equation}
\Psi_\mathtt{A} = \text{restrict}_\mathtt{A}(\Psi_\mathtt{C})
\end{equation}
Clearly
\begin{equation}
\Omega_\mathtt{A|| C} \subseteq \Omega_\mathtt{A}
\end{equation}
since the definition of $\Omega_\mathtt{A\vert C}$ is more restrictive.  The interesting question is whether
\begin{equation}
\Omega_\mathtt{A|| C} = \Omega_\mathtt{A}
\end{equation}
for all $\mathtt{C}$. This might not be true because, even though a solution ${\tilde{\Psi}}_\mathtt{A}$ might satisfy $\text{FieldEqns}_{GR}$ on a patch $\mathscr{M}_\mathtt{A}$, it might be impossible to complete it into a solution over the bigger patch $\mathscr{M}_\mathtt{C}$.

\subsubsection{Mixed solutions}

It is natural to define a notion of   mixed candidate solutions.  These are the union of any number of pure candidate solutions.
\begin{equation}
\Psi_\mathtt{A} = \bigcup_l \Psi_\mathtt{A}[l]
\end{equation}                                    \index{solutions!mixed}
We call these   mixed since the different $\Psi_\mathtt{A}[l]$ correspond to different physical situations.

Given a mixed candidate solution, $\Psi_\mathtt{A}$, we can apply the sort operation defined as \index{sort!for solutions}
\begin{equation}
\text{sort}\left(\bigcup_l \Psi_\mathtt{A}[l] \right) = \left\{ \Psi_\mathtt{A}[l]: \forall l \right\}
\end{equation}
The sort operation separates out the physically different pure candidate solutions.  In practise, it would be very difficult to do this calculation.  However, the operation is well defined.

It is useful to define the inverse sort operation also.  In fact, this is the flatten operation of set theory.
\begin{equation}
\text{flatten}\left( \left\{ \Psi_\mathtt{A}[l]: \forall l \right\}  \right) = \bigcup_l \Psi_\mathtt{A}[l]
\end{equation}          \index{flatten}
When applied to sets of pure states, the flatten operation can be inverted by the sort operation.

The space of   mixed candidate solutions is
\begin{equation}
\Omega_\mathtt{A}[\text{mixed cand}]
\end{equation}
consists of all $\Psi_\mathtt{A}$ such that all elements of $\text{sort}(\Psi_\mathtt{A})$ are in $\Omega_\mathtt{A}[\text{cand}]$ (the space of   pure candidate solutions).

The space of mixed solutions is
\begin{equation}
\Omega_\mathtt{A}[\text{mixed}]
\end{equation}
consists of all $\Psi_\mathtt{A}$ such that all elements of $\text{sort}(\Psi_\mathtt{A})$ are in $\Omega_\mathtt{A}$ (the space of   pure solutions).

\subsubsection{Special solutions}\label{sec:specialsolutions}

\paragraph{The empty solution:}

One special solution is the empty set
\begin{equation}
{\tilde{\Psi}}_\mathtt{A}[\varnothing] = \varnothing
\end{equation}
This is generated if $\mathscr{M}_\mathtt{A}=\varnothing$ for this solution since then there are no points, $p$, corresponding to points in $\mathtt{A}$.  Then we have
\begin{equation}\label{emptysolution}
\Psi_\mathtt{A}[\varnothing] = \{ \varnothing \}
\end{equation}
We call this the \emph{empty solution}. \index{solutions!empty} Given the above definition of $\Omega_\mathtt{A}$, it is clear that
\begin{equation}
\Psi_\mathtt{A}[\varnothing] \in \Omega_\mathtt{A}  ~~\forall ~\mathtt{A}
\end{equation}
since $\text{FieldEqns}_{GR}$ are trivially satisfied for all points in $\mathscr{M}_\mathtt{A}=\varnothing$.  Furthermore, this generates $\Gamma_\mathtt{A}=\varnothing$ which is a subset of $\mathtt{A}$ as required.  Note, incidentally, that this fact does not imply the existence of a non-empty solution $\Psi_\mathtt{U}$ such that $\text{restrict}_\mathtt{A}(\Psi_\mathtt{U})=\Psi_\mathtt{A}[\varnothing]$.  For small enough regions, $\mathtt{A}$ of op-space, empty solutions will be the generic case.

\paragraph{The null solution}

Another special solution is
\begin{equation}\label{nullsolution}
\Psi_\mathtt{A}[\mathbf{0}] = \varnothing
\end{equation}
We call this the \emph{null solutions}.  \index{solutions!null}
This is a very different object from the empty solution (compare \eqref{nullsolution} with \eqref{emptysolution}).  $\Psi_\mathtt{A}[\mathbf{0}]$ corresponds to the empty set of solutions so no solution is given. Since some situation must always pertain (even if it is only the empty solution) the null solution can never actually describe the physical situation.  It is useful to have notation for this solution though.  When we introduce the language of propositions we will see that the null proposition corresponds to the statement that is always false.

\paragraph{Deterministic solution}

Another interesting solution is the solution that is always true.   We call this the deterministic solution.  In fact we can define the deterministic candidate solution and the deterministic solution (these are two different things).

The \emph{deterministic candidate solution} for a region $\mathtt{A}$ of op-space is defined as
\begin{equation}
\Psi_\mathtt{A}[\text{det}, \text{cand}] = \text{flatten}(\Omega_\mathtt{A}[\text{cand}])
=\bigcup_{\Psi_\mathtt{A}\in \Omega_\mathtt{A}[\text{cand}] }\Psi_\mathtt{A}
\end{equation}
This is a mixture over all possible candidate solutions in this region.

The \emph{deterministic solution} \index{solution!deterministic} for region $\mathtt{A}$ is
\begin{equation}
\Psi_\mathtt{A}[\text{det}] = \text{flatten}(\Omega_\mathtt{A})
=\bigcup_{\Psi_\mathtt{A}\in \Omega_\mathtt{A} } \Psi_\mathtt{A}
\end{equation}
This is a property of General Relativity (or whatever diffeomorphism invariant theory we are looking at).  We can be sure that the actual situation is described by a one pure solution in $\text{sort}(\Psi_\mathtt{A}[\text{det}])$.

It is possible that the only solution in some regions of op-space is the empty solution (indeed, for small enough $\mathtt{A}$, this will be the generic situation).  For such regions we have
\begin{equation}
\Psi_\mathtt{A}[\text{det}] = \{\varnothing\} ~~~ \text{iff}~~~ \Omega_\mathtt{A} = \{ \Psi_\mathtt{A}[\varnothing] \}
\end{equation}
In such regions we can be sure that nothing will happen.

\paragraph{Given $\Gamma_\mathtt{A}$ solution}

The most fine-grained observations we can make in a region $\mathtt{A}$ of op-space correspond to sets, $\Gamma_\mathtt{A}\in\mathtt{A}$.  This would correspond so seeing all $\mathbf{S} \in \Gamma_\mathtt{A}$ and \emph{not} seeing any other $\mathbf{S}\in \mathtt{A}$.  We can define the candidate solution consistent with this situation as
\begin{equation}
\Psi_\mathtt{A}[\Gamma_\mathtt{A}, \text{cand}] =\text{flatten}(\Omega_\mathtt{A}[\Gamma_\mathtt{A}, \text{cand}])= \bigcup_{\Psi_\mathtt{A}\in \Omega_\mathtt{A}[\Gamma_\mathtt{A}, \text{cand}]} \Psi_\mathtt{A}
\end{equation}
This is the union of all pure candidate solutions belonging to the set
\begin{equation}
\Omega_\mathtt{A}[\Gamma_\mathtt{A}, \text{cand}] \subseteq \Omega_\mathtt{A}[\text{cand}]
\end{equation}
of pure candidate solutions having the given $\Gamma_\mathtt{A}$.

We can also define the mixed solution consistent with this observation as
\begin{equation}
\Psi_\mathtt{A}[\Gamma_\mathtt{A}] =\text{flatten}(\Omega_\mathtt{A}[\Gamma_\mathtt{A}]) = \bigcup_{\Psi_\mathtt{A}\in \Omega_\mathtt{A}[\Gamma_\mathtt{A}]} \Psi_\mathtt{A}
\end{equation}                   \index{PsiAGamma@$\Psi_\mathtt{A}[\Gamma_\mathtt{A}]$}
where this is the union of all pure solutions belonging to the set
\begin{equation}
\Omega_\mathtt{A}[\Gamma_\mathtt{A}] \subseteq \Omega_\mathtt{A}
\end{equation}
of pure solutions having the given $\Gamma_\mathtt{A}$.  If $\Omega_\mathtt{A}[\Gamma_\mathtt{A}]$ is empty then $\Psi_\mathtt{A}[\Gamma_\mathtt{A}] = \emptyset$ (the null solution).
It can be the case that $\Psi_\mathtt{A}[\Gamma_\mathtt{A}]$ is pure. This will happen if we have a non-degenerate situation so $\Gamma_\mathtt{A}$ has the dimension of the manifold (as discussed in Sec.\ \ref{sec:degeneracyinopspace}) and we have a set of scalars in $\mathbf{S}$ that is sufficient to determine all the physical fields in $\mathtt{A}$ up to diffeomorphisms (as in the Westman Sonego approach).  This would appear to be the world we live which is sufficiently messy that every point can be expected to have different scalar field coincidences.   Thus, in General Relativity, messiness induces purity. We might contrast with the case of Quantum Theory where we normally think of messiness inducing mixed states.  There is, however, a parallel in Quantum Theory with this phenomena.  This is that when we make a maximal measurement we induce a pure state. Likewise here, when we have a fine-grained observation, we induce purity.  We should note that in Quantum Theory we are talking about probabilistic mixtures rather than logical mixtures. Later we will talk about probabilistic mixtures in the context of General Relativity.

\paragraph{Operational solutions}

We may make course-grained observations of observables in $\mathtt{A}$.  Such course-grained observables are associated with a set, $O_\mathtt{A}$, of possible $\Gamma_\mathtt{A}$ sets consistent with the observation.  We define the \emph{candidate operational solution} associated with $O_\mathtt{A}$ as
\begin{equation}
\Psi_\mathtt{A}[O_\mathtt{A}, \text{cand}] =\text{flatten}(\Omega_\mathtt{A}[O_\mathtt{A}, \text{cand}])= \bigcup_{\Psi_\mathtt{A}\in  \Omega_\mathtt{A}[O_\mathtt{A}, \text{cand}]} \Psi_\mathtt{A}
\end{equation}
where
\begin{equation}
\Omega_\mathtt{A}[O_\mathtt{A}, \text{cand}] = \bigcup_{\Gamma_\mathtt{A}\in O_\mathtt{A}} \Omega_\mathtt{A}[\Gamma_\mathtt{A}, \text{cand}]
\end{equation}
is the set of all pure candidate solutions consistent with $O_\mathtt{A}$.  We let
\begin{equation}
\Omega_\mathtt{A}[\text{op cand}]
\end{equation}
be the set of all candidate operational solutions in $\mathtt{A}$.

Similarly, we define the \emph{operational solution} \index{solutions!operational} associated with $O_\mathtt{A}$ as
\begin{equation}
\Psi_\mathtt{A}[O_\mathtt{A}] =\text{flatten}(\Omega_\mathtt{A}[O_\mathtt{A}])= \bigcup_{\Psi_\mathtt{A}\in  \Omega_\mathtt{A}[O_\mathtt{A}]} \Psi_\mathtt{A}
\end{equation}
where
\begin{equation}
\Omega_\mathtt{A}[O_\mathtt{A}] = \bigcup_{\Gamma_\mathtt{A}\in O_\mathtt{A}} \Omega_\mathtt{A}[\Gamma_\mathtt{A}]
\end{equation}
is the set of all pure solutions consistent with $O_\mathtt{A}$.
These are operational solutions consistent with the field equations. We let
\begin{equation}
\Omega_\mathtt{A}[\text{Op}]
\end{equation}
be the set of all operational solutions in $\mathtt{A}$.

Given $\Gamma_\mathtt{A}$ solutions are special cases of operational solutions. We will say that they are \emph{operationally pure} \index{solutions!operationally pure} since they cannot be regarded as a mixture of other solutions in the set of operational solutions.   We let
\begin{equation}\label{Omegaoppure}
\Omega_\mathtt{A}[\text{op pure}]
\end{equation}
be the set of all operationally pure solutions in $\mathtt{A}$.  This set is important as it is the most fine-grained we can expect to observe.  We can define an operation on operational solutions that returns the set of operationally pure solutions that are mixed to give that operational solution
\begin{equation}\label{opsortdefn}
\text{opsort}(\Psi_\mathtt{A}[O_\mathtt{A}]) = \left\{ \Psi_\mathtt{A}[\Gamma_\mathtt{A}]: \forall \Gamma_\mathtt{A}\in O_\mathtt{A} \right\}
\end{equation}
This is clearly similar to the sort operation.

\section{Composition and non-separability}

We are interested in composing two solutions
\begin{equation}
\Psi_\mathtt{A}\in\Omega_\mathtt{A}[\text{mixed}]~~~~ \mathrm{and} ~~~~\Psi_\mathtt{B}\in\Omega_\mathtt{B}[\text{mixed}]
\end{equation}
by imposing conditions at the surface in op-space where $\mathtt{A}$ and $\mathtt{B}$ meet such that we get a solution in $\Omega_{\mathtt{A}\cup\mathtt{B}}[\text{mixed}]$ for the combined region $\mathtt{A}\cup\mathtt{B}$.   We may also be interested in composing solutions for multiple regions in this way.

To join these two solutions we define the notion of a \emph{typing surface} that lives at the boundary between $\mathtt{A}$ and $\mathtt{B}$.  Then we consider matching boundary conditions at this typing surface for each $\tilde{\Psi}_\mathtt{A}\in \Psi_\mathtt{A}$ and $\tilde{\Psi}_\mathtt{B}\in \Psi_\mathtt{B}$.  There are two types of boundary conditions: manifold boundary conditions and field boundary conditions.  The manifold boundary conditions test to see if the appropriate part of the boundary of the manifold patch $\mathscr{M}_\mathtt{A}\subset\mathscr{W}_\mathtt{A}$ associated with $\tilde{\Psi}_\mathtt{A}$ can be identified (in a certain prescribed way) with the appropriate part of the boundary of the manifold patch $\mathscr{M}_\mathtt{B}\subset\mathscr{W}_\mathtt{B}$ associated with $\tilde{\Psi}_\mathtt{B}$ so that we get a manifold patch associated with the two regions.   If the manifold boundary conditions are met then we need a check to see if the fields defined on these manifolds also match appropriately so that we have a solution to the field equations of GR for the composite manifold patch.  We keep only those $\tilde{\Psi}_\mathtt{A}\in \Psi_\mathtt{A}$ and $\tilde{\Psi}_\mathtt{B}\in \Psi_\mathtt{B}$ for which the boundary conditions are met. The set of these objects forms the composed solution for region $\mathtt{A\cup B}$.

We provide what are clearly sufficient boundary conditions for both the manifold and field matching.  It is anticipated that there are more efficient ways of specifying the boundary conditions.

\subsection{Typing surfaces}

We wish to define a notion of \emph{typing surface}. \index{typing surface} We will denote typing surfaces by $\mathtt{a}$, $\mathtt{b}$, \dots.  They are surfaces in op-space equipped with a coordinate system.  We can write
\begin{equation}
\mathtt{a} = \big( \text{set}(\mathtt{a}), \text{coord}(\mathtt{a}) \big)
\end{equation}
The \emph{bounding surface}, $\text{set}(\mathtt{a})$, \index{bounding surface} is a set of points in op-space that comprise a patch of a $K-1$ dimensional surface (here $K$ is the dimension of op-space).  Typically we will consider this bounding surface to be all or part of a bounding surface of some region, $\mathtt{A}$.  For the time being we will take this to be defined by some smooth function, $h_\mathtt{a}(\mathbf{S})=0$ (along with inequalities that restrict to the given patch).  Later we will consider bounding surfaces that are not smoothly defined everywhere (they will be formed by gluing together a finite number of smooth surfaces).
The \emph{coordinate system}, $\text{coord}(\mathtt{a})$, consists of a set of $K$ scalars (smooth functions of $\mathbf{S}$)
\begin{equation}
\mathbf{X}_\mathtt{a} = \left\{ X_\mathtt{a}^l: l=0 ~~\text{to}~~ K-1 \right\}
\end{equation}
such that: (i) the coordinates labeled by $l=1$ to $K-1$ are in the surface so that
\begin{equation}
\frac{\partial h_\mathtt{a}(\mathbf{S})}{\partial X_\mathtt{a}^l}=0  ~~~ \text{for}~l=1~\text{to}~K
\end{equation}
where  $h_\mathtt{a}(\mathbf{S})=0$ is satisfied on the typing surface; and (ii) the $X_\mathtt{a}^0$ coordinate points out of the surface such that
\begin{equation}
\frac{\partial h_\mathtt{a}(\mathbf{S})}{\partial X_\mathtt{a}^0} \not= 0
\end{equation}
We require that $\mathbf{X}_\mathtt{a}$ is an invertible function of $\mathbf{S}$ for all $\mathbf{S}\in\text{set}(\mathtt{a})$.  This means that in the vicinity of the typing surface, we can define the surface by
\begin{equation}
X_\mathtt{a}^0 (\mathbf{S}) = 0
\end{equation}
(i.e.\ we can replace the function $h_\mathtt{a}(\cdot)$ by the function $X_\mathtt{a}^0(\cdot)$.

It is convenient to associate a direction with a typing surfaces, $\mathtt{a}$.  This simply indicates one or the other side of the bounding surface is taken to be the positive direction.  We choose the side of the bounding surface in which the coordinate $X_\mathtt{a}^0$ increases to be the positive direction and indicate this by $\text{dir}(\mathtt{a})$.
We denote by $\text{coord}^\mathtt{R}(\mathtt{a})$ the coordinate system in which we replace $X_\mathtt{a}^0$ by $-X_\mathtt{a}^0$.  Then we write
\begin{equation}
\mathtt{a^R} = \big( \text{set}(\mathtt{a}),  \text{coord}^\mathtt{R}(\mathtt{a}) \big)
\end{equation}
to indicate the typing surface with its direction reversed.

\subsection{Boundary conditions}\label{sec:boundaryconditions}

Now we want to consider boundary conditions at typing surfaces.  We will say that the boundary conditions at typing surface $\mathtt{a}$ are
\begin{equation}
\mathbnd{a}\in \Lambda_\mathtt{a}[\text{spec}]
\end{equation}
where $\text{spec}$ are specifications to be discussed.  Similarly, we associate boundary conditions $\mathbnd{b}$ with typing surface $\mathtt{b}$, $\mathbnd{c}$ with $\mathtt{c}$, and so on.

In this section we will explore the nature of these boundary conditions - they are unusual because of the curious way solutions map into the op-space. We will consider manifold boundary conditions and field boundary conditions separately then put them together.

\subsubsection{Manifold patch at typing surface}

Consider a solution, $\tilde{\Psi}_\mathtt{A}$, for region $\mathtt{A}$ of op-space with typing surface, $\mathtt{a}$, at part (or all) of its boundary.  We define
\begin{equation}
\mathscr{M}_\mathtt{a|A}= \left\{ p:\forall p\in\mathscr{M}_\mathtt{A}~ \text{s.t.} ~\mathbf{S}(p)\in \text{set}(\mathtt{a}) \right\}
\end{equation}
where we calculate $\mathbf{S}(p)$ from $\tilde{\Psi}_\mathtt{A}$.  This is the set of points in $\mathscr{M}_\mathtt{A}$ that correspond, for this solution, to points in the typing surface, $\mathtt{a}$.

It turns out to be convenient to associate a separate chartable space, $\mathscr{W}_\mathtt{a}$ with each typing surface, $\mathtt{a}$.  We can then express boundary conditions by mapping manifolds and fields into this chartable space using the appropriate fiducial identity map.  Thus, we write
\begin{equation}
\mathscr{M}_\mathtt{a|A|a} = \varphi^I_\mathtt{a\leftarrow A}(\mathscr{M}_\mathtt{a|A})\subset \mathscr{W}_\mathtt{a}
\end{equation}
where $\varphi^I_\mathtt{a\leftarrow A}$ is the fiducial identity map from $\mathscr{W}_\mathtt{A}$ to $\mathscr{W}_\mathtt{a}$.  In this notation, the rightmost entry of the subscript on $\mathscr{M}_\mathtt{a|A|a}$ indicates the chartable space.

\subsubsection{Manifold boundary conditions}

\paragraph{The problem}

Consider two regions, $\mathtt{A}$ and $\mathtt{B}$, of op-space that meet at part of their boundary at $\text{set}(\mathtt{a})$.  We wish to join $\mathscr{M}_\mathtt{A}$ to $\mathscr{M}_\mathtt{B}$ by attempting to identifying points in $\mathscr{M}_\mathtt{a|A}\subset \mathscr{M}_\mathtt{A}$ with points in $\mathscr{M}_\mathtt{a|B}\subset \mathscr{M}_\mathtt{B}$ so that we have a manifold patch associated with $\mathtt{A\cup B}$. For this we require three conditions: (i) identification, (ii) direction, and (iii) smoothness.

\paragraph{Identification condition}

We will only allow identification of points that are related by the corresponding fiducial identity map.  In particular, this means we must have
\begin{equation}
\varphi^I_\mathtt{a\leftarrow A}( \mathscr{M}_\mathtt{a|A}) = \varphi^I_\mathtt{a\leftarrow B} (\mathscr{M}_\mathtt{a|B})
\end{equation}
or, equivalently,
\begin{equation}
\mathscr{M}_\mathtt{a|A|a} = \mathscr{M}_\mathtt{a|B|a}
\end{equation}
This is the \emph{identification condition}.

\paragraph{Direction condition}

Second we impose a \emph{direction condition} so that $\mathscr{M}_\mathtt{A}$ and $\mathscr{M}_\mathtt{B}$ are, in an appropriate sense, on opposite sides of the join.  The appropriate sense is that we can continue any chart, $x_{\mathtt{A}i}$, that covers $\mathscr{M}_\mathtt{A}$ in the vicinity of $\mathscr{M}_\mathtt{a|A}$ into a chart $x_{\mathtt{B}i}$ that covers $\mathscr{M}_\mathtt{B}$ in the vicinity of $\mathscr{M}_\mathtt{a|B}$.  We can write the function defining the surface $\text{set}(\mathtt{a})$ as
\begin{equation}
X_\mathtt{a}^0 (x_{\mathtt{A}i}) = 0   ~~~~  X_\mathtt{a}^0 (x_{\mathtt{B}i}) = 0
\end{equation}
with respect to coordinates for the $\mathtt{A}$ side and the $\mathtt{B}$ side respectively.  Here $i$ such that $x_{\mathtt{A}i}\not= -$.
If the typing surface points away from $\mathtt{A}$ (and towards $\mathtt{B}$) then, in the vicinity of the typing surface (but not on the typing surface) the  $X_\mathtt{a}^0$ coordinate will be negative for
$\breve{x}_\mathtt{A}+\delta x_{\mathtt{A}i}\in f_\mathtt{A}(\mathscr{M}_\mathtt{A})$.  Hence, to first order, we can write
\begin{equation}\label{directionA}
X_\mathtt{a}^0 (x_{\mathtt{A}i}+\delta x_{\mathtt{A}i}) = \frac{\partial X_\mathtt{a}^0}{\partial x^\mu_{\mathtt{A}i} } \delta x^\mu_{\mathtt{A}i} < 0
\end{equation}
On the $\mathtt{B}$ side of the boundary the $X_\mathtt{a}^0$ coordinate will be positive so we have
\begin{equation}\label{directionB}
\frac{\partial X_\mathtt{a}^0}{\partial x^\mu_{\mathtt{B}i} } \delta x^\mu_{\mathtt{B}i} > 0
\end{equation}
Thus the necessary condition to ensure that the coordinate system continues across the boundary under the identification is that
\begin{equation}
\frac{\partial X_\mathtt{a}^0}{\partial x^\mu_{\mathtt{A}i} }
\end{equation}
and
\begin{equation}
\frac{\partial X_\mathtt{a}^0}{\partial x^\mu_{\mathtt{B}i} }
\end{equation}
are of the same sign (positive or negative) at the points to be identified.  This condition can be expressed as an equation by mapping both expressions into the chartable space $\mathscr{W}_\mathtt{a}$ using the appropriate fiducial identity maps. In particular, on side $\mathtt{A}$ we map $X_\mathtt{a}^0$ on $\mathscr{M}_\mathtt{A}$ to $\mathscr{M}_\mathtt{A|a}$ where
\begin{equation}
\mathscr{M}_\mathtt{A|a}= \varphi^I_\mathtt{a\leftarrow A}(\mathscr{M}_\mathtt{A})\subset \mathscr{W}_\mathtt{a}
\end{equation}
before evaluating (and similarly for side $\mathtt{B}$).  This condition becomes
\begin{equation}\label{directionKcondition}
\left.\frac{\partial X_\mathtt{a}^0}{\partial x^\mu_{\mathtt{a}i} }\right|_{\tilde{\Psi}_\mathtt{A}} = k^2 \left.\frac{\partial X_\mathtt{a}^0}{\partial x^\mu_{\mathtt{a}i} }\right|_{\tilde{\Psi}_\mathtt{B}}
\end{equation}
where $k^2$ is positive.   This must be true for all $p\in\mathscr{M}_\mathtt{a|A|a}$.  If this condition is satisfied then any given non-zero $\delta x$ can either satisfy \eqref{directionA} or \eqref{directionB} which demonstrates that we can continue a chart across the boundary under the identification.

If we suppose that the scalar fields in $\mathbf{S}$ are differentiable then we can simply write the direction condition as
\begin{equation}\label{directioncondition}
\left.\frac{\partial X_\mathtt{a}^0}{\partial x^\mu_{\mathtt{a}i} }\right|_{\tilde{\Psi}_\mathtt{A}} =  \left.\frac{\partial X_\mathtt{a}^0}{\partial x^\mu_{\mathtt{a}i} }\right|_{\tilde{\Psi}_\mathtt{B}}
\end{equation}
This is a stronger condition but it will help us to use the language of matching boundary conditions.  We will write this condition down in coordinate free notation as
\begin{equation}
\left.\partial_{\mathtt{a}\pmb{\mu}} X_\mathtt{a}^0\right|_{\tilde{\Psi}_\mathtt{A}} = \left.\partial_{\mathtt{b}\pmb{\mu}} X_\mathtt{a}^0\right|_{\tilde{\Psi}_\mathtt{B}}
\end{equation}
this being equivalent to \eqref{directioncondition} when evaluated in any chart.  If this is true in any given coordinate system then the smoothness condition (to be discussed next) guarantees that it is true for all coordinate systems.  Hence, it is sufficient to check the condition in one coordinate system.

\paragraph{Smoothness condition}

The third condition we need to impose is a \emph{smoothness condition}.  For a $C^r$ manifold we also require that these transition maps are $r$ times differentiable.  Hence we also require that the derivatives (up to $r$th derivatives) of these transition maps are numerically equal at the points which are identified.  In fact, because of Whitney's theorem (discussed in Sec.\ \ref{sec:manifoldsinchartablesec}) it is only necessary for the first derivative to match at these points.    It then follows from this theorem that there exists an atlas for which they match for all derivatives. By matching fist derivatives we ensure that the tangent spaces match up so we can transform tensors in the same way on either side of the boundary.  Hence, we need to check that
\begin{equation}
\frac{\partial (x_{\mathtt{A}j}\circ x_{\mathtt{A}i}^{-1})}{\partial x_{\mathtt{A}i}^\mu}
\end{equation}
on the $\mathtt{A}$ side is equal to
\begin{equation}
 \frac{\partial (x_{\mathtt{B}j}\circ x_{\mathtt{B}i}^{-1})}{\partial x_{\mathtt{B}i}^\mu}
\end{equation}
on the $\mathtt{B}$ at each point being identified.   We can map these conditions into $\mathscr{W}_\mathtt{a}$ under the appropriate fiducial identity map so that the condition becomes
\begin{equation}\label{smoothnesscondition}
\left.\frac{\partial (x_{\mathtt{a}j}\circ x_{\mathtt{a}i}^{-1})}{\partial x_{\mathtt{a}i}^\mu}\right|_{\tilde{\Psi}_\mathtt{A}} = \left.\frac{\partial (x_{\mathtt{a}j}\circ x_{\mathtt{a}i}^{-1})}{\partial x_{\mathtt{a}i}^\mu}\right|_{\tilde{\Psi}_\mathtt{B}}
\end{equation}
This condition must be true for all $p\in\mathscr{M}_\mathtt{a|A|a}$ and for all pairs, $i$ and $j$, for which $x_i\not= -$ and $x_j\not=-$.  If we define
\begin{equation}
\pmb{\partial} x_\mathtt{a} = \left\{ \left(ij, \frac{\partial (x_{\mathtt{a}j}\circ x_{\mathtt{A}i}^{-1})}{\partial x_{\mathtt{a}i}^\mu}\right): \forall i,j\right\}
\end{equation}
then we can write the smoothness condition as
\begin{equation}
\left.\pmb{\partial} x_\mathtt{a}\right|_{\tilde{\Psi}_\mathtt{A}} = \left.\pmb{\partial} x_\mathtt{a}\right|_{\tilde{\Psi}_\mathtt{B}}
\end{equation}
This must be true for all the points in the boundaries that are identified.

\paragraph{Resulting manifold}

We write
\begin{equation}
\mathscr{M}_\mathtt{A}\cup_\mathtt{a}\mathscr{M}_\mathtt{B}
\end{equation}
for the manifold patch we obtain by taking the union of $\mathscr{M}_\mathtt{A}$ and $\mathscr{M}_\mathtt{B}$ and identifying any $p\in\mathscr{M}_\mathtt{a|A}\subset\mathscr{M}_\mathtt{A}$ with the corresponding point (mapped by the corresponding fiducial identity map) $p\in\mathscr{M}_\mathtt{a|B}\subset\mathscr{M}_\mathtt{B}$.  If all three conditions above are satisfied then $\mathscr{M}_\mathtt{A}\cup_\mathtt{a}\mathscr{M}_\mathtt{B}$ is itself a manifold patch.

\subsubsection{Field boundary conditions}

If the boundary conditions for the two patches to form a manifold when taken together are met then we can now ask if the physical fields defined on these manifold patches match at the boundary such that we have a solution for the composite patch.    To this end we require that the fields, $\pmb{\Phi}$, themselves match after being mapped into the appropriate subset of $\mathscr{W}_\mathtt{a}$ by the fiducial identity map.  We also require that all tensor fields formed by taking derivatives match.  However, all these fields are related by field equations that have to be true at each point on the manifold (and, in particular, at all points in the boundary).  These field equations relate derivatives. Hence, it is not necessary to match all derivatives separately. If we match enough, then it will follow from the field equations that the remaining derivatives are matched.  We will assume (in the Local Matching Assumption in Sec.\ \ref{sec:joiningsolutions}) that we only need to specify a finite number of derivatives which we put into the set, $\pmb{\pi}$.

The two matching conditions are then
\begin{equation}
\left.\pmb{\Phi}\right|_{\tilde{\Psi}_\mathtt{A}} = \left.\pmb{\Phi}\right|_{\tilde{\Psi}_\mathtt{B}}  ~~~~ \left. \pmb{\pi}\right|_{\tilde{\Psi}_\mathtt{A}} = \left. \pmb{\pi}\right|_{\tilde{\Psi}_\mathtt{A}}
\end{equation}
for all $p_\mathtt{A}\in\mathscr{M}_\mathtt{a|A|a}$.

We construct the tensor fields in the set $\pmb{\pi}_\mathtt{a|A}$ from the tensor fields in $\pmb{\Phi}$, the scalars in $\mathbf{X}_\mathtt{a}$, and derivatives thereof.  For example, if
\begin{equation}
\pmb{\Phi}= \left( g_{\pmb{\mu\nu}}, j^{\pmb{\mu}}[a], F^{\pmb{\mu\nu}} \right)
\end{equation}
we might define
\begin{equation}
\pmb{\pi}= \left( \nabla_{\pmb{\mu}} F^{\pmb{\mu\nu}}, \frac{\partial X^k}{\partial x^{\pmb{\mu}} }, \nabla_{\pmb{\mu}} g_{\pmb{\mu\nu}},
\nabla_{\pmb{\mu}} j_{\pmb{\nu}}[a] \right)
\end{equation}
where these are evaluated from the fields in $\tilde{\Psi}_\mathtt{A}$ for points in $\mathscr{M}_\mathtt{a|A}$.  Of course, we can consider many different choices of functional form for $\pmb{\pi}$.  We need a choice of $\pmb{\pi}$ such that the Local Matching Assumption to be introduced later is true.

\subsubsection{The full boundary conditions}\label{sec:fullboundaryconditions}

We can now collect together all the boundary conditions that must be satisfied.  We define
\begin{equation}\label{vartheta}
\pmb{\Theta} = (\pmb{\Phi}, \pmb{\pi},  \partial_{\mathtt{a}\pmb{\mu}} X_\mathtt{a}^0, \pmb{\partial} x_\mathtt{a} )
\end{equation}    \index{Theta@$\pmb{\Theta}$}
Now we define the \emph{boundary conditions} \index{boundary conditions} at $\mathtt{A}$ induced by the candidate solution ${\tilde{\Psi}}_\mathtt{A}$ to be
\begin{equation}
\tilde{\theta}_\mathtt{a}({\tilde{\Psi}}_\mathtt{A})= \left\{ (p, \pmb{\Theta}): \forall p \in \mathscr{M}_\mathtt{a|A|a} \right\}
\end{equation}
where $\pmb{\Theta}$ (at each point $p$) and $\mathscr{M}_\mathtt{a|A|a}$, are calculated from $\tilde{\Psi}_\mathtt{A}$ (after applying the fiducial identity map $\varphi^I_\mathtt{a\leftarrow A}$).  Hence, although $\tilde{\Psi}_\mathtt{A}$ lives in $\mathscr{W}_\mathtt{A}$, the boundary conditions for a typing surface, $\mathtt{a}$, live in $\mathscr{W}_\mathtt{a}$.

Many of these boundary conditions are physically equivalent as they are related by elements in $G^\text{diffeo}_\mathtt{A}$. Hence, it makes sense to define
\begin{equation}\label{thetawithouttildedefn}
\theta_\mathtt{a}(\Psi_\mathtt{A}) = \left\{ \tilde{\theta}_\mathtt{a}({\tilde{\Psi}}_\mathtt{A}): \forall \tilde{\Psi}_\mathtt{A}\in \Psi_\mathtt{A} \right\}
\end{equation}     \index{thetaa@$\theta_\mathtt{a}$}
This is a diffeomorphism invariant way of presenting the boundary conditions induced by a candidate solution $\Psi_\mathtt{a}$ at $\mathtt{a}$.  Note that this definition works for pure and mixed solutions.

\subsubsection{Composite typing surfaces}

In this section we will consider composite typing surfaces \index{typing surfaces!composite} - that is a typing surface that is regarded as being made up of more than one typing surface.  This is especially important when we want to have typing surfaces that do not join smoothly, or ones where a single coordinate system cannot chart the whole surface.  Further, if we wish to consider many regions of op-space that join together to cover some larger region of op-space then it is impossible that all the boundaries of these smaller regions are smooth.

A composite typing surface corresponds to a list $(\mathtt{a}, \mathtt{b})$.  We will usually denote this by $\mathtt{ab}$.  Similarly, the boundary conditions at such a typing surface correspond to a list $(\mathbnd{a}, \mathbnd{b})$. We will usually denote this by $\mathbnd{ab}$.  We write
\begin{equation}
\tilde{\theta}_\mathtt{ab} (\Psi_\mathtt{A}) = \left(\tilde{\theta}_\mathtt{a}(\Psi_\mathtt{A}), \tilde{\theta}_\mathtt{b}(\Psi_\mathtt{A})\right)
\end{equation}
and
\begin{equation}
\theta_\mathtt{ab} (\Psi_\mathtt{A}) = \left(\theta_\mathtt{a}(\Psi_\mathtt{A}), \theta_\mathtt{b}(\Psi_\mathtt{A})\right)
\end{equation}
We can think of a composite typing surface as a typing surface in its own right.  Thus, we will sometimes denote a composite system by a single letter.  For example, we may write $\mathtt{d}=\mathtt{ab}$.  We allow the components of a composite typing surface to meet (in an $K-2$ dimensional subset of op-space) but not to overlap more than this.  We allow composite typing surfaces to be constructed from any finite number of basic (smooth) typing surfaces.   Henceforth, when when we discuss typing surfaces, we allow them to be composite (and therefore possibly not smooth).

\subsubsection{Joining solutions}\label{sec:joiningsolutions}

Now we will show how to join two solutions, $\Psi_\mathtt{A}$ and $\Psi_\mathtt{B}$, \index{joining solutions} to form a new solution for the region $\mathtt{A}\cup\mathtt{B}$.  The interior of the two regions, $\mathtt{A}$ and $\mathtt{B}$ are non-overlapping but we allow that their boundaries overlap at some $\text{set}(\mathtt{a})$ associated with a typing surface, $\mathtt{a}$.

If the matching condition
\begin{equation}
\tilde{\theta}_\mathtt{a}({\tilde{\Psi}}_\mathtt{A}) = \tilde{\theta}_\mathtt{a}({\tilde{\Psi}}_\mathtt{B})
\end{equation}
then the two solutions match at the typing surface $\mathtt{a}$.  We define
 \begin{equation}
\tilde{\Psi}_\mathtt{A} \cup_\mathtt{a} \tilde{\Psi}_\mathtt{B} =
\left\{ \begin{array}{ll}
\{ (p, \Phi(p)): \forall p\in\mathscr{M}_\mathtt{A}\cup_\mathtt{a}\mathscr{M}_\mathtt{B} \} & \text{if} ~ \tilde{\theta}_\mathtt{a}({\tilde{\Psi}}_\mathtt{A}) = \tilde{\theta}_\mathtt{a}({\tilde{\Psi}}_\mathtt{B}) \\
- & \text{else}
\end{array}
\right.
\end{equation}       \index{cupa@$\cup_\mathtt{a}$}
where here $-$ is the null element having the property that, if it is added to a set, it does not change the set (so $x\cup\{ -\} = x$ for any set $x$).  In the case where we have matching we get, essentially, the union of $\tilde{\Psi}_\mathtt{A}$ and $\tilde{\Psi}_\mathtt{B}$ except that points at the boundary have been identified (and therefore the fields must match at these points).  Now we can define
\begin{equation}\label{joinsolutions}
\Psi_\mathtt{A} \Cup_\mathtt{a} \Psi_\mathtt{B} =
\left\{ {\tilde{\Psi}}_\mathtt{A}\cup_\mathtt{a}{\tilde{\Psi}}_\mathtt{B}: \forall {\tilde{\Psi}}_\mathtt{A}\in \Psi_\mathtt{A}, \forall {\tilde{\Psi}}_\mathtt{B}\in \Psi_\mathtt{B}  \right\}
\end{equation}       \index{Cupa@$\Cup_\mathtt{a}$}
Note that this definition works for joining pure and mixed solutions.  It also works for joining at composite typing surfaces.  We keep pairs ${\tilde{\Psi}}_\mathtt{A}\in \Psi_\mathtt{A}$ and ${\tilde{\Psi}}_\mathtt{B}\in \Psi_\mathtt{B}$ when they match up (have the same boundary conditions) otherwise they contribute the null element.  In the case that the two solutions do not have matching boundary conditions for any pair of elements we get the null solution, $\Psi_\mathtt{A} \Cup_\mathtt{a} \Psi_\mathtt{B}=\emptyset$.

In the case that the two solutions we are joining are pure and they match for some ${\tilde{\Psi}}_\mathtt{A}\in \Psi_\mathtt{A}$ and ${\tilde{\Psi}}_\mathtt{B}\in \Psi_\mathtt{B}$ then there must be a match in $\Psi_\mathtt{B}$ for every element of $\Psi_\mathtt{A}$ since we vary over all diffeomorphisms on each side and hence
\begin{equation}
\theta_\mathtt{a}(\Psi_\mathtt{A})= \theta_\mathtt{a}(\Psi_\mathtt{B})
\end{equation}
(Note that there are no tildes in this expression - here $\theta$ is defined as in \eqref{thetawithouttildedefn}.)  This situation can be quite interesting as we will see shortly.

In the case that the two solutions are mixed then we will get a non-null solution for the join if
\begin{equation}
\theta_\mathtt{a}(\Psi_\mathtt{A})\cap \theta_\mathtt{a}(\Psi_\mathtt{B}) \not=\emptyset
\end{equation}
as then at least some of the pure solutions in the sorts of $\Psi_\mathtt{A}$ and $\Psi_\mathtt{B}$ match up.

\subsubsection{The local matching assumption}

For $\Psi_\mathtt{A} \Cup_\mathtt{a} \Psi_\mathtt{B}$ to actually be a solution we need to make the right choice of $\pmb{\pi}$. This is the purpose of the following assumption: \index{local matching assumption}
\begin{quote}
{\bf Local matching assumption.}
Consider two regions in Op-space, $\mathtt{A}$ and $\mathtt{B}$ whose interiors are non-overlapping and which meet at a typing surface $\mathtt{a}$ (which could be composite).
Further, consider pure solutions, $\Psi_\mathtt{A}\in\Omega_\mathtt{A}$ and $\Psi_\mathtt{B}\in\Omega_\mathtt{B}$.
We assume that, for any physically realistic set, $\text{FieldEqns}_{GR}$, there exists a choice of $\pmb{\pi}$ \emph{with a finite number of elements} such that
\begin{equation}\label{localmatching}
\Psi_\mathtt{A} \Cup_\mathtt{a} \Psi_\mathtt{B} \in \Omega_{\mathtt{A}\cup\mathtt{B}}[\text{mixed}]
\end{equation}
for any pair of regions, $\mathtt{A}$ and $\mathtt{B}$.
\end{quote}
That is, there exist sensible boundary conditions such that when we join two actual solutions then we get an actual solution.
In the first place, this assumption is motivated by the fact that $\text{FieldEqns}_{GR}$ are local field equations (equations pertaining to a point, $p$).  Since the two solutions, ${\tilde{\Psi}}_\mathtt{A}$ and ${\tilde{\Psi}}_\mathtt{B}$, overlap on $\mathtt{a}$ we require that all fields and all derivatives derivatives are equal (coming from the two solutions).  If we choose $\pmb{\pi}$ to be a list of all such derivatives then clearly the matching condition, if satisfied, will entail that ${\tilde{\Psi}}_\mathtt{A}\cup{\tilde{\Psi}}_\mathtt{B}$ is a solution.  However, such a list is infinite. It is possible, however, that we can choose a finite list of such derivatives for $\pmb{\pi}$ such that the relations in $\text{FieldEqns}_{GR}$ can then be used to show that all other derivatives are, in fact, equal.  This is the essential content of the local matching assumption.

There is a twist in this tale however (anticipated by the fact that we put $\Omega_{\mathtt{A}\cup\mathtt{B}}[\text{mixed}]$ in \eqref{localmatching} even though we are considering joining pure solutions).  Since we are actually working with diffeomorphism invariant solutions, we must consider matching every pair of elements from $\Psi_\mathtt{A}$ and $\Psi_\mathtt{B}$.   It is possible then, that there are physically inequivalent matches (i.e.\ not mapped into one another by diffeomorphisms).   This would mean that, even though we are taking $\Psi_\mathtt{A}$ and $\Psi_\mathtt{B}$ to be pure, $\Psi_\mathtt{A} \Cup_\mathtt{a} \Psi_\mathtt{B}$ could be mixed such that
\begin{equation}\label{sortset}
\text{sort}(\Psi_\mathtt{A} \Cup_\mathtt{a} \Psi_\mathtt{B})
\end{equation}
has two or more elements.  An example of how this can happen is provided in Sec.\ \ref{sec:curiousnonseprability}.   One consequence of this is that there exist solutions $\Psi_\mathtt{A\cup B}$ such that
\begin{equation}
\Psi_\mathtt{A\cup B} \not= \text{restrict}_\mathtt{A}(\Psi_\mathtt{A\cup B}) \Cup_\mathtt{a} \text{restrict}_\mathtt{B}(\Psi_\mathtt{A\cup B})
\end{equation}
This will be the case for any element of the set in \eqref{sortset} (assuming this set has more than one element).  Hence, we cannot split a solution up in to two parts by restriction then simply join them back together and necessarily get back the original solution.

We might have taken the attitude that we should only work with pure solutions in any formulation of General Relativity.
However, because of diffeomorphism invariance, we represent even pure solutions as a set of solutions, $\tilde{\Psi}_\mathtt{A}$.  Consequently, the most natural way of joining two pure solutions is with the $\Cup_\mathtt{a}$ operation as in \eqref{joinsolutions}.  Under this operation, joining pure solutions can lead to mixed solutions. We are, then, naturally led to having mixed solutions.

We note that, although the local matching assumption is stated for the special case of joining pure solutions, it clearly follows from this assumption that if we join mixed solutions, $\Psi_\mathtt{A}$ and $\Psi_\mathtt{B}$, we get a solution $\Psi_\mathtt{A\cup B}\in \Omega_\mathtt{A\cup B}[\text{mixed}]$.

\subsubsection{The null join}\label{sec:thenulljoin}

A special case is where $\mathtt{A}\cap\mathtt{B}=\varnothing$. In this case there is no tying surface between $\mathtt A$ and $\mathtt B$.  We will say that the two surfaces are joined by the \emph{null join} \index{null join} denoted by $\mathtt{0}$.  In this case there are no matching conditions to check and we have
\begin{equation}
\Psi_\mathtt{A}\Cup_\mathtt{0}\Psi_\mathtt{B} = \left\{ \tilde{\Psi}_\mathtt{A}\cup\tilde{\Psi}_\mathtt{B}: \forall \tilde{\Psi}_\mathtt{A} \in \Psi_\mathtt{A}, \forall \tilde{\Psi}_\mathtt{B}\in\Psi_\mathtt{B} \right\}
\end{equation}
The null join is useful when we are joining multiple solutions as it enables us to proceed in any order \cite{hardy2013theory}.

\subsubsection{Special cases}\label{sec:specialcases}

If we join two empty solutions at a typing surface $\mathtt{a}$ we clearly get the empty solution for the composite region.
\begin{equation}
\Psi_\mathtt{A}[\varnothing] \Cup_\mathtt{a} \Psi_\mathtt{B}[\varnothing] = \Psi_\mathtt{A\cup B} [\varnothing]
\end{equation}
The boundary conditions clearly match as we have
\begin{equation}
{\theta}_\mathtt{a}({{\Psi}}_\mathtt{A}[\varnothing]) = {\theta}_\mathtt{a}({{\Psi}}_\mathtt{B}[\varnothing])= \{ \varnothing\}
\end{equation}
This boundary condition is discussed below in Sec.\ \ref{sec:specialboundaryconditions} as the special case, $\mathbnd{a}_\varnothing$.  It is useful we can join empty solutions as they are the generic situation for small enough $\mathtt{A}$.

We can also investigate null solutions.  If we combine two null solutions we get a null solution for the combined region.
\begin{equation}\label{ABnulljoin}
\Psi_\mathtt{A}[\mathbf{0}] \Cup_\mathtt{a} \Psi_\mathtt{B}[\mathbf{0}] = \Psi_\mathtt{A\cup B} [\mathbf{0}]
\end{equation}
This follows from the definition of $\Cup_\mathtt{a}$.  In fact, more generally, we have
\begin{equation}
\Psi_\mathtt{A} \Cup_\mathtt{a} \Psi_\mathtt{B}[\mathbf{0}] = \Psi_\mathtt{A\cup B} [\mathbf{0}]
\end{equation}
This also follows from the definition.   Additionally, we have
\begin{equation}
\Psi_\mathtt{A} \Cup_\mathtt{a} \Psi_\mathtt{B} = \Psi_\mathtt{A\cup B} [\mathbf{0}] ~~~\text{if}~~ \theta_\mathtt{a}(\Psi_\mathtt{A})\cap \theta_\mathtt{a}(\Psi_\mathtt{B}) =\varnothing
\end{equation}
since then there can be no matches (there cannot even be matches of empty solutions).

A interesting case is when we join $\Psi_\mathtt{A}[\Gamma_\mathtt{A}]$ and $\Psi_\mathtt{B}[\Gamma_\mathtt{B}]$ at $\mathtt{a}$.  We get
\begin{equation}
\Psi_\mathtt{A}[\Gamma_\mathtt{A}]\Cup_\mathtt{a}\Psi_\mathtt{B}[\Gamma_\mathtt{B}]=
\left\{ \begin{array}{ll}
\Psi_\mathtt{A\cup B} [\mathbf{0}] &  \text{if}~~
                                   \theta_\mathtt{a}(\Psi_\mathtt{A}[\Gamma_\mathtt{A}])\cap \theta_\mathtt{a}(\Psi_\mathtt{B}[\Gamma_\mathtt{B}])  = \varnothing \\
\Psi_\mathtt{A\cup B} [\Gamma_\mathtt{A}\cup\Gamma_\mathtt{B}] & \text{if}~~
                                   \theta_\mathtt{a}(\Psi_\mathtt{A}[\Gamma_\mathtt{A}]) \cap \theta_\mathtt{a}(\Psi_\mathtt{B}[\Gamma_\mathtt{B}]) \not=\varnothing
\end{array} \right.
\end{equation}
This is not completely obvious.  To prove it we work backwards.  Note that
\begin{equation}
\text{restrict}_\mathtt{A}(\Psi_\mathtt{A\cup B}[\Gamma_\mathtt{A}\cup\Gamma_\mathtt{B}]) \subseteq \Psi_\mathtt{A}[\Gamma_\mathtt{A}] ~~~~~
\text{restrict}_\mathtt{B}(\Psi_\mathtt{A\cup B}[\Gamma_\mathtt{A}\cup\Gamma_\mathtt{B}]) \subseteq \Psi_\mathtt{B}[\Gamma_\mathtt{B}]
\end{equation}
Hence, when we join $\Psi_\mathtt{A}[\Gamma_\mathtt{A}]$ and $\Psi_\mathtt{B}[\Gamma_\mathtt{B}]$ then (in the case that there is a match) we have all the necessary pure solutions in the sort of these two solutions to fully reconstruct $\Psi_\mathtt{A\cup B}[\Gamma_\mathtt{A}\cup\Gamma_\mathtt{B}]$.

We can also consider joining two operational solutions associated with sets $O_\mathtt{A}$ and $O_\mathtt{B}$. We obtain
\begin{equation}
\Psi_\mathtt{A}[O_\mathtt{A}]\Cup_\mathtt{a}\Psi_\mathtt{B}[O_\mathtt{B}]=  \Psi_\mathtt{A\cup B}[O_\mathtt{A\cup B}]
\end{equation}
where
\begin{equation}
O_\mathtt{A\cup B} \subseteq O_\mathtt{A} \Cup O_\mathtt{B}
\end{equation}
Here
\begin{equation}\label{OACupOB}
O_\mathtt{A} \Cup O_\mathtt{B} = \left\{ \Gamma_\mathtt{A}\cup \Gamma_\mathtt{B}: \forall \Gamma_\mathtt{A}\in O_\mathtt{A}, \forall \Gamma_\mathtt{B}\in O_\mathtt{B} \right\}
\end{equation}
and
\begin{equation}\label{OAcupB}
O_\mathtt{A\cup B} = \left\{ \Gamma_\mathtt{A}\cup \Gamma_\mathtt{B}: \forall \Gamma_\mathtt{A}\in O_\mathtt{A}, \forall \Gamma_\mathtt{B}\in O_\mathtt{B} ~~\text{s.t.} ~~~ \theta_\mathtt{a}(\Psi_\mathtt{A}[\Gamma_\mathtt{A}]) \cap \theta_\mathtt{a}(\Psi_\mathtt{B}[\Gamma_\mathtt{B}]) \not=\varnothing \right\}
\end{equation}
In other words, $O_\mathtt{A\cup B}$ is the subset for which there is a match.

Finally, consider the case of joining two deterministic solutions.  We have
\begin{equation}
\Psi_\mathtt{A}[\text{det}]\Cup_\mathtt{a} \Psi_\mathtt{B}[\text{det}] = \Psi_\mathtt{A\cup B} [\text{det}]
\end{equation}
because
\begin{equation}
\text{restrict}_\mathtt{A}(\Psi_\mathtt{A\cup B}[\text{det}]) \subseteq \Psi_\mathtt{A}[\text{det}] ~~~~~
\text{restrict}_\mathtt{B}(\Psi_\mathtt{A\cup B}[\text{det}]) \subseteq \Psi_\mathtt{B}[\text{det}]
\end{equation}
and hence we have all the necessary pure solutions in the sort of these two solutions to fully reconstruct $\Psi_\mathtt{A\cup B} [\text{det}]$.

\subsubsection{Sets of boundary conditions}

We are now interested in the set of boundary conditions induced by sets of solutions at $\mathtt{a}$. We define
\begin{equation}
\Lambda_\mathtt{a}[\mathtt{A}] =\left\{ \theta_\mathtt{a}(\Psi_\mathtt{A}): \forall \Psi_\mathtt{A}\in\Omega_\mathtt{A} \right\}
\end{equation}           \index{Lambda@$\Lambda_\mathtt{a}$}
where $\text{set}(\mathtt{a}) \subseteq \mathtt{A}$.
This is the full set of boundary conditions - we will call them \emph{pure boundary conditions} - that can be induced by pure solutions in $\mathtt{A}$.   We define
\begin{equation}
\Lambda_\mathtt{a}[\mathtt{A}, \text{mixed}] =\left\{ \theta_\mathtt{a}(\Psi_\mathtt{A}): \forall \Psi_\mathtt{A}\in\Omega_\mathtt{A}[\text{mixed}] \right\}
\end{equation}
as the full set of boundary conditions - we will call them \emph{mixed boundary conditions} - that can be induced by mixed solutions in $\mathtt{A}$.

Note that
\begin{equation}
\Lambda_\mathtt{a}[\mathtt{A}] \subseteq \Lambda_\mathtt{a}[\mathtt{B}] ~~\text{forall} ~~ \mathtt{B} \subset \mathtt{A} ~~ \text{s.t.} ~\text{set}(\mathtt{a}) \subset \mathtt{B}
\end{equation}
because we always have $\Omega_{\mathtt{B}||\mathtt{A}} \subseteq \Omega_\mathtt{B} $ as discussed in Sec.\ \ref{sec:puresolutions}.
We define
\begin{equation}
\Lambda_\mathtt{a} = \bigcup_{\mathtt{B}\supset \text{set}(\mathtt{a})} \Lambda_\mathtt{a}[\mathtt{B}]
\end{equation}
This is the full set of boundary conditions that can be induced by any solution for any region having $\text{set}(\mathtt{a})$ as a subset.   Similarly, we define
\begin{equation}
\Lambda_\mathtt{a}[\text{mixed}] = \bigcup_{\mathtt{B}\supset \text{set}(\mathtt{a})} \Lambda_\mathtt{a}[\mathtt{B}, \text{mixed}]
\end{equation}
This is the full set of mixed boundary conditions that can be induced by any solution for any region including $\text{set}(\mathtt{a})$.

In the case of candidate solutions we have
\begin{equation}\label{candproperty}
\Omega_{\mathtt{B}||\mathtt{A}}[\text{cand}] = \Omega_\mathtt{B}[\text{cand}] ~~~~~
\Omega_{\mathtt{B}||\mathtt{A}}[\text{mixed cand }] = \Omega_\mathtt{B}[\text{mixed cand}]
\end{equation}
as we are just looking at all possible sufficiently differentiable fields defined over the appropriate parts of the manifolds. Hence we can define
\begin{equation}
\Lambda_\mathtt{a}[\text{cand}] =\left\{ \theta_\mathtt{a}(\Psi_\mathtt{A}): \forall \Psi_\mathtt{A}\in\Omega_\mathtt{A}[\text{cand}] \right\}
\end{equation}
for any $\mathtt{A}$ such that $\text{set}(\mathtt{a})\subset \mathtt{A}$.  By virtue of \eqref{candproperty}, we get the same set, $\Lambda_\mathtt{a}[\text{cand}]$, for any choice $\mathtt{A}\supset \text{set}(\mathtt{a})$ and hence we need not include $\mathtt{A}$ in the specification of this set.
This is the full set of boundary conditions at $\mathtt{a}$ that can be induced by pure candidate solutions.   Similarly, we define
\begin{equation}
\Lambda_\mathtt{a}[\text{mixed cand}] =\left\{ \theta_\mathtt{a}(\Psi_\mathtt{A}): \forall \Psi_\mathtt{A}\in\Omega_\mathtt{A}[\text{mixed cand}] \right\}
\end{equation}
for any $\mathtt{A}$ such that $\text{set}(\mathtt{a})\subset \mathtt{A}$.  This is the full set of boundary conditions that can be induced at $\mathtt{a}$ by mixed candidate solutions.

It is useful to define a sort operation for boundary conditions. \index{sort!for boundary conditions} Thus, if $\mathbnd{a}[l]$ are distinct pure boundary conditions, then
\begin{equation}
\text{sort}(\cup_l \mathbnd{a}[l]) = \{ \mathbnd{a}[l]: \forall l\}
\end{equation}
Thus, the sort operation, when acting on boundary conditions, returns a list of the pure boundary conditions in the given mixed boundary condition.

\subsubsection{Special boundary conditions}\label{sec:specialboundaryconditions}

It is useful to define a number of special boundary conditions.  First, we define $\mathbnd{a}_\varnothing$ as the boundary condition induced at typing surface $\mathtt{a}$ by the empty solution
\begin{equation}
\mathbnd{a}_\varnothing = \theta_\mathtt{a} (\Psi_\mathtt{A}[\varnothing])
\end{equation}
where we assume that $\text{set}(\mathtt{a}) \subset \mathtt{A}$.  Clearly $\mathbnd{a}_\varnothing=\{ \varnothing \}$.

The null boundary conditions  \index{boundary conditions!null} are the conditions that are always false. This is represented by
\begin{equation}
\mathbnd{a}_\mathbf{0}= \varnothing
\end{equation}
Since there must be some boundary conditions on $\mathtt{a}$ (even if only the empty boundary conditions) it is clear that $\mathbnd{a}_\mathbf{0}$ cannot happen.

Another special boundary condition is $\mathbnd{a}_\text{det}[\mathtt{A}]$ defined as
\begin{equation}
\mathbnd{a}_\text{det}[\mathtt{A}]= \theta_\mathtt{a}(\Psi_\mathtt{A} [\text{det}])
\end{equation}
This is the \emph{deterministic boundary condition} \index{boundary conditions!deterministic}  - the boundary condition that is always true.  It is induced at $\mathtt{a}$ by the mixture of all possible solutions in $\mathtt{A}$.

We also define
\begin{equation}
\mathbnd{a}_{\Gamma_\mathtt{A}} = \theta_\mathtt{a} (\Psi_\mathtt{A}[\Gamma_\mathtt{A}])
\end{equation}
as the boundary condition induced by a given $\Gamma_\mathtt{A}$.

\subsection{Composite boundary condition space}

We define $\Lambda_\mathtt{ab}[\text{spec}]$ as
\begin{equation}
\Lambda_\mathtt{ab}[\text{spec}] = \Lambda_\mathtt{a}[\text{spec}]\times \Lambda_\mathtt{b}[\text{spec}]
\end{equation}
where $\text{spec}$ is the specification of what set of boundary conditions we are discussing.  We will write an element of this space as $\mathbnd{ab}$.

\subsection{Given $\mathbnd{a}$ solutions}\label{sec:determinism}

Consider a typing surface, $\mathtt{a}$, corresponding to the full boundary of some region, $\mathtt{A}$, of op-space
\begin{equation}
\text{set}(\mathtt{a}) = \text{boundary}(\mathtt{A})
\end{equation}
Let
\begin{equation}
\Omega_\mathtt{A}[\mathbnd{a}] \subseteq\Omega_\mathtt{A}
\end{equation}
be the set of pure solutions having the property that $\theta_\mathtt{a}(\Psi_\mathtt{A})=\mathbnd{a}$.  Note that if there is no solution in $\Omega_\mathtt{A}$ consistent with this boundary condition then $\Psi_\mathtt{A}[\mathbnd{a}]$ is the empty solution, $\Psi_\mathtt{A}[\varnothing]$.   We define
\begin{equation}
\Psi_\mathtt{a} [\mathbnd{a}] = \text{flatten}(\Omega_\mathtt{A}[\mathbnd{a}])
\end{equation}
This is the mixed solution consisting of all solutions having the given boundary condition.

Since $\mathbnd{a}$ corresponds to the full boundary of $\mathtt{A}$ it is reasonable to assume it contains sufficient information to fully determine the solution: \index{determinism}
\begin{quote}
\emph{Determinism (when no agency):} The solution, $\Psi_\mathtt{A}[\mathbnd{a}]$, associated with a given pure boundary condition, $\mathbnd{a}$, at a typing surface $\mathtt{a}$ having $\text{set}(\mathtt{a})=\text{bound}(\mathtt{A})$, is always pure.
\end{quote}
This assumption will fail to be true if we do not have a complete set of field equations. This might be the case if we so not know, or do not care to solve, for all aspect of the physics.   Furthermore, when we introduce agency into the picture, we will have a phenomena we term \emph{jitter} in which the scalar coincidences associated with the agent's choices may not fully fix the solution.  This will lead to such solutions being mixed.

\subsection{Curious nonseparability}\label{sec:curiousnonseprability}

We will now give an example in which
\begin{equation}
\text{sort}(\Psi_\mathtt{A} \Cup_\mathtt{a} \Psi_\mathtt{B})
\end{equation}
has more than one member (in fact an infinite number of members) when $\Psi_\mathtt{A}$ and $\Psi_\mathtt{B}$ are pure.  This corresponds to a curious kind of nonseparability.  \index{nonseparability, curious} $\Psi_\mathtt{A}$ is a full specification of the beables for region $\mathtt{A}$ and $\Psi_\mathtt{B}$ is a full specification of the beables for region $\mathtt{B}$.   When physics is separable then a full specification of the beables for region $\mathtt{A}$ and for region $\mathtt{B}$ would be sufficient to give a full specification of the beables for the composite region, $\mathtt{A}\cup\mathtt{B}$.  Thus, separability implies
\begin{equation}
B_\mathtt{A\cup B}(\Psi_\mathtt{A\cup B}) = f\left(B_\mathtt{A}(\Psi_\mathtt{A}), B_\mathtt{B}(\Psi_\mathtt{B}) \right)
\end{equation}
This is not the case here. This nonseparability is a consequence of the fact that beables are invariant under diffeomorphisms.  Thus, if we apply a diffeomorphism to a physical situation, it does not count as a physical transformation (it does not effect the beables).

For simplicity (picturability) consider a manifold that is three dimensional (this example could easily be extended to four dimensions).  We set up a coordinate system $x^0=t$, $x^1=r$, and $x^2=\theta$.  We consider non-interacting dust fluids of types $1$ to $3$, these being distinguishable from one another (let us say that they have different colours).   Dust fluid $i$ is described by $(\rho_i(x), U^\mu[i](x))$  (see Appendix \ref{Sec:MatterinGR}).  We define our op-space to be
\begin{equation}
\mathbf{S}= (\rho_1, \rho_2, \rho_3)
\end{equation}

We will describe an initial distribution of fluids by describing three contributions at time $t=0$. We use the function
\begin{equation}
\varphi(x)=
\left\{
\begin{array}{ll}
\exp\left(\frac{-1}{1-x^2}\right) &\text{for} |x|<1 \\
0  & \text{otherwise}
\end{array}
\right.
\end{equation}
This is a smooth function (it is a bump function).

The first contribution is of fluid of type $2$ according to
\begin{equation}
\rho_2=\varphi(r-4), ~~~~~ U^\mu[2] = (a, 0, v_r)
\end{equation}
where $v_r>0$, the radial speed of the dust particles, is a constant and $a$ is chosen so that $g_{\mu\nu}U^\mu[2]U^\nu[2]=1$.

The second contribution is of fluid of type $1$ according to
\begin{equation}
\rho_1=\varphi(r-2), ~~~~~ U^\mu[2] = (b, 0, V_r(r))
\end{equation}
where $V_r(r)\geq 0$ is much greater than $v_r$ in the interval from $r=1$ to $r=3$ and $b$ is chosen so that $g_{\mu\nu}U^\mu[1]U^\nu[1]=1$.  Fluid 1 from this contribution will overtake fluid 2 as the fluids expand.

The third contribution is a mixture of all three fluids such that
\begin{equation}
U^\mu[1]=U^\mu[2]=U^\mu[3] = (b, 0, V_r(r))
\end{equation}
and
\begin{equation}
\rho_1+\rho_2+\rho_3 = f(r)
\end{equation}
(so it is is symmetric under rotations).  For this third contribution we impose that $\rho_1=\rho_2=0$ for all $1\leq r \leq 5$.  We impose that (i) $\rho_1> \rho_2$ for all $r \leq 1$, (ii) $\rho_2 > \rho_1$ for all $r> 5$.  Further, we impose that the distributions of the individual fluids are \emph{not} rotationally symmetric for $0< r< 1$ and $5<r$.
Finally, we impose that $f(r)=k$ up to some value of $r$ much greater than $5$ after which it tails off.  Here $k$ is a constant which we take to be much bigger than $\varphi(0)$.

The over all situation here is that we have three regions at time $t=0$.  The first region for $0<r<1$ is not rotationally symmetric and has $\rho_1>\rho_2$.  The second region, for $1\leq r\leq 5$ is rotationally symmetric.  The third region, for $5< r$, is not rotationally symmetric and has $\rho_1<\rho_2$.  However, if we ignore the colours of the fluids, then we have over all rotational symmetry.
Hence it is not inconsistent to choose a metric at time $t=0$ that is also rotationally symmetric. Further, the metric will remain symmetric under rotations as the system evolves.

As the system evolves the evolution will be dominated by the third contribution.  We assume that the overall densities are such that gravity plays a very small role. Hence the fluid will diffuse tending to zero density after a long time.  In the region $1\leq r\leq 5$ the distributions $\rho_1$ and $\rho_2$ will evolve into one another.  The value of $\rho_1$ at the values $r$ for which $\rho_1=\rho_2$ will, for a while, increase monotonically with time from $\rho_1=0$ at time $t=0$ to the value $\rho_1=q$ at time $t=T$ where we take $T$ to be small enough that the asymmetries have still not propagated into the region $2\leq r\leq 4$.   During this same time, $\rho_3$ will decrease from its maximum value, $\rho_3=k$ at this place to a slightly smaller value, $\rho_3=k^-$. It is possible, after a long time, there will be other places where $\rho_1=\rho_2$.  However, any such places will have a much smaller value of $\rho_3$ and consequently do not happen in the same part of op-space.  Given these considerations, we define the bounding surface, $\mathtt{a}$, where
\begin{equation}
\text{set}(\mathtt{a})=\left\{(\rho_1, \rho_2, \rho_3): \text{s.t.}~ \rho_1=\rho_2, ~ 0 \leq \rho_1 \leq q, ~       k^- \leq \rho_3 \leq k \right\}
\end{equation}
We can choose any suitable coordinates for this typing surface (recall that a typing surface also has a set of coordinates associated with it).
This bounding surface is designed to pick out the circularly symmetric part of the solution above.

Now we wish to choose two regions, $\mathtt{A}$ and $\mathtt{B}$, of op-space where $\mathtt{A}$ corresponds to the interior (inside where $\rho_1=\rho_2$) and $\mathtt{B}$ correspond to the exterior (outside where $\rho_1=\rho_2$).  We will require that $\mathtt{A}$ and $\mathtt{B}$ meet only at $\mathtt{a}$.  We define $\mathtt{A}$ to be
\begin{equation}
\mathtt{A}=\left\{(\rho_1, \rho_2, \rho_3):0\leq \rho_2 \leq \rho_1 \leq q,  ~   h(\rho_1 - \rho_2) \leq \rho_3 \leq k \right\}
\end{equation}
and $\mathtt{B}$ to be
\begin{equation}
\mathtt{B}=\left\{(\rho_1, \rho_2, \rho_3):0\leq \rho_1 \leq \rho_2 \leq q,~  h(\rho_2 - \rho_1) \leq \rho_3 \leq k \right\}
\end{equation}
where
\begin{equation}
h(x) = k^- + \alpha \frac{x}{q}
\end{equation}
and $\alpha$ is chosen to be big enough that the interesting action happening between time $0$ and $T$ is captured in $\mathtt{A}$ and $\mathtt{B}$, but small enough that not much of what happens before or after these times is captured in these regions.

We can solve and find a solution, $\tilde{\Psi}$, in the given coordinate system.  Using this we can we can calculate $\Psi_\mathtt{A}$ and $\Psi_\mathtt{B}$ (by restricting to $\mathtt{A}$ and $\mathtt{B}$ respectively, and then acting with all $\vartheta\in G_\text{diffeo}$). These solutions will be pure.    Now, the interior will be mapped into $\mathtt{A}$ and the exterior will be mapped into $\mathtt{B}$.  We can find another solution, $\tilde{\Psi}'$ by rotating the initial fields in the region $0\leq r\leq 1$ (since these are rotationally asymmetric this will describe a new physical situation).  With this new solution we can calculate $\Psi'_\mathtt{A}$ and $\Psi'_\mathtt{B}$ by the same process as before.  It is clear that we will obtain
\begin{equation}
\Psi'_\mathtt{A}= \Psi_\mathtt{A} ~~~~~ \Psi'_\mathtt{B}= \Psi_\mathtt{B}
\end{equation}
However, the extremal invariant distances between the asymmetric features in the inner and outer regions is different for the two solutions. Hence
\begin{equation}
\Psi'_{\mathtt{A}\cup\mathtt{B}} \not= \Psi_{\mathtt{A}\cup\mathtt{B}}
\end{equation}
The boundary conditions between the two regions is the same in each case
\begin{equation}
\theta_\mathtt{a} (\Psi_\mathtt{A}) = \theta_\mathtt{a} (\Psi_\mathtt{B}) = \theta_\mathtt{a} (\Psi'_\mathtt{A}) = \theta_\mathtt{a} (\Psi'_\mathtt{B})
\end{equation}
Hence, when we form
\begin{equation}
\Psi_\mathtt{A} \Cup_\mathtt{a} \Psi_\mathtt{B}
\end{equation}
we will pick up both solutions (and, indeed, infinitely many more since we get one for each rotation) and hence we obtain a mixed solution.  This means that specifying $\Psi_\mathtt{A}$ and $\Psi_\mathtt{B}$ separately does not specify $\Psi_{\mathtt{A}\cup\mathtt{B}}$ and hence General Relativity is non-separable.

It is worth noting a few things. First, we could use the metric to define more scalars and enlarge our op-space.  However, the metric itself is rotationally symmetric.  Hence, we could not rescue the situation and restore separability this way.  Second, this non-separability is a consequence of insisting that beables are invariant under diffeomorphisms.  In pre-General Relativistic classical field theories, there is a fixed background and the coordinate, $x$, ascribed to a particular point is regarded as part of the physical description.  In such theories, we would have separability even for an example having similar rotationally symmetry.    This is because then the internal rotation described in the previous paragraph would create an ontologically new situation (as the coordinate, $x$, is part of the ontology).

This non-separability of General Relativity demonstrates that it GR has a certain nonclassicality not shared by the earlier classical theories (unless we retrospectively impose something like diffeomorphism invariance on them).

\section{Agency}\label{sec:Agency}

\subsection{Introducing Agency}

\index{agency}

The notion of agency is well incorporated into Quantum Theory.  We can allow a quantum system to pass through some apparatus having knobs on it. For different settings of the knobs we get different evolution.   These different evolutions are associated with different completely positive maps acting on the incoming state.  In General Relativity, on the other hand, the usual picture is that of the block universe - a single solution for all points in space and time given.  Such solutions are usually obtained (in the world of numerical relativity at least) by taking an initial space-like hypersurface with fields defined on it and evolving this according to a canonical formulation of General Relativity.  Such a picture makes it difficult to incorporate agency.  We can, however, think of agency as an effective notion to deal with situations having agents. Here agents are systems that can, in a controlled manner, magnify some property that is below the resolution we are solving the field equations to being above this resolution.

One way to introduce agency in General Relativity is through a force density field, $G^\mu$.  The force density field can be set by varying some other field, $\chi^{\mu\dots\nu}$. We will give an example below.   We will call fields such as $\chi^{\mu\dots\nu}$ \emph{agency fields}. \index{agency fields} We will let $\pmb{\chi}$ be the list of all the agency fields.  The agency fields are in addition to the matter and metric fields already encoded in $\pmb{\Phi}$.

There is a problem though.  Agents can only control what they can observe.  By Assertion 1, we can only observe scalar fields taking certain values in coincidence with one another. The fields in in $\pmb{\chi}$ are (in general) tensors so cannot be measured.  Thus, we will assume that agents can set some quantity
\begin{equation}
\mathbf{Q}= (Q_1, Q_2, \dots Q_L)
\end{equation}            \index{Q@$\mathbf{Q}$}
where $Q_l$ (for $l=1$ to $L$) is a scalar.   We use the fields in $\pmb{\Phi}$ to calculate the scalars in $\mathbf{S}$.  Then we will use the scalars in $\mathbf{S}$ along with the fields in $\pmb{\chi}$ to calculate the scalars in $\mathbf{Q}$.

Once we introduce an effective notion of agency we need to introduce another effective notion - a time direction field, $\pmb{\tau}$.  This is because we suppose that the consequences of a choice can only be felt to the future of where the choice was made. General Relativity, on the other hand, is time symmetric so we need to single out a time direction at each point where a choice might be made.

The full set of fields are given by $(\pmb{\Phi}, \pmb{\chi}, \pmb{\tau})$.  The time direction field is not directly observable (it is not a scalar) but rather can be observed through the behaviour of directly observed quantities.  Thus, the full set of observables are $(\mathbf{S}, \mathbf{Q})$.

An important notion is that of an \emph{agency strategy}. \index{agency strategies} This is where we pre-specify what choice of $\mathbf{Q}$ would be made at each point, $\mathbf{S}$, in some region, $\mathtt{A}$, of op-space should that point be realized.  We specify an agent strategy for $\mathtt{A}$ as
\begin{equation}
\mathbf{Q}_\mathtt{A} = \Big\{ \big({\bf S}, \mathbf{Q}(\mathbf{S}) \big): \forall {\bf S}\in \mathtt{A} \Big\}
\end{equation}   \index{QA@$\mathbf{Q}_\mathtt{A}$}
The agent strategy in some region $\mathtt{A}$ is the analogue of  a  choice of knob setting on an apparatus in Quantum Theory.

Causality imposes that choices cannot effect the past.  This leads to a constraint on solutions.  We will describe how to implement this constraint in terms of boundary conditions.

\subsection{An example: a vast fleet of spaceships}

Imagine a single spaceship with a captain at the helm. He can decide to steer the spaceship to the left or right.  Once he has made his choice the spaceship will move, altering the metric and other fields in $\pmb{\Phi}$ accordingly. So it matters for the physics what choice he makes.   His decision to move left or right originates at the level of the neurons in his brain. Assuming that the captain's behavior is, at the fundamental level, determined by the field equations, we could imagine simply deterministically solving the field equations given some initial information that serves to determine what choice the captain will make.  However, usually when we solve such equations, we work to some resolution and, typically, we would expect to put neurons below the level of that resolution.  One reason for this is computational tractability - solving for the behavior of the brain (even assuming that the brain could be described by the classical equations of General Relativity) would be computationally expensive.  Thus, in situations where we have agents, we would like to pursue a different approach.

We asserted, in Sec.\ \ref{sec:fieldsareeverything}, that everything in General Relativity should be described in terms of fields.  We will seek to apply this to agency as well.  To set this up, imagine a fleet of spaceships so vast that we can treat it as a dust fluid.  In this fluid approximation we can imagine that, at each point $p$, we have a choice of how to steer the ship.  To be more concrete, assume that the ships in the fleet are powered through interaction with another dust fluid - the wind fluid.   Then a complete set of equations is given by
\begin{equation}
\nabla_\mu T^{\mu\nu}[\text{ship}] = G^\nu ~~ \nabla_\mu T^{\mu\nu}[\text{wind}] = - G^\nu ~~~ G^{\mu\nu} = 8\pi T^{\mu\nu}
\end{equation}
where
\begin{equation}
T^{\mu\nu}[\text{ship}] = \rho[\text{ship}] U^\mu[\text{ship}] U^\nu[\text{ship}] ~~~~~  T^{\mu\nu}[\text{wind}] = \rho[\text{wind}] U^\mu[\text{ship}] U^\nu[\text{wind}]
\end{equation}
as long as we choose some form for the force density, $G^\mu$, between the ship and the wind.  One possible force density
\begin{equation}
G^\mu = \chi^\mu \chi^\alpha \rho[\text{wind}] U_\alpha[\text{wind}]
\end{equation}
We can choose $\chi^\mu$ to be spacelike.  We can interpret the direction of $\chi^\mu$ is the direction of the normal to the sail and the magnitude of $\chi^\mu$ to be proportional to the cross-section of the sail.  This equation provides for a force density that is proportional to the component of the current of the wind normal to the sail.

There is a problem, however.  As a field, $\chi^{\pmb{\mu}}(p)$, should be specified at every $p\in\mathscr{M}$.  However, we cannot specify a point, $p$, in a diffeomorphism invariant fashion. Further, the coordinates of $\chi^\mu$, when represented in a coordinate system, are also not invariant.  Choices have to be beables. But, more than this, choices have to be observables. Thus, they must correspond to scalars taking certain values in coincidence (according to Assertion 1).  Let us suppose that we have already nominated a set of scalar fields, $\mathbf{S}$, to form the op-space.  Next we can choose a set of scalars, $\mathbf{X}=(X^1, X^2, \dots X^K)$ that are functions of the scalar fields in $\mathbf{S}$.  Now we can define Westman Sonego scalars
\begin{equation}
\chi^k = \frac{\partial X^k}{\partial x^\mu} \chi^\mu
\end{equation}
This is a list of scalars.   We can calculate these scalars at every point, $\mathbf{S}$, in op-space.  Thus, we can write $\chi^k(\mathbf{S})$.  Although an agent cannot specify $\chi^\mu(p)$ (as this is not invariant under diffeomorphisms), he can specify $\chi^k(\mathbf{S})$.  Indeed, $(\mathbf{S}, \chi^k)$ is an observable (as it corresponds to asserting that certain scalars have certain values in coincidence with one another).   We can then define the agent choice to be $\mathbf{Q}=(\chi^k: k=1~\text{to}~K)$.

\subsection{Agency fields}

If there are no agency fields the field equations of General Relativity take the form given by the matter field equations \eqref{Matterfieldequations} and Einstein field equations \eqref{Einsteinfieldequations}.

\index{agency fields}

If we have agency fields, such as $\chi^{\mu\nu}$ in the above example, then we now have to include these in the field equations.  The left hand side of the Einstein field equation is very constrained by the way in which the equation was obtained. Hence we cannot expect it to depend on agency fields (at least not without modifying General Relativity).   The right hand side could, in principle, depend on the agency fields as long as $\nabla_\nu T^{\mu\nu}=0$ remains true.  However, it seems more reasonable that the total stress-energy tensor does not have a functional dependence on the agency fields since different choices of these fields correspond to small differences below the resolution we are working to.  In fact, for the same reason, it is reasonable to demand that the stress-energy tensors, $T^{\mu\nu}[n]$, associated with the different matter fields do not have a functional dependence on the agency fields.  Rather, we can make the physics depend on the agency fields through the matter field equations alone.  Thus, we leave (\ref{Einsteinfieldequations}) unchanged and we modify (\ref{Matterfieldequations}) as follows
\begin{equation}\label{Matterfieldequationswithagency}
f_l(\vec{\varphi}, \vec{\varphi}_{,\alpha}, \vec{\varphi}_{,{\alpha\beta}}, g_{\mu\nu}, g_{\mu\nu,\gamma}, \pmb{\chi} ) = 0  ~~~~~l=1~\text{to}~L
\end{equation}
where $\pmb{\chi}$ is a list of agency fields (these are tensor field of various types)
\begin{equation}
\pmb{\chi}=
\left(
\begin{array}{c}
\chi^{\mu_1 \dots \mu_{r_1}}[1] \\
\chi^{\mu_1 \dots \mu_{r_2}}[2] \\
\vdots \\
\chi^{\mu_1 \dots \mu_{r_N}}[N]
\end{array}
\right)
\end{equation}
We insist that all the fields in $\pmb{\chi}$ have only superscripts (not subscripts) so that we can convert them to Westman Sonego scalars as in the example above (we can always achieve this by using $g^{\mu\nu}$ to raise lowered subscripts).  The equations (\ref{Matterfieldequationswithagency}) in conjunction with the Einstein field equations (\ref{Einsteinfieldequations}) form a set of coupled field equations that we can solve if we know the agency fields, $\pmb{\chi}$.

\subsection{Agent choices}\label{sec:agentchoices}

The agent choice \index{agent choices} must be specified by a set of scalars, $\mathbf{Q}$. We insist these are calculated from $\pmb{\chi}$ and $\mathbf{S}$.  One possibility is that we calculate Westman Sonego scalars as follows. First we define
\begin{equation}
\mathbf{X} = (X^1, X^2, \dots X^K)
\end{equation}
as some set of scalars, $X^k(\mathbf{S})$ (i.e.\ they are functions of $\mathbf{S}$.  Now we define
\begin{equation}\label{chitoX}
\pmb{\chi}^{\bf X}  :=
\left(
\begin{array}{l}
\partial_{\mu_1} X^{i_1} \dots \partial_{\mu_{r_1}} X^{i_{r_1}}\chi^{\mu_1 \dots \mu_{r_1}}[1] \\
\partial_{\mu_1} X^{i_1} \dots \partial_{\mu_{r_2}} X^{i_{r_2}}\chi^{\mu_1 \dots \mu_{r_2}}[2] \\
~~~~~~~~~~~~~~~~~\vdots \\
\partial_{\mu_1} X^{i_1} \dots \partial_{\mu_{r_N}} X^{i_{r_N}}\chi^{\mu_1 \dots \mu_{r_N}}[N]
\end{array}
\right)
\end{equation}
This is the list of Westman Sonego scalars corresponding to the list of tensors in $\pmb{\chi}$.  We can now set $\mathbf{Q}=\pmb{\chi}^\mathbf{X}$ for the agent choice.  We could imagine other ways of choosing $\mathbf{Q}$ as a function of $\pmb{\chi}$ and $\mathbf{S}$.

In fact we will demand that $\mathbf{Q}$ is specified as a function of $\mathbf{S}$. This means that $\mathbf{Q}(p)$ is constrained to take the same values for all $p\in \mathscr{M}_\mathbf{S}$.   This is imposed by demanding that solutions are consistent with \emph{agent strategies}. \index{agent strategies} An agent strategy is given by providing
\begin{equation}
\mathbf{Q}_\mathtt{A} = \left\{ (\mathbf{S}, \mathbf{Q}): \forall \mathbf{S}\in \mathtt{A}  \right\}
\end{equation}                      \index{QA@$\mathbf{Q}_\mathtt{A}$}
When we have an agent strategy $\mathbf{Q}_\mathbf{A}$ we demand that $(\mathbf{S}(p),\mathbf{Q}(p))\in \mathbf{Q}_\mathtt{A}$ where $(\mathbf{S}(p),\mathbf{Q}(p))$ is calculated from the points $p\in \mathscr{M}_\mathtt{A}$ for every $\tilde{\Psi}_\mathtt{A}\in\Psi_\mathtt{A}$.

\subsection{Time direction field}\label{sec:timedirectionfield}

Agents are time asymmetric as they magnify below resolution properties to being above resolution.  This happens in some given time direction.  The field equations of General Relativity are time-symmetric.  Thus, if we introduce the effective notion of agency then we also need to introduce an effective notion of time direction.  The metric already specifies a double light cone structure at every point in the manifold.  The time direction field must pick out a forward light cone at every point in the manifold.  To do this we introduce an extra effective field - the \emph{time direction field}, \index{time direction field} $\tau^\mu$ \index{tau@$\pmb{\tau}$} and then allow gauge freedom (as any vector in the forward light cone will do the job).   We demand that
\begin{equation}
-g_{\mu\nu}\tau^\mu\tau^\nu > 0
\end{equation}
so that it is time-like (the minus sign is because we are using the $(-, +, \dots, +)$ signature convention).

We will say that a vector, $v^\mu $, is forward pointing if
\begin{equation}
-g_{\mu\nu} v^\mu v^\nu > 0 ~~~~\text{and}~~~~ -g_{\mu\nu} \tau^\mu v^\nu  > 0
\end{equation}
since then it belongs to the same light cone as $\pmb{\tau}$.  At each point, $p\in\mathscr{M}$, we define the set
\begin{equation}
C^+({\bf g}, \pmb{\tau}) = \{ {\bf v}: \forall {\bf v}~~\text{s.t.}~~ -g_{\mu\nu} v^\mu v^\nu > 0 ~~~~\text{and}~~~~ -g_{\mu\nu} \tau^\mu v^\nu  > 0 \}
\end{equation}
This defines the forward cone at $p$.

If $\tau^\mu$ is a time direction field then $\tilde{\tau}^\mu$ is a physically equivalent time direction field iff
\begin{equation}
\tilde{\pmb{\tau}}(p) \in C^+[\pmb{\tau}(p), {\bf g}(p)] ~\forall p
\end{equation}
We will write
\begin{equation}
\tilde{\tau}^\mu=\sigma^{\mu}_\nu \tau^\nu
\end{equation}
and we say that
\begin{equation}
\pmb{\sigma} \in \Sigma^+
\end{equation}
for maps, $\pmb{\sigma}$, that transform between physically equivalent time direction fields.  At each point $p$ this corresponds to a time direction preserving Lorentz boost in a local Lorentz frame (in which the metric is the Minkowski metric $\eta_{\bar{\mu}\bar{\nu}}$) times a positive rescaling factor, $\alpha^2$.  We can write
\begin{equation}\label{sigmaformone}
\sigma^\mu_\nu = \alpha^2 e\indices{^\mu_{\bar{\mu}}} \Lambda^{\bar{\mu}}_{\bar{\nu}} e\indices{_\nu^{\bar{\nu}}}
\end{equation}
where $\Lambda^{\bar{\mu}}_{\bar{\nu}}$ is a lorentz boost that preserves time orientation and where the veilbein, $e^\mu_{\bar{\mu}}$, is a transformation that takes us to a local frame (with coordinates $x^{\bar{\mu}}$) in which we have the Minkowski metric.  It satisfies
\begin{equation}\label{vielbeinmetric}
g^{\mu\nu} = e\indices{^\mu_{\bar{\mu}}} e\indices{^\nu_{\bar{\nu}}} \eta^{\bar{\mu}\bar{\nu}}
\end{equation}
Further, we have
\begin{equation}
e\indices{_\mu^{\bar{\mu}}} = g_{\mu\nu} \eta^{\bar{\mu}\bar{\nu}} e\indices{^\nu_{\bar{\nu}}}
\end{equation}
for the veilbein with indices flipped.  Another way to represent the transformation $\sigma^\mu_\nu$ is to use two different vielbeins
\begin{equation}\label{sigmaformtwo}
\sigma^\mu_\nu = \alpha^2 {e'}\indices{^\mu_{\bar{\mu}}}  e\indices{_\nu^{\bar{\nu}}}    ~~~ \text{where} ~~~ {e'}\indices{^\mu_{\bar{0}}} e\indices{_\mu^{\bar{0}}} < 0
\end{equation}
where ${e'}\indices{^\mu_{\bar{\mu}}}$ also satisfies (\ref{vielbeinmetric}).  The condition ${e'}\indices{^\mu_{\bar{0}}} e\indices{_\mu^{\bar{0}}} <0$ ensures that the time-like axis (associated with $\bar{\mu}=\bar{0}$) has the same direction for the two vielbeins (it must be negative because of our signature convention).  Since different choice of vielbein satisfying this condition are related by a time orientation preserving Lorentz transformation,  (\ref{sigmaformone}) and (\ref{sigmaformtwo}) are equivalent.

This gauge freedom is to be expected as we are using $D$ real numbers in $\tau^\mu$ to specify a single bit of information (which is the forward light cone as opposed to the backward light cone).  Really, we should consider $(\tau^\mu(p), g_{\mu\nu}(p))$ as a pair since, taken together, they specify the forward light cone.  We can cook up other ways of specifying the time direction field and metric jointly.  For example, we could use the veilbein with a forward pointing $e^\mu_0$ component.  For our purposes, however, it suffices to work with the pair, $(\tau^\mu, g_{\mu\nu})$, to represent time direction and metric information.  One reason for making this choice is that can choose $\pmb{\tau}(p)=0$ for points on the manifold where we do not need a time direction field (in particular, for points that are outside the forward light cone of any places where agents can act).

One way to measure $\pmb{\tau}$ is by exploding a \lq\lq causal puffball\rq\rq. \index{causal puffball} This would consist of an explosion fluids out from a small volume where the fluid elements travel at all speeds up to that of light.  If an agent decides to explode a causal puffball at $\bf S$ then coming out from this point, in the op-space, would be all these different fluids marking out the future light cone. Note this only determines $\pmb{\tau}$ up to physically irrelevant aspects - i.e.\ it determines the future light cone.  In the absence of a measurement to measure the time direction field, it is possible that we cannot read $\pmb{\tau}$ from the op space.  Thus, time direction must be considered to live at the hidden variable level.

\subsection{Representing solutions with agency}

Now we have introduced additional effective fields, $\pmb{\chi}$ and $\pmb{\tau}$, we need to revisit how to represent solutions.  We write a solution for region $\mathtt{A}$ as
\begin{equation}
\tilde{\Psi}_\mathtt{A} = \left\{ (p, \pmb{\Phi}, \pmb{\chi}, \pmb{\tau}): \forall p\in\mathscr{M}_\mathtt{A} \right\}
\end{equation}     \index{PsiAtilde@$\tilde{\Psi}_\mathtt{A}$}
where, just like the case without agency,
\begin{equation}
\mathscr{M}_\mathtt{A}= \left\{ p: \forall p ~\text{s.t.}~ \mathbf{S}(p)\in\mathtt{A} \right\}
\end{equation}
This solution is in gauge dependant form.

Now we have two gauge groups acting, $G_\text{diffeo}$ and $\Sigma^+$ (for the time direction field) where $\varphi\in G_\text{diffeo}$ and $\pmb{\sigma}\in\Sigma^+$.  We will combine them into a single gauge group, $G^+$, \index{G@$G^+$} with elements $\vartheta=(\pmb{\sigma}, \varphi)$ which act as
\begin{equation}
\vartheta^*\tilde{\Psi}_\mathtt{A} = \left\{ (p, \varphi^*\pmb{\Phi}, \varphi^*\pmb{\chi}, \pmb{\sigma} \varphi^*\pmb{\tau}): \forall p\in\varphi(\mathscr{M}_\mathtt{A}) \right\}
\end{equation}
Compare this with \eqref{phitildePsiA}.  Note, in particular, that the $\pmb{\sigma}$ part of the transformation only acts on the $\pmb{\tau}$ field.  This means, for example, if there is a field, $J^\mu$ (calculable from $\pmb{\Phi}$) that is parallel to $\tau^\mu$ before the $\vartheta$ transformation, these two vectors may not be parallel afterwards.

We can now give a gauge independent representation of the solution as
\begin{equation}
\Psi_\mathtt{A} = \left\{ \vartheta^*\tilde{\Psi}_\mathtt{A}: \forall \vartheta\in G^+ \right\}
\end{equation}      \index{PsiA@$\tilde{\Psi}_\mathtt{A}$}
This solution is invariant under the gauge group, $G^+$.  We define beables to be quantities that are invariant under $G^+$.  In particular, we can write beables as
\begin{equation}
B_\mathtt{A}(\Psi_\mathtt{A})
\end{equation}
where this is guaranteed to be invariant under $G^+$ because the elements of the set, $\Psi_\mathtt{A}$, come in no particular order.

\subsection{Turning a solution with agency inside out}\label{sec:turningasolutionwithagencyinsideout}

In Sec.\ \ref{sec:turningasolutioninsideout} we discussed how to turn a solution \lq\lq inside out".  We can represent a solution with agency similarly:
\begin{equation}
\Psi_\mathtt{A} = \left\{ (\mathbf{S}, \mathbf{Q}, \tilde{\lambda}):\forall \mathbf{S}\in \Gamma_\mathtt{A} \right\}
\end{equation}   \index{solutions!inside out form}
To define the hidden variables, $\tilde{\lambda}$, we first define $\omega$ such that
\begin{equation}
(\pmb{\Phi}(p), \pmb{\chi}(p))  \leftrightarrow (\mathbf{S}(p),\mathbf{Q}(p), \pmb{\omega}(p)  )
\end{equation}
In words, $\omega$ are some variables that capture the information in $(\pmb{\Phi}(p), \pmb{\chi}(p))$ that is not contained in $(\mathbf{S}(p),\mathbf{Q}(p))$.  Now we define
\begin{equation}
\tilde{\lambda}(\mathbf{S})= \left\{ (p, \omega, \tau): \forall p\in\mathscr{M}_\mathbf{S} \right\}
\end{equation}
These are the hidden variables associated with the point $\mathbf{S}$.  Note that the time direction field gets included with the hidden variables.

We can, further, define
\begin{equation}
\lambda = \{ \vartheta^*\tilde{\lambda}: \forall \vartheta\in G^+ \}
\end{equation}
Given this we can define
\begin{equation}\label{PsiAinsideoutnotildesG}
\Psi_\mathtt{A} = \left\{ (\mathbf{S}, \lambda): \forall \mathbf{S}\in \Gamma_\mathtt{A} \right\}
\end{equation}
This is a $G^+$ invariant way of presenting the solution that separates it into observables and hidden variables.

\subsection{Solving field equations given agent strategies}

When we have agency the field equations consist of the matter field equations with agency \eqref{Matterfieldequationswithagency} and the Einstein field equations \eqref{Einsteinfieldequations}.  We will call these equations
\begin{equation}
\text{FieldEqns}_\text{GR}^\text{agency}
\end{equation}
We have as many independent matter field equations as we have degrees of freedom in $\pmb{\Phi}$.  The agency fields in $\pmb{\chi}$ are extra fields that have to be specified before we can solve the field equations.  However, specifying these fields as a function of $p\in \mathscr{M}$ is not a diffeomorphism invariant way to state the choices of agents.   Rather, we can specify an agent strategy, $\mathbf{Q}_\mathtt{A}$ (see Sec.\ \ref{sec:agentchoices}), for some region, $\mathtt{A}$ of op-space and attempt to find solutions consistent with this.  Additionally, we have the time direction field, $\pmb{\tau}$, that must play a role in solving the field equations since agent choices can only influence the future (defined with respect to $\pmb{\tau}$).

We notate sets of solutions (or candidate solutions) consistent with a given agent strategy as
\begin{equation}
\Omega_\mathtt{A}[\mathbf{Q}_\mathtt{A}, \text{spec}]
\end{equation}
By consistent, we mean that the values of $(\mathbf{S}, \mathbf{Q})$ calculated from these solutions must belong to $\mathbf{Q}_\mathtt{A}$.  For actual solutions sets we need to be sure that choices only influence the future defined with respect to $\pmb{\tau}$.  We will discuss how to implement this in Sec.\ \ref{sec:causality}.  The variable $\text{spec}$ can be set to: $\text{cand}$ for pure candidate solutions consisting of smooth fields); $\text{mixed cand}$ for mixed candidate solutions; omitted for pure solutions that actually solve the field equations; and $\text{mixed}$ for mixed solutions that actually solve the field equations.

If we omit the $\mathbf{Q}_\mathtt{A}$ then we can define sets of solutions $\Omega_\mathtt{A}[\text{spec}]$ that are consistent with the field equations for at least one agent strategy.

\subsection{Boundary conditions}\label{sec:boundaryconditionswithagency}

The definitions for boundary conditions \index{boundary conditions!with agency} we introduced in Sec.\ \ref{sec:fullboundaryconditions} go through as before though we need to redefine $\pmb{\Theta}$ as
\begin{equation}
\pmb{\Theta} = (\pmb{\Phi}, \pmb{\pi}, \partial_{\mathtt{a}\pmb{\mu}} X_\mathtt{a}^0, \pmb{\partial} x_\mathtt{a}, \pmb{\chi}, \pmb{\tau})
\end{equation}
We have appended the agency field, $\pmb{\chi}$, and time direction field, $\pmb{\tau}$, to the original specification of $\pmb{\Theta}$ given in \eqref{vartheta}.

We can define sets of boundary conditions, $\Lambda_\mathtt{a}[\mathtt{A}, \text{spec}]$ as before.  We can also introduce sets,
\begin{equation}
\Lambda_\mathtt{a}[\mathtt{A}, \mathbf{Q}_\mathtt{A}, \text{spec}]
\end{equation}
that are consistent with a given agency strategy.

\subsection{Causality}\label{sec:causality}

We will now provide a condition on
\begin{equation}
\left\{ (\mathbf{Q}_\mathtt{A}, \Omega_\mathtt{A}[\mathbf{Q}_\mathtt{A}]): \forall \mathbf{Q}_\mathtt{A}\in W_\mathtt{A} \right\}
\end{equation}
that encodes causality - namely that agents actions can only influence the future.  Here $W_\mathtt{A}$ is a set of possible agent strategies in $\mathtt{A}$.
Consider $\tilde{\Psi}_\mathtt{A}[\mathbf{Q}_\mathtt{A}] \in \Psi_\mathtt{A}[\mathbf{Q}_\mathtt{A}]\in \Omega_\mathtt{A}[\mathbf{Q}_\mathtt{A}]$.  This solution lives on a manifold patch, $\mathscr{M}_\mathtt{A}$.

\subsubsection{No influence regions}

First we need to define the notion of an \emph{influence region} \index{influence region} and \emph{no influence region} \index{no influence region} of the manifold pertinent to a representation of some solution.
The influence region of region $\mathtt{B}$ of op-space for solution $\tilde{\Psi}_\mathtt{A}$ where $\mathtt{B}\subseteq\mathtt{A}$ is that part of the manifold patch, $\mathscr{M}_\mathtt{A}$, associated with this solution for which there exist forward causal paths within $\mathscr{M}_\mathtt{A}$ from points in the part of the manifold associated with $\mathtt{B}\cap\Gamma_\mathtt{A}$.  Given a solution, $\tilde{\Psi}_\mathtt{A}$ with associated manifold patch, $\mathscr{M}_\mathtt{A}$, and for which $\mathbf{S}$ takes values in $\Gamma_\mathtt{A}\in\mathtt{A}$, we can define
\begin{equation}
\mathscr{M}_{\mathtt{B}|\mathtt{A}} =
\bigcup_{\mathbf{S}\in \mathtt{B}\cap\Gamma_\mathtt{A}} \mathscr{M}_\mathbf{S}
\end{equation}
Then this influence region is given by,
\begin{equation}
\text{Infl}(\mathtt{B}, \tilde{\Psi}_\mathtt{A}) =
\left\{ p : \forall p\in \mathscr{M}_\mathtt{A} ~\text{s.t.}~\exists ~\text{f.c.p. in}~\mathscr{M}_\mathtt{A}~\text{from}~ \mathscr{M}_{\mathtt{B}|\mathtt{A}} \right\}
\end{equation}
where f.c.p. means \lq\lq forward causal path" (defined with respect to the metric and $\pmb{\tau}$ field).  This is the set of points for which it can be established that a signal could pass from the part manifold associated with the region $\mathtt{B}$ of op-space.  Note that we define f.c.p so that $p$ is in the f.c.p. of itself. This means that $\mathscr{M}_\mathtt{B|A}\subseteq \text{Infl}(\mathtt{B}, \tilde{\Psi}_\mathtt{A})$.

The no-influence region is the set of points in $\mathscr{M}_\mathtt{A}$ that (by looking at the solution, $\tilde{\Psi}_\mathtt{A}$) we can be certain are \emph{not} in the future of points in the part of the manifold associated with $\mathtt{B}$.  It is defined
\begin{equation}
\text{NoInfl}(\mathtt{B}, \tilde{\Psi}_\mathtt{A}) =
\left\{ p: \forall p \in \mathscr{M}_\mathtt{A} - \text{Infl}(\mathtt{B}, \tilde{\Psi}_\mathtt{A}) ~\text{s.t.}~\exists~ \text{b.c.p. in} ~\mathscr{M}_\mathtt{A}~\text{from}~
\text{Infl}(\mathtt{B}, \tilde{\Psi}_\mathtt{A})   \right\}
\end{equation}
where b.c.p. means \lq\lq backwards causal path" (defined with respect to the metric and $\pmb{\tau}$ field).  Note that we define this so that $p$ cannot be reached by a b.c.p. from $p$.  It is clear that the union of the influence region and the no-influence region may still be a proper subset of $\mathscr{M}_\mathtt{A}$.  This is because, in the above definitions, we only include paths causal paths within $\mathscr{M}_\mathtt{A}$.  It is possible that there exist causal paths that go from one point in $\mathscr{M}_\mathtt{A}$ to another but passing through points not in $\mathscr{M}_\mathtt{A}$.

\subsubsection{Causality condition}\label{sec:causalitycondition}

Given these definitions we can provide the causality condition.  The key idea of the causality condition is that, for any pair of strategies differing only in some region, $\mathtt{B}$,  there must exist solutions that are the same when restricted to the no-influence region.  \index{causality condition}
\begin{quote}
\emph{Causality:}
Consider any pair of strategies, $\mathbf{Q}_\mathtt{A}$ and $\mathbf{Q}'_\mathtt{A}$ that differ only in some region, $\mathbf{B}\subseteq \mathbf{A}$, of op-space. Now consider any solution
\begin{equation}
\tilde{\Psi}_\mathtt{A}[\mathbf{Q}_\mathtt{A}] \in \Psi_\mathtt{A}[\mathbf{Q}_\mathtt{A}]\in \Omega_\mathtt{A}[\mathbf{Q}_\mathtt{A}]
\end{equation}
We require that there exists a solution
\begin{equation}
\tilde{\Psi'}_\mathtt{A}[\mathbf{Q}'_\mathtt{A}] \in \Psi'_\mathtt{A}[\mathbf{Q}'_\mathtt{A}]\in \Omega_\mathtt{A}[\mathbf{Q}'_\mathtt{A}]
\end{equation}
such that
\begin{equation}
\text{NoInfl}(\mathtt{B}, \tilde{\Psi}'_\mathtt{A}[\mathbf{Q}'_\mathtt{A}])  = \text{NoInfl}(\mathtt{B}, \tilde{\Psi}_\mathtt{A}[\mathbf{Q}_\mathtt{A}])  = \overline{\mathscr{B}}
\end{equation}
and
\begin{equation}
\text{restrict}_{\overline{\mathscr{B}}} (\Psi_\mathtt{A}[\mathbf{Q}_\mathtt{A}]) = \text{restrict}_{\overline{\mathscr{B}}} (\Psi'_\mathtt{A}[\mathbf{Q'}_\mathtt{A}])
\end{equation}
\end{quote}
This is actually a constraint on the collection sets, $\Omega_\mathtt{A}[\mathbf{Q}_\mathtt{A}]$ over different $\mathbf{Q}_\mathtt{A}$.  If this causality condition is satisfied we can tell a story in which different choices agents might make only effect the future.  To understand this constraint a little more it is useful to look at what it imposes on boundary conditions.

\subsubsection{Deterministic input oriented boundary conditions}

Consider a typing surface, $\mathtt{a}$. Let $\mathscr{M}_{\mathtt{a}|\mathtt{A}}$ be the set of points in $\mathscr{M}_\mathtt{A}$ associated with $\Gamma_\mathtt{A}\cap\text{set}(\mathtt{a})$ for the solutions, $\tilde{\Psi}_\mathtt{A}[\mathtt{a}+]$, to be discussed now.  We want these solutions to have the property that $\mathscr{M}_{\mathtt{a}|\mathtt{A}}$ is to the past of all other points in $\mathscr{M}_\mathtt{A}$. For these solutions we can think of the typing surface, $\mathtt{a}$, as being input oriented.   We define such solutions as:
\begin{equation}
\tilde{\Psi}_\mathtt{A}[\mathtt{a}+,\mathbf{Q}_\mathtt{A}]=\left\{ (p, \pmb{\Phi}): p\in \mathscr{M}_\mathtt{A} ~~\text{s.t.} ~ \mathscr{M}_{\mathtt{a}|\mathtt{A}}\subseteq \text{NoInfl}(\mathbf{A}-\text{set}(\mathtt{a}), \tilde{\Psi}_\mathtt{A}[\mathtt{a}+,\mathbf{Q}_\mathtt{A}]) \right\}
\end{equation}
If we act on any such solution with an element of $G^+$ then the new solution will also be an example of this type. Thus, we can form
\begin{equation}
\Psi_\mathtt{A}[\mathtt{a}+,\mathbf{Q}_\mathtt{A}] = \{ \vartheta^* \tilde{\Psi}_\mathtt{A}[\mathtt{a}+,\mathbf{Q}_\mathtt{A}]: \forall \vartheta\in G^+ \}
\end{equation}
We define
\begin{equation}
{\Lambda}_\mathtt{a}[\mathtt{a}+,\mathbf{Q}_\mathtt{U}] = \left\{ {\theta}_\mathtt{a}({\Psi}_\mathtt{A}[\mathtt{a}+, \mathbf{Q}_\mathtt{A}]):\forall~\mathtt{A},~ \forall {\Psi}_\mathtt{A}[\mathtt{a}+,\mathbf{Q}_\mathtt{A}], ~\text{s.t.}~ \mathbf{Q}_\mathtt{A}\subseteq\mathbf{Q}_\mathtt{U}\right\}
\end{equation}
This is the set of boundary conditions induced by such solutions.  The following follows from the causality assumption \index{boundary causality condition}
\begin{quote}
\emph{Boundary causality condition:} We have
\begin{equation}
\Lambda_\mathtt{a}[\mathtt{a}+,\mathbf{Q}_\mathtt{U}] = \Lambda_\mathtt{a}[\mathtt{a}+,\mathbf{Q}'_\mathtt{U}]
\end{equation}
for any pair of agent strategies $\mathbf{Q}_\mathtt{U}$ and $\mathbf{Q}'_\mathtt{U}$ that agree on $\text{set}(\mathtt{a})$.
\end{quote}
In other words, the set of possible boundary conditions does not depend on choices made in the future.   This means we can simply write
\begin{equation}
\Lambda_\mathtt{a}[\mathtt{a}+,\mathbf{Q}_\mathtt{a}]
\end{equation}
where $\mathbf{Q}_\mathtt{a}$ is an agency strategy for the region, $\text{set}(\mathtt{a})$, of op-space.  We can define the deterministic input oriented boundary condition
\begin{equation}
a_\text{det}[\mathtt{a}+, \mathbf{Q}_\mathtt{a}] = \text{flatten}(\Lambda_\mathtt{a}[\mathtt{a}+,\mathbf{Q}_\mathtt{a}])
\end{equation}
This boundary condition is independent of future choices.  This is reminiscent of the Pavia causality condition \cite{chiribella2010informational} which says that we have a unique deterministic effect.

\subsubsection{Causal completeness}

If we think in terms of obtaining a solution by evolving a state across time then we do not want to encounter a situation in which we simply cannot evolve the state any further. This would happen if the state space for the future did not admit the forward evolution of the state we are trying to evolve.   Similarly, if we have an output oriented boundary condition (coming from the past) then we want there to be a match in the input oriented boundary space.  We will call this the \emph{causal completeness condition} \index{causal completeness condition} which we write this condition down as
\begin{equation}
\Lambda_\mathtt{a}[\mathtt{a}-,\mathbf{Q}_\mathtt{A}] \subseteq \Lambda_\mathtt{a}[\mathtt{a}+,\mathbf{Q}_\mathtt{a}]
\end{equation}
Here $\Lambda_\mathtt{a}[\mathtt{a}-,\mathbf{Q}_\mathtt{A}]$ is the set of boundary conditions induced by solutions for some region $\mathtt{A}$ that are to the past of the typing surface.  To state what we mean more precisely, first we define
\begin{equation}
\tilde{\Psi}_\mathtt{A}[\mathtt{a}-,\mathbf{Q}_\mathtt{A}]=\left\{ (p, \pmb{\Phi}): p\in \mathscr{M}_\mathtt{A} ~~\text{s.t.}~ \mathscr{M}_\mathtt{A}- \mathscr{M}_{\mathtt{a}|\mathtt{A}}\subseteq \text{NoInfl}(\text{set}(\mathtt{a}), \tilde{\Psi}_\mathtt{A}[\mathtt{a}-,\mathbf{Q}_\mathtt{A}]) \right\}
\end{equation}
These are solutions for which  $\mathscr{M}_\mathtt{A}- \mathscr{M}_{\mathtt{a}|\mathtt{A}}$ is in the no influence region of $\text{set}(\mathtt{a})$ (so we can think of typing surface, $\mathtt{a}$, as output oriented).  From these we can define $\Psi_\mathtt{A}[\mathtt{a}-,\mathbf{Q}_\mathtt{A}]$ (as the set formed by the action of $G^+$).  Then $\Lambda_\mathtt{a}[\mathtt{a}-,\mathbf{Q}_\mathtt{A}]$ is the set of boundary conditions induced by all such solutions.

An issue with the causal completeness condition is that, as noted in Sec.\ \ref{sec:puresolutions} there may exist solutions in $\Omega_\mathtt{A}$ that do not exist in $\Omega_\mathtt{A||B}$ when $\mathtt{A}\subset \mathtt{B}$.  Hence, it is not completely clear how confident we can be that solutions associated with some smaller region, $\mathtt{A}$, will necessarily extend into solutions for a bigger region.  this condition is true in General Relativity.  Nonetheless, the existence of canonical formulations of General Relativity suggests we can satisfy this condition at least in those circumstances where the canonical formulation works.

\subsubsection{Further discussion}

The above discussion makes most sense when we have no closed causal loops.  Ideally we would be able to rule these out from our causality condition.  We leave determining whether this can be done for future work.

The causality condition is actually a condition on sets of solutions.  A question is how can we go about generating such sets?  One approach would be to use a canonical formulation of General Relativity and simply solve by evolving forward in time from some set of solutions.  There may be other techniques.

Causality is a central issue in an operational reformulation of General Relativity such as this.  Here we have only scratched the surface of this interesting topic.  However, we see that, by allowing agents as part of the picture, we can give a different kind of treatment of this subject.

\subsection{Special solutions}\label{sec:specialsolutionswithagency}

We can define special solutions as before (see Sec.\ \ref{sec:specialsolutions}), adapting them when necessary to take into account agency.

The empty solution, $\Psi_\mathtt{A}[\varnothing]$, is consistent with any agent strategy since its elements consist of the null manifold (a manifold with no points).

We can define a given $\Gamma_\mathtt{A}$ solution consistent with agent strategy $\mathbf{Q}_\mathtt{A}$
\begin{equation}
\Psi_\mathtt{A} [\mathbf{Q}_\mathtt{A},\Gamma_\mathtt{A}] = \text{flatten}(\Omega_\mathtt{A}[\mathbf{Q}_\mathtt{A},\Gamma_\mathtt{A}])
\end{equation}
where
\begin{equation}
\Omega_\mathtt{A}[\mathbf{Q}_\mathtt{A},\Gamma_\mathtt{A}] \subseteq \Omega_\mathtt{A}[\mathbf{Q}_\mathtt{A}]
\end{equation}
is the set of pure solutions consistent with the given $\Gamma_\mathtt{A}$.

Similarly, we define the \emph{operational solution} consistent with a given $\mathbf{Q}_\mathtt{A}$ associated with $O_\mathtt{A}$ as
\begin{equation}
\Psi_\mathtt{A}[\mathbf{Q}_\mathtt{A}, O_\mathtt{A}] =\text{flatten}(\Omega_\mathtt{A}[\mathbf{Q}_\mathtt{A}, O_\mathtt{A}])
\end{equation}
where
\begin{equation}
\Omega_\mathtt{A}[\mathbf{Q}_\mathtt{A}, O_\mathtt{A}] = \bigcup_{\Gamma_\mathtt{A}\in O_\mathtt{A}} \Omega_\mathtt{A}[\mathbf{Q}_\mathtt{A}, \Gamma_\mathtt{A}]
\end{equation}
is the set of all pure solutions consistent with $O_\mathtt{A}$.  We let
\begin{equation}
\Omega_\mathtt{A}[\mathbf{Q}_\mathtt{A}, \text{op}]
\end{equation}
be the set of all operational solutions in $\mathtt{A}$ consistent with $\mathbf{Q}_\mathtt{A}$.

The \emph{deterministic solution} for region $\mathtt{A}$ given $\mathbf{Q}_\mathtt{A}$ is
\begin{equation}
\Psi_\mathtt{A}[\mathbf{Q}_\mathtt{A}, \text{det}] = \text{flatten}(\Omega_\mathtt{A}[\mathbf{Q}_\mathtt{A}])
\end{equation}
This is the union of all pure solutions for region $\mathtt{A}$ consistent with $\mathbf{Q}_\mathtt{A}$.

\subsection{Agent jitter}\label{sec:Agentjitter}

In Sec.\ \ref{sec:determinism} we stated a determinism assumption (in the absence of agency).  It is interesting to re-examine this assumption when we have agency.  Consider a pure boundary condition, \begin{equation}
\mathbnd{a} \in \Lambda_\mathtt{a}[\mathtt{A}, \mathbf{Q}_\mathtt{A}]
\end{equation}
where the typing surface corresponds to the full boundary of $\mathtt{A}$ (so $\text{set}(\mathtt{a})=\text{boundary}(\mathtt{A})$).  Since this is a pure boundary condition fully surrounding the region, $\mathtt{A}$, is it plausible that it fully determines the solution so that $\Psi_\mathtt{A}[\mathbf{Q}_\mathtt{A}, \mathbnd{a}]$ is pure.
We know (from the way in which the field equations are obtained) that there are as many equations as there are real fields in $\pmb{\Phi}$ (minus the four equations $\nabla_\mu G^{\mu\nu}$ associated with diffeomorphism invariance).  Hence, the key question is whether the constraints coming from the agent strategy, $\mathbf{Q}_\mathtt{A}$, provide as many extra equations as there are real fields in $\pmb{\chi}$.  If so, we have enough equations to find a solution given appropriate boundary conditions.  However, as we will see, it is possible that the agent strategy, $\mathbf{Q}_\mathtt{A}$, does not fully determine the agency fields, $\pmb{\chi}$, in the appropriately diffeomorphism invariant sense.  In this case we do not expect to have determinism.

To illustrate this, consider the case where we choose $\mathbf{Q}= \pmb{\chi}^\mathbf{X}$ as in \eqref{chitoX}. If every point $p\in \mathscr{M}_\mathtt{A}$ has distinct $\mathbf{X}$ (call this the $\mathbf{X}$ non-degenerate case) then the matrix $\frac{\partial X^k}{\partial x^\mu}$ has rank equal to the dimension of $\mathscr{M}$. Thus, we can calculate $\pmb{\chi}$ (represented in the coordinate system $x^\mu$) from $\pmb{\chi}^\mathbf{X}$ (which we have set equal to $\mathbf{Q}$).  Hence, we have as many equations constraining the fields $\pmb{\Phi}$ and $\pmb{\chi}$ as we have real fields.   However, if there is some $\mathbf{X}$ degeneracy so that not every point $p\in \mathscr{M}_\mathtt{A}$ has distinct $\mathbf{X}$ then we could not proceed in this way and we would expect that $\Psi_\mathtt{A}[\mathbf{Q}_\mathtt{A}, \mathbnd{a}]$ is mixed.  Whether we have the $\mathbf{X}$ non-degenerate or degenerate case given only $\mathbf{Q}_\mathtt{a}$ and $\mathbnd{a}$ is a matter of calculation. Though we can read some information off from the boundary condition $\mathbnd{a}$.  If $\mathbnd{a}$ corresponds an intersection $\Gamma_\mathtt{A}\cap\text{set}(\mathtt{a})$ that is lower than $\text{dim}(\mathscr{M})-1$ then $\Gamma_\mathtt{A}$ must have intrinsic dimension smaller than the dimension of $\mathscr{M}$ (at least near to this boundary) and hence we must be in the degenerate case.

If the agent strategy fails to provide a complete set of equations then there will be many distinct pure solutions consistent with $\mathbf{Q}_\mathtt{A}$ and boundary condition $\mathbnd{a}$ and then $\Psi_\mathtt{A}[\mathbf{Q}_\mathtt{A}, \mathbnd{a}]$ will be mixed.  We will call the corresponding noise \emph{agent jitter} \index{agent jitter} - this corresponds to the inability of the agent in these situations to exactly set the agency fields (understood in an appropriate diffeomorphism invariant sense).

\subsection{Joining solutions when we have agency}

We join two solutions when we have agency in the same way as before,
\begin{equation}
\Psi_\mathtt{A}[\mathbf{Q}_\mathtt{A}] \Cup_\mathtt{a} \Psi_\mathtt{B}[\mathbf{Q}_\mathtt{B}]
\end{equation}
the only difference being that we have now expanded the boundary conditions to contain the agency field and the time direction field.

The local matching assumption is essentially the same as before. \index{local matching assumption!with agency}
\begin{quote}
{\bf Local matching assumption (with agency).}
Consider two regions in op-space, $\mathtt{A}$ and $\mathtt{B}$ whose interiors are non-overlapping and which meet at a typing surface $\mathtt{a}$ (which could be composite).
Further, consider pure solutions $\Psi_\mathtt{A}[\mathbf{Q}_\mathtt{A}]\in\Omega_\mathtt{A}[\mathbf{Q}_\mathtt{A}]$ and $\Psi_\mathtt{B}[\mathbf{Q}_\mathtt{B}]\in\Omega_\mathtt{B}[\mathbf{Q}_\mathtt{B}]$.
We assume that, for any physically realistic set, $\text{FieldEqns}_{GR}^\text{agency}$, there exists a choice of $\pmb{\pi}$ \emph{with a finite number of elements} such that
\begin{equation}
\Psi_\mathtt{A}[\mathbf{Q}_\mathtt{A}] \Cup_\mathtt{a} \Psi_\mathtt{B}[\mathbf{Q}_\mathtt{B}]  \in \Omega_{\mathtt{A}\cup\mathtt{B}}[\mathbf{Q}_\mathtt{A}\cup\mathbf{Q}_\mathtt{B}, \text{mixed}]
\end{equation}
for any pair of regions, $\mathtt{A}$ and $\mathtt{B}$.
\end{quote}
Thus, when we join two solutions, we obtain a new solution for the composite region, $\mathtt{A}\cup\mathtt{B}$, consistent with the agent strategy $\mathbf{Q}_\mathtt{A}\cup\mathbf{Q}_\mathtt{B}$.  Note that it is possible that strategies $\mathbf{Q}_\mathtt{A}$ and $\mathbf{Q}_\mathtt{B}$ are inconsistent for some $\mathbf{S}\in \text{set}(\mathtt{a})$.  In this case $\Omega_{\mathtt{A}\cup\mathtt{B}}[\mathbf{Q}_\mathtt{A}\cup\mathbf{Q}_\mathtt{B}, \text{mixed}]$ cannot have any solutions which contain these values of $\mathbf{S}$ since the solutions have to match in their $\pmb{\chi}$ fields.

\newpage

\part{Possibilistic formulation: PoAGeR}
\label{part:possibilisticformulation}

\section{Introduction}

In this part of the paper we will provide a possibilistic formulation of General Relativity. We will call this PoAGeR \index{PoAGeR} (this stands for Possibilistic General Relativity with Agency). In the possibilistic formulation we are only concerned in whether certain observations are possible or not.  If we have a set of mutually exclusive possibilities and only one of them is possible then that one must occur.  The possibilistic formulation of General Relativity is what we are seeking if we wish to know whether certain conditions determine the solution.  One example would be when we have some initial conditions (given in a gauge invariant fashion) and we wish to determine the subsequent evolution.

In the probabilistic formulation, to be given in the next part of this paper, we are concerned with calculating probabilities for different observations.  The possibilistic and probabilistic  approaches will have the same structure.  First we define operations. These correspond to choices made and outcomes seen in some region of op-space.  Then we define encapsulated propositions. These provide propositions as the ontological state in the given region.  Next we define boundary propositions which only concern the ontological state at typing surfaces (that might be associated with some region of op-space).  Then we define generalized states.   Finally, we define operational generalized states.  These allow a calculation to be performed.  The structure is
\begin{equation}
\text{operation} \Rightarrow \text{encap.\ \negs prop} \Rightarrow \text{bound.\ \negs prop} \Leftrightarrow \text{gen.\ \negs state} \Leftrightarrow \text{op.\ \negs gen.\ \negs state}
\end{equation}
Any element determines the elements to the right.  For example, given an encapsulated proposition, we can deduce a generalized state. In some cases we can also go to the left.  In the case of the probabilistic formulation, these elements must be \emph{loaded}. That is, we must specify probabilistic information (we will discuss this in the next part of this paper).

These elements can be composed.  For example, we can join two or more operations to form a new operation associated with the union of the corresponding regions of op-space.

\section{Operations}

\subsection{The idea of an operation}\label{sec:whatanoperationis}

We will now introduce the idea of an operation. This corresponds to what choices are made and what outcomes are seen in some region of op-space.  An operation represents our interface with the underlying physics.   We denote an operation, $\mathsf{A}$, as  \index{operations}
\begin{equation}
\mathsf{A}= \left(\text{strat}(\mathsf{A}), \text{outcome}(\mathsf{A}), \text{reg}(\mathsf{A}), \text{type}(\mathsf{A})\right)
\end{equation}
This contains the following elements
\begin{description}
\item[An agent strategy] $\text{strat}(\mathsf{A})$ of the form $\mathbf{Q}_\mathtt{A}$
\item[An outcome] $\text{outcome}(\mathsf{A})$ of the form $O_\mathtt{A}$ where $O_\mathtt{A}$ is some set of $\Gamma_\mathtt{A}$'s.  If $O_\mathtt{A}$ contains a single element, then we have the fine-grained case,  When $O_\mathtt{A}$ contains multiple or even a continuum of $\Gamma_\mathtt{A}$'s we are in the course-grained case.
\item[A region] $\text{reg}(\mathsf{A})=\mathtt{A}$ of op-space (we adopt the convention of using the same letter, so $\text{reg}(\mathsf{A})=\mathtt{A}$, $\text{reg}(\mathsf{B})=\mathtt{B}$, etc).
\item[A typing surface] $\text{type}(\mathcal{A})$ corresponding to some subset of the boundary of $\text{reg}(\mathsf{A})$.  The typing surface is specified as $(\mathtt{a}, \mathtt{bc})$ for example where $\mathtt{a}$ points away from the enclosed region (so the $X_\mathtt{a}^0$ component points away from the enclosed region) and $\mathtt{bc}$ points towards the enclosed region.
\end{description}
Later we will introduce \emph{loaded operations} which contain some extra information pertaining to probabilistic properties associated with this operation.  We will, further, introduce the notion of \emph{free operations}.  These contain some position arguments that allow them to be placed at arbitrary positions in op-space.

\subsection{Symbolic and diagrammatic notation}

It is useful to introduce superscripts and subscripts corresponding to the typing surfaces.  Thus, we write
\begin{equation}
\mathsf{A}^\mathtt{a}_\mathtt{bc}
\end{equation}
in the case that $\text{type}(\mathsf{A})=(\mathtt{a}, \mathtt{bc})$.  We represent this diagrammatically as
\begin{equation}
\mathsf{A}^\mathtt{a}_\mathtt{bc} ~~~ \Longleftrightarrow
\begin{Compose}{0}{0} \setdefaultfont{\mathsf}\setsecondfont{\mathtt}
\Ucircle{A}{0,0}\thispoint{DL}{-120:4} \thispoint{DR}{-60:4} \thispoint{UC}{90:4}
\joincc[above left]{DL}{60}{A}{-120} \csymbolalt{b}
\joincc[above right]{DR}{120}{A}{-60} \csymbolalt{c}
\joincc[left]{A}{90}{UC}{-90} \csymbolalt{a}
\end{Compose}
\end{equation}
Note that superscripts correspond to outward pointing arrows and subscripts correspond to inward pointing arrows.

\subsection{Joining operations}\label{sec:joiningoperations}

We can compose two operations for two regions to form a new operation for the union of these two regions.  We can join $\mathsf{A}^\mathtt{ab}_\mathtt{cd}$ and $\mathsf{B}^\mathtt{ec}_\mathtt{af}$ when
\begin{equation}
\text{reg}(\mathsf{A}) \cap \text{reg}(\mathsf{B}) =  \text{set}(\mathtt{ac})
\end{equation}
That is the two regions overlap only in some part of the typing surface.  We define the composition of the two operations as
\begin{equation} \label{defnencapprop}
\mathsf{A}^\mathtt{ab}_\mathtt{cd}\mathsf{B}^\mathtt{ec}_\mathtt{af}
= \big( \text{strat}(\mathcal{A})\cup\text{strat}(\mathcal{B}), O_\mathtt{A} \Cup  O_\mathtt{B}, \text{reg}(\mathcal{A})\cup\text{reg}(\mathcal{B}), (\mathtt{be}, \mathtt{df}) \big)
\end{equation}
where
\begin{equation}
O_\mathtt{A} \Cup O_\mathtt{B} = \left\{ \Gamma_\mathtt{A}\cup \Gamma_\mathtt{B}: \forall \Gamma_\mathtt{A}\in O_\mathtt{A}, \forall \Gamma_\mathtt{B}\in O_\mathtt{B} \right\}
\end{equation}
(see discussion in Sec.\ \ref{sec:specialcases}).  This gives us a new operation which we can write as $\mathsf{C}^\mathtt{be}_\mathtt{df}$.  Note that the directions of the typing surfaces where the two regions overlap point in opposite directions.  This is why we match superscripts with subscripts.  We can represent composition of two operations by
\begin{equation}
\mathsf{A}^\mathtt{ab}_\mathtt{cd}\mathsf{B}^\mathtt{ec}_\mathtt{af}  ~~~\Longleftrightarrow~~~
\begin{Compose}{0}{-0.7}   \setdefaultfont{\mathsf}\setsecondfont{\mathtt}
\Ucircle{A}{0,0} \Ucircle{B}{45:5}
\thispoint{Ab}{180:4} \thispoint{Ad}{-90:4}
\thispoint{Be}{$(45:5)+(0:4)$} \thispoint{Bf}{$(45:5)+(90:4)$}
\joincc[above left]{A}{60}{B}{-150} \csymbolalt{a}
\joincc[below right]{B}{-120}{A}{30} \csymbolalt{c}
\joincc[above]{A}{-180}{Ab}{0} \csymbolalt{b} \joincc[left]{Ad}{90}{A}{-90} \csymbolalt{d}
\joincc[above]{B}{0}{Be}{-180} \csymbolalt{e} \joincc[left]{Bf}{-90}{B}{90} \csymbolalt{f}
\end{Compose}
\end{equation}
A special case of joining to operations is  when the two regions associated with the two operations do not overlap at all (they do not meet at any bounding surface). Then we will say we have the \emph{null join}.  For example, consider operations, $\mathsf{D}^\mathtt{bd}_\mathtt{g}$ and $\mathsf{E}^\mathtt{a}_{ce}$.  The null join is
\begin{equation}
\mathsf{D}^\mathtt{bd}_\mathtt{g}\mathsf{E}^\mathtt{a}_\mathtt{ce} =
~~~~\Longleftrightarrow~~~
\begin{Compose}{0}{0} \setdefaultfont{\mathsf}\setsecondfont{\mathtt}
\Ucircle{D}{0,0} \thispoint{DL}{-120:4} \thispoint{DR}{-60:4} \thispoint{UC}{90:4}
\joincc[above left]{D}{-120}{DL}{60} \csymbolalt{b}
\joincc[above right]{DR}{120}{D}{-60} \csymbolalt{g}
\joincc[left]{D}{90}{UC}{-90} \csymbolalt{d}
\Ucircle{E}{9,-2} \thispoint{Ea}{$(E)+(0:4)$} \thispoint{Ec}{$(E)+(120:4)$} \thispoint{Ee}{$(E)+(-120:4)$}
\joincc[above]{E}{0}{Ea}{180} \csymbolalt{a}
\joincc[above right]{Ec}{-60}{E}{120} \csymbolalt{c}
\joincc[below right]{Ee}{60}{E}{-120} \csymbolalt{e}
\end{Compose}
\end{equation}

We can join multiple encapsulated propositions to form a new encapsulated proposition. For example
\begin{equation}
\begin{Compose}{0}{0} \setdefaultfont{\mathsf}\setsecondfont{\mathtt}
\Ucircle{A}{0,0} \Ucircle{B}{-5,5} \Ucircle{C}{3,4} \Ucircle{D}{-3, 11} \Ucircle{E}{2,9}
\joincc[below left]{B}{-65}{A}{115} \csymbolalt{a}
\joincc[below]{A}{80}{C}{-90} \csymbolalt{b}
\joincc[below]{C}{170}{B}{-10} \csymbolalt{c}
\joincc[above left]{B}{25}{E}{-110} \csymbolalt{d}
\joincc[left]{B}{80}{D}{-100} \csymbolalt{e}
\joincc[above right]{D}{-15}{E}{170}
\joincc[right]{C}{100}{E}{-80} \csymbolalt{g}
\thispoint{nA}{-2,-2} \joincc[above left]{A}{-135}{nA}{45} \csymbolalt{f}
\thispoint{nB}{-8,5} \joincc[above]{B}{180}{nB}{0} \csymbolalt{h}
\thispoint{nD}{-1,13} \joincc[above left]{D}{45}{nD}{-135} \csymbolalt{i}
\thispoint{nE}{4,11} \joincc[above left]{nE}{-145}{E}{45} \csymbolalt{j}
\end{Compose}
\end{equation}
Note that it does not matter which order we evaluate this expression. We could evaluate in the order $\mathtt{a}$, $\mathtt{b}$, $\mathtt{c}$, $\mathtt{d}$, $\mathtt{e}$, $\mathtt{f}$, $\mathtt{g}$, $\mathtt{h}$, $\mathtt{i}$, then $\mathtt{j}$. Or we could evaluate it in any other order.  We will get the same operation regardless (this is clear from the way in which the composition of a pair of operations is defined).

\section{Encapsulated propositions}

\subsection{Propositions in General Relativity}

A proposition, pertaining to some region $\mathtt{A}$ of op-space takes the form
\begin{equation}
\text{Prop}_\mathtt{A}(\Psi_\mathtt{A})
\end{equation}     \index{propositions}
This is the proposition that the actual solution in region $\mathtt{A}$ is one of the solutions in $\text{sort}(\Psi_\mathtt{A})$.

A \emph{pure proposition} is where $\Psi_\mathtt{A}$ is pure.  An \emph{operational proposition} is one corresponding to an operational solution.  We will write
\begin{equation}
\text{Prop}_\mathtt{A}(\Psi_\mathtt{A}[\mathbf{Q}_\mathtt{A}, O_\mathtt{A}]) = \text{Prop}[\mathbf{Q}_\mathtt{A}, O_\mathtt{A}]
\end{equation}
These are the propositions we can actually observe in $\mathtt{A}$.  Such a proposition is an \emph{operationally pure proposition} if $O_\mathtt{A}$ has a single member, $\Gamma_\mathtt{A}$. Then we can write
\begin{equation}
\text{Prop}_\mathtt{A}(\Psi_\mathtt{A}[\Gamma_\mathtt{A}]) = \text{Prop}[\Gamma_\mathtt{A}]
\end{equation}
These are the most fine-grained propositions we can actually observe.  Operational propositions that are not pure are \emph{operationally mixed propositions}.

There are some special propositions we should consider.  The empty proposition is
\begin{equation}
\text{Prop}[\varnothing_\mathtt{A}]= \text{Prop}_\mathtt{A}(\Psi_\mathtt{A}[\varnothing_\mathtt{A}])
\end{equation}
For sufficiently small regions, $\mathtt{A}$, of op-space, this is likely to be the generic situation as $\Gamma_\mathtt{U}$ will not occupy most of op-space.

The deterministic proposition is
\begin{equation}
\text{Prop}(\Psi_\mathtt{A}[\mathbf{Q}_\mathtt{A}, \text{det}])
\end{equation}
This is the proposition that is always true in the given region, $\mathtt{A}$.

The null proposition is
\begin{equation}
\text{Prop}[\mathbf{0}_\mathtt{A}]= \text{Prop}_\mathtt{A}(\Psi_\mathtt{A}[\mathbf{0}_\mathtt{A}])=\lnot \text{Prop}[\mathbf{1}_\mathtt{A}]
\end{equation}
(where $\lnot$ is the logical not operation).  This is the proposition that is always false.  It is important to note that the null proposition is very different from the empty proposition. The null proposition is never true whereas the empty proposition is generically true for small enough $\mathtt{A}$.

\subsection{Joining of propositions}

The \emph{physical AND} of two propositions is a new proposition defined as follows
\begin{equation}
\text{Prop}_\mathtt{A}(\Psi_\mathtt{A})\wedge_\mathtt{a}\text{Prop}_\mathtt{A}(\Psi_\mathtt{B})
= \text{Prop}_\mathtt{A\cup B}(\Psi_\mathtt{A}\Cup_\mathtt{a}\Psi_\mathtt{B})
\end{equation}  \index{wedgea@$\wedge_\mathtt{a}$}
where $\mathtt{A}$ and $\mathtt{B}$ meet at $\mathtt{a}$.  We call this the \emph{physical} AND as we impose boundary conditions that come from the field equations.

Note that
\begin{equation}
\text{Prop}_\mathtt{A}(\Psi_\mathtt{A}[\varnothing_\mathtt{A}])\wedge_\mathtt{a}\text{Prop}_\mathtt{A}(\Psi_\mathtt{B}[\varnothing_\mathtt{B}])
=\text{Prop}_\mathtt{A\cup B}(\Psi_\mathtt{A\cup B}[\varnothing_\mathtt{A\cup B}])
\end{equation}
because the empty solutions do match at $\mathtt{a}$.

Also note that
\begin{equation}
\text{Prop}_\mathtt{A}(\Psi_\mathtt{A}[\mathbf{0}_\mathtt{A}])\wedge_\mathtt{a}\text{Prop}_\mathtt{A}(\Psi_\mathtt{B}[\mathbf{0}_\mathtt{B}])
=\text{Prop}_\mathtt{A\cup B}(\Psi_\mathtt{A\cup B}[\mathbf{0}_\mathtt{A\cup B}])
\end{equation}
by virtue of \eqref{ABnulljoin}

\subsection{Encapsulated propositions}\label{sec:encapsulatedpropositions}

Typically, when we consider a proposition, we consider it along with other information.   Hence, it is useful to introduce the notion of an \emph{encapsulated proposition}
\begin{equation}
\mathcal{A}= \left( \text{prop}(\mathcal{A}), \text{strat}(\mathcal{A}), \text{reg}(\mathcal{A}), \text{type}(\mathcal{A}) \right)
\end{equation}  \index{encapsulated proposition}
This contains the following elements.
\begin{description}
\item[A proposition] $\text{prop}(\mathcal{A})$ of the form $\text{Prop}(\Psi_\mathtt{A})$.
\item[An agent strategy] $\text{strat}(\mathcal{A})$ of the form $\mathbf{Q}_\mathtt{A}$.
\item[A region] of op-space, $\text{reg}(\mathcal{A})$ of the form $\mathtt{A}$.  We will adopt the convention of associating region $\mathtt{A}$, with encapsulated proposition $\mathcal{A}$, $\mathtt{B}$ with $\mathcal{B}$, and so on.
\item[A typing surface] $\text{type}(\mathcal{A})$.  This is specified as $(\mathtt{a}, \mathtt{bc})$, for example, where $\mathtt{a}$ points away from the enclosed region (so the $X_\mathtt{a}^0$ component points away from the enclosed region) and $\mathtt{bc}$ points towards the enclosed region.  We require that
\begin{equation}
\text{set}(\text{type}(\mathcal{A})) \subseteq \text{boundary}(\text{reg}(\mathcal{A}))
\end{equation}
so that the typing surfaces cover some, or all, of the boundary of the region associated with $\mathcal{A}$.
\end{description}
It is useful to introduce superscripts and subscripts corresponding to the typing surfaces.  Thus, we write
\begin{equation}
\mathcal{A}^\mathtt{a}_\mathtt{bc}
\end{equation}
in the case that $\text{type}(\mathcal{A})=(\mathtt{a}, \mathtt{bc})$.  We can also represent this diagrammatically as
\begin{equation}
\mathcal{A}^\mathtt{a}_\mathtt{bc} ~~~ \Longleftrightarrow
\begin{Compose}{0}{0} \setdefaultfont{\mathcal}\setsecondfont{\mathtt}
\Ucircle{A}{0,0}\thispoint{DL}{-120:4} \thispoint{DR}{-60:4} \thispoint{UC}{90:4}
\joincc[above left]{DL}{60}{A}{-120} \csymbolalt{b}
\joincc[above right]{DR}{120}{A}{-60} \csymbolalt{c}
\joincc[left]{A}{90}{UC}{-90} \csymbolalt{a}
\end{Compose}
\end{equation}
Note that superscripts correspond to outward pointing arrows and subscripts correspond to inward pointing arrows.

\subsection{Reversing and detyping}

We can define two operations on an encapsulated proposition: reversing the direction of a typing surface; and detyping (we can perform these on operations and boundary propositions as well).

We can reverse the direction of a typing surface (corresponding to raising or lowering a index).
We will reserve the symbol, $\mathcal{Z}$, for this purpose.  To lower a superscript we write
\begin{equation}
\mathcal{D}^\mathtt{b}_\mathtt{gd} = \mathcal{Z}_\mathtt{dd_1} \mathcal{D}^\mathtt{bd_1}_\mathtt{g}
~~~~~~~~~~~~
\begin{Compose}{0}{0} \setdefaultfont{\mathcal}\setsecondfont{\mathtt}
\Ucircle{D}{0,0} \thispoint{DL}{-120:4} \thispoint{DR}{-60:4} \thispoint{UC}{90:4}
\joincc[above left]{D}{-120}{DL}{60} \csymbolalt{b}
\joincc[above right]{DR}{120}{D}{-60} \csymbolalt{g}
\joincc[left]{UC}{-90}{D}{90} \csymbolalt{d}
\end{Compose}
~~~=~~~
\begin{Compose}{0}{0} \setdefaultfont{\mathcal}\setsecondfont{\mathtt}
\Ucircle{D}{0,0} \thispoint{DL}{-120:4} \thispoint{DR}{-60:4} \thispoint{UC}{90:6} \ccircle{f}{0.5}{90:3.3}
\joincc[above left]{D}{-120}{DL}{60} \csymbolalt{b}
\joincc[above right]{DR}{120}{D}{-60} \csymbolalt{g}
\joincc[left]{D}{90}{f}{-90} \csymbolalt{d}  \joincc[left]{UC}{-90}{f}{90} \csymbolalt{d}
\end{Compose}
\end{equation}
where we have the symbolic notation on the right and the diagrammatic notation on the left.  Similarly, to raise a subscript we write
\begin{equation}
\mathcal{D}^\mathtt{bdg} = \mathcal{Z}^\mathtt{gg_1} \mathcal{D}^\mathtt{bd}_\mathtt{g_1}
~~~~~~~~~~~~
\begin{Compose}{0}{0} \setdefaultfont{\mathcal}\setsecondfont{\mathtt}
\Ucircle{D}{0,0} \thispoint{DL}{-120:4} \thispoint{DR}{-60:4} \thispoint{UC}{90:4}
\joincc[above left]{D}{-120}{DL}{60} \csymbolalt{b}
\joincc[above right]{D}{-60}{DR}{120} \csymbolalt{g}
\joincc[left]{D}{90}{UC}{-90} \csymbolalt{d}
\end{Compose}
~~~=~~~
\begin{Compose}{0}{0} \setdefaultfont{\mathcal}\setsecondfont{\mathtt}
\Ucircle{D}{0,0} \thispoint{DL}{-120:4} \thispoint{DR}{-60:6} \ccircle{f}{0.5}{-60:3.3} \thispoint{UC}{90:4}
\joincc[above left]{D}{-120}{DL}{60} \csymbolalt{b}
\joincc[above right]{f}{120}{D}{-60} \csymbolalt{g} \joincc[above right]{f}{-60}{DR}{120} \csymbolalt{g}
\joincc[left]{D}{90}{UC}{-90} \csymbolalt{d}
\end{Compose}
\end{equation}
In the diagrammatic notation, we use a small blank circle to denote flipping a typing boundary.

We can detype a typing surface.  Thus, we go from
\begin{equation}
\mathcal{D}^\mathtt{bd}_\mathtt{g} = \big(\text{prop}(\mathcal{D}), \text{strat}(\mathcal{D}),  \text{reg}(\mathcal{D}), (\mathtt{bd},\mathtt{g})\big)
\end{equation}
to
\begin{equation}
\mathcal{D}^\mathtt{b}_\mathtt{g} = \mathcal{T}_\mathtt{d} \mathcal{D}^\mathtt{bd}_\mathtt{g} = \big(\text{prop}(\mathcal{D}), \text{strat}(\mathcal{D}), \text{reg}(\mathcal{D}), (\mathtt{b}, \mathtt{g})\big)
\end{equation}
We will reserve the symbol, $\mathcal T$, to denote the detyping process.  To detype a subscript, we use $\mathcal{T}^\mathtt{g}$.  We represent detyping diagrammatically by
\begin{equation}
\mathcal{D}^\mathtt{b}_\mathtt{g} = \mathcal{T}_\mathtt{d} \mathcal{D}^\mathtt{bd}_\mathtt{g}
~~~ \Longleftrightarrow
\begin{Compose}{0}{-0.09} \setdefaultfont{\mathcal}\setsecondfont{\mathtt}
\Ucircle{D}{0,0} \thispoint{DL}{-120:4}  \thispoint{DR}{-60:4} \crectanglefill{UC}{0.5}{0.17}{90:4}
\joincc[above left]{D}{-120}{DL}{60} \csymbolalt{b}
\joincc[above right]{DR}{120}{D}{-60} \csymbolalt{g}
\joincbnoarrow[left]{D}{90}{UC}{0} \csymbolalt{d}
\end{Compose}
\end{equation}
Note that there is no need to have an arrow in the diagram on the $\mathtt d$ typing surface as we get the same encapsulated proposition if we first reverse the arrow then detype.

\subsection{Joining encapsulated propositions}

We can compose two encapsulated propositions for two regions to form a new encapsulated proposition for the union of these two regions.  The \emph{external composition} of $\mathcal{A}^\mathtt{ab}_\mathtt{cd}$ and $\mathcal{B}^\mathtt{ec}_\mathtt{af}$ is defined when
\begin{equation}
\text{reg}(\mathcal{A}) \cap \text{reg}(\mathcal{B}) =  \text{set}(\mathtt{ac})
\end{equation}
That is the two regions overlap only in some part of the typing surface (this is the reason this type of composition is called \lq\lq external\rq\rq).  When this condition is satisfied we define external AND composition
\begin{equation} \label{defnencappropB}
\mathcal{A}^\mathtt{ab}_\mathtt{cd}\mathcal{B}^\mathtt{ec}_\mathtt{af}
= \big(\text{prop}( \mathcal{A})\wedge_\mathtt{ac} \text{prop}(\mathcal{B}), \text{strat}(\mathcal{A})\cup\text{strat}(\mathcal{B}), \text{reg}(\mathcal{A})\cup\text{reg}(\mathcal{B}), (\mathtt{be}, \mathtt{df}) \big)
\end{equation}
This gives us a new encapsulated proposition which we can write as $\mathcal{C}^\mathtt{be}_\mathtt{df}$.  Note that the directions of the typing surfaces where the two regions overlap point in the same direction.  This is why we match superscripts with subscripts.  We can represent composition of two encapsulated propositions by
\begin{equation}
\mathcal{A}^\mathtt{ab}_\mathtt{cd}\mathcal{B}^\mathtt{ec}_\mathtt{af}  ~~~\Longleftrightarrow~~~
\begin{Compose}{0}{-0.7}   \setdefaultfont{\mathcal}\setsecondfont{\mathtt}
\Ucircle{A}{0,0} \Ucircle{B}{45:5}
\thispoint{Ab}{180:4} \thispoint{Ad}{-90:4}
\thispoint{Be}{$(45:5)+(0:4)$} \thispoint{Bf}{$(45:5)+(90:4)$}
\joincc[above left]{A}{60}{B}{-150} \csymbolalt{a}
\joincc[below right]{B}{-120}{A}{30} \csymbolalt{c}
\joincc[above]{A}{-180}{Ab}{0} \csymbolalt{b} \joincc[left]{Ad}{90}{A}{-90} \csymbolalt{d}
\joincc[above]{B}{0}{Be}{-180} \csymbolalt{e} \joincc[left]{Bf}{-90}{B}{90} \csymbolalt{f}
\end{Compose}
\end{equation}

A special case of the physical AND join is  when the two regions associated with the two encapsulated propositions do not overlap at all (they do not meet at any bounding surface).   For example, consider encapsulated propositions, $\mathcal{D}^\mathtt{bd}_\mathtt{g}$ and $\mathcal{E}^\mathtt{a}_{ce}$.  The null-AND join is
\begin{equation}
\mathcal{D}^\mathtt{bd}_\mathtt{g}\mathcal{E}^\mathtt{a}_\mathtt{ce} =
\big( \text{prop}(\mathcal{D}) \wedge \text{prop}(\mathcal{E}), \mathbf{Q}_\mathcal{DE}, \text{reg}(\mathcal{D})\cup\text{reg}(\mathcal{D}), (\mathtt{bda}, \mathtt{gce}) \big)
\end{equation}
which can be represented diagrammatically as
\begin{equation}
\begin{Compose}{0}{0} \setdefaultfont{\mathcal}\setsecondfont{\mathtt}
\Ucircle{D}{0,0} \thispoint{DL}{-120:4} \thispoint{DR}{-60:4} \thispoint{UC}{90:4}
\joincc[above left]{D}{-120}{DL}{60} \csymbolalt{b}
\joincc[above right]{DR}{120}{D}{-60} \csymbolalt{g}
\joincc[left]{D}{90}{UC}{-90} \csymbolalt{d}
\Ucircle{E}{9,-2} \thispoint{Ea}{$(E)+(0:4)$} \thispoint{Ec}{$(E)+(120:4)$} \thispoint{Ee}{$(E)+(-120:4)$}
\joincc[above]{E}{0}{Ea}{180} \csymbolalt{a}
\joincc[above right]{Ec}{-60}{E}{120} \csymbolalt{c}
\joincc[below right]{Ee}{60}{E}{-120} \csymbolalt{e}
\end{Compose}
\end{equation}
Note that, under such a join, the solutions corresponding to these encapsulated propositions do not meet at a boundary so there is no physical matching.  Hence, we need only take the logical AND.  This corresponds to joining the solutions by $\Psi_\mathtt{A}\Cup_\mathtt{0}\Psi_\mathtt{B}$ as explained in Sec.\ \ref{sec:thenulljoin}

We can join multiple encapsulated propositions to form a new encapsulated proposition. For example
\begin{equation}\label{compecap}
\begin{Compose}{0}{0} \setdefaultfont{\mathcal}\setsecondfont{\mathtt}
\Ucircle{A}{0,0} \Ucircle{B}{-5,5} \Ucircle{C}{3,4} \Ucircle{D}{-3, 11} \Ucircle{E}{2,9}
\joincc[below left]{B}{-65}{A}{115} \csymbolalt{a}
\joincc[below]{A}{80}{C}{-90} \csymbolalt{b}
\joincc[below]{C}{170}{B}{-10} \csymbolalt{c}
\joincc[above left]{B}{25}{E}{-110} \csymbolalt{d}
\joincc[left]{B}{80}{D}{-100} \csymbolalt{e}
\joincc[above right]{D}{-15}{E}{170}
\joincc[right]{C}{100}{E}{-80} \csymbolalt{g}
\thispoint{nA}{-2,-2} \joincc[above left]{A}{-135}{nA}{45} \csymbolalt{f}
\thispoint{nB}{-8,5} \joincc[above]{B}{180}{nB}{0} \csymbolalt{h}
\thispoint{nD}{-1,13} \joincc[above left]{D}{45}{nD}{-135} \csymbolalt{i}
\thispoint{nE}{4,11} \joincc[above left]{nE}{-145}{E}{45} \csymbolalt{j}
\end{Compose}
\end{equation}
Most crucially, it does not matter which order we evaluate this expression. We could evaluate in the order $\mathtt{a}$, $\mathtt{b}$, $\mathtt{c}$, $\mathtt{d}$, $\mathtt{e}$, $\mathtt{f}$, $\mathtt{g}$, $\mathtt{h}$, $\mathtt{i}$, then $\mathtt{j}$. Or we could evaluate it in any other order.  We will get the same answer regardless since this amounts to checking the AND of the boundary conditions are satisfied at each of these typing surfaces.

\section{Boundary propositions}

\subsection{Introducing boundary propositions}\label{sec:boundarypropositions}

Boundary propositions \index{boundary propositions} are propositions concerning the boundary conditions at typing surfaces.  We will represent boundary propositions by, $\mathbpro{A}$, $\mathbpro{B}$, $\mathbpro{C}$, \dots.   The set of pure boundary conditions at a typing surface, $\mathtt{a}$, is $\Lambda_\mathtt{a}$.  Note that it is useful to use superscripts as well as subscripts here so we also define the set $\Lambda^\mathtt{a}$.  The sets $\Lambda^\mathtt{a}[\text{spec}]$ and $\Lambda_\mathtt{a}[\text{spec}]$ have the same elements.

We write the proposition that the boundary condition at typing surface $\mathtt{a}$ is the pure boundary condition, $\mathbnd{a}\in\Lambda^\mathtt{a}$ (or $\mathbnd{a}\in\Lambda_\mathtt{a}$ for the subscript case), by
\begin{equation}
\mathbpro{P}^\mathtt{a}(\mathbnd{a})   ~~~~~~~\text{or}~~~~~~~ \mathbpro{P}_\mathtt{a}(\mathbnd{a})
\end{equation}
(we reserve the letter $\mathbpro{P}$ for this purpose).  We allow subscript and superscript placement of the typing surface index, $\mathtt{a}$, so we can associate boundary propositions with encapsulated propositions.   We denote by $\mathbpro{A}^\mathtt{a}_\mathtt{bc}$ the boundary proposition induced by the encapsulated proposition, $\mathcal{A}^\mathtt{a}_\mathtt{bc}$, and so on.  We write
\begin{equation}
\mathbpro{A}^\mathtt{a}_\mathtt{bc}=\text{bprop}(\mathcal{A}^\mathtt{a}_\mathtt{bc})
\end{equation}
We will later reserve the symbol $\mathbpro{X}$ to represent boundary propositions belonging to a fiducial set.

First we will establish some notation.  We write
\begin{equation}\label{bpropnotation}
\mathbpro{P}^\mathtt{a}(\mathbnd{a}) \mathbpro{P}_\mathtt{b}(\mathbnd{b}) \mathbpro{P}_\mathtt{c}(\mathbnd{c})
\end{equation}
for the proposition
\begin{equation}
\mathbpro{P}^\mathtt{a}(\mathbnd{a}) \wedge \mathbpro{P}_\mathtt{b}(\mathbnd{b}) \wedge \mathbpro{P}_\mathtt{c}(\mathbnd{c})
\end{equation}
(i.e.\ the $\wedge$ is implicit in \eqref{bpropnotation}). Note that it does not matter what order we write the symbols in - e.g.\
\begin{equation}
\mathbpro{P}_\mathtt{c}(\mathbnd{c})\mathbpro{P}_\mathtt{b}(\mathbnd{b})\mathbpro{P}^\mathtt{a}(\mathbnd{a})
\end{equation}
corresponds to the same proposition (because $\wedge$ is commutative and associative).   We will call a proposition like this a \emph{product} boundary proposition.

A composite pure boundary condition will factorize in the sense that it can be written, for example, as $(\mathbnd{ab}, \mathbnd{c})$ where $\mathbnd{a}$ is the boundary condition at $\mathtt{a}$, $\mathbnd{b}$ at $\mathtt{b}$, and $\mathbnd{c}$ at $\mathtt{c}$.  We call such factorized boundary conditions \emph{product} boundary conditions.   Mixed boundary conditions will not, in general, factorize like this.

Consider an encapsulated proposition, $\mathcal{A}^\mathtt{a}_\mathtt{bc}$, having associated solution $\Psi_\mathtt{A}$.  If this solution is pure then this encapsulated proposition induces a product boundary proposition
\begin{equation}
\mathbpro{A}^\mathtt{a}_\mathtt{bc}=\text{bprop}(\mathcal{A}^\mathtt{a}_\mathtt{bc}) =  \mathbpro{P}^\mathtt{a}(\theta_\mathtt{a}(\Psi_\mathtt{A})) \mathbpro{P}_\mathtt{b}(\theta_\mathtt{b}(\Psi_\mathtt{A})) \mathbpro{P}_\mathtt{c}(\theta_\mathtt{c}(\Psi_\mathtt{A}))
\end{equation}
We can write the boundary proposition induced by a mixed solution, $\Psi_\mathtt{A}$, as
\begin{equation}\label{encaptobprop}
\mathbpro{A}^\mathtt{a}_\mathtt{bc}=\text{bprop}(\mathcal{A}^\mathtt{a}_\mathtt{bc}) = \bigvee_{\mathbnd{abc}\in\text{sort}(\theta_\mathtt{abc}(\Psi_\mathtt{A}))}
\mathbpro{P}^\mathtt{a}(\mathbnd{a})\mathbpro{P}_\mathtt{c}(\mathbnd{c})\mathbpro{P}_\mathtt{b}(\mathbnd{b})
\end{equation}
This is the general case.  Recall that $\text{sort}(\mathbnd{a})$ (for some possibly mixed boundary condition, $\mathbnd{a}$) returns the set of pure boundary conditions whose union gives $\mathbnd{a}$.

We can represent a boundary proposition diagrammatically as
\begin{equation}
\mathbpro{A}^\mathtt{a}_\mathtt{bc} ~~~ \Longleftrightarrow
\begin{Compose}{0}{0} \setdefaultfont{\mathbpro}\setsecondfont{\mathtt}
\Ucircle{A}{0,0}\thispoint{DL}{-120:4} \thispoint{DR}{-60:4} \thispoint{UC}{90:4}
\joincc[above left]{DL}{60}{A}{-120} \csymbolalt{b}
\joincc[above right]{DR}{120}{A}{-60} \csymbolalt{c}
\joincc[left]{A}{90}{UC}{-90} \csymbolalt{a}
\end{Compose}
\end{equation}

\subsection{Composition of boundary propositions}\label{sec:compencapboundprop}

We wish to find a way to compose boundary propositions that commutes with the map from encapsulated propositions:
\begin{equation}\label{commutingebprops}
\text{bprop}(\mathcal{A}^\mathtt{ab}_\mathtt{cd}\mathcal{B}^\mathtt{ec}_\mathtt{af}) = \text{bprop}(\mathcal{A}^\mathtt{ab}_\mathtt{cd})\text{bprop}(\mathcal{B}^\mathtt{ec}_\mathtt{af})
\end{equation}
This is possible because, when we compose encapsulated propositions we only check to see if the boundary conditions match or not.

The trick, when joining boundary propositions at a typing surface, is to return the true proposition, $\mathbf{1}$, if the propositions match at this typing surface and to return the false proposition, $\mathbf{0}$, if they do not match.  To do this we apply \emph{possibilistic contraction}
\begin{equation}\label{posscontraction}
\mathbpro{P}^\mathtt{a}(\mathbnd{a})\mathbpro{P}_\mathtt{a}(\mathbnd{a}') =
\left\{ \begin{array}{ll}
\mathbf{0} & \text{if}~ \mathbnd{a}\not= \mathbnd{a}'  \\
\mathbf{1} & \text{if}~ \mathbnd{a}= \mathbnd{a}'
\end{array} \right.
\end{equation}
whenever there is a repeated typing surface with one superscript and one subscript.  So long as there exists at least one match, we return a $\mathbf{1}$.   Now consider applying this to a more complicated example.  Put
\begin{equation}
\mathbpro{P}^\mathtt{ab}_\mathtt{cd}(\mathbnd{ab},\mathbnd{cd}) = \mathbpro{P}^\mathtt{a}(\mathbnd{a})  \mathbpro{P}^\mathtt{b}(\mathbnd{b})  \mathbpro{P}_\mathtt{c}(\mathbnd{c}) \mathbpro{P}_\mathtt{d}(\mathbnd{d})
\end{equation}
Consider
\begin{equation}
\mathbpro{P}^\mathtt{ab}_\mathtt{cd}(\mathbnd{ab},\mathbnd{cd}) \mathbpro{P}^\mathtt{ec}_\mathtt{af}(\mathbnd{ec}',\mathbnd{a}'\mathbnd{f})
= \mathbpro{P}^\mathtt{a}(\mathbnd{a})  \mathbpro{P}^\mathtt{b}(\mathbnd{b})  \mathbpro{P}_\mathtt{c}(\mathbnd{c}) \mathbpro{P}_\mathtt{d}(\mathbnd{d})
\mathbpro{P}^\mathtt{e}(\mathbnd{e})  \mathbpro{P}^\mathtt{c}(\mathbnd{c}')  \mathbpro{P}_\mathtt{a}(\mathbnd{a}') \mathbpro{P}_\mathtt{f}(\mathbnd{f})
\end{equation}
We apply possibilistic contraction (as in \eqref{posscontraction}) to obtain
\begin{equation}
\mathbpro{P}^\mathtt{ab}_\mathtt{cd}(\mathbnd{ab},\mathbnd{cd}) \mathbpro{P}^\mathtt{ec}_\mathtt{af}(\mathbnd{ec}',\mathbnd{a}'\mathbnd{f})
=  \left\{  \begin{array}{ll}
\mathbpro{P}^\mathtt{be}_\mathtt{df}(\mathbnd{be},\mathbnd{df})  & \text{if}~ (\mathbnd{a}= \mathbnd{a}' ) \wedge (\mathbnd{c}= \mathbnd{c}') \\
\mathbf{0} &  \text{else}
\end{array} \right.
\end{equation}
This allows us to combine product boundary propositions.

We can also apply this to general (non-product) boundary propositions.  Consider
\begin{equation}\label{compboundAB}
\mathbpro{A}^\mathtt{ab}_\mathtt{cd}\mathbpro{B}^\mathtt{ec}_\mathtt{af}
= \left(\bigvee_{\mathbnd{abcd}\in\text{sort}(\theta^\mathtt{ab}_\mathtt{cd}(\Psi_\mathtt{A}))}
\mathbpro{P}^\mathtt{ab}_\mathtt{cd}(\mathbnd{ab},\mathbnd{cd}) \right)
\left(  \bigvee_{\mathbnd{ecaf}\in\text{sort}(\theta^\mathtt{ec}_\mathtt{af}(\Psi_\mathtt{A}))}
\mathbpro{P}^\mathtt{ec}_\mathtt{af}(\mathbnd{ec}',\mathbnd{a}'\mathbnd{f}) \right)
\end{equation}
This simplifies to
\begin{equation}
\mathbpro{A}^\mathtt{ab}_\mathtt{cd}\mathbpro{B}^\mathtt{ec}_\mathtt{af}
=  \bigvee_{\mathbnd{bdef}\in\text{sort}(\theta^\mathtt{be}_\mathtt{df}(\Psi_\mathtt{A}\Cup_\mathtt{ac}\Psi_\mathtt{B}))} \mathbpro{P}^\mathtt{be}_\mathtt{df}(\mathbnd{be},\mathbnd{df})
\end{equation}
if $\Psi_\mathtt{A}\Cup_\mathtt{ac}\Psi_\mathtt{B}\not=\varnothing$. If, on the other hand, $\Psi_\mathtt{A}\Cup_\mathtt{ac}\Psi_\mathtt{B}=\varnothing$ then the expression in \eqref{compboundAB} simplifies to $\mathbf{0}$.  We see that this composition satisfies the commutativity requirement \eqref{commutingebprops}.

We can represent composition of two boundary propositions by
\begin{equation}
\mathbpro{A}^\mathtt{ab}_\mathtt{cd}\mathbpro{B}^\mathtt{ec}_\mathtt{af}  ~~~\Longleftrightarrow~~~
\begin{Compose}{0}{-0.7}   \setdefaultfont{\mathbpro}\setsecondfont{\mathtt}
\Ucircle{A}{0,0} \Ucircle{B}{45:5}
\thispoint{Ab}{180:4} \thispoint{Ad}{-90:4}
\thispoint{Be}{$(45:5)+(0:4)$} \thispoint{Bf}{$(45:5)+(90:4)$}
\joincc[above left]{A}{60}{B}{-150} \csymbolalt{a}
\joincc[below right]{B}{-120}{A}{30} \csymbolalt{c}
\joincc[above]{A}{-180}{Ab}{0} \csymbolalt{b} \joincc[left]{Ad}{90}{A}{-90} \csymbolalt{d}
\joincc[above]{B}{0}{Be}{-180} \csymbolalt{e} \joincc[left]{Bf}{-90}{B}{90} \csymbolalt{f}
\end{Compose}
\end{equation}
If we combine two encapsulated propositions so that there are no typing surfaces left over, such as
\begin{equation}
\mathbpro{A}^\mathtt{a}\mathbpro{B}_\mathtt{a}
\end{equation}
then we get either $\mathbf{0}$ or $\mathbf{1}$ for the associated proposition.  If we obtained these boundary propositions from physically allowed encapsulated propositions then we can interpret a $\mathbf{0}$ as meaning that the given arrangement is impossible. A $\mathbf{1}$, on the other hand, indicates that the given arrangement is possible.

We can compose multiple boundary propositions.  It is clear that we can calculate the resulting boundary proposition by contracting over matched typing surfaces in any order and that the generalization of the commutativity condition \eqref{commutingebprops} will still hold.

\section{Duotensors, fiducials, and generalized states}

\subsection{Possibilistic equivalence}\label{sec:possibilisticequivalence}

An operation, encapsulated proposition, or boundary proposition is said to be \emph{closed} if the associated typing surface is $(-,-)$ (i.e.\ it has no typing surface).  In the diagrammatic notation this means there are no open wires.   In this possibilistic formulation, we are interested in whether such elements are possible or impossible.  We can form a closed element such as this by composing these elements so that no open wires are left over.

For closed boundary propositions, $\mathbpro{A}$, we define the function
\begin{equation}
\text{Poss}(\mathbpro{A})=
\left\{
\begin{array}{ll}
0  & \text{if}~ \mathbpro{A}=\mathbf{0}\\
1  & \text{if} ~\mathbpro{A}=\mathbf{1}
\end{array}
\right.
\end{equation}
This is interpreted as the possibilistic value. A 0 indicates an impossible situation and a 1 indicates a possible situation.

We note that, if both $\mathbpro{A}$ and $\mathbpro{B}$ are closed then we have the factorization property
\begin{equation}
\text{poss}(\mathbpro{AB})= \text{poss}(\mathbpro{A})\text{poss}(\mathbpro{B})
\end{equation}
since $\mathbpro{AB}=\mathbf{1}$ is possible if and only if both $\mathbpro{A}=\mathbf{1}$ and $\mathbpro{B}=\mathbf{1}$ are possible.

We are interested in setting up a notion of possibilistic equivalence. For this purpose we define a $p(\cdot)$ \index{p@$p(\cdot)$} function as follows
\begin{equation}
p(\alpha \mathbpro{A} + \beta \mathbpro{B} + \dots ) = \alpha \text{poss}\mathbpro{A} + \beta \text{poss}\mathbpro{B} + \dots
\end{equation}
where $\alpha, \beta, \dots$ take values 0 or 1 and the addition is possibilistic such that
\begin{equation}\label{possaddition}
0+0=0 ~~~0+1=1 ~~~ 1+0= 1~~~ 1+1 = 1
\end{equation}   \index{possibilistic addition}
In general a possibilistic sum of all 0's is equal to 0, and a sum containing one or more 1 is equal to 1.  The $p(\cdot)$ function provides a linear extension of the $\text{poss}(\cdot)$ function. We will use a similar extension in the probabilistic case.

Now we have defined the $p(\cdot)$ function, we can provide a definition of equivalence.  Let us give an example first. We say that
\begin{equation}
\mathbpro{F}^\mathtt{a}_\mathtt{c}+\mathbpro{G}^\mathtt{a}_\mathtt{c} \equiv \mathbpro{H}^\mathtt{a}_\mathtt{c}+\mathbpro{J}^\mathtt{a}_\mathtt{c}
\end{equation}
if
\begin{equation}
p((\mathbpro{F}^\mathtt{a}_\mathtt{c}+\mathbpro{G}^\mathtt{a}_\mathtt{c})\mathbpro{E}_\mathtt{a}^\mathtt{c})
\equiv
p( (\mathbpro{H}^\mathtt{a}_\mathtt{c}+\mathbpro{J}^\mathtt{a}_\mathtt{c}) \mathbpro{E}_\mathtt{a}^\mathtt{c})
\end{equation}
for all $\mathbpro{E}_\mathtt{a}^\mathtt{c}$.  In general, we say that  \index{possibilistic equivalence}
\begin{equation}
\text{expression}_1 \equiv \text{expression}_2
\end{equation}
if
\begin{equation}
p(\text{expression}_1 \mathbpro{E} )=p(\text{expression}_2\mathbpro{E})
\end{equation}
(where we are suppressing typing surface subscripts and superscripts because we are considering the general case)
for all $\mathbpro{E}$ such that
\begin{equation}
\text{expression}_i \mathbpro{E}
\end{equation}
is a sum the form $\alpha \mathbpro{A} + \beta \mathbpro{B} + \dots $ (of closed boundary propositions).

Note that a closed boundary proposition is equivalent to its possibilistic value:
\begin{equation}\label{AequivpossA}
\mathbpro{A}\equiv \text{poss}(\mathbpro{A})
\end{equation}
because
\begin{equation}\label{AequivpossAproof}
p(\mathbpro{AE})= p(\mathbpro{A})p(\mathbpro{E}) = p(\text{poss}(\mathbpro{A})\mathbpro{E})
\end{equation}
for all closed $\mathbpro{E}$.

There is a related notion to possibilistic equivalence. This is \emph{possibilistic entailment}.  We will say that
\begin{equation}
\text{expression}_1 \leqq \text{expression}_2
\end{equation}
if
\begin{equation}
p(\text{expression}_1 \mathbpro{E} ) \leq p(\text{expression}_2\mathbpro{E})
\end{equation}
for all $\mathbpro{E}$.  In this situation we will say that $\text{expression}_1$ entails $\text{expression}_2$.   The notion of entailment we intend here is perhaps better illustrated by an example pertaining to a more familiar context. We may say that if it is possible that a given bag contains apples, then this entails that it is possible the bag contains apples or oranges.  The converse is not the case.  In our context, something is possible if it is a solution to the field equations that define the physical theory.

\subsection{Boundary fiducials}

A fiducial set of boundary propositions \index{boundary fiducials} takes the form
\begin{equation}
\{\mathbpro{X}_\mathtt{a}^\mathbnd{a}: \mathbnd{a}\in \Lambda_\mathtt{a}[\text{fid}]\subseteq\Lambda_\mathtt{a} \}
\end{equation}
for subscript typing surfaces  and
\begin{equation}
\{ \presub{\mathbnd{a}}{\mathbpro{X}}^\mathtt{a}: \mathbnd{a}\in \Lambda^\mathtt{a}[\text{fid}]\subseteq\Lambda^\mathtt{a} \}
\end{equation}
for superscript typing surfaces.  Here the fiducial sets $\Lambda_\mathtt{a}[\text{fid}]$ and $\Lambda^\mathtt{a}[\text{fid}]$ are sets of possible \emph{pure} boundary conditions at $\mathtt{a}$ that are sufficient for expanding those general boundary propositions that are of interest. These two sets are chosen to have the same elements (so we can match at boundaries).

Fiducial sets must be chosen such that
we can write any boundary proposition of interest as being equivalent to a sum over fiducials
\begin{equation}\label{logicaldecomp}
\mathbpro{A}^\mathtt{a}_\mathtt{bc} \equiv  \presup{\mathbnd{a}}A_\mathbnd{bc} ~  \presub{\mathbnd{a}}{\mathbpro{X}}^\mathtt{a} \mathbpro{X}_\mathtt{b}^\mathbnd{b} \mathbpro{X}_\mathtt{c}^\mathbnd{c}
\end{equation}
where we perform possibilistic summation over repeated indices $\mathbnd{a}$, $\mathbnd{b}$, $\mathbnd{c}$ in the sets $\Lambda^\mathtt{a}[\text{fid}]$, $\Lambda_\mathtt{b}[\text{fid}]$, and $\Lambda_\mathtt{c}[\text{fid}]$ respectively.  This can be a continuous sum (if the fiducial sets are continuous).  However, it is not an integral because we are using possibilistic addition (in which 1+1=1).  This sum simply collects all the cases for which $\presup{\mathbnd{a}}A_\mathbnd{bc}=1$.
If we want to be able to write down any boundary proposition as in \eqref{logicaldecomp} then the fiducial sets must contain all pure boundary conditions.

If we choose
\begin{equation}\label{fidchoicepure}
\Lambda_\mathtt{a}[\text{fid}] = \Lambda_\mathtt{a}[\text{cand}] ~~~~~ \Lambda^\mathtt{a}[\text{fid}] = \Lambda^\mathtt{a}[\text{cand}]
\end{equation}
where $\Lambda_\mathtt{a}$ is the set of pure boundary conditions then we can expand any generalized boundary proposition as in \eqref{logicaldecomp}.  We may make an even more general choice and choose pure boundary conditions induced by pure candidate solutions.  \begin{equation}
\Lambda_\mathtt{a}[\text{fid}] = \Lambda_\mathtt{a}[\text{cand}] ~~~~~ \Lambda^\mathtt{a}[\text{fid}] = \Lambda^\mathtt{a}[\text{cand}]
\end{equation}
The advantage of doing this is that we can then look for principles that pick out the physical cases.

The \emph{generalized possibilistic state} \index{generalized possibilistic state}
\begin{equation}
\presup{\mathbnd{a}}A_\mathbnd{bc}
\end{equation}
is an indicator function for the subset of the fiducial boundary conditions associated with the typing surface $(\mathtt{a},\mathtt{bc})$ that make up the given boundary proposition. If the associated boundary proposition is induced by an encapsulated proposition $\mathcal{A}^\mathtt{a}_\mathtt{bc}$ with associated solution, $\Psi_\mathtt{A}$, then
\begin{equation}
\presup{\mathbnd{a}}A_\mathbnd{bc}=
\left\{
\begin{array}{ll}
1 & \text{if}~ \mathbnd{abc}\subset \theta_\mathtt{abc}(\Psi_\mathtt{A}) \\
0 & \text{else}
\end{array}
\right.
\end{equation}
If $\presup{\mathbnd{a}}A_\mathbnd{bc}=0$ for some particular $\mathbnd{abc}$ then
\begin{equation}
\mathbnd{abc}\cup \theta_\mathtt{abc}(\Psi_\mathtt{A}) =\varnothing
\end{equation}
because we have chosen our fiducial sets so that \eqref{logicaldecomp} is possible.  The generalized possibilistic state is a duotensor \cite{hardy2010formalism, hardy2011reformulating}- it has pre and post scripts and there is also a hopping metric (to be introduced below) which can turn subscripts into pre-subscripts and pre-superscripts into superscripts.

It is possible that there are relations between the elements of the fiducial set.  These could be linear equivalences such as
\begin{equation}
\mathbpro{X}_\mathtt{a}^\mathbnd{a} + \mathbpro{X}_\mathtt{a}^{\mathbnd{a}'}\equiv \mathbpro{X}_\mathtt{a}^{\mathbnd{a}'''}
\end{equation}
for some given $\mathbnd{a}$, $\mathbnd{a}'$, and $\mathbnd{a}''$.  Or they could be entailments such as
\begin{equation}
\mathbpro{X}_\mathtt{a}^\mathbnd{a} + \mathbpro{X}_\mathtt{a}^{\mathbnd{a}'} \leqq \mathbpro{X}_\mathtt{a}^{\mathbnd{a}'''}
\end{equation}
In such cases there are corresponding constraints on the elements of the generalized possibilistic state.   Although such linear equivalences and entailments lead to inefficiency in the representation of generalized possibilistic states, they are easily accommodated in the possibilistic case and there may be physical reasons to allow such fiducial sets.

It is useful to use a diagrammatic notation for fiducials.  We put
\begin{equation}
\mathbpro{X}_\mathtt{a}^\mathbnd{a}  ~~\Leftrightarrow~~
\begin{Compose}{0}{0} \setdefaultfont{\mathbpro}\setsecondfont{\mathtt}\setthirdfont{\mathbnd}
\thispoint{pX}{0,-3}
\Scircle{X}{0,0} \blackdotsq{Xa}{0,3} \joincc[left]{pX}{90}{X}{-90} \csymbolalt{a} \joincc[left]{X}{90}{Xa}{-90}\csymbolthird{a}
\end{Compose}
~~~~~~~~
\presub{\mathbnd{a}}{\mathbpro{X}}^\mathtt{a} ~~\Leftrightarrow~~
\begin{Compose}{0}{0} \setdefaultfont{\mathbpro}\setsecondfont{\mathtt}\setthirdfont{\mathbnd}
\thispoint{pX}{0,-3} \Scircle{X}{0,0} \blackdotsq{Xa}{0,3}
\joincc[left]{X}{-90}{pX}{90}\csymbolalt{a}
\joincc[left]{Xa}{-90}{X}{90} \csymbolthird{a}
\end{Compose}
\end{equation}
The black dot must be matched with a white dot.  We can write \eqref{logicaldecomp} in diagrammatic form
\begin{equation}
\mathbpro{A}^\mathtt{a}_\mathtt{bc} \equiv  \presup{\mathbnd{a}}A_\mathbnd{bc} ~  \presub{\mathbnd{a}}{\mathbpro{X}}^\mathtt{a} \mathbpro{X}_\mathtt{b}^\mathbnd{b} \mathbpro{X}_\mathtt{c}^\mathbnd{c}
~~~~~ \Longleftrightarrow ~~~~~
\begin{Compose}{0}{0} \setdefaultfont{\mathbpro}\setsecondfont{\mathtt}\setthirdfont{\mathbnd}
\Ucircle{A}{0,0}\thispoint{DL}{-120:4} \thispoint{DR}{-60:4} \thispoint{UC}{90:4}
\joincc[above left]{DL}{60}{A}{-120} \csymbolalt{b}
\joincc[above right]{DR}{120}{A}{-60} \csymbolalt{c}
\joincc[left]{A}{90}{UC}{-90} \csymbolalt{a}
\end{Compose}
~~~ \equiv ~~~
\begin{Compose}{0}{0} \setdefaultfont{\mathbpro}\setsecondfont{\mathtt}\setthirdfont{\mathbnd}
\ucircle{A}{0,0}\csymbolfourth{A}
\scircle{DLX}{-120:5}\csymbol{X} \scircle{DRX}{-60:5}\csymbol{X} \scircle{UCX}{90:5} \csymbol{X}
\thispoint{DL}{-120:8} \thispoint{DR}{-60:8} \thispoint{UC}{90:8}
\joincc[above left]{DL}{60}{DLX}{-120} \csymbolalt{b}
\joincc[above right]{DR}{120}{DRX}{-60} \csymbolalt{c}
\joincc[left]{UCX}{90}{UC}{-90} \csymbolalt{a}
\joinccbwsq[above left]{DLX}{60}{A}{-120} \csymbolthird{b}
\joinccbwsq[above right]{DRX}{120}{A}{-60} \csymbolthird{c}
\joinccwbsq[left]{A}{90}{UCX}{-90} \csymbolthird{a}
\end{Compose}
\end{equation}
Where the small black and white squares are matched we sum over the corresponding elements of the fiducial set.

We have the following map between the placement indices and black and white squares
\begin{equation}
\prescript{\square}{\blacksquare} A^\blacksquare_\square
\end{equation}
This is because, in symbolic notation we keep to the tradition of matching superscripts with subscripts, and therefore matching pre-superscripts with pre-subscripts whereas in diagrammatic notation, we match black with white squares.

\subsection{The hopping metric}

We define the \emph{hopping metric} \index{hopping metric!possibilsitic} by
\begin{equation}\label{posshoppingdefn}
\presub{\mathbnd{a}'}h^\mathbnd{a}= \text{poss} (\presub{\mathbnd{a}'}X^\mathtt{a} X_\mathtt{a}^\mathbnd{a} )
~~~~~~\Leftrightarrow~~~~~~
\begin{Compose}{0}{-0.1}\setthirdfont{\mathbnd}
\vbbmatrixsq{h}{0,0}\csymbolthird{a}
\end{Compose}
~=~
\text{poss}\left(
\begin{Compose}{0}{-0.5} \setdefaultfont{\mathbpro}\setsecondfont{\mathtt}\setthirdfont{\mathbnd}
\blackdotsq{d1}{0,-3} \scircle{X1}{0,0}\csymbol{X}\scircle{X2}{0,4} \csymbol{X} \blackdotsq{d2}{0,7}
\joincc[left]{d1}{90}{X1}{-90} \csymbolthird{a} \joincc[left]{X1}{90}{X2}{-90}\csymbolalt{a} \joincc[left]{X2}{90}{d2}{-90} \csymbolthird{a}
\end{Compose}
\right)
\end{equation}
Note that in the symbolic notation we must have $\mathbnd{a}$ and $\mathbnd{a}'$ to indicate that these are not equal. However, in the diagrammatic case, we can see that they are different from the fact that we have two black squares. In the diagrammatic notation, we place a $\mathbnd{a}$ next to the hopping metric to remind us what type of boundary condition this is.
We can also write \eqref{posshoppingdefn} as
\begin{equation}
\presub{\mathbnd{a}'}h^\mathbnd{a} \equiv\presub{\mathbnd{a}'}X^\mathtt{a} X_\mathtt{a}^\mathbnd{a}
~~~~~~\Leftrightarrow~~~~~~
\begin{Compose}{0}{-0.1}\setthirdfont{\mathbnd}
\vbbmatrixsq{h}{0,0}\csymbolthird{a}
\end{Compose}
~\equiv~
\begin{Compose}{0}{-0.5} \setdefaultfont{\mathbpro}\setsecondfont{\mathtt}\setthirdfont{\mathbnd}
\blackdotsq{d1}{0,-3} \scircle{X1}{0,0}\csymbol{X}\scircle{X2}{0,4} \csymbol{X} \blackdotsq{d2}{0,7}
\joincc[left]{d1}{90}{X1}{-90} \csymbolthird{a} \joincc[left]{X1}{90}{X2}{-90}\csymbolalt{a} \joincc[left]{X2}{90}{d2}{-90} \csymbolthird{a}
\end{Compose}
\end{equation}
because closed boundary propositions are equivalent to their own possibilistic values (see \eqref{AequivpossA}).

Clearly we have
\begin{equation}
\presub{\mathbnd{a}'}h^\mathbnd{a} =
\left\{
\begin{array}{ll}
0 & \text{if}~ \mathbnd{a}\not= \mathbnd{a}'  \\
1 & \text{if}~ \mathbnd{a}= \mathbnd{a}'
\end{array}
\right.
\end{equation}
We can define an inverse hopping metric, $\presup{\mathbnd{a}'}h_{\mathbnd{a}''}$ (or $\wwdotssq$), such that
\begin{equation}
     \presup{\mathbnd{a}'}h_{\mathbnd{a}''}  \presub{\mathbnd{a}'}h^\mathbnd{a} = \delta_{\mathbnd{a}''}^\mathbnd{a}    ~~~\text{and}~~~~  \presup{\mathbnd{a}}h_{\mathbnd{a}'}  \presub{\mathbnd{a}''}h^{\mathbnd{a}'}  = \prescript{\mathbnd{a}}{\mathbnd{a}''}\delta
\end{equation}
or, diagrammatically,
\begin{equation}
\begin{Compose}{0}{-0.1} \setdefaultfont{\mathbpro}\setsecondfont{\mathtt}\setthirdfont{\mathbnd}
     \wwmatrixsq{hI}{0,0}\csymbolthird{a}\bbmatrixsq{h}{3.9,0}\csymbolthird{a}
\end{Compose}
=
\begin{Compose}{0}{-0.1} \setdefaultfont{\mathbpro}\setsecondfont{\mathtt}\setthirdfont{\mathbnd}
\wbmatrixsq{I}{0,0} \csymbolthird{a}
\end{Compose}
~~~~\text{and}~~~~
\begin{Compose}{0}{-0.1} \setdefaultfont{\mathbpro}\setsecondfont{\mathtt}\setthirdfont{\mathbnd}
\bbmatrixsq{h}{0,0}\csymbolthird{a}\wwmatrixsq{hI}{3.9,0}\csymbolthird{a}
\end{Compose}
=
\begin{Compose}{0}{-0.1} \setdefaultfont{\mathbpro}\setsecondfont{\mathtt}\setthirdfont{\mathbnd}
\bwmatrixsq{I}{0,0} \csymbolthird{a}
\end{Compose}
\end{equation}
where the repeated index is possibilistically summed over according to \eqref{possaddition} (such that $1+1=1$), and
\begin{equation}
\delta_{\mathbnd{a}''}^\mathbnd{a} =
\left\{
\begin{array}{ll}
0 & \text{if}~ \mathbnd{a}\not= \mathbnd{a}''  \\
1 & \text{if}~ \mathbnd{a} = \mathbnd{a}''
\end{array}
\right.
\end{equation}
We are using possibilistic addition and so the inverse must be understood in that context.  In fact we can see that \begin{equation}
\presup{\mathbnd{a}'}h_\mathbnd{a} = \presub{\mathbnd{a}} h^{\mathbnd{a}'} = \delta_\mathbnd{a}^{\mathbnd{a}'}   ~~~~ (!)
\end{equation}
where the $(!)$ because we are indicating numerical equality of the matrix elements without balancing the index placement.  In the quantum case we also have a hopping metric.  However, in that case, it is not equal to the $\delta_\mathbnd{a}^{\mathbnd{a}'}$.  It is for this reason that we give it a special name (rather than just thinking of it as a delta function of some sort).

We can use the hopping metric and its inverse  to hop indices over a symbol or, in diagrammatic notation, change the colour of squares.  For example,
\begin{equation}
\begin{Compose}{0}{-0.5} \setdefaultfont{\mathbpro}\setsecondfont{\mathtt}\setthirdfont{\mathbnd}
\ucircle{A}{0,0}\csymbolfourth{A} \blackdotsq{U}{90:6} \thispoint{DL}{-120:6}\thispoint{DR}{-60:6}
\joincc[left]{A}{90}{U}{-90} \csymbolthird{a}
\joinccdecorate[above left]{DL}{60}{A}{-120}{,->-=0.5, -bsq=0} \csymbolthird{b}
\joinccdecorate[above right]{DR}{120}{A}{-60}{,->-=0.5, -wsq=0} \csymbolthird{c}
\end{Compose}
~~~~=~~~~
\begin{Compose}{0}{-0.5} \setdefaultfont{\mathbpro}\setsecondfont{\mathtt}\setthirdfont{\mathbnd}
\ucircle{A}{0,0}\csymbolfourth{A} \blackdotsq{U}{90:6} \thispoint{DL}{-120:6}\thispoint{DR}{-60:6}
\joinccwbsq[left]{A}{90}{U}{-90} \csymbolthird{a}
\joinccdecorate[above left]{DL}{60}{A}{-120}{,->-=0.25,->-=-0.10,-bsqw-=0.5, -bsq=0} \csymbolthird{b}
\joinccdecorate[above right]{DR}{120}{A}{-60}{,->-=0.5, -wsq=0} \csymbolthird{c}
\end{Compose}
\end{equation}
This shows how to use $\bbdotssq$ to change a white square into a black square.  We can use $\wwdotssq$ to change black squares into white squares.

\subsection{Simple case}

We will now show how this works with the simplest case - where we have a single typing surface
\begin{equation}
\mathbpro{A}_\mathtt{a}\mathbpro{B}^\mathtt{a}
\end{equation}
First consider
\begin{equation}
\mathbpro{A}_\mathtt{a} \equiv A_\mathbnd{a} \mathbpro{X}_\mathtt{a}^\mathbnd{a}
~~~~ \Leftrightarrow~~~~
\begin{Compose}{0}{-1}\setdefaultfont{\mathbpro}\setsecondfont{\mathtt}
\Ucircle{A}{0,0}   \thispoint{pXA}{0,6} \joincc[left]{A}{90}{pXA}{-90} \csymbolalt{a}
\end{Compose}
~\equiv~
\begin{Compose}{0}{-1} \setdefaultfont{\mathbpro}\setsecondfont{\mathtt}
\ucircle{A}{0,0}\csymbolfourth{A} \Scircle{X}{0,6} \joinccwbsq[left]{A}{90}{X}{-90} \csymbolthird{a}  \thispoint{pXA}{0,10} \joincc[left]{X}{90}{pXA}{-90}\csymbolalt{a}
\end{Compose}
\end{equation}
Further, we define
\begin{equation}
  \presub{\mathbnd{a}}B := \text{poss} (\presub{\mathbnd{a}}{\mathbpro{X}}^\mathtt{a} \mathbpro{B}_\mathtt{a})
~~~~ \Leftrightarrow~~~~
\begin{Compose}{0}{-1}\setthirdfont{\mathbnd}
\blackdotsq{pXB}{0,0} \ucircle{B}{0,6}\csymbolfourth{B}  \joincc[left]{pXB}{90}{B}{-90}\csymbolthird{a}
\end{Compose}
~:=~
\text{poss}\left(
\begin{Compose}{0}{-1} \setdefaultfont{\mathbpro}\setsecondfont{\mathtt}\setthirdfont{\mathbnd}
\blackdotsq{pXB}{0,0} \scircle{X}{0,3}\csymbol{X}\Ucircle{B}{0,7}
\joincc[left]{pXB}{90}{X}{-90} \csymbolthird{a}
\joincc[left]{X}{90}{B}{-90}\csymbolalt{a}
\end{Compose}
\right)
\end{equation}
Then we can write
\begin{equation}
\mathbpro{A}_\mathtt{a}\mathbpro{B}^\mathtt{a} \equiv A_a \mathbpro{X}_\mathtt{a}^a \mathbpro{B}^\mathtt{a} \equiv A_a B^a ~~~~ \Leftrightarrow~~~~
\begin{Compose}{0}{-1}\setdefaultfont{\mathbpro}\setsecondfont{\mathtt}\setthirdfont{\mathbnd}
\Ucircle{A}{0,0} \Ucircle{B}{0,6} \joincc[left]{A}{90}{B}{-90}\csymbolalt{a}
\end{Compose}
~\equiv~
\begin{Compose}{0}{-1} \setdefaultfont{\mathbpro}\setsecondfont{\mathtt}\setthirdfont{\mathbnd}
\ucircle{A}{0,0}\csymbolfourth{A}\scircle{X}{0,6}\csymbol{X}\Ucircle{B}{0,10}
\joinccwbsq[left]{A}{90}{X}{-90} \csymbolthird{a}
\joincc[left]{X}{90}{B}{-90}\csymbolalt{a}
\end{Compose}
~\equiv~
\begin{Compose}{0}{-1} \setdefaultfont{\mathbpro}\setsecondfont{\mathtt}\setthirdfont{\mathbnd}
\ucircle{A}{0,0}\csymbolfourth{A}\ucircle{B}{0,6}\csymbolfourth{B}
\joinccwbsq[left]{A}{90}{B}{-90} \csymbolthird{a}
\end{Compose}
\end{equation}
There is an alternative way to do the calculation above.  In diagrammatic form this is
\begin{equation}
\begin{Compose}{0}{-1}\setdefaultfont{\mathbpro}\setsecondfont{\mathtt}\setthirdfont{\mathbnd}
\Ucircle{A}{0,0} \Ucircle{B}{0,6} \joincc[left]{A}{90}{B}{-90}\csymbolalt{a}
\end{Compose}
~\equiv~
\begin{Compose}{0}{-1}\setdefaultfont{\mathbpro}\setsecondfont{\mathtt}\setthirdfont{\mathbnd}
\Ucircle{A}{0,0} \Scircle{X}{0,4} \ucircle{B}{0,10}\csymbolfourth{B}
\joincc[left]{A}{90}{X}{-90}\csymbolalt{a}
\joinccbwsq[left]{X}{90}{B}{-90} \csymbolthird{a}
\end{Compose}
~\equiv~
\begin{Compose}{0}{-1} \setdefaultfont{\mathbpro}\setsecondfont{\mathtt}\setthirdfont{\mathbnd}
\ucircle{A}{0,0}\csymbolfourth{A}\ucircle{B}{0,6}\csymbolfourth{B}
\joinccbwsq[left]{A}{90}{B}{-90} \csymbolthird{a}
\end{Compose}
\end{equation}
There is yet another way to do the calculation - this time more symmetric:
\begin{equation}
\begin{Compose}{0}{-1}\setdefaultfont{\mathbpro}\setsecondfont{\mathtt}\setthirdfont{\mathbnd}
\Ucircle{A}{0,0} \Ucircle{B}{0,6} \joincc[left]{A}{90}{B}{-90}\csymbolalt{a}
\end{Compose}
~\equiv~
\begin{Compose}{0}{-1} \setdefaultfont{\mathbpro}\setsecondfont{\mathtt}\setthirdfont{\mathbnd}
\ucircle{A}{0,0}\csymbolfourth{A}\scircle{X}{0,5}\csymbol{X}\scircle{X2}{0,8}\csymbol{X}\Ucircle{B}{0,13}
\joinccwbsq[left]{A}{90}{X}{-90} \csymbolthird{a}
\joincc[left]{X}{90}{X2}{-90}\csymbolalt{a}
\joinccbwsq[left]{X2}{90}{B}{-90}\csymbolthird{a}
\end{Compose}
~\equiv~
\begin{Compose}{0}{-1} \setdefaultfont{\mathbpro}\setsecondfont{\mathtt}\setthirdfont{\mathbnd}
\ucircle{A}{0,0}\csymbolfourth{A}\ucircle{B}{0,9.5}\csymbolfourth{B}
\joinccwbbwsq[left]{A}{90}{B}{-90} \csymbolthird{a}
\end{Compose}
\end{equation}
This means that $\nbdotssq\hspace{1.8pt}\wndotssq=\nwdotssq\hspace{1.8pt}\bndotssq=\nwdotssq\hspace{1.8pt}\bbdotssq\hspace{1.5pt}\wndotssq$ and hence we can simply write a
full line
\begin{equation}\label{twoways}
\begin{Compose}{0}{-1} \setdefaultfont{\mathbpro}\setsecondfont{\mathtt}\setthirdfont{\mathbnd}
\ucircle{A}{0,0}\csymbolfourth{A}\ucircle{B}{0,6}\csymbolfourth{B}
\joinccwbsq[left]{A}{90}{B}{-90} \csymbolthird{a}
\end{Compose}
~=~
\begin{Compose}{0}{-1} \setdefaultfont{\mathbpro}\setsecondfont{\mathtt}\setthirdfont{\mathbnd}
\ucircle{A}{0,0}\csymbolfourth{A}\ucircle{B}{0,6}\csymbolfourth{B}
\joinccbwsq[left]{A}{90}{B}{-90} \csymbolthird{a}
\end{Compose}
~=~
\begin{Compose}{0}{-1} \setdefaultfont{\mathbpro}\setsecondfont{\mathtt}\setthirdfont{\mathbnd}
\ucircle{A}{0,0}\csymbolfourth{A}\ucircle{B}{0,9.5}\csymbolfourth{B}
\joinccwbbwsq[left]{A}{90}{B}{-90} \csymbolthird{a}
\end{Compose}
~=~
\begin{Compose}{0}{-1} \setdefaultfont{\mathbpro}\setsecondfont{\mathtt}\setthirdfont{\mathbnd}
\ucircle{A}{0,0}\csymbolfourth{A}\ucircle{B}{0,6}\csymbolfourth{B}
\joincc[left]{A}{90}{B}{-90} \csymbolthird{a}
\end{Compose}
\end{equation}
Since an boundary proposition is equivalent to its possibilistic value we have
\begin{equation}
\text{poss}(\mathbpro{A}_\mathtt{a}\mathbpro{B}^\mathtt{a}) = A_\mathbnd{a} B^\mathbnd{a}
~~~~ \Leftrightarrow~~~~
\text{poss}\left(
\begin{Compose}{0}{-1}\setdefaultfont{\mathbpro}\setsecondfont{\mathtt}\setthirdfont{\mathbnd}
\Ucircle{A}{0,0} \Ucircle{B}{0,6} \joincc[left]{A}{90}{B}{-90}\csymbolalt{a}
\end{Compose}
\right)
~=~
\begin{Compose}{0}{-1} \setdefaultfont{\mathnormal}\setsecondfont{\mathtt}\setthirdfont{\mathbnd}
\Ucircle{A}{0,0}\Ucircle{B}{0,6}
\joincc[left]{A}{90}{B}{-90} \csymbolthird{a}
\end{Compose}
\end{equation}
Hence, we can calculate whether a simple composite boundary proposition is possible or not using the generalized possibilistic states, $A_\mathbnd{a}$ and $B^\mathbnd{a}$.

\subsection{General case}\label{sec:generalcasepossibilistic}

In general we wish to calculate the possibilistic value for a more complicated circuit such as
\begin{equation}
\begin{Compose}{0}{0} \setdefaultfont{\mathbpro}\setsecondfont{\mathtt}\setthirdfont{\mathbnd}
\Ucircle{A}{0,0} \Ucircle{B}{-5,5} \Ucircle{C}{3,4} \Ucircle{D}{-3, 11} \Ucircle{E}{2,9}
\joincc[below left]{B}{-65}{A}{115} \csymbolalt{a}
\joincc[below]{A}{80}{C}{-90} \csymbolalt{b}
\joincc[below]{C}{170}{B}{-10} \csymbolalt{c}
\joincc[above left]{B}{25}{E}{-110} \csymbolalt{d}
\joincc[left]{B}{80}{D}{-100} \csymbolalt{e}
\joincc[above right]{D}{-15}{E}{170}\csymbolalt{k}
\joincc[right]{C}{100}{E}{-80} \csymbolalt{g}
\end{Compose}
\end{equation}
To do this we can insert the decomposition of each boundary proposition in terms of fiducials, as in \eqref{logicaldecomp}, then use
\begin{equation}
\begin{Compose}{0}{0} \setdefaultfont{\mathbpro}\setsecondfont{\mathtt}\setthirdfont{\mathbnd}
\thispoint{L}{0,0}\scircle{X1}{6,0}\csymbol{X}\scircle{X2}{10,0}\csymbol{X}\thispoint{R}{16,0}
\joinccwbsq[above]{L}{0}{X1}{180}\csymbolthird{a} \joincc[above]{X1}{0}{X2}{180} \csymbolalt{a} \joinccbwsq[above]{X2}{0}{R}{180} \csymbolthird{a}
\end{Compose}
~\equiv~
\begin{Compose}{0}{0} \setdefaultfont{\mathbpro}\setsecondfont{\mathtt}\setthirdfont{\mathbnd}
\thispoint{L}{0,0} \thispoint{R}{8,0} \joinccwbbwsq[above]{L}{0}{R}{180} \csymbolthird{a}
\end{Compose}
\end{equation}
on each wire to get the equivalent expression
\begin{equation}
\begin{Compose}{0}{0} \setdefaultfont{\mathnormal}\setsecondfont{\mathbnd}
\Ucircle{A}{-2,-2} \Ucircle{B}{-9,5} \Ucircle{C}{6,4} \Ucircle{D}{-6, 16} \Ucircle{E}{3,13}
\joinccwbbwsq[below left]{B}{-65}{A}{115} \csymbolalt{a}
\joinccwbbwsq[below]{A}{80}{C}{-90} \csymbolalt{b}
\joinccwbbwsq[below]{C}{170}{B}{-10} \csymbolalt{c}
\joinccwbbwsq[above left]{B}{25}{E}{-110} \csymbolalt{d}
\joinccwbbwsq[left]{B}{80}{D}{-100} \csymbolalt{e}
\joinccwbbwsq[above right]{D}{-15}{E}{170}\csymbolalt{k}
\joinccwbbwsq[right]{C}{100}{E}{-80} \csymbolalt{g}
\end{Compose}
\end{equation}
We can replace matched black and white squares with a full wire (as in \eqref{twoways}).  Finally, we obtain
\begin{equation}
\begin{Compose}{0}{-1.3} \setdefaultfont{\mathbpro}\setsecondfont{\mathtt}
\Ucircle{A}{0,0} \Ucircle{B}{-5,5} \Ucircle{C}{3,4} \Ucircle{D}{-3, 11} \Ucircle{E}{2,9}
\joincc[below left]{B}{-65}{A}{115} \csymbolalt{a}
\joincc[below]{A}{80}{C}{-90} \csymbolalt{b}
\joincc[below]{C}{170}{B}{-10} \csymbolalt{c}
\joincc[above left]{B}{25}{E}{-110} \csymbolalt{d}
\joincc[left]{B}{80}{D}{-100} \csymbolalt{e}
\joincc[above right]{D}{-15}{E}{170}\csymbolalt{k}
\joincc[right]{C}{100}{E}{-80} \csymbolalt{g}
\end{Compose}
~~\equiv~~
\begin{Compose}{0}{-1.3} \setdefaultfont{\mathnormal}\setsecondfont{\mathbnd}
\Ucircle{A}{0,0} \Ucircle{B}{-5,5} \Ucircle{C}{3,4} \Ucircle{D}{-3, 11} \Ucircle{E}{2,9}
\joincc[below left]{B}{-65}{A}{115} \csymbolalt{a}
\joincc[below]{A}{80}{C}{-90} \csymbolalt{b}
\joincc[below]{C}{170}{B}{-10} \csymbolalt{c}
\joincc[above left]{B}{25}{E}{-110} \csymbolalt{d}
\joincc[left]{B}{80}{D}{-100} \csymbolalt{e}
\joincc[above right]{D}{-15}{E}{170}\csymbolalt{k}
\joincc[right]{C}{100}{E}{-80} \csymbolalt{g}
\end{Compose}
\end{equation}
Hence the calculation for whether a closed composite boundary proposition is given by means of a calculation involving the generalized possibilistic state \emph{having the same compositional structure}. To actually do this calculation we insert black and white squares on the wires. Then we do possibilistic summation over the corresponding fiducial boundary conditions.  The answer is either 0 or 1.

There is a more general case, when we have some open wires left over. For example,
\begin{equation}
\begin{Compose}{0}{0} \setdefaultfont{\mathbpro}\setsecondfont{\mathtt}
\Ucircle{A}{0,0} \Ucircle{B}{-5,5} \Ucircle{C}{3,4} \Ucircle{D}{-3, 11} \Ucircle{E}{2,9}
\joincc[below left]{B}{-65}{A}{115} \csymbolalt{a}
\joincc[below]{A}{80}{C}{-90} \csymbolalt{b}
\joincc[below]{C}{170}{B}{-10} \csymbolalt{c}
\joincc[above left]{B}{25}{E}{-110} \csymbolalt{d}
\joincc[left]{B}{80}{D}{-100} \csymbolalt{e}
\joincc[above right]{D}{-15}{E}{170}\csymbolalt{k}
\joincc[right]{C}{100}{E}{-80} \csymbolalt{g}
\thispoint{nA}{-2,-2} \joincc[above left]{A}{-135}{nA}{45} \csymbolalt{f}
\thispoint{nB}{-8,5} \joincc[above]{B}{180}{nB}{0} \csymbolalt{h}
\thispoint{nD}{-1,13} \joincc[above left]{D}{45}{nD}{-135} \csymbolalt{i}
\thispoint{nE}{4,11} \joincc[above left]{nE}{-145}{E}{45} \csymbolalt{j}
\end{Compose}
\end{equation}
Then the above procedure will give us
\begin{equation}\label{compencapboundprop}
\begin{Compose}{0}{0} \setdefaultfont{\mathbpro}\setsecondfont{\mathtt}\setthirdfont{\mathbnd}
\ucircle{A}{0,0}\csymbolfourth{A} \ucircle{B}{-5,5}\csymbolfourth{B} \ucircle{C}{3,4}\csymbolfourth{C} \ucircle{D}{-3, 11}\csymbolfourth{D} \ucircle{E}{2,9}\csymbolfourth{E}
\joincc[below left]{B}{-65}{A}{115} \csymbolthird{a}
\joincc[below]{A}{80}{C}{-90} \csymbolthird{b}
\joincc[below]{C}{170}{B}{-10} \csymbolthird{c}
\joincc[above left]{B}{25}{E}{-110} \csymbolthird{d}
\joincc[left]{B}{80}{D}{-100} \csymbolthird{e}
\joincc[above right]{D}{-15}{E}{170}\csymbolthird{k}
\joincc[right]{C}{100}{E}{-80} \csymbolthird{g}
\scircle{XA}{-6,-1} \csymbol{X} \scircle{XB}{-11,5}\csymbol{X} \scircle{XD}{2, 15}\csymbol{X} \scircle{XE}{7,13}\csymbol{X}
\joinccwbsq[above]{A}{-170}{XA}{5}\csymbolthird[0,10]{f} \joinccwbsq[above]{B}{180}{XB}{0}\csymbolthird[0,10]{h} \joinccwbsq[above left]{D}{30}{XD}{-160}\csymbolthird{i}
\joinccbwsq[above left]{XE}{-145}{E}{60}\csymbolthird{j}
\thispoint{nA}{-8,-1.2} \joincc[above left]{XA}{-175}{nA}{25} \csymbolalt{f}
\thispoint{nB}{-13,5} \joincc[above]{XB}{180}{nB}{0} \csymbolalt{h}
\thispoint{nD}{4,17} \joincc[above left]{XD}{45}{nD}{-135} \csymbolalt{i}
\thispoint{nE}{9,15} \joincc[above left]{nE}{-145}{XE}{45} \csymbolalt{j}
\end{Compose}
\end{equation}
This is equal to the generalized possibilistic state
\begin{equation}\label{compgenpossstate}
\begin{Compose}{0}{0} \setdefaultfont{\mathbpro}\setsecondfont{\mathtt}\setthirdfont{\mathbnd}
\ucircle{A}{0,0}\csymbolfourth{A} \ucircle{B}{-5,5}\csymbolfourth{B} \ucircle{C}{3,4}\csymbolfourth{C} \ucircle{D}{-3, 11}\csymbolfourth{D} \ucircle{E}{2,9}\csymbolfourth{E}
\joincc[below left]{B}{-65}{A}{115} \csymbolthird{a}
\joincc[below]{A}{80}{C}{-90} \csymbolthird{b}
\joincc[below]{C}{170}{B}{-10} \csymbolthird{c}
\joincc[above left]{B}{25}{E}{-110} \csymbolthird{d}
\joincc[left]{B}{80}{D}{-100} \csymbolthird{e}
\joincc[above right]{D}{-15}{E}{170}\csymbolthird{k}
\joincc[right]{C}{100}{E}{-80} \csymbolthird{g}
\whitedotsq{XA}{-4,-0.5}  \whitedotsq{XB}{-9,4} \whitedotsq{XD}{1, 14} \whitedotsq{XE}{5,12}
\joincc[above]{A}{-170}{XA}{5}\csymbolthird[0,10]{f}
\joincc[above]{B}{180}{XB}{0}\csymbolthird[0,10]{h}
\joincc[above left]{D}{30}{XD}{-160}\csymbolthird{i}
\joincc[above left]{XE}{-145}{E}{60}\csymbolthird{j}
\end{Compose}
\end{equation}
weighted with the fiducial elements shown.  The expression in \eqref{compgenpossstate} is the generalized possibilistic state associated with the composite boundary proposition in \eqref{compencapboundprop}.

\section{Operational possibilistic formulation}\label{sec:operationalpossibilisticformulation}

\subsection{Operational fiducial boundary conditions}\label{sec:operationalfiducialboundaryconditions}

The actual situation we find ourselves in is that we want to know whether some operationally described situation possible or not (according to the field equations of General Relativity) for some composite region $\mathtt{A}\cup\mathtt{B}\cup \dots$. We would like to work entirely in terms of $(\mathbf{Q}_\mathtt{A}, O_\mathtt{A})$, $(\mathbf{Q}_\mathtt{B}, O_\mathtt{B})$, \dots since these are the operationally accessible quantities.  For any given region of op-space, $\mathtt{A}$, we can associate an operational solution,
$\Psi_\mathtt{A}[\mathbf{Q}_\mathtt{A}, O_\mathtt{A}]$, with $(\mathbf{Q}_\mathtt{A}, O_\mathtt{A})$.  Thus, we can ask whether all the operational solutions for the composite region of interest, give rise to a non-null solution when they are joined together.

The problem we face is, till now, joining requires looking at boundary conditions, $\mathbnd{a}$, that are specified in a very abstract way (as described in Sec.\ \ref{sec:boundaryconditions} and \ref{sec:boundaryconditionswithagency}).  We will see here how to specify such boundary conditions in terms of operationally accessible quantities as long as the following assumption is satisfied. \index{induced boundary purity assumption}
\begin{quote}
{\bf Induced boundary purity assumption}. For any pure boundary condition, $\mathbnd{a}\in\Lambda_\mathtt{a}$, there exists some $(Z, \mathbf{Q}_\mathtt{Z}, O_\mathtt{Z})$ such that
\begin{equation}
\theta_\mathtt{a}(\Psi_\mathtt{Z}[\mathbf{Q}_\mathtt{Z}, O_\mathtt{Z}])=\mathbnd{a}
\end{equation}
Here $\mathtt{Z}$ is a region of op-space such that $\text{set}(\mathtt{a})\subseteq \mathtt{Z}$ and $\Psi_\mathtt{Z}[\mathbf{Q}_\mathtt{Z}, O_\mathtt{Z}]$ is an operational solution.  We do not demand that $\text{set}(\mathtt{a})$ is on the boundary of $\mathtt{Z}$ (we allow it to be in the interior).
\end{quote}
To argue that this is a reasonable assumption, consider a smaller region, $\mathtt{Y}\subset \mathtt{Z}$, an agent strategy for this region, $\mathbf{Q}_\mathtt{Y}\subset\mathbf{Q}_\mathtt{Z}$, and an outcome set, $O_\mathtt{Y}$ that is obtained from $O_\mathtt{Z}$ by taking the intersection of each element $\Gamma_\mathtt{Z}\in O_\mathtt{Z}$ with $\mathtt{Y}$.  In this case, it might be the case that there are many pure solutions in the sort of $\Psi_\mathtt{Y}[\mathbf{Q}_\mathtt{Y}, O_\mathtt{Y}]$, and that this solution induces a mixed boundary condition at $\mathtt{a}$. This is because there may not be enough information in $(\mathbf{Q}_\mathtt{Y}, O_\mathtt{Y})$ to fix a pure boundary condition at $\mathtt{a}$.   However, it is reasonable to assume that for a big enough region (i.e.\ $\mathtt{Z}$) we can have enough information to fix a pure boundary condition.  Hence, we can associate a set of triples, $(\mathtt{Z}, \mathbf{Q}_\mathtt{Z}, O_\mathtt{Z})$ with the elements of a fiducial set, $\Lambda_\mathtt{a}[\text{fid}]$ of boundary conditions.

Every join at some typing surface, such as $\mathtt{a}$, \index{fiducial boundary conditions!operational} happens at the boundary between two regions of op-space, such as $\mathtt{A}$ and $\mathtt{B}$.   For the $\mathtt{A}$ side we choose sets of triples,
\begin{equation}
(\mathtt{Z}, \mathbf{Q}_\mathtt{Z}, O_\mathtt{Z})\in \Upsilon[\mathtt{A}]_\mathtt{a}, ~~~~ (\mathtt{Z}, \mathbf{Q}_\mathtt{Z}, O_\mathtt{Z})\in \Upsilon[\mathtt{A}]^\mathtt{a}
\end{equation}  \index{UpsilonAa@$\Upsilon[\mathtt{A}]_\mathtt{a}$}
where these two sets are equal.  Similarly, for the $\mathtt{B}$ side, we have
\begin{equation}
(\mathtt{Z}, \mathbf{Q}_\mathtt{Z}, O_\mathtt{Z})\in \Upsilon[\mathtt{B}]_\mathtt{a}, ~~~~ (\mathtt{Z}, \mathbf{Q}_\mathtt{Z}, O_\mathtt{Z})\in \Upsilon[\mathtt{B}]^\mathtt{a}
\end{equation}
Note that the $\mathtt{A}$ and $\mathtt{B}$ here are just serving as labels. At this stage we impose no particular relationship between the $\mathtt{Z}$'s and $\mathtt{A}$ or $\mathtt{B}$ (we will discuss such possibilities in Sec.\ \ref{sec:choicesofUpsilonsets}).  We can choose $\Upsilon[\mathtt{A}]_\mathtt{a}=\Upsilon[\mathtt{B}]_\mathtt{a}$ but it is interesting to allow these sets to be different.

We intend to use the elements of these $\Upsilon$ sets as subscripts and superscripts so it is cumbersome to write $(\mathtt{Z}, \mathbf{Q}_\mathtt{Z}, O_\mathtt{Z})$. Instead, we will write
\begin{equation}
a_\mathtt{A}\in \Upsilon[\mathtt{A}]_\mathtt{a}, ~~ a_\mathtt{A}\in \Upsilon[\mathtt{A}]^\mathtt{a},~~ a_\mathtt{B}\in \Upsilon[\mathtt{B}]_\mathtt{a}, ~~a_\mathtt{B}\in\Upsilon[\mathtt{B}]^\mathtt{a}.
\end{equation}
We require that there exists an invertible one-to-one map
\begin{equation}
a_\mathtt{A}    = q[\mathtt{A}]_\mathtt{a}(\mathbnd{a})
\end{equation}
between the element of $\Lambda_\mathtt{a}[\text{fid}]$ and the corresponding elements of $\Upsilon[\mathtt{A}]_\mathtt{a}$.  Specifically, the inverse of $q[\mathtt{A}]_\mathtt{a}(\cdot)$ is
\begin{equation}\label{invertiblemapforUpsilon}
\mathbnd{a} = \theta_\mathtt{a}(\Psi_\mathtt{Z}[\mathbf{Q}_\mathtt{Z}, O_\mathtt{Z}]) ~~~\text{for}~  a_\mathtt{A}=(\mathtt{Z}, \mathbf{Q}_\mathtt{Z}, O_\mathtt{Z})
\end{equation}
Similarly, there is a invertible one-to-one map between the elements of $\Lambda^\mathtt{a}[\text{fid}]$ and the corresponding elements of $\Upsilon^\mathtt{a}[\mathtt{A}]$.
\begin{equation}
a_\mathtt{A} = q[\mathtt{A}]^\mathtt{a}(\mathbnd{a})
\end{equation}
so
\begin{equation}
\Upsilon[\mathtt{A}]_\mathtt{a} = q[\mathtt{A}]_\mathtt{a}( \Lambda_\mathtt{a}[\text{fid}])  ~~~~~  \Upsilon[\mathtt{A}]^\mathtt{a} = q[\mathtt{A}]^\mathtt{a}( \Lambda^\mathtt{a}[\text{fid}])
\end{equation}
There are similar invertible functions, $q[\mathtt{B}]_\mathtt{a}$ and $q[\mathtt{B}]^\mathtt{a}$, for the $\mathtt{B}$ side.

Now we can define
\begin{equation}
\begin{Compose}{0}{0}\setfourthfont{\mathbnd}\setsecondfont{\mathtt}
\whitedot{A}{0,0}\csymbolthird[-26,0]{a_\mathtt{A}} \blackdotsq{B}{2.5,0}\csymbolfourth[20,0]{a} \joincc{A}{0}{B}{180} \csymbolalt{A}
\end{Compose}
~~~~ \Leftrightarrow~~~~
h[\mathtt{A}]^\mathbnd{a}_{a_\mathtt{A}} =
\left\{
\begin{array}{ll}
0  & \text{if} ~ a_\mathtt{A}\not= q[\mathtt{A}]_\mathtt{a}(\mathbnd{a}) \\
1  & \text{if} ~ a_\mathtt{A}    = q[\mathtt{A}]_\mathtt{a}[\mathtt{A}](\mathbnd{a})
\end{array}
\right.
\end{equation}
having inverse
\begin{equation}
\begin{Compose}{0}{0}\setfourthfont{\mathbnd}\setsecondfont{\mathtt}
\blackdotsq{B}{0,0}\csymbolfourth[-20,0]{a} \whitedot{A}{2.5,0}\csymbolthird[26,0]{a_\mathtt{A}}\joincc{B}{0}{A}{180}
\end{Compose}
~~~~~\Leftrightarrow~~~~~ h_\mathbnd{a}^{a_\mathtt{A}}
\end{equation}
for $\mathbnd{a}\in\Lambda_\mathtt{a}[\text{fid}]$ and $a_\mathtt{A}\in\Upsilon[\mathtt{A}]_\mathtt{a}$.  We can also define
\begin{equation}
\begin{Compose}{0}{0}\setfourthfont{\mathbnd}\setsecondfont{\mathtt}
\blackdotsq{A}{0,0}\csymbolfourth[-20,0]{a} \whitedot{B}{2.5,0}\csymbolthird[26,0]{a_\mathtt{A}} \joincc{A}{0}{B}{180}
\end{Compose}
~~~~ \Leftrightarrow
\prescript{a_\mathtt{A}}{\mathbnd{a}}h =
\left\{
\begin{array}{ll}
0  & \text{if} ~ a_\mathtt{A}\not= q[\mathtt{A}]_\mathtt{a}(\mathbnd{a}) \\
1  & \text{if} ~ a_\mathtt{A}    = q[\mathtt{A}]_\mathtt{a}(\mathbnd{a})
\end{array}
\right.
\end{equation}
with inverse
\begin{equation}
\begin{Compose}{0}{0}\setfourthfont{\mathbnd}\setsecondfont{\mathtt}
\whitedot{B}{0,0}\csymbolfourth[-20,0]{a} \blackdotsq{A}{2.5,0}\csymbolthird[26,0]{a_\mathtt{A}} \joincc{B}{0}{A}{180}
\end{Compose}
~~~~~\Leftrightarrow ~~~~~\prescript{\mathbnd{a}}{a_\mathtt{A}}h
\end{equation}
for $\mathbnd{a}\in\Lambda^\mathtt{a}[\text{fid}]$ and $a_\mathtt{A}\in\Upsilon^\mathtt{a}$.
We can match black squares with white squares.  Then we perform possibilistic summation over the elements of the corresponding fiducial set, $\Lambda_\mathtt{a}[\text{fid}]$ or $\Lambda^\mathtt{a}[\text{fid}]$.    We can also match black dots with white dots.  Then we perform possibilistic summation over the elements of the corresponding operational fiducial set, $\Upsilon[\mathtt{A}]_\mathtt{a}$ or $\Upsilon[\mathtt{A}]^\mathtt{a}$.

\subsection{Operational generalized possibilistic states}

We can use the invertible maps just discussed to change squares to dots and vice versa.  Thus, for every calculation involving black and white squares, we can write down the corresponding one involving black and white dots.

Alternatively, we can simply start with fiducial propositions labeled by elements of $\Upsilon[\mathtt{A}]_\mathtt{a}$ and $\Upsilon[\mathtt{A}]^\mathtt{a}$ as follows
\begin{equation}
\mathbpro{X}_\mathtt{a}^{a_\mathtt{A}}  ~~\Leftrightarrow~~
\begin{Compose}{0}{0} \setdefaultfont{\mathbpro}\setsecondfont{\mathtt}
\thispoint{pX}{0,-3}
\Scircle{X}{0,0} \blackdot{Xa}{0,3} \joincc[left]{pX}{90}{X}{-90} \csymbolalt{a} \joincc[left]{X}{90}{Xa}{-90}\csymbolthird{a_\mathtt{A}}
\end{Compose}
~~~~~~~~
\presub{a_\mathtt{A}}{\mathbpro{X}}^\mathtt{a} ~~\Leftrightarrow~~
\begin{Compose}{0}{0} \setdefaultfont{\mathbpro}\setsecondfont{\mathtt}
\thispoint{pX}{0,-3} \Scircle{X}{0,0} \blackdot{Xa}{0,3}
\joincc[left]{X}{-90}{pX}{90}\csymbolalt{a}
\joincc[left]{Xa}{-90}{X}{90} \csymbolthird{a_\mathtt{A}}
\end{Compose}
\end{equation}
These are the boundary propositions at $\mathtt{a}$ associated with the boundary condition induced by $a_\mathtt{A}$ at $\mathtt{a}$ as given by \eqref{invertiblemapforUpsilon}.

Then we can decompose an boundary proposition in an equivalent form involving these fiducials giving
\begin{equation}
\mathbpro{A}^\mathtt{a}_\mathtt{bc} \equiv  \presup{a_\mathtt{A}}A_{b_\mathtt{A}c_\mathtt{A}} ~
\presub{a_\mathtt{A}}{\mathbpro{X}}^\mathtt{a} \mathbpro{X}_\mathtt{b}^{b_\mathtt{A}} \mathbpro{X}_\mathtt{c}^{c_\mathtt{A}}
~~~~~ \Longleftrightarrow ~~~~~
\begin{Compose}{0}{0} \setdefaultfont{\mathbpro}\setsecondfont{\mathtt}
\Ucircle{A}{0,0}\thispoint{DL}{-120:4} \thispoint{DR}{-60:4} \thispoint{UC}{90:4}
\joincc[above left]{DL}{60}{A}{-120} \csymbolalt{b}
\joincc[above right]{DR}{120}{A}{-60} \csymbolalt{c}
\joincc[left]{A}{90}{UC}{-90} \csymbolalt{a}
\end{Compose}
~~~ \equiv ~~~
\begin{Compose}{0}{0} \setdefaultfont{\mathbpro}\setsecondfont{\mathtt}
\ucircle{A}{0,0}\csymbolthird{A}
\scircle{DLX}{-120:5}\csymbol{X} \scircle{DRX}{-60:5}\csymbol{X} \scircle{UCX}{90:5} \csymbol{X}
\thispoint{DL}{-120:8} \thispoint{DR}{-60:8} \thispoint{UC}{90:8}
\joincc[above left]{DL}{60}{DLX}{-120} \csymbolalt{b}
\joincc[above right]{DR}{120}{DRX}{-60} \csymbolalt{c}
\joincc[left]{UCX}{90}{UC}{-90} \csymbolalt{a}
\joinccbw[above left]{DLX}{60}{A}{-120} \csymbolthird{b_\mathtt{A}}
\joinccbw[above right]{DRX}{120}{A}{-60} \csymbolthird{c_\mathtt{A}}
\joinccwb[left]{A}{90}{UCX}{-90} \csymbolthird{a_\mathtt{A}}
\end{Compose}
\end{equation}
The object, $\presup{a_\mathtt{A}}A_{b_\mathtt{A}c_\mathtt{A}}$, is the operational generalized possibilistic state. \index{generalized possibilistic state!operational}

\subsection{Operational hopping metric}

We can join generalized operational possibilistic states together by use of the operational hopping metric. This is defined for a typing surface $\mathtt{a}$ bordering operational regions $\mathtt{A}$ and $\mathtt{B}$ in terms of operational fiducial sets $\Upsilon[\mathtt{A}]^\mathtt{a}$ and $\Upsilon[\mathtt{B}]_\mathtt{a}$ as follows:
\begin{equation}
\presub{a_\mathtt{A}}h^{a_\mathtt{B}}= \text{poss} (\presub{a_\mathtt{A}}{\mathbpro{X}}^\mathtt{a} \mathbpro{X}_\mathtt{a}^{a_\mathtt{B}} )
~~~~~~\Leftrightarrow~~~~~~
\begin{Compose}{0}{-0.1}
\vbbmatrix{h}{0,0}\csymbolthird[32,-62]{a_\mathtt{A}} \csymbolthird[32,62]{a_\mathtt{B}}
\end{Compose}
~=~
\text{poss}\left(
\begin{Compose}{0}{-0.5} \setdefaultfont{\mathbpro}\setsecondfont{\mathtt}
\blackdot{d1}{0,-3} \scircle{X1}{0,0}\csymbol{X}\scircle{X2}{0,4} \csymbol{X} \blackdot{d2}{0,7}
\joincc[left]{d1}{90}{X1}{-90} \csymbolthird[32,-60]{a_\mathtt{A}} \joincc[left]{X1}{90}{X2}{-90}\csymbolalt{a}
\joincc[left]{X2}{90}{d2}{-90} \csymbolthird[32,60]{a_\mathtt{B}}
\end{Compose}
\right)
\end{equation}  \index{hopping metric!operational}
To actually calculate the operational hopping metric, $\bbdots$, we use the hopping metric, $\bbdotssq$, as follows
\begin{equation}
a_\mathtt{A}\bbdots a_\mathtt{B}
= a_\mathtt{A} \customdots[,->-=0.5, *-, -wsq=1]\hspace{-5.2pt} \placesymbol[0,0.35]{\mathbnd{a}}\hspace{-5.3pt} \bbdotssq \hspace{-5.3pt} \placesymbol[0,0.35]{\mathbnd{a}} \hspace{-5.2pt} \customdots[,->-=0.5, -wsq=0, -*]a_\mathtt{B}
\end{equation}
We can invert the operational hopping metric to obtain $\wwdots$ if and only if the inverse, $\wwdotssq$, exists.

\subsection{Choices of $\Upsilon$ sets}\label{sec:choicesofUpsilonsets}

Here we will discuss different strategies for choosing the $\Upsilon$ sets.  The constraints are that $\text{set}(\mathtt{a})\subset \mathtt{Z}$, the operational solution associated with each member of an $\Upsilon$ set must induce a pure boundary condition at $\mathtt{a}$, and we must get the full set of boundary conditions in the corresponding $\Lambda_\mathtt{a}[\text{fid}]$.

One way to choose the $\Upsilon$ sets is to have
\begin{equation}
\Upsilon[\mathtt{A}]_\mathtt{a} = \Upsilon[\mathtt{B}]_\mathtt{a} = \Upsilon_\mathtt{a},  ~~~~~ \Upsilon[\mathtt{A}]^\mathtt{a} = \Upsilon[\mathtt{B}]^\mathtt{a} = \Upsilon^\mathtt{a}
\end{equation}
Then these sets are associated with the typing surface rather than a region of op-space.  If we do this we can remove the $\mathtt{A}$ and $\mathtt{B}$ subscripts on the labels, $a_\mathtt{A}$ and $a_\mathtt{B}$.

Another choice is to choose the $\mathtt{Z}$'s from $\Upsilon[\mathtt{A}]_\mathtt{a}$ to be non-overlapping with the $\mathtt{Z}$'s from $\Upsilon[\mathtt{B}]_\mathtt{a}$ except at $\text{set}(\mathtt{a})$.  To do this we can choose two regions, $\mathtt{Z}_\mathtt{A}$ and $\mathtt{Z}_\mathtt{B}$, such that
\begin{equation}
\mathtt{Z}_\mathtt{A}\cap\mathtt{Z}_\mathtt{B}=\text{set}\mathtt{a}
\end{equation}
Further, we have
\begin{equation}
\text{set}(\mathtt{a})\subset \mathtt{Z}\subseteq \mathtt{Z}_\mathtt{A}
\end{equation}
for every $\mathtt{Z}$ from a triple $(\mathtt{Z}, \mathbf{Q}_\mathtt{Z}, O_\mathtt{Z})\in \Upsilon[\mathtt{A}]_\mathtt{a}$ and
\begin{equation}
\text{set}(\mathtt{a})\subset \mathtt{Z}\subseteq \mathtt{Z}_\mathtt{B}
\end{equation}
for every $\mathtt{Z}$ from a triple $(\mathtt{Z}, \mathbf{Q}_\mathtt{Z}, O_\mathtt{Z})\in \Upsilon[\mathtt{B}]_\mathtt{a}$.  It does not follow from the induced boundary purity assumption (from Sec.\ \ref{sec:operationalfiducialboundaryconditions}) that we can choose the $\Upsilon$ sets in this way.  However, it is reasonable to assume that we can (for the same reasons that this assumption is reasonable).  We will call such such a choice of $\Upsilon$ sets the \emph{natural} choice.  It has the great advantage that the fiducial circuits used to define the hopping metric,
\begin{equation}
\presub{a_\mathtt{A}}h^{a_\mathtt{B}}= \text{poss} (\presub{a_\mathtt{A}}{\mathbpro{X}}^\mathtt{a} \mathbpro{X}_\mathtt{a}^{a_\mathtt{B}} )
~~~~~~\Leftrightarrow~~~~~~
\begin{Compose}{0}{-0.1}
\vbbmatrix{h}{0,0}\csymbolthird[32,-62]{a_\mathtt{A}} \csymbolthird[32,62]{a_\mathtt{B}}
\end{Compose}
~=~
\text{poss}\left(
\begin{Compose}{0}{-0.5} \setdefaultfont{\mathbpro}\setsecondfont{\mathtt}
\blackdot{d1}{0,-3} \scircle{X1}{0,0}\csymbol{X}\scircle{X2}{0,4} \csymbol{X} \blackdot{d2}{0,7}
\joincc[left]{d1}{90}{X1}{-90} \csymbolthird[32,-60]{a_\mathtt{A}} \joincc[left]{X1}{90}{X2}{-90}\csymbolalt{a}
\joincc[left]{X2}{90}{d2}{-90} \csymbolthird[32,60]{a_\mathtt{B}}
\end{Compose}
\right)
\end{equation}
can be naturally understood as physical circuits corresponding to some bigger region $\mathtt{Z}\cup\mathtt{Z}'$ with agent strategy and outcome sets
\begin{equation}
(\mathbf{Q}_\mathtt{Z}\cup\mathbf{Q}_{\mathtt{Z}'}, O_\mathtt{Z}\cup O_{\mathtt{Z}'})
\end{equation}
We adopt this choice for the operational approach to be described in Sec.\ \ref{sec:operationalmanifestlyinvariantformulation}.

In this natural choice, we can choose $\mathtt{Z}_\mathtt{A}$ to be on the $\mathtt{A}$ side of $\mathtt{a}$ and $\mathtt{Z}_\mathtt{B}$ to be on the $\mathtt{B}$ side (though we are free to choose things the other way round).

There is one final possibility that may be possible in some circumstances. This is a special case of the natural choice. For some choices of fiducial set, $\Lambda_\mathtt{a}[\text{fid}]$, we will be able to choose $\mathtt{Z}_\mathtt{A}=\mathtt{A}$ and $\mathtt{Z}_\mathtt{B}=\mathtt{B}$.  This will be true if the operational solutions we want to consider in $\mathtt{A}$ and $\mathtt{B}$ are sufficiently \lq\lq messy" that they induce pure boundary conditions at $\mathtt{a}$.

\subsection{Manifestly diffeomorphism invariant calculations}\label{manifestinvcalcs}

We are interested in performing a calculation to know whether some operationally described situation is possible or not.  To do this we start with the description as a composite operation, then map to the description as a composite encapsulated proposition, then map to the description as a composite boundary proposition, then to an composite operational generalized state.  This looks like
\begin{equation}
\begin{array}{lccc}
{} &
\begin{Compose}{0}{0} \setdefaultfont{\mathsf}\setsecondfont{\mathtt}
\Ucircle{A}{0,0} \Ucircle{B}{-5,5} \Ucircle{C}{3,4} \Ucircle{D}{-3, 11} \Ucircle{E}{2,9}
\joincc[below left]{B}{-65}{A}{115} \csymbolalt{a}
\joincc[below]{A}{80}{C}{-90} \csymbolalt{b}
\joincc[below]{C}{170}{B}{-10} \csymbolalt{c}
\joincc[above left]{B}{25}{E}{-110} \csymbolalt{d}
\joincc[left]{B}{80}{D}{-100} \csymbolalt{e}
\joincc[above right]{D}{-15}{E}{170}
\joincc[right]{C}{100}{E}{-80} \csymbolalt{g}
\end{Compose}
& \longrightarrow &
\begin{Compose}{0}{0} \setdefaultfont{\mathcal}\setsecondfont{\mathtt}
\Ucircle{A}{0,0} \Ucircle{B}{-5,5} \Ucircle{C}{3,4} \Ucircle{D}{-3, 11} \Ucircle{E}{2,9}
\joincc[below left]{B}{-65}{A}{115} \csymbolalt{a}
\joincc[below]{A}{80}{C}{-90} \csymbolalt{b}
\joincc[below]{C}{170}{B}{-10} \csymbolalt{c}
\joincc[above left]{B}{25}{E}{-110} \csymbolalt{d}
\joincc[left]{B}{80}{D}{-100} \csymbolalt{e}
\joincc[above right]{D}{-15}{E}{170}
\joincc[right]{C}{100}{E}{-80} \csymbolalt{g}
\end{Compose}
\\
\longrightarrow &
\begin{Compose}{0}{0} \setdefaultfont{\mathbpro}\setsecondfont{\mathtt}
\Ucircle{A}{0,0} \Ucircle{B}{-5,5} \Ucircle{C}{3,4} \Ucircle{D}{-3, 11} \Ucircle{E}{2,9}
\joincc[below left]{B}{-65}{A}{115} \csymbolalt{a}
\joincc[below]{A}{80}{C}{-90} \csymbolalt{b}
\joincc[below]{C}{170}{B}{-10} \csymbolalt{c}
\joincc[above left]{B}{25}{E}{-110} \csymbolalt{d}
\joincc[left]{B}{80}{D}{-100} \csymbolalt{e}
\joincc[above right]{D}{-15}{E}{170}
\joincc[right]{C}{100}{E}{-80} \csymbolalt{g}
\end{Compose}
& \longrightarrow &
\begin{Compose}{0}{0} \setdefaultfont{\mathnormal}\setsecondfont{\mathnormal}
\Ucircle{A}{-2,-2} \Ucircle{B}{-9,5} \Ucircle{C}{6,4} \Ucircle{D}{-6, 16} \Ucircle{E}{3,13}
\joinccwbbw[below left]{B}{-65}{A}{115}  \csymbolwbunder{a_\mathtt{B}} \csymbolbwunder{a_\mathtt{A}}
\joinccwbbw[below]{A}{80}{C}{-90} \csymbolwbunder{b_\mathtt{A}}\csymbolbwunder{b_\mathtt{C}}
\joinccwbbw[below]{C}{170}{B}{-10} \csymbolwbunder[0,-6.5]{c_\mathtt{C}}\csymbolbwunder[0,-6.5]{c_\mathtt{B}}
\joinccwbbw[above left]{B}{25}{E}{-110} \csymbolwb{d_\mathtt{B}}\csymbolbw{d_\mathtt{E}}
\joinccwbbw[left]{B}{80}{D}{-100} \csymbolwb{e_\mathtt{B}}\csymbolbw{e_\mathtt{D}}
\joinccwbbw[above right]{D}{-15}{E}{170}\csymbolwb{k_\mathtt{D}}\csymbolbw{k_\mathtt{E}}
\joinccwbbw[right]{C}{100}{E}{-80} \csymbolwb{g_\mathtt{C}}\csymbolbwunder{g_\mathtt{E}}
\end{Compose}
\end{array}
\end{equation}
We assume that the encapsulated propositions arising after the first step correspond to actual solutions (rather than candidate solutions). Hence, this is the step in which the field equations of General Relativity are used.  Any operation (e.g.\ $\mathsf{A}^\mathtt{a}$)  for which there are no actual solutions will lead to an operational generalized possibilistic state equal to $0$ ( $A^a=0$ for all $a$ in our example).

\subsection{Operational manifestly invariant formulation}\label{sec:operationalmanifestlyinvariantformulation}

We can extend our $p(\cdot)$ function so that it can also act on linear sums of operations
\begin{equation}
p(\alpha \mathsf{A} + \beta\mathsf{B} +\dots) = \alpha \text{Poss}(\mathsf{A}) + \beta \text{Poss}(\mathsf{B})+\dots
\end{equation}
where $\mathsf{A}$, $\mathsf{B}$, \dots are closed operations, $\alpha$, $\beta$, \dots are equal to 0 or 1 and possibilistic addition is used.
We could also allow it to act on mixed sums like $\alpha \mathsf{A} + \beta\mathbpro{B}$ though we will not have any need to use such mixed expressions in this paper.  We can use this to set up a notion of equivalence for sums of operations like that established for sums of boundary propositions in Sec.\ \ref{sec:possibilisticequivalence}.  For example, we would say that
\begin{equation}
\alpha\mathsf{C}^\mathtt{a}_\mathtt{c}+ \beta\mathsf{D}^\mathtt{a}_\mathtt{c} \equiv \gamma\mathsf{F}^\mathtt{a}_\mathtt{c} + \delta\mathsf{G}^\mathtt{a}_\mathtt{c}
\end{equation}
if
\begin{equation}
p((\alpha\mathsf{C}^\mathtt{a}_\mathtt{c}+ \beta\mathsf{D}^\mathtt{a}_\mathtt{c})\mathsf{E}^\mathtt{c}_\mathtt{a}) \equiv
p((\gamma\mathsf{F}^\mathtt{a}_\mathtt{c} + \delta\mathsf{G}^\mathtt{a}_\mathtt{c}\mathsf{E}^\mathtt{c}_\mathtt{a})
\end{equation}
We can do this in general as in Sec. \ref{sec:possibilisticequivalence}.  In particular, note that
\begin{equation}
\mathsf{A}\equiv\text{Poss}(\mathsf{A})
\end{equation}
for the same reason as given in \eqref{AequivpossAproof} for boundary propositions.

We choose $\Upsilon$ sets according to the natural choice discussed in Sec.\ \ref{sec:choicesofUpsilonsets}.  We define fiducial sets of operations
\begin{equation}
\mathsf{X}_\mathtt{a}^{a_\mathtt{A}}  ~~\Leftrightarrow~~
\begin{Compose}{0}{0} \setdefaultfont{\mathsf}\setsecondfont{\mathtt}
\thispoint{pX}{0,-3}
\Scircle{X}{0,0} \blackdot{Xa}{0,3} \joincc[left]{pX}{90}{X}{-90} \csymbolalt{a} \joincc[left]{X}{90}{Xa}{-90}\csymbolthird{a_\mathtt{A}}
\end{Compose}
~~~~~~~~
\presub{a_\mathtt{A}}{\mathsf{X}}^\mathtt{a} ~~\Leftrightarrow~~
\begin{Compose}{0}{0} \setdefaultfont{\mathsf}\setsecondfont{\mathtt}
\thispoint{pX}{0,-3} \Scircle{X}{0,0} \blackdot{Xa}{0,3}
\joincc[left]{X}{-90}{pX}{90}\csymbolalt{a}
\joincc[left]{Xa}{-90}{X}{90} \csymbolthird{a_\mathtt{A}}
\end{Compose}
\end{equation}
where $\mathsf{X}_\mathtt{a}^{a_\mathtt{A}}$ is the operation described by the triple $(\mathtt{Z}, \mathbf{Q}_\mathtt{Z}, O_\mathtt{Z})=a_\mathtt{a}$ having type $(-,\mathtt{a})$, and similarly for $\presub{a_\mathtt{A}}{\mathsf{X}}^\mathtt{a}$ which has typing surface $(\mathtt{a}, -)$.  This means that
\begin{equation}
\text{poss}\left(
\begin{Compose}{0}{-0.5} \setdefaultfont{\mathsf}\setsecondfont{\mathtt}
\blackdot{d1}{0,-3} \scircle{X1}{0,0}\csymbol{X}\scircle{X2}{0,4} \csymbol{X} \blackdot{d2}{0,7}
\joincc[left]{d1}{90}{X1}{-90} \csymbolthird[32,-60]{a_\mathtt{A}} \joincc[left]{X1}{90}{X2}{-90}\csymbolalt{a}
\joincc[left]{X2}{90}{d2}{-90} \csymbolthird[32,60]{a_\mathtt{B}}
\end{Compose}
\right)
~=~
\text{poss}\left(
\begin{Compose}{0}{-0.5} \setdefaultfont{\mathbpro}\setsecondfont{\mathtt}
\blackdot{d1}{0,-3} \scircle{X1}{0,0}\csymbol{X}\scircle{X2}{0,4} \csymbol{X} \blackdot{d2}{0,7}
\joincc[left]{d1}{90}{X1}{-90} \csymbolthird[32,-60]{a_\mathtt{A}} \joincc[left]{X1}{90}{X2}{-90}\csymbolalt{a}
\joincc[left]{X2}{90}{d2}{-90} \csymbolthird[32,60]{a_\mathtt{B}}
\end{Compose}
\right)
~=~
\begin{Compose}{0}{-0.1}
\vbbmatrix{h}{0,0}\csymbolthird[32,-62]{a_\mathtt{A}} \csymbolthird[32,62]{a_\mathtt{B}}
\end{Compose}
\end{equation}
This, in turn, implies we have decomposition locality \index{decomposition locality} for any operation:
\begin{equation}
\mathsf{A}^\mathtt{a}_\mathtt{bc} \equiv  \presup{a_\mathtt{A}}A_{b_\mathtt{A}c_\mathtt{A}} ~
\presub{a_\mathtt{A}}{\mathsf{X}}^\mathtt{a} \mathsf{X}_\mathtt{b}^{b_\mathtt{A}} \mathsf{X}_\mathtt{c}^{c_\mathtt{A}}
~~~~~ \Longleftrightarrow ~~~~~
\begin{Compose}{0}{0} \setdefaultfont{\mathsf}\setsecondfont{\mathtt}
\Ucircle{A}{0,0}\thispoint{DL}{-120:4} \thispoint{DR}{-60:4} \thispoint{UC}{90:4}
\joincc[above left]{DL}{60}{A}{-120} \csymbolalt{b}
\joincc[above right]{DR}{120}{A}{-60} \csymbolalt{c}
\joincc[left]{A}{90}{UC}{-90} \csymbolalt{a}
\end{Compose}
~~~ \equiv ~~~
\begin{Compose}{0}{0} \setdefaultfont{\mathsf}\setsecondfont{\mathtt}
\ucircle{A}{0,0}\csymbolthird{A}
\scircle{DLX}{-120:5}\csymbol{X} \scircle{DRX}{-60:5}\csymbol{X} \scircle{UCX}{90:5} \csymbol{X}
\thispoint{DL}{-120:8} \thispoint{DR}{-60:8} \thispoint{UC}{90:8}
\joincc[above left]{DL}{60}{DLX}{-120} \csymbolalt{b}
\joincc[above right]{DR}{120}{DRX}{-60} \csymbolalt{c}
\joincc[left]{UCX}{90}{UC}{-90} \csymbolalt{a}
\joinccbw[above left]{DLX}{60}{A}{-120} \csymbolthird{b_\mathtt{A}}
\joinccbw[above right]{DRX}{120}{A}{-60} \csymbolthird{c_\mathtt{A}}
\joinccwb[left]{A}{90}{UCX}{-90} \csymbolthird{a_\mathtt{A}}
\end{Compose}
\end{equation}
This is clear because, if we substitute such decompositions into any operation expression and use expressions such as
\begin{equation}
\begin{Compose}{0}{0} \setdefaultfont{\mathsf}\setsecondfont{\mathtt}
\thispoint{L}{0,0}\scircle{X1}{6,0}\csymbol{X}\scircle{X2}{10,0}\csymbol{X}\thispoint{R}{16,0}
\joinccwb[above]{L}{0}{X1}{180}\csymbolthird{a_\mathtt{B}} \joincc[above]{X1}{0}{X2}{180} \csymbolalt{a} \joinccbw[above]{X2}{0}{R}{180} \csymbolthird{a_\mathtt{A}}
\end{Compose}
~\equiv~
\begin{Compose}{0}{0} \setdefaultfont{\mathbpro}\setsecondfont{\mathtt}\setthirdfont{\mathbnd}
\thispoint{L}{0,0} \thispoint{R}{8,0} \joinccwbbw[above]{L}{0}{R}{180} \csymbolthird{a}
\end{Compose}
\end{equation}
then we can convert directly from the operation diagram to a generalized state calculation as, for  example, in
\begin{equation}
\begin{Compose}{0}{0} \setdefaultfont{\mathsf}\setsecondfont{\mathtt}
\Ucircle{A}{0,0} \Ucircle{B}{-5,5} \Ucircle{C}{3,4} \Ucircle{D}{-3, 11} \Ucircle{E}{2,9}
\joincc[below left]{B}{-65}{A}{115} \csymbolalt{a}
\joincc[below]{A}{80}{C}{-90} \csymbolalt{b}
\joincc[below]{C}{170}{B}{-10} \csymbolalt{c}
\joincc[above left]{B}{25}{E}{-110} \csymbolalt{d}
\joincc[left]{B}{80}{D}{-100} \csymbolalt{e}
\joincc[above right]{D}{-15}{E}{170}
\joincc[right]{C}{100}{E}{-80} \csymbolalt{g}
\end{Compose}
~~~\longrightarrow~~~
\begin{Compose}{0}{0} \setdefaultfont{\mathnormal}\setsecondfont{\mathnormal}
\Ucircle{A}{-2,-2} \Ucircle{B}{-9,5} \Ucircle{C}{6,4} \Ucircle{D}{-6, 16} \Ucircle{E}{3,13}
\joinccwbbw[below left]{B}{-65}{A}{115}  \csymbolwbunder{a_\mathtt{B}} \csymbolbwunder{a_\mathtt{A}}
\joinccwbbw[below]{A}{80}{C}{-90} \csymbolwbunder{b_\mathtt{A}}\csymbolbwunder{b_\mathtt{C}}
\joinccwbbw[below]{C}{170}{B}{-10} \csymbolwbunder[0,-6.5]{c_\mathtt{C}}\csymbolbwunder[0,-6.5]{c_\mathtt{B}}
\joinccwbbw[above left]{B}{25}{E}{-110} \csymbolwb{d_\mathtt{B}}\csymbolbw{d_\mathtt{E}}
\joinccwbbw[left]{B}{80}{D}{-100} \csymbolwb{e_\mathtt{B}}\csymbolbw{e_\mathtt{D}}
\joinccwbbw[above right]{D}{-15}{E}{170}\csymbolwb{k_\mathtt{D}}\csymbolbw{k_\mathtt{E}}
\joinccwbbw[right]{C}{100}{E}{-80} \csymbolwb{g_\mathtt{C}}\csymbolbwunder{g_\mathtt{E}}
\end{Compose}
\end{equation}
This is the basic expression of the sought after operational possibilistic formulation of General Relativity - PoAGeR.

\subsection{Comments on PoAGeR}

This calculation depends on (i) the operational generalized possibilistic state for each operation of interest and (ii) the operational hopping metric for each typing surface, $\mathtt{a}$ and pair of op-regions of interest.  We note that these objects are \emph{manifestly invariant} under diffeomorphisms and time orientation preserving transformations on the $\tau^\mu$ field (i.e. they are invariant under the action of $G^+$).  Furthermore, they do not achieve such manifest invariance in a heavy-handed way (they do not rely on forming sets of objects generated by the group $G^+$).  This is because these objects depend on the manifestly invariant objects, $\Gamma_\mathtt{A}$ and $\mathbf{Q}_\mathtt{A}$ rather than on $\Psi_\mathtt{A}$ and $\theta_\mathtt{a}(\Psi_\mathtt{A})$. These latter objects are invariant under $G^+$ but only by virtue of the fact that they are formed by taking the action of $G+$ on $\tilde{\Psi}_\mathtt{A}$ and $\tilde{\theta}_\mathtt{a}(\Psi_\mathtt{A})$.  For the sake of having useful terminology, we will reserve the term \lq\lq manifestly invariant" for the situation where we do not make use of objects (like $\Psi_\mathtt{A}$ and $\theta_\mathtt{a}(\Psi_\mathtt{A})$) constructed by taking the action of $G^+$.

It is much more appealing to express our calculations in terms of manifestly invariant objects.  To actually calculate these objects, however, it appears that we need to use objects like $\Psi_\mathtt{A}$ and $\theta_\mathtt{a}(\Psi_\mathtt{A})$.  We can take an attitude inspired by Quantum Theory. In quantum theory every operation, $\mathsf{A}^\mathsf{a}_\mathsf{b}$ is associated with a operator $\hat{A}^\mathsf{a}_\mathsf{b}$ (see Part \ref{part:operatortensorQT} of this paper).   There exist theorems that any operator satisfying the physicality constraints (see Sec.\ \ref{sec:physicality}) can be realized by some operation (and, in general, there are multiple ways of doing this). In Quantum Theory people are in the habit of referring to an operation by the operator associated with it because this is a good way to label the corresponding equivalence class and the operator is needed to calculate the probability in any case. Thus, it is of interest simply to explore the properties of operators in this space. The space of possible operators in Quantum Theory is constrained by some very general principles (that probabilities should  be between 0 and 1 and that we cannot signal backward in time). Similarly, it may be possible to prove that the space of operational generalized possibilistic states (the analogue of operators) can be constrained by some very general principles and that all instances in this space are realizable.  Then we can explore the properties of General Relativity with respect to these spaces.

We also note that the approach here is consistent with (and indeed motivated by) the principle of general compositionality described in Sec.\ \ref{sec:introductionatbeginning}.

\subsection{Black and white dots: external and internal points of view}\label{sec:blackandwhitedots}

We can change white dots into black dots using the operational hopping metric
\begin{equation}
\begin{Compose}{0}{0}\setdefaultfont{\mathnormal}
\Ucircle{A}{0,0}
\blackdot{U}{90:6} \csymbol[90:25]{a_\mathtt{B}}
\blackdot{DL}{-150:6}\csymbol[-150:25]{d_\mathtt{D}}
\blackdot{DR}{-30:6} \csymbol[-30:30]{e_\mathtt{C}}
\joinccwb{A}{90}{U}{-90} \csymbolwb{a_\mathtt{A}}
\joinccwb{A}{-150}{DL}{30}\csymbolwb{d_\mathtt{A}}
\joinccbw{DR}{150}{A}{-30}\csymbolbw{e_\mathtt{A}}
\end{Compose}
~~~=~~~
\begin{Compose}{0}{0}\setdefaultfont{\mathnormal}
\Ucircle{A}{0,0}
\blackdot{U}{90:6} \csymbol[90:25]{a_\mathtt{B}}
\blackdot{DL}{-150:6}\csymbol[-150:25]{d_\mathtt{C}}
\blackdot{DR}{-30:6} \csymbol[-30:30]{e_\mathtt{D}}
\joincc{A}{90}{U}{-90}
\joincc{A}{-150}{DL}{30}
\joincc{DR}{150}{A}{-30}
\end{Compose}
\end{equation}
Using the inverse, $\wwdots$, we can change black dots into white dots.  Paying attention to the subscripts on the summation labels we see that the object with all white dots represents an \lq\lq internal" point of view while the object with all black dots corresponds to an \lq\lq external" point of view
\begin{equation}
\text{\lq\lq internal":}
\begin{Compose}{0}{0}\setdefaultfont{\mathnormal}
\Ucircle{A}{0,0}
\whitedot{U}{90:6} \csymbol[90:25]{a_\mathtt{A}}
\whitedot{DL}{-150:6}\csymbol[-150:25]{d_\mathtt{A}}
\whitedot{DR}{-30:6} \csymbol[-30:30]{e_\mathtt{A}}
\joincc{A}{90}{U}{-90}
\joincc{A}{-150}{DL}{30}
\joincc{DR}{150}{A}{-30}
\end{Compose}
~~~~~~~~
\text{\lq\lq external":}
\begin{Compose}{0}{0}\setdefaultfont{\mathnormal}
\Ucircle{A}{0,0}
\blackdot{U}{90:6} \csymbol[90:25]{a_\mathtt{B}}
\blackdot{DL}{-150:6}\csymbol[-150:25]{d_\mathtt{C}}
\blackdot{DR}{-30:6} \csymbol[-30:30]{e_\mathtt{D}}
\joincc{A}{90}{U}{-90}
\joincc{A}{-150}{DL}{30}
\joincc{DR}{150}{A}{-30}
\end{Compose}
\end{equation}
We put inverted commas around internal and external since the op-space regions because although $a_\mathtt{A}$, for example, is labeled by $\mathtt{A}$, the corresponding $\mathtt{Z}$ may or may not be a subset of $\mathtt{A}$.   We note that we can also have objects with both black and white dots.

\subsection{Formalism locality}\label{sec:formalismlocalityPoss}

\index{formalism locality}

In Sec.\ \ref{manifestinvcalcs} we considered only circuits with no wires left open.  This means there are no typing surfaces left unmatched. Then the calculation for an composite operation yields 0 or 1 (not possible or possible). If we have wires left over then we perform a calculation such as
\begin{equation}
\begin{Compose}{0}{0} \setdefaultfont{\mathsf}\setsecondfont{\mathtt}
\Ucircle{A}{0,0} \Ucircle{B}{-5,5} \Ucircle{C}{3,4} \Ucircle{D}{-3, 11} \Ucircle{E}{2,9}
\joincc[below left]{B}{-65}{A}{115} \csymbolalt{a}
\joincc[below]{A}{80}{C}{-90} \csymbolalt{b}
\joincc[below]{C}{170}{B}{-10} \csymbolalt{c}
\joincc[above left]{B}{25}{E}{-110} \csymbolalt{d}
\joincc[left]{B}{80}{D}{-100} \csymbolalt{e}
\joincc[above right]{D}{-15}{E}{170}
\joincc[right]{C}{100}{E}{-80} \csymbolalt{g}
\thispoint{nA}{-2,-2} \joincc[above left]{A}{-135}{nA}{45} \csymbolalt{f}
\thispoint{nB}{-8,5} \joincc[above]{B}{180}{nB}{0} \csymbolalt{h}
\thispoint{nD}{-1,13} \joincc[above left]{D}{45}{nD}{-135} \csymbolalt{i}
\thispoint{nE}{4,11} \joincc[above left]{nE}{-145}{E}{45} \csymbolalt{j}
\end{Compose}
~~~\longrightarrow~~~
\begin{Compose}{0}{0} \setdefaultfont{\mathnormal}\setsecondfont{\mathnormal}
\Ucircle{A}{-2,-2} \Ucircle{B}{-9,5} \Ucircle{C}{6,4} \Ucircle{D}{-6, 16} \Ucircle{E}{3,13}
\joinccwbbw[below left]{B}{-65}{A}{115}  \csymbolwbunder{a_\mathtt{B}} \csymbolbwunder{a_\mathtt{A}}
\joinccwbbw[below]{A}{80}{C}{-90} \csymbolwbunder{b_\mathtt{A}}\csymbolbwunder{b_\mathtt{C}}
\joinccwbbw[below]{C}{170}{B}{-10} \csymbolwbunder[0,-6.5]{c_\mathtt{C}}\csymbolbwunder[0,-6.5]{c_\mathtt{B}}
\joinccwbbw[above left]{B}{25}{E}{-110} \csymbolwb{d_\mathtt{B}}\csymbolbw{d_\mathtt{E}}
\joinccwbbw[left]{B}{80}{D}{-100} \csymbolwb{e_\mathtt{B}}\csymbolbw{e_\mathtt{D}}
\joinccwbbw[above right]{D}{-15}{E}{170}\csymbolwb{k_\mathtt{D}}\csymbolbw{k_\mathtt{E}}
\joinccwbbw[right]{C}{100}{E}{-80} \csymbolwb{g_\mathtt{C}}\csymbolbwunder{g_\mathtt{E}}
\whitedot{nA}{-4,-4} \csymbolalt[-20,-20]{f_\mathtt{A}}    \joincc[above left]{A}{-135}{nA}{45}
\whitedot{nB}{-12,5} \csymbolalt[-25,0]{h_\mathtt{A}}\joincc[above right]{B}{180}{nB}{0}
\whitedot{nD}{-4,18}\csymbolalt[15,20]{i_\mathtt{A}} \joincc[above left]{D}{45}{nD}{-135}
\whitedot{nE}{5,15} \csymbolalt[15,20]{j_\mathtt{A}} \joincc[above left]{nE}{-145}{E}{45}
\end{Compose}
\end{equation}
That is, we obtain an operational generalized possibilistic state (with open wires on it).

If there is at least one choice of labels on the open wires for which this state is equal to 1 then the composite operation is possible (there exists at least one solution to the field equations giving rise to the corresponding outcomes at each operation with the given settings).  We can represent this mathematically.  Let us represent this composite operation by $\mathsf{G}^\mathtt{fhi}_\mathtt{j}$ and the corresponding operational state by
\begin{equation}
\presup{f_\mathtt{A}h_\mathtt{B}i_\mathtt{D}}G_{j_\mathtt{E}}
\end{equation}
We define $\presub{f_\mathtt{A}}T$ to be equal to 1 for all $f_\mathtt{A}\in\Upsilon[\mathtt{A}]_\mathtt{a}$ (we reserve the symbol $T$ for this purpose).  Similarly, we define  $T^{j_\mathtt{E}}$ to be equal to 1 for all $j_\mathtt{E}\in\Upsilon[\mathtt{E}]^\mathtt{a}$.  Then we can write
\begin{equation}\label{Gtracing}
G:= \presup{f_\mathtt{A}h_\mathtt{B}i_\mathtt{D}}G_{j_\mathtt{E}} \presub{f_\mathtt{A}}T \smallspace \presub{h_\mathtt{B}}T \smallspace \presub{i_\mathtt{D}}T \smallspace T^{j_\mathtt{E}}
\end{equation}
If $G=1$ then the operation is possible and if $G=0$ the operation is not possible.

We may, however, be interested in more refined questions where we make use of the information that would be lost were we to trace over the indices as in \eqref{Gtracing}.  For example, we might be interested in whether some condition in op-region $\mathtt{A}$ is sufficient to fix a certain property in op-region $\mathtt{B}$ independently of what happens and what is done in rest of op-space.  Consider some operation, $\mathsf{A}^\mathtt{a}_\mathtt{b}$, in region $\mathtt{A}$. This will be the condition. Consider some set of possible operations, $\mathsf{B}_\mathtt{a}^\mathtt{c}[n]$, in region $\mathtt{B}$ where we have a fixed $\mathbf{Q}_\mathtt{B}$ and $n$ labels a set, $O_\mathtt{B}[n]$, of possible outcomes.  Assume, further, that these outcome sets are disjoint and they cover all possible outcomes for region $\mathtt{B}$ (including the empty ourcome where $\Gamma_\mathtt{B}=\varnothing$).  Then if
\begin{equation}
A^{a_\mathtt{A}}_{b_\mathtt{A}}\presup{c_\mathtt{B}}B_{a_\mathtt{A}}[1] = \sum_n A^{a_\mathtt{A}}_{b_\mathtt{A}}\presup{c_\mathtt{B}}B_{a_\mathtt{A}}[n] ~~~\text{and}~~~ A^{a_\mathtt{A}}_{b_\mathtt{A}}\presup{c_\mathtt{B}}B_{a_\mathtt{A}}[n]= 0 ~\forall  n\geq 2
\end{equation}
then, if the condition associated with operation $\mathsf{A}^\mathtt{a}_\mathtt{b}$ is observed in region $\mathtt{A}$ and we have setting $\mathbf{Q}_\mathtt{B}$ in region $\mathtt{B}$, then outcome $O_\mathtt{A}[1]$ must occur in region $\mathtt{B}$ regardless of what is seen and done outside $\mathtt{A}\cup\mathtt{B}$.  Thus we see that we can, in certain circumstances, make predictions for a region without attention to what happens elsewhere.  Elsewhere \cite{hardy2010bformalism} this has been called \emph{formalism locality}.  This sort of calculation is the manifestly diffeomorphism invariant analogue of writing down some initial conditions and evolving the state.  Rather than having initial conditions on an initial spacelike hypersurface in manifold and evolving, we specify conditions in some region of op-space and see what this implies for some other region.

\subsection{Causality in possibilistic formulation}\label{sec:causalityinpossbilisticformulation}

When we choose our sets of allowed pure solutions (such as $\Omega_\mathtt{A}[\mathbf{Q}_\mathtt{A}]$) we can insist that they satisfy the causality condition in Sec.\ \ref{sec:causalitycondition}.  This guarantees that, when we match solutions for different op-space regions, the causality condition is satisfied for the resulting composite region and, hence, we can tell a causal story as outlined in Sec.\ \ref{sec:causality} for each pure solution in any resulting mixture. \index{causality}

The operational generalized state, $A^{a_\mathtt{A}}_{b_\mathtt{A}c_\mathtt{A}}$ for example, associated with an operation $\mathsf{A}^\mathtt{a}_\mathtt{cb}$ may have non-zero entries for more than one of the possible $(a_\mathtt{A}, b_\mathtt{A}, c_\mathtt{A})$.  These map to distinct boundary conditions $(\mathbnd{a}, \mathbnd{b},\mathbnd{c})$ at the surface of the associated region, $\mathtt{A}$, of op-space.  Hence, more than one possible boundary condition is associated with such an operation.  These distinct boundary conditions may correspond to different causal structures at the boundary.  Thus, we cannot necessarily think of some particular typing surface (or even some part of some particular typing surface) as corresponding either to an input (in which the time direction is inward) or an output (in which the time direction is outward).  This means we have fuzzy causal structure.   Hence, when we plug operations together we are not simply matching outputs with inputs (as, for example, in the circuit framework for quantum theory).  Rather, causal matching is taken care of at a more detailed level in the matching of the $\pmb{\tau}$ field when we do possibilistic summation over the different boundary conditions.

Ideally we would be able to impose a simple condition on operational generalized states that tells us that causality is respected. In the case of those theories having fixed causal structure, such a condition exists in the operational framework.  This condition is that the deterministic effect is unique and is sometimes called the Pavia causality condition (as it was invented by Chribella, D'Ariano, and Perinotti \cite{chiribella2010informational}.  A similar condition for the case where the causal structure is not fixed is a worthy target of future research.

\newpage

\part{Probabilistic formulation: PAGeR}\label{part:probabilisticformulation}

\section{Introduction}

We will provide a probabilistic formulation.  We will call this PAGeR (Probabilistic General Relativity with Agency). \index{PAGeR}  We can assume that these probabilities arise, simply, from lack of knowledge.
The main difference with the possibilistic case is that we introduce \emph{loading}. This corresponds to extra specification used to determine certain local probabilities. We use the word loading in the sense of \emph{loading the dice}. In classical physics probabilities are put on top of the ontology (they are subjective).  The values of the probabilities may come from some beliefs, from statistical considerations (such as a Maxwell distribution), from well characterized noise, from a manufacturer's guarantee (in the case of some instrument), from symmetry considerations or elsewhere.   We will introduce the notion of a load abstractly only specifying what this actually means when we have reached the appropriate point in the discussion (and, even then, we will only define this concept for the cases we need to in order to do the relevant calculations).

\section{Objects in formalism}

We are, again, motivated by the principle of general compositionality.  To this end, the structure of the approach here will be similar (though not identical) to that for the possibilistic case.  We will have loaded operations, loaded encapsulated propositions, loaded boundary propositions, and operational generalized states. The notion of a load is different for these different objects.   We will only use the notion of a loaded encapsulated proposition when it is closed (and accordingly, only define what we mean by a load for the closed case).   We will define a map directly from loaded operations to loaded boundary propositions.

We will use fiducials, and a hopping metric in a similar way to the possibilistic case.  Here the addition will be regular addition (rather than possibilistic addition).

\subsection{Loaded Operations}\label{sec:loadedoperationsPAGeR}

We can associate a \emph{loaded operation}\index{loaded operation} with a region of op-space.   This corresponds to
\begin{equation}
\mathsf{A}= \left(\text{strat}(\mathsf{A}), \text{outcome}(\mathsf{A}), \text{load}(\mathsf{A}), \text{reg}(\mathsf{A}), \text{type}(\mathsf{A})\right)
\end{equation}
The new element we have introduced here (compared with operations in the possibilistic case) is the loading description $\text{load}(\mathsf{A})$.  We will sometimes write this as $L_\mathsf{A}$.  The loading for a loaded operation can be different from the loading for a loaded encapsulated proposition.  In particular, we assume that the loading for a loaded operation is some operationally accessible quantity - this could correspond to beliefs, be determined from observables determined by the management beables, instrument manufactures specifications, or some other information. The role of loading will be clarified below.  The other elements of an operation are the same as in the possibilistic case as discussed in Sec.\ \ref{sec:whatanoperationis}.

As before, we will introduce subscripts and superscripts and corresponding diagrammatic notation
\begin{equation}
\mathsf{A}^\mathtt{a}_\mathtt{bc} ~~~ \Longleftrightarrow
\begin{Compose}{0}{0} \setdefaultfont{\mathsf}\setsecondfont{\mathtt}
\Ucircle{A}{0,0}\thispoint{DL}{-120:4} \thispoint{DR}{-60:4} \thispoint{UC}{90:4}
\joincc[above left]{DL}{60}{A}{-120} \csymbolalt{b}
\joincc[above right]{DR}{120}{A}{-60} \csymbolalt{c}
\joincc[left]{A}{90}{UC}{-90} \csymbolalt{a}
\end{Compose}
\end{equation}
This enables us to provide diagrams for calculations.

We compose two loaded operations in the same way that two (unloaded) operations are composed (see Sec.\ \ref{sec:joiningoperations}) though now we need a rule for combining the loading.  Since the loading is part of the description of the operation we could simply assume that
\begin{equation}
\text{load}(\mathsf{AB})=(\text{load}(\mathsf{A}), \text{load}(\mathsf{B}))
\end{equation}
(or we can write $L_\mathsf{AB}= L_\mathsf{A}L_\mathsf{B}$ where the cartesian product is implied).  On the other hand, if we are only interested in equivalence classes (operations that behave the same when joined to other operations) then we may find that different combinations $L_\mathsf{A}L_\mathsf{B}$ behave the same.  In such cases we might want to represent the loading more efficiently by some function $L_\mathsf{AB}(L_\mathsf{A}, L_\mathsf{B})$.

We represent joining loaded operations as before by
\begin{equation}
\mathsf{A}^\mathtt{ab}_\mathtt{cd}\mathsf{B}^\mathtt{ec}_\mathtt{af}  ~~~\Longleftrightarrow~~~
\begin{Compose}{0}{-0.7}   \setdefaultfont{\mathsf}\setsecondfont{\mathtt}
\Ucircle{A}{0,0} \Ucircle{B}{45:5}
\thispoint{Ab}{180:4} \thispoint{Ad}{-90:4}
\thispoint{Be}{$(45:5)+(0:4)$} \thispoint{Bf}{$(45:5)+(90:4)$}
\joincc[above left]{A}{60}{B}{-150} \csymbolalt{a}
\joincc[below right]{B}{-120}{A}{30} \csymbolalt{c}
\joincc[above]{A}{-180}{Ab}{0} \csymbolalt{b} \joincc[left]{Ad}{90}{A}{-90} \csymbolalt{d}
\joincc[above]{B}{0}{Be}{-180} \csymbolalt{e} \joincc[left]{Bf}{-90}{B}{90} \csymbolalt{f}
\end{Compose}
\end{equation}
When we join multiple loaded operations we obtain a new loaded operation.

An \emph{outcome complete set of loaded operations} is a set
\begin{equation}
\left\{ \mathsf{A}^\mathtt{a}_\mathtt{bc}[k_\mathtt{A}]:k_\mathtt{A}\in S_\mathtt{A}\right\}
\end{equation}
where each loaded operation has the same $\mathbf{Q}_\mathtt{A}$ and where the associated outcome sets are disjoint
\begin{equation}
O_\mathtt{A}[k_\mathtt{A}] \cap O_\mathtt{A}[k'_\mathtt{A}] = \varnothing ~~\text{for} ~ k_\mathtt{A}\not=k'_\mathtt{A}
\end{equation}
and the union of all outcome sets,
\begin{equation}
\bigcup_{k_\mathtt{A}\in S_\mathtt{A}} O_\mathtt{A}[k_\mathtt{A}],
\end{equation}
is the set of all possible outcomes (i.e. the set of all possible $\Gamma_\mathtt{A}\in\mathtt{A}$).  In fact we can restrict to the set of all physically possible $\Gamma_\mathtt{A}$ if this set has been characterized. Or we can restrict to the set of all $\Gamma_\mathtt{A}$ that have appropriate smoothness properties as dictated by the fact that they must arise from solutions, $\Psi_\mathtt{A}$.

\subsection{Probabilities for closed loaded operations}\label{sec:probabilitiesforclosed}

A closed loaded operation is one having no open typing surfaces left over.  We make the following assumption: \index{probability assumption}
\begin{quote}
\textbf{Probability assumption.} We can associate a probability
\begin{equation}
\text{prob}(\mathsf{A})= \text{Prob}(O_\mathtt{A}|\mathbf{Q}_\mathtt{A}, L_\mathtt{A})
\end{equation}
with any \emph{closed} loaded operation that depends only on the specification of this loaded operation (and is independent of what outcomes are seen and what choices are made elsewhere):
\end{quote}
Rather than thinking of this as an assumption, we can regard it as definition of what we mean by a closed operation.

One consequence of the probability assumption is that probabilities factorize when we have two closed operations
\begin{equation}
\text{Prob}(\mathsf{AB}) = \text{Prob}(\mathsf{A})\text{Prob}(\mathsf{B})
\end{equation}
where $\mathsf{A}$ and $\mathsf{B}$ are closed because
\begin{equation}
\begin{split}
\text{Prob}(O_\mathtt{A}, O_\mathtt{B}|\mathbf{Q}_\mathtt{A}, &\mathbf{Q}_\mathtt{B}, L_\mathtt{A}, L_\mathtt{B}) \\
&=
\text{Prob}(O_\mathtt{A}| O_\mathtt{B}\mathbf{Q}_\mathtt{A}, \mathbf{Q}_\mathtt{B}, L_\mathtt{A}, L_\mathtt{B}) \text{Prob}(O_\mathtt{B}|\mathbf{Q}_\mathtt{A}, \mathbf{Q}_\mathtt{B}, L_\mathtt{A}, L_\mathtt{B})  \\
&= \text{Prob}(O_\mathtt{A}|\mathbf{Q}_\mathtt{A}, L_\mathtt{A}) \text{Prob}(O_\mathtt{B}|\mathbf{Q}_\mathtt{B}, L_\mathtt{B})
\end{split}
\end{equation}
The last step follows because the probability of $O_\mathtt{A}$ only depends on $\mathbf{Q}_\mathtt{A}$ and $L_\mathtt{A}$ by the probability assumption (and similarly in the $\mathtt{B}$ case).

It is possible that the set, $O_\mathtt{A}$, is of measure zero.  Then we should, instead, use a probability density. Then we have
\begin{equation}
\text{ProbDensity}(\mathsf{A}[k_\mathtt{A}])
\end{equation}
for the loaded operations in an outcome complete set of loaded operations.  We must have
\begin{equation}
\SumInt_{k_\mathtt{A}\in S_\mathtt{A}} \text{ProbDensity}(\mathsf{A}[k_\mathtt{A}]) \mathrm{d}k_\mathtt{A} = 1
\end{equation}
so that the total probability is equal to 1.

There is an issue with the probability assumption above.  A closed operation has no typing surface but it may have some boundary that is not not associated with a typing surface. In this case we can expect there to be correlations across this boundary.  We can take one of two attitudes towards this.  Either we can demand that, at the untyped part of the boundary, the operation corresponds to some pure boundary conditions. This would prevent there being correlations.  Or we can assume that, for the purposes at hand, we are simply not going to be making comparisons across any untyped boundary.

\subsection{loaded encapsulated propositions}\label{sec:loadedencapprops}

We define a loaded encapsulated proposition as follows
\begin{equation}\label{defnloadedencap}
\mathcal{A}= \left( \text{prop}(\mathcal{A}), \text{strat}(\mathcal{A}), \text{load}(\mathcal{A}), \text{reg}(\mathcal{A}), \text{type}(\mathcal{A}) \right)
\end{equation}  \index{loaded encapsulated propositions}
This is the same as in the possibilistic case (as discussed in Sec.\ \ref{sec:encapsulatedpropositions}) except that we have added $\text{load}(\mathcal{A})$ (sometimes denoted $L_\mathcal{A}$).  This is an extra piece of information we can add to specify appropriate probabilistic information (as this cannot be captured by the other elements).   We will only require a definition of what we mean by the load for closed loaded encapsulated propositions.  In this case (as explained below) the load corresponds to a subnormalized probability distribution over those pure solutions consistent with $\text{prop}(\mathcal{A})$.

We will add the typing surfaces as superscripts and subscripts, or as wires in diagrammatic notation. For example
\begin{equation}
\mathcal{A}^\mathtt{a}_\mathtt{bc} ~~~ \Longleftrightarrow
\begin{Compose}{0}{0} \setdefaultfont{\mathcal}\setsecondfont{\mathtt}
\Ucircle{A}{0,0}\thispoint{DL}{-120:4} \thispoint{DR}{-60:4} \thispoint{UC}{90:4}
\joincc[above left]{DL}{60}{A}{-120} \csymbolalt{b}
\joincc[above right]{DR}{120}{A}{-60} \csymbolalt{c}
\joincc[left]{A}{90}{UC}{-90} \csymbolalt{a}
\end{Compose}
\end{equation}
We can join encapsulated loaded propositions
\begin{equation}
\mathcal{A}^\mathtt{ab}_\mathtt{cd}\mathcal{B}^\mathtt{ec}_\mathtt{af}  ~~~\Longleftrightarrow~~~
\begin{Compose}{0}{-0.7}   \setdefaultfont{\mathcal}\setsecondfont{\mathtt}
\Ucircle{A}{0,0} \Ucircle{B}{45:5}
\thispoint{Ab}{180:4} \thispoint{Ad}{-90:4}
\thispoint{Be}{$(45:5)+(0:4)$} \thispoint{Bf}{$(45:5)+(90:4)$}
\joincc[above left]{A}{60}{B}{-150} \csymbolalt{a}
\joincc[below right]{B}{-120}{A}{30} \csymbolalt{c}
\joincc[above]{A}{-180}{Ab}{0} \csymbolalt{b} \joincc[left]{Ad}{90}{A}{-90} \csymbolalt{d}
\joincc[above]{B}{0}{Be}{-180} \csymbolalt{e} \joincc[left]{Bf}{-90}{B}{90} \csymbolalt{f}
\end{Compose}
\end{equation}
to give a new encapsulated loaded proposition.   The rules for combining two encapsulated loaded propositions are the same as for two encapsulated propositions except that now we need some rules to say what $\text{load}(\mathcal{AB})$ is.  We can simply take this to be
\begin{equation}
\text{load}(\mathcal{AB})=(\text{load}(\mathcal{A}), \text{load}(\mathcal{B}))
\end{equation}
However, if we are interested in equivalence classes (encapsulated loaded propositions that have the same behavior when connected to others) then we may use a more efficient loading description for the new composite encapsulated loaded proposition.

\subsection{Probabilities for closed loaded encapsulated propositions}\label{sec:probsforclosedlencapprops}

A closed loaded encapsulated proposition is one having no typing surfaces (so $\text{type}(\mathcal{A})=(-,-)$.   We will associate a probability, $\text{Prob}(\mathcal{A})$ with any closed loaded encapsulated proposition, $\mathcal{A}$.  We understand this as being due to some subnormalized probability distribution over the underlying pure solutions.   Let $\Psi_\mathtt{A}[\mathcal{A}]$ be the solution (this will be mixed in general) associated with the proposition $\text{prop}(\mathcal{A})$.  Consider the pure solutions
\begin{equation}
\Psi_\mathtt{A}\in \text{sort}(\Psi_\mathtt{A}[\mathcal{A}])
\end{equation}
We associate with any such closed $\mathcal{A}$ a \emph{subnormalized} probability distribution over these pure solutions
\begin{equation}
\rho^{L_\mathcal{A}}_\mathtt{A}(\Psi_\mathtt{A})  ~~~\text{for}~~~ \Psi_\mathtt{A}\in \text{sort}(\Psi_\mathtt{A}[\mathcal{A}])
\end{equation}
such that the probability associated with the loaded encapsulated proposition is given by
\begin{equation}
\SumInt_{\Psi_\mathtt{A}\in\text{sort}(\Psi_\mathtt{A}[\mathcal{A}])} \rho^{L_\mathcal{A}}_\mathtt{A}(\Psi_\mathtt{A}) d\Psi_\mathtt{A} = \text{Prob}(\mathcal{A})
\end{equation}
This means that, in the case of closed loaded encapsulated propositions, we will say that the loading is given by this subnormalized distribution over pure states (this is the content of the \emph{Probability assumption} to be introduced more carefully in Sec.\ \ref{sec:probabilitiesforclosed}).   We use the $\SumInt$ symbol here to indicate the necessary kind of summation/integration.  We should bear in mind that  $\Psi_\mathtt{A}$ is a set of $\tilde{\Psi}_\mathtt{A}$ where these $\tilde{\Psi}_\mathtt{A}$ are given by specifying fields on a variable manifold.  It is not entirely clear how to perform integration of this nature - it is clearly at least as difficult as functional integration.   Later we will replace integration over the space of $\Psi_\mathtt{A}$ with integration over an operational space which will be more clearly defined.

\subsection{Map from loaded operations to loaded encapsulated propositions}\label{sec:maplopstolencaps}

We want a map from closed loaded operations to closed loaded encapsulated propositions
\begin{equation}
\mathsf{A} \rightarrow \mathcal{A}  ~~~\text{when}~~~ \text{type}(\mathcal{A})=\text{type}(\mathsf{A}) =(-,-)
\end{equation}
so that we can calculate loaded encapsulated propositions given a loaded operation.   This means we  need to specify the elements in \eqref{defnloadedencap} of the loaded encapsulated proposition.  We put
\begin{equation}
\text{strat}(\mathcal{A})=\text{strat}(\mathsf{A})=\mathbf{Q}_\mathtt{A}, ~~~~~
\text{reg}(\mathcal{A})=\text{reg}(\mathsf{A})=\mathtt{A}, ~~~~~
\end{equation}
Then we choose
\begin{equation}
\text{prop}(\mathcal{A})=\text{Prop}(\mathbf{Q}_\mathtt{A}, O_\mathtt{A})
\end{equation}
where $O_\mathtt{A}=\text{outcome}(\mathsf{A})$ and $\text{Prop}(\mathbf{Q}_\mathtt{A}, O_\mathtt{A})$ is the proposition associated with the operational solution $\Psi_\mathtt{A}[\mathbf{Q}_\mathtt{A}, O_\mathtt{A}]$.  Finally, we need some map,
\begin{equation}
L_\mathsf{A} \rightarrow L_\mathcal{A}
\end{equation}
We require (for the special case of closed objects) that we can calculate the subnormalized probability distribution associated with $\mathcal{A}$ from the loading description, $L_\mathsf{A}$, for the operation.  We also require
such that
\begin{equation}
\text{prob}(\mathcal{A})=\text{Prob}(\mathsf{A})
\end{equation}

\subsection{Loaded boundary propositions}

Now we wish to introduce the idea of loaded boundary propositions. \index{loaded boundary propositions} A loaded boundary proposition pertains to a typing surface and is denoted by $\mathlbpro{A}^\mathtt{a}$, $\mathlbpro{B}^\mathtt{a}_\mathtt{b}$, etc.  A loaded boundary proposition will be written as
\begin{equation}
\mathlbpro{A}^\mathtt{a}_\mathtt{bc}
\end{equation}
and it contains information about the boundary proposition, $\mathbpro{A}^\mathtt{a}_\mathtt{bc}$, in the sense of Sec.\ \ref{sec:boundarypropositions} and also has some loading information, $L_\mathlbpro{A}$, built in.  We are not yet in a position to specify exactly what we mean by a loaded boundary proposition.  Once we have developed the appropriate theory, we will show how to represent them as a sum over fiducials.

We will associate such loaded boundary propositions with loaded encapsulated propositions using the notation $\mathlbpro{A}$ is associated with $\mathcal{A}$, $\mathlbpro{B}$ is associated with $\mathcal{B}$, etc.  We reserve the symbol $\mathlbpro{X}$ for fiducials.

We can represent a loaded boundary proposition diagrammatically as
\begin{equation}
\mathlbpro{A}^\mathtt{a}_\mathtt{bc} ~~~ \Longleftrightarrow
\begin{Compose}{0}{0} \setdefaultfont{\mathlbpro}\setsecondfont{\mathtt}
\Ucircle{A}{0,0}\thispoint{DL}{-120:4} \thispoint{DR}{-60:4} \thispoint{UC}{90:4}
\joincc[above left]{DL}{60}{A}{-120} \csymbolalt{b}
\joincc[above right]{DR}{120}{A}{-60} \csymbolalt{c}
\joincc[left]{A}{90}{UC}{-90} \csymbolalt{a}
\end{Compose}
\end{equation}
We could attempt to provide composition rules at this stage.  However, we have not specified what we mean by the loading. Hence, we will first introduce probabilities and the associated machinery as this will play an important role in understanding what it means to combining loaded boundary propositions.

\subsection{Probabilities for closed loaded boundary propositions}

We demand that a closed loaded boundary proposition takes the form
\begin{equation}
\mathlbpro{A} = \text{Prob}(\mathlbpro{A}) \mathbf{1}
\end{equation}
where $\mathbf{1}$ is the proposition that is always true.  Compare this with the case of a closed boundary proposition (in the possiblistic case, see Sec.\ \ref{sec:boundarypropositions}) which can be equal to $\mathbf{1}$ or $\mathbf{0}$.

\section{The simple case}

\subsection{Composite region: simple case}\label{sec:compositeregion:simplecase}

Now we consider the simple case of a composite loaded operation
\begin{equation}
\mathsf{A}^\mathtt{a}\mathsf{B}_\mathtt{a} ~~~~~ \Leftrightarrow~~~~~
\begin{Compose}{0}{-0.7}\setdefaultfont{\mathsf}\setsecondfont{\mathtt}
\Ucircle{A}{0,0} \Ucircle{B}{0,5} \joincc[left]{A}{90}{B}{-90} \csymbolalt{a}
\end{Compose}
\end{equation}
Since this is a closed loaded operation it, and the associated loaded encapsulated proposition, will have a probability associated with it. Further, there will be a subnormalized distribution $\rho^{L_\mathtt{A}L_\mathtt{B}}_\mathtt{A\cup B}(\Psi_\mathtt{A\cup B})$ over pure states
$\Psi_\mathtt{A\cup B}\in \text{sort}(\Psi_\mathtt{A}[\mathcal{A}]\Cup_\mathtt{a} \Psi_\mathtt{B}[\mathcal{B}])$.

This distribution will induce a distribution over pure boundary conditions, $\mathbnd{a}\in \Lambda_\mathtt{a}$, associated with the typing surface $\mathtt{a}$
\begin{equation}
\rho_\mathtt{a}(\mathbnd{a}) ~~~~\text{such that} ~~~~ \SumInt_{\mathbnd{a}\in\Lambda_\mathtt{a}} \rho_\mathtt{a}(\mathbnd{a}) d\mathbnd{a} = \text{Prob}(\mathsf{A}^\mathtt{a}\mathsf{B}_\mathtt{a})
\end{equation}
where
\begin{equation}
\rho_\mathtt{a}(\mathbnd{a})= \SumInt_{\Psi_\mathtt{A}\in\Omega_\mathtt{A}[\mathbnd{a}]} \rho(\Psi_\mathtt{A}) d\Psi_\mathtt{A}
\end{equation}
where $\Omega_\mathtt{A}[\mathbnd{a}]$ is the set of pure solutions, $\Psi_\mathtt{A\cup B}$, for which $\theta_\mathtt{a}(\Psi_\mathtt{A\cup B}) =\mathbnd{a}$.  What we learn from this is that we can calculate the probability for the composite loaded operation from a function defined at the typing surface.   This tells us that the associated loaded boundary proposition can be written as
\begin{equation}
\mathlbpro{A}^\mathtt{a} \mathlbpro{B}_\mathtt{a} = \left( \SumInt \rho_\mathtt{a}(\mathbnd{a}) \mathrm{d}\mathbnd{a}\right) \mathbf{1}
\end{equation}
We can gain extra insight into this by writing it as
\begin{equation}
\mathlbpro{A}^\mathtt{a} \mathlbpro{B}_\mathtt{a}
= p_{\mathlbpro{AB}}(\theta_\mathtt{a}(\Psi_\mathtt{A}[\mathcal{A}])\cap\theta_\mathtt{a}(\Psi_\mathtt{B}[\mathcal{B}]) ) \mathbf{1}
\end{equation}
where
\begin{equation}
p_{\mathlbpro{AB}}(\mathbnd{a}') = \SumInt_{\mathbnd{a}\in\text{sort}(\mathbnd{a}')} \rho_\mathtt{a}(\mathbnd{a}) \mathrm{d}\mathbnd{a}
\end{equation}
Note that
\begin{equation}
p_{\mathlbpro{AB}}(\varnothing)=0
\end{equation}
This tells us how to calculate the probability associated with a closed loaded proposition for the simple composite case.

\subsection{Introducing fiducials}\label{sec:introducingfiducialsprob}

We can proceed as follows.  We can write
\begin{equation}
\text{Prob}(\mathsf{A}^\mathtt{a}\mathsf{B}_\mathtt{a}) = \text{Prob}(\mathcal{A}^\mathtt{a}\mathcal{B}_\mathtt{a})
= \text{Prob}(\mathlbpro{A}^\mathtt{a}\mathlbpro{B}_\mathtt{a})
\end{equation}
We introduce a set of fiducial loaded boundary propositions
\begin{equation}
\mathlbpro{X}_\mathtt{a}^{\mathlbnd{a}}   ~~~~~  \mathlbnd{a} \in \bar{\Lambda}_\mathtt{a}[\text{fid}]
\end{equation}     \index{fiducial loaded boundary propositions}
where $\mathlbnd{a}$ labels elements of the set, $\bar{\Lambda}_\mathtt{a}[\text{fid}]$ of fiducial loaded propositions.  This set may be continuous or discrete depending on the situations we wish to treat.   Given the fiducial set we can write
\begin{equation}
\text{Prob}(\mathlbpro{A}^\mathtt{a}\mathlbpro{B}_\mathtt{a}) = \text{Prob}(\mathlbpro{A}^\mathtt{a}\mathlbpro{X}^{\mathlbnd{a}}_\mathtt{a}) B_{\mathlbnd{a}}
= A^{\mathlbnd{a}} B_{\mathlbnd{a}}
\end{equation}
where summation and/or integration over $\mathlbnd{a}\in \bar{\Lambda}_\mathtt{a}[\text{fid}]$ is implicit and
\begin{equation}
 A^{\mathlbnd{a}} := \text{Prob}(\mathlbpro{A}^\mathtt{a}\mathlbpro{X}^\mathbnd{a}_\mathtt{a})
\end{equation}
If we write the summation/integration in explicitly we have
\begin{equation}
\text{Prob}(\mathlbpro{A}^\mathtt{a}\mathlbpro{B}_\mathtt{a}) = \SumInt_{\mathlbnd{a}\in \bar{\Lambda}[\text{fid}]} A^{\mathlbnd{a}} B_{\mathlbnd{a}} \mathrm{d} \mathlbnd{a}
\end{equation}
The summation (and/or integration) is now by the standard rules of arithmetic (not possibilistic summation).  Note that we can always choose a fiducial set big enough that it is possible to write the probability in this way because the set of fiducial loaded boundary propositions could consist of all loaded boundary propositions.  However, in general, the probabilities for different situations will depend on one another. Hence, we would expect a much more efficient fiducial sets to be possible.  We assume that the set chosen is, in some appropriate sense, minimal.  We assume linear dependence because probabilities add linearly convex combinations (see Appendix B of \cite{hardy2011reformulating} for more discussion).  We use the $\SumInt$ and $\mathrm{d}\mathlbnd{a}$ symbols to indicate the appropriate type of summatin/integration (as per the discussion at the end of Sec.\ \ref{sec:probsforclosedlencapprops}).f

The objects $A^\mathlbnd{a}$ and $B_\mathlbnd{a}$ are \emph{generalized probabilistic states}. These will be discussed in more detail in Sec. \ref{sec:generalprobcase}.

The choice of fiducial set here will depend on the situation we want to model.  If any probability distribution over pure solutions is possible then a possible choice of fiducial set is where the loaded boundary propositions correspond to pure boundary conditions (the loading will then, effectively, be a delta function distribution centered on this pure boundary condition).   In general, we can have arbitrary mixtures of any given probability distributions. This is the reason that we use linear compression above (indeed, in a very general sense, if we allow arbitrary mixtures, we must use linear compression \cite{hardy2011reformulating}).

\subsection{Equivalence, equality, and the $p(\cdot)$ function}\label{sec:equivalenceequality}

We can set up a notion of probabilistic equivalence \index{probabilistic equivalence} (in a similar fashion to the possibilistic case).  First we define a $p(\cdot)$ function
\begin{equation}
p(\alpha \mathlbpro{A} + \beta \mathlbpro{B} + \dots ) =  \alpha \text{prob}(\mathlbpro{A}) + \beta \text{prob}(\mathlbpro{B}) + \dots
\end{equation}
where $\mathlbpro{A}$, $\mathlbpro{B}$, \dots are \emph{closed} loaded boundary propositions and $\alpha$, $\beta$, \dots are real numbers.  Note this is a different $p(\cdot)$ function to that used in the possibilistic case.

Now we have defined a $p(\cdot)$ function, we can provide a definition of equivalence.  Let us give an example first. We say that
\begin{equation}
\mathlbpro{F}^\mathtt{a}_\mathtt{c}+\mathlbpro{G}^\mathtt{a}_\mathtt{c} \equiv \mathlbpro{H}^\mathtt{a}_\mathtt{c}+\mathlbpro{J}^\mathtt{a}_\mathtt{c}
\end{equation}
if
\begin{equation}
p((\mathlbpro{F}^\mathtt{a}_\mathtt{c}+\mathlbpro{G}^\mathtt{a}_\mathtt{c})\mathlbpro{E}_\mathtt{a}^\mathtt{c})
=
p( (\mathlbpro{H}^\mathtt{a}_\mathtt{c}+\mathlbpro{J}^\mathtt{a}_\mathtt{c}) \mathlbpro{E}_\mathtt{a}^\mathtt{c})
\end{equation}
for all $\mathlbpro{E}_\mathtt{a}^\mathtt{c}$.  In general, we say that
\begin{equation}
\text{expression}_1 \equiv \text{expression}_2
\end{equation}
if
\begin{equation}
p(\text{expression}_1 \mathlbpro{E} )=p(\text{expression}_2\mathlbpro{E})
\end{equation}
(where we are suppressing typing surface subscripts and superscripts because we are considering the general case)
for all $\mathlbpro{E}$ such that
\begin{equation}
\text{expression}_i \mathlbpro{E}
\end{equation}
is a sum the form $\alpha \mathlbpro{A} + \beta \mathlbpro{B} + \dots $ (of closed loaded boundary propositions).

Note that a closed boundary proposition is equivalent to its own probability:
\begin{equation}\label{AequivprobA}
\mathlbpro{A}\equiv \text{poss}(\mathlbpro{A})
\end{equation}
because
\begin{equation}\label{AequivprobAreason}
p(\mathlbpro{AE})= p(\mathlbpro{A})p(\mathlbpro{E}) = p(\text{poss}(\mathlbpro{A})\mathlbpro{E})
\end{equation}
for all closed $\mathlbpro{E}$.  In the first step we use the fact that probabilities factorize for two closed loaded operations (and, consequently, two closed loaded boundary propositions).  In the second step we use the properties of the $p(\cdot)$ function.

We will say that two expressions are \emph{equal} if they are equivalent and each term is of the same type.  Thus, in the example above, we will actually say
\begin{equation}
\mathlbpro{F}^\mathtt{a}_\mathtt{c}+\mathlbpro{G}^\mathtt{a}_\mathtt{c} = \mathlbpro{H}^\mathtt{a}_\mathtt{c}+\mathlbpro{J}^\mathtt{a}_\mathtt{c}
\end{equation}
because each term has the same type.  This will enable us to define loaded boundary propositions as sums over fiducials. \index{loaded boundary propositions} On the other hand, we cannot insert an equals sign in \eqref{AequivprobA}.

\subsection{The simple case again}

Now we can write
\begin{equation}
\text{Prob}(\mathlbpro{A}^\mathtt{a}\mathlbpro{B}_\mathtt{a}) = \text{Prob}(\mathlbpro{A}^\mathtt{a}\mathlbpro{X}^{\mathlbnd{a}}_\mathtt{a}) B_{\mathlbnd{a}}
= p(\mathlbpro{A}^\mathtt{a}\mathlbpro{X}^{\mathlbnd{a}}_\mathtt{a}B_{\mathlbnd{a}})
\end{equation}
Since this is true for any $\mathlbpro{A}^\mathtt{a}$ it follows from the remarks at the end of Sec.\ \ref{sec:equivalenceequality} that
\begin{equation}
\mathlbpro{B}_\mathtt{a} =\mathlbpro{X}^{\mathlbnd{a}}_\mathtt{a} B_{\mathlbnd{a}}
\end{equation}
where summation over $\mathlbnd{a}\in\bar{\Lambda}_\mathtt{a}[\text{fid}]$ is implied.  Here we that a loaded boundary proposition corresponds to a weighted sum over fiducials.  This will be generalized for arbitrary loaded boundary propositions in Sec.\ \ref{sec:generalprobcase}.

We can do the same thing for $\mathlbpro{A}^\mathtt{a}$.  First we introduce a fiducial set of loaded boundary propositions
\begin{equation}
\presub{\mathlbnd{a}}
{\mathlbpro{X}}^\mathtt{a} ~~~~ \text{for}~ \mathlbnd{a}\in \bar{\Lambda}^\mathtt{a}[\text{fid}]
\end{equation}
Now we can write
\begin{equation}
\text{Prob}(\mathlbpro{A}^\mathtt{a}\mathlbpro{B}_\mathtt{a})
= \presup{\mathlbnd{a}}A \smallspace \text{Prob}(\presub{\mathlbnd{a}}{\mathlbpro{X}}^\mathtt{a}\mathlbpro{B}_\mathtt{a})
= p(\presup{\mathlbnd{a}}A\presub{\mathlbnd{a}}{\mathlbpro{X}}^\mathtt{a}\mathlbpro{B}_\mathtt{a})
\end{equation}
Since this is true for any $\mathlbpro{B}_\mathtt{a}$ we have
\begin{equation}
\mathlbpro{A}^\mathtt{a} = \presup{\mathlbnd{a}}A \presub{\mathlbnd{a}}{\mathlbpro{X}}^\mathtt{a}
\end{equation}
where summation is over $\mathlbnd{a}\in\bar{\Lambda}^\mathtt{a}[\text{fid}]$.

We can introduce diagrammatic notation for fiducials:
\begin{equation}
\mathlbpro{X}_\mathtt{a}^{\mathlbnd{a}}  ~~\Leftrightarrow~~
\begin{Compose}{0}{0} \setdefaultfont{\mathlbpro}\setsecondfont{\mathtt}\setthirdfont{\mathlbnd}
\thispoint{pX}{0,-3}
\Scircle{X}{0,0} \blackdotsq{Xa}{0,3} \joincc[left]{pX}{90}{X}{-90} \csymbolalt{a} \joincc[left]{X}{90}{Xa}{-90}\csymbolthird{a}
\end{Compose}
~~~~~~~~
\presub{\mathlbnd{a}}{\mathlbpro{X}}^{\mathtt{a}} ~~\Leftrightarrow~~
\begin{Compose}{0}{0} \setdefaultfont{\mathlbpro}\setsecondfont{\mathtt}\setthirdfont{\mathlbnd}
\thispoint{pX}{0,-3} \Scircle{X}{0,0} \blackdotsq{Xa}{0,3}
\joincc[left]{X}{-90}{pX}{90}\csymbolalt{a}
\joincc[left]{Xa}{-90}{X}{90} \csymbolthird{a}
\end{Compose}
\end{equation}
Then we can write
\begin{equation}
\mathlbpro{A}^\mathtt{a}\mathlbpro{B}_\mathtt{a}=\mathlbpro{A}^\mathtt{a}\mathlbpro{X}^{\mathlbnd{a}}_\mathtt{a}B_{\mathlbnd{a}} \equiv A^{\mathlbnd{a}}B_{\mathlbnd{a}}
~~~~ \Leftrightarrow~~~~
\begin{Compose}{0}{0}\setdefaultfont{\mathlbpro}\setsecondfont{\mathtt}\setthirdfont{\mathlbnd}
\Ucircle{A}{0,0} \Ucircle{B}{0,6} \joincc[left]{A}{90}{B}{-90}\csymbolalt{a}
\end{Compose}
~=~
\begin{Compose}{0}{0}\setdefaultfont{\mathlbpro}\setsecondfont{\mathtt}\setthirdfont{\mathlbnd}
\Ucircle{A}{0,0} \Scircle{X}{0,4} \ucircle{B}{0,10}\csymbolfourth{B}
\joincc[left]{A}{90}{X}{-90}\csymbolalt{a}
\joinccbwsq[left]{X}{90}{B}{-90} \csymbolthird{a}
\end{Compose}
~\equiv~
\begin{Compose}{0}{0} \setdefaultfont{\mathlbpro}\setsecondfont{\mathtt}\setthirdfont{\mathlbnd}
\ucircle{A}{0,0}\csymbolfourth{A}\ucircle{B}{0,6}\csymbolfourth{B}
\joinccbwsq[left]{A}{90}{B}{-90} \csymbolthird{a}
\end{Compose}
\end{equation}
and
\begin{equation}
\mathlbpro{A}^\mathtt{a}\mathlbpro{B}_\mathtt{a} = \presup{\mathlbnd{a}}A \presub{\mathlbnd{a}}{\mathlbpro{X}}^\mathtt{a} \mathlbpro{B}_\mathtt{a}
\equiv   \presup{\mathlbnd{a}}A  \presub{\mathlbnd{a}}B ~~~~ \Leftrightarrow~~~~
\begin{Compose}{0}{0}\setdefaultfont{\mathlbpro}\setsecondfont{\mathtt}\setthirdfont{\mathlbnd}
\Ucircle{A}{0,0} \Ucircle{B}{0,6} \joincc[left]{A}{90}{B}{-90}\csymbolalt{a}
\end{Compose}
~=~
\begin{Compose}{0}{0} \setdefaultfont{\mathlbpro}\setsecondfont{\mathtt}\setthirdfont{\mathlbnd}
\ucircle{A}{0,0}\csymbolfourth{A}\scircle{X}{0,6}\csymbol{X}\Ucircle{B}{0,10}
\joinccwbsq[left]{A}{90}{X}{-90} \csymbolthird{a}
\joincc[left]{X}{90}{B}{-90}\csymbolalt{a}
\end{Compose}
~\equiv~
\begin{Compose}{0}{0} \setdefaultfont{\mathlbpro}\setsecondfont{\mathtt}\setthirdfont{\mathlbnd}
\ucircle{A}{0,0}\csymbolfourth{A}\ucircle{B}{0,6}\csymbolfourth{B}
\joinccwbsq[left]{A}{90}{B}{-90} \csymbolthird{a}
\end{Compose}
\end{equation}

\subsection{The hopping metric}

We can write
\begin{equation}
\mathlbpro{A}^\mathtt{a}\mathlbpro{B}_\mathtt{a} =
\presup{\mathlbnd{a}}A \presub{\mathlbnd{a}}{\mathlbpro{X}}^\mathtt{a}\mathlbpro{X}^{\mathlbnd{a'}}_\mathtt{a}B_{\mathlbnd{a'}} \equiv
\presup{\mathlbnd{a}}A \presub{\mathlbnd{a}}h^{\mathlbnd{a'}} B_{\mathlbnd{a'}}
\end{equation}
where we define the hopping metric
\begin{equation}
\presub{\mathlbnd{a}'}h^{\mathlbnd{a}}= \text{prob} (\presub{\mathlbnd{a}'}X^\mathtt{a} X_\mathtt{a}^\mathlbnd{a} )
~~~~~~\Leftrightarrow~~~~~~
\begin{Compose}{0}{-0.1}\setthirdfont{\mathlbnd}
\vbbmatrixsq{h}{0,0}\csymbolthird{a}
\end{Compose}
~=~
\text{prob}\left(
\begin{Compose}{0}{-0.5} \setdefaultfont{\mathlbpro}\setsecondfont{\mathtt}\setthirdfont{\mathlbnd}
\blackdotsq{d1}{0,-3} \scircle{X1}{0,0}\csymbol{X}\scircle{X2}{0,4} \csymbol{X} \blackdotsq{d2}{0,7}
\joincc[left]{d1}{90}{X1}{-90} \csymbolthird{a} \joincc[left]{X1}{90}{X2}{-90}\csymbolalt{a} \joincc[left]{X2}{90}{d2}{-90} \csymbolthird{a}
\end{Compose}
\right)
\end{equation}  \index{hopping metric!probabilistic}
We can also write
\begin{equation}
\presub{\mathlbnd{a}'}h^\mathlbnd{a} \equiv\presub{\mathlbnd{a}'}X^\mathtt{a} X_\mathtt{a}^\mathlbnd{a}
~~~~~~\Leftrightarrow~~~~~~
\begin{Compose}{0}{-0.1}\setthirdfont{\mathlbnd}
\vbbmatrixsq{h}{0,0}\csymbolthird{a}
\end{Compose}
~\equiv~
\begin{Compose}{0}{-0.5} \setdefaultfont{\mathlbpro}\setsecondfont{\mathtt}\setthirdfont{\mathlbnd}
\blackdotsq{d1}{0,-3} \scircle{X1}{0,0}\csymbol{X}\scircle{X2}{0,4} \csymbol{X} \blackdotsq{d2}{0,7}
\joincc[left]{d1}{90}{X1}{-90} \csymbolthird{a} \joincc[left]{X1}{90}{X2}{-90}\csymbolalt{a} \joincc[left]{X2}{90}{d2}{-90} \csymbolthird{a}
\end{Compose}
\end{equation}
We can calculate the hopping metric from the induced distribution over boundary conditions as in Sec.\ \ref{sec:compositeregion:simplecase}.

We can use the hopping metric to represent the calculation for the simple case in a more symmetric way as follows:
\begin{equation}
\begin{Compose}{0}{0}\setdefaultfont{\mathlbpro}\setsecondfont{\mathtt}\setthirdfont{\mathlbnd}
\Ucircle{A}{0,0} \Ucircle{B}{0,6} \joincc[left]{A}{90}{B}{-90}\csymbolalt{a}
\end{Compose}
~=~
\begin{Compose}{0}{0} \setdefaultfont{\mathlbpro}\setsecondfont{\mathtt}\setthirdfont{\mathlbnd}
\ucircle{A}{0,0}\csymbolfourth{A}\scircle{X}{0,5}\csymbol{X}\scircle{X2}{0,8}\csymbol{X}\Ucircle{B}{0,13}
\joinccwbsq[left]{A}{90}{X}{-90} \csymbolthird{a}
\joincc[left]{X}{90}{X2}{-90}\csymbolalt{a}
\joinccbwsq[left]{X2}{90}{B}{-90}\csymbolthird{a}
\end{Compose}
~\equiv~
\begin{Compose}{0}{0} \setdefaultfont{\mathlbpro}\setsecondfont{\mathtt}\setthirdfont{\mathlbnd}
\ucircle{A}{0,0}\csymbolfourth{A}\ucircle{B}{0,9.5}\csymbolfourth{B}
\joinccwbbwsq[left]{A}{90}{B}{-90} \csymbolthird{a}
\end{Compose}
\end{equation}
This means that $\nbdotssq\hspace{1.8pt}\wndotssq=\nwdotssq\hspace{1.8pt}\bndotssq=\nwdotssq\hspace{1.8pt}\bbdotssq\hspace{1.5pt}\wndotssq$ and hence we can simply write a
full line:
\begin{equation}
\begin{Compose}{0}{0} \setdefaultfont{\mathlbpro}\setsecondfont{\mathtt}\setthirdfont{\mathlbnd}
\ucircle{A}{0,0}\csymbolfourth{A}\ucircle{B}{0,6}\csymbolfourth{B}
\joinccwbsq[left]{A}{90}{B}{-90} \csymbolthird{a}
\end{Compose}
~=~
\begin{Compose}{0}{0} \setdefaultfont{\mathlbpro}\setsecondfont{\mathtt}\setthirdfont{\mathlbnd}
\ucircle{A}{0,0}\csymbolfourth{A}\ucircle{B}{0,6}\csymbolfourth{B}
\joinccbwsq[left]{A}{90}{B}{-90} \csymbolthird{a}
\end{Compose}
~=~
\begin{Compose}{0}{0} \setdefaultfont{\mathlbpro}\setsecondfont{\mathtt}\setthirdfont{\mathlbnd}
\ucircle{A}{0,0}\csymbolfourth{A}\ucircle{B}{0,9.5}\csymbolfourth{B}
\joinccwbbwsq[left]{A}{90}{B}{-90} \csymbolthird{a}
\end{Compose}
~=~
\begin{Compose}{0}{0} \setdefaultfont{\mathlbpro}\setsecondfont{\mathtt}\setthirdfont{\mathlbnd}
\ucircle{A}{0,0}\csymbolfourth{A}\ucircle{B}{0,6}\csymbolfourth{B}
\joincc[left]{A}{90}{B}{-90} \csymbolthird{a}
\end{Compose}
\end{equation}
Since a closed loaded boundary proposition is equivalent to its probability we have
\begin{equation}
\text{prob}(\mathlbpro{A}_\mathtt{a}\mathlbpro{B}^\mathtt{a}) = A_\mathlbnd{a} B^\mathlbnd{a}
~~~~ \Leftrightarrow~~~~
\text{prob}\left(\negs
\begin{Compose}{0}{-0.75}\setdefaultfont{\mathlbpro}\setsecondfont{\mathtt}\setthirdfont{\mathlbnd}
\Ucircle{A}{0,0} \Ucircle{B}{0,6} \joincc[left]{A}{90}{B}{-90}\csymbolalt{a}
\end{Compose}
\smallspace\right)
~=~
\begin{Compose}{0}{-0.75} \setdefaultfont{\mathnormal}\setsecondfont{\mathtt}\setthirdfont{\mathlbnd}
\ucircle{A}{0,0}\csymbolthird{A}\ucircle{B}{0,6}\csymbolthird{B}
\joincc[left]{A}{90}{B}{-90} \csymbolthird{a}
\end{Compose}
\end{equation}
Hence, we can calculate whether a simple composite boundary proposition is possible or not using the generalized probabilistic states, $A_\mathlbnd{a}$ and $B^\mathlbnd{a}$.

In some cases we can define an inverse hopping metric, $\presup{\mathlbnd{a}'}h_{\mathlbnd{a}''}$ (or $\wwdotssq$), such that
\begin{equation}
     \presup{\mathlbnd{a}'}h_{\mathlbnd{a}''}  \presub{\mathlbnd{a}'}h^\mathlbnd{a} = \delta_{\mathlbnd{a}''}^\mathlbnd{a}    ~~~\text{and}~~~~  \presup{\mathlbnd{a}}h_{\mathlbnd{a}'}  \presub{\mathlbnd{a}''}h^{\mathlbnd{a}'}  = \prescript{\mathlbnd{a}}{\mathlbnd{a}''}\delta
\end{equation}
or, diagrammatically,
\begin{equation}
\begin{Compose}{0}{-0.1} \setdefaultfont{\mathlbpro}\setsecondfont{\mathtt}\setthirdfont{\mathlbnd}
     \wwmatrixsq{hI}{0,0}\csymbolthird{a}\bbmatrixsq{h}{3.9,0}\csymbolthird{a}
\end{Compose}
=
\begin{Compose}{0}{-0.1} \setdefaultfont{\mathlbpro}\setsecondfont{\mathtt}\setthirdfont{\mathlbnd}
\wbmatrixsq{I}{0,0} \csymbolthird{a}
\end{Compose}
~~~~\text{and}~~~~
\begin{Compose}{0}{-0.1} \setdefaultfont{\mathlbpro}\setsecondfont{\mathtt}\setthirdfont{\mathlbnd}
\bbmatrixsq{h}{0,0}\csymbolthird{a}\wwmatrixsq{hI}{3.9,0}\csymbolthird{a}
\end{Compose}
=
\begin{Compose}{0}{-0.1} \setdefaultfont{\mathlbpro}\setsecondfont{\mathtt}\setthirdfont{\mathlbnd}
\bwmatrixsq{I}{0,0} \csymbolthird{a}
\end{Compose}
\end{equation}
We need to provide an interpretation for the object $\delta_{\mathlbnd{a}''}^\mathlbnd{a}$.  If we have countable fiducial sets then we can interpret this as a Kronnecker delta function.  If the fiducial sets are not countable then the obvious thing to try is to identify it with the Dirac delta function $\delta(\mathlbnd{a}-\mathlbnd{a}'')$. However, this is problematic because, in general, the set of $\mathlbnd{a}$ is not specified by a finite number of real parameters.  Thus, instead we will say $\delta_{\mathlbnd{a}''}^\mathlbnd{a}$ is the identity substitution operator defined in Appendix \ref{sec:thesubstitutionoperator}.

An inverse will only exist if the fiducial sets are chosen appropriately.  We do not need there to exist an inverse though it is useful if one does exist.

\subsection{Possible choices of fiducials}\label{sec:possiblechoicesoffiducuals}

Here we review some choices we can make for the sets, $\bar{\Lambda}_\mathtt{a}[\text{fid}]$ and $\bar{\Lambda}^\mathtt{a}[\text{fid}]$, of fiducial loaded boundary propositions.

We can choose fiducial loaded boundary propositions associated with pure boundary conditions.  Each such fiducial loaded boundary proposition can then be thought of as corresponding to a delta function probability distribution centered on some pure boundary condition, $\mathbnd{a}$.  In this case we can label the boundary conditions by $\mathlbnd{a}=\mathbnd{a}\in \Lambda_\mathtt{a}$ and
\begin{equation}
\presub{\mathlbnd{a}'}h^{\mathlbnd{a}}=
\text{Prob}(\presub{\mathlbnd{a}'}{\mathlbpro{X}}^\mathtt{a} \mathlbpro{X}_\mathtt{a}^{\mathlbnd{a}} ) =
\presub{\mathlbnd{a}'}\delta^{\mathlbnd{a}}
\end{equation}
where $\presub{\mathlbnd{a}'}\delta^{\mathlbnd{a}}$ is the identity substitution operator defined in Appendix \ref{sec:thesubstitutionoperator}.

Another choice is where the fiducial loaded boundary propositions are associated with op-pure boundary conditions so that $\mathlbnd{a}=\mathbnd{a}\in \Lambda_\mathtt{a}[\text{op-pure}]$
and
\begin{equation}
\presub{\mathlbnd{a}'}h^{\mathlbnd{a}}=
\text{Prob}(\presub{\mathlbnd{a}'}{\mathlbpro{X}}^\mathtt{a} \mathlbpro{X}_\mathtt{a}^{\mathlbnd{a}} ) =
 \presub{\mathlbnd{a}'}w^{\mathlbnd{a}}
\end{equation}
where $\presub{\mathlbnd{a}'}w^{\mathlbnd{a}}$ is the weighted substitution operator defined in Appendix \ref{sec:thesubstitutionoperator}.

\section{The General Case}\label{sec:generalprobcase}

\subsection{Decomposition locality}\label{sec:decompositionpossibilisticoperations}

We write a general loaded proposition as
\begin{equation}\label{loadedlocaldecomp}
\mathlbpro{A}^\mathtt{a}_\mathtt{bc} = \presup{\mathlbnd{a}}A_{\mathlbnd{b}\mathlbnd{c}} ~  \presub{\mathlbnd{a}}{\mathlbpro{X}}^\mathtt{a} \mathlbpro{X}_\mathtt{b}^\mathlbnd{b} \mathlbpro{X}_\mathtt{c}^\mathlbnd{c}
~~~~~ \Longleftrightarrow ~~~~~
\begin{Compose}{0}{0} \setdefaultfont{\mathlbpro}\setsecondfont{\mathtt}\setthirdfont{\mathlbnd}
\Ucircle{A}{0,0}\thispoint{DL}{-120:4} \thispoint{DR}{-60:4} \thispoint{UC}{90:4}
\joincc[above left]{DL}{60}{A}{-120} \csymbolalt{b}
\joincc[above right]{DR}{120}{A}{-60} \csymbolalt{c}
\joincc[left]{A}{90}{UC}{-90} \csymbolalt{a}
\end{Compose}
~~~ = ~~~
\begin{Compose}{0}{0} \setdefaultfont{\mathlbpro}\setsecondfont{\mathtt}\setthirdfont{\mathlbnd}
\ucircle{A}{0,0}\csymbolfourth{A}
\scircle{DLX}{-120:5}\csymbol{X} \scircle{DRX}{-60:5}\csymbol{X} \scircle{UCX}{90:5} \csymbol{X}
\thispoint{DL}{-120:8} \thispoint{DR}{-60:8} \thispoint{UC}{90:8}
\joincc[above left]{DL}{60}{DLX}{-120} \csymbolalt{b}
\joincc[above right]{DR}{120}{DRX}{-60} \csymbolalt{c}
\joincc[left]{UCX}{90}{UC}{-90} \csymbolalt{a}
\joinccbwsq[above left]{DLX}{60}{A}{-120} \csymbolthird{b}
\joinccbwsq[above right]{DRX}{120}{A}{-60} \csymbolthird{c}
\joinccwbsq[left]{A}{90}{UCX}{-90} \csymbolthird{a}
\end{Compose}
\end{equation}
We call this property \emph{decomposition locality} \index{decomposition locality!probabilistic case} - it is closely related to tomographic locality (and equivalent in the case that the hopping metric is invertible).  The duotensor, $\presup{\mathlbnd{a}}A_{\mathlbnd{b}\mathlbnd{c}}$, is the \emph{generalized probabilistic state}.  Clearly our choice of fiducial sets determines how much probabilistic information at the boundary we can represent in this way.  If we choose the fiducial loaded boundary propositions associated with pure boundary conditions such that $\mathlbnd{a}=\mathbnd{a}\in \Lambda_\mathtt{a}$ as discussed in Sec.\ \ref{sec:possiblechoicesoffiducuals} then it is clear that we can represent any amount of information at the boundary (effectively we can get to any induced probability distribution at the boundary) in this way.  However, if our choice of fiducial sets has any sort of course graining then we will not be able to represent so wide a variety of situations.

\subsection{Loading for loaded operations}\label{sec:mapstoloaded}

We need to say what the loading is for a non-closed loaded operation.  This loading, $L_\mathtt{A}$, supplies the appropriate probabilistic information so that we can calculate the loaded boundary proposition.  \index{loading}

One choice is to say that the loading for an operation, $\mathsf{A}_\mathtt{bc}^\mathtt{a}$, say, simply is the generalized state:
\begin{equation}
L_\mathtt{A} =  \presup{\mathlbnd{a}}A_{\mathlbnd{b}\mathlbnd{c}}
\end{equation}
In this case we see that, for a composite operation, $L_\mathtt{AB}\not=(L_\mathtt{A},L_\mathtt{B})$ in general (this possibility was discussed in  Sec.\ \ref{sec:loadedoperationsPAGeR}).

Another choice, we may want $L_\mathtt{A}$ to be specified by some other parameters (such as by sets of scalars associated with management beables).  In this case insist that the generalized state is given by some function of the loading.  In this case we may have $L_\mathtt{AB}=(L_\mathtt{A},L_\mathtt{B})$.

For either choice of how we specify the loading we have a map
\begin{equation}\label{loadedmaps}
\text{loaded operation} \rightarrow \text{loaded boundary proposition}
\end{equation}
and we are in a position to calculate probabilities for general composite operations as we will now see.

\subsection{Calculation for general case}

We start with some composite loaded operation. For example
\begin{equation}
\begin{Compose}{0}{0} \setdefaultfont{\mathsf}\setsecondfont{\mathtt}\setthirdfont{\mathtt}
\Ucircle{A}{0,0} \Ucircle{B}{-5,5} \Ucircle{C}{3,4} \Ucircle{D}{-3, 11} \Ucircle{E}{2,9}
\joincc[below left]{B}{-65}{A}{115} \csymbolalt{a}
\joincc[below]{A}{80}{C}{-90} \csymbolalt{b}
\joincc[below]{C}{170}{B}{-10} \csymbolalt{c}
\joincc[above left]{B}{25}{E}{-110} \csymbolalt{d}
\joincc[left]{B}{80}{D}{-100} \csymbolalt{e}
\joincc[above right]{D}{-15}{E}{170}\csymbolalt{k}
\joincc[right]{C}{100}{E}{-80} \csymbolalt{g}
\end{Compose}
\end{equation}
Here we consider a closed composite loaded operation. We consider the open case in Sec.\ \ref{sec:formalismlocalityinPAGeR}.
We can now map each of the operations in this composite operation to the corresponding loaded boundary proposition. In our example this gives
\begin{equation}
\begin{Compose}{0}{0} \setdefaultfont{\mathlbpro}\setsecondfont{\mathtt}\setthirdfont{\mathlbnd}
\Ucircle{A}{0,0} \Ucircle{B}{-5,5} \Ucircle{C}{3,4} \Ucircle{D}{-3, 11} \Ucircle{E}{2,9}
\joincc[below left]{B}{-65}{A}{115} \csymbolalt{a}
\joincc[below]{A}{80}{C}{-90} \csymbolalt{b}
\joincc[below]{C}{170}{B}{-10} \csymbolalt{c}
\joincc[above left]{B}{25}{E}{-110} \csymbolalt{d}
\joincc[left]{B}{80}{D}{-100} \csymbolalt{e}
\joincc[above right]{D}{-15}{E}{170}\csymbolalt{k}
\joincc[right]{C}{100}{E}{-80} \csymbolalt{g}
\end{Compose}
\end{equation}
To simplify this we can insert the local decomposition of each boundary proposition in terms of fiducials, as in \eqref{loadedlocaldecomp},then use
\begin{equation}
\begin{Compose}{0}{0} \setdefaultfont{\mathlbpro}\setsecondfont{\mathtt}\setthirdfont{\mathlbnd}
\thispoint{L}{0,0}\scircle{X1}{6,0}\csymbol{X}\scircle{X2}{10,0}\csymbol{X}\thispoint{R}{16,0}
\joinccwbsq[above]{L}{0}{X1}{180}\csymbolthird{a} \joincc[above]{X1}{0}{X2}{180} \csymbolalt{a} \joinccbwsq[above]{X2}{0}{R}{180} \csymbolthird{a}
\end{Compose}
~\equiv~
\begin{Compose}{0}{0} \setdefaultfont{\mathbpro}\setsecondfont{\mathtt}\setthirdfont{\mathbnd}
\thispoint{L}{0,0} \thispoint{R}{8,0} \joinccwbbwsq[above]{L}{0}{R}{180} \csymbolthird{a}
\end{Compose}
\end{equation}
on each wire to get the equivalent expression
\begin{equation}
\begin{Compose}{0}{0} \setdefaultfont{\mathnormal}\setsecondfont{\mathlbnd}
\Ucircle{A}{-2,-2} \Ucircle{B}{-9,5} \Ucircle{C}{6,4} \Ucircle{D}{-6, 16} \Ucircle{E}{3,13}
\joinccwbbwsq[below left]{B}{-65}{A}{115} \csymbolalt{a}
\joinccwbbwsq[below]{A}{80}{C}{-90} \csymbolalt{b}
\joinccwbbwsq[below]{C}{170}{B}{-10} \csymbolalt{c}
\joinccwbbwsq[above left]{B}{25}{E}{-110} \csymbolalt{d}
\joinccwbbwsq[left]{B}{80}{D}{-100} \csymbolalt{e}
\joinccwbbwsq[above right]{D}{-15}{E}{170}\csymbolalt{k}
\joinccwbbwsq[right]{C}{100}{E}{-80} \csymbolalt{g}
\end{Compose}
\end{equation}
We can replace matched black and white squares with a full wire. Hence we obtain
\begin{equation}
\begin{Compose}{0}{-1.5} \setdefaultfont{\mathlbpro}\setsecondfont{\mathtt}
\Ucircle{A}{0,0} \Ucircle{B}{-5,5} \Ucircle{C}{3,4} \Ucircle{D}{-3, 11} \Ucircle{E}{2,9}
\joincc[below left]{B}{-65}{A}{115} \csymbolalt{a}
\joincc[below]{A}{80}{C}{-90} \csymbolalt{b}
\joincc[below]{C}{170}{B}{-10} \csymbolalt{c}
\joincc[above left]{B}{25}{E}{-110} \csymbolalt{d}
\joincc[left]{B}{80}{D}{-100} \csymbolalt{e}
\joincc[above right]{D}{-15}{E}{170}\csymbolalt{k}
\joincc[right]{C}{100}{E}{-80} \csymbolalt{g}
\end{Compose}
~~\equiv~~
\begin{Compose}{0}{-1.5} \setdefaultfont{\mathnormal}\setsecondfont{\mathbnd}
\Ucircle{A}{0,0} \Ucircle{B}{-5,5} \Ucircle{C}{3,4} \Ucircle{D}{-3, 11} \Ucircle{E}{2,9}
\joincc[below left]{B}{-65}{A}{115} \csymbolalt{a}
\joincc[below]{A}{80}{C}{-90} \csymbolalt{b}
\joincc[below]{C}{170}{B}{-10} \csymbolalt{c}
\joincc[above left]{B}{25}{E}{-110} \csymbolalt{d}
\joincc[left]{B}{80}{D}{-100} \csymbolalt{e}
\joincc[above right]{D}{-15}{E}{170}\csymbolalt{k}
\joincc[right]{C}{100}{E}{-80} \csymbolalt{g}
\end{Compose}
\end{equation}
Black and white dots can be reinserted if we want to actually do the calculation.  This means
\begin{equation}
\text{prob}\left(
\begin{Compose}{0}{-1.5} \setdefaultfont{\mathsf}\setsecondfont{\mathtt}
\Ucircle{A}{0,0} \Ucircle{B}{-5,5} \Ucircle{C}{3,4} \Ucircle{D}{-3, 11} \Ucircle{E}{2,9}
\joincc[below left]{B}{-65}{A}{115} \csymbolalt{a}
\joincc[below]{A}{80}{C}{-90} \csymbolalt{b}
\joincc[below]{C}{170}{B}{-10} \csymbolalt{c}
\joincc[above left]{B}{25}{E}{-110} \csymbolalt{d}
\joincc[left]{B}{80}{D}{-100} \csymbolalt{e}
\joincc[above right]{D}{-15}{E}{170}\csymbolalt{k}
\joincc[right]{C}{100}{E}{-80} \csymbolalt{g}
\end{Compose}
\right)
~~=~~
\begin{Compose}{0}{-1.5} \setdefaultfont{\mathnormal}\setsecondfont{\mathbnd}
\Ucircle{A}{0,0} \Ucircle{B}{-5,5} \Ucircle{C}{3,4} \Ucircle{D}{-3, 11} \Ucircle{E}{2,9}
\joincc[below left]{B}{-65}{A}{115} \csymbolalt{a}
\joincc[below]{A}{80}{C}{-90} \csymbolalt{b}
\joincc[below]{C}{170}{B}{-10} \csymbolalt{c}
\joincc[above left]{B}{25}{E}{-110} \csymbolalt{d}
\joincc[left]{B}{80}{D}{-100} \csymbolalt{e}
\joincc[above right]{D}{-15}{E}{170}\csymbolalt{k}
\joincc[right]{C}{100}{E}{-80} \csymbolalt{g}
\end{Compose}
\end{equation}
The probability for a closed composite loaded operation is given by an expression having the same compositional structure.

\section{Operational probabilistic formulation}\label{sec:operationalprobabilisticformulation}

\subsection{Operational loaded boundary propositions}\label{sec:operationalloadedboundarypropositions}

In the possibilistic case (in Sec. \ref{sec:operationalfiducialboundaryconditions}) we noted that if a certain assumption holds (the induced boundary purity assumption) we can chose a set, $\Upsilon[\mathtt{A}]_\mathtt{a}$, of fiducial boundary propositions labeled by $a_\mathtt{A}=(\mathtt{Z}, \mathbf{Q}_\mathtt{Z}, O_\mathtt{Z})$.  We required that there exists an invertible map between $\Upsilon[\mathtt{A}]_\mathtt{a}$ and $\Lambda_\mathtt{a}[\text{fid}]$.  Here we will consider a set, $\bar{\Upsilon}[\mathtt{A}]_\mathtt{a}$, of fiducial \emph{loaded} boundary propositions labeled by $a_\mathtt{A}=(\mathtt{Z}, \mathbf{Q}_\mathtt{Z}, O_\mathtt{Z}, L_\mathsf{Z})$ (where $L_\mathsf{Z}$ is the loading for the associated loaded operation). We have
\begin{equation}
a_\mathtt{A}=(\mathbf{Q}_\mathtt{A}, O_\mathtt{A}, L_\mathsf{A})\in \bar{\Upsilon}[\mathtt{A}]_\mathtt{a}, ~~~~ a_\mathtt{A}=(\mathbf{Q}_\mathtt{A}, O_\mathtt{A}, L_\mathsf{A})\in \bar{\Upsilon}[\mathtt{A}]^\mathtt{a}
\end{equation}
for the fiducial sets.  We can think of these fiducial sets labeling fiducial loaded boundary propositions or fiducial loaded operations.

We require that there exists a one-to-one map between the elements of  $\bar{\Upsilon}[\mathtt{A}]_\mathtt{a}$ and the elements of $\bar{\Lambda}_\mathtt{a}[\text{fid}]$ - the labels for the fiducial sets in our previous notation.  We have a similar situation for the superscript case.
\begin{equation}
\bar{\Upsilon}[\mathtt{A}]_\mathtt{a} = \bar{q}[\mathtt{A}]_\mathtt{a}(    \bar{\Lambda}_\mathtt{a}[\text{fid}]) ~~~~~
\bar{\Upsilon}[\mathtt{A}]^\mathtt{a} = \bar{q}[\mathtt{A}]^\mathtt{a}(    \bar{\Lambda}^\mathtt{a}[\text{fid}])
\end{equation}

For $\mathlbnd{a}\in\bar{\Lambda}_\mathtt{a}[\text{fid}]$ and $a_\mathtt{A}\in\bar{\Upsilon}[\mathtt{A}]_\mathtt{a}$ we define
\begin{equation}
\begin{Compose}{0}{0}\setfourthfont{\mathlbnd}\setsecondfont{\mathtt}
\whitedot{A}{0,0}\csymbolthird[-30,0]{a_\mathtt{A}} \blackdotsq{B}{2.5,0}\csymbolfourth[20,0]{a} \joincc{A}{0}{B}{180} \csymbolalt{A}
\end{Compose}
~~~~ \Leftrightarrow~~~~
s[\mathtt{A}]^\mathlbnd{a}_{a_\mathtt{A}}
\end{equation}
This is the substitution operator discussed in Appendix \ref{sec:thesubstitutionoperator} which implements a change of variables (from $\mathlbnd{a}$ to $a_\mathtt{A}$).
The inverse is
\begin{equation}
\begin{Compose}{0}{0}\setfourthfont{\mathlbnd}\setsecondfont{\mathtt}
\blackdotsq{B}{0,0}\csymbolfourth[-20,0]{a} \whitedot{A}{2.5,0}\csymbolthird[30,0]{a_\mathtt{A}}\joincc{B}{0}{A}{180}
\end{Compose}
~~~~~\Leftrightarrow~~~~~ s_\mathlbnd{a}^{a_\mathtt{A}}
\end{equation}
This is also a substitution operator.

For $\mathlbnd{a}\in\Lambda^\mathtt{a}[\text{fid}]$ and $a_\mathtt{A}\in\bar{\Upsilon}^\mathtt{a}$ we define
\begin{equation}
\begin{Compose}{0}{0}\setfourthfont{\mathlbnd}\setsecondfont{\mathtt}
\blackdotsq{A}{0,0}\csymbolfourth[-20,0]{a} \whitedot{B}{2.5,0}\csymbolthird[28,0]{a_\mathtt{A}} \joincc{A}{0}{B}{180}
\end{Compose}
~~~~ \Leftrightarrow
\prescript{a_\mathtt{A}}{\mathlbnd{a}}s
\end{equation}
with inverse
\begin{equation}
\begin{Compose}{0}{0}\setfourthfont{\mathlbnd}\setsecondfont{\mathtt}
\whitedot{B}{0,0}\csymbolfourth[-20,0]{a} \blackdotsq{A}{2.5,0}\csymbolthird[30,0]{a_\mathtt{A}} \joincc{B}{0}{A}{180}
\end{Compose}
~~~~~\Leftrightarrow ~~~~~\prescript{\mathlbnd{a}}{a_\mathtt{A}}s
\end{equation}
These are also substitution operators.

We can justify the existence of these $\bar{\Upsilon}$ sets in the same way as we did for $\Upsilon$ sets in the possiblistic case (in Sec.\ \ref{sec:operationalfiducialboundaryconditions}). We can choose the elements of $\bar{\Lambda}_\mathtt{a}[\text{fid}]$ to correspond to delta function probability distributions centered on pure boundary conditions.  Then it follows from the induced boundary purity assumption is true then there exist operations that give rise to these elements of $\bar{\Lambda}_\mathtt{a}[\text{fid}]$.  We choose loaded operations corresponding to delta function probability distributions centered at such operations.

We can use the invertible maps above to convert squares to dots and vice versa (this is really just a change of variables).  Alternatively, we can do everything from scratch using fiducials
\begin{equation}
\mathlbpro{X}_\mathtt{a}^{a_\mathtt{A}}  ~~\Leftrightarrow~~
\begin{Compose}{0}{0} \setdefaultfont{\mathlbpro}\setsecondfont{\mathtt}
\thispoint{pX}{0,-3}
\Scircle{X}{0,0} \blackdot{Xa}{0,3} \joincc[left]{pX}{90}{X}{-90} \csymbolalt{a} \joincc[left]{X}{90}{Xa}{-90}\csymbolthird{a_\mathtt{A}}
\end{Compose}
~~~~~~~~
\presub{a_\mathtt{A}}{\mathlbpro{X}}^\mathtt{a} ~~\Leftrightarrow~~
\begin{Compose}{0}{0} \setdefaultfont{\mathlbpro}\setsecondfont{\mathtt}
\thispoint{pX}{0,-3} \Scircle{X}{0,0} \blackdot{Xa}{0,3}
\joincc[left]{X}{-90}{pX}{90}\csymbolalt{a}
\joincc[left]{Xa}{-90}{X}{90} \csymbolthird{a_\mathtt{A}}
\end{Compose}
\end{equation}
Then we can decompose a loaded boundary proposition in an equivalent form involving these fiducials giving
\begin{equation}
\mathlbpro{A}^\mathtt{a}_\mathtt{bc} \equiv  \presup{a_\mathtt{A}}A_{b_\mathtt{A}c_\mathtt{A}} ~
\presub{a_\mathtt{A}}{\mathlbpro{X}}^\mathtt{a} \mathlbpro{X}_\mathtt{b}^{b_\mathtt{A}} \mathlbpro{X}_\mathtt{c}^{c_\mathtt{A}}
~~~~~ \Longleftrightarrow ~~~~~
\begin{Compose}{0}{0} \setdefaultfont{\mathlbpro}\setsecondfont{\mathtt}
\Ucircle{A}{0,0}\thispoint{DL}{-120:4} \thispoint{DR}{-60:4} \thispoint{UC}{90:4}
\joincc[above left]{DL}{60}{A}{-120} \csymbolalt{b}
\joincc[above right]{DR}{120}{A}{-60} \csymbolalt{c}
\joincc[left]{A}{90}{UC}{-90} \csymbolalt{a}
\end{Compose}
~~~ \equiv ~~~
\begin{Compose}{0}{0} \setdefaultfont{\mathlbpro}\setsecondfont{\mathtt}
\ucircle{A}{0,0}\csymbolthird{A}
\scircle{DLX}{-120:5}\csymbol{X} \scircle{DRX}{-60:5}\csymbol{X} \scircle{UCX}{90:5} \csymbol{X}
\thispoint{DL}{-120:8} \thispoint{DR}{-60:8} \thispoint{UC}{90:8}
\joincc[above left]{DL}{60}{DLX}{-120} \csymbolalt{b}
\joincc[above right]{DR}{120}{DRX}{-60} \csymbolalt{c}
\joincc[left]{UCX}{90}{UC}{-90} \csymbolalt{a}
\joinccbw[above left]{DLX}{60}{A}{-120} \csymbolthird{b_\mathtt{A}}
\joinccbw[above right]{DRX}{120}{A}{-60} \csymbolthird{c_\mathtt{A}}
\joinccwb[left]{A}{90}{UCX}{-90} \csymbolthird{a_\mathtt{A}}
\end{Compose}
\end{equation}      \index{loaded boundary propositions!operational} \index{decomposition locality}
The object, $\presup{a_\mathtt{A}}A_{b_\mathtt{A}c_\mathtt{A}}$, is the operational generalized possibilistic state.

\subsection{Fiducial loaded operations}\label{sec:fiducialloadedoperations}

As in the possibilistic case, we can extend our $p(\cdot)$ function so that
\begin{equation}
p(\alpha \mathsf{A} + \beta\mathsf{B} +\dots) = \alpha \text{Poss}(\mathsf{A}) + \beta \text{Poss}(\mathsf{B})+\dots
\end{equation}
where $\mathsf{A}$, $\mathsf{B}$, \dots are closed loaded operations, and $\alpha$, $\beta$, \dots are real numbers.  Now regular (rather than possibilistic) addition is used.  This allows us to set up an equivalence notion for loaded operations.  For example, we would say that
\begin{equation}
\alpha\mathsf{C}^\mathtt{a}_\mathtt{c}+ \beta\mathsf{D}^\mathtt{a}_\mathtt{c} \equiv \gamma\mathsf{F}^\mathtt{a}_\mathtt{c} + \delta\mathsf{G}^\mathtt{a}_\mathtt{c}
\end{equation}
if
\begin{equation}
p((\alpha\mathsf{C}^\mathtt{a}_\mathtt{c}+ \beta\mathsf{D}^\mathtt{a}_\mathtt{c})\mathsf{E}^\mathtt{c}_\mathtt{a}) =
p((\gamma\mathsf{F}^\mathtt{a}_\mathtt{c} + \delta\mathsf{G}^\mathtt{a}_\mathtt{c}\mathsf{E}^\mathtt{c}_\mathtt{a})
\end{equation}
We can do this in general as in Sec. \ref{sec:equivalenceequality}.  In particular, note that
\begin{equation}
\mathsf{A}\equiv\text{Prob}(\mathsf{A})
\end{equation}
for the same reason as given in \eqref{AequivprobAreason} for loaded boundary propositions.

We now define fiducial sets of loaded operations  \index{fiducial loaded operations}
\begin{equation}
\mathsf{X}_\mathtt{a}^{a_\mathtt{A}}  ~~\Leftrightarrow~~
\begin{Compose}{0}{0} \setdefaultfont{\mathsf}\setsecondfont{\mathtt}
\thispoint{pX}{0,-3}
\Scircle{X}{0,0} \blackdot{Xa}{0,3} \joincc[left]{pX}{90}{X}{-90} \csymbolalt{a} \joincc[left]{X}{90}{Xa}{-90}\csymbolthird{a_\mathtt{A}}
\end{Compose}
~~~~~~~~
\presub{a_\mathtt{A}}{\mathsf{X}}^\mathtt{a} ~~\Leftrightarrow~~
\begin{Compose}{0}{0} \setdefaultfont{\mathsf}\setsecondfont{\mathtt}
\thispoint{pX}{0,-3} \Scircle{X}{0,0} \blackdot{Xa}{0,3}
\joincc[left]{X}{-90}{pX}{90}\csymbolalt{a}
\joincc[left]{Xa}{-90}{X}{90} \csymbolthird{a_\mathtt{A}}
\end{Compose}
\end{equation}
We will see in Sec.\ \ref{sec:decompositionlocalityforloadedoperations} that we can use these fiducials to expand general loaded operations.

\subsection{Operational hopping metric}

We can define the operational hopping metric in terms of the fiducial loaded boundary propositions or in terms of the fiducial loaded operations
\begin{equation}\label{samehoppingloadedcase}
\text{prob}\left(
\begin{Compose}{0}{-0.5} \setdefaultfont{\mathsf}\setsecondfont{\mathtt}
\blackdot{d1}{0,-3} \scircle{X1}{0,0}\csymbol{X}\scircle{X2}{0,4} \csymbol{X} \blackdot{d2}{0,7}
\joincc[left]{d1}{90}{X1}{-90} \csymbolthird[32,-60]{a_\mathtt{A}} \joincc[left]{X1}{90}{X2}{-90}\csymbolalt{a}
\joincc[left]{X2}{90}{d2}{-90} \csymbolthird[32,60]{a_\mathtt{B}}
\end{Compose}
\right)
~=~
\text{prob}\left(
\begin{Compose}{0}{-0.5} \setdefaultfont{\mathlbpro}\setsecondfont{\mathtt}
\blackdot{d1}{0,-3} \scircle{X1}{0,0}\csymbol{X}\scircle{X2}{0,4} \csymbol{X} \blackdot{d2}{0,7}
\joincc[left]{d1}{90}{X1}{-90} \csymbolthird[32,-60]{a_\mathtt{A}} \joincc[left]{X1}{90}{X2}{-90}\csymbolalt{a}
\joincc[left]{X2}{90}{d2}{-90} \csymbolthird[32,60]{a_\mathtt{B}}
\end{Compose}
\right)
~=~
\begin{Compose}{0}{-0.1}
\vbbmatrix{h}{0,0}\csymbolthird[32,-62]{a_\mathtt{A}} \csymbolthird[32,62]{a_\mathtt{B}}
\end{Compose}
\end{equation}  \index{hopping metric!operational probabilistic}
To actually calculate the operational hopping metric, $\bbdots$, we use the hopping metric, $\bbdotssq$, as follows
\begin{equation}
a_\mathtt{A}\bbdots a_\mathtt{B}
= a_\mathtt{A} \customdots[,->-=0.5, *-, -wsq=1]\hspace{-5.2pt} \placesymbol[0,0.35]{\mathlbnd{a}}\hspace{-5.3pt} \bbdotssq \hspace{-5.3pt} \placesymbol[0,0.35]{\mathlbnd{a}} \hspace{-5.2pt} \customdots[,->-=0.5, -wsq=0, -*]a_\mathtt{B}
\end{equation}
We can invert the operational hopping metric to obtain $\wwdots$ if and only if the inverse, $\wwdotssq$, exists.

\subsection{Decomposition locality for loaded operations}\label{sec:decompositionlocalityforloadedoperations}

It follows from the fact that we have a notion of equivalence for operations and the fact that we get the same hopping metric whether we use fiducial operations or fiducial loaded boundary propositions (as in \eqref{samehoppingloadedcase}) that we have decomposition locality \index{decomposition locality} for a general loaded operation
\begin{equation}
\mathsf{A}^\mathtt{a}_\mathtt{bc} \equiv  \presup{a_\mathtt{A}}A_{b_\mathtt{A}c_\mathtt{A}} ~
\presub{a_\mathtt{A}}{\mathsf{X}}^\mathtt{a} \mathsf{X}_\mathtt{b}^{b_\mathtt{A}} \mathsf{X}_\mathtt{c}^{c_\mathtt{A}}
~~~~~ \Longleftrightarrow ~~~~~
\begin{Compose}{0}{0} \setdefaultfont{\mathsf}\setsecondfont{\mathtt}
\Ucircle{A}{0,0}\thispoint{DL}{-120:4} \thispoint{DR}{-60:4} \thispoint{UC}{90:4}
\joincc[above left]{DL}{60}{A}{-120} \csymbolalt{b}
\joincc[above right]{DR}{120}{A}{-60} \csymbolalt{c}
\joincc[left]{A}{90}{UC}{-90} \csymbolalt{a}
\end{Compose}
~~~ \equiv ~~~
\begin{Compose}{0}{0} \setdefaultfont{\mathsf}\setsecondfont{\mathtt}
\ucircle{A}{0,0}\csymbolthird{A}
\scircle{DLX}{-120:5}\csymbol{X} \scircle{DRX}{-60:5}\csymbol{X} \scircle{UCX}{90:5} \csymbol{X}
\thispoint{DL}{-120:8} \thispoint{DR}{-60:8} \thispoint{UC}{90:8}
\joincc[above left]{DL}{60}{DLX}{-120} \csymbolalt{b}
\joincc[above right]{DR}{120}{DRX}{-60} \csymbolalt{c}
\joincc[left]{UCX}{90}{UC}{-90} \csymbolalt{a}
\joinccbw[above left]{DLX}{60}{A}{-120} \csymbolthird{b_\mathtt{A}}
\joinccbw[above right]{DRX}{120}{A}{-60} \csymbolthird{c_\mathtt{A}}
\joinccwb[left]{A}{90}{UCX}{-90} \csymbolthird{a_\mathtt{A}}
\end{Compose}
\end{equation}
This is because we have
\begin{equation}
\begin{Compose}{0}{0} \setdefaultfont{\mathlbpro}\setsecondfont{\mathtt}
\thispoint{L}{0,0}\scircle{X1}{6,0}\csymbol{X}\scircle{X2}{10,0}\csymbol{X}\thispoint{R}{16,0}
\joinccwb[above]{L}{0}{X1}{180}\csymbolthird{a_\mathtt{B}} \joincc[above]{X1}{0}{X2}{180} \csymbolalt{a} \joinccbw[above]{X2}{0}{R}{180} \csymbolthird{a_\mathtt{A}}
\end{Compose}
~\equiv~
\begin{Compose}{0}{0} \setdefaultfont{\mathbpro}\setsecondfont{\mathtt}\setthirdfont{\mathbnd}
\thispoint{L}{0,0} \thispoint{R}{8,0} \joinccwbbw[above]{L}{0}{R}{180} \csymbolthird{a}
\end{Compose}
\end{equation}
and
\begin{equation}
\begin{Compose}{0}{0} \setdefaultfont{\mathsf}\setsecondfont{\mathtt}
\thispoint{L}{0,0}\scircle{X1}{6,0}\csymbol{X}\scircle{X2}{10,0}\csymbol{X}\thispoint{R}{16,0}
\joinccwb[above]{L}{0}{X1}{180}\csymbolthird{a_\mathtt{B}} \joincc[above]{X1}{0}{X2}{180} \csymbolalt{a} \joinccbw[above]{X2}{0}{R}{180} \csymbolthird{a_\mathtt{A}}
\end{Compose}
~\equiv~
\begin{Compose}{0}{0} \setdefaultfont{\mathbpro}\setsecondfont{\mathtt}\setthirdfont{\mathbnd}
\thispoint{L}{0,0} \thispoint{R}{8,0} \joinccwbbw[above]{L}{0}{R}{180} \csymbolthird{a}
\end{Compose}
\end{equation}
and hence the probability for both
\begin{equation}
\begin{Compose}{0}{-1.5} \setdefaultfont{\mathsf}\setsecondfont{\mathtt}
\Ucircle{A}{0,0} \Ucircle{B}{-5,5} \Ucircle{C}{3,4} \Ucircle{D}{-3, 11} \Ucircle{E}{2,9}
\joincc[below left]{B}{-65}{A}{115} \csymbolalt{a}
\joincc[below]{A}{80}{C}{-90} \csymbolalt{b}
\joincc[below]{C}{170}{B}{-10} \csymbolalt{c}
\joincc[above left]{B}{25}{E}{-110} \csymbolalt{d}
\joincc[left]{B}{80}{D}{-100} \csymbolalt{e}
\joincc[above right]{D}{-15}{E}{170}
\joincc[right]{C}{100}{E}{-80} \csymbolalt{g}
\end{Compose}
~~~~~ \text{and}~~~~~
\begin{Compose}{0}{-1.5} \setdefaultfont{\mathlbpro}\setsecondfont{\mathtt}
\Ucircle{A}{0,0} \Ucircle{B}{-5,5} \Ucircle{C}{3,4} \Ucircle{D}{-3, 11} \Ucircle{E}{2,9}
\joincc[below left]{B}{-65}{A}{115} \csymbolalt{a}
\joincc[below]{A}{80}{C}{-90} \csymbolalt{b}
\joincc[below]{C}{170}{B}{-10} \csymbolalt{c}
\joincc[above left]{B}{25}{E}{-110} \csymbolalt{d}
\joincc[left]{B}{80}{D}{-100} \csymbolalt{e}
\joincc[above right]{D}{-15}{E}{170}
\joincc[right]{C}{100}{E}{-80} \csymbolalt{g}
\end{Compose}
\end{equation}
is given by
\begin{equation}
\begin{Compose}{0}{0} \setdefaultfont{\mathnormal}\setsecondfont{\mathnormal}
\Ucircle{A}{-2,-2} \Ucircle{B}{-9,5} \Ucircle{C}{6,4} \Ucircle{D}{-6, 16} \Ucircle{E}{3,13}
\joinccwbbw[below left]{B}{-65}{A}{115}  \csymbolwbunder{a_\mathtt{B}} \csymbolbwunder{a_\mathtt{A}}
\joinccwbbw[below]{A}{80}{C}{-90} \csymbolwbunder{b_\mathtt{A}}\csymbolbwunder{b_\mathtt{C}}
\joinccwbbw[below]{C}{170}{B}{-10} \csymbolwbunder[0,-6.5]{c_\mathtt{C}}\csymbolbwunder[0,-6.5]{c_\mathtt{B}}
\joinccwbbw[above left]{B}{25}{E}{-110} \csymbolwb{d_\mathtt{B}}\csymbolbw{d_\mathtt{E}}
\joinccwbbw[left]{B}{80}{D}{-100} \csymbolwb{e_\mathtt{B}}\csymbolbw{e_\mathtt{D}}
\joinccwbbw[above right]{D}{-15}{E}{170}\csymbolwb{k_\mathtt{D}}\csymbolbw{k_\mathtt{E}}
\joinccwbbw[right]{C}{100}{E}{-80} \csymbolwb{g_\mathtt{C}}\csymbolbwunder{g_\mathtt{E}}
\end{Compose}
\end{equation}
Hence, given any composite loaded operation we can calculate a probability using generalized probabilistic states.

\subsection{Manifest invariance of PAGeR}

The specification of a load, $L_\mathsf{A}$, for an operation, $\mathsf{A}$, must be given in \lq operationally accessible" terms.  For example, we could get this loading from some observables extracted from management beables.  An example of how this might happen is if there are extra scalar fields constructed from fields that do not enter into $\pmb{\Phi}$ but, that can be used at the management level.  Observation of these could provide some probabilistic information about the fields in $\pmb{\Phi}$. Alternatively the loading could depend on beliefs of the observer. Arguably, such beliefs are operationally accessible at least to the observer in question (and to everybody once written down).

In any case, it is reasonable to assert that $L_\mathsf{A}$ is manifestly invariant under the action of $G^+$.  Hence, the labels  $a_\mathtt{A}=(\mathtt{Z}, \mathbf{Q}_\mathtt{Z}, O_\mathtt{Z}, L_\mathsf{Z})$, are manifestly invariant under $G^+$.  Operational generalized probabilistic states and the hopping metric are manifestly invariant under $G^+$.  This means that this operational probabilistic formulation of General Relativity is manifestly invariant under diffeomorphisms (and time-orientation preserving transformations).

\subsection{Black and white dots}

The remarks in Sec.\ \ref{sec:blackandwhitedots} for the possibilistic case go through exactly as written there for the probabilistic case also.  In particular, we can change white dots to black dots using the hopping metric $\bbdots$ and, if the inverse hopping metric $\wwdots$ exists, we can use it to change black dots into white dots.  Additionally we have \lq\lq internal" and \lq\lq external" points of view
\begin{equation}
\text{\lq\lq internal":}
\begin{Compose}{0}{0}\setdefaultfont{\mathnormal}
\Ucircle{A}{0,0}
\whitedot{U}{90:6} \csymbol[90:25]{a_\mathtt{A}}
\whitedot{DL}{-150:6}\csymbol[-150:25]{d_\mathtt{A}}
\whitedot{DR}{-30:6} \csymbol[-30:30]{e_\mathtt{A}}
\joincc{A}{90}{U}{-90}
\joincc{A}{-150}{DL}{30}
\joincc{DR}{150}{A}{-30}
\end{Compose}
~~~~~~~~
\text{\lq\lq external":}
\begin{Compose}{0}{0}\setdefaultfont{\mathnormal}
\Ucircle{A}{0,0}
\blackdot{U}{90:6} \csymbol[90:25]{a_\mathtt{B}}
\blackdot{DL}{-150:6}\csymbol[-150:25]{d_\mathtt{C}}
\blackdot{DR}{-30:6} \csymbol[-30:30]{e_\mathtt{D}}
\joincc{A}{90}{U}{-90}
\joincc{A}{-150}{DL}{30}
\joincc{DR}{150}{A}{-30}
\end{Compose}
\end{equation}
We also note that if a generalized state has all black dots it gives a list of probabilities for each fiducial arrangement.

\subsection{Formalism locality in PAGeR}\label{sec:formalismlocalityinPAGeR}

The formalism, as described so far, tells us how to calculate probabilities for closed composite loaded operations.  However, a more typical situation we might find ourselves in is where the part of the world we are interested in is not closed off from the rest of the world - in particular, it may be correlated with the rest of the world.  One way to proceed is to find a larger part of the world that is closed, do our calculations for that, and then make inferences for the smaller part of the world we are actually interested in.  This is not a very satisfactory way to proceed however. Thus, instead, we will set up a way of doing calculations directly for the part of the world (i.e.\ op-space) we are interested in without having to refer to parts of the world that we are not interested in.  We call the ability to do this formalism locality \cite{hardy2010bformalism}. \index{formalism locality}

Imagine we are doing a calculation for some op-space region
\[
\mathtt{G=A\cup B\cup C\cup D\cup E}
\]
A loaded operation for this region having open wires looks like the following
\begin{equation}
\begin{Compose}{0}{0} \setdefaultfont{\mathsf}\setsecondfont{\mathtt}
\Ucircle{A}{0,0} \Ucircle{B}{-5,5} \Ucircle{C}{3,4} \Ucircle{D}{-3, 11} \Ucircle{E}{2,9}
\joincc[below left]{B}{-65}{A}{115} \csymbolalt{a}
\joincc[below]{A}{80}{C}{-90} \csymbolalt{b}
\joincc[below]{C}{170}{B}{-10} \csymbolalt{c}
\joincc[above left]{B}{25}{E}{-110} \csymbolalt{d}
\joincc[left]{B}{80}{D}{-100} \csymbolalt{e}
\joincc[above right]{D}{-15}{E}{170}
\joincc[right]{C}{100}{E}{-80} \csymbolalt{g}
\thispoint{nA}{-2,-2} \joincc[above left]{A}{-135}{nA}{45} \csymbolalt{f}
\thispoint{nB}{-8,5} \joincc[above]{B}{180}{nB}{0} \csymbolalt{h}
\thispoint{nD}{-1,13} \joincc[above left]{D}{45}{nD}{-135} \csymbolalt{i}
\thispoint{nE}{4,11} \joincc[above left]{nE}{-145}{E}{45} \csymbolalt{j}
\end{Compose}
\end{equation}
We can calculate the operational generalized probabilistic state associated with this:
\begin{equation}
\begin{Compose}{0}{0} \setdefaultfont{\mathnormal}\setsecondfont{\mathnormal}
\Ucircle{A}{-2,-2} \Ucircle{B}{-9,5} \Ucircle{C}{6,4} \Ucircle{D}{-6, 16} \Ucircle{E}{3,13}
\joinccwbbw[below left]{B}{-65}{A}{115}  \csymbolwbunder{a_\mathtt{B}} \csymbolbwunder{a_\mathtt{A}}
\joinccwbbw[below]{A}{80}{C}{-90} \csymbolwbunder{b_\mathtt{A}}\csymbolbwunder{b_\mathtt{C}}
\joinccwbbw[below]{C}{170}{B}{-10} \csymbolwbunder[0,-6.5]{c_\mathtt{C}}\csymbolbwunder[0,-6.5]{c_\mathtt{B}}
\joinccwbbw[above left]{B}{25}{E}{-110} \csymbolwb{d_\mathtt{B}}\csymbolbw{d_\mathtt{E}}
\joinccwbbw[left]{B}{80}{D}{-100} \csymbolwb{e_\mathtt{B}}\csymbolbw{e_\mathtt{D}}
\joinccwbbw[above right]{D}{-15}{E}{170}\csymbolwb{k_\mathtt{D}}\csymbolbw{k_\mathtt{E}}
\joinccwbbw[right]{C}{100}{E}{-80} \csymbolwb{g_\mathtt{C}}\csymbolbwunder{g_\mathtt{E}}
\whitedot{nA}{-4,-4} \csymbolalt[-20,-20]{f_\mathtt{A}}    \joincc[above left]{A}{-135}{nA}{45}
\whitedot{nB}{-12,5} \csymbolalt[-25,0]{h_\mathtt{A}}\joincc[above right]{B}{180}{nB}{0}
\whitedot{nD}{-4,18}\csymbolalt[15,20]{i_\mathtt{A}} \joincc[above left]{D}{45}{nD}{-135}
\whitedot{nE}{5,15} \csymbolalt[15,20]{j_\mathtt{A}} \joincc[above left]{nE}{-145}{E}{45}
\end{Compose}
\end{equation}
We can represent this operational generalized probabilistic state by
\begin{equation}
\presup{f_\mathtt{A}h_\mathtt{B}i_\mathtt{D}}G_{j_\mathtt{E}}
\end{equation}
where this has the same number of open wires.

We may be interested in different possible outcomes in each of the component regions.  Then we can consider different loaded operations pertaining to region $\mathtt{G}$.  Consider a set of loaded operations $\mathsf{G}^\mathtt{fhi}_\mathtt{j}[l]$ where $l$ labels the different outcome sets for the composite region.  Associated with each of these loaded operations is an operational generalized probabilistic state
$\presup{f_\mathtt{A}h_\mathtt{B}i_\mathtt{D}}G_{j_\mathtt{E}}[l]$.  Now we can ask what the relative probability is for two such operations (i.e. what is the ratio of the number of times we see outcomes $l$ versus $l'$ when we repeat the experiment very many times?).  The problem is that this probability may depend on what happens outside of $\mathtt{G}$.  For the moment, assume that what happens outside is described by the operation $\mathsf{F}_\mathtt{fhi}^\mathtt{j}$.  Then the relative probability is given by
\begin{equation}\label{Grelprob}
\frac{\text{prob}(\mathsf{G}^\mathtt{fhi}_\mathtt{j}[l] \mathsf{F}_\mathtt{fhi}^\mathtt{j})}{\text{prob}(\mathsf{G}^\mathtt{fhi}_\mathtt{j}[l'] \mathsf{F}_\mathtt{fhi}^\mathtt{j})}
= \frac{
\presup{f_\mathtt{A}h_\mathtt{B}i_\mathtt{D}}G_{j_\mathtt{E}}[l] \presub{f_\mathtt{A}h_\mathtt{B}i_\mathtt{D}}F^{j_\mathtt{E}}
}{
\presup{f_\mathtt{A}h_\mathtt{B}i_\mathtt{D}}G_{j_\mathtt{E}}[l'] \presub{f_\mathtt{A}h_\mathtt{B}i_\mathtt{D}}F^{j_\mathtt{E}}
}
\end{equation}
Now, in the special case where
\begin{equation}\label{formlocproportionalcond}
\presup{f_\mathtt{A}h_\mathtt{B}i_\mathtt{D}}G_{j_\mathtt{E}[l]}  \propto \presup{f_\mathtt{A}h_\mathtt{B}i_\mathtt{D}}G_{j_\mathtt{E}[l']}
\end{equation}
we see that the dependence on what happens outside $\mathtt{G}$ drops out.  On the other hand, if this proportionality does not hold then, at least mathematically, the relative probability depends on $\presub{f_\mathtt{A}h_\mathtt{B}i_\mathtt{D}}F^{j_\mathtt{E}}$.  We can make a stronger statement. If the hoping metric is invertible then the fiducial elements are not over complete (for our present purposes we assume that this is the case).  Then we can find pairs of allowed operational generalized probabilistic states,  $\presub{f_\mathtt{A}h_\mathtt{B}i_\mathtt{D}}F^{j_\mathtt{E}}$ and $\presub{f_\mathtt{A}h_\mathtt{B}i_\mathtt{D}}F'^{j_\mathtt{E}}$, such that the relative probability in \eqref{Grelprob} will be be different.

Hence we can adopt a two prong attack. First we ask the question of whether the relative probability between $l$ and $l'$ is well conditioned (independent of conditions outside $\mathtt{G}$).  This will be the case if the proportionality condition in \eqref{formlocproportionalcond} holds.  If this condition does hold then, second, we can calculate the relative probability as
\begin{equation}
\text{relprob}\left(\frac{(\mathsf{G}^\mathtt{fhi}_\mathtt{j}[l] )}{(\mathsf{G}^\mathtt{fhi}_\mathtt{j}[l'] )}\right)
= \frac{
\presup{f_\mathtt{A}h_\mathtt{B}i_\mathtt{D}}G_{j_\mathtt{E}}[l]
}{
\presup{f_\mathtt{A}h_\mathtt{B}i_\mathtt{D}}G_{j_\mathtt{E}}[l']
}
\end{equation}
where
\begin{equation}
\frac{
\presup{f_\mathtt{A}h_\mathtt{B}i_\mathtt{D}}G_{j_\mathtt{E}}[l]
}{
\presup{f_\mathtt{A}h_\mathtt{B}i_\mathtt{D}}G_{j_\mathtt{E}}[l']
}
\end{equation}
is equal to the proportionality constant in \eqref{formlocproportionalcond}.

For a generic region of op-space we would expect that such relative probabilities will not be well conditioned (they will fail the proportionality test). This two prong attack is a way of picking out those situations where we can say something.

The notion of a relative probability is more general than that of a probability. In particular, if $l'$ corresponds to the set of all possible outcomes, then the relative probability is equal to the probability of $l$.  It is possible, however, that such probabilities will not be well conditioned even though some relative probabilities (where $l'$ is not the set of all possible outcomes) are well conditioned.

\section{Causality in PAGeR}

In Sec.\ \ref{sec:causalityinpossbilisticformulation} we discussed causality \index{causality} in the possibilistic formulation of General Relativity.  Similar considerations apply here.  In particular, we should choose the underlying set of allowed pure solutions, $\Omega_\mathtt{A}[\mathbf{Q}_\mathtt{A}]$, so that they satisfy the causality condition of Sec.\ \ref{sec:causalitycondition} so that we can tell a causal story.  Generalized states correspond to a probabilistic mixture of different causal situations at the typing surfaces.  Ideally, we would like to have a causality condition on these generalized states that imposed that it is not possible to signal to the past in this situation where we have probabilistic causal structure.

\newpage

\part{Operator Tensor Quantum Theory}\label{part:operatortensorQT}

\section{Introduction}

As a prelude to discussing Quantum Gravity we will provide a review of the operator tensor formulation of Quantum Theory developed in \cite{hardy2011reformulating, hardy2012operator, hardy2015quantum}. We will extend this approach a little so it also applies to continuous Hilbert spaces and also to Quantum Field Theory.

\section{Finite dimensional Hilbert spaces}

First we will discuss the case of finite dimensional Hilbert spaces - this is the case covered in previous work.

\subsection{Operations}

Here we will review the operator tensor approach to Quantum Theory \cite{hardy2011reformulating, hardy2012operator, hardy2015quantum}. We will slightly modify the diagrammatic notation from that used in earlier work so as to be consistent with the notation in this paper (in particular, we use circles rather than boxes and indicate time direction by arrows).  Consider some experiment that can be modeled as being built out of operations.  An operation \index{operations!quantum case}corresponds to one use of an apparatus.  It has a set of outcomes associated with it.  The elements of this set might be detector clicks, a pointer reading, or some other classical outcomes on the apparatus.  We say that the operation \lq\lq happens" if the actual outcome is in this set.  The operation also has inputs and outputs associated with it. The symbolic and diagrammatic notation for an operation is
\begin{equation}
\mathsf{A}_\mathsf{a_1b_2}^\mathsf{a_3c_4}  ~~~~~~\Leftrightarrow~~~~~~
\begin{Compose}{0}{-0.1}\setdefaultfont{\mathsf}
\Ucircle{A}{0,0}
\thispoint{ALL}{-3,-3}\thispoint{ALR}{3,-3} \thispoint{AUL}{-3,3}\thispoint{AUR}{3,3}
\joincc{ALL}{90}{A}{-110} \csymbol{a} \joincc{ALR}{90}{A}{-70} \csymbol{b}
\joincc{A}{110}{AUL}{-90} \csymbol{a} \joincc{A}{70}{AUR}{-90}\csymbol{c}
\end{Compose}
\end{equation}
Here $\mathsf{a}$, $\mathsf{b}$, \dots are different types \index{types} of quantum system (e.g. electrons, photons, \dots).  In the symbolic notation subscripts indicate inputs and superscripts indicate outputs. In the diagrammatic notation this is indicated with arrows.  If we have a different outcome set associated with some apparatus then we use a different symbol (e.g.\ $\mathsf{B}_\mathsf{a_1b_2}^\mathsf{a_3c_4}$ instead of $\mathsf{A}_\mathsf{a_1b_2}^\mathsf{a_3c_4}$). Alternatively, we can represent different outcome sets in square brackets (e.g.\ $\mathsf{A}_\mathsf{a_1b_2}^\mathsf{a_3c_4}[1]$ and $\mathsf{A}_\mathsf{a_1b_2}^\mathsf{a_3c_4}[2]$).  In the symbolic notation we need include integer subscripts on these system types to indicate the wiring.  For example, we can consider circuits such as
\begin{equation}\label{QTcircuitB}
\mathsf{A^{a_1a_2} B_{a_1c_3}^{a_5d_4} C_{a_2}^{c_3b_6} D_{a_5c_7} E_{d_4b_6}^{c_7}} ~~~~~~\Leftrightarrow~~~~~~
\begin{Compose}{0}{-2.2} \setdefaultfont{\mathsf}\setsecondfont{\mathsf}\setthirdfont{\mathsf}
\Ucircle{A}{0,0} \Ucircle{B}{-5,9} \Ucircle{C}{3,4} \Ucircle{D}{-3, 15} \Ucircle{E}{2,13}
\joincc[below left]{A}{115}{B}{-65} \csymbolalt{a}
\joincc[below]{A}{80}{C}{-90} \csymbolalt{a}
\joincc[below]{C}{170}{B}{-10} \csymbolalt{c}
\joincc[above left]{B}{25}{E}{-110} \csymbolalt{d}
\joincc[left]{B}{80}{D}{-100} \csymbolalt{a}
\joincc[above right]{E}{170}{D}{-15}\csymbolalt{c}
\joincc[right]{C}{100}{E}{-80} \csymbolalt{g}
\end{Compose}
\end{equation}
The integer subscripts on the type labels (in the symbolic notation) label the wires.  Note that this circuit is a directed acyclic graph (there are no close loops in it).  This is necessary here because the interpretation is that systems move forward in time in the direction of the arrows.

Our objective is to calculate the joint probability that the outcome for each operation in some circuit is in the outcome set associated with that operation.

\subsection{Decomposition locality}

We assume that any circuit (this must be closed) has a probability that is independent of settings and outcomes elsewhere.  We can use this to define a $p(\cdot)$ function which induces a notion of equivalence as in Sec.\ \ref{sec:fiducialloadedoperations} (also see \cite{hardy2011reformulating}.  In Quantum Theory it turns out that operations are decomposition local \index{decomposition locality!quantum case} so we can write
\begin{equation}
\mathsf{A}_\mathsf{a_1b_2}^\mathsf{a_3c_4} \equiv
\presup{a_3c_4}A_{a_1b_2}  \mathsf{X}_\mathsf{a_1}^{a_1} \mathsf{X}_\mathsf{b_2}^{b_2} \presub{a_3}{\mathsf{X}}^\mathsf{a_3} \presub{c_4}{\mathsf{X}}^\mathsf{c_4}
\end{equation}
\begin{equation}
\begin{Compose}{0}{0}\setdefaultfont{\mathsf}
\Ucircle{A}{0,0}
\thispoint{ALL}{-3,-3}\thispoint{ALR}{3,-3} \thispoint{AUL}{-3,3}\thispoint{AUR}{3,3}
\joincc{ALL}{90}{A}{-110} \csymbol{a} \joincc{ALR}{90}{A}{-70} \csymbol{b}
\joincc{A}{110}{AUL}{-90} \csymbol{a} \joincc{A}{70}{AUR}{-90}\csymbol{c}
\end{Compose}
~\equiv~
\begin{Compose}{0}{0}\setdefaultfont{\mathsf}\setsecondfont{\mathnormal}
\ucircle{A}{0,0}\csymbolalt{A}
\scircle{XLL}{-4,-4}\csymbol{X}\thispoint{ALL}{-7,-7}\joincc{ALL}{90}{XLL}{-135} \csymbol{a} \joinccbw{XLL}{45}{A}{-110}\csymbolalt{a}
\scircle{XLR}{4,-4}\csymbol{X}\thispoint{ALR}{7,-7}  \joincc{ALR}{90}{XLR}{-45} \csymbol{b}  \joinccbw{XLR}{135}{A}{-70}\csymbolalt{b}
\scircle{XUL}{-4,4}\csymbol{X}\thispoint{AUL}{-7,7}  \joinccwb{A}{110}{XUL}{-45}  \csymbolalt{a}   \joincc{XUL}{135}{AUL}{-90} \csymbol{a}
\scircle{XUR}{4,4}\csymbol{X}\thispoint{AUR}{7,7}    \joinccwb{A}{70}{XUR}{-135} \csymbolalt{c} \joincc{XUR}{45}{AUR}{-90}\csymbol{c}
\end{Compose}
\end{equation}
where
\begin{equation}
\presub{a}{\mathsf{X}}^\mathsf{a} ~~\Leftrightarrow~~
\begin{Compose}{0}{0} \setdefaultfont{\mathsf}\setsecondfont{\mathsf}
\thispoint{pX}{0,3} \Scircle{X}{0,0} \blackdot{Xa}{0,-3}
\joincc[left]{X}{90}{pX}{-90}\csymbolalt{a}
\joincc[left]{Xa}{90}{X}{-90} \csymbolthird{a}
\end{Compose}
\end{equation}
are fiducial preparations \index{fiducial preparations} for a system of type $\mathsf{a}$ and
\begin{equation}
\mathsf{X}_\mathsf{a_1}^{a_1}  ~~\Leftrightarrow~~
\begin{Compose}{0}{0} \setdefaultfont{\mathsf}\setsecondfont{\mathsf}
\thispoint{pX}{0,-3}
\Scircle{X}{0,0} \blackdot{Xa}{0,3} \joincc[left]{pX}{90}{X}{-90} \csymbolalt{a} \joincc[left]{X}{90}{Xa}{-90}\csymbolthird{a}
\end{Compose}
\end{equation}
are fiducial results \index{fiducial results} (measurement plus outcome) for systems of type $\mathsf{a}$.  The object
\begin{equation}
\presup{a_3c_4}A_{a_1b_2}  ~~~~~~\Leftrightarrow~~~~~~
\begin{Compose}{0}{-0.1}\setdefaultfont{\mathnormal}
\Ucircle{A}{0,0}
\whitedot{ALL}{-3,-3}\whitedot{ALR}{3,-3} \whitedot{AUL}{-3,3}\whitedot{AUR}{3,3}
\joincc{ALL}{90}{A}{-110} \csymbol{a} \joincc{ALR}{90}{A}{-70} \csymbol{b}
\joincc{A}{110}{AUL}{-90} \csymbol{a} \joincc{A}{70}{AUR}{-90}\csymbol{c}
\end{Compose}
\end{equation}
is the \emph{generalized state}. \index{generalized state!quantum case}  We will use this to construct operator tensors later.

\subsection{Hopping metric}

We define the hopping metric as follows
\begin{equation}
\presub{a'}h^{a}= \text{prob} (\presub{a'}X^\mathtt{a} X_\mathtt{a}^a )
~~~~~~\Leftrightarrow~~~~~~
\begin{Compose}{0}{-0.1}
\vbbmatrix{h}{0,0}\csymbolthird{a}
\end{Compose}
~=~
\text{prob}\left(
\begin{Compose}{0}{-0.5} \setdefaultfont{\mathsf}\setsecondfont{\mathsf}\setthirdfont{\mathnormal}
\blackdot{d1}{0,-3} \scircle{X1}{0,0}\csymbol{X}\scircle{X2}{0,4} \csymbol{X} \blackdot{d2}{0,7}
\joincc[left]{d1}{90}{X1}{-90} \csymbolthird{a} \joincc[left]{X1}{90}{X2}{-90}\csymbolalt{a} \joincc[left]{X2}{90}{d2}{-90} \csymbolthird{a}
\end{Compose}
\right)
\end{equation}   \index{hopping metric!quantum case}
This has the same role as the hopping metric discussed in previous parts of the paper.

We can replace each operation in the circuit in \eqref{QTcircuitB} and then substitute in the hopping metric to get the equivalent expression
\begin{equation}\label{QTgenstatecircuit}
\begin{Compose}{0}{0} \setdefaultfont{\mathnormal}\setsecondfont{\mathnormal}
\Ucircle{A}{0,0} \Ucircle{B}{-8,11} \Ucircle{C}{5,6} \Ucircle{D}{-5, 19} \Ucircle{E}{2,15}
\joinccwbbw[below left]{A}{115}{B}{-65} \csymbolalt{a}
\joinccwbbw[below]{A}{80}{C}{-90} \csymbolalt{a}
\joinccwbbw[below]{C}{170}{B}{-10} \csymbolalt{c}
\joinccwbbw[above left]{B}{25}{E}{-110} \csymbolalt{d}
\joinccwbbw[left]{B}{80}{D}{-100} \csymbolalt{a}
\joinccwbbw[above right]{E}{170}{D}{-15}\csymbolalt{c}
\joinccwbbw[right]{C}{100}{E}{-80} \csymbolalt{g}
\end{Compose}
\end{equation}
This gives us the probability associated with the circuit in terms of generalized probabilistic states and hopping metrics.  However, for Quantum Theory, we can write this in a more convenient form using operator tensors.

\subsection{Operator tensors}

We associate Hilbert spaces $\mathcal{H}^\mathsf{a_1}$ (for outputs) and $\mathcal{H}_\mathsf{a_1}$ (for inputs) with a system of type $\mathsf{a}$.  These Hilbert spaces are of dimension $N_\mathsf{a}$.   Hermitian operators acting on $\mathcal{H}^\mathsf{a_1}$ are written $\hat{A}^\mathtt{a_1}$, $\hat{B}^\mathsf{a_1}$, \dots. Hermitian operators acting on $\mathcal{H}_\mathsf{a_1}$ are written $\hat{A}_\mathsf{a_1}$, $\hat{B}_\mathsf{a_1}$, \dots.  We denote by
\begin{equation}\label{QTtrace}
\hat{A}^\mathsf{a_1} \hat{B}_\mathsf{a_1}
\end{equation}
the trace of the product of these two operators.  The repeated type label indicates taking the trace (this is analogous to Einstein's summation convention).

An operator $\hat{A}_\mathsf{a_1b_2}^\mathsf{a_3c_4}$ is an Hermitian operator acting on Hilbert space $\mathcal{H}_\mathsf{a_1}\otimes\mathcal{H}_\mathsf{b_2}\otimes\mathcal{H}^\mathsf{a_3}\otimes\mathcal{H}^\mathsf{c_4}$.  We can write this Hilbert space as  $\mathcal{H}_\mathsf{a_1}\mathcal{H}_\mathsf{b_2}\mathcal{H}^\mathsf{a_3}\mathcal{H}^\mathsf{c_4}$ or, simply, as $\mathcal{H}_\mathsf{a_1b_2}^\mathsf{a_3c_4}$ Note that there none of the integers on the indices are repeated.  We notate this operator diagrammatically by
\begin{equation}
\begin{Compose}{0}{0}\setdefaultfont{\mathsf}
\ucircle{A}{0,0} \csymbolthird{\hat{A}}
\thispoint{ALL}{-3,-3}\thispoint{ALR}{3,-3} \thispoint{AUL}{-3,3}\thispoint{AUR}{3,3}
\joincc{ALL}{90}{A}{-110} \csymbol{a} \joincc{ALR}{90}{A}{-70} \csymbol{b}
\joincc{A}{110}{AUL}{-90} \csymbol{a} \joincc{A}{70}{AUR}{-90}\csymbol{c}
\end{Compose}
\end{equation}  \index{operator tensors}
Since we also have subscripts and superscripts we call these \emph{operator tensors}.  By
\begin{equation}
\hat{A}^\mathsf{a_1} \hat{C}^\mathsf{b_2}_\mathsf{c_3}
\end{equation}
we mean the tensor product of these operators.  Note that the order is not important - $\hat{C}^\mathsf{b_2}_\mathsf{c_3}\hat{A}^\mathsf{a_1}$ means the same thing as all salient ordering information is contained in the type labels.  By
\begin{equation}
\hat{A}^\mathtt{a_1} \hat{B}_\mathtt{a_2}
\end{equation}
we mean the tensor product of these operators (compare this with \eqref{QTtrace} where the integer subscript on the type label is repeated and so the trace is taken).

We can find fiducial operators notated by
\begin{equation}
\presub{a}{\hat{X}}^\mathsf{a} ~~\Leftrightarrow~~
\begin{Compose}{0}{0} \setdefaultfont{\hat}\setsecondfont{\mathsf}
\thispoint{pX}{0,3} \Scircle{X}{0,0} \blackdot{Xa}{0,-3}
\joincc[left]{X}{90}{pX}{-90}\csymbolalt{a}
\joincc[left]{Xa}{90}{X}{-90} \csymbolthird{a}
\end{Compose}
\end{equation}
and   \index{fiducial operators}
\begin{equation}
\hat{X}_\mathsf{a_1}^{a_1}  ~~\Leftrightarrow~~
\begin{Compose}{0}{0} \setdefaultfont{\hat}\setsecondfont{\mathsf}
\thispoint{pX}{0,-3}
\Scircle{X}{0,0} \blackdot{Xa}{0,3} \joincc[left]{pX}{90}{X}{-90} \csymbolalt{a} \joincc[left]{X}{90}{Xa}{-90}\csymbolthird{a}
\end{Compose}
\end{equation}
such that
\begin{equation}
\begin{Compose}{0}{-0.5} \setdefaultfont{\hat}\setsecondfont{\mathsf}\setthirdfont{\mathnormal}
\blackdot{d1}{0,-3} \scircle{X1}{0,0}\csymbol{X}\scircle{X2}{0,4} \csymbol{X} \blackdot{d2}{0,7}
\joincc[left]{d1}{90}{X1}{-90} \csymbolthird{a} \joincc[left]{X1}{90}{X2}{-90}\csymbolalt{a} \joincc[left]{X2}{90}{d2}{-90} \csymbolthird{a}
\end{Compose}
~=~
\text{prob}\left(
\begin{Compose}{0}{-0.5} \setdefaultfont{\mathsf}\setsecondfont{\mathsf}\setthirdfont{\mathnormal}
\blackdot{d1}{0,-3} \scircle{X1}{0,0}\csymbol{X}\scircle{X2}{0,4} \csymbol{X} \blackdot{d2}{0,7}
\joincc[left]{d1}{90}{X1}{-90} \csymbolthird{a} \joincc[left]{X1}{90}{X2}{-90}\csymbolalt{a} \joincc[left]{X2}{90}{d2}{-90} \csymbolthird{a}
\end{Compose}
\right)
~=~
\begin{Compose}{0}{-0.1}
\vbbmatrix{h}{0,0}\csymbolthird{a}
\end{Compose}
\end{equation}
So the hopping matrix obtained from fiducial operators is the same as that obtained from fiducial operations.  The label $a_1$ runs from 1 to $N_\mathsf{a}^2$ since this is how many linearly independent Hermitian operators there are acting on a Hilbert space of dimension $N_\mathsf{a}$.

\subsection{Operator decomposition locality}

Now
\begin{equation}\label{fidproduct}
\hat{X}_\mathsf{b_2}^{b_2} \presub{a_3}{\hat{X}}^\mathsf{a_3} \presub{c_4}{\hat{X}}^\mathsf{c_4}
\end{equation}
is the tensor product of these fiducial operators.  It is a property of Hermitian operators acting on complex Hilbert spaces that the set of operators in \eqref{fidproduct} taken over all values of the labels $a_1, b_2, a_3, c_4$ span the space of Hermitian operators acting on $\mathcal{H}_\mathsf{a_1b_2}^\mathsf{a_3c_4}$.  Consequently, we can write a general Hermitian operator for this Hilbert space as
\begin{equation}
\hat{A}_\mathsf{a_1b_2}^\mathsf{a_3c_4} =
\presup{a_3c_4}A_{a_1b_2}  \hat{X}_\mathsf{a_1}^{a_1} \hat{X}_\mathsf{b_2}^{b_2} \presub{a_3}{\hat{X}}^\mathsf{a_3} \presub{c_4}{\hat{X}}^\mathsf{c_4}
\end{equation}
\begin{equation}\label{decomplocaloperatorexpression}
\begin{Compose}{0}{0}\setdefaultfont{\hat}\setsecondfont{\mathsf}
\Ucircle{A}{0,0}
\thispoint{ALL}{-3,-3}\thispoint{ALR}{3,-3} \thispoint{AUL}{-3,3}\thispoint{AUR}{3,3}
\joincc{ALL}{90}{A}{-110} \csymbolalt{a} \joincc{ALR}{90}{A}{-70} \csymbolalt{b}
\joincc{A}{110}{AUL}{-90} \csymbolalt{a} \joincc{A}{70}{AUR}{-90}\csymbolalt{c}
\end{Compose}
~=~
\begin{Compose}{0}{0}\setdefaultfont{\hat}\setsecondfont{\mathnormal}\setthirdfont{\mathsf}
\Ucircle{A}{0,0}
\scircle{XLL}{-4,-4}\csymbol{X}\thispoint{ALL}{-7,-7}\joincc{ALL}{90}{XLL}{-135} \csymbol{a} \joinccbw{XLL}{45}{A}{-110}\csymbolalt{a}
\scircle{XLR}{4,-4}\csymbol{X}\thispoint{ALR}{7,-7}  \joincc{ALR}{90}{XLR}{-45} \csymbol{b}  \joinccbw{XLR}{135}{A}{-70}\csymbolalt{b}
\scircle{XUL}{-4,4}\csymbol{X}\thispoint{AUL}{-7,7}  \joinccwb{A}{110}{XUL}{-45}  \csymbolalt{a}   \joincc{XUL}{135}{AUL}{-90} \csymbol{a}
\scircle{XUR}{4,4}\csymbol{X}\thispoint{AUR}{7,7}    \joinccwb{A}{70}{XUR}{-135} \csymbolalt{c} \joincc{XUR}{45}{AUR}{-90}\csymbol{c}
\end{Compose}
\end{equation} \index{decomposition locality}
It is interesting to note that operators acting on real or quaternionic Hilbert spaces cannot be decomposed in this way.

\subsection{Operator Circuits}

It is now simple to see that the expression for the probability in \eqref{QTgenstatecircuit} is equal to
\begin{equation}\label{QTopcircuit}
\hat{A}^\mathsf{a_1a_2} \hat{B}_\mathsf{a_1c_3}^\mathsf{a_5d_4} \hat{C}_\mathsf{a_2}^\mathsf{c_3b_6} \hat{D}_\mathsf{a_5c_7} \hat{E}_\mathsf{d_4b_6}^\mathsf{c_7} ~~~~~~\Leftrightarrow~~~~~~
\begin{Compose}{0}{-2.2} \setdefaultfont{\hat}\setsecondfont{\mathsf}\setthirdfont{\mathsf}
\Ucircle{A}{0,0} \Ucircle{B}{-5,9} \Ucircle{C}{3,4} \Ucircle{D}{-3, 15} \Ucircle{E}{2,13}
\joincc[below left]{A}{115}{B}{-65} \csymbolalt{a}
\joincc[below]{A}{80}{C}{-90} \csymbolalt{a}
\joincc[below]{C}{170}{B}{-10} \csymbolalt{c}
\joincc[above left]{B}{25}{E}{-110} \csymbolalt{d}
\joincc[left]{B}{80}{D}{-100} \csymbolalt{a}
\joincc[above right]{E}{170}{D}{-15}\csymbolalt{c}
\joincc[right]{C}{100}{E}{-80} \csymbolalt{g}
\end{Compose}
\end{equation}
because if we insert the local decomposition form (such as in \eqref{decomplocaloperatorexpression}) in this we get back to \eqref{QTgenstatecircuit}.
To evaluate this expression we can either simply reverse the process by expanding out each operator in terms of fiduials and do the calculation in terms of generalized states. Or we can interpret any repeated index, one raise and one lowered, (or any wire in the diagrammatic notation) as corresponding to taking the partial trace over the product in the relevant part of the space.  To illustrate this we note that
\begin{equation}
\hat{A}^\mathsf{a_1} \hat{B}_\mathsf{a_1} \hat{C}^\mathsf{b_2}
\end{equation}
is the trace of the product of $\hat{A}^\mathsf{a_1}$ and $\hat{B}_\mathsf{a_1}$ times the operator $\hat{C}^\mathsf{b_2}$.  We are effectively taking the partial trace in the $\mathcal{H}_\mathsf{a_1}$ part of the Hilbert space.  If we have an expression like
\begin{equation}
\hat{A}^\mathsf{a_1} \hat{D}_\mathsf{a_1}^\mathsf{b_2}
\end{equation}
we can evaluate this by the partial trace in the $\mathcal{H}_\mathsf{a_1}$ Hilbert space.

\subsection{Physicality}\label{sec:physicality}

We have shown that we can determine the probability for a quantum circuit using operator tensors.  The real usefulness of this approach, however, is that it is simple to write down  conditions on operator tensors that must be satisfied if the corresponding operator can be physically realized.  An equivalent characterization in terms of the generalized probabilistic states (such as $\presub{a_1b_2}A^{a_3c_4}$) is possible in principle but it is much easier to provide conditions on the operator tensors.

Any directed acyclic graph (such as the quantum circuits we consider here) can be \emph{completely foliated} \index{foliation!of circuit} into \emph{space-like hypersurfaces}.   A space-like hypersurface is a set of wires in the graph that partition the graph into two pieces (i.e. were these wires to be deleted we would have two disjoint parts of the graph) and such that there does not exist a forward directed path through the circuit from any wire to any other wire in this space-like hypersurface.  To visualize this we can think of a drawing a line through the graph intersecting wires. Two space-like hypersurfaces \emph{cross} if some wires in one are to the future of the other and some are to the past.   A complete foliation is a set of space-like hypersurfaces that do not cross and where every wire is included in at least one space-like hypersurface (some wires may be in more than one such hypersurface).  There always exists at least one such complete foliation for a directed acyclic graph.  Generically (for big enough graphs) there exist multiple such complete foliations.  We can think of the space-like hypersurfaces in such a foliation as providing a time parameter.

The most general circuit containing the operation, $\mathsf{B}_\mathsf{a_1}^\mathsf{b_2}$, can always be written as
\begin{equation}\label{Bisolated}
\mathsf{A}^\mathsf{a_1c_3} \mathsf{B}_\mathsf{a_1}^\mathsf{b_2} \mathsf{C}_\mathsf{b_2c_3}
~~~~\Leftrightarrow~~~~
\begin{Compose}{0}{-1.43}\setdefaultfont{\mathsf}\setsecondfont{\mathsf}
\Ucircle{A}{0,0} \Ucircle{B}{-3,4} \Ucircle{C}{0,8}
\joincc{A}{135}{B}{-90} \csymbolalt{a} \joincc{A}{45}{C}{-45} \csymbolalt{c} \joincc{B}{90}{C}{-135} \csymbolalt{b}
\end{Compose}
\end{equation}
because we can lump together all the operations before $\mathsf{B}_\mathsf{a_1}^\mathsf{b_2}$ and call them $\mathsf{A}^\mathsf{a_1c_3}$ and all the operations after and call them $\mathsf{C}_\mathsf{b_2c_3}$.  The notion of before is with respect to a space-like hypersurface that includes wire $\mathsf{a_1}$ and the notion of after is with respect to another space-like hypersurface differing from the previous one in that it includes wire $\mathsf{b_2}$ rather than $\mathsf{a_1}$ (we can always \lq\lq push" a space-like hypersurface over an operation to obtain another spacelike hypersurface).

We give the following definition \index{physicality!definition}
\begin{quote}
\textbf{Physicality definition}: An operator, $\hat{B}_\mathsf{a_1}^\mathsf{b_2}$, is \emph{physical} if and only if both
\begin{equation}
0 \leq \hat{A}^\mathsf{a_1c_3} \hat{B}_\mathsf{a_1}^\mathsf{b_2} \hat{C}_\mathsf{b_2c_3}
\end{equation}
and
\begin{equation}\label{physcondtwo}
\hat{A}^\mathsf{a_1c_3} \hat{B}_\mathsf{a_1}^\mathsf{b_2} \hat{I}_\mathsf{b_2c_3} \leq 1
\end{equation}
for all rank one projectors $\hat{A}^\mathsf{a_1c_3}$ and $\hat{C}_\mathsf{b_2c_3}$ and all system types $\mathsf{c}$.
Here $\hat{I}_\mathsf{b_2c_3}$ is the identity operator on $\mathcal{H}_\mathsf{b_2c_3}$
\end{quote}
We can then consistently associate operators with operations such that we obtain probabilities when we evaluate the operator circuit corresponding to any operation circuit.

The following theorem is of central importance: \index{physicality!theorem}
\begin{quote}
\textbf{Physicality theorem}: An operator, $\hat{B}_\mathsf{a_1}^\mathsf{b_2}$, is physical if and only if
\begin{equation}\label{physicalityconditions}
0\leq  \hat{B}_\mathsf{a^T_1}^\mathsf{b_2}    ~~~~\text{and}~~~  \hat{B}_\mathsf{a_1}^\mathsf{b_2} \hat{I}_\mathsf{b_2}\leq \hat{I}_\mathsf{a_1}
\end{equation}
where the superscript $\mathsf{T}$ indicates we are taking the partial transpose in the $\mathcal{H}_\mathsf{a_1}$ space.
\end{quote}
These conditions are equivalent to the usual conditions imposed on operators (density matrices are positive, maps are completely positive and trace non-increasing, and measurement outcomes are associated with positive operators).   Any circuit built out of operators satisfying the two conditions in \eqref{physicalityconditions} will necessarily have a value between 0 and 1.

\subsection{Causality}

The motivation for the physicality definition uses notions that have a bearing on causality in various ways.   Since the physicality conditions in \eqref{physicalityconditions} come from this definition we see that the constraints on operators come from these notions of causality.

First, the fact that the circuit is taken to be a directed acyclic graph means we can isolate the given operation ($\mathsf{B}_\mathsf{a_1}^\mathsf{b_2}$) by space-like hypersurfaces and hence write a general circuit in the form given in \eqref{Bisolated}.  This is consistent with interpreting the arrows on the wires as corresponding to a time direction (forward, let's say) and so it is consistent with a certain notion of definite causal structure.

Second, the condition \eqref{physcondtwo} is motivated by the idea that we associate the deterministic operation (where the outcome set is the set of all possible outcomes) having only inputs with the identity operator, $\hat{I}_\mathsf{b_2c_3}$ (since then we cannot have probability greater than one).   There are, however, many ways to realize a deterministic operation having only inputs.  This assumption entails that we associate the same operator with all of them. In the language of Chribella, D'Ariano, and Perinotti, there is a unique deterministic effect associated with any given system type.   This is the content of the \emph{Pavia causality condition} \index{Pavia causality condition} that encodes the principle that we cannot signal backwards in time (if the deterministic effect depended on knob settings for example, then the probability for earlier outcomes would depend on these knob settings so we would be able to signal backwards in time).

Third, we assume enough structure that we can signal into the future.  This is clear because we assume we can prepare states corresponding to rank one projection operators.  Consider a set of possible preparations $\hat{D}^\mathsf{a_1}[n]$ (where $n=1$ to $N$) equal to projection operators onto a basis of the Hilbert space.  We can read off whether we send the $n=1$ case or not using a $\hat{E}_\mathsf{a_1}[1]$ projection operator.  In this way we can send information to the future (as indicated by the arrows on the wires).

\section{Continuous dimensional Hilbert spaces}\label{sec:continuousdimensionalHilbertspace}

\index{continuous dimensional Hilbert spaces}

We will now look at how to adapt this approach to continuous dimensional Hilbert space. For the sake of definiteness consider a Hilbert space with a basis set of states $|q\rangle$ where $q\in\mathbb{R}$ such that
\begin{equation}
\langle q|q'\rangle = \delta(q-q')~~~~\text{and}~~~~ \int \mathrm{d}q|q\rangle\langle q| = \hat{I}
\end{equation}
where $\hat{I}$ is the identity operator on this Hilbert space.  We define
\begin{equation}\label{Pqdefn}
\hat{P}_q = |q\rangle\langle q| ~~~~\text{for}~q\in\mathbb{R}
\end{equation}
and
\begin{equation}\label{Pqqprimedefn}
\hat{P}_{qq'-} = |qq'-\rangle\langle qq'-|, ~~~ \hat{P}_{qq'i} = |qq'i\rangle\langle qq'i|~~~~\text{for}~q> q'~~ \text{where} ~ q,q'\in\mathbb{R}
\end{equation}
where
\begin{equation}
|qq'-\rangle= \frac{1}{\sqrt{2}} (|q\rangle-|q'\rangle) ~~~~~~ |qq'i\rangle=\frac{1}{\sqrt{2}}(|q\rangle+i|q'\rangle)
\end{equation}
We will show that we can write down an arbitrary Hermitian operator as
\begin{equation}\label{genAoperator}
\hat{A} = \int \mathrm{d}q  a(q)\hat{P}_q + \iint_{q> q'}\mathrm{d}q \mathrm{d}q' b(q,q')\hat{P}_{qq'-} + \iint_{q> q'}\mathrm{d}q \mathrm{d}q' c(q,q') \hat{P}_{qq'i}
\end{equation}
where we need to impose $c(q,q)=0$.  As we will see, this last condition can be imposed in a natrual fashion by demanding that $c(q,q')$ is antisymmetric (so $c(q,q')=-c(q',q)$).
We need to treat the $q=q'$ contributions to this integral with a little care.  We make sense of integrating with the constraint $q>q'$ by imposing that, at $q=q'$, the integrand vanishes as follows (and, hence, this is equivalent to integrating with the constraint $q\geq q'$).  First we note that
\begin{equation}
\hat{P}_{qq-}= 0
\end{equation}
Hence the $(q,q)\hat{P}_{qq-}$ term does not contribute.  Now note that
\begin{equation}
\hat{P}_{qqi} = 2 \hat{P}_q
\end{equation}
Hence $c(q,q') \hat{P}_{qq'i}$ term would contribute to the integral as $2c(q,q) \hat{P}_{q}$.  This would lead to a curious double counting with the $a(q)\hat{P}_q$ term.  To avoid this we impose that $c(q,q)=0$.  We will see that this happens naturally in any case.

We can write any Hermitian operator as a weighted sum (or integral) over rank one projection operators:
\begin{equation}\label{Aprojectorcontinuous}
\hat{A} = \SumInt_\alpha \mathrm{d}\alpha |\psi_\alpha\rangle\langle\psi_\alpha|
\end{equation}
We can write
\begin{equation}\label{Psicontinuouscase}
|\psi_\alpha\rangle = \int \mathrm{d}q \psi(q) |q\rangle
\end{equation}
It is easy to show
\begin{align}\label{continuousprojector}
|\psi_\alpha\rangle\langle \psi_\alpha| = &\iint \mathrm{d}q\mathrm{d}q'
\frac{1}{2} \text{Re}(\psi_\alpha(q)\psi_\alpha^*(q')) \left( |q\rangle\langle q'| + |q'\rangle\langle q| \right) \nonumber \\
&+ \iint \mathrm{d}q\mathrm{d}q' \frac{1}{2} \text{Im}(\psi_\alpha(q)\psi_\alpha^*(q')) \left( i|q\rangle\langle q'| - i|q'\rangle\langle q| \right)
\end{align}
It follows from the fact that $\psi_\alpha(q)\psi_\alpha^*(q') = (\psi_\alpha(q')\psi_\alpha^*(q))^*$ that $\text{Re}(\psi_\alpha(q)\psi_\alpha^*(q'))$ is symmetric in $q$ and $q'$ and $\text{Im}(\psi_\alpha(q)\psi_\alpha^*(q'))$ is antisymmetric.   Consider the two integrals on the right hand side of \eqref{continuousprojector} separately.  The integrand in each case is a symmetric function.   We can write the first integral as
\begin{align}\label{firstterm}
& \iint \mathrm{d}q\mathrm{d}q' \text{Re}(\psi_\alpha(q)\psi_\alpha^*(q'))\left(\frac{1}{2} \hat{P}_q +\frac{1}{2}\hat{P}_{q'} - \hat{P}_{qq'-}\right) \nonumber \\
&=\iint \mathrm{d}q \mathrm{d}q' \text{Re}(\psi_\alpha(q)\psi_\alpha^*(q')) \hat{P}_q
           - 2\iint_{q\geq q'} \mathrm{d}q\mathrm{d}q' \text{Re}(\psi_\alpha(q)\psi_\alpha^*(q'))\hat{P}_{qq'-}
\end{align}
Note that $\hat{P}_{qq-}=0$ and so the $q=q'$ line contributes nothing to the $\hat{P}_{qq'-}$ term in this integral.  To manipulate the second integral on the right hand side of \eqref{continuousprojector} we can integrate only over $q\geq q'$ and double the integral (taking advantage of the above mentioned symmetry property for the integrand) to obtain
\begin{align}\label{secondterm}
&2\iint_{q\geq q'} \mathrm{d}q\mathrm{d}q' \frac{1}{2} \text{Im}(\psi_\alpha(q)\psi_\alpha^*(q')) \left( i|q\rangle\langle q'| - i|q'\rangle\langle q| \right) \nonumber\\
&= 2 \iint_{q\geq q'} \mathrm{d}q \mathrm{d}q' \text{Im}(\psi_\alpha(q)\psi_\alpha^*(q'))( \frac{1}{2}\hat{P}_q +\frac{1}{2}\hat{P}_{q'}- \hat{P}_{qq'i})  \nonumber\\
&= \iint \mathrm{d}q \mathrm{d}q' \text{Sym}\left(\text{Im}(\psi_\alpha(q)\psi_\alpha^*(q'))\right) \hat{P}_q
       -  2\iint_{q\geq q'} \mathrm{d}q\mathrm{d}q' \text{Im}(\psi_\alpha(q)\psi_\alpha^*(q')) \hat{P}_{qq'i}
\end{align}
where, for any antisymmetric function, $g(q, q')$, we define
\begin{equation}
\text{Sym}(g(q,q')) = \left\{ \begin{array}{ll}
                              g(q,q') & \text{for} q\geq q' \\
                              -g(q,q') & \text{for} q'\geq q
                              \end{array}  \right.
\end{equation}
Note that $\text{Im}(\psi_\alpha^*(q)\psi_\alpha(q))=0$ and so the $q=q'$ line contributes nothing to the integral \eqref{secondterm}.

Using \eqref{continuousprojector}, \eqref{firstterm}, and \eqref{secondterm} in \eqref{Aprojectorcontinuous}, we see that we can write any Hermitian operator as in \eqref{genAoperator}. In particular, note that the $q'=q$ cases in the $b(q,q')$ and $c(q,q')$ terms do not contribute for the reasons discussed so we can legitimately claim that we are using $\hat{P}_{qq'-}$ and $\hat{P}_{qq'i}$ only when $q>q'$.

We can choose the fiducial operators to be equal to the projection operators in \eqref{Pqdefn} and \eqref{Pqqprimedefn}. We write these as
\begin{equation}
\presub{a_1}{\hat{X}}^\mathsf{a_1}    ~~~~~~~~   {\hat{X}}_\mathsf{a_1}^{a_1}
\end{equation}
where
\begin{equation}
a_1\in \{q:\forall~ q\in \mathbb{R}\}\cup \{qq'-:\forall~ q,q'\in \mathbf{R} ~\text{s.t.} ~q>q' \}\cup \{qq'i:\forall~ q,q'\in \mathbf{R} ~\text{s.t.} ~q>q' \}
\end{equation}
Now we can write a general operator as a weighted $\SumInt$ over these fiducial operators.  For example,
\begin{equation}
\hat{A}_\mathsf{a_1} = A_{a_1} \hat{X}^{a_1}_\mathsf{a_1}
\end{equation}
where, written out explicitly, this means
\begin{equation}
\hat{A}_\mathsf{a_1} = \int \mathrm{d}q A_{q} \hat{X}^q_\mathsf{a_1}
+ \iint_{q>q'}\mathrm{d}q \mathrm{d}q' A_{qq'-} \hat{X}^{qq'-}_\mathsf{a_1}
+ \iint_{q>q'}\mathrm{d}q \mathrm{d}q' A_{qq'i} \hat{X}^{qq'i}_\mathsf{a_1}
\end{equation}
where we require $A_{qqi}=0$. We are justified in regarding the integration for the last two terms on the right as being for $q>q'$ (rather than $q\geq q'$) because the integrands vanish when $q=q'$ (the first one vanishes because $\hat{X}^{qq-}_\mathsf{a_1}=0$, and the second because $A_{qqi}=0$).

There is nothing in the above that particularly depends on having $q$ as an element of $\mathbb{R}$.  We could run the above argument for some other space, $\Lambda$, in place of $\mathbb{R}$ as long as we can represent states by the expression \eqref{Psicontinuouscase}.

We are now able to define the hopping matrix for a system with a continuous dimensional Hilbert space as
\begin{equation}
\presub{a'_1}h^{a_1} = \presub{a_1}{\hat{X}}^\mathsf{a_1} {\hat{X}}_\mathsf{a_1}^{a_1}
\end{equation}
If we can find fiducial operations having the same hopping matrix then we can set up a correspondence between operations and operators as before.

Now consider a composite system.  Any Hermitian operator can be regarded as a weighted sum of product operators. Hence, for the continuous dimensional case also, we have decomposition locality.  For example
\begin{equation}
\hat{A}_\mathsf{a_1b_2}^\mathsf{a_3c_4} =
\presup{a_3c_4}A_{a_1b_2}  \hat{X}_\mathsf{a_1}^{a_1} \hat{X}_\mathsf{b_2}^{b_2} \presub{a_3}{\hat{X}}^\mathsf{a_3} \presub{c_4}{\hat{X}}^\mathsf{c_4}
\end{equation}
or, in diagrammatic notation,
\begin{equation}
\begin{Compose}{0}{0}\setdefaultfont{\hat}\setsecondfont{\mathsf}
\Ucircle{A}{0,0}
\thispoint{ALL}{-3,-3}\thispoint{ALR}{3,-3} \thispoint{AUL}{-3,3}\thispoint{AUR}{3,3}
\joincc{ALL}{90}{A}{-110} \csymbolalt{a} \joincc{ALR}{90}{A}{-70} \csymbolalt{b}
\joincc{A}{110}{AUL}{-90} \csymbolalt{a} \joincc{A}{70}{AUR}{-90}\csymbolalt{c}
\end{Compose}
~=~
\begin{Compose}{0}{0}\setdefaultfont{\hat}\setsecondfont{\mathnormal}\setthirdfont{\mathsf}
\Ucircle{A}{0,0}
\scircle{XLL}{-4,-4}\csymbol{X}\thispoint{ALL}{-7,-7}\joincc{ALL}{90}{XLL}{-135} \csymbol{a} \joinccbwsq{XLL}{45}{A}{-110}\csymbolalt{a}
\scircle{XLR}{4,-4}\csymbol{X}\thispoint{ALR}{7,-7}  \joincc{ALR}{90}{XLR}{-45} \csymbol{b}  \joinccbwsq{XLR}{135}{A}{-70}\csymbolalt{b}
\scircle{XUL}{-4,4}\csymbol{X}\thispoint{AUL}{-7,7}  \joinccwbsq{A}{110}{XUL}{-45}  \csymbolalt{a}   \joincc{XUL}{135}{AUL}{-90} \csymbol{a}
\scircle{XUR}{4,4}\csymbol{X}\thispoint{AUR}{7,7}    \joinccwbsq{A}{70}{XUR}{-135} \csymbolalt{c} \joincc{XUR}{45}{AUR}{-90}\csymbol{c}
\end{Compose}
\end{equation}
where, now, we can interpret the Hilbert spaces corresponding to the system types as being continuous.  We can also treat the situation where some systems have continuous dimensional Hilbert spaces and some have finite dimensional Hilbert spaces in this way.

We can define Physicality for operators in as in the finite dimensional case and obtain a physicality theorem as in Sec.\ \ref{sec:physicality}.  Hence we can set up Quantum Theory when some (or all) of the systems are continuous dimensional in the same way as above.

\section{Operator Tensor Quantum Field Theory}\label{sec:operatortensorquantumfieldtheory}

\subsection{Introduction}

\index{operator tensor quantum field theory}

The operator tensor formulation of Quantum Theory suggests an approach to Quantum Field Theory in which we associated operators with arbitrary regions of space-time.  This was suggested in \cite{hardy2011reformulating}.  A similar approach has since been developed in some detail by Oeckl (called the positive formalism) in \cite{oeckl2013positive} motivated by the ideas in \cite{hardy2011reformulating} and his own general boundary formalism (which goes back to 2003 and previously formulated for the pure state case only). There is a certain convergence (as Oeckl put it in \cite{oeckl2013positive}) between these two approaches. There are, however, some substantial differences between the operator tensor formulation of Quantum Field Theory suggested in \cite{hardy2011reformulating} and Oeckl's positive formalism.  The most significant of these is that Oeckl's operators are positive whereas in the operator tensor approach the operators are positive under input transpose (for this to be possible we need to specify a tensor product factorization of the Hilbert space at the boundary into input and output parts).  Oreshkov and Cerf have also made proposals along related lines \cite{oreshkov2014operational}.  An alternative approach to Quantum Field Theory regarded as a theory of cellular automata has been pursued by D'Ariano, Perinotti and collaborators \cite{d2014derivation, bisio2015quantum}.

We will now spell out in greater detail how the operator tensor approach to Quantum Field Theory works.

\subsection{The discrete case}

Consider a big circuit on a grid composed of left moving quantum systems and right moving quantum systems (these could be qubits).
\begin{equation}
\begin{Compose}{0}{0}
\Cgrid{0.5}{11}{11}{0,0}
\end{Compose}
\end{equation}
At each vertex, there is an operation.  We can set up a coordinate system such that these vertices are at $x=(m\delta,n\delta)$ where $\delta$ is the grid spacing (this coordinate system is aligned with the grid itself).  The operation at each vertex has some setting (agent choice) which we can denote by $Q(x)$ and some outcomes which we denote by $O(x)$. Let the operation at $x$ be $\mathsf{A}[Q(x), O(x)]$.  Each vertex has two wires going into it and two wires going out of it.  We can label the wires by $x$ corresponding to their midpoints.  Hence, the wires are at positions $( m\delta +\delta/2, n\delta + \delta/2)$.

Although we are illustrating this with a 2 dimensional grid the same ideas will go through in any dimension.  We think of the horizontal direction as space and the vertical direction as time.

An operation at $x$ has incoming wires $x^-_l=x-\frac{\delta_l}{2}$ and $x^-_r=x-\frac{\delta_r}{2}$ and outgoing wires $x^+_l=x-\frac{\delta_l}{2}$ and $x^+_r=x-\frac{\delta_r}{2}$. Here $\delta_l$ ($\delta_r$) is a vector connecting adjacent vertices by a left (right) going wire.  We label leftward going wires with $\mathsf{l}$ and rightward going wires with $\mathsf{r}$.  The operation at $x$ can be written as
\begin{equation}
\mathsf{A}_{ \mathsf{l}_{x^-_l} \mathsf{r}_{x^-_r} }^{ \mathsf{l}_{x^+_l} \mathsf{r}_{x^+_r} }[Q(x),O(x)]
\end{equation}
Associated with this operation is an operator
\begin{equation}
\hat{A}_{ \mathsf{l}_{x^-_l} \mathsf{r}_{x^-_r} }^{ \mathsf{l}_{x^+_l} \mathsf{r}_{x^+_r} }[Q(x),O(x)]
\end{equation}
We can take an arbitrary fragment of this circuit such as
\begin{equation}
\begin{Compose}{0}{0}
\Cobject{\thetooth}{C}{1}{1}{1,-0.07}
\Cgrid{0.5}{11}{11}{0,0}
\end{Compose}
\end{equation}
Let $\mathtt{A}$ be the arbitrary region shown where we take $\mathtt{A}$ is the set of $x$'s of the vertices in this region.

Associated with $\mathtt{A}$ is an operation
\begin{equation}
\mathsf{A}_{\mathsf{l}^-_\mathtt{A}\mathsf{r}_\mathtt{A}^-}^{\mathsf{l}^+_\mathtt{A}\mathsf{r}_\mathtt{A}^+}
=  \prod_{x\in \mathtt{A}} \mathsf{A}_{ \mathsf{l}_{x^-_l} \mathsf{r}_{x^-_r} }^{ \mathsf{l}_{x^+_l} \mathsf{r}_{x^+_r} }[Q(x),O(x)]
\end{equation}
and an operator
\begin{equation}
\hat{A}_{\mathsf{l}^-_\mathtt{A}\mathsf{r}_\mathtt{A}^-}^{\mathsf{l}^+_\mathtt{A}\mathsf{r}_\mathtt{A}^+}
=  \prod_{x\in \mathtt{A}} \hat{A}_{ \mathsf{l}_{x^-_l} \mathsf{r}_{x^-_r} }^{ \mathsf{l}_{x^+_l} \mathsf{r}_{x^+_r} }[Q(x),O(x)]
\end{equation}
Here we are taking the partial trace over all the repeated wires as before.  The indices remaining (not traced over) are those associated with the wires that intersect the boundary.  The inward coming wires are
\begin{equation}
\mathsf{l}_{\mathtt{A}^-} = \prod_{x\in \partial\mathtt{A}^{\mathsf{l}-}} \mathsf{l}_x  ~~~~~~  \mathsf{r}_{\mathtt{A}^-} = \prod_{x\in \partial\mathtt{A}^{\mathsf{r}-}} \mathsf{r}_x
\end{equation}
from the left and the right respectively, and the outward going wires are
\begin{equation}
\mathsf{l}_{\mathtt{A}^+} = \prod_{x\in \partial\mathtt{A}^{\mathsf{l}+}} \mathsf{l}_x  ~~~~~~  \mathsf{r}_{\mathtt{A}^+} = \prod_{x\in \partial\mathtt{A}^{\mathsf{r}+}} \mathsf{r}_x
\end{equation}
where $\partial\mathsf{A}^{\mathsf{l}\pm}$ is the set of $x$ labels for all the outgoing/incoming leftward wires of $\mathtt{A}$, and $\partial\mathsf{A}^{\mathsf{r}\pm}$ is the set of $x$ labels for the outgoing/incomng rightward wires of $\mathtt{A}$.

\subsection{Physicality conditions}\label{sec:physicalityconditionsQFToldnotation}

The operation shown has the possibility of feeding outputs at the side back into inputs.  However, we can \lq\lq pull" all the input wires down and all the output wires up and apply the physicality condition. This leads to the physicality theorem. In particular, we can show that
\begin{equation}\label{QFTinputTcond}
0\leq \hat{A}_{(\mathsf{l}^-_\mathtt{A})^\mathsf{T}(\mathsf{r}_\mathtt{A}^-)^\mathsf{T}}^{\mathsf{l}^+_\mathtt{A}\mathsf{r}_\mathtt{A}^+}
\end{equation}
We also get
\begin{equation}\label{QFTcausalityI}
\hat{A}_{\mathsf{l}^-_\mathtt{A}\mathsf{r}_\mathtt{A}^-}^{\mathsf{l}^+_\mathtt{A}\mathsf{r}_\mathtt{A}^+} \hat{I}_{\mathsf{l}^+_\mathtt{A}\mathsf{r}_\mathtt{A}^+}
\leq \hat{I}_{\mathsf{l}^-_\mathtt{A}\mathsf{r}_\mathtt{A}^-}
\end{equation} \index{physicality conditions! discrete QFT}
However, it is possible to get many more conditions of the type in \eqref{QFTcausalityI}.  In particular, in their paper on quantum combs, Chiribella, D'Ariano, and Perinotti \cite{chiribella2009theoretical} provide a recursive method to place constraints on general circuit fragments of this type (there is a dictionary between the operator tensor approach and the quantum combs approach explained in \cite{hardy2011reformulating}).  We need to include these additional constraints to get a complete set of conditions.

\subsection{Change in notation convention}

So far, in this part of the paper on Quantum Theory, we have used the convention that inputs correspond to subscripts and outputs correspond to superscripts.  This has the drawback that, in those parts of the boundary where we have both input wires and output wires, we have to separate them into subscripts and superscripts.   Now we will relax the notation to bring the approach in line with the other parts of this paper. We will allow wires associated with inputs to appear in superscripts, and wires associated with outputs to appear as subscripts.  However, we will still demand, when matching, that inputs are matched with outputs.  To this end, note that for any wire, $\mathsf{w}_x$ at some position, $x$, can be though of as an output (indicated with a $+$) or an input (indicated with a $-$).  We denote a typing surface, $\mathtt{c}$ to be a set of wires with \emph{either} input \emph{or} output indicated
\begin{equation}
\mathtt{c} = \left\{ (x, \pm): \forall x\in \text{set}(\mathtt{c}) \right\}
\end{equation}
For each $x\in\text{set}(\mathtt{c})$ we have either a $+$ or a $-$.  This information can be thought of as the direction of the typing surface \index{typing surface! discrete QFT} at that value of $x$ (in this discrete analogy).
The typing surface $\mathtt{c}$ is associated with
\begin{equation}
\prod_{x\in \text{set}(\mathtt{c})} \mathsf{w}^\pm_x
\end{equation}
in the old notation where $\mathsf{w}^\pm_x$ is the wire at $x$ (this can be left moving or right moving and the $\pm$ indicated output/input.  We obtain $\mathtt{c}^R$ by reversing each direction
\begin{equation}
\mathtt{c}^R = \left\{ (x, \mp): \forall x\in \text{set}(\mathtt{c}) \right\}
\end{equation}
We adopt the convention that, when a typing surface $\mathtt{c}$ appears as a subscript, all its wires are reversed (but not when it appears as a superscript). This means that we can write down expressions like
\begin{equation}
\mathsf{A}_\mathtt{a b}^\mathtt{c} \mathsf{B}_\mathtt{c}
\end{equation}
as now the wires in $\mathtt{c}$ are matched (as every output is matched with an input).

We can write $\mathtt{c}=\mathtt{c}_+\cup \mathtt{c}_-$ (or $\mathtt{c}=\mathtt{c}_+ \mathtt{c}_-$ for notational convenience) where
\begin{equation}
\mathtt{c}_+ = \left\{ (x, +): \forall ~\text{output wires}~ x\in \text{set}(\mathtt{c}) \right\}
\end{equation}
and similarly for $\mathtt{c}_-$.

When $\mathtt{c}$ appears as a superscript we associate with it a Hilbert space
\begin{equation}
\mathcal{H}^\mathtt{c} = \mathcal{H}^\mathtt{c_+}\otimes \mathcal{H}^{c_-}
\end{equation}
Similarly, when $\mathtt{c}$ appears as a subscript we associate the Hilbert space
\begin{equation}
\mathcal{H}_\mathtt{c} = \mathcal{H}_\mathtt{c_+}\otimes \mathcal{H}_{c_-}
\end{equation}
with it.

With this new notational convention in place, we can write an operator as
\begin{equation}
\hat{A}_\mathtt{a b}^\mathtt{c}
\end{equation}
for example. This is an Hermitian operator acting on the Hilbert space
\begin{equation}
\mathcal{H}_\mathtt{a}\otimes \mathcal{H}_\mathtt{b} \otimes \mathcal{H}^\mathtt{c}
\end{equation}
In this example we require that
\begin{equation}\label{boundaryofA}
\text{set}(\mathtt{a}) \cup \text{set}(\mathtt{b}) \cup \text{set}(\mathtt{c})
\end{equation}
is the set of wires going into and coming out of the boundary of $\mathtt{A}$.  In principle, we could close some wires by putting preparation or measurement boxes on them.  This closed part of the boundary of $\mathtt{A}$ would not have any wires sticking out of it (and so would not contribute to $\partial\mathtt{A}$).

We will write $\mathtt{a}^{\mathsf{T}_-}=\mathtt{a}_+(\mathtt{a}_-)^T$ to indicate taking the input transpose so that the condition in \eqref{QFTinputTcond} can be written as
\begin{equation}
0\leq \hat{A}_{\mathtt{a^{\mathsf{T}_-}} b^{\mathsf{T}_-}}^\mathtt{c^{\mathsf{T}_-}}
\end{equation}
We write \eqref{QFTcausalityI} as
\begin{equation}
\hat{A}_\mathtt{a b}^\mathtt{c} \hat{I}^\mathtt{a_+b_+}_\mathtt{c_+} \leq \hat{I}_\mathtt{a_- b_-}^\mathtt{c_-}
\end{equation}         \index{physicality conditions!discrete QFT}
where $\mathtt{a} = \mathtt{a_-}\mathtt{a_+}$ and $-$ indicates the input wires and $+$ the output wires.  Again, as mentioned in Sec.\ \ref{sec:physicalityconditionsQFToldnotation}, we need additional conditions (equivalent to the recursive conditions of Chiribella et al.)  to have a complete set of conditions.

We can compose these objects by joining them at their boundaries. For example,
\begin{equation}
\begin{Compose}{0}{0}
\cobjectwhite{\theblob}{A}{2}{2}{-0.2,-5} \csymbol[0,-90]{A}
\cobjectwhite{\theflag}{B}{2}{1}{-4.75,0} \csymbol[-50,0]{B}
\cobjectwhite{\theblob}{E}{1.5}{1.5}{2.15,3} \csymbol[0,50]{E}
\cobjectwhite{\thetooth}{C}{1}{1}{1.4,-1.75} \csymbol{C}
\cobjectwhite{\theblob}{D}{0.95}{0.75}{-3.8,8} \csymbol{D}
\Cgrid[->, ultra thin]{0.5}{20}{20}{0,-1}
\end{Compose}
\end{equation}
shows the composition of five such objects.

\subsection{The continuous limit}\label{sec:thecontinuouslimit}

The next step is to take the limit as $\delta \rightarrow 0$ while keeping the outline of the region of interest unchanged.  As we take this limit we have more and more vertices inside the region of interest. In the limit: (i) the boundary becomes a continuous set; (ii) the typing surfaces, $\mathtt{a}$, $\mathtt{b}$, \dots become continuous and can be thought of as directed areas;  (iii) the Hilbert spaces, $\mathcal{H}^\mathtt{a}$ and $\mathcal{H}_\mathtt{a}$, associated with any type, $\mathtt{a}$, becomes infinite dimensional - we will treat them as continuous Hilbert spaces; (iiii) the setting $Q(x)$ and outcome $O(x)$ become fields.

We can use the same notation in the continuous case as in the discrete case.  An operation is represented by $\mathsf{A}^\mathsf{c}_\mathsf{ab}$, is associated with some region, $\mathtt{A}$, and has setting and outcomes
\begin{equation}
(Q_\mathsf{A}, O_\mathsf{A}) = \left\{ (x, Q(x), O(x)):\forall x \in \mathtt{A} \right\}
\end{equation}
Although we motivated this with a 2 dimensional grid, we can think of this in any number of dimensions.  We could model this by some fixed manifold, $\mathcal{M}$.

We require that the typing surfaces associated with the operation $\mathsf{A}^\mathsf{c}_\mathsf{ab}$ cover the whole of the typing surface (this may or may not be the whole of the boundary of $\mathtt{A}$).  These typing surfaces can meet but not overlap except at their edges.

We are assuming we have a fixed background causal structure. In particular, assume that we have some metric, $g_{\mu\nu}(x)$, and also a time direction field, $\tau^\mu(x)$ (see Sec.\ \ref{sec:timedirectionfield}).  We represent the typing surface, $\mathtt{a}$, by
\begin{equation}
\mathtt{a} = \left\{ (x, n^\mu): \forall x\in\text{set}(\mathtt{a}) \right\}
\end{equation}
where $\text{set}(\mathtt{a})$ is a surface (having dimension one less than that of the space) and $n^\mu$ is a normal to the surface at $x$.

We cannot decompose a typing surface into the union of output and input sets, $\mathtt{a}=\mathtt{a}_+\cup\mathtt{a}_-$, as before. However, we can still decompose the Hilbert space.  In particular, when $\mathtt{a}$ appears as a superscript we write
\begin{equation}
\mathcal{H}^\mathtt{a} = \mathcal{H}^\mathtt{a_+}\otimes \mathcal{H}^{a_-}
\end{equation}
Similarly, when $\mathtt{a}$ appears as a subscript we have Hilbert space
\begin{equation}
\mathcal{H}_\mathtt{a} = \mathcal{H}_\mathtt{a_+}\otimes \mathcal{H}_{a_-}
\end{equation}
We can deduce something about the nature of the Hilbert spaces in these decompositions from the normal, $n^\mu$, to the surface.
\begin{description}
\item[Future pointing:]   If $n^\mu$ is future pointing (so $-g_{\mu\nu} n^\mu n^\nu \geq 0$ and $g_{\mu\nu} n^\mu \tau^\nu> 0$) then
\begin{equation}
\text{dim}(\mathcal{H}^\mathtt{a_-}) = \text{dim}(\mathcal{H}_\mathtt{a_-}) = 1 ~~~\text{dim}(\mathcal{H}^\mathtt{a_+}) = \text{dim}(\mathcal{H}_\mathtt{a_+}) = \infty
\end{equation}
as there can be no non-trivial information leaving this surface in the direction of $n^\mu$ that is traveling to the past (but information can travel to the future).
\item[Past pointing:] If $n^\mu$ is past pointing (so $-g_{\mu\nu} n^\mu n^\nu \geq 0$ and $g_{\mu\nu} n^\mu \tau^\nu < 0$) then
\begin{equation}
\text{dim}(\mathcal{H}^\mathtt{a_+}) = \text{dim}(\mathcal{H}_\mathtt{a_+}) = 1 ~~~ \text{dim}(\mathcal{H}^\mathtt{a_-}) = \text{dim}(\mathcal{H}_\mathtt{a_-}) = \infty
\end{equation}
as there can be no non-trivial information leaving this surface in the direction of $n^\mu$ that is traveling to the future (but information can come in from the past).
\item[Spacelike:] If $n^\mu$ is spacelike (so $-g_{\mu\nu} n^\mu n^\nu < 0$) then there can be information leaving this surface in the direction of $n^\mu$ coming from the past and also information going to the future.  Hence
\begin{equation}
\text{dim}(\mathcal{H}^\mathtt{a_+}) = \text{dim}(\mathcal{H}_\mathtt{a_+}) = \text{dim}(\mathcal{H}^\mathtt{a_-}) = \text{dim}(\mathcal{H}_\mathtt{a_-})= \infty
\end{equation}
\end{description}
These different cases are best understood by looking at the discrete case (and thinking about the limit).  Rather than imposing the above constraints on the Hilbert spaces associated with different parts of the boundary we could alow
\begin{equation}
\text{dim}(\mathcal{H}^\mathtt{a_+}) = \text{dim}(\mathcal{H}_\mathtt{a_+}) = \text{dim}(\mathcal{H}^\mathtt{a_-}) = \text{dim}(\mathcal{H}_\mathtt{a_-})= \infty
\end{equation}
regardless of the direction of $n^\mu$ and, instead, impose constraints on the space of allowed operators coming from the relationship between this normal vector and the metric.  This would amount to the same thing but may be a better way to look for generalizations of the structure discussed here.

An operator is represented by
\begin{equation}
\hat{A}_{\mathtt{a}\mathtt{b}}^\mathtt{c}
\end{equation}
This is a Hermitian operator acting on a Hilbert space $\mathcal{H}_\mathtt{a}\otimes\mathcal{H}_\mathtt{b}\otimes\mathcal{H}^\mathtt{c}$.  Physicality demands that this operator is positive under input transpose
\begin{equation}
0\leq \hat{A}_\mathtt{a^{T_-}b^{T_-}}^\mathtt{c}
\end{equation}   \index{physicality conditions!QFT}
We also have normalization conditions such as
\begin{equation}
\hat{A}_\mathtt{a b}^\mathtt{c} \hat{I}^\mathtt{a_+b_+}_\mathtt{c_+} \leq \hat{I}_\mathtt{a_- b_-}^\mathtt{c_-}
\end{equation}
An important problem that would complete this formulation of QFT is to write down the complete set of normalization condition.  These conditions must (i) guarantee that we cannot signal outside the forward light cone and (ii) that probabilities are bounded between 0 and 1.  Finding these conditions is an important problem for future work.  Here we will simply give these conditions a name
\begin{equation}\label{QFTphysicalityCond}
\text{QFTphysicalityCond}(\hat{A}_\mathtt{a b}^\mathtt{c})
\end{equation}
If this condition is satisfied then the operator tensor is physically possible.  It is not clear how to extend the recursive techniques of Chribella et al. \cite{chiribella2009theoretical} condition because this is not a discrete situation.

Another condition that we should impose for physical reasons, in setting up a calculation, is that the $Q(x)$ and $O(x)$ fields match up at boundaries between operators (though, from a mathematical point of view, it may be interesting to consider cases with discontinuities.

\subsection{Calculations}\label{sec:calculationsQFT}

We can use operator tensors to extract probabilities as before.  If we have a closed circuit then the operator tensor will give a probability.  If we have open circuits then we can use the ideas of formalism locality.  We will illustrate how to do a calculation for the open case.

We can associate an operator diagram with the situation where we have several regions that border one another. For example,
\begin{equation}
\begin{Compose}{0}{0}
\cobjectwhite{\theblob}{A}{2}{2}{-0.2,-5} \csymbol[0,-90]{A}
\cobjectwhite{\theflag}{B}{2}{1}{-4.75,0} \csymbol[-50,0]{B}
\cobjectwhite{\theblob}{E}{1.5}{1.5}{2.15,3} \csymbol[0,50]{E}
\cobjectwhite{\thetooth}{C}{1}{1}{1.4,-1.75} \csymbol{C}
\cobjectwhite{\theblob}{D}{0.95}{0.75}{-3.8,8} \csymbol{D}
\end{Compose}
\Leftrightarrow ~~
\begin{Compose}{0}{-1.1} \setdefaultfont{\mathsf}\setsecondfont{\mathtt}
\Ucircle{A}{1,-1} \Ucircle{B}{-5,5} \Ucircle{C}{-0.3,4} \Ucircle{D}{-3, 11} \Ucircle{E}{2,9}
\joincc[below left]{B}{-65}{A}{130} \csymbolalt{a}
\joincc[right]{A}{100}{C}{-90} \csymbolalt{b}
\joincc[below]{C}{170}{B}{-10} \csymbolalt{c}
\joincc[above left]{B}{25}{E}{-140} \csymbolalt{d}
\joincc[left]{B}{80}{D}{-100} \csymbolalt{e}
\joincc[above right]{D}{-15}{E}{170}
\joincc[below right]{C}{70}{E}{-95} \csymbolalt{g}
\joincc[right]{E}{-60}{A}{40}\csymbolalt{k}
\thispoint{nA}{-2,-2} \joincc[above left]{A}{-135}{nA}{45} \csymbolalt{f}
\thispoint{nB}{-8,5} \joincc[above]{B}{180}{nB}{0} \csymbolalt{h}
\thispoint{nD}{-1,13} \joincc[above left]{D}{45}{nD}{-135} \csymbolalt{i}
\thispoint{nE}{4,11} \joincc[above left]{nE}{-145}{E}{45} \csymbolalt{j}
\end{Compose}
\end{equation}
This corresponds to the operator
\begin{equation}
\begin{Compose}{0}{0} \setdefaultfont{\hat}\setsecondfont{\mathtt}
\Ucircle{A}{1,-1} \Ucircle{B}{-5,5} \Ucircle{C}{-0.3,4} \Ucircle{D}{-3, 11} \Ucircle{E}{2,9}
\joincc[below left]{B}{-65}{A}{130} \csymbolalt{a}
\joincc[right]{A}{100}{C}{-90} \csymbolalt{b}
\joincc[below]{C}{170}{B}{-10} \csymbolalt{c}
\joincc[above left]{B}{25}{E}{-140} \csymbolalt{d}
\joincc[left]{B}{80}{D}{-100} \csymbolalt{e}
\joincc[above right]{D}{-15}{E}{170}
\joincc[below right]{C}{70}{E}{-95} \csymbolalt{g}
\joincc[right]{E}{-60}{A}{40}\csymbolalt{k}
\thispoint{nA}{-2,-2} \joincc[above left]{A}{-135}{nA}{45} \csymbolalt{f}
\thispoint{nB}{-8,5} \joincc[above]{B}{180}{nB}{0} \csymbolalt{h}
\thispoint{nD}{-1,13} \joincc[above left]{D}{45}{nD}{-135} \csymbolalt{i}
\thispoint{nE}{4,11} \joincc[above left]{nE}{-145}{E}{45} \csymbolalt{j}
\end{Compose}
\end{equation}
We can calculate different operators for different outcomes in each of these regions.  When two operators of this sort are proportional then the relative probability is well conditioned (i.e.\ independent of what settings are chosen and what outcomes are seen outside this composite region) and is given by the constant of proportionality.  If the two operators are not proportional then the probability is not well conditioned so - it may depend on what is happening outside the region of interest.  Here we are simply applying the ideas of formalism locality as discussed in Sec.\ \ref{sec:formalismlocalityPoss} (for PoAGeR) and Sec.\ \ref{sec:formalismlocalityinPAGeR} (for PAGeR).   One example to consider the two operations
\begin{equation}
\begin{Compose}{0}{-1.4} \setdefaultfont{\mathsf}\setsecondfont{\mathtt}
\Ucircle{A}{1,-1} \Ucircle{B}{-5,5} \Ucircle{C}{-0.3,4} \Ucircle{D}{-3, 11} \Ucircle{E}{2,9}
\joincc[below left]{B}{-65}{A}{130} \csymbolalt{a}
\joincc[right]{A}{100}{C}{-90} \csymbolalt{b}
\joincc[below]{C}{170}{B}{-10} \csymbolalt{c}
\joincc[above left]{B}{25}{E}{-140} \csymbolalt{d}
\joincc[left]{B}{80}{D}{-100} \csymbolalt{e}
\joincc[above right]{D}{-15}{E}{170}
\joincc[below right]{C}{70}{E}{-95} \csymbolalt{g}
\joincc[right]{E}{-60}{A}{40}\csymbolalt{k}
\thispoint{nA}{-2,-2} \joincc[above left]{A}{-135}{nA}{45} \csymbolalt{f}
\thispoint{nB}{-8,5} \joincc[above]{B}{180}{nB}{0} \csymbolalt{h}
\thispoint{nD}{-1,13} \joincc[above left]{D}{45}{nD}{-135} \csymbolalt{i}
\thispoint{nE}{4,11} \joincc[above left]{nE}{-145}{E}{45} \csymbolalt{j}
\end{Compose}
~~~ \text{and}~~~
\begin{Compose}{0}{-1.4} \setdefaultfont{\mathsf}\setsecondfont{\mathtt}
\Ucircle{A}{1,-1} \Ucircle{B}{-5,5} \ucircle{C}{-0.3,4}\csymbol{C'}\Ucircle{D}{-3, 11} \Ucircle{E}{2,9}
\joincc[below left]{B}{-65}{A}{130} \csymbolalt{a}
\joincc[right]{A}{100}{C}{-90} \csymbolalt{b}
\joincc[below]{C}{170}{B}{-10} \csymbolalt{c}
\joincc[above left]{B}{25}{E}{-140} \csymbolalt{d}
\joincc[left]{B}{80}{D}{-100} \csymbolalt{e}
\joincc[above right]{D}{-15}{E}{170}
\joincc[below right]{C}{70}{E}{-95} \csymbolalt{g}
\joincc[right]{E}{-60}{A}{40}\csymbolalt{k}
\thispoint{nA}{-2,-2} \joincc[above left]{A}{-135}{nA}{45} \csymbolalt{f}
\thispoint{nB}{-8,5} \joincc[above]{B}{180}{nB}{0} \csymbolalt{h}
\thispoint{nD}{-1,13} \joincc[above left]{D}{45}{nD}{-135} \csymbolalt{i}
\thispoint{nE}{4,11} \joincc[above left]{nE}{-145}{E}{45} \csymbolalt{j}
\end{Compose}
\end{equation}
where, in this example, the operation in the centre is $C$ in one case and $C'$ in the other case (but all the other operations are the same).  The relative probability for these two scenarios is well conditioned if the operators
\begin{equation}
\begin{Compose}{0}{-1.4} \setdefaultfont{\hat}\setsecondfont{\mathtt}
\Ucircle{A}{1,-1} \Ucircle{B}{-5,5} \Ucircle{C}{-0.3,4} \Ucircle{D}{-3, 11} \Ucircle{E}{2,9}
\joincc[below left]{B}{-65}{A}{130} \csymbolalt{a}
\joincc[right]{A}{100}{C}{-90} \csymbolalt{b}
\joincc[below]{C}{170}{B}{-10} \csymbolalt{c}
\joincc[above left]{B}{25}{E}{-140} \csymbolalt{d}
\joincc[left]{B}{80}{D}{-100} \csymbolalt{e}
\joincc[above right]{D}{-15}{E}{170}
\joincc[below right]{C}{70}{E}{-95} \csymbolalt{g}
\joincc[right]{E}{-60}{A}{40}\csymbolalt{k}
\thispoint{nA}{-2,-2} \joincc[above left]{A}{-135}{nA}{45} \csymbolalt{f}
\thispoint{nB}{-8,5} \joincc[above]{B}{180}{nB}{0} \csymbolalt{h}
\thispoint{nD}{-1,13} \joincc[above left]{D}{45}{nD}{-135} \csymbolalt{i}
\thispoint{nE}{4,11} \joincc[above left]{nE}{-145}{E}{45} \csymbolalt{j}
\end{Compose}
~~~ \text{and}~~~
\begin{Compose}{0}{-1.4} \setdefaultfont{\hat}\setsecondfont{\mathtt}
\Ucircle{A}{1,-1} \Ucircle{B}{-5,5} \ucircle{C}{-0.3,4}\csymbol{C'}\Ucircle{D}{-3, 11} \Ucircle{E}{2,9}
\joincc[below left]{B}{-65}{A}{130} \csymbolalt{a}
\joincc[right]{A}{100}{C}{-90} \csymbolalt{b}
\joincc[below]{C}{170}{B}{-10} \csymbolalt{c}
\joincc[above left]{B}{25}{E}{-140} \csymbolalt{d}
\joincc[left]{B}{80}{D}{-100} \csymbolalt{e}
\joincc[above right]{D}{-15}{E}{170}
\joincc[below right]{C}{70}{E}{-95} \csymbolalt{g}
\joincc[right]{E}{-60}{A}{40}\csymbolalt{k}
\thispoint{nA}{-2,-2} \joincc[above left]{A}{-135}{nA}{45} \csymbolalt{f}
\thispoint{nB}{-8,5} \joincc[above]{B}{180}{nB}{0} \csymbolalt{h}
\thispoint{nD}{-1,13} \joincc[above left]{D}{45}{nD}{-135} \csymbolalt{i}
\thispoint{nE}{4,11} \joincc[above left]{nE}{-145}{E}{45} \csymbolalt{j}
\end{Compose}
\end{equation}
are proportional and then the relative probability is equal to the constant of proportionality.

\subsection{Discussion}

The operator tensor formulation of Quantum Field Theory provides an operational approach that is intimately wedded to the spacetime structure.  We can think in terms of inputting and outputting data (i.e.\ settings and outcomes) during the course of an experiment.  We are not required to use the scattering paradigm in which we send in particles from $-\infty$ and measure them at $+\infty$.

There are some technical problems that need to be solved for this to represent a full formulation of QFT.  First, we need a full characterization of the set of operators associated with an given space-time region (this is the QFT physicality condition \eqref{QFTphysicalityCond}).  Second, the Hilbert spaces required may be \lq\lq bigger" than the Hilbert space consider in the continuous Hilbert spaces considered in Sec.\ \ref{sec:continuousdimensionalHilbertspace} in that they associate an amplitude with arbitrary field configurations.  We can define an Dirac delta function for Hilbert spaces with a basis described by a point in a finite dimensional space, $\mathbb{R}^N$.  However, arbitrary field configurations require a continuous infinity of real numbers for their specification.  Third, we need to be sure that, in spite of these problems, we can define the trace operation and also have a useful notion of the identity operator.  The substitution operator defined in Appendix \ref{sec:thesubstitutionoperator} may play a role here.

There is also a deeper conceptual question - are the coordinates, $x$, operational?  We could be in a situation in which we can actually read $x$ off some fixed physical reference frame.  This reference frame would have to be classical and there could be no back-reaction between it and the degrees of freedom we are interested in measuring.  This assumption would ultimately have to break down. The approach in the other parts of this paper, in which we define an operational space, offers a way forward even in the limited context of Quantum Field Theory.

\newpage

\part{Quantum Gravity: QuAGeR}\label{part:QuantumGravity}

\section{The problem of Quantum Gravity}

We will sketch three possible routes to Quantum Gravity - the abstract, the ontological, and the principled route.  In the abstract route, we consider a kind of quantization in which we construct a candidate hopping metric for Quantum Gravity and suitably enlarge the space of generalized probabilistic states.   In the ontological approach we explore the possibility that PAGeR actually has Quantum Theory as a limit (where unusual quantum effects might have to do with having unusual mixtures of different causal scenarios).  In the principled route, we suggest that we might be able to write down some set of principles that apply to a suitably general framework and obtain a theory of Quantum Gravity.   This part of the paper is mostly speculative. However, we will make some inroads into the first approach.

First, it is worth making some remarks about what, exactly, the problem of Quantum Gravity is.  \index{Quantum Gravity!the problem of} We take it to be to find a theory which reproduces the predictions of Quantum Theory on the one hand, and General Relativity on the other hand at least in those circumstances where those theories have been shown to be successful.  Given that it can reproduce the predictions of Quantum Theory and of General Relativity we expect there to be some limit that can be applied to Quantum Gravity that gives these less fundamental theories.  We may also expect there will be situations which neither General Relativity or Quantum Theory are able to account for - these will be genuinely new Quantum Gravitational effects.

One way to look at this is to note that Quantum Theory and General Relativity are each conservative and radical in complementary respects.   Quantum theory is inherently probabilistic (radical feature) while General Relativity is deterministic (conservative feature).  General Relativity, on the other hand, has dynamical causal structure (radical feature) while, in Quantum Theory the causal structure is fixed and given in advance (conservative feature).  It is most likely that a theory of Quantum Gravity will take the radical road in each case. Hence, it will have dynamical causal structure and be inherently probabilistic. In fact, we an expect it to be a little more radical still for the following reason. In Quantum Theory any dynamical quantity is also subject to quantum indefiniteness. There is, in some sense, no-matter-of-the-fact as to which slit a particle going through an interfometer takes.  Similarly, we expect that there may be no matter of the fact as to what the causal structure is in Quantum Gravity. That is we expect there to be indefinite causal structure.  Indefinite causal structure has been a subject of much study in recent years (see discussion in Sec.\ \ref{sec:thisresearchprogramandotherwork}.)

We will pursue an approach consistent with the principle of general compositionality (see Sec.\ \ref{sec:introductionatbeginning}).  In fact, once we have indefinite causal structure, we are more-or-less forced to such an approach.  If we have definite causal structure then we can demand that describe the world in terms of parts that are circumscribed by this causal structure.  The most common example of this is the idea that our components should be time slices.  This leads to a particular way of doing physics that does not respect the afore mentioned principle.  Once we have indefinite causal structure this avenue is effectively blocked.

The theory of Quantum Gravity itself may end up being formulated in an entirely new mathematical language.  Indeed, we should be willing to countenance the possibility that the mathematics of Quantum Gravity is as different to the mathematics of Quantum Theory or General Relativity as the mathematics of either of those theories is from Newtonian mechanics.

There are various approaches to Quantum Gravity that are active areas of research - string theory, loop quantum gravity, spin foams, dynamical triangulations, causal sets, and others (see discussion in Sec.\ \ref{sec:thisresearchprogramandotherwork}.  The approach being pursued here is distinct in that it comes from the tradition of operational approaches to physics and general probability theories (GPTs).

\section{An abstract approach - Quantization}\label{sec:anabstractapproach}

Now we will explore an approach to finding a theory of Quantum Gravity based on the operator tensor formulation of quantum theory and PAGeR as developed earlier.  This is still at a speculative stage. However, we will push the approach further than the other two approaches to be suggested.

\subsection{Quantization and GRization}

The approach to obtaining a theory of Quantum Gravity explored in this section is, perhaps, best summed up in the diagram below
\begin{displaymath}
\xymatrix{ \text{OpQT} \ar[r]^{\text{GRize}} & \text{QuAGeR}  \\
\text{CProbT} \ar[u]^{\text{quantize}} \ar[r]_{\text{GRize}} & \text{PAGeR} \ar[u]_{\text{quantize}} }
\end{displaymath}
CProbT is classical probability theory. This is the theory of dice and coins and concerns probability simplices and stochastic maps.  OpQT is Operational Quantum Theory.  This is the rather practical formulation of Quantum Theory used extensively in Quantum Information.  This is usually expressed in terms of density matrices, completely positive maps, and POVMs (positive operator valued measures).  We will use the operator tensor formulation of Quantum Theory (see Part \ref{part:operatortensorQT} here.  This is an alternative (though equivalent) operational formulation of Quantum theory.   \index{Quantum Gravity!abstract approach}   \index{QuAGeR}  \index{quantization} \index{GRization}

It is worth mentioning an alternative possible strategy (which we will not pursue here) summed up as below
\begin{displaymath}
\xymatrix{ \text{OpQT} \ar[r]^{\text{GRize}} & \text{QuAGeR}  \\
\text{CPossT} \ar[u]^{\text{quantize}} \ar[r]_{\text{GRize}} & \text{PoAGeR} \ar[u]_{\text{quantize}} }
\end{displaymath}
Here CPossT is classical possibilistic theory.   That is it might be better to base a strategy on possibilistic formulations of the classical theories with probabilities coming in at the Quantum level.

\subsection{Square root and square approach to quantization}

\index{quantization!square root and square}

Consider a classical system described consisting of a single particle. Its ontological state is described by a point, $a=(q, p)$, in phase space.  To illustrate the following remarks, we will imagine that these points are discretized where $q=(1, 2, \dots, N)$ and similarly $p=(1, 2, \dots, N)$.  Then the number of distinguishable ontic states is $N^2$.  If we have a classical probabilistic mixture then we would describe the state by the list of $N^2$ probabilities
\begin{equation}\label{classicalaA}
\presup{a}A =\text{prob}(a)
\end{equation}
In quantum theory we would first \lq\lq take the square root" and associate the basis vectors of a Hilbert space with $q$ (we could have chosen $p$).  This gives us a set of $N$ basis vectors $\{|q\rangle\}$.  This is the number of perfectly distinguishable states (in a single shot measurement).   But then we \lq\lq square" because, to represent a general density matrix we need to form a space of linearly independent Hermitian operators.  An example of such a set is
\begin{equation}\label{hatPqqxy}
\hat{P}_q = |q\rangle\langle q|, ~~~ \hat{P}_{qq'x}= |qq'x\rangle \langle qq'x|, ~~~ \hat{P}_{qq'y}= |qq'x\rangle \langle qq'y|
\end{equation}
where
\begin{equation}
|qq'x\rangle = \frac{1}{\sqrt{2}}(|q\rangle+ |q'\rangle) ~~~~~~ |qq'y\rangle = \frac{1}{\sqrt{2}}(|q\rangle+ i|q'\rangle)
\end{equation}
for $q\not=q'$.  There are $N^2$ projectors in \eqref{hatPqqxy}. We write these $N^2$ linearly independent projectors as
\begin{equation}
\presub{a}{\hat{X}}^\mathtt{a}~~~~~~ \text{where}~~ a\in\{q\}\cup \{qq'x\}\cup \{qq'y\}
\end{equation}
This is our fiducial set (for the case with raised subscript, $\mathtt{a}$).
Now we can write a general density matrix, $\hat{A}^\mathtt{a}$, as
\begin{equation}
\hat{A}^\mathtt{a} = \presub{a}{\hat{X}}^\mathtt{a} \smallspace \presup{a}A
\end{equation}
We can represent a quantum state by $\hat{A}^\mathtt{a}$, or equivalently by the coefficients, $\presup{a}A$.  These coefficients are different from the classical state in \eqref{classicalaA} - in particular, they can be negative. However, it is striking that we have $N^2$ entries in each case.

The classical limit must re-emerge.  One way to take a classical limit (often used in Quantum Information) is to restrict ourselves to diagonal density matrices. However, this only gives us $N$ distinct possibilities rather than the $N^2$ we need for the full phase space.  It would be interesting to understand, from this point of view, how the classical phase space emerges in the limit.

\subsection{Classical level of description}\label{sec:classicallevel}

When we provide an operational interpretation of a situation described by Quantum Theory, we must nominate certain quantities as being described classically (corresponding to settings, outcomes, and the operational description of the composite structure of the experiment in question).  We call these the \emph{nominated classical quantities}. \index{classical quantities in QG}  We want to have enough classicality to support our notion of an operation.  This requires that we have outcome sets, $O_\mathtt{A}$ built out of $\Gamma_\mathtt{A}$ sets which are, in turn built out of points, $\mathbf{S}$, in op-space.  We also need the notion of a op-space region, $\mathtt{A}$, and typing surfaces, $\mathtt{a}$. These are also built out of points, $\mathbf{S}$, in op-space.  Hence, it is sufficient to nominate the points, $\mathbf{S}$, to be classical to support the notion of an operation.  We can imagine that the scalars, $S_k$, that make up $\mathbf{S}$ correspond only to large enough systems that it is reasonable to treat them classically.  We could, indeed, allow these scalars to be course-grainings over more fine-grained scalars.  As long as we are able to define an op-space to describe our operations then we can proceed.  We also need to be able to describe settings, $\mathbf{Q}_\mathtt{A}$.  Since these also correspond to lists of scalars, similar remarks apply.  We may also want to describe some loading, $L_\mathtt{A}$.  This loading may depend on additional scalars treated at the management level.   Hence, at the level of operational description, we are assuming the same mathematics as in the classical case - we have op-space spanned by scalars.  The nature of the description of the systems that are not directly observed (what we called the hidden variables in the classical case) may be very different in the case of Quantum Gravity.

\subsection{Loaded operations}

Given the above remarks about having a classical level of description, we can define loaded operations in exactly the same way as for PAGeR in Sec.\ \ref{sec:loadedoperationsPAGeR}.  Thus, an operation has the following components
\begin{equation}
\mathsf{A}=( \text{strat}(\mathsf{A}), \text{outcome}(\mathsf{A}), \text{load}(\mathsf{A}), \text{reg}(\mathsf{A}), \text{type}(\mathsf{A}) )
\end{equation}          \index{loaded operations!Quantum Gravity}
We give loaded operations subscripts and superscripts as before and we can also represent them diagrammatically.
As before, we will introduce subscripts and superscripts and corresponding diagrammatic notation
\begin{equation}
\mathsf{A}^\mathtt{a}_\mathtt{bc} ~~~ \Longleftrightarrow
\begin{Compose}{0}{0} \setdefaultfont{\mathsf}\setsecondfont{\mathtt}
\Ucircle{A}{0,0}\thispoint{DL}{-120:4} \thispoint{DR}{-60:4} \thispoint{UC}{90:4}
\joincc[above left]{DL}{60}{A}{-120} \csymbolalt{b}
\joincc[above right]{DR}{120}{A}{-60} \csymbolalt{c}
\joincc[left]{A}{90}{UC}{-90} \csymbolalt{a}
\end{Compose}
\end{equation}
We can join operations together to form composite operations covering multiple regions of op-space.

\subsection{Choosing the Hilbert space: taking the \lq\lq square root"}

As a first step in this approach to constructing QuAGeR we need to take the square root - that is find the object that is analogous to $q$ rather than $(q, p)$.
We had previously defined our boundary conditions in Sec.\ \ref{sec:fullboundaryconditions} as
\begin{equation}
\theta_\mathtt{a}(\Psi_\mathtt{A})
= \left\{ \tilde{\theta}_\mathtt{a}({\tilde{\Psi}}_\mathtt{A}): \forall \tilde{\Psi}_\mathtt{A}\in \Psi_\mathtt{A} \right\}
= \left\{ \tilde{\theta}_\mathtt{a}(\vartheta^*{\tilde{\Psi}}_\mathtt{A}): \forall \varphi_\mathtt{A}\in G^+_\mathtt{A} \right\}
\end{equation}
where
\begin{equation}
\tilde{\theta}_\mathtt{a}({\tilde{\Psi}}_\mathtt{A})= \left\{ (p, \pmb{\Theta}): \forall p \in \mathscr{M}_\mathtt{a|A|a} \right\}
\end{equation}
and (in the agency case - see Sec.\ \ref{sec:boundaryconditionswithagency}) we have
\begin{equation}
\pmb{\Theta} = (\pmb{\Phi}, \pmb{\pi}, \partial_{\mathtt{a}\pmb{\mu}} X_\mathtt{a}^0, \pmb{\partial} x_\mathtt{a}, \pmb{\chi}, \pmb{\tau})
\end{equation}
These boundary conditions are analogous to $(q,p)$.

We would like to provide boundary conditions analogous to $q$.  One proposal is that we simply drop the $\pmb{\pi}$. Thus, we define
\begin{equation}\label{oneproposalforQ}
\pmb{\Theta}_Q = (\pmb{\Phi}, \partial_{\mathtt{a}\pmb{\mu}} X_\mathtt{a}^0, \pmb{\partial} x_\mathtt{a}, \pmb{\chi}, \pmb{\tau})
\end{equation}   \index{ThetaQ@$\pmb{\Theta}_Q$}
Then our boundary conditions are given by
\begin{equation}\label{qboundaryconditions}
q_\mathtt{a}= \left\{ \vartheta^*\tilde{q}_\mathtt{a} :\forall \vartheta\in G^+_\mathtt{A} \right\}
\end{equation}
where
\begin{equation}
\tilde{q}_\mathtt{a} = \left\{ (\pmb{\Theta}_Q, p): \forall p \in \mathscr{M}_\mathtt{a} \right\}
\end{equation}
Previously we regarded these boundary conditions as being induced at the boundary by solutions to the field equations.  However, since we are working towards a new theory, we cannot assume that there are any field equations (in the classical sense).  Hence, we simply consider all possible boundary conditions of this nature.  Note that we could try different proposals other than \eqref{oneproposalforQ}. The strategy outlined here will go through the same way.  As we will see shortly, these boundary conditions get mapped to operational specifications on the interior of regions and it is the latter that are ultimately more important.

Let $\Lambda_\mathtt{a}[\text{basis}]$ be the set of possible $q_\mathtt{a}$.  We will regard these as the labels for a basis $|q_\mathtt{a}\rangle$ of the Hilbert space.  A technical problem here is that this is a space of fields on manifolds.  This is parameterized by an infinite number of real parameters.  Hence it appears that the Hilbert space will be non-separable and there may also be issues defining the Dirac delta function for
$\langle q_\mathtt{a}|q'_\mathtt{a}\rangle$.  We may be able to solve these technical problems by some kind of regularization or by using the substitution operator defined in Sec.\ \ref{sec:thesubstitutionoperator}.  However, given that the approach to quantum gravity we are outlining here is rather speculative in the first place, we will not embark on attempting to resolve this here.

\subsection{Choosing the fiducials: taking the \lq\lq square"}

We will now assume that we can construct fiducials corresponding to superpositions of these basis states.  Thus, we assume that is meaningful to discuss
\begin{equation}
|q_\mathtt{a}q'_\mathtt{a}-\rangle = \frac{1}{\sqrt{2}} (|q_\mathtt{a}\rangle -  |q'_\mathtt{a}\rangle) ~~~~~~
|q_\mathtt{a}q'_\mathtt{a}i\rangle = \frac{1}{\sqrt{2}} (|q_\mathtt{a}\rangle + i|q'_\mathtt{a}\rangle)
\end{equation}
We consider the fiducials
\begin{equation}
\hat{X}_\mathtt{a}^{\mathqbnd{a}}  ~~\Leftrightarrow~~
\begin{Compose}{0}{0} \setdefaultfont{\mathnormal}\setsecondfont{\mathtt}\setthirdfont{\mathqbnd}
\thispoint{pX}{0,-3}
\Scircle{X}{0,0} \blackdotsq{Xa}{0,3} \joincc[left]{pX}{90}{X}{-90} \csymbolalt{a} \joincc[left]{X}{90}{Xa}{-90}\csymbolthird{a}
\end{Compose}
~~~~~~~~
\presub{\mathqbnd{a}}{\hat{X}}^{\mathtt{a}} ~~\Leftrightarrow~~
\begin{Compose}{0}{0} \setdefaultfont{\mathnormal}\setsecondfont{\mathtt}\setthirdfont{\mathqbnd}
\thispoint{pX}{0,-3} \Scircle{X}{0,0} \blackdotsq{Xa}{0,3}
\joincc[left]{X}{-90}{pX}{90}\csymbolalt{a}
\joincc[left]{Xa}{-90}{X}{90} \csymbolthird{a}
\end{Compose}
\end{equation}  \index{fiducial operators in QG}
where the fiducial set (same for raised and lowered subscripts) is taken to be
\begin{equation}\label{QGfiducialset}
\mathqbnd{a}\in \{q_\mathtt{a}:\forall~ q_\mathtt{a}\}
\cup \{q_\mathtt{a}q'_\mathtt{a}-:\forall~ q_\mathtt{a},q'_\mathtt{a} ~\text{s.t.} ~q_\mathtt{a}\not= q'_\mathtt{a} \}
\cup \{q_\mathtt{a}q'_\mathtt{a}i:\forall~ q_\mathtt{a}, q'_\mathtt{a} ~\text{s.t.} ~q_\mathtt{a}\not= q'_\mathtt{a} \}
\end{equation}
where $q_\mathtt{a}$ and $q'_\mathtt{a}$ are in $\Lambda_\mathtt{a}[\text{basis}]$.
These fiducial operators correspond to the projection operators $|a\rangle\langle a|$.  We are double counting in the $q_\mathtt{a}q'_\mathtt{a}-$ case as $q'_\mathtt{a}q_\mathtt{a}-$ also appears (and similarly in the $q_\mathtt{a}q'_\mathtt{a}i$ case).  If we want the associated hopping metric to be invertible then we should remove this double counting.  We will write $\Lambda^\mathtt{a}[\text{fid}]$ and $\Lambda_\mathtt{a}[\text{fid}]$ for the fiducial sets (in \eqref{QGfiducialset}) of $\mathqbnd{a}$'s.  We could, or course, choose different fiducial sets.

We can use these fiducials to define a hopping  metric
\begin{equation}
\presub{\mathqbnd{a}'}h^{\mathqbnd{a}}= \text{prob} (\presub{\mathqbnd{a}'}X^\mathtt{a} X_\mathtt{a}^\mathqbnd{a} )
~~~~~~\Leftrightarrow~~~~~~
\begin{Compose}{0}{-0.1}\setthirdfont{\mathqbnd}
\vbbmatrixsq{h}{0,0}\csymbolthird{a}
\end{Compose}
~=~
\begin{Compose}{0}{-0.5} \setdefaultfont{\hat}\setsecondfont{\mathtt}\setthirdfont{\mathlbnd}
\blackdotsq{d1}{0,-3} \scircle{X1}{0,0}\csymbol{X}\scircle{X2}{0,4} \csymbol{X} \blackdotsq{d2}{0,7}
\joincc[left]{d1}{90}{X1}{-90} \csymbolthird{a} \joincc[left]{X1}{90}{X2}{-90}\csymbolalt{a} \joincc[left]{X2}{90}{d2}{-90} \csymbolthird{a}
\end{Compose}
\end{equation}

We can write a general operator in fully decomposed form. For example
\begin{equation}
\hat{A}_\mathsf{a_1b_2}^\mathsf{a_3c_4} =
\presup{\mathqbnd{a_3c_4}}A_{\mathqbnd{a_1b_2}}  \hat{X}_\mathsf{a_1}^\mathqbnd{a_1} \hat{X}_\mathsf{b_2}^\mathqbnd{b_2} \presub{\mathqbnd{a_3}}{\hat{X}}^\mathsf{a_3} \presub{\mathqbnd{c_4}}{\hat{X}}^\mathsf{c_4}
\end{equation}
\begin{equation}
\begin{Compose}{0}{0}\setdefaultfont{\hat}\setsecondfont{\mathtt}
\Ucircle{A}{0,0}
\thispoint{ALL}{-3,-3}\thispoint{ALR}{3,-3} \thispoint{AUL}{-3,3}\thispoint{AUR}{3,3}
\joincc{ALL}{90}{A}{-110} \csymbolalt{a} \joincc{ALR}{90}{A}{-70} \csymbolalt{b}
\joincc{A}{110}{AUL}{-90} \csymbolalt{a} \joincc{A}{70}{AUR}{-90}\csymbolalt{c}
\end{Compose}
~=~
\begin{Compose}{0}{0}\setdefaultfont{\hat}\setsecondfont{\mathqbnd}\setthirdfont{\mathtt}
\Ucircle{A}{0,0}
\scircle{XLL}{-4,-4}\csymbol{X}\thispoint{ALL}{-7,-7}\joincc{ALL}{90}{XLL}{-135} \csymbol{a} \joinccbwsq{XLL}{45}{A}{-110}\csymbolalt{a}
\scircle{XLR}{4,-4}\csymbol{X}\thispoint{ALR}{7,-7}  \joincc{ALR}{90}{XLR}{-45} \csymbol{b}  \joinccbwsq{XLR}{135}{A}{-70}\csymbolalt{b}
\scircle{XUL}{-4,4}\csymbol{X}\thispoint{AUL}{-7,7}  \joinccwbsq{A}{110}{XUL}{-45}  \csymbolalt{a}   \joincc{XUL}{135}{AUL}{-90} \csymbol{a}
\scircle{XUR}{4,4}\csymbol{X}\thispoint{AUR}{7,7}    \joinccwbsq{A}{70}{XUR}{-135} \csymbolalt{c} \joincc{XUR}{45}{AUR}{-90}\csymbol{c}
\end{Compose}
\end{equation}
Here the repeated indexes for the fiducial label ($\mathqbnd{a}_1$ for example) indicated integration/summation over that quantity.  We need a measure, $\mathrm{d}\mathqbnd{a}_1$, to accomplish this integration.

\subsection{Operational Quantum Gravity}\label{sec:OperationalQuantumGravity}

So far we have set this up this theory in terms of boundary conditions $\mathqbnd{a}$. These are inconvenient quantities to use because they are formed by acting with $G^+$ on $\tilde{\mathqbnd{a}}$.  However, as we noted in Sec.\ \ref{sec:classicallevel}, we can nominate some scalar fields to provide a classical level of description.  Each operation must be described in terms of $(\mathtt{Z}, \mathbf{Q}_\mathtt{Z}, O_\mathtt{Z}, L_\mathsf{Z})$ (strictly the loading, $L_\mathsf{Z}$, may depend on beliefs as well as scalar quantities but it is something we can write down and so operational in this sense).  Thus, we assume that there exists an invertible map from every element, $\mathqbnd{a}\in \Lambda^\mathtt{a}[\text{fid}]$, to some operational procedure $a_\mathtt{A}\in \Upsilon[\mathtt{A}]^\mathtt{a}$ where $a_\mathtt{A}=(\mathtt{Z}, \mathbf{Q}_\mathtt{Z}, O_\mathtt{Z}, L_\mathsf{Z})$ gives the fiducial operation associated with the boundary condition $\mathqbnd{a}$.   Similarly, for the subscript case, we assume $\mathbf{a}\in \Lambda_\mathtt{a}[\text{fid}]$ can be mapped to $a\in \Upsilon[\mathtt{A}]_\mathtt{a}$.

The map from $\mathqbnd{a}\in\Lambda_\mathtt{a}[\text{fid}]$ to $a_\mathtt{A}\in\Upsilon[\mathtt{A}]_\mathtt{a}$ can be written
\begin{equation}
\begin{Compose}{0}{0}\setfourthfont{\mathqbnd}\setsecondfont{\mathtt}
\whitedot{A}{0,0}\csymbolthird[-30,0]{a_\mathtt{A}} \blackdotsq{B}{2.5,0}\csymbolfourth[20,0]{a} \joincc{A}{0}{B}{180} \csymbolalt{A}
\end{Compose}
~~~~ \Leftrightarrow~~~~
s^\mathqbnd{a}_{a_\mathtt{A}}
\end{equation}
with inverse
\begin{equation}
\begin{Compose}{0}{0}\setfourthfont{\mathqbnd}\setsecondfont{\mathtt}
\blackdotsq{B}{0,0}\csymbolfourth[-20,0]{a} \whitedot{A}{2.5,0}\csymbolthird[30,0]{a_\mathtt{A}}\joincc{B}{0}{A}{180}
\end{Compose}
~~~~~\Leftrightarrow~~~~~ s_\mathqbnd{a}^{a_\mathtt{A}}
\end{equation}
Here $s^\mathqbnd{a}_{a_\mathtt{A}}$ and $s_\mathqbnd{a}^{a_\mathtt{A}}$ are substitution operators (as defined in Appendix \ref{sec:thesubstitutionoperator}).

The map from $\mathqbnd{a}\in\Lambda^\mathtt{a}[\text{fid}]$ to $a_\mathtt{A}\in\Upsilon^\mathtt{a}$ can be written
\begin{equation}
\begin{Compose}{0}{0}\setfourthfont{\mathqbnd}\setsecondfont{\mathtt}
\blackdotsq{A}{0,0}\csymbolfourth[-20,0]{a} \whitedot{B}{2.5,0}\csymbolthird[30,0]{a_\mathtt{A}} \joincc{A}{0}{B}{180}
\end{Compose}
~~~~ \Leftrightarrow
\prescript{a_\mathtt{A}}{\mathqbnd{a}}s
\end{equation}
with inverse
\begin{equation}
\begin{Compose}{0}{0}\setfourthfont{\mathqbnd}\setsecondfont{\mathtt}
\whitedot{B}{0,0}\csymbolfourth[-20,0]{a} \blackdotsq{A}{2.5,0}\csymbolthird[30,0]{a_\mathtt{A}} \joincc{B}{0}{A}{180}
\end{Compose}
~~~~~\Leftrightarrow ~~~~~\prescript{\mathqbnd{a}}{a_\mathtt{A}}h
\end{equation}
where these are substitution operators as well.

\subsection{A calculation in this framework}

In this framework we start with a circuit or a fragment of a circuit built out of operations.  The notation for the calculation would look just like the case of Quantum Field Theory as discussed in Sec.\ \ref{sec:calculationsQFT} (although the interpretation is a little different).  We start with a bunch of operations wired together corresponding to some composite region in op-space.  For example
\begin{equation}
\begin{Compose}{0}{0}
\cobjectwhite{\theblob}{A}{2}{2}{-0.2,-5} \csymbol[0,-90]{A}
\cobjectwhite{\theflag}{B}{2}{1}{-4.75,0} \csymbol[-50,0]{B}
\cobjectwhite{\theblob}{E}{1.5}{1.5}{2.15,3} \csymbol[0,50]{E}
\cobjectwhite{\thetooth}{C}{1}{1}{1.4,-1.75} \csymbol{C}
\cobjectwhite{\theblob}{D}{0.95}{0.75}{-3.8,8} \csymbol{D}
\end{Compose}
\Leftrightarrow ~~
\begin{Compose}{0}{-1.1} \setdefaultfont{\mathsf}\setsecondfont{\mathtt}
\Ucircle{A}{1,-1} \Ucircle{B}{-5,5} \Ucircle{C}{-0.3,4} \Ucircle{D}{-3, 11} \Ucircle{E}{2,9}
\joincc[below left]{B}{-65}{A}{130} \csymbolalt{a}
\joincc[right]{A}{100}{C}{-90} \csymbolalt{b}
\joincc[below]{C}{170}{B}{-10} \csymbolalt{c}
\joincc[above left]{B}{25}{E}{-140} \csymbolalt{d}
\joincc[left]{B}{80}{D}{-100} \csymbolalt{e}
\joincc[above right]{D}{-15}{E}{170}
\joincc[below right]{C}{70}{E}{-95} \csymbolalt{g}
\joincc[right]{E}{-60}{A}{40}\csymbolalt{k}
\thispoint{nA}{-2,-2} \joincc[above left]{A}{-135}{nA}{45} \csymbolalt{f}
\thispoint{nB}{-8,5} \joincc[above]{B}{180}{nB}{0} \csymbolalt{h}
\thispoint{nD}{-1,13} \joincc[above left]{D}{45}{nD}{-135} \csymbolalt{i}
\thispoint{nE}{4,11} \joincc[above left]{nE}{-145}{E}{45} \csymbolalt{j}
\end{Compose}
\end{equation}
For ease of graphical representation, we are representing op-space in two dimensions on the left above.  Of course, a more realistic situation will have many more dimensions than this.  Then we map this to operators
\begin{equation}
\begin{Compose}{0}{0} \setdefaultfont{\hat}\setsecondfont{\mathtt}
\Ucircle{A}{1,-1} \Ucircle{B}{-5,5} \Ucircle{C}{-0.3,4} \Ucircle{D}{-3, 11} \Ucircle{E}{2,9}
\joincc[below left]{B}{-65}{A}{130} \csymbolalt{a}
\joincc[right]{A}{100}{C}{-90} \csymbolalt{b}
\joincc[below]{C}{170}{B}{-10} \csymbolalt{c}
\joincc[above left]{B}{25}{E}{-140} \csymbolalt{d}
\joincc[left]{B}{80}{D}{-100} \csymbolalt{e}
\joincc[above right]{D}{-15}{E}{170}
\joincc[below right]{C}{70}{E}{-95} \csymbolalt{g}
\joincc[right]{E}{-60}{A}{40}\csymbolalt{k}
\thispoint{nA}{-2,-2} \joincc[above left]{A}{-135}{nA}{45} \csymbolalt{f}
\thispoint{nB}{-8,5} \joincc[above]{B}{180}{nB}{0} \csymbolalt{h}
\thispoint{nD}{-1,13} \joincc[above left]{D}{45}{nD}{-135} \csymbolalt{i}
\thispoint{nE}{4,11} \joincc[above left]{nE}{-145}{E}{45} \csymbolalt{j}
\end{Compose}
\end{equation}
We can calculate different operators for different outcomes in each of these regions.  We can use the ideas of formalism locality as discussed Sec.\ \ref{sec:formalismlocalityPoss} (for PoAGeR), Sec.\ \ref{sec:formalismlocalityinPAGeR} (for PAGeR), and in Sec.\ \ref{sec:calculationsQFT} (for QFT).   When two of these operators are proportional then the relative probability is well conditioned and is given by the constant of proportionality.  If the two operators are not proportional then the probability is not well conditioned so - it may depend on what is happening outside the region of interest.   One example (which looks, from a notational point of view, the same as the example considered in Sec.\ \ref{sec:calculationsQFT} (for Quantum Field Theory) is to consider the two operations
\begin{equation}
\begin{Compose}{0}{-1.4} \setdefaultfont{\mathsf}\setsecondfont{\mathtt}
\Ucircle{A}{1,-1} \Ucircle{B}{-5,5} \Ucircle{C}{-0.3,4} \Ucircle{D}{-3, 11} \Ucircle{E}{2,9}
\joincc[below left]{B}{-65}{A}{130} \csymbolalt{a}
\joincc[right]{A}{100}{C}{-90} \csymbolalt{b}
\joincc[below]{C}{170}{B}{-10} \csymbolalt{c}
\joincc[above left]{B}{25}{E}{-140} \csymbolalt{d}
\joincc[left]{B}{80}{D}{-100} \csymbolalt{e}
\joincc[above right]{D}{-15}{E}{170}
\joincc[below right]{C}{70}{E}{-95} \csymbolalt{g}
\joincc[right]{E}{-60}{A}{40}\csymbolalt{k}
\thispoint{nA}{-2,-2} \joincc[above left]{A}{-135}{nA}{45} \csymbolalt{f}
\thispoint{nB}{-8,5} \joincc[above]{B}{180}{nB}{0} \csymbolalt{h}
\thispoint{nD}{-1,13} \joincc[above left]{D}{45}{nD}{-135} \csymbolalt{i}
\thispoint{nE}{4,11} \joincc[above left]{nE}{-145}{E}{45} \csymbolalt{j}
\end{Compose}
~~~ \text{and}~~~
\begin{Compose}{0}{-1.4} \setdefaultfont{\mathsf}\setsecondfont{\mathtt}
\Ucircle{A}{1,-1} \Ucircle{B}{-5,5} \ucircle{C}{-0.3,4}\csymbol{C'}\Ucircle{D}{-3, 11} \Ucircle{E}{2,9}
\joincc[below left]{B}{-65}{A}{130} \csymbolalt{a}
\joincc[right]{A}{100}{C}{-90} \csymbolalt{b}
\joincc[below]{C}{170}{B}{-10} \csymbolalt{c}
\joincc[above left]{B}{25}{E}{-140} \csymbolalt{d}
\joincc[left]{B}{80}{D}{-100} \csymbolalt{e}
\joincc[above right]{D}{-15}{E}{170}
\joincc[below right]{C}{70}{E}{-95} \csymbolalt{g}
\joincc[right]{E}{-60}{A}{40}\csymbolalt{k}
\thispoint{nA}{-2,-2} \joincc[above left]{A}{-135}{nA}{45} \csymbolalt{f}
\thispoint{nB}{-8,5} \joincc[above]{B}{180}{nB}{0} \csymbolalt{h}
\thispoint{nD}{-1,13} \joincc[above left]{D}{45}{nD}{-135} \csymbolalt{i}
\thispoint{nE}{4,11} \joincc[above left]{nE}{-145}{E}{45} \csymbolalt{j}
\end{Compose}
\end{equation}
where the operation in the centre is $C$ in one case and $C'$ in the other case (but all the other operations are the same).  The relative probability for these two scenarios is well conditioned if the operators
\begin{equation}
\begin{Compose}{0}{-1.4} \setdefaultfont{\hat}\setsecondfont{\mathtt}
\Ucircle{A}{1,-1} \Ucircle{B}{-5,5} \Ucircle{C}{-0.3,4} \Ucircle{D}{-3, 11} \Ucircle{E}{2,9}
\joincc[below left]{B}{-65}{A}{130} \csymbolalt{a}
\joincc[right]{A}{100}{C}{-90} \csymbolalt{b}
\joincc[below]{C}{170}{B}{-10} \csymbolalt{c}
\joincc[above left]{B}{25}{E}{-140} \csymbolalt{d}
\joincc[left]{B}{80}{D}{-100} \csymbolalt{e}
\joincc[above right]{D}{-15}{E}{170}
\joincc[below right]{C}{70}{E}{-95} \csymbolalt{g}
\joincc[right]{E}{-60}{A}{40}\csymbolalt{k}
\thispoint{nA}{-2,-2} \joincc[above left]{A}{-135}{nA}{45} \csymbolalt{f}
\thispoint{nB}{-8,5} \joincc[above]{B}{180}{nB}{0} \csymbolalt{h}
\thispoint{nD}{-1,13} \joincc[above left]{D}{45}{nD}{-135} \csymbolalt{i}
\thispoint{nE}{4,11} \joincc[above left]{nE}{-145}{E}{45} \csymbolalt{j}
\end{Compose}
~~~ \text{and}~~~
\begin{Compose}{0}{-1.4} \setdefaultfont{\hat}\setsecondfont{\mathtt}
\Ucircle{A}{1,-1} \Ucircle{B}{-5,5} \ucircle{C}{-0.3,4}\csymbol{C'}\Ucircle{D}{-3, 11} \Ucircle{E}{2,9}
\joincc[below left]{B}{-65}{A}{130} \csymbolalt{a}
\joincc[right]{A}{100}{C}{-90} \csymbolalt{b}
\joincc[below]{C}{170}{B}{-10} \csymbolalt{c}
\joincc[above left]{B}{25}{E}{-140} \csymbolalt{d}
\joincc[left]{B}{80}{D}{-100} \csymbolalt{e}
\joincc[above right]{D}{-15}{E}{170}
\joincc[below right]{C}{70}{E}{-95} \csymbolalt{g}
\joincc[right]{E}{-60}{A}{40}\csymbolalt{k}
\thispoint{nA}{-2,-2} \joincc[above left]{A}{-135}{nA}{45} \csymbolalt{f}
\thispoint{nB}{-8,5} \joincc[above]{B}{180}{nB}{0} \csymbolalt{h}
\thispoint{nD}{-1,13} \joincc[above left]{D}{45}{nD}{-135} \csymbolalt{i}
\thispoint{nE}{4,11} \joincc[above left]{nE}{-145}{E}{45} \csymbolalt{j}
\end{Compose}
\end{equation}
are proportional and then the relative probability is equal to the constant of proportionality.

\subsection{What are the physicality conditions?}\label{sec:whatarethephysicalityconditions}

Although the calculations look similar to those in the operator tensor formulation of Quantum Field Theory, something different is happening in this case.  The boundaries between regions in the QFT case correspond to well defined causal situations (they are either time-like future pointing, time-like past pointing, or space like pointing). However, in this case the boundaries do not necessarily correspond to a well defined causal situation. In fact they may correspond to indefinite causal structure in which there is no-matter-of-the-fact as to what the causal structure is (something like a quantum superposition of different causal structures).  This is because the classical description of the boundary (as given by the typing surface) need not fix the metric and time direction field along the boundary.

\index{physicality conditions!in QG}

In the operator tensor formulation of Quantum Theory we were able to give a clear characterization of the operators. In the operator tensor formulation of Quantum Field Theory such a characterization should exist also (though we only partially provided it).  The characterization of operators in these cases follows from the Pavia causality condition (that the deterministic effect is unique \cite{chiribella2010informational}.  This condition is possible because we have fixed background causal structure in these cases.  We are seeking a condition that still encodes some kind of causality but does not depend on having fixed (or even definite) causal structure.  The best way to illustrate what we are seeking is by analogy with the transition from Special Relativity to General Relativity
\begin{equation}
\begin{array}{lcl}
g_{\bar{\mu}\bar{\nu}}=\eta_{\bar{\mu}\bar{\nu}} ~~\text{in SR} & \longrightarrow & G_{\mu\nu} = 8\pi T_{\mu\nu}  \\
\text{QFT physicality cond.} & \longrightarrow & \text{missing QG cond.}
\end{array}
\end{equation}
The geometrical equation of Special Relativity (that $g_{\bar{\mu}\bar{\nu}}=\eta_{\bar{\mu}\bar{\nu}}$ where $\bar{\mu}$ indicates the components of inertial coordinates) is analogous to the QFT physicality condition (discussed in Sec.\ \ref{sec:thecontinuouslimit} - see \eqref{QFTphysicalityCond}).  The QFT physicality condition was not actually given but it is the QFT generalization of the physicality condition discussed in Sec.\ \ref{sec:physicality} and Sec.\ \ref{sec:physicalityconditionsQFToldnotation}.   A theory of Quantum Gravity is, in this framework, specified by providing dynamical conditions that specify the space of allowed operators. The principle goal then is to find this missing condition (an easier goal would be to, first, find the QFT physicality condition).

It is worth emphasizing that a we are suggesting taking the same attitude towards dynamics here as in operational Quantum Theory - namely that it is sufficient to specify the constraints on possible operators to have captured the essential aspects of the theory.  In operational Quantum Theory any operator that is physical (or completely positive and trace non-increasing in usual language) can be realized by some Hamiltonian acting on the given systems and appropriately chosen ancilla systems).

We can pose the question of what these missing equations are in a more mathematical way.  The question is what are the most general constraints we can write down on operators such that closed circuits have values between 0 and 1 (so they can be interpreted as probabilities).  The constraints we need for Quantum Gravity may be just these or there may be further constraints coming from more physical considerations (such as causality).  This is analogous to the fact that we need to express the equations of General Relativity in terms of tensors, but the actual equations for General Relativity are more specialized - in particular, the equation that depends on second derivatives of the metric is the Einstein field equation.  

One issue that we have not addressed so far is whether and how the Planck length, 
\begin{equation}
\ell_P=\sqrt{\frac{\hbar G}{c^3}} 
\end{equation}
enters into this formalism of Quantum Gravity.  One possibility is that a length scale crops up naturally in finding physicality conditions. In other words, it might be impossible, in the quantum case, to find physicality conditions that do not involve a length scale.  

We motivated this version of QuAGeR by finding a map from a set of fiducial operators, $\Lambda_\mathtt{a}[\text{fid}]$, built from some set of basis states, $\Lambda_\mathtt{a}[\text{basis}]$, whose elements are of the form in \eqref{qboundaryconditions} whose definition depends on some choice as, for example, in \eqref{oneproposalforQ}.  Once we have this map we obtain a sets of fiducial operations such as $\Upsilon[\mathtt{A}]_\mathtt{a}$.  The elements of these $\Upsilon$ sets are described in entirely operational terms.  According to this procedure, whatever choice we make for basis (such as \eqref{oneproposalforQ}), we obtain a operationally described $\Upsilon$ sets.  If we are able to find physicality conditions directly on operations then this should suggest directly what sets will serve for the fiducial $\Upsilon$ sets. In such a scenario, we would not need to concern ourselves with finding the right choice of boundary conditions corresponding to a basis (as in \eqref{oneproposalforQ}).

\section{Ontological approach to Quantum Gravity}

\subsection{Ontology}

Classical General Relativity has a clear ontology - beables are quantities that are invariant under diffeomorphisms.  This is, consequently, also true in PoAGeR and PAGeR (the diffeomorphism group being enlarged to allow the gauge symmetry on the time direction field introduced because we have agency in the picture).  However, the abstract approach to Quantun Gravity outlined above in Sec.\ \ref{sec:anabstractapproach} is quite different. There is no clear ontology beyond the operational level.  This is, of course, also true in Quantum Theory (as outlined in Part \ref{part:operatortensorQT}) as well.  Our highest priority is to find a theory of Quantum Gravity. However, it would be great if we also found a new ontology along the way.  In this section we discuss the possibility that the ontology of Quantum Gravity might be closer to that of General Relativity.  This is, however, very speculative. We are not able to present any really compelling reasons supporting this idea - just a few suggestive hints.\index{Quantum Gravity!ontological approach}

\subsection{Maybe PAGeR is Quantum Gravity}

The first possibility is that General Relativity, when construed in an operational and probabilistic way, actually has Quantum Theory as a limit.  Were this the case then we could assert that General Relativity (in the guise of PAGeR) is actually the sought after theory of Quantum Gravity.

At first sight it would appear that this is immediately ruled out.  General Relativity is a local field theory and local field theories cannot violate Bell inequalities.  A theory of Quantum Gravity must have at least that part of Quantum Theory that has been verified in the laboratory as a limit.  Bell inequalities have actually been violated. Consequently, we might argue, we have to go beyond classical General Relativity and, indeed, local field theories in general (see Kent's discussion in \cite{kent2009proposed}).

However, when we say the field equations of General Relativity are local, we mean this with respect to the manifold.  But we know that we cannot localize beables on the manifold.  The space we directly observe is op-space (spanned by scalar fields according to the assertion in Sec.\ \ref{sec:observables}).  So the real question is whether we have a theory that is local in op-space. Certainly we see the curious form of nonseparability as discussed in Sec.\ \ref{sec:curiousnonseprability}.  The solution associated with a composite region is not fixed by the solutions associated with the component regions.  At an intuitive level this is a bit like Quantum Theory. However, this kind of nonseparability does not imply a violation of any Bell inequalities.

Nevertheless, we are clearly in a new ballpark.  While PAGeR has a clear ontology it does not obviously fit inside the standard hidden variable frameworks normally considered in Quantum Foundations which associate a hidden variable at each moment in time.  This is effectively the case in the ontological model of Harrigan and Spekkens \cite{harrigan2010einstein} for example.  Is it conceivable that PAGeR could violate Bell inequalities, give rise to interference, contextuality and all the other interesting effects of Quantum Theory?

One possibility is that, although we have local field equations on the manifold, when viewed from the vantage point of op-space there are some effective nonlocal effects that are masked by having a statistical mixture.  The idea of statistical masking appears in the de Broglie Bohm approach to Quantum Theory (see, particularly the work of Valentini \cite{valentini1991signal}).  A major challenge of such an approach would be to achieve such masking in a natural way without fine tuning \cite{wood2015lesson}.   The de Broglie Bohm model is, however, nonlocal at the level of the partial differential equations.  The idea that we might have locality at the fundamental level but violations at the effective level has been discussed in a different guise by Markoupoulou and Smolin \cite{markopoulou2004quantum}.  Another related idea, discussed by Edwards \cite{edwards2014non} is that the metric, at the fundamental level, allows an particle to be close to each of two distant objects.

\subsection{More general ontological models}\label{sec:moregeneralontologicalmodels}

We could, instead, consider more general ontological models that reduce to PAGeR in one limit and to Quantum Theory in another.  Such a model could have solutions
\begin{equation}\label{genontmodelsoln}
\Psi_\mathtt{A} = \left\{ (\mathbf{S}, \lambda): \forall \mathbf{S}\in \Gamma_\mathtt{A} \right\}
\end{equation}
where $\lambda$ are hidden variables associated with the point $\mathbf{S}$.  Indeed, in Sec.\ \ref{sec:turningasolutioninsideout} (and Sec.\ \ref{sec:turningasolutionwithagencyinsideout} for the case with agency) we showed that solutions can be represented just like this in General Relativity.   In that case, $\lambda$ had a particular interpretation in terms of fields on a manifold and points $\mathbf{S}$, correspond to nonlocal regions on the manifold. However, we can also think of a solution such as \eqref{genontmodelsoln} without reference to a manifold. In the case of a general ontological model we can allow $\lambda$ to be any kind of hidden variables.  While this way of representing the solution is local in op-space (as we attach hidden variables to each point in $\Gamma$) these hidden variables may contain very nonlocal information about any underlying fundamental space (such as the manifold).  In particular, the causal structure on points $\mathbf{S}\in\Gamma$ can be dynamically determined. For one solution a certain point $\mathsf{S}$ might be earlier than another point, $\mathbf{S}'$, while on another solution the temporal order may be reversed. We could have a mixture of these two situations if there is no way of determining from the observable information which pertains. It is even possible that the parts of the underlying fundamental space associated with $\mathsf{S}$ and $\mathsf{S}'$ are intertwined such that there is no clear sense in which one is before or after the other.   In such models the causal structure could play a very non-trivial role quite distinct from the role it plays in standard hidden variable models (wherein the causal structure is taken to be fixed).

\subsection{Discussion}

What is really required to make this kind of model interesting is a clear demonstration that some key features of Quantum Theory could be reproduced.  The most compelling such effect would be a violation of Bell inequalities.  The above discussion makes it clear that it is at least worth investigating the possibility that we might be able to get Quantum Theory from General Relativity or some other similar ontological model.

\section{Principles, axioms and postulates}

\subsection{Introduction}

There has been significant work in recent years on deriving Quantum Theory from sets of postulates in an operational setting (see discussion in Sec.\ \ref{sec:thisresearchprogramandotherwork}).  In the case of Quantum Theory these are \emph{reconstructions} since we already have Quantum Theory.  However, it is possible that we could use this approach to \emph{construct} a theory of Quantum Gravity from scratch.  The general idea is that we take some framework for physical theories, write down some postulates, and obtain the theory in this way.    \index{Quantum Gravity!principled approach}

\subsection{Postulates for Quantum Theory}\label{sec:postulatesforQT}

\index{postulates!for Quantum Theory}

There are various sets of postulates from which we can obtain Quantum Theory.   In 2001 in \cite{hardy2001quantum} the present author provided the following postulates
\begin{description}
\item[Information.]  Systems having, or constrained to have a given information carrying capacity have the same properties.
\item[Information locality.] The information carrying capacity (measured in bits) of a composite system is the sum of the information carrying capacities of the components (or $N_\mathsf{ab}=N_\mathsf{a}N_\mathsf{b}$).
\item[Tomographic locality.] We can determine the state of any composite system by making local measurements on its components (or $K_\mathsf{ab}=K_\mathsf{a}K_\mathsf{b}$)
\item[Continuity.]  There exists a continuous reversible transformation between any pair of pure states.
\item[Simplicity.] States are described by the smallest number of probabilities consistent with the other postulates.
\end{description}
Here $N_\mathsf{a}$ is the maximum number of states that can be distinguished in a single shot for a system of type $\mathsf{a}$ and $K_\mathsf{a}$ is the number of probabilities required to specify a general state for a system of type $\mathsf{a}$.

The above set of postulates has an unnatural \lq\lq simplicity postulate".  Various people showed how to avoid such a postulate in subsequent work \cite{dakic2009quantum, masanes2010derivation, chiribella2010probabilistic, chiribella2010informational, hardy2011reformulating}.  The present author provided another set of 2011 \cite{hardy2011reformulating} that do not have a simplicity postulate
\begin{description}
\item[Logical sharpness.] There is a one-to-one map between pure states and maximal effects such that we get probability one.
\item[Information locality.] The information carrying capacity (measured in bits) of a composite system is the sum of the information carrying capacities of the components (or $N_\mathsf{ab}=N_\mathsf{a}N_\mathsf{b}$).
\item[Tomographic locality.] We can determine the state of any composite system by making local measurements on its components (or $K_\mathsf{ab}=K_\mathsf{a}K_\mathsf{b}$)
\item[Permutability.] There exists a reversible transformation on any system effecting any permutation of any given set of maximally distinguishable set of states for that system.
\item[Sturdiness.]  Filters are non-flattening.
\end{description}
A filter is a transformation which blocks some states in some given maximal set of distinguishable states and passes unchanged all states that are not attenuated at all.  A non-flat set of states is a set of states that span the space of states that are passed by some filter.  The last postulate means that when any non-flat set of states is sent into a filter, a non-flat set of states comes out.  This set of postulates give Classical Probability Theory and Quantum Theory as the only two possibilities.  To single out Quantum Theory we need only add any property that is inconsistent with Classical Probability theory and consistent with Quantum Theory.  For example, we can replace the \emph{Permutability} axiom with the following
\begin{description}
\item[Compound permutability.] There exists a compound reversible transformation on any system effecting any permutation of any given set of maximally distinguishable set of states for that system.
\end{description}
Here a compound reversible transformation is one that can be built out of two sequential transformations where neither is equal to the identity.

These sets of postulates apply to a given framework. For the second set the actual framework is spelled out in great detail in \cite{hardy2011reformulating} (which has much in common with the framework developed by Chiribella, D'Ariano, and Perinotti \cite{chiribella2010probabilistic, chiribella2010informational}).  In particular it applies to circuits that are directed acyclic graphs (so there are no closed loops).  If we have indefinite causal structure (as we expect in Quantum Gravity) then we would not want to restrict our attention to such graphs.  Additionally, many of the objects used in these systems of postulates require, for their definition, the input/output structure.  For example, a state is associated with an operation having outputs only (a preparation).  An effect is associated with an operation having inputs only.  These play a different role in the system of postulates.  If we have indefinite causal structure then we do not necessarily have inputs and outputs so we cannot use concepts in our postulates that rely on these concepts.

It would be interesting to attempt to write down postulates for operator tensor Quantum Field Theory as presented in Sec.\ \ref{sec:operatortensorquantumfieldtheory}. Indeed, such postulates would be more relevant to the task of finding a theory of Quantum Gravity.  Some of the postulates we have for Quantum Theory would go through as they are. In particular, we can understand the tomographic locality postulate as saying that it is always possible to fully characterize an operation (now pertaining to an arbitrary region of space-time) by local fiducials around the boundary.  The principle of general compositionality (see Sec.\ \ref{sec:introductionatbeginning}) is clearly satisfied for the operator tensor formulation of Quantum Field Theory and may be a useful principle for a reconstruction attempt.

\subsection{Postulates for General Relativity?}\label{sec:postulatesforGR}

Einstein's route to General Relativity was driven by the equivalence principle and also by the need to reproduce Special Relativity and Newtonian Gravitation as special cases. This is discussed in great detail in Appendix \ref{appendix:standarformulationofGR}.  From the Equivalence principle he motivated the idea that the equations of Special Relativity should apply in a local inertial reference frame.  He also motivated the idea that the equations should be written in general coordinates and, from this, the principle of general covariance - that the equations should be written in such a way that they take the same form in any coordinate system.  To achieve the latter he invoked the machinery of tensor fields on manifolds.  Also, from the equivalence principle, he equates the metric with the gravitational field and seeks field equations for this new gravitational field that are similar to Poison's equation (as this is the appropriate form of Netwon's law of gravitation for this purpose) and, further, actually reduces to the Poison equation in an appropriate limit.  He was further guided in this by the conservation of energy-momentum in Special Relativity.

There appears to be a certain inevitability to Einstein's derivation of General Relativity.  Alternatives to General Relativity do exist (such as Brans-Dicke \cite{brans1961mach} and  Lovelock gravity \cite{lovelock1971einstein}).  However, these alternatives accept as their starting point the idea that we have a manifold with fields on it.  A priori this is quite a reasonable starting point. However, the principle of general covariance leads to the hole argument (as detailed in Appendix \ref{Sec:principleequivcovar}) and from this we deduce that beables cannot be localized on any subset of the manifold.  This suggests that, maybe, we should not be starting with a framework in which there are manifolds with fields defined on them.

Further evidence for this point of view comes from the argument in Appendix \ref{sec:firstderivative} in which it is seen that we have exactly the right number of parameters (40 in four dimensional space-time) to set the first derivatives of the metric to zero.  Hence, by the time we have written down the metric as a tensor on a manifold, we have implicitly assumed enough structure that the existence of a locally flat reference frame is guaranteed at every point on the manifold. It would be more compelling to take, as a postulate, that there must exist a locally flat reference frame, and from that derive some properties.  For this strategy to make sense we would need to start by assuming less.

Here we make two suggestions.  We can pursue an operational approach or an operational approach supplemented by hidden variables.

For the operational approach we need an operational space of some kind to start. In principle this could be anything. However, to give ourselves a little more traction we assume that it is defined by a number of variables, $\mathbf{S}=(S_1, S_2, \dots, S_K)$ where each of the variables are real valued.  We can consider regions, $\mathtt{A}$, $\mathtt{B}$, \dots of this space. In any such region we can form observables, $O_\mathtt{A}$, which are sets of sets, $\Gamma_\mathtt{A}\in\mathtt{A}$.  We can very quickly arrive at the operational possibilistic framework detailed in Sec.\ \ref{sec:operationalpossibilisticformulation} or the operational probabilistic framework detailed in Sec.\ \ref{sec:operationalprobabilisticformulation} in which the fiducials are labeled by operational specifications.  This does not give us General Relativity of course. However, it is a suitable setting for General Relativity.  To actually get General Relativity we would need to impose some sort of postulates that give the correct hopping metric and the correct space of states.

For the operational approach supplemented by hidden variables framework was already described (in Sec.\ \ref{sec:moregeneralontologicalmodels}).  The main idea is that we introduce hidden variables, $\lambda$, at each point $\mathbf{S}$.  We can also seek postulates that refer to this hidden variable structure.

We will not actually provide a set of postulates for General Relativity but simply look at a few ideas.

\index{postulates!for General Relativity?}

First, we have causality. To define what we mean by causality in an operational framework we need to introduce a notion of agency and a notion of time direction.  Agent choices can be described by some variables $\mathbf{Q}(\mathbf{S})$.  It is less clear how to define a notion of time direction in this general operational framework without introducing an underling manifold. This remains a challenge. An appropriate definition of causality would have to formalize the idea assumption that different choices can only influence what happens in the future.

Another property of General Relativity is that the local dimension of $\Gamma$ at some point $\mathbf{S}$ is less than or equal to some constant (the dimension of the manifold).  If we include the hidden variables then there we may be able to write an equality
\begin{equation}\label{dimensionpostulate}
\text{dim}_\mathbf{S}(\Gamma) + \text{dim}_\mathsf{S}(\lambda) = D
\end{equation}
where $\text{dim}_\mathsf{S}(\lambda)$ is some function on $\lambda$. If we define $\lambda$ as in Sec.\ \ref{sec:turningasolutioninsideout} (or Sec.\ \ref{sec:turningasolutionwithagencyinsideout} in the case with agency) then this is the dimension of $\mathscr{M}_\mathsf{S}$ (i.e. the dimension of the part of the manifold that gets mapped to $\mathsf{S}$.  We can think of \eqref{dimensionpostulate} as the basis for a postulate.

General Relativity satisfies the assumption of decomposition locality for operations as explained in Sec.\ \ref{sec:decompositionpossibilisticoperations} (and also in Sec.\ \ref{sec:decompositionlocalityforloadedoperations} for loaded operations).  This is equivalent to the assumption of tomographic locality (used in reconstructions of Quantum Theory as mentioned in Sec.\ \ref{sec:postulatesforQT}) as long as the hopping metric is invertible \cite{hardy2011reformulating}.  Hence, it is reasonable to elevate tomographic locality or decomposition locality to the level of a postulate for General Relativity.  There is a deep connection between this property and the even deeper principle of general compositionality (see the discussion in \cite{hardy2013theory} pertaining to the composition principle).  The principle of general compositionality might be a good starting point for such a reconstruction of General Relativity.

The field equations of General Relativity act as a constraint on generalized states in the operational frameworks presented earlier.   These comprise matter field equations and Einstein's field equation.  We propose the following attitude. We do not want to attempt to derive the matter field equations - they are contingent on rather different considerations from the ones we have been thinking about here.  They are analogous to choosing some particular Hamiltonian in Quantum Theory.  In the operational formulation of Quantum Theory we consider the integrated effect of Hamiltonians for some operation.  In fact, further, we allow non-unitary evolution (corresponding to completely positive maps).  We are interested in characterizing the full set of evolutions.  The constraint on the space of such evolutions comes from the physicality condition which is intimately related to causality.  Thus, we suggest that the role of the Einstein field equations is analogous to the role of causality in Quantum Theory.   If this intuition is correct, then by characterizing the space of generalized states in our formulation of General Relativity using causality, we are effectively characterizing the Einstein's field equations.

\subsection{Postulates for Quantum Gravity?}

There are three frameworks we might consider within which we could impose postulates. These are the probabilistic operational framework (as discussed in Sec.\ \ref{sec:postulatesforGR}), the operational/ontological framework (as discussed in Sec.\ \ref{sec:moregeneralontologicalmodels}) and the operator tensor framework (as discussed in Sec.\ \ref{sec:OperationalQuantumGravity}).  The latter framework takes more of the structure of standard Quantum Theory as given.

\index{postulates!for Quantum Gravity?}

In one of these frameworks we can attempt to impose postulates to narrow down to a theory of Quantum Gravity.   These postulates might come from General Relativity, Quantum Theory, or be novel to the theory of Quantum Gravity.

Tomographic locality (or the related property of decomposition locality) is a very natural property to assume as it appears to be true in both Quantum Theory and General Relativity.  Causality is also a property that both theories share. In this case we have to provide a definition of causality that works for Quantum Theory, General Relativity, and most particularly, for the nascent theory of Quantum Gravity.  This is a challenge as a theory of Quantum Gravity will, most likely, have indefinite causal structure.  The deeper principle of general compositionality might be a good starting point here.

The constraint that probabilities are between 0 and 1 will impose some mathematical structure on the sets of allowed generalized states (or operator tensors).  If we take, as given, the operator tensor framework then we need to find constraints on operators tensors such that, when combined together, we get probabilities between 0 and 1.  In the context of Quantum Theory we adopted a certain definition of physicality.  This included assuming that we can prepare arbitrary rank one projectors for preparations and for effects (associated with measurement outcomes) and, implicitly, that the identity operator is associated with the deterministic effect (which has all outcomes associated with it).  This assumption imposes certain background causal structure (we need to be able to talk about preparations and measurements and we need to have some sort of definite causal structure to do this).  If we relax the definition of physicality then we may be able to obtain more general constraints on operators.

Indefinite causal structure would put strain on some of the notions used in the axioms for Quantum Theory.  For example, assuming that there is a reversible transformation suggests that time is treated on a different footing from space (as this transformation is reversible in the time direction). The idea of perfectly distinguishable states also suggests a time direction.  We can use such states to signal reliably from the past to the future.  The integer $N_\mathsf{a}$ relies on the notion of perfectly distinguishable states.

\newpage

\part{Discussions}\label{part:discussions}

\section{Other Tools}

\subsection{Free operations}\label{sec:freeoperations}

To do physics, it is useful to have objects that we can put in different places.  However, encapsulated propositions pertain to a particular region of op-space.  Thus, it is useful to invent a notion of \emph{free operation} \index{free operations}(by analogy with free vectors) along with associated notions for the generalized states.   Here we provide some preliminary steps in this direction.

Recall that an operation (possibly loaded) has the following form:
\begin{equation}\label{operationdefinitionothertools}
\mathsf{A}= \left(\text{strat}(\mathsf{A}), \text{outcome}(\mathsf{A}), \text{load}(\mathsf{A}), \text{reg}(\mathsf{A}), \text{type}(\mathsf{A})\right)
\end{equation}
The load element is only necessary if we are considering a loaded operation.

Now consider a number of basic operations $\mathsf{A}$, $\mathsf{B}$, \dots.  We will assume that we can build all operations of interest by moving these basic operations around.  We can imagine moving the basic operation, $\mathsf{A}$, by applying some transformation, $\pmb{\alpha}\in F_\mathsf{A}$, acting on op-space and also on the space of strategies and the space of loadings.  We write
\begin{equation}
\pmb{\alpha}^\star \mathsf{A}
\end{equation}
for the new operation (this is in imitation of the notation for diffeomorphisms). Each of the elements of $\mathsf{A}$ shown in \eqref{operationdefinitionothertools} will be transformed under $\pmb{\alpha}$. We will use the $\star$ notation for all such transformations (thus we will write $\pmb{\alpha}^\star\mathbf{S}$, etc).  We assume that this transformation is smooth and invertible like a diffeomorphism (however, it is not clear that the space $\pmb{\alpha}$ acts on is a manifold so we will stop short of assuming that it is actually a diffeomorphism).  The set, $F_\mathsf{A}$, of possible transformations can be chosen to suit our needs. It might include only translations, translations and rotations, or be equal to the full set of possible such invertible smooth transformations.  We define
\begin{equation}
\mathsf{A}[\pmb{\alpha}] = \pmb{\alpha}^\star \mathsf{A}, ~~~~~~ \mathsf{B}[\pmb{\beta}] = \pmb{\alpha}^\star \mathsf{B}, \dots
\end{equation}
In this way we can move the basic operations around.

Each basic operation has some typing surfaces.  Hence, these typing surfaces can also be moved around.  Imagine that we have three basic operations $\mathsf{A}^\mathtt{a}_\mathtt{bc}$, $\mathsf{B}^\mathtt{cf}$, and $\mathsf{C}_\mathtt{ef}$.  It might turn out that there exists $\alpha\in F_\mathsf{A}$ and $\beta\in F_\mathsf{C}$ such that $\alpha^\star \mathtt{a}= \gamma^\star\mathtt{e}$.  Then we will say that $\mathtt{a}$ and $\mathtt{e}$ belong to the same equivalence class, which we will denote by $\mathsf{a}$ (note this is a different font, we use $\backslash$mathtt for typing surfaces and $\backslash$mathsf for types).  If none of the other typing surfaces on these three basic operations can be mapped into one another then the set of transformations that generate this equivalence class is $F_\mathsf{a}= F_\mathsf{A}\cup F_\mathsf{C}$.

In general, we put all the typing surfaces that can be generated by transformations on the basic operations into appropriate equivalence classes and give them names $\mathsf{a}$, $\mathsf{b}$, \dots. We will call these equivalence classes \emph{types}. They are analogous to the system types used in the operator tensor formulation of Quantum Theory.  Let us assume, in our example, that we have only three such types, $\mathsf{a}$, $\mathsf{b}$, and $\mathsf{c}$.  We can put these as subscripts and superscripts on the basic operations replacing the old typing surfaces so we have, say, $\mathsf{A}^\mathsf{a_1}_\mathsf{b2c_3}$, $\mathsf{B}^\mathsf{c_1b_2}$, and $\mathsf{C}_\mathsf{a_1b_2}$.  The integers are necessary as we can now have many repetitions of the same type when we wire together some operations.

We can move these basic operations around and join them to each other. For example,
\begin{equation}
\mathsf{A}_\mathsf{b_1c_2}^\mathsf{a_4}[\pmb{\alpha}] \mathsf{B}^\mathsf{c_2b_3}[\pmb{\beta}] \mathsf{C}_\mathsf{a_4b_5}[\pmb{\gamma}] \mathsf{B}^\mathsf{c_6b_5}[\pmb{\beta}']
~~~~\Leftrightarrow~~~~
\begin{Compose}{0}{0} \setdefaultfont{\mathsf}\setsecondfont{\mathsf}
\ccircle{A}{1.65}{0,0} \csymbol{\mathsf{A}\mathnormal{[\pmb{\alpha}]}}  \thispoint{Al}{-4,0}
\ccircle{Blow}{1.65}{6,-6} \csymbol{\mathsf{B}\mathnormal{[\pmb{\beta}]}}  \thispoint{Blowl}{3, -9}
\ccircle{C}{1.65}{5,5} \csymbol{\mathsf{C}\mathnormal{[\pmb{\gamma}]}}
\ccircle{Bup}{1.65}{10,10} \csymbol{\mathsf{B}\mathnormal{[\pmb{\beta}']}}  \thispoint{Bupl}{7,13}
\joincc[above right]{Blow}{140}{A}{-50} \csymbolalt{c}
\joincc[above left]{A}{50}{C}{-140} \csymbolalt{a}
\joincc[below right]{Bup}{-140}{C}{50}\csymbolalt{b}
\joincc[above right]{Bup}{140}{Bupl}{-50} \csymbolalt{c}
\joincc[below right]{Blow}{-140}{Blowl}{50} \csymbolalt{b}
\joincc[above]{Al}{0}{A}{-180} \csymbolalt{b}
\end{Compose}
\end{equation}
Note that we can have multiple copies of each operation (though with different transformations applied) and multiple copies of each type.  For this to actually make sense it is necessary that the typing surfaces actually match.  We can assume that this calculation is done in advance so we take it as given.

We can associate a generalized state with each operation. For example
\begin{equation}
\mathsf{A}_\mathsf{b_1c_2}^\mathsf{a_4}[\pmb{\alpha}] \rightarrow    A_{b_1c_2}^{a_4}[\pmb{\alpha}]
\end{equation}
The generalized states are a function of the transformation, $\pmb{\alpha}$.  If we choose our basis operations well then we may be able to find cases where the generalized state depends on $\pmb{\alpha}$ in a simple way.  Then we can turn the operational description into a calculation for the generalized state for the composite operation. For the above example we obtain generalized state
\begin{equation}
{A}_{b_1c_2}^{a_4}[\pmb{\alpha}] {B}^{c_2b_3}[\pmb{\beta}] {C}_{a_4b_5}[\pmb{\gamma}] {B}^{c_6b_5}[\pmb{\beta}']
~~~~\Leftrightarrow~~~~
\begin{Compose}{0}{0} \setdefaultfont{\mathnormal}\setsecondfont{\mathnormal}
\ccircle{A}{1.65}{0,0} \csymbol{\mathnormal{A}\mathnormal{[\pmb{\alpha}]}}  \thispoint{Al}{-4,0}
\ccircle{Blow}{1.65}{6,-6} \csymbol{\mathnormal{B}\mathnormal{[\pmb{\beta}]}}  \thispoint{Blowl}{3, -9}
\ccircle{C}{1.65}{5,5} \csymbol{\mathnormal{C}\mathnormal{[\pmb{\gamma}]}}
\ccircle{Bup}{1.65}{10,10} \csymbol{\mathnormal{B}\mathnormal{[\pmb{\beta}']}}  \thispoint{Bupl}{7,13}
\joincc[above right]{Blow}{140}{A}{-50} \csymbolalt{c}
\joincc[above left]{A}{50}{C}{-140} \csymbolalt{a}
\joincc[below right]{Bup}{-140}{C}{50}\csymbolalt{b}
\joincc[above right]{Bup}{140}{Bupl}{-50} \csymbolalt{c}
\joincc[below right]{Blow}{-140}{Blowl}{50} \csymbolalt{b}
\joincc[above]{Al}{0}{A}{-180} \csymbolalt{b}
\end{Compose}
\end{equation}
This will work for the possibilistic or probabilistic formulations. For the operator tensor case, we have operator tensors in place of generalized states
\begin{equation}
\mathsf{A}_\mathsf{b_1c_2}^\mathsf{a_4}[\pmb{\alpha}] \rightarrow    \hat{A}_\mathsf{b_1c_2}^\mathsf{a_4}[\pmb{\alpha}]
\end{equation}
and we combine them as before.

\subsection{Infinitesimal approach}

The general philosophy in this paper is that we should take a  compositional approach to physics.  We can make predictions concerning bigger regions of op-space by joining together objects corresponding to smaller regions.  This is a useful approach if the smaller regions are simpler.  This suggests we should take the limit of considering infinitesimal elements of op-space.  Here we will outline first steps in this direction.
\index{infinitesimal approach}

Let
\begin{equation}
\{\pmb{\delta}_k: \forall k=1 ~\text{to}~K\}
\end{equation}
be set of $K$ linearly independent small vectors (that we will let tend to zero) in op-space.  Let
\begin{equation}
\mathtt{A}_\text{discrete}= \{ \sum_k m_k \pmb{\delta}_k: \forall~ \text{integers} ~ m_k ~ \text{such that} ~ \sum_k m_k \pmb{\delta}_k \in \mathtt{A}\}
\end{equation}
These are the points on the grid generated by the $\pmb{\delta}_k$ that lie inside $\mathtt{A}$.  We will take these points to be the centre of our elements. Each element is a $K$-dimensional parallelepiped  centered at at some $\mathbf{S}\in\mathtt{A}_\text{discrete}$.   Midway between any pair of such points separated by $\pmb{\delta}_k$ (for some $k$) is a face which constitutes a typing surface.   We can refer to this typing surface as
\begin{equation}
\mathtt{a}_{\mathbf{S}^\pm_k}  ~~~~ \text{where}~~ \mathbf{S}^\pm_k = \mathbf{S} \pm \frac{1}{2} \pmb{\delta}_k
\end{equation}
where the subscript indicates the position of the centre of the typing surface in op-space.  The direction associated with the typing surface is the $\pmb{\delta}_k$ direction.

The element of op-space centred at $\mathbf{S}$ will be written as $\delta\mathtt{A}(\mathbf{S})$.  Then
\begin{equation}
\mathtt{A} = \bigcup_{\mathbf{S}\in \mathtt{A}_\text{discrete} } \delta\mathtt{A}(\mathbf{S})
\end{equation}
We can write an operation associated with $\delta\mathtt{A}(\mathbf{S})$ as
\begin{equation}\label{elementoperator}
\delta\mathsf{A}^{\mathtt{a}_{\mathbf{S}^+_1} \dots \mathtt{a}_{\mathbf{S}^+_K} }_{\mathtt{a}_{\mathbf{S}^-_1} \dots \mathtt{a}_{\mathbf{S}^-_K} } [\mathbf{S}]
\end{equation}
Now we can write the corresponding operation associated with region $\mathtt{A}$ as
\begin{equation}
\mathsf{A}^\mathtt{b}_\mathtt{c} = \lim_{\pmb{\delta}_k \rightarrow 0} \prod_{\mathbf{S}\in \mathtt{A}_\text{discrete} }
\delta\mathsf{A}^{\mathtt{a}_{\mathbf{S}^+_1} \dots \mathtt{a}_{\mathbf{S}^+_K} }_{\mathtt{a}_{\mathbf{S}^-_1} \dots \mathtt{a}_{\mathbf{S}^-_K} } [\mathbf{S}]
\end{equation}
where it is understood that we are taking the limit as the length of $\pmb{\delta}_k$ tends to zero for all $k$.
All the interior typing surfaces get matched leaving only the exterior typing surfaces unmatched. We denote the exterior typing surface as $(\mathtt{b},\mathtt{c})$.

We can associate a generalized state (possibilistic or probabilistic) or an operator tensor with the operator in \eqref{elementoperator} which we denote as
\begin{equation}
\delta{A}^{{a}_{\mathbf{S}^+_1} \dots {a}_{\mathbf{S}^+_K} }_{{a}_{\mathbf{S}^-_1} \dots {a}_{\mathbf{S}^-_K} } [\mathbf{S}]
\end{equation}
for the generalized state or
\begin{equation}\label{elementoperatortensor}
\delta\hat{A}^{\mathtt{a}_{\mathbf{S}^+_1} \dots \mathtt{a}_{\mathbf{S}^+_K} }_{\mathtt{a}_{\mathbf{S}^-_1} \dots \mathtt{a}_{\mathbf{S}^-_K} } [\mathbf{S}]
\end{equation}
for the operator tensor.

We can now write down the corresponding calculation to obtain the generalized state associated with the operation $\mathsf{A}^\mathtt{b}_\mathtt{c}$
as
\begin{equation}
{A}^{b}_{c} = \lim_{\pmb{\delta}_k \rightarrow 0} \prod_{\mathbf{S}\in \mathtt{A}_\text{discrete} }
\delta\mathsf{A}^{{a}_{\mathbf{S}^+_1} \dots {a}_{\mathbf{S}^+_K} }_{{a}_{\mathbf{S}^-_1} \dots {a}_{\mathbf{S}^-_K} } [\mathbf{S}]
\end{equation}
for the generalized state case and
\begin{equation}
\hat{A}^\mathtt{b}_\mathtt{c} = \lim_{\pmb{\delta}_k \rightarrow 0} \prod_{\mathbf{S}\in \mathtt{A}_\text{discrete} }
\delta\hat{A}^{\mathtt{a}_{\mathbf{S}^+_1} \dots \mathtt{a}_{\mathbf{S}^+_K} }_{\mathtt{a}_{\mathbf{S}^-_1} \dots \mathtt{a}_{\mathbf{S}^-_K} } [\mathbf{S}]
\end{equation}
for the operator tensor case.

Assuming that such calculations can be performed by appropriate techniques, the question now is whether we can write down, in simple terms, an expression for the generalized state (or operator tensor) associated with such infinitesimal elements.  We leave this for future work.

\section{Conclusions}

In this paper we considered theories in which solutions are given by defining fields on a manifold and whose equations are invariant under diffeomorphisms.  General Relativity is such a theory.  We have seen how to formulate General Relativity (and, indeed, all such theories) in an operational way.  We did this by setting up an operational space.  The particular choice of operational space we made was where we nominated a set of scalars to form the axes of the space (this was motivated by related ideas of Westman and Sonego).  We could investigate other choices of operational space.  We showed how to set up compositional calculations for the possibilistic and probabilistic cases in terms of operations, generalized states, and other related objects.  This provides a formulation of General Relativity similar to the operator tensor formulation of quantum theory.  We described a route to formulating Quantum Field Theory in this kind of framework and then we made proposals on how to develop a theory of Quantum Gravity in operational terms.

One interesting aspect of General Relativity which becomes evident when formulated in this way is that it exhibits a curious non-separability.  When we provide pure solution descriptions of two regions of op-space, the combined region may turn out to be described by a mixed state.

The route we took to operationalism here was to define our operations out of the fields appearing in the field equations of General Relativity. We could have attempted to take a different approach in which we take certain instruments (such as clocks, light rays, fluxometers, \dots) as primitives and then attempt to formulate the equations of General Relativity in terms of quantities measured by these instruments.  One problem with such an approach is that instruments will, in general, only function in a limited range of physical circumstances. They may break up under very strong gravitational fields for example.  By defining operational space directly in terms of fields we avoid this.

We could, nevertheless, attempt to construct instruments out of the basic fields within the operational framework discussed in this paper.  One example of how we might do this is the two-fluid clock explained in Sec.\ \ref{sec:fieldsareeverything} (wherein two fluids mix over time in a spherical blob).  Such instruments could, possibly, be associated with free operations (the latter were discussed in Sec.\ \ref{sec:freeoperations}).  Developing this aspect of the theory would be important in making this framework useful for more practical calculations.

There remain a number of challenges, both technical and conceptual, in pushing this program forward.  The most important of these is to write down physicality conditions for generalized states (and, in the quantum cases, for operator tensors) that capture the causality condition. This is especially challenging as we have fuzzy causal structure (indefinite causal structure in the case of Quantum Gravity). In the quantum case there is the possibility (as discussed in Sec.\ \ref{sec:whatarethephysicalityconditions}) that a candidate for the Planck length will emerge naturally in finding the appropriate physicality condition.  Another kind of technical challenge is to deal with fact that we have sets of boundary conditions with infinite cardinality. We need to be more careful in ensuring that the integrals are properly defined in such cases.

This framework has the advantage that we need only consider the part of the world we are interested in (because we have formalism locality).  This could be advantageous in numerical relativity. Standard numerical relativity techniques use a canonical formulation of General Relativity in which a state defined on a given space-like slice is evolved in time.  This means we may have to consider parts of the universe that are not directly relevant to the phenomena we are investigating.  The approach here, on the other hand, may alow us to simulate a much smaller part of the universe when investigating specific questions.

General Relativity is a strange place.  By formulating it in operational terms we can hope to get a better handle on how the theory relates to our own experience of the world. Further, Quantum Theory can be readily understood in operational terms (indeed, it may be argued that there is no consensus on how to understand Quantum Theory beyond this).  By formulating the two theories in similar terms we can hope to open up the way to a theory of Quantum Gravity.  Ultimately we would like to move beyond a purely operational understanding.  We can hope that the odd mix of conceptual tensions at play in the problem of Quantum Gravity may lead us to an understanding of the deeper reality underlying such a theory.  From this point of view we should regard operationalism as a methodology aimed at making progress in physics. Not an end in itself.

\section*{Acknowledgements}

I am especially grateful to Joy Christian for convincing me many years ago that people working in Quantum Foundations need to take the problem of Quantum Gravity seriously and for teaching me Einstein's hole argument.   I am grateful to Christopher Fuchs for getting me on the path of thinking about axioms and postulates for physics and to Antony Valentini for frequently reminding me that \lq\lq reality is real". I am grateful to Bob Coecke for introducing me to diagrammatic approaches to physics and to Markus Mueller for discussions on Quantum Field Theory which influenced the section of this paper on that subject.  I am also grateful to Hans Westman for taking time to explain his scalar fields approach which is so important for this present project.  Eventually I was convinced.

I have benefited enormously from conversations with the \lq\lq Pavia group" Giulio Chribella (now in Hong Kong), Mauro D'Ariano, and Paulo Perinotti, with Perimeter Institute colleagues Robert Spekkens, Matthew Pusey, Rafael Sorkin, Bianca Ditrich, Laurent Freidel, Lee Smolin, and from discussions with Caslav Brukner, Ognyan Oreshkov, Robert Oeckl, Adrian Kent, Samson Abransky, Matthew Leifer, Scott Aaronson, Achim Kempf, and many other people. I am grateful to Neil Turok for discussions and for the quote from de Witt I used in Sec.\ \ref{sec:thisresearchprogramandotherwork}.  I am also grateful to Paul Busch for information on the history of operational quantum theory and Ka\'ca Bradonji\'c for permission to use the cover art.

I am very grateful to everyone at Seven Shores Cafe where most of this work was done.

I am deeply grateful to Zivy for her support, encouragement, and editing help during this project, and to Vivienne and Helen for teaching me about the world we live in in ways no adult could. Me: \lq\lq How do you know where you are?"  Vivienne: \lq\lq By looking and seeing what is around you,"  which just about sums up this paper.

Research at Perimeter Institute is supported by the Government of Canada through Industry Canada and by the Province of Ontario through the Ministry
of Economic Development and Innovation.  This project was made possible in part through the support of a grant from the John Templeton Foundation. The opinions expressed in this publication are those of the author and do not necessarily reflect the views of the John Templeton Foundation.  I am grateful also to FQXi for support through grant 2015-145578 (entitled "Categorical Compositional Physics").

The term PAGeR is homage to the Blackberry Pager and, thereby, to Mike Lazaridis for founding the Perimeter Institute - a place where foundational research is encouraged.

\newpage

\part{Appendices}

\appendix

\section{Standard formulation of General Relativity}\label{appendix:standarformulationofGR}

In this Appendix we will review the standard formulation of General Relativity. We will go into more detail than might be expected in a research paper of this nature. We do this for two reasons. First, we want to put ourselves in a good position to make some of the conceptual and mathematical moves that happen in the main text.  And second, because it is anticipated that many of the people reading this paper will come from a Quantum Foundations, Quantum Information, or Computer Science background and so a review of the standard formulation of General Relativity might be quite useful.  This review borrows from many textbooks (especially Schutz \cite{schutz2009first}, Dirac \cite{dirac1996general}, Misner, Thorne, and Wheeler \cite{misner1973gravitation},  Wald \cite{wald2010general}, Malament \cite{malament2006classical}, Poisson \cite{poisson2004relativist}, Hawking and Ellis \cite{hawking1973large}) with a few pedagogical forays that may be novel.  There are also a number of very good online video courses available on PIRSA by Turok \cite{Turok2015relativity}, Sorkin \cite{Sorkin2015elements}, Kempf \cite{Kempf2015general}, Gregory \cite{Gregory2015gravitational}, and Poisson \cite{Poisson2012advanced}.

General Relativity, like Quantum Theory, is a radical conceptual innovation on the classical theories that preceded it.  Two particular radical features stand out.  First,  the causal structure (as conveyed by the light cone structure inferred from the metric) is dynamical. Unlike the classical physics of Newton and Maxwell, we do not have a fixed causal background.   Second, the coordinate $x^\mu$ used in General Relativity cannot be regarded objectively real because the physics is invariant under general coordinate transformations.   General Relativity is really a theory of fields on top of fields.  These two facts make it much more challenging to formulate General Relativity in an operational manner.

In this review I will emphasize a few things that are important for the purposes of the main text. First, except in the special case where we have no matter, Einstein's equation is only one of a set of coupled field equations that have to be solved to obtain a solution (or \lq\lq model") representing the physical situation.  The best way to come at this issue is start with Special Relativity.  From Special Relativity we get the matter field equations.  At this stage (before we go to General Relativity) we have as many equations as we have degrees of freedom and hence these equations can be solved.  The principe of equivalence suggests these equations must be true locally but not globally.  We apply a technique sometimes called minimal substitution to convert the Special Relativistic matter field equations into General Relativistic matter field equations in accord with the principle of equivalence and the demand that the equations must be true in general coordinates.  This involves replacing the constant Minkowski metric, $\eta_{\bar\mu\bar\nu}$, with the variable metric $g_{\mu\nu}$.  It follows that we need some extra equations so we have enough equations to solve for both the matter fields and the metric.  The extra equations are given by the Einstein field equation.  Second, we will emphasize that General Relativity is a theory of fields. Devices such as test particles, clocks, rulers, and even physical reference frames must be regarded as corresponding to particular configurations of fields and so are effective notions.  Consequently we will not discuss test particles and geodesics (except in the context of a dust fluid where we actually have fields). Finally, we emphasize the ontology of General Relativity.  Einstein's hole argument tells us that physically real quantities are those that are invariant under diffeomorphisms.

\subsection{The principle of equivalence and the principle of general covariance}\label{Sec:principleequivcovar}

Einstein considered a man in an elevator. \index{elevator}  As long as the elevator is not too big there is no way from inside the elevator to distinguish the situation where the elevator is falling freely under gravity from the situation where it is floating out in space.  All objects will fall with the same acceleration and so appear to be moving inertially inside the elevator in both situations.  The \emph{weak equivalence principle} \index{equivalence principle} states that different bodies fall in the same way. This follows from the equivalence of inertial and gravitational mass as famously (and probably apocryphally) demonstrated by Galileo at the Leaning Tower of Pisa.  The \emph{strong equivalence principle} states that the laws of physics are the same locally in any falling frame of reference.  This principle plays an essential role in obtaining the matter field equations for General Relativity from the matter field equations of Special Relativity as we will see in Sec.\ \ref{Sec:minimalsubstitution}.

Since falling elevators are accelerating relative to one another, it is clear that we cannot provide coordinates for a single reference frame that covers every locally inertial situation with a locally inertial coordinate system. Thus, any reference frame that covers any finite part of spacetime will, generically, have to be non-inertial (i.e.\ accelerating) in most places.  If we have to use general coordinate systems then there is no special coordinate system. Further, we can consider general transformations between such coordinate systems. We are forced to the point of view that any coordinate system is as good as any other.  The \emph{principle of general covariance} \index{principle of general covariance} says that the laws of physics should be written in a way that they take the same form in any coordinate system.  To actually implement this we will set up the mathematical machinery of manifolds and tensor fields.

\subsection{Manifolds}\label{Sec:Manifolds}

Manifolds are essential for formulating General Relativity. \index{manifolds}  A manifold is set of points, $\mathscr M$, that can be covered by a family of \emph{charts} \index{charts} (labeled here by $i$) all having the same dimension, $D$ (the dimension of the space).  Each chart consists of a one-to-one mapping, $\omega_i$, from some open set $\mathscr{O}_i\subseteq \mathscr{M}$ onto an open region, $V_i \subseteq\mathbb{R}^D$.  We require
\begin{equation}
\bigcup_i \mathscr{O}_i = \mathscr{M}
\end{equation}
so these open sets cover the manifold. We can, thus, represent the points in the region $O_i$ of the manifold by an $D$-tuple of coordinates, $\{ x^0, x^1, \dots, x^{D-1} \}$ corresponding to points in the region $V_i$.   Every point in the manifold is covered by at least one such \emph{coordinate system} as well as points in the vicinity of that point (since we have open regions).  We will often denote such a point by $x$ (by which we really mean $\{x^\mu|\mu=0, 1, \dots D-1 \}$).

There are some technical requirements for such a cover of charts to constitute a manifold.  Where they overlap, pairs of charts must be smoothly \lq\lq sewn together" in an appropriate technical sense. Thus, in the overlap $\mathscr{O}_i\cap\mathscr{O}_j$ we require that the map
\begin{equation}
              \omega_j \circ \omega^{-1}_i(x)
\end{equation}
from points, $x$ in $V_i$ (via the manifold) to $V_j$ is smooth (infinitely differentiable).  An additional technical requirement is that $\omega_i(\mathscr{O}_i\cap\mathscr{O}_j)$ must be an open set in $\mathbb{R}^D$. See \cite{wald2010general, malament2006classical} for further discussion of the definition of a manifold.

In General Relativity a particular solution, once found, will \lq\lq live on" a particular manifold.  A different solution may live on the same manifold or a different manifold. Thus, we are interested in the space of manifolds (an approach to looking at the space of manifolds is introduced in Sec.\ \ref{sec:diffeos}).

\subsection{Tensors}

To define a tensor we need a vector space $V$ spanned by some basis vectors $\{ {\bf e}_\mu: \mu=0 ~\text{to}~D-1\}$.  We also need the dual space $V^*$ consisting of the linear functionals on $V$. The dual space also forms a vector space and is be spanned by the basis $\{ {\bf e}^\mu:\mu=0 ~\text{to}~D-1\}$ where
 \begin{equation}
{\bf e}^\mu ({\bf e_\nu})= \delta^\mu_\nu
\end{equation}
We write a $(q,r)$-type tensor as
\begin{equation}\label{qrtensor}
{\bf T}= T^{\mu_1\dots \mu_q}_{\nu_1\dots \nu_r} \ {\bf e}_{\mu_1}\otimes \dots \otimes {\bf e}_{\mu_q} \otimes {\bf e}^{\nu_1} \otimes \dots \otimes {\bf e}^{\nu_r}
\end{equation}         \index{tensor}
We can use Penrose's abstract index notation
\begin{equation}\label{Penrosenotation}
{\bf T}= T^{a_1\dots a_q}_{b_1\dots b_r}
\end{equation}
to indicate indicate that we have an $(q,r)$-type tensor \cite{penrose1971angular}. In this case, the indices $a$, $b$, are just place holders indicating what space the tensor lives in.  This notation also takes care of contractions when we write down abstract expressions between tensors.  A tensor is a basis independent object. If we write it in a different basis then the object itself is unchanged.  However, the coefficients
\begin{equation}
T^{\mu_1\dots \mu_q}_{\nu_1\dots \nu_r}
\end{equation}
will transform.  Let $\{ {\bf e}_{\mu'}: \mu'=0 ~\text{to}~D-1\}$ and $\{ {\bf e}^{\mu'}:\mu'=0 ~\text{to}~D-1\}$ be the new basis and new dual basis respectively satisfying
 \begin{equation}
{\bf e}^{\mu' }({\bf e_{\nu'}})= \delta^{\mu'}_{\nu'}
\end{equation}
Note that the primes on the indices inform us that we have new bases.  We can write the old bases in terms of the new bases
\begin{equation}\label{newbases}
{\bf e}_{\mu} = e\indices{^{\mu'}_\mu} {\bf e}_{\mu'}  ~~~~~ {\bf e}^\mu = e\indices{_{\mu'}^\mu} {\bf e}^{\mu'}
\end{equation}
Then we can write
\begin{equation}\label{qrprimetensor}
{\bf T}= T^{{\mu'}_1\dots {\mu'}_q}_{{\nu'}_1\dots {\nu'}_r} \ {\bf e}_{{\mu'}_1}\otimes \dots \otimes {\bf e}_{{\mu'}_q} \otimes {\bf e}^{{\nu'}_1} \otimes \dots \otimes {\bf e}^{{\nu'}_r}
\end{equation}
Now comparing (\ref{qrtensor}) and (\ref{qrprimetensor}) and using (\ref{newbases}) we obtain
\begin{equation}
T^{{\mu'}_1\dots {\mu'}_q}_{{\nu'}_1\dots {\nu'}_r} =  e\indices{^{{\mu'}_1}_{\mu_1}} \dots e\indices{^{{\mu'}_q}_{\mu_q}} e\indices{_{{\mu'}_1}^{\mu_1}} \dots e\indices{_{{\mu'}_r}^{\mu_r} } T^{\mu_1\dots \mu_q}_{\nu_1\dots \nu_r}
\end{equation}
Thus the components of a tensor must transform in this way so that the tensor can be regarded as a basis independent object.

\subsection{Tensor fields}

So far we have only spoken of tensors. In General Relativity we are actually interested in \emph{tensor fields}. \index{tensor fields} To define a tensor field on a manifold, $\mathscr M$, we must specify a tensor at each point in the manifold.  To represent a tensor field in terms of its components at each point we must define a basis (and dual basis) at every point on the manifold.  The standard way of doing this is quite curious. The bases are formed from differential objects associated with the coordinate system pertaining to a chart covering the manifold at each point. The basis so obtained is called a \emph{coordinate basis}. These differential objects transform under a general coordinate transformation and this is how we transform to a new basis.  There are other ways of getting a basis.  The tetrad approach, for example, uses the metric field to form a basis corresponding to a local inertial reference frame at $p$.  We will outline only the coordinate basis approach here.

At every point, $p$, in a manifold we can form the \emph{tangent space}. \index{tangent space} A vector at $p$ lives in this tangent space. For a sphere we can think of the tangent space at $p$ as being like a tangent plane which touches the sphere at $p$.  This picture, however, entails thinking of the sphere as being embedded in three dimensions.  It is possible, using differential structure, to define the notion of a tangent space intrinsically without reference to any such embedding.  We would like a basis set, $e_\mu$, for the tangent space at $p$ obtained in an intrinsic way.   To obtain such a basis set, first we note that there is a one-to-one correspondence between a vector $(A^0, A^1, \dots A^{D-1})$ and the directional derivative operator  ${\bf A}:=A^\mu \partial_\mu$ (where $\partial_\mu := \frac{\partial}{\partial x^\mu}$).   The derivative operator, $\partial_\mu$, acts on real valued functions defined on the manifold and, hence, can be thought of as object intrinsic to the manifold.  Hence, a basis for the tangent space is provided by the operators $\{\partial_\mu: \mu=0 ~\text{to}~ D-1\}$.  We also need a dual space whose elements are linear functional on the vectors.  For this we use the \emph{one-form} $df$, \index{one form} associated with a function $f$ on the manifold, defined by
\begin{equation}\label{oneformdefinition}
df( {\bf A}) := {\bf A} f
\end{equation}
The notation $df$ is suggestive of an infinitesimal difference.  Note, however, that $df$ is defined through the above equation.
For the basis vectors, $\partial_\mu$, we have
\begin{equation}
df (\partial_\mu) = \partial_\mu f
\end{equation}
Then a basis for the dual space is given by $\{ dx^\mu : \mu=0~\text{to}~D-1\}$ since
\begin{equation}\label{deltadual}
dx^\mu (\partial_\nu) = \partial_\nu x^\mu = \frac{\partial x^\mu}{\partial x^\nu} = \delta^\mu_\nu
\end{equation}
Hence we can expand a tensor field as
\begin{equation}\label{coordbasisqrtensor}
{\bf T}= T^{\mu_1\dots \mu_q}_{\nu_1\dots \nu_r} \ {\partial}_{\mu_1}\otimes \dots \otimes {\partial}_{\mu_q} \otimes {dx}^{\nu_1} \otimes \dots \otimes {dx}^{\nu_r}
\end{equation}
by plugging our new bases into (\ref{qrtensor}).

We are interested in how the components of tensor fields, when expressed in this way, transform when we go to a new basis.  We induce a transformation of the bases $\{\partial_\mu\}$ and $\{ dx^\mu\}$ when we perform a general invertible coordinate transformation
\begin{equation}
x^{\mu} \longrightarrow x^{\mu'} = h^{\mu'}(\{x^\mu\})
\end{equation}
where we only consider smooth (infinitely differentiable) functions, $h(\cdot)$.  Under such a transformation we have
\begin{equation}\label{chainrule}
\partial_{\mu} = \frac{ \partial x^{\mu'}}{\partial {x^{\mu}}} \partial_{\mu'}
\end{equation}
by the chain rule.  For the dual basis we have
\begin{equation}\label{dxtransformation}
dx^{\mu} = \frac{ \partial x^{\mu}}{\partial {x^{\mu'}}} dx^{\mu'}
\end{equation}
since, for an arbitrary vector, ${\bf A}$, using (\ref{chainrule},\ref{oneformdefinition}) we have
\begin{equation}
dx^\mu({\bf A}) = A^\nu \partial_\nu x^\mu = A^\nu \frac{\partial x^{\nu'}}{\partial x^{\nu}} \partial_{\nu'} x^\mu
= A^\nu dx^{\nu'}(\partial_\nu)  \frac{ \partial x^{\mu}}{\partial {x^{\nu'}}}
= \frac{ \partial x^{\mu}}{\partial {x^{\nu'}}} dx^{\nu'} ({\bf A})
\end{equation}
Note that (\ref{dxtransformation}) is relationship we would expect under the interpretation that $df$ is an an infinitesimal (which motivates this notation).

Putting (\ref{chainrule},\ref{dxtransformation}) into (\ref{coordbasisqrtensor}) we obtain the transformation equation
\begin{equation}
T^{{\mu'}_1\dots {\mu'}_q}_{{\nu'}_1\dots {\nu'}_r} =  \frac{ \partial x^{{\mu'}_1}}{\partial {x^{{\mu}_1}}}\dots \frac{ \partial x^{{\mu'}_q}}{\partial {x^{{\mu}_q}}} \frac{ \partial x^{{\nu}_1}}{\partial {x^{{\nu'}_1}}} \dots \frac{ \partial x^{{\nu}_r}}{\partial {x^{\nu'}_r}} T^{\mu_1\dots \mu_q}_{\nu_1\dots \nu_r}
\end{equation}
for the components of the tensors.  Sometimes the components of the tensor are referred to as being a tensor so long as they satisfy the above transformation equation.  We will use this language below when it is convenient.

An important trick in tensorial analysis is to find a statement relating tensorial quantities that is true in a given coordinate system.  Since these are tensorial quantities the statement then becomes true in any coordinate system.  Another important trick it the quotient theorem. This says, for example, if $A^{\mu\nu}_{\alpha}$ is a tensor and $A^{\mu\nu}_{\alpha} B^\alpha_\mu$ is a tensor then $B^\alpha_\mu$ is a tensor.  This is easy to prove (see \cite{dirac1996general}).

\subsection{The metric}

A metric field is given by a $(0,2)$-type tensor
\begin{equation}
{\bf g}= g_{\mu\nu} dx^\mu \otimes dx^\nu
\end{equation}  \index{metric}
where the matrix $g_{\mu\nu}$ is symmetric ($g_{\nu\mu}=g_{\mu\nu}$) and invertible. We define
\begin{equation}
g^{\mu\nu} := (g_{\mu\nu})^{-1}
\end{equation}
A metric provides a scalar product between vectors
\begin{equation}
{\bf U}\cdot {\bf V} = g_{\mu\nu} U^\mu V^\nu
\end{equation}
for any pair of vectors in the tangent space at any point, $x$.

A special metric is the \emph{Minkowski metric}. \index{Minkowski metric}This has the form
\begin{equation}
\eta_{\mu\nu}= \text{diag}(-1, 1, 1, \dots, 1)
\end{equation}
and is the metric of Special Relativity (in the four dimensional case the coordinates are $(x^0=ct, x^1=x, x^2=y, x^3=z)$).

In General Relativity the metric is further taken to be Lorentzian - it has signature $(-, +, +, \dots +)$ \index{Lorentzian signature} where the signature is the list of signs of the eigenvalues.  Furthermore, in General Relativity, the metric is given a certain physical significance -  it can be used to calculate the invariant distance between points that are infinitesimally close.  Define
\begin{equation}
\delta {\bf x} = \delta x^\mu \partial_\mu
\end{equation}
where we take the $\delta x^\mu$'s to be very small.  This is a small vector in the tangent space (we can think of it as an infinitesimal).  Then the square of the distance along $\delta{\bf x}$ (in the limit as $\delta x^\mu \rightarrow 0$) is
\begin{equation}\label{metricfords}
\begin{split}
\delta s^2 & = {\bf g}(\delta {\bf x} \otimes \delta {\bf x}) = g_{\mu\nu} dx^\mu \otimes dx^\nu ( \delta x^\alpha \partial_\alpha \otimes \delta x^\beta \partial_\beta )
= g_{\mu\nu} \delta^\mu_\alpha \delta^\nu_\beta \delta x^\alpha \delta x^\beta \\
&= g_{\mu\nu} \delta x^\mu \delta x^\nu
\end{split}
\end{equation}
If $g_{\mu\nu}=\eta_{\mu\nu}$ then we see immediately that $\delta s^2$ is the invariant infinitesimal square interval of Special Relativity.    We will see in Sec.\ \ref{Sec:Zerothderivative} that we can always find a coordinate system such that the metric is equal to the Minkowski metric.  Since $\delta s^2$ is a scalar it takes the same value in all coordinate systems.  Hence,  by appealing to Special Relativity, we can always interpret this as the invariant infinitesimal square interval with the following properties.  If $\delta s^2$ is negative then we have a time-like separation along $\delta {\bf x}$ and $\sqrt{|\delta s^2|}$ is equal to the proper time elapsed between the end points of the vector.  If $\delta s^2$ is positive we have a space-like separation and $\sqrt{\delta s^2}$ is the spacial distance between the end points of the vector.  If  $\delta x^2$ is zero we have a null separation (light-like).  The metric provides causal structure.  It also provides a scale (measured in some appropriate units).

The fact that the metric is invertible means that we can use it to raise and lower subscripts. For example,
\begin{equation}
R\indices{_\alpha_\beta_\mu_\nu} = g_{\alpha\sigma}R\indices{^\sigma_\beta_\mu_\nu}, ~~~ A^{\mu\nu} = g^{\nu\sigma} A\indices{^\mu_\sigma}
\end{equation}
This is a useful notational convention.

\subsection{The covariant derivative}

Physical theories are expressed in terms of derivatives.  Since we are building our physical theory out of tensors we are interested in things like
\begin{equation}
\partial_\gamma T^{\mu_1\dots \mu_q}_{\nu_1\dots \nu_r}, ~~~~~~ \partial_\gamma\partial_\delta T^{\mu_1\dots \mu_q}_{\nu_1\dots \nu_r}
\end{equation}
and so on.  A natural question is whether these kind of objects are tensors.  Let us consider the simplest case first.  Consider differentiating a scalar, $\partial_\mu S$.  This is a tensor because
\begin{equation}\label{scalartrans}
\partial_{\mu'} S = \frac{\partial x^{\mu'}}{\partial x^\mu} \partial_\mu S
\end{equation}
by the chain rule for partial differentiation.  Next consider the case of the derivative of a vector, $\partial_\nu V^\mu$.  We can show that this is not a tensor by differentiating the transformation equation for this vector (and using the chain rule)
\begin{equation}
\begin{split}
\partial_{\nu'} V^{\mu'} &= \frac{\partial}{\partial x^{\nu'}}\left( \frac{\partial x^{\mu'}}{\partial x^\mu} V^\mu\right) \\
&= \frac{\partial x^{\mu'}}{\partial x^\mu} \frac{\partial x^\nu}{\partial x^{\nu'}}  \partial_\nu V^\mu + \frac{\partial x^{\mu'}}{\partial x^{\nu'} \partial x^\mu} V^\mu
\end{split}
\end{equation}
The presence of the second term on the right shows that $\partial_\nu V^\mu$ does not transform as a tensor.  In general, except for the special case of a scalar, differentiation of a tensor does not return a tensor.  However, physical laws make heavy use of differentiation and so we need a notion of derivative that is suited to a curved manifold.  For this purpose a \emph{covariant derivative}, \index{covariant derivative} denoted $\nabla_\mu$, is introduced.

To get a handle on the properties of the covariant derivative we will, as we go along, demand they have a number of properties motivated by the analogy with normal derivatives.  Since $\partial_\mu S$ is a tensor (as we saw in (\ref{scalartrans})), we demand
\begin{equation}\label{covscalar}
\nabla_\mu S = \partial_\mu S
\end{equation}
for any scalar field, $S$.  The covariant derivative of a scalar field is equal to the partial derivative.

The next case to consider is a vector.  We write
\begin{equation}\label{covupperindices}
\nabla_\mu V^\alpha = \partial_\mu V^\alpha + \Gamma\indices{^\alpha_\nu_\mu} V^\nu
\end{equation}
The second term on the right is the \lq\lq correction" that makes $\nabla_\mu V^\alpha$ tensorial even though $\partial_\mu V^\alpha$ is not. The notation $V\indices{^\nu_{,\mu}}:= \partial_\mu V^\nu$ and $V\indices{^\nu_{;\mu}}:= \nabla_\mu V^\nu$ is frequently used so then we write
\begin{equation}
V\indices{^\alpha_{;\mu}} = V\indices{^\alpha_{,\mu}} + \Gamma\indices{^\alpha_\nu_\mu} V^\nu
\end{equation}
The object, $\Gamma\indices{^\nu_\gamma_\mu}$, is called the \emph{connection}. \index{connection}  It is not a tensor as it does not transform appropriately.  It can easily be shown that it transforms as follows
\begin{equation}
\Gamma\indices{^{\alpha'}_{\nu'}_{\mu'}} = \frac{\partial x^{\alpha'}}{\partial x^{\alpha}} \frac{\partial x^{\nu}}{\partial x^{\nu'}} \frac{\partial x^{\mu}}{\partial x^{\mu'}}
 \Gamma\indices{^{\alpha}_{\nu}_{\mu}}
- \frac{\partial^2 x^{\alpha'}}{\partial x^{\nu}\partial x^{\mu}}
\frac{\partial x^{\nu}}{\partial x^{\nu'}}\frac{\partial x^{\mu}}{\partial x^{\mu'}}
\end{equation}
We see from the presence of the second term that the connection does not transform as a tensor.

To calculate the covariant derivative of arbitrary tensors we proceed as follows. First consider a $(0,1)$ type tensor.  If we put
\begin{equation}
\nabla_\mu U_\alpha = \partial_\mu U_\alpha + Q\indices{^\nu_\alpha_\mu} U_\nu
\end{equation}
then by imposing $\nabla_{\mu}(U_\alpha V^\alpha) = \partial_\mu(U_\alpha V^\alpha)$ (since $U_\alpha V^\alpha$ is a scalar field) we quickly obtain that $Q\indices{^\nu_\alpha_\nu} = - \Gamma\indices{^\nu_\alpha_\nu}$.  So we can write
\begin{equation}\label{covlowerindice}
\nabla_\mu U_\alpha = \partial_\mu U_\alpha - \Gamma\indices{^\nu_\alpha_\mu} U_\nu
\end{equation}
Now we demand that $\nabla_\mu$ satisfies the same product rule as normal differentiation.  Then
\begin{equation}
\begin{split}
\nabla_\mu (U_\alpha & V^\beta W^\gamma) = \nabla_\mu( U_\alpha) V^\beta W^\gamma + U_\alpha \nabla_\mu (V^\beta) W^\gamma + U_\alpha V^\beta \nabla_\mu (W^\gamma) \\
&= \partial_\mu (U_\alpha V^\beta W^\gamma) - \Gamma\indices{^\delta_\alpha_\mu}U_\delta V^\beta W^\gamma + \Gamma\indices{^\beta_\delta_\mu}U_\alpha V^\delta W^\gamma + \Gamma\indices{^\gamma_\delta_\mu}U_\alpha V^\beta W^\delta
\end{split}
\end{equation}
We further demand that $\nabla_\mu$ acts linearly.  Since a general $(2,1)$ type tensor can be written as a sum of tensors like $U_\alpha V^\beta W^\gamma$ we have, by linearity,
\begin{equation}
\nabla_\mu T_\alpha^{\beta\gamma} = \partial_\mu  T_\alpha^{\beta\gamma}
- \Gamma\indices{^\delta_\alpha_\mu}T_\delta^{\beta\gamma}
+ \Gamma\indices{^\beta_\delta_\mu} T_\alpha^{\delta\gamma}
+ \Gamma\indices{^\beta_\delta_\mu} T_\alpha^{\beta\delta}
\end{equation}
This approach clearly generalizes to arbitrary tensors. We pick up a minus sign correction for each subscript and a plus sign correction for each superscript.

Partial derivatives commute so $\partial_\mu\partial_\nu=\partial_\nu\partial_\mu$.  We cannot demand that covariant derivatives commute in general (there is no way to get the $\Gamma$ terms to cancel). However, we can demand that they commute when acting on a scalar field.  Thus we demand
\begin{equation}\label{torsionfreedefinition}
\nabla_\mu \nabla_\nu S= \nabla_\nu\nabla_\mu S
\end{equation}
where $S$ is a scalar field. Using (\ref{covscalar}, \ref{covlowerindice}) in (\ref{torsionfreedefinition}) we obtain
\begin{equation}
\Gamma\indices{^\alpha_\mu_\nu} = \Gamma\indices{^\alpha_\nu_\mu}
\end{equation}
i.e.\ the connection is symmetric in its two subscripts. In this case the covariant derivative is said to be \emph{torsion free}. \index{torsion free} It is possible to relax this requirement and work with covariant derivatives with torsion.  In General Relativity, however, the covariant derivative is taken to be torsion free.

The notions of connection and metric are logically independent.  However, as we will see in Sec.\ \ref{sec:firstderivative}, we can make an explicit choice for the connection in terms of first derivatives of the metric.  This choice matters for the physics (as we will see Sec.\ \ref{Sec:minimalsubstitution}) because, in General Relativity, we substitute the covariant derivative for the partial derivative of the Special Relativistic physical laws.

\subsection{Locally flat and local inertial reference frames}

We define a \emph{locally flat reference frame} \index{locally flat reference frame} at a point $p$ to be a choice of coordinates, $x^{\ubar{\mu}}$ such that, at $p$,
\begin{enumerate}
\item The metric is in Minkowski form,
\begin{equation}
\left. g_{\ubar{\mu}\ubar{\nu}}\right|_p=\eta_{\ubar{\mu}\ubar{\nu}} := \text{diag}(-1, +1, +1, \dots , +1)
\end{equation}
\item The first derivatives of the metric vanish
\begin{equation}
 \left.g_{\ubar{\mu}\ubar{\nu},\ubar{\alpha}} \right|_p = 0
\end{equation}
\end{enumerate}
By a Taylor expansion (see Sec.\ \ref{sec:derivatives}) this means that the metric is Minkowski to first order in a small region around $p$.
We use a \lq\lq bar" under the index ($\ubar{\mu}$) as this has a connotation of flatness.  We will show below that, at any point $p$, we can find a frame in which the metric is in Minkowski form (this is proved in Sec.\ \ref{Sec:Zerothderivative}) and the first derivatives of the metric vanish (this is shown in Sec.\ \ref{sec:firstderivative}).  Hence, we can find a locally flat reference frame at any point, $p$.

We define a \emph{local inertial reference frame} \index{local inertial reference frame} at a point $p$ to be a choice of coordinates, $x^{\bar\mu}$, such that, at $p$
\begin{enumerate}
\item The metric is in Minkowski form,
\begin{equation}
\left. g_{\bar\mu\bar\nu}\right|_p=\eta_{\bar\mu\bar\nu} := \text{diag}(-1, +1, +1, \dots , +1)
\end{equation}
\item The connection vanishes
\begin{equation}
\Gamma\indices{^{\bar{\alpha}}_{\bar\mu}_{\bar\nu}} = 0
\end{equation}
\end{enumerate}
This means that, in a local inertial reference frame, the covariant derivative acts as a simple partial derivative $\nabla_{\bar\mu}=\partial_{\bar\mu}$.

This is interesting because it provides the link between the equations of Special Relativity (now taken to apply locally in a local inertial reference frame) and the equations of General Relativity (which must be written in a general coordinate system but can reduce to the equations of Special Relativity in a local inertial reference frame).   We will discuss this in Sec.\ \ref{Sec:minimalsubstitution}).

In fact, it is reasonable to demand that the local frame of reference in which the equations of Special Relativity apply is also locally flat (as well as being locally inertial) as this is true in Special Relativity.  We can  show that the equivalence of locally flat reference frames and local inertial reference frames everywhere implies
\begin{equation}\label{metriccompatibility}
g_{\mu\nu;\alpha} = 0
\end{equation}
This is known as the \emph{metric compatibility condition}. \index{metric compatibility}  To prove this consider a reference frame, $x^{\bar{\mu}}$, which is both a locally flat reference frame and a locally inertial reference frame at some point $p$. Hence, at $p$, we have
\begin{equation}
g_{\bar\mu\bar\nu,\bar\alpha}=0 ~~~\text{and}~~~ \Gamma\indices{^{\bar{\alpha}}_{\bar\mu}_{\bar\nu}} = 0
\end{equation}
then, in this reference frame we have
\begin{equation}
g_{\bar\mu\bar\nu;\bar\alpha}=0
\end{equation}
at $p$.  However, $g_{\bar\mu\bar\nu;\bar\alpha}$ is a tensor and hence we have $g_{\mu\nu;\alpha} = 0$ in any frame at $p$.  Equivalence of locally flat and local inertial reference frames everywhere implies metric compatibility everywhere.

It is also worth noting that, if we assume metric compatibility, then we can prove equivalence of locally flat and local inertial reference frames.
To see this, first note that in a local inertial reference frame metric compatibility implies
\begin{equation}
g_{\bar\mu\bar\nu,\bar\alpha}=0
\end{equation}
as then the connection vanishes.  But this is the condition for being in a locally flat reference frame.  Second note that if we are in a locally flat reference frame then metric compatibility implies
\begin{equation}
\Gamma\indices{^{\ubar{\alpha}}_{\ubar{\mu}}_{\ubar{\nu}}} = 0
\end{equation}
as then the derivative of the metric vanishes. But this is the condition for being in a local inertial reference frame.  Hence we have proved the equivalence of locally flat and locally inertial reference frames under metric compatibility.

The equation for metric compatibility \eqref{metriccompatibility} can be expanded out in terms of the connection.  Then, after a little index manipulation, we obtain
\begin{equation}\label{Gammausingmetric}
\Gamma\indices{^\alpha_\mu_\nu} =\frac{1}{2} g^{\alpha\gamma}(g_{\gamma\mu,\nu} + g_{\gamma\nu,\mu} - g_{\mu\nu,\gamma})
\end{equation}
where we have used the symmetry of the connection in its subscripts (since the covariant derivative is taken to be torsion free).

Since we can always satisfy metric compatibility (by choosing $\Gamma\indices{^\alpha_\mu_\nu}$ according to \eqref{Gammausingmetric}), and we can always find coordinates that specify a locally flat reference frame at any point, it follows that we can always find coordinates that specify a local inertial reference frame at any point.   The mathematical fact that we can always find a local inertial reference frame is essential for the physical reasoning used in setting up General Relativity.

\subsection{Derivatives of the metric at  a point}\label{sec:derivatives}

One way to understand the various objects that appear in General Relativity is to consider derivatives of the transformation equation for the metric starting with the zeroth derivative and going up to the third derivative.   The discussion here is motivated by (though different from) Schultz's discussion of the local flatness theorem \cite{schutz2009first} (see also Poisson's discussion \cite{poisson2004relativist}).

In a locally flat reference frame, the metric is Minkowski to first order in a small region about $p$ since, by a Taylor expansion of the metric about $p$
\begin{equation}\label{Taylormetric}
g_{\ubar{\mu}\ubar{\nu}}(x) = \eta_{\ubar{\mu}\ubar{\nu}} + (x^{\ubar{\alpha}}-x^{\ubar{\alpha}}(p)) \left. g_{\ubar{\mu}\ubar{\nu},\ubar{\alpha}}\right|_p + O( (x^{\ubar{\alpha}}- x^{\ubar{\alpha}}(p))^2)
\end{equation}
but, by our definition of a locally flat reference frame, the second term on the right vanishes.  We can try to go further and choose our transformation so that second derivatives of the metric vanish.  It turns out that we cannot make all of them vanish and this gives rise to the idea of curvature.

From here on, we will impose that we have metric compatibility (by making the choice in \eqref{Gammausingmetric} for the connection).  Hence a locally flat coordinate system is also a local inertial coordinate system (and vice versa).  We will use the notation $x^{\bar\mu}$ for the corresponding coordinates.  We will simply refer to this frame as a local inertial frame (local flatness being taken as given).

We wish to find tensors built out of the metric and its derivatives and identities between them.  Our main strategy is to find statements that are true in a local inertial reference frame, find a tensor that gives rise to this statement in this frame, and then this statement is true in all frames.  We will use the symmetry of the metric $g_{\mu\nu}=g_{\nu\mu}$, the symmetry of the connection in its subscripts ($\Gamma\indices{^\alpha_\mu_\nu}= \Gamma\indices{^\alpha_\nu_\mu}$) and the symmetry of partial differentiation
\begin{equation}
\begin{split}
&\frac{\partial^2}{\partial x^\mu\partial x^\nu}= \frac{\partial^2}{\partial x^\nu\partial x^\mu}\\
\frac{\partial^3}{\partial x^\mu \partial x^\nu \partial x^\gamma}& = \frac{\partial^3}{\partial x^\gamma \partial x^\mu \partial x^\nu }
= \frac{\partial^3}{\partial x^\nu \partial x^\mu\partial x^\gamma } = \dots
\end{split}
\end{equation}
These symmetries imply certain identities, for example,
\begin{equation}
g_{\mu\nu,\alpha\beta} - g_{\nu\mu,\beta\alpha}=0
\end{equation}
Such identities are only useful to us if they can be \lq\lq lifted" into tensorial identities.   The most important example of such identities are the Bianchi identities (see Sec.\ \ref{Sec:Bianchi}) which are in terms of third derivatives of the metric (i.e.\ $g_{\nu\mu,\beta\alpha\gamma}$).

We will employ parameter counting arguments below.  For ease of discussion, we will consider the case where we have a four dimensional manifold but similar conclusions go through for arbitrary dimension.

\subsubsection{Zeroth derivative and the Minkowski metric}\label{Sec:Zerothderivative}

\index{metric!zeroth derivative}

First we look at the metric at $p$ (zeroth derivative).  We will show that we can always transform this to the Minkowski metric. If we transform to new coordinates, $x^{\bar\mu}$ (we use this notation because we anticipate that we are transforming to a locally inertial flat frame), then the metric at $p$ becomes
\begin{equation}
g_{\bar\mu\bar\nu}(p) = \left. \frac{\partial x^{\mu}}{\partial x^{\bar\mu}}\right|_p \left. \frac{\partial x^{\nu}}{\partial x^{\bar\nu}}\right|_p g_{\mu\nu}(p)
\end{equation}

The metric, $g_{\mu\nu}(p)$, has ten free parameters (as it is symmetric). The matrix $\left. \frac{\partial x^{\mu}}{\partial x^{\bar\mu}}\right|_p$ provides us with sixteen free parameters. We can use ten of these free parameters to set the metric to diagonal form.  In General Relativity this always has signature $(-, +, +, +)$.  Thus, we obtain
\begin{equation}
g_{\bar\mu\bar\nu}(p) = \eta_{\bar\mu\bar\nu} := \text{diag}(-1, +1, +1, +1)
\end{equation}
Note that we do not have complete freedom with this transformation - we cannot use it to change the signature of the metric.  We have six free parameters left over.  These are the six parameters of the local Lorentz transformation.  Thus, the sixteen parameters of the first derivatives $\left. \frac{\partial x^{\mu}}{\partial x^{\bar\mu}}\right|_p $ can be chosen to put the metric at $p$ in Minkowski form along with providing a particular choice of Lorentz frame.  We can, however, go a bit further and chose a transformation puts the metric in a small region around $p$ in Minkowski form (to first order) as well as choosing a Lorentz frame that covers this small region.  To do this we need to consider first derivatives of the metric (as is clear from (\ref{Taylormetric})).

\subsubsection{First derivatives and local flatness}\label{sec:firstderivative}

\index{metric!first derivatives} \index{local flatness}

Now look at the first derivatives of the metric at $p$.  We will show we can always transform these to be equal to zero.  We can differentiate the transformation equation for the metric.
\begin{equation}
g_{\bar\mu\bar\nu,\bar\alpha} = \frac{\partial}{\partial x^{\bar\alpha}} \left(  \frac{\partial x^{\mu}}{\partial x^{\bar\mu}}\frac{\partial x^{\nu}}{\partial x^{\bar\nu}}g_{\mu\nu} \right)
\end{equation}
This gives
\begin{equation}\label{firstderivativemetric}
g_{\bar\mu\bar\nu,\bar\alpha} = \frac{\partial x^{\mu}}{\partial x^{\bar\mu}}\frac{\partial x^{\nu}}{\partial x^{\bar\nu}}\frac{\partial x^{\alpha}}{\partial x^{\bar\alpha}} g_{\mu\nu,\alpha}
+ \frac{\partial^2 x^\mu}{\partial x^{\bar\alpha} \partial x^{\bar\mu}} \frac{\partial x^\nu}{\partial x^{\bar\nu} } g_{\mu\nu}
+ \frac{\partial x^\mu}{\partial x^{\bar\mu} }  \frac{\partial^2 x^\nu}{\partial x^{\bar\alpha}\partial x^{\bar\nu}} g_{\mu\nu}
\end{equation}
Consider this expression at point $p$.  Assume we have fixed already the sixteen parameters $\left.\frac{\partial x^{\mu}}{\partial x^{\bar\mu}}\right|_p$ to transform the metric into Minkowski form and provide a particular choice of Lorentz frame at $p$.  Then the only degrees of freedom we have left are the 40 free parameters in the second derivatives,
\begin{equation}\label{fortysecondderivs}
\left. \frac{\partial^2 x^\mu}{\partial x^{\bar\alpha} \partial x^{\bar\mu}}\right|_p
\end{equation}
We also have 40 parameters $g_{\mu\nu,\alpha}(p)$ which we wish to transform to zero.  We have just enough free parameters to do this.  Thus, we can choose the second derivatives in (\ref{fortysecondderivs}) such that
\begin{equation}
g_{\bar\mu\bar\nu,\bar\alpha}(p) = 0
\end{equation}
Hence, we can always set up a locally flat reference frame at any given point, $p$ (and, since we impose metric compatibility via \eqref{Gammausingmetric}, this frame is also locally inertial).

One interesting observation follows from the fact that we can make all the first derivatives of the metric vanish.  This is that
\begin{quote}
Any tensor for which, when written out, every term contains a first derivative of the metric ($g_{\gamma\mu,\nu}$) or a covariant derivative of the metric ($g_{\gamma\mu;\nu}$) must be identically equal to zero.
\end{quote}
The reason for this is that, in a local inertial reference frame, such tensors are equal to zero.  Note, incidentally, that the connection is written in terms of first derivatives of the metric (\ref{Gammausingmetric}) but it is not a tensor and hence can vanish in a local inertial reference frame while not vanishing for other choices of coordinate system.  Also note that it is possible that a non-vanishing tensor, written out explicitly in terms of derivatives, may have some terms that depend on first derivatives.  An example of this is the curvature tensor to be discussed in the next section. These terms are necessary to make the whole object into a tensor even though, in a local inertial reference frame, they vanish.

\subsubsection{Second derivatives and the curvature tensor}

\index{metric!second derivatives}

If we differentiate (\ref{firstderivativemetric}) again we obtain an expression that looks like
\begin{equation}\label{secondderivativemetric}
\begin{split}
g_{\bar\mu\bar\nu,\bar\alpha\bar\beta}= & \frac{\partial^3 x^\mu}{\partial x^{\bar\beta} \partial x^{\bar\alpha} \partial x^{\bar\mu}} \frac{\partial x^\nu}{\partial x^{\bar\nu} } g_{\mu\nu}
+ \frac{\partial x^\mu}{\partial x^{\bar\mu} }  \frac{\partial^3 x^\nu}{\partial x^{\bar\beta} \partial x^{\bar\alpha}\partial x^{\bar\nu}} g_{\mu\nu}\\
& + \text{terms involving only 1st and 2nd derivatives of}~x^\mu~\text{wrt} ~ x^{\bar\nu}
\end{split}
\end{equation}
Consider this at the point $p$.  Assume we have already fixed the 16 parameters $\left.\frac{\partial x^{\mu}}{\partial x^{\bar\mu}}\right|_p$ to transform to a local inertial reference frame at $p$ and, further, we have fixed the 40 parameters $\left. \frac{\partial^2 x^\mu}{\partial x^{\bar\alpha} \partial x^{\bar\mu}}\right|_p$ so that we have transformed to a locally flat region.  Is it possible to transform the 100 second derivatives of the metric so that they vanish?  That is, can we set $g_{\bar\mu\bar\nu,\bar\alpha\bar\beta}=0$? The only free parameters we have are the third derivatives of $x$
\begin{equation}
\left. \frac{\partial^3 x^\mu}{\partial x^{\bar\beta} \partial x^{\bar\alpha} \partial x^{\bar\mu}} \right|_p
\end{equation}
But there are only 80 free parameters here.  Hence, we are 20 parameters short of being able to set all the second derivatives of the metric to zero.  This means that there can be some nontrivial curvature of the manifold so far as the metric is concerned and, further, that this nonzero curvature is measured by 20 parameters.  In particular,  consider the following linear combination of second derivatives of the metric
\begin{equation}\label{inertialframeR}
R_{\bar\alpha\bar\beta\bar\mu\bar\nu} = \frac{1}{2}(g_{\bar\alpha\bar\nu,\bar\beta\bar\mu} -g_{\bar\alpha\bar\mu,\bar\beta\bar\nu} + g_{\bar\beta\bar\mu,\bar\alpha\bar\nu} - g_{\bar\beta\bar\nu,\bar\alpha\bar\mu})
\end{equation}
If we substitute in (\ref{secondderivativemetric}) into (\ref{inertialframeR}) then all terms containing third derivatives of $x$ cancel.  This means we cannot, in general, set $R_{\bar\alpha\bar\beta\bar\mu\bar\nu}$ to zero (as these third derivatives of $x$ are the only degrees of freedom we would have to do this).  Hence $R_{\bar\alpha\bar\beta\bar\mu\bar\nu}$ it can be thought of as measuring the curvature - the departure from flatness that occurs when we get to second derivatives of the metric.  Now note from (\ref{inertialframeR}) that
\begin{equation}\label{inertialRidentone}
R_{\bar\alpha\bar\beta\bar\mu\bar\nu}= -R_{\bar\beta\bar\alpha\bar\mu\bar\nu}=-R_{\bar\alpha\bar\beta\bar\nu\bar\mu}=R_{\bar\mu\bar\nu\bar\alpha\bar\beta}
\end{equation}
\begin{equation}\label{inertialRidenttwo}
R_{\bar\alpha\bar\beta\bar\mu\bar\nu}+R_{\bar\alpha\bar\nu\bar\beta\bar\mu}+R_{\bar\alpha\bar\mu\bar\nu\bar\beta}=0
\end{equation}
By taking into account these identities, we can show that $R_{\bar\alpha\bar\beta\bar\mu\bar\nu}$ contains exactly 20 independent degrees of freedom.  This means it is the full list of combinations of $g_{\bar\mu\bar\nu,\bar\alpha\bar\beta}$ that cannot be made to vanish and therefore a good measure of curvature.

We have defined $R_{\bar\alpha\bar\beta\bar\mu\bar\nu}$ in (\ref{inertialframeR}) in a local inertial reference frame.  We would like to define a tensor that reduces to this given form in a local inertial reference frame.  To do this note that the covariant derivative commutes when acting on scalars (assuming it is torsion free) but not when acting on vectors. Thus, it is natural to consider $R\indices{^\mu_\nu_\alpha_\beta}$ defined through
\begin{equation}
\nabla_\beta\nabla_\alpha V^\mu - \nabla_\alpha \nabla_\beta V^\mu = R\indices{^\mu_\nu_\alpha_\beta} V^\nu
\end{equation}
The object $R\indices{^\mu_\nu_\alpha_\beta}$ is a tensor.  This follows from the fact that $g_{\mu\nu}$ is a tensor, covariant differentiation returns tensors, and the quotient theorem. Further, as we will see, it contains second derivatives of the metric (but not higher derivatives) so it possible it reduces to $R_{\bar\alpha\bar\beta\bar\mu\bar\nu}$ in the locally inertial reference frame (once the first index has been lowered).  To see that this is true first note, using (\ref{covupperindices}), that
\begin{equation}\label{RusingGammas}
R\indices{^\alpha_\beta_\mu_\nu} = \Gamma\indices{^\alpha_{\beta\nu,\mu}} - \Gamma\indices{^\alpha_{\beta\mu,\nu}}
+ \Gamma\indices{^\alpha_{\sigma\mu}}\Gamma\indices{^\sigma_{\beta\nu}} - \Gamma\indices{^\alpha_{\sigma\mu}}\Gamma{^\sigma_{\beta\mu}}
\end{equation}
We can write (\ref{RusingGammas}) in any indices we want (as these are the components of a tensor).  If we take these indices to be the barred indices of the local inertial reference frame  then $\Gamma\indices{^{\bar\alpha}_{\bar\beta\bar\mu}}=0$.  Now only the first two terms on the right hand side of (\ref{RusingGammas}) survive (expressed in the barred indices). Using (\ref{Gammausingmetric}), we obtain (\ref{inertialframeR}).  Hence $R\indices{^\alpha_\beta_\mu_\nu}$ has the properties we looking for.  It is called the \emph{curvature tensor}. \index{curvature tensor} Now we know that this is a tensor we can lift the identities (\ref{inertialRidentone}, \ref{inertialRidenttwo}) to a general reference frame.  We have the identities
\begin{equation}\label{Ridentone}
R_{\alpha\beta\mu\nu}= -R_{\beta\alpha\mu\nu}=-R_{\alpha\beta\nu\mu}=R_{\mu\nu\alpha\beta}
\end{equation}
\begin{equation}\label{Ridenttwo}
R_{\alpha\beta\mu\nu}+R_{\alpha\nu\beta\mu}+R_{\alpha\mu\nu\beta}=0
\end{equation}
true in any frame.

We can use the curvature tensor to define further important objects.  The \emph{Ricci tensor} \index{Ricci tensor} is defined as
\begin{equation}
R_{\mu\nu} := R\indices{^\lambda_\mu_\lambda_\nu}
\end{equation}
Note that it follows from the above identities that (i) the Ricci tensor is symmetric and (ii) all other contractions of the curvature tensor either vanish or are equal to $\pm R_{\mu\nu}$.  The Ricci scalar is defined as \index{Ricci scalar}
\begin{equation}
R := g^{\mu\nu} R_{\mu\nu}
\end{equation}

We can transform to a coordinate system such that any given point, we have a local inertial reference frame.  However, we cannot find a transformation that does this for all points simultaneously. This is clear since the curvature tensor will not vanish in general. Hence according to (\ref{Taylormetric}) there will, in general, be $O( (x^{\bar\alpha}- x^{\bar\alpha}(p))^2)$ correction terms for the Minkowski metric near any point we have an inertial reference frame.  We can, of course, impose flatness everywhere through the condition
\begin{equation}
R\indices{^\alpha_\beta_\mu_\nu}(x)=0  ~~~ \forall x
\end{equation}
This returns us to the flat world of Special Relativity.

\subsubsection{Third derivatives and the Bianchi identity}\label{Sec:Bianchi}

\index{metric!third derivatives}

We could continue the exercise of differentiating the metric and looking for tensors in terms of higher derivatives of the metric.  However, General Relativity does not use any such higher derivatives with the exception of one important example.  If we differentiate (\ref{RusingGammas}) then evaluate in a local inertial reference frame we obtain
\begin{equation}
R_{\bar\alpha\beta\bar\mu\bar\nu,\bar\lambda} = \frac{1}{2}
(g_{\bar\alpha\bar\nu,\bar\beta\bar\mu\bar\lambda} - g_{\bar\alpha\bar\mu,\bar\beta\bar\nu\bar\lambda}
+g_{\bar\beta\bar\mu,\bar\alpha\bar\mu\bar\lambda} - g_{\bar\beta\bar\nu,\bar\alpha\bar\mu\bar\lambda})
\end{equation}
From the symmetry of the metric and the symmetry of partial differentiation we see that
\begin{equation}
R_{\bar\alpha\bar\beta\bar\mu\bar\nu,\bar\lambda} + R_{\bar\alpha\bar\beta\bar\lambda\bar\mu,\bar\nu} + R_{\bar\alpha\bar\beta\bar\nu\bar\lambda,\bar\mu} = 0
\end{equation}
Thus, by replacing partial derivatives with covariant derivatives we get the tensorial identity
\begin{equation}
R_{\alpha\beta\mu\nu;\lambda} + R_{\alpha\beta\lambda\mu;\nu} + R_{\alpha\beta\nu\lambda;\mu} = 0
\end{equation}    \index{Bianchi identity}
valid in any reference frame. This is called the Bianchi identity. It plays an important role in General Relativity.  In particular it follows from the Bianchi identity and the identities (\ref{Ridentone}, \ref{Ridenttwo}) that the Einstein tensor, defined as
\begin{equation}
G^{\mu\nu} := R^{\mu\nu} - \frac{1}{2} g^{\mu\nu} R
\end{equation}
satisfies the identity
\begin{equation} \label{Gconserved}
\nabla_\nu G^{\mu\nu} = 0
\end{equation}

\subsection{Matter}\label{sec:equivalenceprinc}

So far we have, essentially, only been concerned with setting up the mathematics for General Relativity.  We have introduced a metric field.  This plays the role of the gravitational field in General Relativity.  We have also chosen particular connection motivated by physical reasons.  However, we have not yet introduced matter fields nor have we introduced any physical equations that the various fields (gravitational and matter) must obey.   The so called \emph{matter fields} \index{matter fields}  is the name given to all the other fields apart from the gravitational field, $g_{\mu\nu}$. \index{gravitational field} To obtain the appropriate matter field equations in General Relativity we use a technique called minimal substitution on the field equations from Special Relativity.  Hence we will start by reviewing the matter field equations from Special Relativity.

\subsubsection{Matter in Special Relativity}\label{Sec:matterinSR}

In Special Relativity space-time is flat so that there is a global inertial reference frame with a global coordinate system $x^{\bar\mu}$ (which can be subject to Lorentz transformations). In four dimensions these coordinates correspond to $(ct, x, y, z)$.  We can consider many types of matter in Special Relativity and write down field equations for them.  These field equations pertain to the inertial coordinate system.  \index{matter field equations!Special Relativity}

Before doing this, however, it is useful to introduce the stress-energy tensor. \index{stress energy tensor} Each type of matter, $n$, will have associated with it a stress-energy tensor, $T^{\mu\nu}[n]$.  This is defined such that, in a local inertial reference frame, the $\bar\mu\bar\nu$ component of this is equal to the flux of $\bar\mu$ momentum across a surface of constant $x^{\bar\nu}$.

The stress energy tensor is symmetric ($T^{\bar\mu\bar\nu}[n]=T^{\bar\nu\bar\mu}[n]$).    The equalities $T^{\bar 0 \bar k}=T^{\bar k \bar 0}$ (where $k=1,2,\dots$) follow virtue of the way the stress energy is defined.  To show $T^{kl}=T^{lk}$ (where $k,l=1,2,\dots$) we can make a physical argument, namely that symmetry follows from fact that the torque on an element must tend to zero at least as fast as the moment of inertia as we decrease the fluid element size.  See \cite{misner1973gravitation} for details.

The energy current associated with matter of type $n$ as seen in a local inertial rest frame in which the four velocity, $v^{\bar\mu}$, is given by
\begin{equation}
j^{\bar\mu}=g_{\bar\alpha\bar\nu} T^{\bar\mu\bar\alpha}[n]v^{\bar\nu}
\end{equation}
This follows from the definition of the stress-energy tensor.

For each field we have
\begin{equation}\label{Twithforce}
\partial_{\bar\nu} T^{\bar\mu\bar\nu}[n] = G^{\bar\mu}
\end{equation}
where $G^{\bar\nu}$ is the force density acting on the fluid element at $x$.

The total stress-energy tensor, $T^{\bar\mu\bar\nu}=\sum_n T^{\bar\mu\bar\nu}[n]$, is found to satisfy the following conservation equation in Special Relativity
\begin{equation}\label{Tconservation}
\partial_{\bar\nu} T^{\bar\mu\bar\nu} =0
\end{equation}
since all the force densities cancel at each $x$ by Newton's second law.  This corresponds to conservation of $\bar\mu$ momentum.

It must be possible to derive the equations (\ref{Twithforce}, \ref{Tconservation}) from the matter field equations since these field equations are supposed to fully describe the behaviour of the fields.  For a dust the converse is also true.  The matter field equations for dust fluid can actually be derived from  (\ref{Twithforce}, \ref{Tconservation}).

The types of matter in Special Relativity include
\begin{description}
\item[Dust] \index{dust!Special Relativity} is a fluid which is fully described by fields $(U^{\bar\mu}, \rho)$ where $U^{\bar\mu}$ is the four-velocity of a fluid cell at $x$ (four-velocities satisfy $\eta_{\alpha\beta} U^\alpha U^\beta =1$) and $\rho$ is the energy density in the rest frame of the fluid.  If we think of a fluid as being a course-grained way of treating a bunch of particles moving in various directions, then for a dust, all the particles in a fluid cell are moving with the same four-velocity.  The stress energy tensor for dust is given by
    \begin{equation}
    T^{\bar\mu\bar\nu}[\text{dust}] = \rho U^{\bar\mu} U^{\bar \nu}
    \end{equation}
    as $U^{\bar\nu}$ is the flux with which four momentum, $\rho U^{\bar\mu}$, flows in the $\bar\nu$ direction.   For a dust, the matter field equations are given by (\ref{Twithforce})
    \begin{equation}
    \partial_{\bar\nu} (\rho U^{\bar\mu} U^{\bar\nu}) = G^{\bar\mu}[\text{dust}]
    \end{equation}
    where $G^{\bar\mu}[\text{dust}]$ is the force density acting on the dust.
\item[Perfect fluids] \index{perfect fluids!Special Relativity} are fluids which, everywhere, look isotropic in their rest frame.  They are described by fields $(U^{\bar\mu}, \rho, P)$ where $U^{\bar\mu}$ and $\rho$ are defined as for a dust and $P(x)$ is the pressure at $x$ as measured in the rest frame of the fluid. The stress energy tensor for a perfect fluid is
    \begin{equation}
    T^{\bar\mu\bar\nu}[\text{perfect fluid}] = (\rho + P) \rho U^{\bar\mu} U^{\bar\nu} + P \eta^{\bar\mu\bar\mu}
    \end{equation}
    Here $\eta^{\bar\mu\bar\nu}=\text{diag}(-1, 1, 1, \dots)$ is the inverse of $\eta_{\bar\mu\bar\nu}$.  The equations of motion are given by
    \begin{equation}
    \partial_{\bar\nu} ((\rho + P) \rho U^{\bar\mu} U^{\bar\nu} + P \eta^{\bar\mu\bar\mu}) = G^{\bar\mu}[\text{perfect fluid}]
    \end{equation}
    and the fact that $\partial_\mu (\eta_{\alpha\beta} U^\alpha U^\beta)=0 $ (since $\eta_{\alpha\beta} U^\alpha U^\beta =1$).  These do not constitute a full set of equations.  To provide these we also need an \emph{equation of state} which provides some relationship between $\rho$ and $P$. For example, for a radiation fluid (consisting only of electromagnetic radiation) $\rho=3P$.
\item[Electromagnetic fields] \index{electromagnetic fields!Special Relativity} denoted by an antisymmetric matrix $F_{\bar\mu\bar\nu}$. In Special Relativity this satisfies Maxwell's equations
    \begin{equation}
    \partial_{\bar\mu} F^{\bar\nu\bar\mu} = 4\pi J^{\bar\nu}  ~~~~~
    \partial_{\bar\mu} F_{\bar\nu\bar\lambda} + \partial_{\bar\lambda} F_{\bar\mu\bar\nu} + \partial_{\bar\nu} F_{\bar\lambda\bar\mu} = 0
    \end{equation}
    Here $J^{\bar\nu}$ is the charge current density associated with a charged fluid.
    The stress-energy tensor for an electromagnetic field is given by
    \begin{equation}\label{Telectromagnetic}
    T^{\bar\mu\bar\nu}[\text{electromagnetic}]=
    F\indices{^{\bar\mu}_{\bar\gamma}} F^{\bar\nu\bar\gamma}
    - \frac{1}{4} \eta^{\bar\mu\bar\nu} F_{\bar\gamma\bar\delta} F^{\bar\gamma\bar\delta}
    \end{equation}
    It can be shown from Maxwell's equations that this is conserved if the charge current is zero everywhere.
\item[Charged fluid] \index{charged fluid!Special Relativity} A dust or perfect fluid can be charged.  In this case we have an additional field, $q$. This is a scalar and defined to be equal to the charge density in the rest frame of the fluid.   The charge current associated with the fluid is given by
    \begin{equation}
    J^{\bar\mu} = q U^{\bar\mu}
    \end{equation}
    The fluid will experience a force density
    \begin{equation}
    G^{\bar\mu}= \eta^{\bar\mu\bar\gamma} F_{\bar\gamma\bar\nu} J^{\bar\nu}
    \end{equation}
    Assuming there are no other forces present, we have
    \begin{equation}
    \partial_{\bar\nu} (T^{\bar\mu\bar\nu}[\text{charged fluid}] + T^{\bar\mu\bar\nu}[\text{electromagnetic}]) = 0
    \end{equation}
    Indeed, one way to motivate the expression in (\ref{Telectromagnetic}) is to impose conservation of the total stress energy tensor in the presence of a charged fluid.
\end{description}
There are many other types of fields considered (general fluids, Klein-Gordon fields, Yang-Mills fields, scalar fields \dots).

\subsubsection{Minimal substitution technique}\label{Sec:minimalsubstitution}

We get field equations for the matter fields in General Relativity by taking the field equations from Special Relativity and adopting them to General Relativity.  This gives us matter field equations for General Relativity.  We are guided in this process by a few principles.  First we adopt the \emph{principle of general covariance} - that the laws of physics should be written in a form which is invariant under general coordinate transformations (the equations look the same before and after a general coordinate transformation).  This is guaranteed by writing them in terms of tensors.  Second, we adopt the \emph{principle of equivalence}.  The  equivalence principle says that, if we are in a local inertial reference frame, the laws of physics should be those given by Special Relativity.   Assume we have the Special Relativistic laws written down in tensorial form but where we are constrained to local inertial coordinates, $x^{\bar \mu}$ (in four dimensions these are$(ct, x, y, z)$).  The laws are then invariant under Lorentz transformations.  Then a prescription for implementing the equivalence principle is by using the following \emph{minimal substitution rule}:  \index{minimal substitution}
\begin{enumerate}
\item Replace the Minkowski metric, $\eta_{\bar\mu\bar\nu}$, by $g_{\mu\nu}$.
\item Replace partial derivatives, $\partial_{\bar\mu}$, by covariant derivatives, $\nabla_{\mu}$.
\item Replace all remaining indexes associated with inertial coordinates (represented by $\bar\mu$, $\bar\nu$, \dots) with indexes associated with general coordinates (indexed by $\mu$, $\nu$, \dots).
\end{enumerate}
When we re-specialize to a local inertial reference frame this will give us back the Special Relativistic equations by virtue of our choice of covariant derivative using the connection in (\ref{Gammausingmetric}) above.

In some cases the minimal substitution prescription above is ambiguous.  For example, we could replace $\partial_{\bar\mu}\partial_{\bar\nu} A^{\bar\gamma}$ with $\nabla_\mu\nabla_\nu A^\gamma$ or we could note that it is equal to $\frac{1}{2}(\partial_{\bar\mu}\partial_{\bar\nu} A^{\bar\gamma} + \partial_{\bar\nu}\partial_{\bar\mu} A^{\bar\gamma})$ and replace it with $\frac{1}{2}(\nabla_\mu\nabla_\nu A^\gamma + \nabla_\nu\nabla_\mu A^\gamma)$ which is different (since the covariant derivative does not commute).  In such cases we have to find other physical reasons to make the right choice.  For the examples of matter fields given in Sec.\ \ref{Sec:matterinSR} there is no such ambiguity.

The minimal substitution rule is not the only way of guaranteeing that we are consistent with Special Relativity.  We could, for example, also add terms depending on the curvature tensor to the Special Relativistic equations.  This would give us back Special Relativity for flat space as then the curvature tensor vanishes. It would, however, be inconsistent with the equivalence principle.

\subsubsection{Matter field equations in General Relativity}\label{Sec:MatterinGR}

We can now use the minimal substitution technique to obtain the matter field equations for General Relativity along with the stress-energy tensor for each matter field. Importantly, after minimal substitution, we are in general coordinates, $x^\mu$ (and not restricted to inertial coordinates, $x^{\bar\mu}$).  The stress energy tensor for each field, $n$, now obeys
\begin{equation}
\nabla_\nu T^{\mu\nu}[n]= G^\mu
\end{equation}
The total stress energy tensor now obeys the \lq\lq conservation" equation
\begin{equation}
\nabla_{\nu} T^{\mu\nu} = \partial_\nu T^{\mu\nu} + \Gamma\indices{^\mu_\sigma_\nu} T^{\sigma\nu} + \Gamma\indices{^\nu_\nu_\sigma} T^{\mu\sigma} = 0
\end{equation}
We put inverted commas around the word conservation since there is actually a correction arising from the connection.  The physical reason for this is that the gravitational field can actually put energy momentum into (or out of) the matter degrees of freedom.

For the examples we discussed earlier we obtain
\begin{description}
\item[Dust.] \index{dust!General Relativity} The stress energy tensor for dust is given by
    \begin{equation}
    T^{\mu\nu}[\text{dust}] = \rho U^{\mu} U^{ \nu}
    \end{equation}
    as before (but in general coordinates).  The matter field equations are now
    \begin{equation}\label{dusteqnofmotion}
    \nabla_{\nu} (\rho U^{\mu} U^{\nu}) = G^{\mu}[\text{dust}]
    \end{equation}
    where $G^\mu[\text{dust}]$ is the external force density acting on the dust.
\item[Perfect fluids.] \index{perfect fluids!General Relativity} The stress energy tensor for a perfect fluid is
    \begin{equation}
    T^{\mu\nu}[\text{perfect fluid}] = (\rho + P) \rho U^{\mu} U^{\nu} + P g^{\mu\mu}
    \end{equation}
    The equations of motion are given by
    \begin{equation}
    \nabla_{\nu} ((\rho + P) \rho U^{\mu} U^{\nu} + P g^{\mu\mu}) = G^{\mu}[\text{perfect fluid}]
    \end{equation}
    and an appropriate equation of state.
\item[Electromagnetic fields] \index{electromagnetic fields!General Relativity} Maxwell's equations in General Relativity are
    \begin{equation}
    \nabla_{\mu} F^{\nu\mu} = 4\pi J^{\nu}  ~~~~~
    \nabla_\mu F_{\nu\lambda} + \nabla_{\lambda} F_{\mu\nu} + \nabla_{\nu} F_{\lambda\mu} = 0
    \end{equation}
    The stress-energy tensor for an electromagnetic field is given by
    \begin{equation}\label{TelectromagneticGR}
    T^{\mu\nu}[\text{electromagnetic}]=
    F\indices{^{\mu}_{\gamma}} F^{\nu\gamma}
    - \frac{1}{4} g^{\mu\nu} F_{\gamma\delta} F^{\gamma\delta}
    \end{equation}
\item[Charged fluid] \index{charged fluids!General Relativity} The fluid will experience a force density
    \begin{equation}
    G^{\mu}= g^{\mu\gamma} F_{\gamma\nu} J^{\nu}
    \end{equation}
    Assuming there are no other forces present, we have
    \begin{equation}
    \nabla_{\nu} (T^{\mu\nu}[\text{charged fluid}] + T^{\mu\nu}[\text{electromagnetic}]) = 0
    \end{equation}
\end{description}

\subsubsection{Small fluid blobs move along geodesics}\label{Sec:fluidflowalonggeodesics}

In his original formulation of General Relativity, Einstein included the additional postulate that test particles (these have vanishingly small rest frame energy) move along geodesics.  However, it turns out that we do not need this bolt-on postulate.  Rather, it can be proven, under very weak assumptions, that small fluid blobs subject to no external forces move along geodesics (at least in the limit as the fluid blob has arbitrarily small rest frame energy and size). \index{blobs} \index{geodesics}

Let $U^\alpha$ be the four velocity.   The covariant derivative along the direction of motion is $U^\alpha \nabla_\alpha$.  For geodesic motion, we demand that the velocity does not change in the direction of motion (with respect to this covariant derivative).  This gives us the following definition for geodesic motion:
\begin{equation}\label{geodesiccondition}
U^\alpha \nabla_\alpha U^\beta = 0
\end{equation}

For a dust fluid it follows immediately from the equations of motion (\ref{dusteqnofmotion}) that the fluid elements flow along geodesics.  Consider
\begin{equation}
\begin{split}
   0= \nabla_\nu T^{\mu\nu}[\text{dust}] & = \nabla_\nu (\rho U^{\mu} U^{ \nu}) \\
    & = \rho U^\nu \nabla_\nu U^\mu + \rho U^\mu \nabla_\nu U^\nu +  U^\mu U^\nu \partial_\nu \rho
\end{split}
\end{equation}
Now we note that the second two terms, $\rho U^\mu \nabla_\nu U^\nu +  U^\mu U^\nu \partial_\nu \rho$, are parallel to $U^\mu$.  We can see that the first term, on the other hand, is perpendicular to $U^\mu$ (by contracting it it by $g_{\alpha\mu} U^\alpha$ and using the fact that the covariant derivative of $g_{\alpha\mu} U^\alpha U^\mu=1$ must vanish).  Hence (\ref{geodesiccondition}) is satisfied and we have geodesic motion.

In the case of more general matter have been a number of attempts to show that small blobs follow geodesics \cite{einstein1938gravitational, geroch1975motion, ehlers2004equation}. The most compelling is due to Geroch and Jang \cite{geroch1975motion}.  They show that if (i) we have $\nabla_\nu T^{\mu\nu}=0$ and (ii) the \emph{strengthened dominant energy condition} (SDEC) on $T^{\mu\nu}$ holds, then any world tube outside of which $T^{\mu\nu}$ vanishes must centre around a geodesic (in the limit of small $\rho$ and small radius for this world tube (see \cite{malament2006classical} for a good discussion of this.  See also the discussion by Weatherall \cite{weatherall2011status} who raises some concerns).  The SDEC on $T^{\mu\nu}$ is that, for any time like $v^\mu$, we have (a) $v^\mu v^\nu T_{\mu\nu} \geq 0$ everywhere and (b) $T^\mu_\nu v^\nu$ is time-like where ever $T^{\mu\nu}\not=0$.
It is worth noting that blobs will not always follow geodesics if the appropriate conditions are not met.  For example, spinning blobs, or blobs with non-zero quadrupole moment may not follow geodesics.

\subsection{The Einstein field equations}\label{Sec:Einsteinfieldeqns}

The Special Relativistic field equations form a simultaneous set of equations that are just sufficient to find a solution given appropriate boundary data.  I.e. we have the same number of equations as unknowns.  When we perform the minimal substitution procedure, we introduce additional unknowns, namely the components of the metric (there are 10 of these in four dimensional space time).  However, the minimal substitution procedure does not, by itself, increase the number of equations.  Hence, we need some additional equations. For example, we need 10 additional equations in four dimensional space time (there is an important subtly which actually reduces this to 6 independent equations - we will discuss this below in Sec.\  \ref{Sec:Bianchiresolution}).  It is also worth noting that the matter field equations obtained by minimal substitution only depend on zeroth and first derivatives of the metric (as required by the equivalence principle).  The Einstein field equations also depend on second derivatives (but not higher).  This distinguishes them among the field equations.

Newton's law of gravitation can be expressed as a Poisson equation
\begin{equation}
(\partial_x^2 +\partial_y^2 + \partial_z^2) \phi = 4\pi G \rho
\end{equation}
where $\phi$ is the gravitational potential, $\rho$ is the mass density and $G$ is a constant.  On the left we have gravitational degrees of freedom.  Note in particular, that we have second derivatives of the potential. On the right we have matter degrees of freedom.  In relativity the analogue of $\rho$ is $T^{\mu\nu}$.  Hence, it is reasonable to posit that the General Relativistic equations replacing the Poisson equation above have the form
\begin{equation}
Q^{\mu\nu} = k T^{\mu\nu}
\end{equation}
where $Q^{\mu\nu}$ is a $(2,0)$ tensor depending only on the metric up to second derivatives.
By the equivalence principle we also require $\nabla_\nu T^{\mu\nu} = 0$.  Hence we require $\nabla_\nu Q^{\mu\nu} =0$.  From (\ref{Gconserved}), we see that
\begin{equation}\label{EinsteinsQchoice}
Q^{\mu\nu} = G^{\mu\nu} + \Lambda g^{\mu\nu}
\end{equation}
has the required property.  Here $\Lambda$  is a constant and we use metric compatibility to see that the second term vanishes when we calculate $\nabla_\nu Q^{\mu\nu}$.   This gives
\begin{equation}
G^{\mu\nu} + \Lambda g^{\mu\nu} = 8\pi T^{\mu\nu}
\end{equation}
This is \emph{Einstein's field equation}. \index{Einstein's field equation}  The constant $k$ has been fixed (in units where $G=1$ and $c=1$) by comparison with Newton's law.  $G$ and $\Lambda$ are constants of nature - they must be constant across the entire universe. $G$ is Netwon's constant, and $\Lambda$ is called the \emph{cosmological constant}.  Modern measurements suggest $\Lambda$ is small but non-zero.  It can be ignored except at cosmological scales.  The Einstein field equation provides the necessary extra 10 equations (in four dimensions) so that we can solve for the matter fields and the metric (modulo the important subtly we will discuss below).  It is worth noting that there is no tensor, $Q^{\mu\nu}$, that depends only on the metric and first derivatives (other than functions of the metric itself) as we noted in Sec.\ \ref{sec:firstderivative}.  Hence, we have to go upto second derivatives.  There is an interesting theorem due to Lovelock \cite{lovelock1972four} that $G_{\mu\nu}$ and $g_{\mu\nu}$ are the only divergence free tensors on a four dimensional manifold formed from the metric and its first two derivatives.  This suggests that, for the four dimensional case at least, the choice in  (\ref{EinsteinsQchoice}) is necessary.  For higher dimensions a similar statement is true if we insist that $Q^{\mu\nu}$ is linear in second derivatives of the metric \cite{cartan1922equations, weyl1922space, vermeil1917notiz}.

\subsection{General Relativity as a system of coupled field equations}\label{Sec:GRascoupledfieldeqns}

A general physical situation in General Relativity involves solving a system of coupled field equations.  These comprise the matter field equations (derived from the Special Relativistic matter field equations by the minimal substitution rule) along with Einstein's field equation.

Let us denote the matter fields by $\vec{F} = \{ F[nj] :j=1 ~\text{to} ~ J_n, n= 1 ~\text{to}~ N \}$ where $F[nj]$ is a tensor field (could be a scalar, a vector, or a higher rank tensor of some type). We have labeled the different types of matter by $n$.  For each type of matter we may have one or more fields (labeled by $j$).  For example, for a perfect fluid we have $(U^\mu, \rho, P)$ - i.e. a vector field and two scalar fields.  Let $a_{nj}$ be the number of real degrees of freedom in $F[nj]$.  For example, if $F[nj]$ is a symmetric rank two tensor for some given $(n, j)$ then $a_{nj}=10$ in four dimensional spacetime.   The total number of matter degrees of freedom is
\begin{equation}
a=\sum_{n=1}^N \sum_{j=1}^{J_n} a_{nj}
\end{equation}
In addition to these degrees of freedom we also have $g_{\mu\nu}$ (which has $D/2(D+1)$ degrees of freedom where $D$ is the dimension of spacetime).

We have a number of matter field equations
\begin{equation}\label{GeneralMatterfieldeqn}
f_l(\vec{F}, \vec{F}_{,\alpha}, \vec{F}_{,{\alpha\beta}}, g_{\mu\nu}, g_{\mu\nu,\gamma} ) = 0  ~~~~~l=1~\text{to}~L
\end{equation}
in addition to Einstein's field equation
\begin{equation}\label{GeneralEinsteinfieldeqn}
G^{\alpha\beta}(g_{\mu\nu}, g_{\mu\nu,\gamma}, g_{\mu\nu,\gamma\delta}) = 8\pi T^{\alpha\beta}(\vec{F}, \vec{F}_{,\gamma}, g_{\mu\nu}, g_{\mu\nu,\gamma})
\end{equation}
We have indicated the dependencies on the matter and gravitational fields and their derivatives.  These functional forms cover all the examples of matter fields that are commonly considered.  It is possible that some matter fields require more general functional forms.  The equivalence principle, however, prohibits $f_l$ in (\ref{GeneralMatterfieldeqn}) from depending on $g_{\mu\nu,\gamma\delta}$ or higher derivatives.  The form of the functions $f_l(\cdot)$ are constrained by the fact that the equations (\ref{GeneralMatterfieldeqn}) must be written in tensorial form (as in the examples we gave in Sec.\ \ref{Sec:MatterinGR}).  Consequently they each have a form that is invariant under general coordinate transformations.   Assuming that all the equations in (\ref{GeneralMatterfieldeqn}) are independent and that equation $l$ contains $b_l$ independent equations when written out in full, then we require that
\begin{equation}
\sum_{l=1}^L b_l = a
\end{equation}
so that we have the same number of independent matter field equations is the same as the number of degrees of freedom for the matter fields.

In the examples usually considered, the stress energy tensor for a given matter type only depends on the matter fields associated with that type of matter. Then we can write the functional dependence as
\begin{equation}
T^{\mu\nu}[n]= (\{F[ni], F_{,\gamma}[ni]: i=1~\text{to}~ I_n\}, g_{\mu\nu}, g_{\mu\nu,\gamma} )
\end{equation}
Note that we still require dependence on $g_{\mu\nu}$.  In some cases (e.g.\ a Klein Gordon field) there will also be dependence on $g_{\mu\nu,\gamma}$.  The total stress energy is given by
\begin{equation}
T^{\mu\nu} = \sum T^{\mu\nu}[n]
\end{equation}
In principle it is possible that there are contributions to the stress energy arising from interactions between the fields.  However, for the examples discussed in the literature this is not the case.

If the different types of matter do not interact with one another then there will be separate equations for each type of matter. Then we will have separate equations for each type of matter. \begin{equation}
f_{nj}(\{F[ni], F_{,\gamma}[ni], F_{,\gamma\delta}: i=1~\text{to}~ I_n\}, g_{\mu\nu}, g_{\mu\nu,\gamma} ) =0
\end{equation}
for $j=1$ to $J_n$ and $n=1$ to $N$.  If, however, they interact (for example a charged fluid interacts with an electromagnetic field) then there will be terms corresponding to the interaction.

All these equations can be obtained by general techniques by writing down a Lagrangian density and then extremizing an action.

All the field equations relate matter and gravitational degrees of freedom.  In this sense the Einstein field equation is not special.  However, when we look a little closer Einsten's field equation is special among the  field equations in General Relativity for various reasons.
\begin{enumerate}
\item It is the only equation to depend on the second derivative of the metric, $g_{\mu\nu,\gamma\delta}$ (this is a consequence of the equivalence principle). Correspondingly, it is the only equation that concerns the curvature.
\item It only depends on the matter fields through the total stress energy tensor. This is specified by ten components. Thus the immediate effect of all the possible complexity of matter on the curvature is, at each point, $x$, reduced to only ten parameters.  Of course, the matter field equations do depend on the metric and its first derivative so the metric does \lq\lq see" all the complexity of matter beyond the filter of the stress-energy tensor.
\item The left hand side is completely geometrical (depends only on the metric).  Apart from the cosmological constant (which is often set to zero), it has no free parameters.  If we change the left hand side then we are in a different physical theory for gravity.
\item The Einstein field equation satisfies the property that the divergence of the left hand side is identically equal to zero (with the consequence that the divergence of the stress energy tensor must also have zero divergence).  We will see in Sec.\ \ref{Sec:Bianchiresolution} that this odd property corresponds to something especially curious:  among all the field equations, it is Einstein's field equation that shoulders the burden resulting from invariance in form (of all the field equations) under general coordinate transformations.
\end{enumerate}

\subsection{The hole argument and diffeomorphism invariance}\label{Sec:HoleArgument}

We will now discuss the \emph{hole argument}. \index{hole argument} This was originally developed by Einstein in 1913 and revived in the modern context by philosophers John Earman \cite{earman1986space}, John Stachel \cite{stachel1989einstein}, and John Norton \cite{norton1993general} (so called John${}^3$).  This has had a big impact on the development of theories of Quantum Gravity (especially in the loop quantum gravity community \cite{rovelli2007quantum}). Stachel has written a recent review \cite{stachel2014hole}.  Much has been written on it by both philosophers and physicists (for example, \cite{macdonald2001einstein, rickles2005new, smolin2006case, christian200114}).

A solution in General Relativity is given by specifying metric and matter fields at all points on some manifold, that is:
\begin{equation}
{\tilde{\Psi}}=\bigg\{\big(p, {\bf g}(p), \vec{\mathbf{F}}(p)\big): \forall p\in\mathscr{M}\bigg\}
\end{equation}
That is we must specify a manifold and the metric and matter fields at all points on the manifold. Here ${\bf g}$ is the metric field tensor (in basis independent notation) and $ \vec{\mathbf{F}}(p)$ is the list of matter fields (in basis independent notation).  This way of writing the solution is a useful abstraction for reasoning about General Relativity.  However, to actually write down a solution we need to cover the manifold by charts and write down the components of the tensors in the given coordinate systems.  If $\cup_i \mathscr{O}_i=\mathscr{M}$ where $\mathscr{O}_i$ are open regions and $V_i\subseteq \mathbb{R}^D$ are a set of charts for which there exists an one-to-one smooth map such that $\omega_i(p) = x\in V_i$ for $p\in \mathscr{O}_i$ (as described in Sec.\ \ref{Sec:Manifolds}) then we can represent a solution as
\begin{equation}\label{solutioncharted}
{\tilde{\Psi}}[\{V_i\}]= \bigg\{\Big(i, \Big\{\big(x, g_{\mu\nu}(x), \vec{F}(x)\big):\forall x\in V_i\Big\}\Big): \forall i\bigg\}
\end{equation}
This is a set of $i$-indexed sets that cover the manifold.  Where these sets correspond to overlapping regions on the manifold there are obvious consistency constraints (so that the fields are related by the same map as the charts).

\subsubsection{The argument}

Many solutions to the field equations are possible.  However, if we specify sufficient boundary data then we expect to have a unique solution.
One way to set up this intuition is to imagine an open set of the manifold $\mathscr{ H}\subset \mathscr{M}$ having the same dimension as the manifold.  We call this the \emph{hole}.  Further, let this be fully enclosed by a thick \emph{shell}, $\mathscr{S}\subseteq \mathscr{M}$.  We impose that $\mathscr S$ and $\mathscr H$ are disjoint and that the boundary of $\mathscr H$ is everywhere adjacent to $\mathscr S$ (there are no gaps between $\mathscr S$ and $\mathscr H$).  We suppose that $\mathscr S$ forms a thick shell around the hole.  For simplicity, we will further assume that $\mathscr{H}\cup \mathscr{S}$ is covered by a single chart.  Now we can imagine imagine specifying $g_{\mu\nu}(x)$ and $\vec{F}(x)$ in a given coordinate system at all points in the shell.  Given the thickness of the shell, it is reasonable to expect that such boundary data would fully determine the solution in the hole.

However, consider the following argument.  Let
\begin{equation}\label{holesolutionA}
\big\{(x, g_{\mu\nu}(x), \vec{F}(x)): x\in C_\mathscr{H}\big\}
\end{equation}
be a solution in the hole given a full specification of boundary data
\begin{equation}
\big\{(x, g_{\mu\nu}(x), \vec{F}(x)): x\in C_\mathscr{S}\big\}
\end{equation}
in the shell.  Here $C_\mathscr{H}$ are the coordinates corresponding to points in $\mathscr{H}$ for the hole and similarly for the shell.  We obtain this solution for the hole (in \eqref{holesolutionA}) by solving the field equations (\ref{GeneralMatterfieldeqn}, \ref{GeneralEinsteinfieldeqn}).  Now consider a general invertible transformation to a new coordinate system, $x'=x'(x)$, which leaves the coordinates for points in $\mathscr S$ unchanged (so we have the same boundary data). We can regard this as a passive transformation, and we obtain
\begin{equation}
\big\{ (x', {g'}_{\mu'\nu'}(x'), \vec{F}'(x')): x\in C_\mathscr{H}\big\}
\end{equation}
where we have put primes on the $g'$ and $\vec{F}'$ to indicate that the functions have changed (because we have performed tensor transformations).  Now the form of the field equations (\ref{GeneralMatterfieldeqn}, \ref{GeneralEinsteinfieldeqn}) are invariant under general coordinate transformations.  The field equations have the same form when written down explicitly in either the $x$ or the $x'$ coordinates.  Hence
\begin{equation}
\big\{(x,{g'}_{\mu\nu}(x), \vec{F}'(x)): x\in C_\mathscr{H}\big\}
\end{equation}
must also be a solution in the original $x$ coordinate system.  Here $g'_{00}(\cdot)=g'_{0'0'}(\cdot)$, $g'_{01}(\cdot)=g'_{0'1'}(\cdot)$, \dots (with similar equations for the matter fields).  In fact there are infinitely many such general transformations. It now looks like we have infinitely many different solutions in the hole in the same coordinate system given the same boundary data.   This appears to be a radical breakdown of determinism - the field equations simply fail to determine the solution uniquely.

\subsubsection{An example}

\paragraph{Smooth transformations.}

One objection might be that we cannot have smooth transformations that leave the coordinates in the shell unchanged while transforming points in the hole.  Certainly it is true that such a transformation cannot be analytic (since, if it was, we would be able to determine the coordinates in the vicinity of the boundary by a Taylor expansion).  However, we can use \emph{bump functions}.  A bump function is is equal to zero outside any given open set, smooth everywhere, and can be equal to any function inside any compact subset of the aforementioned open set.  We can use bump functions to implement arbitrary smooth transformations on any compact subset of an arbitrary open set and which equal the identity outside this open set.  An example of a bump function is
\begin{equation}
\varphi(x)=
\left\{
\begin{array}{ll}
\exp\left(\frac{-1}{1-x^2}\right) &\text{for} |x|<1 \\
0  & \text{otherwise}
\end{array}
\right.
\end{equation}
This looks like a bump between $x=-1$ and $x=1$.  It is smooth (all its derivatives are continuous).  We will consider a transformation from $x$ to $x'$ coordinates.   We use the bump function to define the inverse transformation from $x'$ to $x$
\begin{equation}\label{bumptransformation}
x^{0} =  x^{0'} + a \varphi(x^{0'}),  ~~ x^{1} =  x^{1'} + a \varphi(x^{1'}), ~~\dots
\end{equation}
where we choose $|a|$ small enough that this function is invertible.  These transformations act only inside the cube bounded by $|x^\mu|<1$ and this constitutes our hole.  Now consider a simple case where we have no matter fields and the metric in the shell is everywhere just $g_{\mu\nu}(x)=\eta_{\mu\nu}$.

\paragraph{Vacuum solutions.}

Then, in appropriate coordinates, a solution to the Einstein equations in vacuum, $G^{\mu\nu}=0$, is simply
\begin{equation}\label{Minkowskihole}
g_{\mu\nu}(x)=\eta_{\mu\nu}
\end{equation}
Now we can apply the transformation above. We obtain
\begin{equation}\label{primedmetricinhole}
\begin{split}
g_{\mu'\nu'} &= \frac{\partial x^\mu}{\partial x^{\mu'} }\frac{\partial x^\nu}{\partial x^{\nu'}} \eta_{\mu\nu} \\
&= \text{diag}\left( -\left(1 - \frac{2ax^{0'}}{1-(x^{0'})^2} \varphi\left(x^{0'}\right)\right)^2, \left(1 - \frac{2ax^{1'}}{1-(x^{1'})^2} \varphi\left(x^{1'}\right)\right)^2, \dots \right)
\end{split}
\end{equation}
Now this is a solution to $G^{\mu'\nu'}=0$ with the given boundary conditions.  However, the equation $G^{\mu'\nu'}=0$ written out in full with respect to $x'$ has exactly the same form as the equation $G^{\mu\nu}=0$ written out with respect to $x$. The boundary conditions are not affected. Hence
\begin{equation}\label{unprimedmetricinhole}
g_{\mu\nu}= \text{diag}\left(- \left(1 - \frac{2ax^{0}}{1-(x^{0})^2} \varphi\left(x^{0}\right)\right)^2, \left(1 - \frac{2ax^{1}}{1-(x^{1})^2} \varphi\left(x^{1}\right)\right)^2, \dots \right)
\end{equation}
is a solution to $G^{\mu\nu}=0$.  But this is a different solution from the one in (\ref{Minkowskihole}).  Indeed, we can vary $a$ to obtain an infinite number of different solutions.

\paragraph{Add a little matter.}

Let us go back to solution in (\ref{Minkowskihole}) where we have $g_{\mu\nu}(x)=\eta_{\mu\nu}$ everywhere in the shell and hole.  We can add a little matter to this picture. Thus, imagine adding lots of small \lq\lq blobs" of different types of noninteracting fluids.  We assume that the blobs do not spread while in the shell or hole.  We label the types (and therefore the blobs) by $i=1,2,\dots$.  The energy densities (in the rest frame) are $\rho[i](x)$ and the velocities are $U^\mu[i](x)$. We assume that $\rho[i](x)$ is small enough that gravitational effects can be ignored and hence this matter will not effect the metric appreciably in the shell or hole.  Given that the blobs are small they will follow geodesics (see Sec.\ \ref{Sec:fluidflowalonggeodesics}).  We assume that the paths of the blobs in the shell are given as part of the boundary conditions.

\paragraph{Intersection graph.}

The blobs are noninteracting - they pass right through each other.  We assume that, if they coincide, they do so only momentarily.  We can associate a graph with these blobs where the nodes correspond to intersections of two or more blobs (for example, $\text{node}(1,5,7)$, for a point where blobs $1$, $5$, and $7$ interact) and edges labeled by the type of fluid associated with the blob passing between any two nodes (for example $\text{edge}(5)$).  There may be more than one node or edge with the same name.  Each node would have a position, for example $x[\text{node}(1,5,7),n]$ where $n=1, 2, 3, \dots$ labels different nodes having the same blobs interacting.

\paragraph{A solution in the $x$ coordinate system.}

We can solve the field equations for this situation.  First we do this in the $x$ coordinate system as before.  We are assuming that gravitational effects can be ignored and hence a  solution for the metric in the hole (given that we have the Minkowski metric in the shell) is just that $g_{\mu\nu}(x)=\eta_{\mu\nu}$.  With this metric the geodesics will be straight lines in the $x$ coordinate system.  Hence the blobs will all follow straight lines.  The intersections graph can be calculated from this and we can also attach a position, $x[\text{node}-]$, to each node on the graph.  The edges of the graph will correspond to straight lines in the $x$ coordinate system.

\paragraph{A solution in the $x'$ coordinate system.}

Now we can transform to the $x'$ coordinate system.  The metric in the new coordinate system is given by (\ref{primedmetricinhole}).  The positions of the nodes, $x'[\text{node}-]$, in the new coordinate system are given by solving
\begin{equation}\label{bumptransformationnodes}
\begin{split}
x^{0}[\text{node}-] &=  x^{0'}[\text{node}-] + a \varphi(x^{0'})[\text{node}-], \\
x^{1}[\text{node}-] &=  x^{1'}[\text{node}-] + a \varphi(x^{1'})[\text{node}-], \\
\dots &
\end{split}
\end{equation}
for the primed coordinates.

\paragraph{Another solution in the $x$ coordinate system.}

But now we note that the field equations take the same form in both the $x$ and the $x'$ coordinate systems. Hence, in the original $x$ coordinate system we have another solution for the same field equations (with the same boundary conditions).  Namely where the metric is that in (\ref{primedmetricinhole}) but with the primes removed (i.e.\ as given in \ref{unprimedmetricinhole})) and the node positions are just take the same numerical values in the $x$ coordinate system as the $x'[\text{node}-]$ in the $x'$ coordinate system. Thus, for this second solution in the $x$ coordinate system, the nodes are in different positions.  Since the metric is no longer in Minkowski form the geodesics no longer correspond to straight lines and so the edges of the intersection graph will no longer correspond to straight lines in the $x$ coordinate system.   Since we can vary the parameter $a$ in the transformation in (\ref{bumptransformation}), we actually generate an infinite number of different solutions (all in the same coordinate system and with the same boundary conditions) for the metric, the values of the coordinates at the nodes, and the shape of the paths of the blobs.  It appears we have a radical break down of determinism in that the field equations fail to determine unique solution even when given ample boundary conditions.    However, \dots

\subsubsection{The resolution}\label{Sec:resolution}

\dots the intersection graph will be unchanged.  The key idea in the resolution is that we identify all apparently different solutions that can occur through active transformations (as above) in a given coordinate system as corresponding to the same physical situation. Those things that are the same for different solutions are physical.  Those things that are different correspond only to \lq\lq gauge".  Thus, there is no reality to the position, $x[\text{node}(1, 5, 7)]$ at which a particular event (such as the intersection of blobs $1$, $5$, and $7$) happens even within a given coordinate system.  Einstein said many times (for example in a letter to Ehrenfest, 5th January 1916) that, by virtue of this argument, the space-time coordinates \lq\lq thereby lose the last vestige of physical reality" \cite{sep-einstein-philscience}.  That we have a particular intersection graph is something real, however, since the intersection graph remains the same for different solutions.     The coordinates we attach to the nodes are not physically meaningful, neither are the shapes of the curve associated with the edges in the given coordinate system.

The extremal invariant distance between any pair of intersections is another meaningful quantity (it is the same for different solutions given by an active transformation in a given coordinate system). To see this note that we can write the extremal invariant distance between $\text{node}$ and $\text{node}'$ as
\begin{equation}\label{extremaldistance}
s= \underset{\text{all~paths}}{\text{extremum}} \int_{\lambda(\text{node})}^{\lambda(\text{node}')}\sqrt{\left| g_{\mu\nu} \frac{dx^\mu}{d\lambda}\frac{dx^\nu}{d\lambda}   \right|} d\lambda
\end{equation}
where the paths are parameterized by $\lambda$ through functions $x^\mu(\lambda)$.  The term inside the square root comes from (\ref{metricfords}).  It is a scalar quantity depending on the path.  Different solutions in a given coordinate system are related by an active transformation of the coordinates. We can let this act on the path as well. Since the metric field is also transformed by the same active transformation, the value of the integral along a path will remain unchanged.  Hence, when we extremise over all paths, we will get the same answer (indeed, given that we are extremising over all paths, we do not actually have to transform the paths when we perform the active transformation).

\subsubsection{Diffeomorphisms}\label{Sec:diffeomorphims}

It is instructive to think about the hole argument in coordinate free language.   The different (and physically equivalent) solutions are mapped into one another by an active transformation of the points on the manifold. This is how the nodes come to have different coordinates for the same coordinate system.   To understand this we need the notion of a diffeomorphism and we need to understand the effect of diffeomorphisms on tensor fields.

\index{diffeomorphism}
First consider a map, $\varphi$ between manifolds $\mathscr M$ and $\mathscr N$ ($\varphi:\mathscr{M}\rightarrow \mathscr{N}$).  This takes the point $p\in\mathscr{M}$ to the point $\varphi(p)\in\mathscr{N}$.  If the map is smooth, onto, one-to-one, and the inverse map is also smooth then we say it is a diffeomorphism.  If there exists a diffeomorphism between two manifolds then they have the same manifold structure.  Thus we can have a diffeomorphism that maps a manifold onto itself  ($\varphi: \mathscr{M}\rightarrow \mathscr{M}$).

\paragraph{Push forward and pull back.}

First we will consider the general case where $\varphi$ might not be a diffeomorphism. Since the map $\varphi$ is from $\mathscr M$ to $\mathscr N$ we think of this direction as being the \lq\lq forward" direction.

Consider a real valued function, $f$, defined on $\mathscr N$.  We can use the map, $\varphi$, to \emph{pull back} \index{pull back} this function so we get a real valued function on $\mathscr M$.  We define the pull back, $\varphi_* f$, of $f$ as
\begin{equation}
\varphi_* f = f \circ \varphi
\end{equation}
This clearly defines a function on $\mathscr M$.

Now we have this notion, we can define the \emph{push forward} \index{push forward} of a (1, 0) type vector field, $V(p)$, defined on $\mathscr{M}$ to $\mathscr N$ through the following equation
\begin{equation}\label{pushforwardV}
\varphi^* V f =      V \varphi_* f  ~~ \forall ~ f
\end{equation}
Where $\varphi^* V(p)$ is a vector at $\varphi(p)$ in $\mathscr N$.  Let us introduce the coordinate system $\{ x^\mu: \mu=0~\text{to}~ D_\mathscr{M}-1\}$ for $\mathscr M$ and $\{ y^\alpha: \alpha=0~\text{to} ~ D_\mathscr{N}-1\}$ for $\mathscr N$ (where, just in this subsubsection, we use $\mu, \nu, \dots$ for $\mathscr M$ and $\alpha, \beta, \dots$ for $\mathscr N$).  Then we can expand $V=V^\mu \partial_\mu$ and $\varphi^* V = (\varphi^* V)^\alpha \partial_\alpha$.  Putting this into (\ref{pushforwardV}) and choosing $f=y^\beta$, we obtain
\begin{equation}
(\varphi^* V)^\alpha \partial_\alpha y^\beta  =V^\mu \partial_\mu (\varphi_* y^\beta)
\end{equation}
Now $\varphi_* y^\beta$ is the pull back of the coordinate, $y^\beta$, into the $\mathscr M$ manifold.  Since it has been pulled back to $\mathscr M$, we can differentiate it with respect to $x^\mu$.  We write
\begin{equation}
\frac{\partial y^\beta}{\partial x^\mu} := \partial_\mu (\varphi_* y^\beta)
\end{equation}
for shorthand.  Using $\partial_\alpha y^\beta=\delta^\beta_\alpha$, we obtain
\begin{equation}\label{pushforwardVcomponents}
(\varphi^* V)^\beta =   \frac{\partial y^\beta}{\partial x^\mu}  V^\mu
\end{equation}
for the components of the push forward vector in terms of the original vector.

We can use the push forward of a vector to define the pull back  of a one form, $df$ (defined in (\ref{oneformdefinition})) through the equation
\begin{equation}
\varphi_* df (V) = df(\varphi^* V) ~~ \forall V
\end{equation}
This takes a one form, $df$, at $\varphi(p)$ on $\mathscr N$ and returns a pull back one form, $\varphi_* df$, at $p$ on $\mathscr M$.  Using (\ref{pushforwardVcomponents}) we obtain
\begin{equation}\label{pullbackdfcomponents}
(\varphi_* df)_\mu = \frac{\partial y^\alpha}{\partial x^\mu} (df)_\alpha
\end{equation}
for the components of the push back one form in $\mathscr M$.

The situation is summed up as follows
\begin{flalign}
\mathscr{M} &\overset{\varphi}\longrightarrow  \mathscr{N}  \\
\varphi_* f    &  \longleftarrow      f     \\
V           & \longrightarrow      \varphi^*V  \\
\varphi_* df     & \longleftarrow       df
\end{flalign}
Note we cannot pull back a $(1,0)$ type vector and we cannot push forward a one form (a $(0,1)$ type vector).

\paragraph{Push forward with a diffeomorphism.}

If we have a diffeomorphism then $\varphi^{-1}$ exists and we can pull and push objects in both directions. Since we have a diffeomorphism, the two manifolds have the same manifold structure.

The simplest example is a scalar field, $S(p)$, which gets pushed forward to $\varphi^*S(p)=S(\varphi(p))$.   We can also push forward  a one form by using the inverse map, $\varphi^{-1}$.
\begin{equation}
\varphi^* df = (\varphi^{-1})_* df
\end{equation}
This pushes forward $df$ at $p\in\mathscr{M}$ to $\varphi^* df$ at point $\varphi(p)\in \mathscr{N}$.

In coordinates we obtain
\begin{equation}
(\varphi^* df)_\alpha = \frac{ \partial ([\varphi^{-1}]_* x^\mu)}{\partial y^\alpha } (df)_\mu
\end{equation}
(we get this from (\ref{pullbackdfcomponents}) - we need to swap the positions of $x$ and $y$ in the partial derivative as we are pushing forward rather than pulling back).
The main thing about the partial derivative here is that it compares $[\varphi^{-1}]_* x^\mu$ with $y^\alpha$ which is the same thing as comparing $x^\mu$ with $\varphi_* y^\alpha$.  Hence,
\begin{equation}
\frac{ \partial ([\varphi^{-1}]_* x^\mu)}{\partial y^\alpha } =  \frac{\partial x^\mu}{\partial (\varphi_* y^\alpha)} =: \frac{\partial x^\mu}{\partial y^\alpha}
\end{equation}
and therefore,
\begin{equation}\label{phidfcomponents}
(\varphi^* df)_\alpha = \frac{\partial x^\mu}{\partial y^\alpha} (df)_\mu
\end{equation}

We are now in a position to see how to push forward a general tensor.  The push forward of a general tensor, $T^{a_1\dots a_q}_{b_1\dots b_r}$ (we are using Penrose notation to indicate the type of tensor we have (see (\ref{Penrosenotation})) is defined through requiring
\begin{equation}\label{phiTdefinition}
\varphi^*T^{a_1\dots a_q}_{b_1\dots b_r} df_{a_1}  \dots df_{a_q} V^{b_1}\dots V^{b_r} = T^{a_1\dots a_q}_{b_1\dots b_r} \varphi^*df_{a_1}  \dots \varphi^*df_{a_q} \varphi^*V^{b_1}\dots \varphi^*V^{b_r}
\end{equation}
for all $df_{a_i}$ and $V^{b_j}$.  It is instructive to give the effect of a diffeomorphism on the components of the tensor.  This is given by
\begin{equation}
\varphi^*  T^{\alpha_1 \dots \alpha_q}_{\beta_1\dots \beta_r} =
\frac{\partial y^{\alpha_1}}{\partial x^{\mu_1}}\dots  \frac{\partial y^{\alpha_q}}{\partial x^{\mu_q}}
\frac{\partial x^{\beta_1}}{\partial y^{\nu_1}}  \dots \frac{\partial x^{\beta_r}}{\partial y^{\nu_r}}
T^{\mu_1\dots \mu_q}_{\nu_1\dots \nu_r}
\end{equation}
as is seen by substituting (\ref{pushforwardVcomponents}, \ref{phidfcomponents}) into (\ref{phiTdefinition}).  In the case where both $p$ and $\varphi(p)$ are covered by the same coordinate system, the partial derivatives are simply comparing two different regions in this coordinate system that are related by $\varphi$.

A particular application of a diffeomorphism is when we have $\mathscr{N}=\mathscr{M}$. In this case diffeomorphisms simply push the fields around on $\mathscr{M}$.

\paragraph{The hole argument using diffeomorphisms.}

We can represent a solution as
\begin{equation}
{\tilde{\Psi}}= \Big\{ \big(p, {\bf g}(p), \vec{\bf F}(p)\big):\forall p\in \mathscr{M}\big\}
\end{equation}
Consider a diffeomorphism, $\varphi:\mathscr{M}\rightarrow \mathscr{N}$ on the manifold.  If we act this diffeomorphism on on ${\tilde{\Psi}}$ we get
\begin{equation}
\begin{split}
\varphi^*{\tilde{\Psi}}= & \Big\{ \big(\varphi(p), \varphi^*{\bf g}(\varphi(p)), \varphi^*\vec{\bf F}(\varphi(p))\big):\forall \varphi(p)\in \varphi(\mathscr{M})\big\} \nonumber\\
= &\Big\{ \big(p, \varphi^*{\bf g}(p), \varphi^*\vec{\bf F}(p)\big):\forall p\in \varphi(\mathscr{M})\big\}
\end{split}
\end{equation}
From the above arguments, it is clear that $\varphi^*{\tilde{\Psi}}$ is also a solution to the field equations (now understood in coordinate free terms) for any diffeomorphism, $\varphi$.  This means that the physical content of a solution is represented by something like
\begin{equation}
\underset{\text{all}~ \varphi}{\text{mod}} \{ \big(p, \varphi^*{\bf g}({p}), \varphi^*\vec{\bf F}({p})\big):\forall p\in \varphi(\mathscr{M})\}
\end{equation}
where we read this equation to mean that we \lq\lq mod out" over the diffeomorphism group.   We can represent this by the following object
\begin{equation}
\Psi = \big\{ \varphi^*  \tilde{\Psi}: \forall \varphi \big\}
\end{equation}
We can think of this as a diffeomorphism invariant way of representing the solution.  Although we need the manifold to represent each element in $\Psi$, there is no physical meaning to where the fields are on this manifold.  If we move all the fields at once (as we do with a diffeomorphism) then the relations between the fields remain as they were (in an appropriate sense) and so the physics is unchanged.

In the main text we discuss \emph{beables} as functions of a solution that are invariant under diffeomorphisms (see Sec.\ \ref{sec:candidatesolutionsfieldequationsandbeables}).

\subsubsection{Role of the Bianchi identities}\label{Sec:Bianchiresolution}

We noted in Sec.\ \ref{Sec:Einsteinfieldeqns} and Sec.\ \ref{Sec:GRascoupledfieldeqns} that we have the same number of field equations as degrees of freedom resulting from the fields modulo an important subtlety.  The point is that if we really have the same number equations as degrees of freedom, then we should be able to solve the equations and find a unique solution given appropriate boundary conditions. However, we know from the hole argument that this is not possible.  In four dimensions we have four arbitrary degrees of freedom associated with general transformations (or general diffeomorphisms) and any solution must be arbitrary up to these four transformations.  Consequently, we need four fewer equations.  Fortunately, we have exactly four fewer equations since the left hand side of the Einstein field equations satisfies $\nabla_\nu G^{\mu\nu}=0$.  Note that the right hand side must, consequently, satisfy $\nabla_\nu T^{\mu\nu}=0$.  However, this does not constitute an additional constraint on the matter degrees of freedom beyond the constraint already given by the matter field equations (since it must be possible to derive $\nabla_\nu T^{\mu\nu}=0$ from the matter field equations).

We see that the Einstein field equation shoulders the burden of diffeomorphism invariance.  Further, given that the matter field equations are obtained by the minimal substitution technique, they could not shoulder this burden - the Einstein field equations have to do this themselves.





\subsection{Causality}\label{sec:causalityone}

A big motivation for setting up the framework of General Relativity is causality.  Influences should not travel outside the light cone structure given by the metric.  Given this, it would be good to have a simple proof that General Relativity is, indeed, causal in an appropriate sense.

\index{causality!in standard GR}

An immediate problem, when addressing this question, is to come up with a suitable definition of causality.  The intuitive notion is that we should only be able to signal from point $p$ to point $q$ if $q$ is in the future of $p$ (so $q$ should not be outside the forward light cone at $p$). There are four issues arise when trying to define causality in General Relativity.
\begin{description}
\item[Agency.] To even consider signalling, we need the possibility of different choices.  However, in General Relativity, we have a deterministic set of field equations (up to physically irrelevant coordinate transformations)so we cannot have different choices.
\item[Dynamical causal structure.]  The light-cone structure is not fixed in advance but is, rather, something we solve for (by determining $g_{\mu\nu}$).  Hence, it may depend on what \lq\lq choice" we make at $p$.
\item[Signalling with $g_{\mu\nu}$.] The metric, $g_{\mu\nu}$, determines the causal structure, but it can also be used to signal (by sending gravity waves for example).  We need to be sure that the latter respects the former.
\item[Identifying $p$ and $q$.] The hole argument makes it clear that $p$ is not really physical object since it is not invariant under diffeomorphisms so there is a problem in defining causality with respect to points on the manifold.
\end{description}
These points are considered in great detail in the main text.  In particular, in Sec.\ \ref{sec:Agency} we will introduce a way to include agency so that different choices can be formally considered. In Sec.\ \ref{sec:observables} we will introduce a diffeomorphism invariant \emph{operational space} regarded as the local arena in which observe the world.  In Sec.\ \ref{sec:causality} we give a detailed discussion of causality using operational space and the notion of agency.

In the absence a formal way of considering agency, one way to look at causality is to consider functional dependencies.  Hawking and Ellis define what they call \emph{local causality} (this is the same thing we have been calling \lq\lq causality") in terms of the Cauchy problem (where we time evolve fields specified on a space-like hypersurface):
\begin{quote}
{\bf Local causality (Hawking Ellis):} \lq\lq Let $p\in\mathscr{U}$ be such that every non-spacelike curve through $p$ intersects the spacelike surface $x^0=0$ within $\mathscr{U}$.  Let $\mathscr{F}$ be the set of points in the surface $x^4=0$ which can be reached by non-spacelike curves in $\mathscr U$ from $p$. Then we require that the values of the matter fields at $p$ must be uniquely determined by the values of the fields and their derivatives up to some finite order on $\mathscr F$, and that they are not uniquely determined by the values on any proper subset of $\mathscr F$ to which it can be continuously retracted."
\end{quote}
Here $\mathscr U$ is a \emph{convex normal neighbourhood} of $p$.  Roughly speaking, a normal neighbourhood about $p$ is one which is small enough that it can be coordinatized by non-crossing geodesics.  Such a neighbourhood is also convex if it there exists a unique geodesic between any pair of points in it (see \cite{hawking1973large} for definitions).

That General Relativity is locally causal in the sense of Hawking and Ellis (when certain assumptions are true) can be seen by setting up the Cauchy problem.  This problem is particularly interesting in the context of General Relativity because of diffeomorphism invariance.  An elementary discussion of the Cauchy problem can be found in \cite{weinberg1972gravitation} (more detailed discussions can be found in many advanced texts including, of course, \cite{hawking1973large}).

Whatever definition of causality one adopts, the intuitive reason that General Relativity is causal is that it inherits this from Special Relativity.  If, at a point $p$, we transform to a local inertial reference frame, then the matter field equations look locally like those of Special Relativity.  By assumption, these equations do not allow propagation of signals outside the light cone at $p$ (assuming we rule out tacyons and other such exotic Special Relativistic possibilities).  This is true at any point, $p$, and hence we should not be able to use these matter fields to signal outside the light cone.

In Genera Relativity it is possible, in principle at least, that the manifold has a global property that does allow information to travel backward in time (since it concerned only matter fields).  Indeed, solutions do exist that have closed timelike curves. In particular, G\"odel exhibited such a solution  \cite{godel1949example}.  It is common to rule out such solutions (for example it is commonly assumed that the solutions we seek should be foliable into spacelike hypersurfaces indexed by some time $t$).

\subsection{General Relativity is about fields}\label{Sec:GRaboutfields}

General Relativity consists of (i) the minimal substitution technique for getting General Relativity matter field equations from the Special Relativistic counterparts and (ii) The Einstein field equations.

General Relativity should be regarded as a theory of fields (where we mod out over diffeomorphisms) and nothing more.  This point is worth making since, in setting up Relativity (Special and General) much use is made of clocks, measuring rods, light rays, test particles, and physical reference frames.   These are useful notions that Einstein employed for the purposes of theory construction. However, once we have General Relativity, it provides its own ontology. Although it may be difficult to explicitly do so, we can imagine clocks, measuring rods, light rays, and test particles all being built out of matter and metric fields.  The extent to which we are unable to construct such objects is simply a measure of the extent to which we should consider such objects as only having an approximate role in General Relativity.   In Appendix \ref{Sec:fluidflowalonggeodesics} we discussed how it follows from the field equations of General Relativity (actually matter field equations) that small blobs of fluid should move on geodesics.  This allows us to recover (approximately) the notion of test particles.  Similarly, physical reference frames require that we have some physical features against which we can determine space-time points.  These must be provided by physical fields also.

\section{The substitution operator}\label{sec:thesubstitutionoperator}

In this appendix we define the substitution operator
\begin{equation}
s_{x}^{y}
\end{equation}
where $x\in S$ and $y\in R$ may correspond to a discrete quantity, a continuous quantity, or something more general (like boundary conditions such as $\mathlbnd{a}$ considered in the main text).  The substitution operator effects a change of variables (a common procedure in calculus). It has an effect somewhat similar to a Dirac delta function but can be more generally defined.

Consider an expression
\begin{equation}
A^x B_x
\end{equation}
The notational convention here is that we integrate/sum  w.r.t.\ $x$
\begin{equation}
A^x B_x = \SumInt \mathrm{d}x A(x) \bar{B}(x)
\end{equation}
where $A^x=A(x)$ and $B_x=\bar{B}(x)$ (the bar over the $B$ marks that $B_x$ is associated with a white dot).
However we could change variables to $y$ where $x=s(y)$.  We take $s(\cdot)$ to be invertible.  We require that
\begin{equation}
S=s(R)
\end{equation}
We have
\begin{equation}
\SumInt_{x\in S} \mathrm{d}x A(x) \bar{B}(x) = \SumInt_{y\in R} s'(y)\mathrm{d}y A(s(y)) \bar{B}(s(y)) = A^y B_y
\end{equation}
where $s'$ denotes the Jacobian of the transformation.  We need to define quantities so that this holds. We can write
\begin{equation}
A^y = s_x^y A^x  :=  f(y) A(s(y))   ~~~~~~       B_y =  s_y^x B_x := g(y) \bar{B}(s(y))
\end{equation}
where
\begin{equation}
f(y)g(y) = s'(y)
\end{equation}
We will put $g(y)=1$ for the application in the application in the main text so that
\begin{equation}\label{substitutionoperationdefn}
A^y = s_x^y A^x  :=  s'(y) A(s(y))   ~~~~~~       B_y =  s_y^x B_x := f(y) \bar{B}(s(y))
\end{equation}
In this case the $s'(y)$ ix on the $A^y$ associated with a black dot (when represented diagrammatically as a duotensor).  This is appropriate as the Jacobian modifies the probability density for the application in the main text.  In the case that we simply have summation (not integration) then the Jacobian is unity (equals 1).

Note that the repeated index $x$ in $s_x^y A^x$ does not indicate integration but, rather, substitution.
The interpretation of repeated indices is allowed to depend on the objects the indices are attached to.  In other words, we have a \emph{handshake}.  Whenever we have an expression like $D^x E_x$ we can decide how to interpret the repeated index.  Each quantity $D^y$ and $E_y$ can have priority rules associated with it.  If we have a substitution operator, $s_x^y$, then it places priority $1$ for a substitution operation as in \eqref{substitutionoperationdefn}.   Usual quantities like $A^x$ in \eqref{substitutionoperationdefn} have priority $2$ for the integration operation.  The handshake takes the highest priority operation.  We have the freedom to have a handshake because we use the more general notation $D^x E_x$ rather than always writing this down as an integral.

A special case is where we substitute an $x$ for itself.  Then the Jacobian is the identity.  We write the identity substitution operator as
\begin{equation}
\delta_{x}^{x'}
\end{equation}
We can write
\begin{equation}
A^x B_x = A^y B_y = A^x s_x^y  s^{x'}_y B_{x'}
\end{equation}
Since this must be true for all $A^x$ and $B_x$ we can have
\begin{equation}\label{substhenbackagain}
s_x^y  s^{x'}_y =\delta_x^{x'}
\end{equation}
This corresponds to substituting $y$ for $x$ then substituting back again.  Note that, in our notation, $s'(x) = (s'(y))^{-1}$ it is consistent with \eqref{substitutionoperationdefn} to interpret \eqref{substhenbackagain} as the identity.

We can define substitution operators for pre-subscripts and pre-superscripts.
\begin{equation}
A^xB_x = \int_{x\in S} \mathrm{d}x A(x) \bar{B}(x) = \int_{y\in R} s'(y) \mathrm{d}y A(s(y)) B(s(y)) =  \presup{y}A \presub{y}B
\end{equation}
To make this work we need
\begin{equation}
\presup{y}A= \presup{y}s_x A^x := \alpha(y) A(s(y)) ~~~~~~  \presub{y}B = \presub{y}s^x B_x := \beta(y) \bar{B}(s(y))
\end{equation}
where
\begin{equation}\label{getJac}
\alpha(y)\beta(y) = s'(y)
\end{equation}
We can do this by putting
\begin{equation}
\presup{y}A=\prescript{y}{x}s \presup{x}A := \bar{A}(s(y))  ~~~~~~ \presub{y}B=\prescript{x}{y}s \presub{x}B := s'(y) B(s(y))
\end{equation}
where, again, we put the Jacobian on the case associated with a black dot.

We have
\begin{equation}
\presup{y}s_x \presub{y}s^{x'} = \delta_x^{x'}
\end{equation}
where $\delta_x^{x'}$ is the identity substitution operator.   We also define the identity substitution operator
\begin{equation}
\presub{x'}\delta^x
\end{equation}
such that
\begin{equation}
\presup{x}A= \presup{x}\delta_{x'} A^{x'} :=  A(x) ~~~~~~  \presub{x}B = \presub{x}\delta^{x'} B_{x'} :=  \bar{B}(x)
\end{equation}
This is where we substitute $x$ for itself (hence the Jacobian is unity).  This object is a little unnatural as it hops an index over the main symbol - it is not really an identity operator.

Finally, we can define the weighted substitution operator for self-substitution of a variable $x$
\begin{equation}
\presup{x}A= \presup{x}w_{x'} A^{x'} := \alpha(x) A(x) ~~~~~~  \presub{x}B = \presub{x}s^{x'} B_{x'} :=  \beta{x}\bar{B}(x)
\end{equation}
where we do not impose that $\alpha(x)\beta(x)=1$.

Rather than using the substitution operator we can use the Dirac delta function.  We can write
\begin{equation}
A^y B_y= \iint \mathrm{d} x \mathrm{d}y  A(x) B(x) s'(y) \delta(x-s(y))
\end{equation}
This achieves substitution of $x$ by $s(y)$.  However it is not clear how to define the Dirac delta function for an infinite dimensional space.  Further, from a computational point of view, it is easier to perform a change of variables than to integrate.  The handshake protocol allows us to avoid using the Dirac delta function.  The Dirac delta function does have properties that are useful when evaluating integrals that are not true of the substitution operator (for example, $x\delta'(x) = -\delta(x)$).  However, the substitution operator is sufficient for the purposes we have in mind.

\newpage

\bibliography{QGbibJuly2016}
\bibliographystyle{plain}

\newpage

\printindex

\end{document}